\documentclass[12pt,a4paper,oneside]{report} 
\usepackage[margin=30mm]{geometry}

\usepackage{setspace}
\usepackage{zref-totpages}

\usepackage{amsmath,amssymb,amsfonts,amsthm}
\usepackage{mathrsfs} 
\usepackage{caption,subcaption}

 \usepackage{fancyhdr}
 
 \DeclareMathOperator\arctanh{arctanh}

\usepackage{microtype}

\usepackage{graphicx,color}
\usepackage{accents}
\usepackage{mathtools}
\usepackage[nocompress]{cite}
 \usepackage{tikz,tikz-3dplot}
 \usetikzlibrary{shapes.geometric}
 \usetikzlibrary{matrix}
\usetikzlibrary{shadings}
\usetikzlibrary{arrows.meta}
\usetikzlibrary{calc}
\usetikzlibrary{decorations.markings}
\usepackage{enumerate}
\usepackage{lipsum}
\usepackage{cancel}
\usepackage{physics}
\usepackage{pgfplots}
\usepackage{longtable}
\usepackage{afterpage}
\newcommand\myemptypage{
    \null
    \thispagestyle{empty}
    \newpage
    }

\tikzset{>=latex} 
\tikzstyle{vector}=[->, thick]
\tikzstyle{thin arrow}=[dashed,thin,-{Latex[length=4,width=3]}]

\newcommand\pgfmathsinandcos[3]{%
\pgfmathsetmacro#1{sin(#3)}%
\pgfmathsetmacro#2{cos(#3)}%
}
\newcommand\LongitudePlane[3][current plane]{%
\pgfmathsinandcos\sinEl\cosEl{#2} 
\pgfmathsinandcos\sint\cost{#3} 
\tikzset{#1/.estyle={cm={\cost,\sint*\sinEl,0,\cosEl,(0,0)}}}
}
\newcommand\LatitudePlane[3][current plane]{%
\pgfmathsinandcos\sinEl\cosEl{#2} 
\pgfmathsinandcos\sint\cost{#3} 
\pgfmathsetmacro\yshift{\cosEl*\sint}
\tikzset{#1/.style={cm={\cost,0,0,\cost*\sinEl,(0,\yshift)}}} %
}

\newcommand\DrawLongitudeCircleredL[2][1]{
    \LongitudePlane{\angEl}{#2}
    \tikzset{current plane/.estyle={cm={\cost,\sint*\sinEl,0,\cosEl,(0,0)},scale=#1}}
     \pgfmathsetmacro\angVis{atan(sin(#2)*cos(\angEl)/sin(\angEl))} %
    \draw[current plane,black,semithick] (90:1) arc (90:180:1);   

    \draw[current plane,black!50,ultra thick,->] (120:1) -- ([turn]-110:0.4);
    \draw[current plane,black!50,ultra thick,->] (150:1) -- ([turn]-140:0.4);
    \draw[current plane,black!50,ultra thick,->] (180:1) -- ([turn]-180:0.55);
}

\newcommand\DrawLongitudeCircleredR[2][1]{
    \LongitudePlane{\angEl}{#2}
    \tikzset{current plane/.estyle={cm={\cost,\sint*\sinEl,0,\cosEl,(0,0)},scale=#1}}
     \pgfmathsetmacro\angVis{atan(sin(#2)*cos(\angEl)/sin(\angEl))} %
    \draw[current plane,black,semithick] (90:1) arc (90:180:1);   
    
       \draw[current plane,black!70,ultra thick,->] (150:1) -- ([turn]0:0.6);
        \draw[current plane,black!70,ultra thick,->] (120:1) -- ([turn]30:0.6);
    \draw[current plane,black!70,ultra thick,->] (90:1) -- ([turn]60:0.5);
}

\newcommand\DrawLatitudeCircle[2][1]{
    \LatitudePlane{\angEl}{#2}
    \tikzset{current plane/.prefix style={scale=#1}}
    \pgfmathsetmacro\sinVis{sin(#2)/cos(#2)*sin(\angEl)/cos(\angEl)}
    \pgfmathsetmacro\angVis{asin(min(1,max(\sinVis,-1)))}
    \draw[current plane,thin,black,dashed] (\angVis:1) arc (\angVis:-\angVis-180:1);
    \draw[current plane,very thin,dashed,black!80] (180-\angVis:1) arc (180-\angVis:\angVis:1);
}

\newcommand\DrawLatitudeCirclered[2][1]{
    \LatitudePlane{\angEl}{#2}
    \tikzset{current plane/.prefix style={scale=#1}}
    \pgfmathsetmacro\sinVis{sin(#2)/cos(#2)*sin(\angEl)/cos(\angEl)}
    \pgfmathsetmacro\angVis{asin(min(1,max(\sinVis,-1)))}
    \draw[current plane,black,semithick] (\angPhiTwo:1) node[below right] {$$} arc (\angPhiTwo:\angPhiOne:1) node[below left] {$$}; 

    \foreach \r in {-130,-90,...,-50}{
    \draw[current plane,black,ultra thick,->] (\r:1) -- ([turn]90:.5);
    }
}

\usepackage{quotchap}

\usepackage{pdflscape}

\newcommand{\ourG}{\mathbf{G}}
\newcommand{\ourB}{\mathbf{B}}
\newcommand{\ourC}{\mathbf{C}}

\numberwithin{equation}{section}

\theoremstyle{plain}
\newtheorem{theorem}{Theorem}
\theoremstyle{definition}

\usepackage[nocompress]{cite}
\usepackage{hyperref}
\hypersetup{colorlinks=true,allcolors=blue,urlcolor=cyan}

\title{A Unified Approach to Geometric Modifications of Gravity}
\author{Erik Jensko}

\begin{document}
\pagecolor{white}

\emergencystretch 3em

\pagenumbering{roman} 
\setcounter{page}{0}

\newgeometry{margin=40mm}

    \newpage%
    \null%
    \vspace*{5em}%

    \begin{center}%
        {\Huge A Unified Approach to Geometric}\\[1em]
        {\Huge Modifications of Gravity }\\[5em]%
        {\Large \itshape Erik Jensko}\\%
    \end{center}%
    \vfill%
    \begin{center}%
    A thesis submitted in partial fulfilment \\
    of the requirements for the degree of \\
    \textbf{Doctor of Philosophy} \\
    of \\
    \textbf{University College London}
    \end{center}%
    \vspace{2em}%
    \begin{center}%
    Department of Mathematics \\
    University College London\\
    \end{center}%
    \vspace{2em}%
    \begin{center}%
    December 7, 2023%
    \end{center}%

\restoregeometry

\renewcommand{\chapterheadstartvskip}{\vspace*{-2\baselineskip}} 

\pagenumbering{arabic} 
\setcounter{page}{2}

\chapter*{Declaration}
I, Erik Jensko, confirm that the work presented in this thesis is my own.
         Where information has been derived from other sources, I confirm that this has been indicated in the work.

\chapter*{Abstract}

This thesis studies modified theories of gravity from a geometric viewpoint. The primary motivation for considering alternatives to General Relativity comes from cosmological observations and theoretical arguments, which are discussed from the outset. Moreover, we explore the various different approaches to modifying gravity. We begin with an introduction of the mathematical foundations of General Relativity and gravitational theories in non-Riemannian geometries. 
Then, starting from the decomposition of the Einstein-Hilbert action into bulk and boundary terms, we construct new modifications of General Relativity.
These modifications  break diffeomorphism invariance or local Lorentz invariance, allowing one to bypass Lovelock's theorem while remaining second-order and without introducing additional fields. We work in both the Levi-Civita framework of General Relativity and the non-Riemannian metric-affine framework; the latter leads to a new Einstein-Cartan-type theory with propagating torsion. Important comparisons are also made with the modified teleparallel theories, and we construct a unified framework that encompasses all of these theories. The equivalence between theories that break fundamental symmetries in the Riemannian setting and non-Riemannian theories of gravity is explored in detail. This leads to a dual interpretation of teleparallel gravity, one in terms of geometric quantities and the other in terms of non-covariant objects.
 We then study the cosmological applications of these modified theories, making use of dynamical systems techniques. Strict conditions that models must satisfy in order to give rise to accelerated expansion (de Sitter solutions) are derived using these methods. Another key result is that the modified Einstein-Cartan theories can drive inflation in the early universe, replacing the initial cosmological singularity of General Relativity. To conclude, we discuss the viability of these modifications and possible future directions, examining their significance and relevance to the broader field of gravitational physics as a whole.

\chapter*{Impact statement}
This thesis investigated novel theories of gravity, with a focus on their mathematical and theoretical properties. These modifications open up a number of routes for constructing new theories of gravity, and also give us new ways of understanding our current ones. This broadens the current knowledge within this domain of physics and allows future researchers to develop new methods based on the results presented here.

Most of the work originated from the different decompositions of the Einstein-Hilbert action into bulk and boundary terms. Making use of the tetrad formalism, it was shown how working in the coordinate basis led to an equivalence with the symmetric teleparallel theories, whilst working in the orthonormal basis led to an equivalence with the metric teleparallel theories. Moreover, this equivalence also extended to their modifications as well. This gives an entirely new perspective from which to view these popular modified theories of gravity. 

In the cosmological setting, it was shown that these different gravitational theories have the same background equations of motion. This allowed for a systematic study of all possible models, making use of dynamical systems theory techniques. A key finding was the ubiquitous existence of late-time, stable, de Sitter fixed points, corresponding to an accelerating expanding universe. The necessary conditions for these solutions were also derived. 

Lastly, it was found that the modified Einstein-Cartan theories led to propagating torsion, which was shown to give rise to inflation. These models have the potential to answer fundamental questions relating to the early universe and the origin of the cosmos. More studies are needed to fully investigate the feasibility of these types of theories and to connect this theoretical work with observational data and constraints.

Overall, this thesis contributes to the global community of researchers working on questions relating to gravitation and cosmology by enhancing our understanding of modified theories of gravity. The mathematical techniques developed can also be applied in a wide range of contexts within theoretical physics.

\chapter*{Acknowledgements}
A number of people deserve thanks for their contributions toward this thesis and my work, and also for their help and friendship during the past four years of my PhD. It is only fitting that I attempt to include many of their names here below.

Firstly, I would like to thank my PhD advisor Christian B\"{o}hmer, for his constant support and encouragement. His dedication and passion for our projects has made it a truly enjoyable experience. Without his guidance, this work would not have been possible. As we always seem to come up with no end of new ideas to work on, I look forward to our continued collaborations. I count myself lucky to have had such a brilliant advisor.

I am also grateful to my collaborators and the people I have met during conferences and research visits. I would like to thank Tomi Koivisto for hosting me at the University of Tartu, as well as the theory group, including Laur J\"{a}rv, Sebastian Bahamonde, Jo\~{a}o Rosa and many others.
The lively group discussions were fun and informative, and the trips and events organised by Laur were a pleasure to attend.
Also to the friends made there, particularly Yufeng and Linda, the visit would not have been the same without their company.

For my visit to the University of Science and Technology of China, I have to thank Yifu Cai, for being an amazing host and making me feel exceptionally welcome. The trips around Anhui and to Yifu's hometown, with the company of Wentao Luo and Chris Miller, was a truly memorable experience. Likewise, I appreciate the opportunity to meet and get to know the other visitors such as Manos Saridakis, Misao Sasaki, and Masahide Yamaguchi. From all of these people I feel I have learnt something valuable.
 A special thanks goes to members of the particle cosmology group, Yumin Hu, Yaqi Zhao, Amara Ilyas, and many of the students, for taking care of me during my visit (especially when it came to translations). The ongoing discussions with Yumin have been extremely useful, and I'm glad to have gained a close collaborator and friend. Trips like these have been a real highlight of my PhD.

Specifically related to the work on this thesis, I want to thank Lucía Menendez-Pidal de Cristina for help on topics that she knows far better than I. Moreover, insightful discussions with Tomi Koivisto and Ruth Lazkoz have been invaluable for my work.
 It's also safe to say that without the support of my undergraduate and postgraduate advisors, Pete Millington and Ed Copeland, I would not even have begun this PhD. Their teachings and enthusiasm have stuck with me throughout my academic career.

To my friends and family, all of whom I am immensely grateful for, I owe the biggest thanks. The support of my parents, Joelle \& Duncan, has been essential. I've always been able to count on the both of them. An additional thanks goes to my mum, for proofreading all {\ztotpages} pages of this thesis. The friends I've made and grown closer to during the PhD have made my time in London thoroughly enjoyable, and I want to thank Holly and Eleanor in particular. The UCL Barbell Club also deserves a mention, and I'm fortunate to have had great people to train with over the past four years. 

Most of all, I would like to thank Vicky. It is no understatement to say that without you by my side throughout these years, none of this would have been possible -- the proofreading of this thesis (\textit{and every previous dissertation}) being yet another thing I'm thankful for. I am most appreciative for the never-ending patience, support and love you have given me over the past eight years. This journey has been achievable only thanks to you, and so I am eternally grateful.


\clearpage

\tableofcontents

\clearpage

\chapter*{Notation}
\addcontentsline{toc}{chapter}{Notation}
A general overview of the notation used throughout the thesis is presented here. Greek letters $\mu, \nu, \lambda, ...$ refer to spacetime indices whilst Latin letters $a, b, c, ...$ refer to tangent space indices, also called flat-space or Lorentz indices. Symmetry and anti-symmetry brackets are defined as $A_{(\mu \nu)}  = (A_{\mu \nu} + A_{\nu \mu})/2$ and $A_{[\mu \nu]}  = (A_{\mu \nu} - A_{\nu \mu})/2$ respectively. The commutator, or Lie bracket, is defined as $[A,B] = AB - BA$. In general, objects with an overbar $\bar{A}_{\mu}$ are with respect to the affine connection, with a tilde $\tilde{A}_{\mu}$ are with respect to the Riemann-Cartan connection (with vanishing non-metricity), and without any decoration $A_{\mu}$ are with respect to the Levi-Civita connection. Notable exceptions to this rule are the tensors of torsion, contortion and non-metricity and the spin connection. 
A comprehensive list of commonly used symbols, in index notation, is given below. 
{
\begin{longtable}{l r  p{10cm} c }
\hline
$g_{\mu \nu}$ & $\qquad \qquad     $ & metric tensor \\
$g^{\mu \nu}$ & $\qquad \qquad   $ & inverse metric \\
$g$ &  & metric determinant \\
$\eta_{ab}$ &  & Minkowski metric \\
$e_{a}{}^{\mu}$ &  & tetrad or frame field \\
$e^{a}{}_{\mu}$ &  & co-tetrad \\
$e$ &  & tetrad determinant \\
$\varPhi^{A}$ &  & matter fields \\
$\Lambda^{a}{}_{b}$ &  & Lorentz matrix \\
$\Lambda_{a}{}^{b}$ &  & inverse Lorentz matrix \\
$ \bar{\Gamma}^{\mu}_{\nu \lambda} $ && affine connection \\
$ \tilde{\Gamma}^{\mu}_{\nu \lambda} $ && Riemann-Cartan connection \\
$ \Gamma^{\mu}_{\nu \lambda} $ && Levi-Civita connection \\
$ \omega_{\mu}{}^{a}{}_{b} $ && affine spin connection \\
$ \tilde{\omega}_{\mu}{}^{a}{}_{b} $ && Riemann-Cartan spin connection \\
$ \mathring{\omega}_{\mu}{}^{a}{}_{b} $ && Levi-Civita spin connection \\
$ \bar{\nabla}_{\mu}$ &  & affine covariant derivative \\
$ \tilde{\nabla}_{\mu}$ &  & Riemann-Cartan covariant derivative \\
$ \nabla_{\mu}$ &  & Levi-Civita covariant derivative \\
$ T^{\mu}{}_{\nu \lambda} $ && torsion tensor \\
$ Q_{\mu \nu \lambda} $ && non-metricity tensor \\
$ K_{\mu \nu}{}^{\lambda} $ && contortion tensor \\
$ \bar{R}_{\mu \nu \lambda}{}^{\gamma} $ && affine Riemann curvature tensor \\
$ \bar{R}_{\mu \nu} $ && affine Ricci tensor
 \\
$ \bar{R} $ && affine Ricci scalar 
\\
$ R_{\mu \nu \lambda}{}^{\gamma} $ && Levi-Civita Riemann curvature tensor \\
$ R_{\mu \nu} $ && Levi-Civita  Ricci tensor 
\\
$ R $ && Levi-Civita  Ricci scalar 
\\
$ \bar{G}_{\mu \nu} $ && affine Einstein tensor \\
$ \tilde{G}_{\mu \nu} $ && Riemann-Cartan Einstein tensor \\
$ G_{\mu \nu} $ && Levi-Civita Einstein tensor \\
$ P^{\mu \nu}{}_{\lambda}$ &  & Palatini tensor \\
$ \tilde{P}^{\mu \nu}{}_{\lambda}$ &  & Riemann-Cartan Palatini tensor  \\
$T_{\mu \nu}$ &  & metric energy-momentum tensor 
\\
${}^{(\bar{\Gamma})}T_{\mu \nu}$ &  & metric energy-momentum tensor (\textit{affine framework}) 
\\
$\Omega^{a}{}_{\mu}$ &  & tetradic energy-momentum tensor 
\\
$\Sigma^{a}{}_{\mu}$ &  & canonical energy-momentum tensor (\textit{affine framework})  
\\
$\Delta^{\mu \nu}{}_{\lambda}$ &  & hypermomentum tensor  
\\
$S^{\mu}{}_{a}{}^{b}$ &  & intrinsic spin tensor 
\\
$T$ &  & torsion scalar \\
$Q$ &  & non-metricity scalar \\
$B_T$ &  & torsion boundary term \\
$B_Q$ &  & non-metricity boundary term \\
$b_{\Lambda}$ &  & teleparallel gauge boundary term \\
$b_{\xi}$ &  & symmetric teleparallel gauge boundary term \\
$\ourG$ &  & Levi-Civita bulk term \\
$\mathfrak{G}$ &  & tetradic Levi-Civita bulk term \\
$\bar{\ourG}$ &  & affine bulk term \\
$\ourB$ &  & Levi-Civita boundary term \\
$\mathfrak{B}$ &  & tetradic Levi-Civita boundary term \\
$\bar{\ourB}$ &  & affine boundary term \\
$M^{\mu \nu}{}_{\lambda}$ &  & Levi-Civita rank-three object 
\\
$\mathfrak{M}^{\mu}{}_{a}{}^{b}$ &  & tetradic Levi-Civita rank-three object 
\\
$E^{\mu \nu \lambda}$ &  & Levi-Civita superpotential 
\\
$\bar{E}^{\mu \nu \lambda}$ &  & affine superpotential 
\\
$\mathbb{B}$ &  & metric-tetrad boundary term \\
\hline
\end{longtable}}


\clearpage

\myemptypage

 \pagestyle{fancy}%
 \renewcommand{\chaptermark}[1]{\markboth{#1}{#1}}
 \fancyhead{}
\fancyhf[LOH]{\textit{\chaptername\ \thechapter . \ \leftmark}}
\fancyhf[REH]{\textit{\nouppercase{\rightmark}}}

\renewcommand{\chapterheadstartvskip}{\vspace*{-2\baselineskip}}


\begin{savequote}[70mm]
The supreme task of the physicist is to arrive at those universal elementary laws from which the cosmos can be built up by pure deduction. There is no logical path to these laws; only intuition, resting on sympathetic understanding of experience, can reach them.
\qauthor{`Motive des Forschens' (1918) \\ Albert Einstein}
\end{savequote}
\chapter{Introduction}
\label{chapter1}

The best current theory of gravitation and spacetime is Einstein's 1915 theory of General Relativity (GR), a classical field theory where geometry and curvature play the role of gravity. Whilst being over 100 years old, the theory of General Relativity has passed every experimental test to date, from solar systems tests to astrophysical and cosmological experiments~\cite{Will:2014kxa}. The language used to describe the mathematics of GR is differential geometry, and gravity is interpreted as the curvature of spacetime itself. It is somewhat miraculous that such a rich and varied array of phenomena and accurate predictions can come from such a conceptually simple and mathematically elegant theory.

The equations governing the dynamics of General Relativity are known as the Einstein field equations
\begin{equation} \label{EFE0}
R_{\mu \nu} - \frac{1}{2} R\,  g_{\mu \nu} = \frac{8 \pi G}{c^4} T_{\mu \nu} \, ,
\end{equation}
where $R_{\mu \nu}$ and $R$ are related to the curvature of spacetime, $g_{\mu \nu}$ is the metric tensor describing the local geometry of spacetime, and $T_{\mu \nu}$ is the energy-momentum tensor of matter. The left-hand side represents `geometry and curvature', while the right-hand side is `energy and momentum'. This is summarised in the famous quote by Wheeler, ``Space tells matter how to move, matter tells space how to curve”~\cite{Misner:1973prb}. 

The constants appearing, $G$ and $c$, are Newton's gravitational constant and the speed of light, giving equation~(\ref{EFE0}) units of inverse length squared. The fact that there is only \textit{one} fundamental parameter, Einstein's gravitational coupling constant $\kappa = 8 \pi G / c^4$, that needs to be experimentally measured and determined (in contrast to the roughly 19 free parameters in the Standard Model~\cite{Uzan:2010pm}) is also a testament to the simplicity of Einstein's theory.

After more than a century since its birth, the predictions of General Relativity are still being confirmed. This is especially relevant over the past few decades, where we have entered the age of `precision cosmology', see~\cite{Ishak:2018his} and references therein. The 2015 LIGO detection of gravitational waves from the black hole merger GW150914 is a prime example of the ongoing successful predictions of GR~\cite{LIGOScientific:2016aoc,LIGOScientific:2016lio}. Along with quantum theory and the Standard Model of particle physics, General Relativity represents one of the two pillars of modern physics.

However, modifications of Einstein's original theory of General Relativity can be traced back almost as far as GR itself. In the late 1910s and early 1920s Cartan, Weyl, Eddington, Einstein and a host of others made significant work in this direction~\cite{Goenner:2004se,Goenner:2014mka}. Many of these `modified theories' were naturally born from the developments and extensions of the geometric framework on which GR rests: Riemannian geometry. These extensions were in part motivated by an effort to unify the fundamental forces in a unified field theory~\cite{Goenner:2004se,Goenner:2014mka}. For example, Weyl's gauge theoretic approach to gravity came from his goal of “geometrizing" the electromagnetic interaction and unifying it with gravity~\cite{Weyl1919}. This also led to the birth of gauge theory itself, which underpins the Standard Model.

As a whole, these developments gave rise to a broader geometric framework from which one can attempt to formulate and understand gravity; one based not just around the spacetime metric but also the affine connection, an integral component of differential geometry. In General Relativity, the affine connection is \textit{chosen} to be the Levi-Civita connection, which is completely determined by the metric tensor.  A richer structure is revealed by considering an a priori independent affine connection, known as \textit{metric-affine} or \textit{non-Riemannian} geometry.

In this thesis, we will study modified theories of gravity based in both the traditional Riemannian setting of GR and the broader non-Riemannian setting. In the Riemannian setting, we propose a novel modification of General Relativity based on the non-covariant, first-order \textit{Einstein action}. We then go on to generalise this theory to the metric-affine case with an independent affine connection. A particular focus, and an important result in itself, is the comparison between our newly proposed models and the modified theories based on \textit{teleparallel geometries}\footnote{Teleparallel geometries are a sub-class of the metric-affine geometries with vanishing affine curvature.}. This is achieved through the study of boundary terms. These boundary terms play an important role in other areas of physics, such as being related to the entropy of black holes. In our modified theories, we see that these boundary terms also play a prominent role. One surprising result is that they can give rise to inflation in the early universe.

The other aspect that links our modifications and the teleparallel theories is symmetry. The proposed theories based on the Einstein action break fundamental symmetries of nature. By restoring the covariance of these models, we in fact find an equivalence with the modified teleparallel models. This serves as an important link between symmetry and non-Riemannian geometry.

In the remainder of this chapter, we motivate the need to modify General Relativity, taking a look at some of the observational and theoretical challenges GR faces. We also give an outline of the different ways that one can go about constructing modifications of gravity. Many of the theories introduced share parallels with our modifications, and we will reference them again in the later chapters. 

\section{Why modify General Relativity?}
\label{section1.1}
Despite its overwhelming observational successes and its elegant mathematical structure, there are important reasons one would look for modifications of General Relativity or alternative theories of gravity. These can broadly be divided into observational and theoretical issues. The observational challenges are related to the so-called `dark sector', i.e., dark matter and dark energy. The theoretical issues are concerned with topics such as singularities and the need for a quantum theory of gravity.
Perhaps the most infamous problem, relating both to observations and theory, is the \textit{cosmological constant problem}. A brief outline of these challenges will be given, as well as some potential hints about how modified gravity may solve these issues.

\subsection{Observational challenges}
\label{section1.1.1}
The current benchmark cosmological model is the Lambda-Cold-Dark-Matter ($\Lambda$CDM) model, also known as the concordance model, where $\Lambda$ stands for the cosmological constant added to the Einstein field equations~(\ref{EFE0}). The equations then take the form
\begin{equation} \label{EFE01}
R_{\mu \nu} - \frac{1}{2}R \, g_{\mu \nu} + \Lambda \, g_{\mu \nu} = \kappa T_{\mu \nu} \, .
\end{equation}
Dark matter is needed to explain galaxy rotation curves, observations in the CMB, and a host of other astrophysical and cosmological phenomena~\cite{Bertone:2016nfn}, whilst dark energy is required to account for the observed accelerating expansion of the Universe~\cite{SupernovaSearchTeam:1998fmf,SupernovaCosmologyProject:1998vns}. Both of these require us to go beyond the Standard Model in some way. As of yet, we currently lack a microscopic description for dark matter, despite accounting for roughly 85\% of the matter in the Universe, and the cosmological constant has its own issues.
Within the $\Lambda$CDM framework, two recent and important observational discrepancies that need addressing are the \textit{Hubble tension} and the \textit{S$_8$ tension}.

The Hubble constant is the present-day value for the rate of expansion of the Universe $H_0 = H(t_{\textrm{today}})$, where $H(t):= \dot{a}(t)/a(t)$, dot denotes the time derivative, and $a(t)$ is the scale factor of the Universe appearing in the Friedmann-Lema\^{i}tre-Robertson-Walker (FLRW) metric~\cite{Dodelson:2003ft}. The Hubble tension is the disagreement between the inferred value of the Hubble constant $H_0$ from early-time observations versus the late-time value of $H_0$ obtained from local (mostly model-independent) measurements~\cite{weinberg2013observational,DiValentino:2020zio,Riess:2019qba}. These early-time (high redshift $z \gg 100$) measurements come from the Planck analysis of the Cosmic Microwave Background (CMB), giving a value\footnote{These values vary slightly with the inclusion of other data sets, e.g., with Baryon Acoustic Oscillations one can find $H_0 = 67.9 \pm 1.1$km s$^{-1}$ Mpc$^{-1}$~\cite{ivanov2020cosmological}. For a recent and thorough review on the Hubble tension and possible alleviations, see~\cite{DiValentino:2021izs}.} of $H_0 = 67.27 \pm 0.60$km s$^{-1}$ Mpc$^{-1}$~\cite{Planck:2018vyg}. It is important to note that these measurements are model-dependent.

Late-time (low redshift $z \leq 0.01$) measurements come from the `distance ladders' constructed from Supernovae calibrated by Cepheids, with recent observations giving a value of $H_0 = 73.04 \pm 1.04$km s$^{-1}$ Mpc$^{-1}$~\cite{riess2022comprehensive}. Low redshift measurements using other methods give similarly large values for $H_0$ compared to the early-time values.
Over the last few years, this tension has only grown and become more statistically significant, with most current values being in disagreement at the level of at least $5\sigma$~\cite{DiValentino:2020zio,Riess:2019qba}. Moreover, this tension cannot be easily removed by systematics, which would in fact require multiple independent systematic errors~\cite{DiValentino:2020zio}, also see~\cite{Bidenko:2023cgm}.

Modified theories can change the effective strength of gravity at different points in the Universe's history and so have many possible ways to address these issues, though many of the simplest theories (such as certain early dark energy models~\cite{Poulin:2023lkg}) are not satisfactory. This often necessitates studying more radical modifications of gravity, and some of these have been demonstrated to be consistent with measurements from both late and early times. See~\cite{DiValentino:2021izs} for a comprehensive overview.

The $S_8$ tension is related to the amplitude of growth of structure formation and the observed matter clustering in the late Universe, see~\cite{DiValentino:2020vvd,Abdalla:2022yfr}. The inferred value of matter clustering from early Universe measurements such as the CMB is in disagreement with the late-time, local measurements from gravitational lensing and galaxy clustering~\cite{Heymans:2020gsg}.

 This is typically quantified by $S_{8} = \sigma_{8}  \sqrt{\Omega_{m}/0.3}$, where $\sigma_{8}$ is the root mean square of the amplitude of matter perturbations smoothed over the scales $8 h^{-1} \textrm{Mpc}^{-1}$~\cite{Poulin:2022sgp}. Here, $h$ is the dimensionless Hubble constant $h = H_0/ (100$ km s$^{-1}$ Mpc$^{-1})$ and $\Omega_{m}$ is the current matter density parameter. For a survey of recent values of $S_{8}$ from various different datasets, see~\cite{DiValentino:2021izs}. In general, early versus late values are in tension at the level of around $2-3\sigma$.
An added complexity is that most proposed solutions for the $H_0$ tension tend to worsen the $S_8$ tension, or vice-versa. For example, late-time modifications of $\Lambda$CDM have to meet a number of stringent constraints to alleviate both of these tensions simultaneously~\cite{Heisenberg:2022gqk}.

Both of these tensions may be pointing towards new physics beyond General Relativity. Within the $\Lambda$CDM framework, it should be noted that there are further anomalies and tensions besides those presented here. For example, on smaller scales there are some long-standing discrepancies relating to dark matter and galactic observations. These are the `cuspy halo' problem~\cite{deBlok:2009sp} and the `missing satellite' or `dwarf galaxy' problem~\cite{mateo1998dwarf}. The first refers to the density profiles of low-mass galaxies and the second to the number and distribution of dwarf galaxies. In both these cases, the predictions of $\Lambda$CDM are in conflict with observation and there is currently no agreed-upon resolution. However, it should be noted that these discrepancies appear to be less significant than those of $H_0$ and $S_8$. For an overview and review of $\Lambda$CDM tensions, including all those mentioned above, see~\cite{Perivolaropoulos:2021jda,Abdalla:2022yfr}.

\subsection{Theoretical challenges}
\label{section1.1.2}
\subsubsection{Quantum gravity}

The observational challenges outlined above are concerned with the infrared (IR) structure of gravity; this means at low energies and on large scales. At the opposite end of the spectrum in the ultraviolet (UV), on small scales and at high energies, General Relativity also faces problems. These are primarily to do with quantum gravity, and we refer to the works~\cite{Kiefer:2004xyv,Carlip:2001wq} for an overview of these topics. Here we briefly mention a few of these issues, and why modifications of GR could be useful in these regards.

It is well known that GR as a quantum theory is non-renormalizable~\cite{Stelle:1976gc,Goroff:1985sz}. This can be inferred from the fact that the coupling constant of the Einstein-Hilbert action has negative mass dimension\footnote{We will usually work in natural units $\hbar = c =1$ throughout this thesis.} $[G] = -2$, see for instance~\cite{zee2013einstein}. Consequently, we cannot use our usual perturbative techniques for gravity up to and above the Planck scale $M_{p} \propto  (G)^{-1/2}$, where new physics might come into play.  We will discuss the implications of these limitations shortly, and whether or not this is an issue.

One solution to non-renormalizability is to modify the gravitational action by including higher powers of curvature (and hence higher derivatives), which can lead to a non-negative coupling constant. A key example is quadratic gravity~\cite{Stelle:1976gc}, see also~\cite{Salvio:2018crh,Alvarez-Gaume:2015rwa}. As famously shown by Stelle, the resulting theory is in fact partially renormalizable~\cite{Stelle:1976gc,Stelle:1977ry}. However, these modifications can also introduce ghost degrees of freedom~\cite{Salvio:2018crh}, making them potentially unsuitable in the IR. For example, in the usual quantization prescription, quadratic curvature terms are known to result in the loss of unitarity\footnote{Note that if alternative approaches are taken~\cite{Donoghue:2019fcb}, unitarity may be preserved at the cost of violating micro-causality.}~\cite{Stelle:1977ry,Stelle:1976gc,Alvarez:1988tb}. In Sec.~\ref{section1.2.1} we study these actions in more detail and discuss some of these ghostly instabilities.

Other problems in quantizing gravity are related to the diffeomorphism invariance and background independence of General Relativity. These topics are discussed in Chapter~\ref{chapter2}, but see~\cite{Carlip:2001wq} for a review focusing on the quantum aspects. The crux of the matter is identifying the physical degrees of freedom, and distinguishing these from the gauge degrees of freedom~\cite{Wald:1984}. One way that this presents itself as an obstacle for quantum gravity is the overcounting of states in the path-integral formulation, leading to unwanted infinities~\cite{Loll:2022ibq}. Another major challenge is the `problem of time', which is again a consequence of diffeomorphism invariance~\cite{Isham:1992ms}. This is because coordinates lose their physical meaning in background independent theories such as GR, making reconciliation with standard quantum theories, where time takes a special role, especially difficult. Lastly, the notion of observables becomes less clear in quantum gravity, where diffeomorphism invariance implies that observables must be non-local~\cite{Loll:2022ibq,Rovelli:1990ph}.

In light of some of these issues, the two main approaches to quantum gravity take the following different routes: either assuming that the spacetime metric is not the fundamental field (as in string theory), or assuming that the classical, smooth background of GR should instead be fundamentally quantum\footnote{Another approach to quantizing geometry is noncommutative geometry, see~\cite{beggs2020quantum}.} (in loop quantum gravity)~\cite{Carlip:2001wq,Kiefer:2004xyv}. The former gives up on background independence and primarily uses perturbative techniques, whilst the latter maintains background independence but is non-perturbative~\cite{Rovelli:2004tv}. However, problems still remain in both of these theories~\cite{Carlip:2001wq}, and it is safe to say that we do not yet have a fully agreed-upon theory of quantum gravity.

We should also mention two other proposals, the asymptotic safety paradigm~\cite{Niedermaier:2006wt} and Ho\v{r}ava-Lifshitz gravity~\cite{Horava:2009uw}. The former approaches the problem of non-renormalizability from a non-perturbative perspective, using techniques based on the renormalization group flow to control the UV-behaviour of the theory~\cite{Niedermaier:2006wt,Bonanno:2020bil}. The latter includes higher spatial derivates to ensure renormalizability, similar to the quadratic gravity theories, but breaks Lorentz covariance to maintain lower-order time derivatives, see Sec.~\ref{section1.2.4}. Again, both of these theories have their own shortcomings and setbacks~\cite{Donoghue:2019clr,Bonanno:2020bil,Padilla:2010ge,Wang:2017brl}, and there is still no consensus on the correct approach.

On the other hand, the lack of a UV-complete theory of quantum gravity does not mean that we cannot use perturbative techniques in quantum General Relativity. Instead, one takes the perspective of treating quantum GR as an effective field theory (EFT), valid up to some cutoff scale (e.g., at or below the Planck scale)~\cite{Donoghue:1994dn,Donoghue:2012zc}. In the EFT treatment of quantum gravity, one can still make meaningful predictions at ordinary energy scales. Non-renormalizable theories can be seen in the same way as renormalizable ones, they are just more sensitive to the (integrated out) microscopic physics~\cite{Burgess:2003jk}. These techniques have been successfully employed in a variety of gravitational contexts~\cite{Gubitosi:2012hu,Carrasco:2012cv}, the EFT of inflation~\cite{Cheung:2007st} being a particularly notable case. We will later make reference to studies using these techniques applied to the modified gravity theories of this thesis in Sec.~\ref{section5.2.2}.

The first step of constructing an EFT is to write down the most general action consistent with the low-energy symmetries of the theory, ordered in an energy expansion. In GR, the symmetries are diffeomorphism and local Lorentz invariance, and the action takes the form
\begin{equation} \label{EFT}
S_{\textrm{grav}}[g] = \int \sqrt{-g} d^4x \Big[ \Lambda + \frac{1}{2\kappa} R + c_1 R^2 + c_2 R_{\mu \nu} R^{\mu \nu} + \ldots \Big] \, ,
\end{equation}
where all terms (which will be formally defined in the following chapters) are Lorentz and coordinate scalars. Moreover, the action is organised in an energy expansion, i.e., by powers of derivatives of the metric~\cite{Donoghue:1994dn,Burgess:2003jk}.
The first term is the cosmological constant, which must be experimentally determined. This is discussed in more detail below, but observationally we know $\Lambda$ to be very small, so it is often dropped from the expansion. The leading-order term is then the Einstein-Hilbert action~(\ref{EH}) with $\kappa = 8 \pi G$, which gives the standard Einstein field equations~(\ref{EFE0}). We then have the quadratic curvature terms with dimensionless parameters $c_1$ and $c_2$, and higher-order terms represented by the ellipsis in~(\ref{EFT}). There are an infinite number of such terms, and the free parameters must be determined experimentally.  However, these higher-order terms are suppressed by inverse powers of the cutoff scale. In other words, these higher-order correction terms do not affect the low-energy physics. 
Consequently, at the energy scales of everyday physics, including those reached in the most powerful particle colliders, only the Einstein-Hilbert term is important.

 The next-to-leading-order quadratic terms are extremely weak if $c_1$ and $c_2$ are of order unity. In fact, they are only bound to be less than around $10^{74}$~\cite{Stelle:1977ry}. One may also wonder about the previously mentioned pathologies associated with these quadratic terms, but it has been shown that these issues occur at scales beyond the Planck scale~\cite{Donoghue:2012zc,simon1990higher}. At these energies, all operators become equally important. This again demonstrates the use of effective field theory within its domain of validity.

Despite these successes, there are reasons to go beyond the effective field theory of GR.  Firstly, the EFT of quantum GR does not represent a fundamental theory of quantum gravity~\cite{Steinwachs:2020jkj}. At the high-energy and large-curvature limits, the EFT treatment breaks down and a UV-complete theory is needed. These regimes are crucial for answering questions about the early universe or the interior of black hole solutions, where GR predicts singularities~\cite{Coley:2018mzr}. Another interesting limitation is in the extreme IR limit, where non-local quantum gravitational effects could build up over large distances, even if the local curvature is small~\cite{Donoghue:2022eay}. This could be problematic in spacetimes with horizons and singularities, where the EFT is sensitive to the metric components that diverge at these points.

Instead, it is also possible to consider that the standard perturbative expansion is still valid but that Einstein's theory is an incomplete description of gravity~\cite{Steinwachs:2020jkj}. This motivates searches for alternative gravitational actions, such as those in different geometric settings~\cite{Hehl:1994ue} or with additional fields~\cite{Clifton:2011jh}. In contrast to the EFT treatment of GR, these modified actions typically lead to new additional propagating degrees of freedom. The quadratic curvature theories mentioned previously serve as a typical example. Another important example is Ho\v{r}ava-Lifshitz gravity~\cite{Horava:2009uw}, which we discuss in Sec.~\ref{section1.2.4}.

Within the EFT framework, an interesting perspective was put forward by Donoghue in~\cite{Donoghue:2009mn}: if one relaxes the assumption of gauge invariance, it could be the case that gauge symmetries are simply a low-energy emergent phenomena. This means allowing operators into the effective Lagrangian that violate these gauge symmetries, which in the case of GR is diffeomorphism invariance\footnote{A very similar situation occurs in Ho\v{r}ava Gravity, where Lorentz invariance is broken in the UV but
recovered in the IR, see Sec.~\ref{section1.2.4}.}. This is explored explicitly in~\cite{Anber:2009qp}, where non-covariant actions are proposed and tested. The theories we study in Chapter~\ref{chapter5} take a very similar form, and can thus be motivated with similar arguments. In Sec.~\ref{section1.2.4} we further discuss symmetry-breaking theories, as well as the concept of a minimal length scale~\cite{Garay:1994en}, which is not too far removed from the topics discussed above. A natural choice of minimum length scale could be related to the Planck length $l_{p} = (\hbar G/c^3)^{1/2} \approx 1.6 \times 10^{-35}$m, where quantum gravitational field fluctuations are thought to become important~\cite{Wheeler:1957mu}. In summary, modified gravitational theories that break covariance may be a promising starting point. In Chapter~\ref{chapter6} we also show how this can lead to positive outcomes for the classical singularities of GR.

\subsubsection{The cosmological constant problem}

Perhaps the most infamous challenge for General Relativity is the \textit{cosmological constant problem}~\cite{Weinberg:1988cp}. Using a conventional quantum field theory (QFT) treatment, the zero-point vacuum energy of quantum matter fields can be calculated. However, General Relativity and the equivalence principle (see Sec.~\ref{section3.3}) tell us that all forms of energy density should gravitate, and this includes the energy of the vacuum. In fact, the Colella, Overhauser and Werner (COW) experiments~\cite{colella1975observation} seem to verify that the weak equivalence principle holds at the quantum mechanical level~\cite{martin2012everything}.
The vacuum energy can then be directly equated with the cosmological constant $\Lambda$ in the Einstein field equations~(\ref{EFE01}).  A naive calculation leads to the famous 120 orders of magnitude difference between the QFT prediction and the observed value of the cosmological constant\footnote{The observed value of $\rho_{\Lambda}^{\textrm{obs}} \approx 10^{-47} \textrm{GeV}^4$ whereas the zero-point energy in QFT with a cutoff at the Planck scale is around $\rho_{\Lambda}^{\textrm{bare}} \approx 10^{74} \textrm{GeV}^4$. Taking the cutoff instead at the QCD scale still gives a value of $\rho_{\Lambda}^{\textrm{bare}} \approx 10^{-3} \textrm{GeV}^4$, requiring significant fine-tuning~\cite{martin2012everything}.}. 

In the context of QFT and effective field theory, this may not seem to be such an issue. There, it is commonplace to use renormalisation techniques to tame divergences, so one could expect the renormalised cosmological constant $\rho_{\Lambda}^{\textrm{obs}}$ to be obtained by simply subtracting a counterterm 
\begin{equation}
\rho_{\Lambda}^{\textrm{obs}} = \rho_{\Lambda}^{\textrm{bare}} + \rho_{\Lambda}^{\textrm{counter}} \,, 
\end{equation}
where $\rho_{\Lambda}^{\textrm{obs}} \approx 10^{-47} \textrm{GeV}^4$ is the observed cosmological constant of our universe. Here $ \rho_{\Lambda}^{\textrm{bare}}$ is the vacuum energy of quantum fields, calculated up to some particular energy scale, and is typically many orders of magnitude greater than $\rho_{\Lambda}^{\textrm{obs}}$. Hence $\rho_{\Lambda}^{\textrm{counter}} $ must be significantly fine-tuned to match observation, and the non-zero value of $\rho_{\Lambda}^{\textrm{obs}}$ makes symmetry arguments harder to construct.

The  precision fine-tuning of the counter term $\rho_{\Lambda}^{\textrm{counter}}$ poses a question of naturalness, but a more detailed treatment reveals the problem is actually far worse: this fine-tuning must be repeated at each higher-order in the perturbation theory, making the theory highly sensitive to the (unknown) UV physics. This is known as `radiative instability', for details see Weinberg~\cite{Weinberg:1988cp} or, for a modern treatment, the lectures of Padilla~\cite{Padilla:2015aaa}. In the language of effective field theory, the radiative instability of the cosmological constant means that new, low-energy counter-terms need to be chosen whenever the cutoff energy scale is changed, which goes against the whole point of effective field theory.

This problem has been notoriously hard to solve. Further related issues include phase transitions, which would again ruin the fine-tuning of the cosmological constant~\cite{martin2012everything}. Moreover, there is a very powerful no-go theorem due to 
Weinberg~\cite{Weinberg:1988cp} that shows that self-tuning is not possible without fine-tuning, given some fairly weak assumptions such as local translation invariance of the vacuum.
More recent approaches in modified theories of gravity may show some promise though, for a review see~\cite{Bernardo:2022cck}. In particular, models that break symmetries, such as Lorentz invariance or diffeomorphism invariance~\cite{Khoury:2018vdv}, have been shown to lead to `degravitation', another promising avenue for addressing the cosmological constant problem, see~\cite{Arkani-Hamed:2002ukf,niedermann2017gravitational}.

Finally, a related potential issue is the \textit{coincidence problem}~\cite{steinhardt1997cosmological}, where the current ratio of the dark energy density parameter $\Omega_{\Lambda ,0}$ and matter density parameter $\Omega_{m,0}$ is of order one. This can be seen as a special epoch of the universe's history, one where we are in the privileged position to observe the accelerated expansion of the universe. However, this is really just another problem of fine-tuning, specifically, of the initial conditions in the early universe and of the cosmological constant such that they lead to an evolution compatible with both structure formation and our current observations~\cite{jenskoscalar}. This is also pointed out in~\cite{Velten:2014nra}, where it is explained that the real coincidence is that $\Omega_{\Lambda} \approx 0.7$.

Both the coincidence problem and cosmological constant problem are theoretical issues, though they concern observation. It is therefore useful to explore theories beyond General Relativity and the $\Lambda$CDM model. For a review of dynamical approaches to dark energy as well as its relation to the cosmological constant and coincidence problems, see~\cite{Copeland:2006wr} and references therein.

\section{Theories beyond General Relativity}
\label{section1.2}
The challenges outlined above motivate us to look for gravitational theories beyond General Relativity. It should be mentioned that alternative theories were studied long before many of these issues became apparent, largely as a result of the extended geometric framework that was developed around the time of GR itself~\cite{Goenner:2004se,Goenner:2014mka}. From a modern perspective, modifications of gravity can largely be classified by which of the fundamental assumptions of GR they break. 

 An important theorem due to Lovelock gives the precise conditions that need to be met in order to have field equations which differ from General Relativity~\cite{Lovelock:1971yv,lovelock1972four}
\begin{theorem}[Lovelock's Theorem] \label{Theorem1}
The only local, divergence-free tensor constructed from the metric and its first two derivatives in four dimensions is
\begin{equation} 
E^{\mu \nu} = \alpha G^{\mu \nu} + \beta g^{\mu \nu} \, ,
\end{equation}
where $G^{\mu \nu} \equiv R^{\mu \nu} - \frac{1}{2} R g^{\mu \nu}$ is the Einstein tensor and $\alpha$ and $\beta$ are constants.
\end{theorem}
\noindent Note that the symmetry of $E^{\mu \nu}$ and its linearity in second derivatives follow directly from the other assumptions~\cite{Lovelock:1971yv,lovelock1972four}.
The resulting tensor is just the left-hand side of the Einstein field equations with a cosmological constant term~(\ref{EFE01}). In terms of action principles, we see that General Relativity is the unique second-order theory of gravity that one can obtain from a four-dimensional diffeomorphism and locally Lorentz invariant action constructed from the metric alone $S[g]$.

It then follows that modified theories must break some of these assumptions -- for example, by considering: 
\begin{enumerate}[(i)]
\item higher than second-order derivatives of the metric,
\item fields other than the metric,
\item  non-Riemannian geometries,
\item breaking local Lorentz invariance or diffeomorphism invariance,
\item non-locality,
\item extra spacetime dimensions.
\end{enumerate}
More radical approaches, such as theories not derived from a Lagrangian (e.g., emergent theories) or quantum approaches, may not strictly fall into these categories.

The focus of this thesis will be on modifications that are geometric (non-Riemannian) and those that break fundamental symmetries. Hence, modifications (i)-(iv) will be particularly important. We will not discuss non-local theories (v), alternative dimensions (vi), or other approaches to modified gravity. For these reasons, we will briefly give an introduction to these types of modifications (i)-(iv) and provide more detailed references throughout. For a more thorough review of modified theories of gravity, including many not mentioned here, we refer the reader to the works~\cite{Clifton:2011jh,CANTATA:2021ktz}.

\subsection{Invariants}
\label{section1.2.1}
The action of General Relativity is the Einstein-Hilbert action~(\ref{EH}), constructed from the Ricci scalar $R$ of the Levi-Civita connection~(\ref{Levi-Civita}). The Ricci scalar includes second-order derivatives of the metric tensor, but only the first-order terms contribute to the equations of motion\footnote{See Chapter~\ref{chapter4} for details, and equation~(\ref{RGB}) in particular, for the explicit decomposition into first and second-order terms.}. Hence, the resulting field equations are of second-order. For non-linear functions of the Ricci scalar, e.g., $R^2$, the field equations will generically be fourth-order. In fact, these non-linear modified terms were written down as early as 1918 by Weyl~\cite{Weyl:1918ib,schimming2004history} and shortly after in 1923 by Eddington~\cite{eddington1923mathematical}. 
The equations of motion will also be fourth-order for any invariant contractions of the curvature tensor $R_{\mu \nu \lambda}{}^{\gamma}$ that cannot be rewritten as the sum of first-order terms and second-order terms confined to the boundary, see for example~\cite{ruzmaikina1970quadratic}.

It is also well known that higher-derivative gravity, being constructed from actions with additional geometric invariants beyond the Ricci scalar, leads to favourable UV properties~\cite{Biswas:2005qr}. In particular, Stelle famously showed that fourth-order modifications of GR leads to a renormalizable theory~\cite{Stelle:1976gc,Stelle:1977ry}. However, as we will touch upon at the end of this discussion, higher-derivative theories have the disadvantage of being prone to instabilities~\cite{Biswas:2011ar}. Here we briefly review some of these modifications based on invariants constructed from the metric tensor, beginning with the Ricci scalar.

The class of theories based on an arbitrary function of the Ricci scalar is known as $f(R)$ gravity. In this case, it is generally not possible to write the action solely in terms of first derivatives of the metric and boundary terms, therefore breaking assumption (i).
Some of the earliest works on $f(R)$ gravity with $f$ being completely arbitrary include~\cite{Buchdahl:1970ynr,Barrow:1983rx}. In the following years these theories became much more popular, see~\cite{Sotiriou:2008rp,DeFelice:2010aj,Nojiri:2010wj} for reviews.

The action and resulting field equations, which will be useful to compare with the modified theories studied later in the thesis, are given by
\begin{equation} \label{f(R)_action}
S_{f(R)} = \frac{1}{2\kappa} \int f(R) \sqrt{-g} d^4x \, ,
\end{equation}
\begin{equation}
  \label{f(R)field_equation}
   f'(R) R_{\mu \nu} + \big[g_{\mu \nu} \Box - \nabla_{\mu} \nabla_{\nu}\big] f'(R) - \frac{1}{2} g_{\mu \nu} f(R) = \kappa T_{\mu \nu} \, ,
\end{equation}
where $\Box = \nabla^{\mu} \nabla_{\mu}$ is the d'Alembert operator, $\nabla_{\mu}$ is the covariant derivative,
and a prime denotes the derivative with respect to the dependent variable.
 Further details will be given in Chapter~\ref{chapter2}, and in Chapter~\ref{chapter5} we look at $f(R)$ gravity explicitly. For any non-linear function $f$ such that $f''(R) \neq 0$, the resulting field equations are generically fourth-order. And it is precisely the linear choice $f = R + \Lambda$ that uniquely leads to the Einstein field equations~(\ref{EFE01}), as predicted by Lovelock's theorem. 

One particular choice that is relevant in the context of the early universe is $f = R + \alpha R^2$, known as Starobinsky inflation~\cite{Starobinsky:1980te}, leading to early-time de-Sitter exponential expansion. This inflationary model, despite being proposed over forty years ago and taking such a simple form, is still in good agreement with all observational data such as from the CMB~\cite{Rodrigues-da-Silva:2021jab}. We will also see that this model corresponds to a minimally coupled scalar field with an exponential-type potential. In Section~\ref{section6.1} and~\ref{section6.3} we also study modified theories leading to inflationary scenarios, with an action taking a similar form to this choice. For further details of Starobinsky inflation, such as quantum implications of the model, see~\cite{Vilenkin:1985md}. We again refer to the reviews of $f(R)$ gravity~\cite{Sotiriou:2008rp,DeFelice:2010aj,Nojiri:2010wj,Clifton:2011jh} and references therein for details regarding the general properties and solutions of these theories. For works focusing on cosmology and viable $f(R)$ models, see~\cite{Amendola:2006we}, and for experimental tests, see~\cite{Will:2014kxa}.

Another particularly interesting invariant is the Gauss-Bonnet term
\begin{equation} \label{GaussBonnet}
\mathcal{G} = R^2 - 4 R^{\mu \nu} R_{\mu \nu} + R_{\mu \nu \lambda \gamma} R^{\mu \nu \lambda \gamma}  \, .
\end{equation}
It turns out that this term is topological in $n=4$ dimensions, and hence has vanishing equations of motion~\cite{Lanczos:1938sf}. Note that with arbitrary factors multiplying each term in~(\ref{GaussBonnet}), the action is no longer topological and generically leads to fourth-order equations of motion, see for example~\cite{Magnano:1990qu}. Explicitly, its variation with respect to the metric tensor leads to the Lanczos tensor~\cite{lanczos1932elektromagnetismus}
\begin{equation}
A_{\mu \nu} = 2 R_{\mu}{}^{\lambda \gamma \rho} R_{\nu \lambda \gamma \rho} - 4 R^{\lambda \gamma} R_{\mu \lambda \nu \gamma} - 4 R_{\mu \lambda} R^{\lambda}{}_{\nu} + 2 R R_{\mu \nu} -\frac{1}{2} \mathcal{G} g_{\mu \nu} \, ,
\end{equation}
which can be seen to be second-order in metric derivatives, see also~\cite{TP:2010} for details. Crucially, this tensor vanishes identically in dimension less than five\footnote{Recently there has been an interest in attempts to extract a limit of $n \rightarrow 4$ Gauss-Bonnet gravity~\cite{Glavan:2019inb}, for a review see~\cite{fernandes20224d}. In our work~\cite{Boehmer:2023lpb} we look at a similar $n \rightarrow 2$ limit of Einstein gravity by breaking diffeomorphism invariance, based on the decomposition of Chapter~\ref{chapter4}.}.

However, just like in the case of the Ricci scalar, a non-linear function of a boundary term is no longer a boundary term. We can then take an arbitrary function of this invariant as our starting action~\cite{Nojiri:2005jg,Nojiri:2005vv}, known as modified Gauss-Bonnet gravity,
\begin{equation}
S_{f(\mathcal{G})} = \frac{1}{2\kappa} \int f(\mathcal{G}) \sqrt{-g} d^4x \, .
\end{equation}
The field equations are generically fourth-order, containing terms of the form $\nabla_{\mu} \nabla_{\nu} f'(\mathcal{G})$. For the full equations of motion, see~\cite{Nojiri:2005jg,Nojiri:2005vv}. In the case when $f$ is linear, the field equations vanish identically in four dimensions. Further details of cosmological solutions can also be found in~\cite{Li:2007jm}. 

Both the Ricci scalar and Gauss-Bonnet term are examples of \textit{Lovelock invariants}~\cite{Lovelock:1971yv}. More generally one can consider \textit{Lovelock gravity} where the action is given by the sum of dimensionally extended Euler densities
\begin{equation} \label{LovelockI}
\mathcal{L}_{\textrm{Lovelock}}= \sqrt{-g} \sum_{i} c_{i} \frac{1}{2^i} \delta^{\mu_{1} \nu_{1} ... \mu_{i} \nu_{i} }_{\rho_{1} \sigma_{1} ... \rho_{i} \sigma_{i} }  \prod_{j}^{i} R^{\rho_{j} \sigma_{j}}{}_{\mu_{j} \nu_{j}} \, ,
\end{equation}
where $c_i$ are the coupling constants and the generalised Kronecker delta is defined as $\delta^{\mu_{1} \mu_{2} ... \mu_{r}}_{\nu_{1} \nu_{2} ... \nu_{r}} = (2r)! \delta^{\mu_{1} \mu_{2} ... \mu_{r}}_{[\nu_{1} \nu_{2} ... \nu_{r}]}$. Each term with index $i$ represents the Euler density in dimension $n=2i$. For example, the $i=1$ term is the Euler density in $n=2$ dimensions, which is the Ricci scalar $R$. The $i=2$ term is the Euler density for $n=4$, which is the Gauss-Bonnet term $R^2 - 4 R^{\mu \nu} R_{\mu \nu} + R_{\mu \nu \lambda \gamma} R^{\mu \nu \lambda \gamma}$, and so on. These terms vanish identically for $i > n/2$ when $n$ is even, or for $i > (n-1)/2$ when $n$ is odd. For further details, see the extensive review by Padmanabhan~\cite{Padmanabhan:2013xyr}. 

These Lagrangians have a number of special and interesting properties. Namely, they lead to second-order, divergence-free, symmetric field equations. In fact, they uniquely satisfy the constraints of Lovelock's theorem in an arbitrary number of dimensions $n$. In our $n=4$ universe, the sole non-trivial contribution is the Ricci scalar.
Another interesting property is that the Lovelock invariants can be decomposed into a so-called `bulk' and `boundary' part~\cite{Mukhopadhyay:2006vu,yale2011structure}. The bulk part leads to the equations of motion whilst the boundary part takes the form of a total derivative. In this thesis, we will mainly be interested in this decomposition for the $i=1$ case of the Ricci scalar, but we will briefly comment on the decomposition of higher-dimensional Euler densities in Chapter~\ref{chapter4}.

Generalised versions of Lovelock gravity, taking the form $f(L_{\textrm{Lovelock}})$ have been studied in~\cite{Bueno:2016dol}. The first two cases, $f(R)$ and $f(\mathcal{G})$ have already been mentioned above. These $f(L_{\textrm{Lovelock}})$ theories generically lead to fourth-order equations of motion, due to the non-linear functions of boundary terms. However, they are also known to not introduce any additional spin-2 degrees of freedom~\cite{DeFelice:2010aj}. Aside from the Lovelock invariants, arbitrary functions of more general scalars constructed from the Riemann tensor of the type $f(R_{\mu \nu \lambda}{}^{\gamma})$ have also been studied, see for example~\cite{Deruelle:2009zk}. These equations of motion are also fourth-order but include the modified Lovelock theories in limiting cases. For a review of higher-order gravitational theories see~\cite{Bueno:2016ypa}.

Beyond these types of theories, there are a class of theories known as \textit{infinite derivative gravity}, where an infinite number of derivatives of curvature scalars are studied~\cite{Biswas:2013cha}. These models can also be inspired from a string theory perspective~\cite{Biswas:2014tua}. These types of theories are \textit{non-local}, but have a number of interesting properties, such as being Ostrogradsky ghost-free (which will be discussed below, see~\cite{Biswas:2011ar,Biswas:2013cha} for details), and potentially resolving the classical singularities that appear in GR, e.g.,~\cite{Conroy:2014dja,Kolar:2023gqi}. However, the field equations of such a theory are clearly highly complex, making studying solutions very difficult.

We should also make an important point regarding the stability of these modifications. It is well known that higher-order derivative theories can become prone to having ghost-like instabilities of the Ostrogradsky kind\footnote{Note that there are other issues relating to stability aside from Ostrogradsky ghosts, such as the Frolov or Dolgov-Kawasaki instabilities. We will not discuss these here, but see~\cite{Clifton:2011jh} for details.}~\cite{2015arXiv150602210W}. For example, as previously discussed, Stelle showed that the renormalizable higher-order action $R + R^2 + R_{\mu \nu}R^{\mu \nu}$ also inevitably gives rise to a massive spin-2 ghost carrying negative energy~\cite{Stelle:1976gc,Stelle:1977ry}. Whether or not this renders higher-derivative theories unsuitable or not depends on the perspective taken\footnote{We focus here on the classical analysis, but should note that these ghosts do not necessarily have negative energy states after quantization~\cite{Donoghue:2021cza}. Moreover, it has been pointed out that classical Ostragoadsky instabilities may be avoided in quantum theories~\cite{Donoghue:2021eto}.}; in an effective theory of gravitation, these issues occur beyond the Planck scale and so one can safely consider these terms within the EFT formulation, see~\cite{Simon:1990ic,Donoghue:2012zc} for further details. 

On the other hand, it is also known that theories constructed from the Lovelock scalars, such as the Ricci scalar $R$ or Gauss-Bonnet term $\mathcal{G}$, evade these ghostly instabilities\cite{Woodard:2006nt}. This also extends to arbitrary functions of those invariants, the
$f(R)$, $f(\mathcal{G})$ and $f(L_{\textrm{Lovelock}})$ theories~\cite{Bueno:2016ypa}. These issues are avoided altogether if the equations of motion are at most second-order in (time) derivatives, such as in Ho\v{r}ava-Lifshitz gravity~\cite{Horava:2009uw}. Most of the modified theories studied in this thesis likewise avoid these issues by being second-order. This is because they often take the form `$f$ of something', similar to in $f(R)$ gravity, but where that `something' is only first-order in derivatives.

\subsection{Additional fields}
\label{section1.2.2}
The next way to modify gravity is to include additional, non-geometric fields beyond the metric tensor. These fields are treated as dynamical, and the gravitational action can generally be written as $S[g,\phi^{A}]$, where $\phi^{A}$ represents these additional fields. Variations of the action are taken with respect to both the metric and the independent $\phi^{A}$ fields. Moreover, unlike matter fields, these $\phi^{A}$ couple to geometry non-minimally, usually via a direct coupling between the Einstein-Hilbert action $R$ and some function of the fields $f(\phi^{A})$. Additionally, the action often contains terms for the dynamics of these fields (e.g., kinetic and potential terms).

The simplest types of fields that can be added to the action are spin-0 scalars $\phi$, which lead to scalar-tensor theories. Just like the modified theories based on invariants, scalar-tensor theories also have a long and rich history, see for example~\cite{Bergmann:1968ve,fujii2003scalar}. In fact, the progenitor of these theories can be seen in the works of Dirac~\cite{Dirac:1937ti,Dirac:1938mt}, where he considered that the fundamental constants of nature may, in fact, not be constant. Jordan then showed that these `varying constants' have to become dynamical fields for a consistent field theory treatment~\cite{jordan1937physikalischen}. The works of Brans and Dicke follow from these investigations, resulting in their seminal work on a varying gravitational constant $G$ and scalar-tensor theory~\cite{Brans:1961sx}. For a review and overview of these theories and their developments, see~\cite{Uzan:2010pm}.

 The action for a general scalar-tensor theory can be written as~\cite{Clifton:2011jh}
\begin{equation} \label{ST_Jordan0}
S_{\phi}[g,\phi] = \frac{1}{16 \pi } \int  \Big( R f(\phi) - \omega(\phi) (\nabla \phi)^2 - 2 V(\phi) \Big) \sqrt{-g} d^4x \, ,
\end{equation}
where $(\nabla \phi)^2 = \nabla_{\mu} \phi \nabla^{\mu} \phi$ and $f$ and $\omega$ are arbitrary functions of $\phi$. The minimally coupled matter Lagrangian takes the standard form $S_{\textrm{M}}[g, \varPhi^{A}]$ with $\varPhi^{A}$ representing matter fields. A simple field redefinition~\cite{Bergmann:1968ve} 
\begin{equation}
f(\phi) \rightarrow \phi \, , \qquad  \omega(\phi) \rightarrow \omega(\phi) \frac{f(\phi)}{f'(\phi)^2} \, ,
\end{equation}
allows us to rewrite the gravitational action as
\begin{equation} \label{ST_Jordan}
S_{\phi}[g,\phi] = \frac{1}{16 \pi} \int  \Big( R \phi - \frac{\omega(\phi)}{\phi}  (\nabla \phi)^2 - 2 V(\phi) \Big) \sqrt{-g} d^4x \, .
\end{equation}
The form of the actions in~(\ref{ST_Jordan0}) and~(\ref{ST_Jordan}) are written in what is known as the \textit{Jordan frame}: matter couples to the metric in the usual way and follows (metric) geodesics, and the equivalence principle is upheld.
 In the limit $\omega \rightarrow \textrm{constant}$ and $V \rightarrow 0$ we obtain Brans-Dicke theory~\cite{Brans:1961sx}. General Relativity is obtained in the limit $\omega \rightarrow \infty$, $\omega'/\omega^2 \rightarrow 0$ and $V \rightarrow \textrm{constant}$.

The equations of motion of~(\ref{ST_Jordan}) are~\cite{Clifton:2011jh}
\begin{align} \label{ST_1}
G_{\mu \nu} + \frac{g_{\mu \nu}}{\phi} \Big( \Box \phi + \frac{\omega}{2\phi} (\nabla \phi)^2 + V \Big) - \frac{1}{\phi} \nabla_{\mu} \nabla_{\nu} \phi - \frac{\omega}{\phi^2} \nabla_{\mu} \phi \nabla_{\nu} \phi &= \frac{8 \pi}{\phi} T_{\mu \nu} \, , \\
(2 \omega + 3) \Box \phi +(\nabla \phi)^2  \omega' + 4 V - 2 \phi V' &= 8 \pi T \, , \label{ST_2}
\end{align}
where $T$ is the trace of the energy-momentum tensor and the Ricci scalar $R$ has been eliminated from the scalar field equation using the trace of~(\ref{ST_1}). These equations of motion are second-order in derivatives of the metric, but now there is the presence of an additional scalar degree of freedom $\phi$. The right-hand side of the Einstein equation~(\ref{ST_1}) shows how this theory can be interpreted as a varying effective gravitational constant $G_{\textrm{eff}}$~\cite{Uzan:2010pm}.
Also note that matter is covariantly conserved $\nabla_{\mu} T^{\mu \nu} =0$, and all other aspects of Lovelock's theorem are upheld.

It is well known that a conformal rescaling\footnote{To be precise, we are performing a \textit{Weyl rescaling} of the metric tensor as opposed to a conformal transformation~\cite{Karananas:2015ioa}.} of the metric can be applied 
\begin{equation}
g_{\mu \nu} \rightarrow \hat{g}_{\mu \nu} = e^{-2\Omega(x)} g_{\mu \nu}  \, ,
\end{equation}
such that the theory is written in a frame where the scalar $\phi$ is minimally coupled to the Ricci scalar (of the new metric). This is accomplished by choosing the conformal factor $e^{-2\Omega} = \phi$ so that $\hat{g}_{\mu \nu} = \phi g_{\mu \nu}$~\cite{Dicke:1961gz}. The frame with respect to the conformal metric $\hat{g}$ is known as the \textit{Einstein frame}. Using the standard conformal transformation properties~\cite{Wald:1984}, the action can be rewritten as
\begin{equation} \label{ST_Einstein}
S_{\hat{\phi}}[\hat{g},\hat{\phi}] =  \int  \Big( \frac{1}{16 \pi}\hat{R} - \frac{1}{2} (\hat{\nabla} \hat{\phi})^2 - \hat{V}(\hat{\phi}) \Big) \sqrt{-
\hat{g}} d^4x \, ,
\end{equation}
where an appropriate field redefinition $\phi \rightarrow \hat{\phi}$ has been made, with $\partial \Omega / \partial \hat{\phi} = \sqrt{4\pi / (3+2 \omega)}$, and $8 \pi \hat{V}(\hat{\phi}) = e^{4 \Omega} V$ (see~\cite{Clifton:2011jh,magnano1994physical} for technical details).
The transformation also affects the matter Lagrangian $S_{\textrm{M}}[g, \varPhi^{A}] \rightarrow S_{\textrm{M}}[e^{2\Omega} \hat{g}, \varPhi^{A}]$ and causes a violation of the equivalence principle (see Sec.~\ref{section2.2.1}). The matter fields $\varPhi^{A}$ no longer follow geodesics of the new metric due to the coupling with $\hat{\phi}$.

The equations of motion for the scalar-tensor theory in the Einstein frame take the form
\begin{align} \label{ST_E1}
\hat{G}_{\mu \nu} &= 8 \pi \Big[ \hat{T}_{\mu \nu} + \hat{\nabla}_{\mu} \hat{\phi} \hat{\nabla}_{\nu} \hat{\phi} - \frac{1}{2} \hat{g}_{\mu \nu} \big( (\hat{\nabla} \hat{\phi})^2 + \hat{V} \big) \Big] \, , \\
\hat{\Box} \hat{\phi} - \hat{V}' &= - \frac{\sqrt{4 \pi}}{(3+2 \omega)^2} \hat{T} \, , \label{ST_E2}
\end{align}
where all terms with a tilde are defined with respect to $\hat{g}_{\mu \nu}$. Similarly, we define the energy-momentum tensor $\hat{T}^{\mu \nu} = e^{6 \Omega} T^{\mu \nu}$. The Einstein equations~(\ref{ST_E1}) take their usual form, but the conservation equation is modified with a new source term $\hat{\nabla}_{\mu} \hat{T}^{\mu \nu} = \sqrt{4 \pi}/(3+2 \omega)^2 \hat{T} \hat{\nabla}^{\nu} \hat{\phi}$, again see~\cite{Clifton:2011jh} for details. 

The equivalences (or lack thereof) between the two formulations in the Einstein and Jordan frame have been well studied; we refer to the works~\cite{Faraoni:1999hp,magnano1994physical,cho1992reinterpretation} for further discussion on this topic. On a related note, we briefly mention that $f(R)$ gravity theories can be rewritten in the form of a scalar-tensor theory by performing a Legendre transformation~\cite{Teyssandier:1983zz,Amendola:2006kh}. Consider the Lagrangian $L = f(\chi) + f'(\chi)(R-\chi)$, with $\chi$ a dynamical field. 
Variations with respect to $\chi$ lead to $f''(\chi)(R-\chi)=0$. Provided $f'' \neq 0$, which reduces to GR, we have $\chi = R$ on-shell. This is then equivalent to $f(R)$ gravity defined in~(\ref{f(R)_action}). Defining the field $\phi = f'(\chi)$ and potential $V(\phi) = (\phi \chi - f)/2$ leads to a resulting Lagrangian $L = \phi R - V(\phi)$. This is just the scalar-tensor theories in the Jordan frame~(\ref{ST_Jordan}) with $\omega =0$. Note that we have assumed $\phi(\chi)$ to be an invertible function. A similar conformal transformation and analysis in the Einstein frame can then be studied~\cite{Amendola:2006kh}.

This hints that in $f(R)$ gravity we have one additional scalar degree of freedom, which indeed is the case~\cite{DeFelice:2010aj}. However, once again some care must be taken with regards to the equivalence between formulations. Here we merely wish to express how seemingly different modifications can in fact turn out to be much more closely related. This is a theme that will be prominent throughout the thesis.

The most general scalar-tensor theories with a single scalar field leading to second-order field equations in $n=4$ dimensions belong to the class of theories known as Hordenski theories~\cite{Kobayashi:2019hrl}. One can then go on to consider more scalar fields~\cite{Damour:1992we}, or additional vector or tensor fields~\cite{Clifton:2011jh}. In order to keep these modifications consistent with General Relativity, within the regimes where it has been successfully tested\cite{Will:2014kxa}, various techniques have been developed to regulate the couplings to the beyond-GR terms. For example, screening mechanisms, such as the Chameleon mechanism~\cite{Khoury:2003rn}, allow the effective mass of the additional scalars to depend on the local density of matter. This allows the modifications to satisfy solar system constraints but still play a prominent role on cosmological scales~\cite{Burrage:2017qrf}. Adding additional field content therefore represents a promising and viable way to modify gravity, though the freedom in the amount of theories that can be considered makes pinpointing a specific model somewhat challenging.

\subsection{Non-Riemannian geometry}
\label{section1.2.3}

The previous modifications have been within the setting of Riemannian geometry. This is based on the unique, torsion-free and metric-compatible Levi-Civita connection. This geometric assumption is also implicit in Lovelock's theorem by requiring the field equations to be constructed solely from the metric. We can abandon this assumption and consider a more general geometric framework, which we call \textit{non-Riemannian} geometry. In the following chapter, we will introduce the mathematical structure in detail and study some of the main non-Riemannian gravitational theories. For this reason, we only give a condensed outline of some aspects of non-Riemannian theories here. For a historical overview, see~\cite{Goenner:2004se,Goenner:2014mka}.

A class of theories of particular interest are the \textit{metric-affine} theories~\cite{Hehl:1994ue}. Here, the metric and affine connection are treated as independent dynamical variables. In some sense, this can be thought of as introducing additional fields, but, like the metric tensor, they are intrinsically geometric in nature. The affine connection has $64$ independent components, resulting in a total of $64+10$ independent components in general. An affine space contains curvature, torsion and non-metricity. These are all geometric quantities related to the connection, introduced in more detail in Sec.~\ref{section2.1}. In General Relativity, torsion and non-metricity are assumed to vanish. 

The prototypical example of a non-Riemannian theory of gravity is Einstein-Cartan theory~\cite{cartan2001riemannian}. These spacetimes possess both curvature and torsion. It is found that torsion couples to the intrinsic spin current of matter\footnote{It is interesting to note that Cartan proposed his theory before the discovery of (quantum mechanical) intrinsic spin.}. The field equations take the form
\begin{align}  \label{ec_01}
\tilde{G}_{\mu \nu} &= \kappa \Sigma_{\mu \nu} \, , \\
T^{\mu}{}_{\rho \sigma} + \delta^{\mu}_{\rho} T^{\nu}{}_{\sigma \nu} - \delta^{\mu}_{\sigma} T^{\nu}{}_{\rho \nu}  &= 2\kappa S^{\mu}{}_{\rho \sigma} \label{ec_02} \, ,
\end{align}
where $\tilde{G}_{\mu \nu}$ denotes the affine Einstein tensor in a Riemann-Cartan space, $T^{\mu}{}_{\nu \lambda}$ is the torsion tensor, $\Sigma_{\mu \nu}$ is the (canonical) energy-momentum tensor and $S^{\mu}{}_{\rho \sigma}$ is the intrinsic spin tensor. Importantly, it is the latter equation that comes from the presence of an independent connection. In standard General Relativity, there is no direct analogue for~(\ref{ec_02}). 
The torsional field equation is algebraic and therefore torsion does not propagate when $S_{\mu \nu \lambda}=0$. It follows that the theory reduces to General Relativity~(\ref{EFE0}) in the vacuum. Einstein-Cartan theory is studied in  Sec.~\ref{section3.2.1}. It is a limiting case of the more general Poincar\'{e} gauge gravity theories~\cite{Hayashi:1979wj,Obukhov:2018bmf,hehl1980four}.

One particularly interesting class of theories are the \textit{teleparallel} gravities~\cite{Bahamonde:2021gfp,Aldrovandi:2013wha,BeltranJimenez:2019esp}. These theories assume that curvature vanishes, but torsion and non-metricity do not. The additional geometric structure that they possess allows for alternative, equivalent formulations of General Relativity. These are known as the \textit{teleparallel equivalents of General Relativity} (TEGRs)~\cite{Hohmann:2021fpr,BeltranJimenez:2019esp,Nester:1998mp,Aldrovandi:2013wha,Maluf:2013gaa}. In the case where non-metricity vanishes, it has long been known that the original (metric-compatible) TEGR action is not invariant under local Lorentz transformations~\cite{Cho:1975dh,Krssak:2018ywd}. Instead, it is \textit{pseudo-invariant}, by which we mean it is invariant up to a boundary term. This makes studying their modifications particularly interesting, where significant deviations from GR can arise. We will return to this point shortly. 

Despite the fact that the teleparallel theories admit equivalents of General Relativity, they are fundamentally different: the former describes gravity with the geometric quantities of torsion and non-metricity, whilst GR describes gravity by the curvature of spacetime. That such an equivalence between these distinct formulations exists is quite remarkable. In Chapter~\ref{chapter3} we explore the nature of this equivalence, and see that is it intimately linked with fundamental symmetries and boundary terms.

Another interesting facet of these geometric, metric-affine theories, is that they naturally arise from a gauge-theoretic perspective. This should not be too surprising, given many of their origins are related to attempts to unify gravity with the other gauge theories of matter~\cite{Blagojevic:2013xpa}. This serves as additional motivation to study these types of theories. In Sec.~\ref{section3.2.3} we discuss the relationships with gauge theories and the gauge gravitational approach.

Beyond the metric-affine framework, one can consider more general geometries such as non-commutative geometry or Finsler geometries~\cite{CANTATA:2021ktz,Pfeifer:2011xi,aschieri2006noncommutative}. Moreover, the previous modifications (i)-(ii) can also be applied to these geometric theories. For example, 
considering non-linear functions of these affine-geometric quantities or coupling them to additional fields leads to even more general theories. In Section~\ref{section5.3.3} we look at the metric-affine version of $f(\bar{R})$ gravity as a limiting case of our generalised theories. When the affine connection is treated as independent from the metric, the Ricci scalar is only first-order in derivatives of the fundamental variables. It follows that the field equations are second-order. Moreover, the standard metric $f(R)$ and the non-Riemannian metric-affine $f(\bar{R})$ theories are inequivalent. Interestingly, metric-affine $f(\bar{R})$ gravity can also be recast as a scalar-tensor theory, which is again inequivalent to its metrical counterpart, see~\cite{Sotiriou:2006hs}.

It is worth drawing special attention to the modified teleparallel gravity models known as $f(T)$ and $f(Q)$ gravity, where $T$ and $Q$ are scalars representing torsion and non-metricity respectively. These modified theories were first proposed in~\cite{Bengochea:2008gz,BeltranJimenez:2017tkd} and have been particularly popular as of late, see also~\cite{Linder:2010py,Krssak:2018ywd,Bahamonde:2021gfp,Hohmann:2022mlc,BeltranJimenez:2019tjy,Cai:2015emx}. However, there is still some confusion and disagreement in the literature regarding their theoretical properties and interpretation~\cite{Blixt:2023kyr,Golovnev:2021lki,Golovnev:2022bfm,Golovnev:2020zpv,Maluf:2018coz}. In Sec.~\ref{section3.3} we will study the teleparallel theories, and then focus on their modifications $f(T)$ \& $f(Q)$ gravity in Sec.~\ref{section5.2.2}. We also aim to clarify some of the issues and disagreements regarding the covariance of these theories. Early expositions of these issues can be found in~\cite{Li:2010cg,Tamanini:2012hg}, while the covariant formulation is discussed in~\cite{krvsvsak2016covariant}. More recent considerations can be found in \cite{Golovnev:2021omn,Hohmann:2021dhr,Maluf:2018coz,Bahamonde:2021gfp}.

In Chapter~\ref{chapter5} we propose a new modified theory within the metric-affine framework, which we call $f(\bar{\ourG})$ gravity~\cite{Boehmer:2023fyl}. The term $\bar{\ourG}$ can be directly related to curvature, torsion and non-metricity, and it possesses intriguing non-covariant properties. We then go on to look at the Einstein-Cartan limit of this theory, showing how our modifications can be written in a similar form to Einstein-Cartan theory. One interesting result, that differs from standard Einstein-Cartan theory, is that the torsional equations are now dynamical instead of algebraic. We can therefore study the effects of propagating torsion. In Chapter~\ref{chapter6}, we look at the cosmological consequences of these modifications.

\subsection{Symmetry breaking}
\label{section1.2.4}

A final route to modifying gravity that we will explore is to consider theories that break fundamental symmetries. The symmetries inherent to General Relativity are the invariance under diffeomorphisms and local Lorentz transformations. These symmetries will be fully explained in Sec.~\ref{section2.3}. In short, diffeomorphisms can be thought of as coordinate transformations and local Lorentz transformations as frame transformations in the tangent space. They bear resemblance to the gauge invariance of gauge theories, with the symmetries representing a redundancy in the mathematical description. The invariance of GR can be seen as following from the action being constructed from a scalar with respect to both of these transformations, along with being background independent.

Many theories that break local Lorentz or diffeomorphism invariance can be motivated by quantum gravity considerations, see Sec.~\ref{section1.1.2}. For example, if we accept that classical field theories like GR cannot be applied at length scales where quantum effects
dominate, then it becomes natural to consider models that break these symmetries at small scales. The existence of a minimal length scale has been considered explicitly in~\cite{Garay:1994en}. One realisation of this idea can be found in the \textit{doubly special relativity} theories~\cite{Amelino-Camelia:2000cpa}, which can be viewed as a low-energy limit of a theory of quantum gravity.
The Born-Infeld scheme~\cite{Born:1934gh} also serves as a simple approach to implement a fundamental length scale; this has been applied in many contexts, from string theory~\cite{Tseytlin:1999dj} to the $f(T)$ gravity theories~\cite{ferraro2008born}, also see~\cite{jimenez2018born} for further applications.
In $f(T)$ gravity this has also been shown to have the potential to regularise the singularities occurring in black hole spacetimes~\cite{Bohmer:2019vff} and cosmological spacetimes~\cite{fiorini2009type}.

An alternative approach is to view General Relativity as an emergent phenomena, in which case diffeomorphisms may again be broken on small enough scales. This has been explored in~\cite{Anber:2009qp,Bluhm:2014oua,Bluhm:2016dzm}. In some of these scenarios, the symmetry breaking is taken to be explicit, whilst in others it is spontaneous. This bears resemblance to the theories known as `spatially covariant gravity'~\cite{Gao:2014fra,khoury2012spatially}. As the name suggests, these theories are only invariant under spatial diffeomorphisms, with time diffeomorphisms broken.
As is well known, the invariance of General Relativity under coordinate transformations leads to a large redundancy in its description, making the degrees of freedom difficult to isolate. This is also known to make the canonical quantization of GR challenging~\cite{Ashtekar:1974uu}, so breaking this redundancy may be a promising avenue to explore.

There are many more theories that break parts of these symmetries or abandon them altogether. On the other hand, some modified theories can be thought of as `gauge-fixed' by construction, meaning that they appear to break these symmetries but are dynamically equivalent to fully covariant theories~\cite{Gao:2014soa}. Perhaps the simplest example is \textit{unimodular gravity}, which is (classically) equivalent to General Relativity with a gauge-fixing condition~\cite{Padilla:2014yea}. This gauge was actually originally used by Einstein as a computational simplification for his field equations~\cite{einstein1922grundlage}, but has continued to be studied as an ``alternative'' to GR ever since (see~\cite{Carballo-Rubio:2022ofy} for a recent review).

Let us briefly sketch the outline of unimodular gravity. The action is given by 
\begin{equation} \label{unimodular}
S_{\textrm{unimodular}}[g,\lambda] = \frac{1}{2\kappa} \int \Big( \sqrt{-g} R  - \lambda (\sqrt{-g} - \epsilon_0) \Big) d^4 x \, ,
\end{equation}
where $ \epsilon_0 $ is a fixed, non-dynamical scalar density and $\lambda(x)$ is a Lagrange multiplier. The theory is only invariant under transverse (volume preserving) diffeomorphisms\footnote{Transverse diffeomorphisms satisfy $\delta_{\xi} \sqrt{-g} = 0$, where $\delta_{\xi}$ is the infinitesimal diffeomorphism given by the Lie derivative in equation~(\ref{Lie metric}). The equation for $\xi$ is simply $\nabla_{\mu}\xi^{\mu}=0$.}, as opposed to the full group of diffeomorphisms.
The variations with respect to the Lagrange multiplier $\lambda(x)$ give the unimodular constraint $\sqrt{-g} = \epsilon_0$, where the constant $\epsilon_0$ was traditionally chosen such that $g =-1$. The metric variations lead to $G_{\mu \nu} + g_{\mu \nu} \lambda/2 = \kappa T_{\mu \nu}$. 

To eliminate the Lagrange multiplier, we first assume the matter action is a scalar. The covariant conservation of the energy-momentum tensor then follows (see Sec.~\ref{section2.2.2}). The Einstein tensor is identically divergence-free, and so from the field equations we obtain $\nabla_{\mu} \lambda(x) = 0$. It follows that the Lagrange multiplier term must be a constant $\lambda(x)= \lambda_0$, and the Einstein field equations take their standard form with a cosmological constant~(\ref{EFE01}). 

It is often claimed that because $\lambda_0$ enters the field equations as an integration constant, this somehow alleviates the usual issues that plague the cosmological constant. As clarified in the work of Padilla and Saltas~\cite{Padilla:2014yea}, these claims do not hold any weight, because the nature of the cosmological constant problem is related to its radiative instability. With regards to its quantum properties, which \textit{can} differ from GR, we refer to the works~\cite{Smolin:2009ti,Bufalo:2015wda,Padilla:2014yea}. 

The above formulation and action is manifestly non-covariant. Alternatively, we can make use of the so-called Stueckelberg trick\footnote{The procedure essentially involves restoring a gauge symmetry by performing the associated gauge transformation and then promoting those gauge parameters to fields, see~\cite{Ruegg:2003ps,deRham:2014zqa}. This is also known as `parameterising the spacetime coordinates'.} to restore the full symmetry of the theory, leading to a fully covariant treatment. The fixed, non-dynamical term needs to be supplemented by Stueckelberg fields so that it transforms correctly. This is simply the Jacobian factor that appears when applying a coordinate transformation $x^{\mu} \rightarrow \hat{x}^{\mu}(x)$. The Stueckelberg  fields are then introduced via the replacement $\hat{x}^{\mu} \rightarrow \xi^{\mu}(x)$, where $\xi^{\mu}(x) = \{ \xi^{1}(x), \xi^{2}(x),\xi^{3}(x), \xi^{4}(x) \}$.

Following~\cite{Padilla:2014yea,kuchavr1991does}, the unimodular action becomes
\begin{equation} \label{unimodular2}
S_{\textrm{Stueckelberg}}[g,\lambda,\xi] = \frac{1}{2\kappa} \int \Big( \sqrt{-g} R  -  \lambda \big(\sqrt{-g} - \epsilon_0 \big| J^{\mu}{}_{\nu} \big| \big) \Big) d^4 x \, ,
\end{equation}
where the Jacobian matrix is defined in terms of the new Stueckelberg fields $J^{\mu}{}_{\nu} = \partial_{\nu} \xi^{\mu}(x)$. The Stueckelberg fields transform as scalars $\xi^{\mu}(x) \rightarrow \hat{\xi}^{\mu}(\hat{x}) = \xi^{\mu}(x)$ so that the action is invariant under the full group of diffeomorphisms. The action reduces to the original one~(\ref{unimodular}) when the `privileged' unimodular coordinates are used $\xi^{\mu}(x) = x^{\mu}$.

One might then wonder about the dynamics associated with the new fields $\xi^{\mu}$. It turns out that for the unimodular action~(\ref{unimodular2}) the Stueckelberg terms take the form of a total derivative~\cite{kuchavr1991does}
\begin{equation} \label{unimod_relation}
| J^{\mu}{}_{\nu}| = 4! \delta^{[\alpha}_{\mu} \delta^{\beta}_{\nu} \delta^{\gamma}_{\rho} \delta^{\lambda]}_{\sigma} \partial_{\alpha} \xi^{\mu} \partial_{\beta} \xi^{\nu} \partial_{\gamma}\xi^{\rho} \partial_{\lambda} \xi^{\sigma} = 4! \partial_{\alpha}\big( \delta^{[\alpha}_{\mu} \delta^{\beta}_{\nu} \delta^{\gamma}_{\rho} \delta^{\lambda]}_{\sigma} \xi^{\mu} \partial_{\beta} \xi^{\nu} \partial_{\gamma}\xi^{\rho} \partial_{\lambda} \xi^{\sigma} \big) \, ,
\end{equation}
and their variations yield the constraint $\partial_{\mu} \lambda(x) =0$. This equation was originally derived from invariance arguments, so it makes sense that it is now a consequence of the Stueckelberg procedure.
Hence, the dynamics are completely equivalent to those in the previous case. As the field equations are equivalent to GR with a cosmological constant, the number of degrees of freedom should be too. This is indeed confirmed in a Hamiltonian analysis~\cite{Henneaux:1989zc}.

From the relation above~(\ref{unimod_relation}), one can also see that the action~(\ref{unimodular2}) is just a special case of the \textit{Henneaux-Teitelboim} action~\cite{Henneaux:1989zc}. There, the action is the same as in~(\ref{unimodular2}) but with the final term replaced by $\partial_{\mu} \tau^{\mu}$, where $\tau^{\mu}$ is a vector density. The variations lead to the same equations of motion as in the unimodular action~(\ref{unimodular2}), except with the unimodular constraint replaced with $\sqrt{-g} = \partial_{\mu} \tau^{\mu}$.

Another very important application of the Stueckelberg trick is in massive gravity~\cite{deRham:2014zqa,Hinterbichler:2011tt}. Massive gravity can be traced back to the work of Fierz and Pauli~\cite{Fierz:1939ix}, and involves adding a mass term to the action of the spin-2 field propagating on Minkowski space (see equation~(\ref{FP}) for the massless Lagrangian). In recent years, much work has been done to cure the pathologies~\cite{vanDam:1970vg,Zakharov:1970cc,Boulware:1972yco} associated with the massive Feirz-Pauli theory, see~\cite{deRham:2014zqa,Babichev:2013usa} and references therein.

The addition of the mass term to the theory breaks the (linearised) diffeomorphism invariance.
 As pointed out by Hinterbichler~\cite{Hinterbichler:2011tt}, the Stueckelberg trick can be used to make any Lagrangian diffeomorphism invariant. This can be seen in the examples above. In massive gravity, the gauge invariance (under coordinate transformations) is broken, and the Stueckelberg trick is used to restore the full diffeomorphism invariance of the theory.
This will turn out to be surprisingly similar to the covariantisation methods that can be applied to our non-covariant theories introduced in Chapters~\ref{chapter4} and~\ref{chapter5} . 
The similarities between massive gravity and symmetric teleparallel $f(Q)$ gravity have already been pointed out in~\cite{BeltranJimenez:2019tme}. We also note that some of the theoretical issues seem to be shared by both these types of theories, namely, that of strong coupling. These topics, whilst very important, are outside of the scope of this work, so we refer to~\cite{Arkani-Hamed:2002bjr,Deffayet:2005ys} and~\cite{BeltranJimenez:2021auj}  for further details in massive gravity and $f(T)$ gravity respectively. We will again mention these issues where relevant, but will not go into technical details.

The other key theory worth mentioning again is Ho\v{r}ava-Lifshitz gravity~\cite{Horava:2009uw} (or Ho\v{r}ava gravity). There, the breaking of invariances also plays an important, physical role. In Ho\v{r}ava gravity, the full diffeomorphism symmetry is broken down into the subgroups consisting of spatial diffeomorphisms and time reparameterizations. As we mentioned previously, the (power-counting) non-renormalizability of GR can be fixed by including higher derivatives. However, we also mentioned that these higher derivatives come with extra degrees of freedom prone to instabilities and pathologies. In Ho\v{r}ava gravity, by breaking diffeomorphism invariance, higher-order spatial derivatives are able to be introduced while maintaining first-order time derivatives, evading Ostrogradsky's theorem.
Due to this attractive feature, there has been a lot of research in this area, see~\cite{Wang:2017brl,Visser:2011mf} for reviews.

The breaking of diffeomorphism invariance brings the possibility of additional degrees of freedom. These can be made explicit by again restoring the invariance with the Stueckelberg trick. In this case, unlike the unimodular example, there are genuine new propagating degrees of freedom. However, these Stueckelberg fields can become strongly coupled~\cite{Charmousis:2009tc,Blas:2009yd}, an issue we mentioned earlier. The authors of~\cite{Charmousis:2009tc} point out that the strong coupling limit is highly reminiscent of the famous vDVZ discontinuity in massive gravity~\cite{Arkani-Hamed:2002bjr}. Using this language, of broken diffeomorphisms and additional degrees of freedom represented by Stueckelberg fields, will turn out to be useful when we consider the geometric modifications of gravity in Chapter~\ref{chapter5}. We also refer to~\cite{Kimpton:2010xi} for a nice summary of these topics in Ho\v{r}ava gravity and its extensions.

Further insight on the dual interpretations of symmetry-breaking gravitational theories can be gained from the work of Jackiw, where Lorentz and diffeomorphism breaking is studied in Chern-Simons modifications of the Einstein-Hilbert action~\cite{RJackiw_2006,RJackiw:2007br}. There, it is found that the full invariance of the theory is reinstated dynamically as a requirement for consistency. The one-sentence abstract of~\cite{RJackiw:2007br} summarises this succinctly: ``\textit{In a diffeomorphism invariant theory, symmetry breaking may be a mask for coordinate choice.}''

For a review of Lorentz violating theories, as well as tests to search for these violations, see~\cite{Mattingly:2005re}. Other notable theories relating to broken symmetries include Einstein-Aether theories~\cite{Jacobson:2007veq,Jacobson:2000xp}, Bumblebee gravity~\cite{Bluhm:2010mi} and string field theory~\cite{Kostelecky:1988zi}. In Einstein-Aether theory, for example, the theory can be written covariantly but preferred reference frame effects are present. It is known that these types of modifications can lead to MOND-like dynamics in regions of low acceleration~\cite{Milgrom:1983ca}. We will return to the MOND theories in Chapter~\ref{chapter5}, where a direct comparison can be made with our modified theories.

Finally, let us turn to the modified theories that will be the focus of this thesis: we will also be interested in non-covariant theories, but using the Einstein-Hilbert action as our starting point. In Section~\ref{section1.2.1} we noted that the Lovelock invariants can be decomposed into a bulk and boundary term. The simplest case was the Ricci scalar. The associated bulk term has the benefit of being first-order in derivatives but is no longer covariant. The symmetries that this decomposition breaks crucially depend on the way in which it is decomposed; on the one hand, it can break diffeomorphism invariance, and on the other, local Lorentz invariance. 

We use this non-covariant decomposition as the starting point to consider more general modifications. What is quite surprising is that these decomposed bulk and boundary terms are fundamentally related to the non-Riemannian theory of teleparallel gravity. Furthermore, by restoring the invariances of the modified actions, we introduce Stueckelberg fields that have a direct relationship with the teleparallel connections. In fact, we will explicitly show that the Stueckelberg fields act to transform the non-covariant bulk terms into the teleparallel scalars.
A similar result was recently shown by Milgrom in the context of modified symmetric teleparallel gravity and MOND~\cite{Milgrom:2019rtd}. The interconnected nature of all of these theories is quite evident, and we aim to explore these relations throughout this work.

\subsection{Outline}
\label{section1.2.5}

Many of the different routes to modifying gravity end up, in some shape or form, being equivalent -- at least at the classical level. This is because, for the most part, they can be framed as General Relativity coupled to extra fields: `\textit{modifying gravity means changing its degrees of freedom}'~\cite{Hinterbichler:2011tt}.
We saw this to be the case with $f(R)$ gravity, which can be recast as a scalar-tensor theory. Moreover, the additional degrees of freedom introduced in the metric-affine theories can also be seen as new fields; in the special case where they do not couple to matter, they often take the form of auxiliary (non-geometric) fields. A key example is in Einstein-Cartan theory, which Nester showed to be dynamically equivalent to standard Einstein gravity coupled to additional sources~\cite{Nester:1977zz}. This is also well explained in the work by Sotiriou~\cite{Sotiriou:2014yhm}, where they show that breaking assumptions means more degrees of freedom, which can often be written in terms of additional scalar fields. 

We see the same to be true for the symmetry-breaking theories. For example, Ho\v{r}ava gravity can be written in a fully covariant manner with additional dynamical auxiliary fields~\cite{Jacobson:2010mx}, linking it to Einstein-Aether theories. Manifestly non-covariant theories are often just gauge-fixed versions of fully invariant theories, perhaps in a `special' choice of frame or coordinates. 
In this thesis, we will show the same to be true for the non-covariant modifications such as $f(\ourG)$ gravity and $f(\mathfrak{G})$ gravity, where $\ourG$ and $\mathfrak{G}$ are different bulk decompositions of the Ricci scalar. These will be compared with the modified teleparallel theories. Before studying these modifications, we return to General Relativity and cover its geometric foundations.

In Chapter~\ref{chapter2} we give a thorough review of the background mathematics necessary to study these non-Riemannian theories of gravity. In particular, the metric-affine structure and the tetrad formalism are introduced. Then, the principles and postulates of General Relativity are examined, showing how one arrives at GR from physical considerations. We then cover the Lagrangian formulation with the Einstein-Hilbert action in both the metric and tetrad formalism and derive the equations of motion. As an aside, we motivate the choice of action from the perspective of spin-2 gravitons. To conclude the chapter, we discuss isometries and symmetries, including the important topics of diffeomorphism invariance and local Lorentz invariance.  

In Chapter~\ref{chapter3} we look at the metric-affine theories of gravitation. This begins with the Palatini formalism, where the metric and connection are treated as independent but matter only couples to the metric. We then study the fully metric-affine theories, introducing the matter currents that couple to the connection. A key example is Einstein-Cartan theory, studied in Sec.~\ref{section3.2.1}. Gauge theoretic motivations are also briefly discussed. Finally, we look at  the teleparallel theories of gravity, studying their foundations and recent developments. 

Chapter~\ref{chapter4} focuses on the different decompositions of the Einstein-Hilbert action into its bulk and boundary parts, which serve as the starting point for the modified theories of the following chapters. This decomposition is studied in both the metric and the tetrad frameworks, leading to distinctly different results. The actions based on these decompositions are shown to lead to the standard Einstein field equations, and we look at the role of diffeomorphisms and local Lorentz transformations for these theories. We then make a crucial comparison with the teleparallel theories in Sec.~\ref{section4.3}, making use of the Stueckelberg formalism previously discussed. In the last section we look at the decomposition in the non-Riemannian setting.

The modified theories of gravity are introduced and studied in Chapter~\ref{chapter5}. These are known as $f(\ourG,\ourB)$ and $f(\bar{\ourG}, \bar{\ourB})$ gravity, the former being in the Levi-Civita framework and the latter being its metric-affine generalisation~\cite{Boehmer:2021aji,Boehmer:2023fyl}. In Sec.~\ref{section5.2} we compare our theories to the modified teleparallel theories. In particular, we wish to draw attention to the diagram in Fig.~\ref{fig:unified}, where we show how all of these modified theories relate to one another. This serves as a basis for a unified approach.

In Chapter~\ref{chapter6} we study the cosmological implications of our modified theories, as well as the modified teleparallel theories $f(T)$ and $f(Q)$ gravity. As an introductory example, we study the Born-Infeld cosmologies and show how the initial big-bang singularity of GR is replaced by an inflationary de-Sitter stage.
In Sec.~\ref{section6.2} a dynamical systems approach is used to study the cosmologies of all of these theories, based on our works~\cite{Boehmer:2022wln,Boehmer:2023knj}. Specifically, we find the necessary conditions for any second-order modified gravity model to possess stable de-Sitter points, giving rise to the accelerated expansion of the universe. We then look at propagating torsion in the metric-affine theories, again finding an early inflationary period replacing the $t \rightarrow 0$ cosmological singularity of standard GR. 

Finally, in Chapter~\ref{chapter7} we summarise the results of this thesis. Extensions based on this work are discussed, such as a new two-dimensional limit of General Relativity, which we proposed in~\cite{Boehmer:2023lpb}. Moreover, we give an outline of possible future studies and investigations that could be conducted, stating their relevance for the wider gravity community.

\begin{savequote}[70mm]
That the knowledge at which geometry aims is knowledge of the eternal, and
not of aught perishing and transient.
\qauthor{`The Republic, Book VII' \\
Plato}
\end{savequote}
\chapter{Geometric foundations of gravity}
\label{chapter2}

The theory of General Relativity is best described using the language of differential geometry. In General Relativity, the dynamical equations of motion are most easily written down in terms of geometric quantities such as the curvature of spacetime. The key dynamical variable is the spacetime metric itself, the geometric quantity used to measure distances and describe causal structure. This is in contrast to most other field theories, where the background spacetime is fixed, usually to be the Minkowski metric of Special Relativity. This can be understood by the fact that the gravitational interaction is best interpreted as a geometric phenomenon and a consequence of geometry itself.

Most alternative theories of gravity can also be described in the same geometric language, though often this structure is much richer than the Riemannian setting of GR. As such, it will be beneficial to provide a short and concise overview of the mathematical language necessary to study the modified theories covered in this thesis. The mathematical formalisms will mainly be introduced using tensor index notation, though some reference to the more compact differential form notation will be shown. The intuition sometimes gained from the differential form calculus notation should be kept in mind, and appropriate references will be given throughout. It is also worth noting that the definitions here will only be as rigorous as required by a physicist performing coordinate-based calculations in theories of gravity.

The main standard sources, and the ones used for this section, are the following textbooks~\cite{JS1954,Lee00,Wald:1984,Misner:1973prb,Frankel:1997ec}. Firstly, the metrical structure on the manifold will be introduced, followed by the affine structure endowed by an independent affine connection. (The reason for this ordering is that metrical structure is more commonly known than affine geometry, despite the fact that the latter is more fundamental and does not require a metric.)
The tetrad formalism will also be introduced, which is needed for studying gravity in non-coordinate bases. This is in fact
essential for coupling spinors to gravity, as well as understanding the gauge structure of gravity and the teleparallel theories.

Next, the fundamental physical principles and assumptions of General Relativity are studied, before reviewing the Einstein-Hilbert action and field equations in both the metric and tetrad formalisms. Finally, topics relating to symmetries in gravitational theories, such as covariance, diffeomorphisms, local Lorentz invariance, isometries and projective transformations are briefly reviewed. All of these topics will be used in the following chapters of this thesis, so it is useful to introduce them here in a pedagogical manner with references for further reading.

\section{Differential geometry and mathematical preliminaries} 
\label{section2.1}
General Relativity is a geometric theory; it can be formulated in terms of geometric, invariant objects on a smooth differentiable manifold $\mathcal{M}$, which we will later take to represent spacetime. The manifold structure allows us to construct local coordinate systems and apply differential calculus at points on the manifold.
The types of manifold of interest are called \textit{Lorentzian manifolds}\footnote{A Lorentzian manifold is the set $(\mathcal{M}, g)$ where $g$ is a metric of Lorentzian signature $(1,n-1)$ with $n$ the number of dimensions. For a more thorough definition of differentiable manifolds, we refer the reader to~\cite{Lee00,Frankel:1997ec}.}, which locally look like flat Minkowski space. In General Relativity the dimension of the manifold is fixed to be four $n=4$, but for generality we work in an arbitrary number of dimensions. 

The geometric objects of importance are tensors, which are multilinear maps that live in the tensor product space built from the tangent space and cotangent space at each point $p$ on the manifold. For a rank $(k,l)$ tensor, this space is
 \begin{equation}
 T_{p}^{(k,l)}(\mathcal{M}) = \overbrace{T_p(\mathcal{M}) \otimes ...  \otimes T_p(\mathcal{M})}^{k \ \textrm{copies} } \otimes \overbrace{T^{*}_p(\mathcal{M}) \otimes ...  \otimes T^{*}_p(\mathcal{M})}^{l  \ \textrm{copies}} \ ,
 \end{equation}
 with $T_{p}(\mathcal{M})$ and $T^{*}_{p}(\mathcal{M})$ the tangent and cotangent space at $p$ respectively. The set of tensor product spaces at all points on the manifold $p \in \mathcal{M}$  is the tensor bundle $T^{(k,l)}(\mathcal{M})$, from which we get tensor fields as sections of the tensor bundle. Hence we see that the tensor fields are indeed geometric objects and do not depend on coordinates. (For brevity, we will often refer to tensor fields simply as tensors, and the same for scalar and vector fields.)

However, as we like to work in coordinates when performing calculations, it is useful to have a coordinate expression for these objects too. A rank $(k,l)$ tensor can be written in local coordinates $x^{\mu}$ as
\begin{equation} \label{Tensor}
T^{(k,l)} = T^{\mu_1...\mu_k}{}_{\nu_1...\nu_l} \frac{\partial}{\partial x^{\mu_1}} \otimes ... \otimes \frac{\partial}{\partial x^{\mu_k}} \otimes dx^{\nu_{1}} \otimes ... \otimes  dx^{\nu_{l}} \ ,
\end{equation}
where $\partial_{\mu} = \partial / \partial x^{\mu}$ and $dx^{\mu}$ are the vector and one-form \textit{coordinate bases} respectively. We will initially work in a coordinate (or holonomic) basis $\mathbf{e}_{\mu} = \partial_{\mu}$ such that the Lie bracket\footnote{The Lie bracket is the commutator of vector fields $U,V \in T(\mathcal{M})$, defined as $[U,V] f := U(V(f)) - V(Y(f))$ where $f$ is a smooth (scalar) function $f \in C^{\infty} (\mathcal{M})$. The expression in local coordinates is $[U^{\mu} \partial_{\mu},V^{\mu} \partial_{\mu}]f = (U^{\mu} \partial_{\mu} V^{\nu} -V^{\mu} \partial_{\mu} U^{\nu})\partial_{\nu}f$, hence the coordinate bases have vanishing Lie bracket as partial derivatives commute. This result is sometimes known as Frobenius’ theorem. \label{footnote:Lie}} vanishes $[\mathbf{e}_{\mu}
,\mathbf{e}_{\mu}]=0$. Later, when introducing the tetrad formulation, we will make use of anholonomic bases. We will stick to the coordinate basis in the equations that follow, dropping terms of anholonomy.

An important property of geometric objects is how their components change when moving from one set of coordinates $\{x^{\mu}\}$ to another $\{ y^{\mu} \}$.
The components of tensors satisfy the following transformation rule under a change of coordinates $x^{\mu} \rightarrow \hat{x}^{\mu}(x)$
\begin{align} \label{Tensor transformation}
 T^{\mu_1...\mu_k}{}_{\nu_1...\nu_l}(x) &\rightarrow \hat{T}^{\mu_1...\mu_k}{}_{\nu_1...\nu_l}(\hat{x}) \nonumber \\  &= \frac{\partial \hat{x}^{\mu_1}}{\partial x^{\rho_1}} ... \frac{\partial \hat{x}^{\mu_k}}{\partial x^{\rho_k}} \, \frac{\partial x^{\sigma_1} }{\partial \hat{x}^{\nu_1}} ... \frac{\partial x^{\sigma_l} }{\partial \hat{x}^{\nu_l}} T^{\rho_1 ...\rho_k}{}_{\sigma_1 ... \sigma_l}(x) \ .
\end{align}
Along with the corresponding transformation rules for the bases, the object $T^{(k,l)}$ is invariant.

Notice that we will often refer to geometric objects solely by their components in local coordinates. For example, we would say that under a change of coordinates (the components of) a tensor transforms \textit{covariantly} according to (\ref{Tensor transformation}). We will also often drop the implicit tensor products in terms like (\ref{Tensor}), as well as the dependence on spacetime coordinates.

\subsubsection{Metric tensor}
The metric $g$ is a symmetric $(0,2)$ tensor that defines an inner product on the tangent space $T_p(\mathcal{M})$ at each point on the manifold $p \in \mathcal{M}$. Given two vectors $U,V \in T_{p} (\mathcal{M})$ the metric is the map \begin{equation}
g (U,V) = g (V,U) \in \mathbb{R} \, .
\end{equation}
This can be thought of as the generalisation of the dot product in Euclidean space and gives a notion of distances on the manifold.
 The metric is also known as the \textit{spacetime interval}, which we write in local coordinates as 
\begin{equation} \label{line element}
{ds}^2 =  g_{\mu \nu}(x) dx^{\mu} \otimes dx^{\nu} \ .
\end{equation}
It describes the local causal structure of spacetime.

The components of the inverse metric are $g^{\mu \nu}(x)$ where $g^{\mu \nu}g_{\mu \lambda} = \delta^{\nu}_{\lambda}$ is the Kronecker delta symbol. Using the metric and its inverse we can raise and lower the indices of 
other objects, e.g. $g_{\mu \nu} v^{\mu} = v_{\nu}$ and  $g^{\mu \nu} v_{\mu} = v^{\nu}$. In other words, the metric acts as an isomorphism between $T_{p} (\mathcal{M})$ and its dual $T^*_{p} (\mathcal{M})$.

Treating the components $g_{\mu \nu}$ as a matrix, the determinant is denoted by $g = \textrm{det}(g_{\mu \nu})$.
We will assume the metric has a Lorentzian signature, meaning it has an index of $(1,n-1)$, where the first item is the number of negative eigenvalues and the second is the number of positive eigenvalues. A spacetime vector $V$ is classified as \textit{spacelike} if $g_{\mu \nu} V^{\mu} V^{\nu} > 0$, \textit{null} if $g_{\mu \nu} V^{\mu} V^{\nu} = 0$ and 
\textit{timelike} if $g_{\mu \nu} V^{\mu} V^{\nu} < 0$.
The physical significance of the metric will be covered in more detail in Sec.~\ref{section2.2.1}.

\subsubsection{Differential forms}
It is worth briefly introducing the notion of \textit{differential forms}. The simplest case are the zero-forms, which are just scalar functions. The next simplest example are the one-forms, which are dual to vectors and live in the dual space or cotangent space $T^*_{p} (\mathcal{M})$ at each point $p$. Taking the collection of each cotangent space at every point on the manifold leads to the cotangent bundle $T^*(\mathcal{M})$, with one-form (fields) being sections of the bundle. Again, we will drop the superfluous use of the word `fields' when describing differential forms.

An example of a differential form already introduced is the coordinate basis one-form $dx^{\mu}$. Taking the coordinate basis as an example, the inner product of a one-form $\omega = \omega_{\mu} dx^{\mu}$ and a vector $v = v^{\mu} \partial_{\mu}$ is 
\begin{equation} \label{innerproduct}
\langle \omega , v \rangle = \omega_{\mu} v^{\nu} \langle dx^{\mu} , \partial_{\nu} \rangle = \omega_{\mu} v^{\nu} \delta_{\nu}^{\mu} =  \omega_{\mu} v^{\mu} \, .
\end{equation}
In other words, the inner product is the map $\langle . \rangle : T^*_{p} (\mathcal{M}) \times T_{p} (\mathcal{M}) \rightarrow \mathbb{R}$.

A differential $k$-form is a totally antisymmetric rank $(0,k)$ tensor 
\begin{equation}
\omega_{\mu_{1} \mu_{2} ...\mu_{k}} = \omega_{[\mu_{1} \mu_{2}...\mu_{k}]} \, .
\end{equation}
 The space of all $k$-forms on the $n$-dimensional manifold is $\Lambda^k(\mathcal{M})$. The set  $\Lambda^k(\mathcal{M})$ is empty if $k > n$, and elements of $\Lambda^n (\mathcal{M})$ are known as top-forms.
 
  Two operations that are useful to introduce for differential forms are the wedge product and the exterior derivative. The wedge product of a $p$-form $\omega$ and $q$-form $\eta$ is defined as
 \begin{equation}
( \omega \wedge \eta )_{\mu_{1}...\mu_{p} \nu_{1}...\nu_{q} } = \frac{(p+q)!}{p! q!} \omega_{[\mu_1...\mu_p} \eta_{\nu_1...\nu_q]} \, ,
 \end{equation}
 where the resulting object is a $(p+q)$-form. This is simply the antisymmetrised tensor product.
 An immediate result is that 
 \begin{equation} 
 \omega \wedge \eta  = (-1)^{pq} \eta \wedge \omega \, ,
 \end{equation}
 and that an odd-degree form wedged with itself is zero. An arbitrary $p$-form $\omega$ can be written in the coordinate basis as
 \begin{equation}\label{pform}
 \omega = \frac{1}{p!} \omega_{\mu_{1}...\mu_{p}} dx^{\mu_{1}} \wedge ... \wedge dx^{\mu_{p}} \, .
 \end{equation}
 
 The second operation is the exterior derivative $d$, which maps $p$-forms to $(p+1)$-forms $d : \Lambda ^{p} (\mathcal{M}) \rightarrow \Lambda ^{p+1} (\mathcal{M})$. For the $p$-form $\omega$ this is
  \begin{equation} \label{exterior}
(d\omega)_{\mu_{1}...\mu_{p+1}} =(p+1) \partial_{[\mu_{1}} w_{\mu_{2}...\mu_{p+1}]} \, ,
 \end{equation}
or in the coordinate basis~(\ref{pform}) equivalently 
\begin{equation} \label{exterior2}
d \omega =  \frac{1}{p!} \partial_{\nu}( \omega_{\mu_{1}...\mu_{p}}) dx^{\nu} \wedge dx^{\mu_{1}} \wedge ... \wedge dx^{\mu_{p}} \, .
\end{equation}

An important result known as the Poincaré lemma is that repeating the exterior derivative on any form twice always gives zero, $d^2=0$. A differential form satisfying $d \omega = 0$ is called \textit{closed}, whilst a form is called \textit{exact} if $\omega = d \eta$ for some form $\eta$. These concepts will be revisited at the end of the following section.

\subsection{Affine structure}
\label{section2.1.1}
A key geometric object is the linear or \textit{affine connection} $\bar{\Gamma}$, which connects nearby tangent spaces on the manifold. More technically, the connection is a map $\bar{\Gamma}: \mathcal{X}(\mathcal{M}) \times \mathcal{X}(\mathcal{M}) \rightarrow \mathcal{X}(\mathcal{M})$, where $\mathcal{X}(\mathcal{M})$ is the space of vector fields on $\mathcal{M}$. In other words, it takes two vectors as an input and gives a vector as an output. By specifying a connection we can \textit{parallel transport} vectors at different points on the manifold in order to compare them and give us a well-defined notion of taking derivatives of vectors. This introduces the concept of the \textit{covariant derivative} $\bar{\nabla}$ of a given connection
\begin{equation}
\bar{\nabla}_{\mu} \mathbf{e}_{\nu} := \bar{\nabla}_{\mathbf{e}_{\mu}} \mathbf{e}_{\nu} = \mathbf{e}_{\lambda} \bar{\Gamma}^{\lambda}_{\mu \nu} \, ,
\end{equation}
where $\bar{\Gamma}^{\nu}_{\mu \lambda}(x)$ are the components of the connection in the coordinate basis $\mathbf{e}_{\mu} = \partial_{\mu}$.

This covariant derivative acts on the components of a vector and one-form by
\begin{equation} \label{Covariant derivative}
\bar{\nabla}_{\mu}v^{\nu} = \partial_{\mu} v^{\nu} + \bar{\Gamma}^{\nu}_{\mu \lambda} v^{\lambda} \quad \quad , \quad \quad \bar{\nabla}_{\mu}v_{\nu} = \partial_{\mu} v_{\nu} - \bar{\Gamma}^{\lambda}_{\mu \nu} v_{\lambda} \, .
\end{equation}
 The covariant derivative can be generalised for arbitrary rank tensors and it maps rank $(k,l)$ tensors to rank $(k,l+1)$ tensors
\begin{multline} \label{Covariant derivative2}
\bar{\nabla}_{\lambda}  T^{\mu_1...\mu_k}{}_{\nu_1...\nu_l}  = \partial_{\lambda} T^{\mu_1...\mu_k}{}_{\nu_1...\nu_l} 
+ \sum^{k}_{i=1} \bar{\Gamma}^{\mu_{i}}_{\lambda \rho} T^{...\mu_{i-1} \rho \mu_{i+1} ...}{}_{\nu_1...\nu_l} \\
- \sum^{l}_{i=1} \bar{\Gamma}^{\rho}_{\lambda \nu_{i}} T^{\mu_1...\mu_k}{}_{...\nu_{i-1} \rho \nu_{i+1} ...}
 \, .
\end{multline}
It is linear and obeys the Leibniz rule.

The affine connection, however, is not a tensor.
Using the formula for the covariant derivative~(\ref{Covariant derivative}) and the knowledge of how tensors transform, we can find the transformation rule for the connection coefficients under a change of coordinates $x^{\mu} \rightarrow \hat{x}^{\mu}(x)$
\begin{equation} \label{Christoffel transformation}
 \bar{\Gamma}^{\gamma}_{\mu \nu} (x) \rightarrow \hat{\bar{\Gamma}}^{\gamma}_{\mu \nu}(\hat{x}) = \frac{\partial x^{\alpha}}{\partial \hat{x}^{\mu}} \frac{\partial x^{\beta}}{\partial \hat{x}^{\nu}} \frac{\partial \hat{x}^{\gamma}}{\partial x^{\sigma}} \bar{\Gamma}^{\sigma}_{\alpha \beta}(x) + \frac{\partial^2 x^{\sigma}}{\partial \hat{x}^{\mu} \partial \hat{x}^{\nu} } \frac{\partial \hat{x}^{\gamma}}{\partial x^{\sigma}} \ .
 \end{equation}

From the derivative operator $\bar{\nabla}$ we can define the notion of \textit{parallel transport}. A vector $v^{\mu}$ is parallel transported along a curve $\gamma:  I \rightarrow \mathcal{M}$ parameterised by $\lambda$ in the interval $I \subset \mathbb{R}$, with tangent vector $u^{\mu} = d \gamma(\lambda) / d \lambda$, if it satisfies the following equation
\begin{equation}\label{Parallel}
\bar{\nabla}_{u} v^{\mu} := u^{\nu} \bar{\nabla}_{\nu} v^{\mu}= 0 \, ,
\end{equation}
along the curve $\lambda \in I$. An \textit{autoparallel} curve is one where its tangent vector parallel transported with respect to itself vanishes $\bar{\nabla}_{u} u^{\mu} = 0$. This can be written for the curve $x^{\mu}(\tau)$ as
\begin{equation}\label{Autoparallel}
\frac{d^2 x^{\mu}}{ d \tau^2} + \bar{\Gamma}^{\mu}_{\nu \rho} \frac{d x^{\nu}}{d \tau} \frac{d x^{\rho}}{d \tau} =0 \,.
\end{equation}
This gives a sense of the straightest possible line through space, and curves satisfying this equation are called autoparallel \textit{geodesics}. The topic of geodesics will be revisited when studying the Levi-Civita connection and General Relativity.

\subsubsection{Torsion, non-metricity and curvature}
The introduction of the affine connection to our geometric framework, now given by the set $(\mathcal{M}, g, \bar{\Gamma})$, naturally gives rise to a number of other important invariant objects. The first is torsion
\begin{equation}\label{Torsion v0}
T(U,V) := \bar{\nabla}_{U} V - \bar{\nabla}_{V} U - [U,V] \, ,
\end{equation}
where $U$ and $V$ are two vector fields $U, V \in T(\mathcal{M})$. In a holonomic basis
the components of torsion can be written as
\begin{equation} \label{Torsion}
T^{\lambda}{}_{\mu \nu} := 2 \bar{\Gamma}^{\lambda}_{[\mu \nu]} = \bar{\Gamma}^{\lambda}_{\mu \nu} - \bar{\Gamma}^{\lambda}_{\nu \mu} \, ,
\end{equation}
which is simply the antisymmetric part of the connection. From~(\ref{Torsion v0}) we see that for scalar functions, torsion measures the amount by which the covariant derivative fails to commute
\begin{equation} 
[\bar{\nabla}_{\mu}, \bar{\nabla}_{\nu}]  f=  -T^{\lambda}{}_{\mu \nu} \bar{\nabla}_{\lambda} f \, .
\end{equation}
Note that torsion is a tensor, which can be seen by looking at the antisymmetric part of the connection transformation rule and noting that the inhomogeneous part vanishes.

\begin{figure}[!hbt]
\centering

\begin{tikzpicture}[line cap=round, scale=1.5]
  \coordinate (O) at (0,0);
  \coordinate (A) at ( 3,0.1);
  \coordinate (B) at ( 0.18,3);
  \coordinate (C) at (3.2,2.7);
  \coordinate (D) at (2.9,3.1);
   \coordinate (O1) at (-0.4,-0.1);
   \coordinate (O2) at (+0.05,-0.5);
    \coordinate (A1) at (3.5,0.0);
    \coordinate (A2) at (2.85,-0.5);
    \coordinate (B1) at ( 0.,3.45);
    \coordinate (B2) at ( -0.3 ,2.9);
  
  \draw[vector,black!85] (B) -- (D) node[midway,above] {$\vb{U+\delta U}$};
  \draw[vector,black!85] (A) -- (C) node[midway,below right=-2] {$\vb{V + \delta V}$};
  
  \filldraw [black] (O) circle (1pt)node[below left] {$p$};
  \draw[thin,dashed,black]     (O1) to[out=40,in=-1] (O) ;
  \draw[thin,dashed,black]     (O) to[out=-90,in=-50] (O2);
  \draw[thin,dashed,black]     (A) to[out=-5,in=10]  (A1);
  \draw[thin,dashed,black]     (A) to[out=-100,in=70]  (A2);
   \draw[thin,dashed,black]     (B) to[out=-80,in=70]  (B1);
    \draw[thin,dashed,black]     (B2) to[out=40,in=0]  (B);
    
  \draw[vector,black] (O) -- (A) node[midway,below] {$\vb{U}$};
  \draw[vector,black] (O) -- (B) node[midway,above left=-2] {$\vb{V}$};
  \draw[vector,dashed,red] (C) -- (D) node[ right=5,black] {$T(\vb{U},\vb{V})$};
\end{tikzpicture}
\caption{Torsion as the closure of infinitesimal parallelograms.}
\label{fig.tor}
\end{figure}
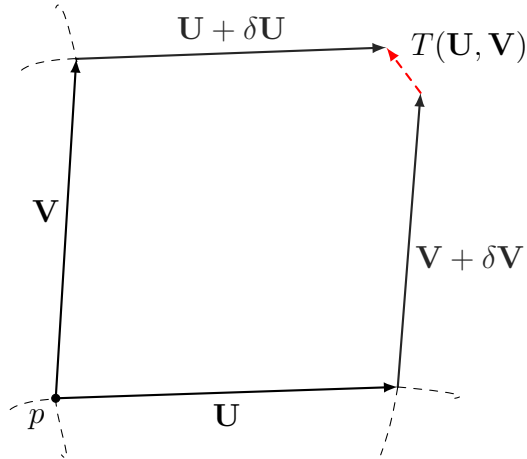

Torsion measures the twisting of tangent spaces upon parallel transport along a curve. This can be viewed geometrically as the closure of the quadrilateral formed from two infinitesimal vectors parallel transported with respect to each other~\cite{Misner:1973prb,Blagojevic:2002du,JS1954}. This is illustrated in Fig.~\ref{fig.tor}, where $U$ and $V$ refer to infinitesimal vectors at $p$ on a space with torsion. Note that the picture in Fig.~\ref{fig.tor} is the same as the visualisation of the commutator (of basis vectors) as the closure of quadrilaterals in the absence of geometric quantities, which is immediately obvious from equation~(\ref{Torsion v0}), again see~\cite{Misner:1973prb}. For more on the geometric interpretation of torsion, we refer to~\cite{hehl2007elie}.

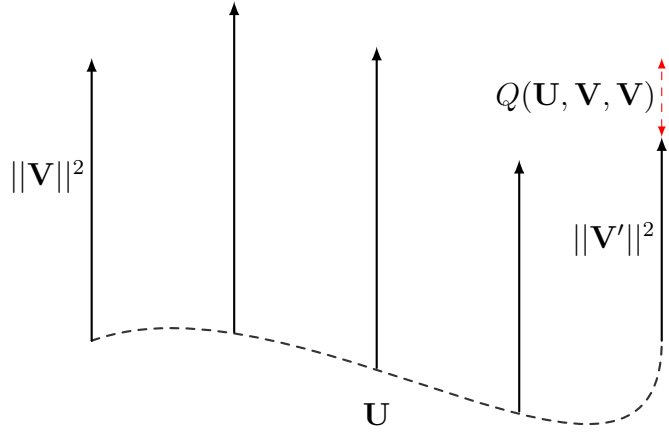
\begin{figure}[!hbt]
\centering
\begin{tikzpicture}[line cap=round, scale=1.5]
  \coordinate (O) at (0,0);
  \coordinate (O1) at (0,2.5);
  \coordinate (A) at (5,0);
  \coordinate (A1) at (5,1.8);
  \coordinate (A2) at (5,2.5);
  
   \coordinate (B) at (1.25,0.075);
  \coordinate (B1) at (1.25,3);
   \coordinate (C) at (2.5,-0.24);
  \coordinate (C1) at (2.5,2.6);
   \coordinate (D) at (3.75,-0.625);
  \coordinate (D1) at (3.75,1.6);
  \coordinate (U) at (2.5,-0.8);

  \draw[dashed,black!80,thick] (O) to[out=20,in=-90] (A);
         \draw[<->,red,dashed] (A1) -- (A2)  node[midway, left=-2,black] {$ Q(\vb{U},\vb{V}, \vb{V})$} ;
    \draw[vector,black] (O) -- (O1) node[midway,above left=-2] {$||\vb{V}||^2$};
     \draw[vector,black] (A) -- (A1) node[midway, left=-2] {$||\vb{V}'||^2$};
     \draw[vector,black] (B) -- (B1);
     \draw[vector,black] (C) -- (C1);
      \draw[vector,black] (D) -- (D1);
     \node (U) at (2.5,-0.65) {$\vb{U}$};
\end{tikzpicture}
\vspace{-7mm}
\caption{Lengths of parallel transported vectors changing due to non-metricity.}
\label{fig.met}
\end{figure}

The second key object in our metric-affine framework is the tensor of non-metricity
\begin{align}
Q(U,V,X) :=  - (\bar{\nabla}_U g)(V,X) \, ,
\end{align}
with $U, V, X \in T(\mathcal{M})$.
In local coordinates the components are
\begin{equation} \label{nonmetricity}
Q_{\lambda \mu \nu} := - \bar{\nabla}_{\lambda}  g_{\mu \nu} \, ,
\end{equation}
which is simply the covariant derivative of the metric tensor.
A geometric interpretation can be given to this tensor, which is the variation in lengths of parallel transported vectors. This can be easily seen using the definition of parallel transport~(\ref{Parallel}) applied to the norm of a vector $||v|| = \sqrt{v^{\mu} v_{\mu}}$ along with the properties of the covariant derivative
\begin{align}
\bar{\nabla}_{u} (||v||^2) =  \bar{\nabla}_{u} (v^{\mu} v^{\nu} g_{\mu \nu}) &= 2v_{\nu} \bar{\nabla}_{u} v^{\nu} + v^{\mu} v^{\nu} \frac{d \gamma^{\rho}}{d \lambda} \bar{\nabla}_{\rho} g_{\mu \nu} \nonumber \\
&=  - v^{\mu} v^{\nu} \frac{d \gamma^{\rho}}{d \lambda} Q_{\rho \mu \nu} \neq 0 \, .
\end{align}
The non-metricity tensor therefore measures the amount by which the magnitude of inner products of parallel transported vectors changes along a curve. A simple illustration is given in Fig.~\ref{fig.met}, where the length of the vector $V$ parallel transported along a curve with tangent $U$ changes proportional to the non-metricity tensor. For detailed figures of all the geometric quantities in this section, we also refer to the textbook~\cite{hurley2000geometry}.
Note that non-metricity requires the existence of a metric tensor, whereas torsion and curvature (introduced below) do not.

Finally, we have the Riemann curvature tensor, defined in terms of the commutator of second covariant derivatives
\begin{equation}
R(U,V)X := \bar{\nabla}_{U} \bar{\nabla}_{V} X - \bar{\nabla}_{V} \bar{\nabla}_{U} X - \bar{\nabla}_{[U,V]} X \, ,
\end{equation}
with $U,X,V \in T(\mathcal{M})$.
Using the definition of torsion we can write this in local coordinates as
\begin{equation} \label{Riemann covar}
\big[ \bar{\nabla}_{\mu} , \bar{\nabla}_{\nu} \big]  v^{\lambda}= \bar{R}_{\mu \nu \gamma}{}^{\lambda} v^{\gamma}  - T^{\rho}{}_{\mu \nu} \bar{\nabla}_{\rho} v^{\lambda} \ ,
\end{equation}
where the curvature tensor has components
\begin{equation} \label{Riemann tensor}
\bar{R}_{\mu \nu \gamma}{}^{\lambda} := 2 \bar{\Gamma}^{\lambda}_{[\mu| \rho|} \bar{\Gamma}^{\rho}_{\nu] \gamma} + 2 \partial_{[\mu} \bar{\Gamma}^{\lambda}_{\nu] \gamma} \ . 
\end{equation}
Equation~(\ref{Riemann covar}) can be generalised to tensors of arbitrary rank $(k,l)$ in a straightforward way
\begin{multline} \label{Riemann covar2}
\big[ \bar{\nabla}_{\mu} , \bar{\nabla}_{\nu} \big] T^{\lambda_1...\lambda_k}{}_{\gamma_1...\gamma_l} = \sum^k_{i=1} \bar{R}_{\mu \nu \rho}{}^{\lambda_i}  T^{...\lambda_{i-1} \rho \lambda_{i+1} ...}{}_{\gamma_1...\gamma_l} \\ -
\sum^l_{i=1} \bar{R}_{\mu \nu \gamma_i}{}^{\rho}  T^{\lambda_1...\lambda_k}{}_{...\gamma_{-1} \rho \gamma_{i+1}...} 
-T^{\rho}{}_{\mu \nu} \bar{\nabla}_{\rho}  T^{\lambda_1...\lambda_k}{}_{\gamma_1...\gamma_l} \, .
\end{multline}
Notice that, unlike torsion and non-metricity, the Riemann tensor contains derivatives of the affine connection. This is because the Riemann tensor measures the integrability of the connection, and a connection is said to be integrable if $\bar{R}_{\mu \nu \gamma}{}^{\lambda}$ vanishes at all points~\cite{JS1954}.

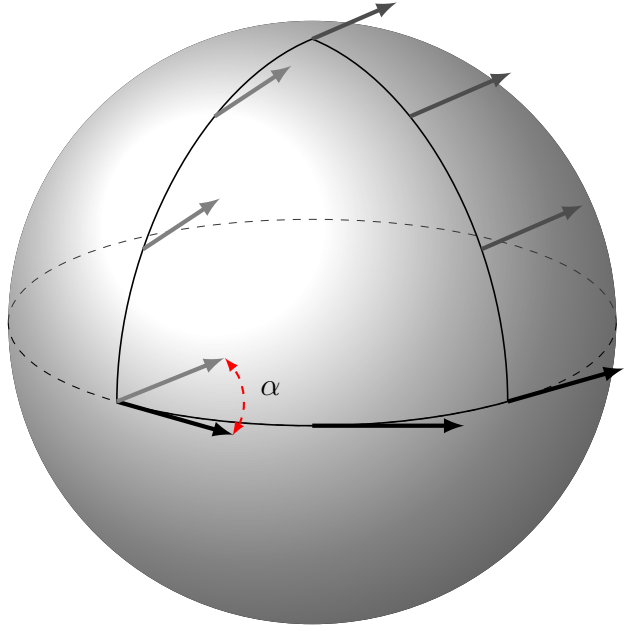
\begin{figure}[!hbt]
\centering
\begin{tikzpicture}[scale=1,every node/.style={minimum size=1cm}]

\def\R{4} 

\def\angEl{20} 
\def\angAz{-100} 
\def\angPhiOne{-130} 
\def\angPhiTwo{-50} 
\def\angBeta{30} 

\fill[ball color=black!0] (0,0) circle (\R); 

\DrawLatitudeCircle[\R]{0} 

\DrawLatitudeCirclered[\R]{0}
\DrawLongitudeCircleredL[\R]{180+\angPhiOne}
\DrawLongitudeCircleredR[\R]{180+\angPhiTwo}

\draw [red,thick,<->,dashed]   (-1.05,-1.5) to[out=60,in=-50] (-1.15,-0.47) node[midway,left ,below=10,black] {$\alpha \quad \quad$ \ \  };

\end{tikzpicture}

\caption{Parallel transported vectors changing around a closed loop by an angle $\alpha$ due to curvature.}
\label{fig.riem1}
\end{figure}

Geometrically, the Riemann tensor measures the curvature of spacetime. Where torsion describes the twisting of tangent spaces parallel transported along a curve, curvature describes how tangent spaces \textit{roll} along a curve. To see the effects of curvature alone, we will temporarily ignore torsion and non-metricity and use the un-barred notation $R_{\mu \nu \gamma}{}^{\lambda}$ to refer to the torsion and non-metricity-free Riemann tensor.
An illustration of the change in angle of a vector parallel transported around a closed loop due to curvature is given in Fig.~\ref{fig.riem1}. This is quite intuitive when thinking about the \textit{extrinsic} curvature of a sphere embedded in a higher-dimensional space, where the same illustration can be drawn.

To be more concrete, consider the infinitesimal parallelogram defined by two (infinitesimal) vectors $U$ and $V$ with origin at the point $p$. Parallel transport a vector $X$ from $p$ in the direction of $U$ then $V$. Next, parallel transport the vector in the reverse direction of $U$ followed by the reverse direction of $V$. The change in the vector $X$ after being parallel transported around this closed loop is given by
\begin{equation}
\delta X^{\mu} = X^{\mu}|_{\textrm{end}} -  X^{\mu}|_{\textrm{start}} = R_{\nu \lambda \gamma}{}^{\mu} U^{\nu} V^{\lambda} X^{\gamma} \, .
\end{equation}
A formal derivation can be found in most textbooks, see for example \cite{Wald:1984}~and~\cite{Misner:1973prb}.

The Riemann tensor is also related to tidal forces. This can be understood by studying the relative deviation between neighbouring geodesics. 
First, let us define a one-parameter family of geodesics $\gamma_s (t)$ with $\Sigma$ a submanifold spanned by the curves of $\gamma_s(t)$. Choosing coordinates $s$ and $t$ with $x^{\mu}(s,t) \in \Sigma$, and defining the tangent vector field $T^{\mu} = \frac{\partial x^{\mu}}{\partial t}$ satisfying $\nabla_{T}T^{\mu} =0$ and the deviation vector field $S^{\mu} = \frac{\partial x^{\mu}}{\partial s}$, we can calculate the `relative acceleration' between nearby geodesics 
\begin{align} \label{geodesicdev}
a^{\mu} &= \nabla_{T} (\nabla_{T} S^{\mu}) = T^{\lambda} \nabla_{\lambda} ( T^{\nu} \nabla_{\nu} S^{\mu} ) 
= T^{\lambda} \nabla_{\lambda} S^{\nu} \nabla_{\nu} T^{\mu} + T^{\lambda} S^{\nu} \nabla_{\lambda} \nabla_{\nu} T^{\mu} \nonumber \\
&= S^{\lambda} \nabla_{\lambda} T^{\nu}  \nabla_{\nu} T^{\mu} + T^{\lambda} S^{\nu} ({\nabla}_{\nu} {\nabla}_{\lambda} T^{\mu} + R_{\lambda \nu \gamma}{}^{\mu} T^{\gamma} )
\nonumber \\
&=  S^{\lambda} \nabla_{\lambda} T^{\nu}  \nabla_{\nu} T^{\mu}  + S^{\nu} \nabla_{\nu}(T^{\lambda} \nabla_{\lambda} T^{\mu}) - S^{\nu} \nabla_{\nu} T^{\lambda} \nabla_{\lambda} T^{\mu} +  T^{\lambda} S^{\nu}R_{\lambda \nu \gamma}{}^{\mu} T^{\gamma}  \nonumber \\
&= R_{\lambda \nu \gamma}{}^{\mu} T^{\lambda} S^{\nu} T^{\gamma} \, .
\end{align}
In the above derivation we used that torsion is zero and that $T^{\mu}$ and $S^{\mu}$ commute such that $T^{\mu} \nabla_{\mu}S^{\nu} = S^{\mu} \nabla_{\mu}T^{\nu}$.

For non-vanishing torsion the above calculation and result is modified slightly. We instead have an additional torsion term coming from the commutator of covariant derivatives~(\ref{Riemann covar}). We also have that the tangent vector fields instead satisfy $T^{\mu} \bar{\nabla}_{\mu}S^{\nu} = S^{\mu} \bar{\nabla}_{\mu}T^{\nu} + T^{\nu}{}_{\mu \lambda} T^{\mu} S^{\lambda}$ coming from the definition of torsion~(\ref{Torsion v0}). The calculation for the (autoparallel) geodesic deviation is then 
\begin{align} \label{geodesicdev2}
a^{\mu} &= \bar{\nabla}_{T} (\bar{\nabla}_{T} S^{\mu}) = T^{\lambda} \bar{\nabla}_{\lambda} (S^{\nu} \bar{\nabla}_{\nu} T^{\mu} + T^{\mu}{}_{\nu \gamma} T^{\nu} S^{\gamma} ) \nonumber \\ 
& \begin{multlined}[t]
=T^{\lambda} \bar{\nabla}_{\lambda} S^{\nu} \bar{\nabla}_{\nu} T^{\mu} 
+T^{\lambda} \bar{\nabla}_{\lambda}(T^{\mu}{}_{\nu \gamma} T^{\nu} S^{\gamma}) \\ 
+ T^{\lambda} S^{\nu} \big( \bar{R}_{\lambda \nu \alpha}{}^{\mu} T^{\alpha} + \bar{\nabla}_{\nu} \bar{\nabla}_{\lambda} T^{\mu} - T^{\alpha}{}_{\lambda \nu} \bar{\nabla}_{\alpha} T^{\mu} \big) 
\end{multlined}
\nonumber \\
& \begin{multlined}[t] = \bar{\nabla}_{\nu} T^{\mu} \big( T^{\lambda} \bar{\nabla}_{\lambda} S^{\nu} - S^{\lambda} \bar{\nabla}_{\lambda} T^{\nu} - T^{\nu}{}_{\lambda \gamma} T^{\lambda} S^{\gamma} \big) + \\ \bar{R}_{\lambda \nu \alpha}{}^{\mu} T^{\lambda} S^{\nu} T^{\alpha}  +  T^{\lambda} \bar{\nabla}_{\lambda}(T^{\mu}{}_{\nu \gamma} T^{\nu} S^{\gamma}) \nonumber
\end{multlined}  \\
&= \bar{R}_{\lambda \nu \alpha}{}^{\mu} T^{\lambda} S^{\nu} T^{\alpha}  +  T^{\lambda} T^{\nu} \bar{\nabla}_{\lambda}(T^{\mu}{}_{\nu \gamma}  S^{\gamma}) \, ,
\end{align}
which can be found in~\cite{Dereli:2019ips}. We therefore see that the presence of torsion also influences the relative deviation of geodesics, even in the absence of curvature.

From contractions of the Riemann tensor we define the Ricci tensor $\bar{R}_{\mu \nu}$ and Ricci scalar $\bar{R}$, which have components 
\begin{align}\label{Ricci tensor}
\bar{R}_{\mu \nu} := \bar{R}_{\lambda \mu \nu}{}^{\lambda} &=  \bar{\Gamma}^{\lambda}_{\lambda \rho} \bar{\Gamma}^{\rho}_{\mu \nu} - \bar{\Gamma}^{\lambda}_{\mu \rho} \bar{\Gamma}^{\rho}_{\lambda \nu} 
+ \partial_{\lambda} \bar{\Gamma}^{\lambda}_{\mu \nu} - \partial_{\mu} \bar{\Gamma}^{\lambda}_{\lambda \nu}    \, , \\
\label{Ricci scalar}
\bar{R} := g^{\mu \nu} \bar{R}_{\mu \nu} &= g^{\mu \nu} (  \bar{\Gamma}^{\lambda}_{\lambda \rho} \bar{\Gamma}^{\rho}_{\mu \nu} - \bar{\Gamma}^{\lambda}_{\mu \rho} \bar{\Gamma}^{\rho}_{\lambda \nu} + \partial_{\lambda} \bar{\Gamma}^{\lambda}_{\mu \nu} - \partial_{\mu} \bar{\Gamma}^{\lambda}_{\lambda \nu} ) \, .
\end{align}
Note that the Riemann and Ricci tensors depend only on the connection whilst the Ricci scalar necessarily requires a metric tensor $\bar{R} = \bar{R}(g,\bar{\Gamma})$. It is also useful to state the symmetry properties of the curvature tensors in relation to torsion and non-metricity
\begin{align}
\bar{R}_{(\mu \nu) \gamma}{}^{\lambda} &=0 \\
\bar{R}_{[\mu \nu \gamma]}{}^{\lambda} &= \bar{\nabla}_{[\mu} T^{\lambda}{}_{\nu \gamma]} - T^{\rho}{}_{[\mu \nu} T^{\lambda}{}_{\gamma] \rho} \\
\bar{R}_{\mu \nu (\gamma \lambda)} &= \bar{\nabla}_{[\mu} Q_{\nu] \gamma \lambda} + \frac{1}{2} T^{\rho}{}_{\mu \nu} Q_{\rho \gamma \lambda} \, .
\end{align}

Using these properties one can show that the geometric tensors satisfy the so-called \textit{Bianchi identities}
\begin{equation} \label{Bianchi}
\bar{\nabla}_{[\rho} \bar{R}_{\mu \nu] \gamma}{}^{\lambda} = T^{\sigma}{}_{[\rho \mu}  \bar{R}_{\nu] \sigma \gamma}{}^{\lambda} \, .
\end{equation}
Contracting over the first and last index and the last two indices respectively gives the contracted Bianchi identities
\begin{align} \label{Bianchi_2}
\bar{\nabla}_{\lambda} \bar{R}_{\mu \nu \gamma}{}^{\lambda} &= 2 \bar{\nabla}_{[\mu} \bar{R}_{\nu] \gamma } 
+ T^{\sigma}{}_{\mu \nu} \bar{R}_{\sigma \gamma}
+ 2 T^{\sigma}{}_{\lambda [ \mu}  \bar{R}_{\nu] \sigma \gamma}{}^{\lambda} \\
 \label{Bianchi_3}
\bar{\nabla}_{[\lambda} \bar{V}_{\mu \nu]} &= T^{\rho}{}_{[\lambda \mu} \bar{V}_{\nu] \rho} \, , 
\end{align}
where we introduced the new notation for the contraction $\bar{V}_{\mu \nu} :=  \bar{R}_{\mu \nu \lambda}{}^{\lambda}$.

\subsubsection{Levi-Civita connection}
A connection of specific importance is the \textit{Levi-Civita connection}. This is the unique torsion-free $\bar{\Gamma}^{\lambda}_{[\mu \nu]} =0$ and metric compatible $\bar{\nabla}_{\lambda} g_{\mu \nu} =0 $ connection. It can be written in local (holonomic) coordinates completely in terms of the metric tensor
\begin{equation} \label{Levi-Civita}
\Gamma^{\lambda}_{\mu \nu} = \frac{1}{2} g^{\lambda \gamma} \big(g_{\mu \gamma , \nu} + g_{\nu \gamma , \mu} - g_{\mu \nu , \gamma} \big) \ ,
\end{equation} 
where we will use Gamma without a bar to refer to this specific connection. Similarly, the covariant derivative associated with the Levi-Civita connection will use the notation $\bar{\nabla}(\Gamma) := \nabla$ without a bar, as will all other torsion-free and metric-compatible quantities. 

The Levi-Civita connection is the one chosen in General Relativity and is of fundamental importance. From the metric compatibility of the Levi-Civita connection we can now raise and lower indices through covariant derivatives, such that $g^{\mu \nu} \nabla_{\lambda} v_{\nu} = \nabla_{\lambda} v^{\mu}$ and  $\nabla^{\mu} \nabla_{\mu} = \nabla_{\mu} \nabla^{\mu}$. The Riemann tensor symmetries are also simplified
\begin{align}
R_{\mu \nu (\gamma \lambda)} &= 0 \quad \quad , \quad \quad R_{[\mu \nu \gamma]}{}^{ \lambda} = 0 \, ,
\end{align}
as are the Bianchi identities
\begin{equation}
\nabla_{[\rho} R_{\mu \nu] \gamma}{}^{\lambda} = 0 \quad \quad , \quad \quad \nabla_{[\rho} R_{\mu \nu]}=0 \, .
\end{equation}
Further contraction with the metric $g^{\mu \nu}$ leads to the twice-contracted Bianchi identity 
\begin{equation} \label{Bianchi2}
\nabla_{\mu} G^{\mu}{}_{\nu} = 0 \, ,
\end{equation}
where we define the \textit{Einstein tensor} as
\begin{equation}
G_{\mu \nu} := R_{\mu \nu} - \frac{1}{2} g_{\mu \nu} R \, .
\end{equation}

Revisiting the topic of geodesics, the extremal curves calculated from minimising the distance interval $ds$ gives another notion of the shortest path on a manifold with a metric. Contrast this with the autoparallel geodesics of equation~(\ref{Autoparallel}), which does not require a metric structure. Minimising this action with appropriate boundary conditions for the path leads to
\begin{equation}
\delta \int ds = \delta \int \sqrt{-g_{\mu \nu} \dot{x}^{\mu} \dot{x}^{\nu}} d \tau = 0 \, ,
\end{equation}
where $\dot{x}^{\mu}(\tau) = d x^{\mu}(\tau)/d\tau$ and an affine path parameterisation has been used. The solution to the above equation is
\begin{align} \label{GR geodesic}
\frac{d^2 x^{\mu}}{ d \tau^2} + \frac{1}{2} g^{\mu \gamma} \big(g_{\nu \gamma , \rho} + g_{\rho \gamma , \nu} - g_{\nu \rho , \gamma} \big) \frac{d x^{\nu}}{d \tau} \frac{d x^{\rho}}{d \tau} &=0 \nonumber \\
\frac{d^2 x^{\mu}}{ d \tau^2}  + \Gamma^{\mu}_{\nu \gamma} \frac{d x^{\nu}}{d \tau} \frac{d x^{\gamma}}{d \tau} &=0 \, .
\end{align}
The combination of metric derivatives have been identified with the Levi-Civita connection in the final line. This is just the equation for autoparallel curves~(\ref{Autoparallel}) with this choice of connection, so both notions of geodesic coincide. Geometries with torsion and non-metricity do not share this property in general\footnote{We emphasise that torsion \textit{does} contribute to the autoparallel geodesic equation despite the affine connection being symmetrised, see equation~(\ref{geodesic contortion2}).}.

It should be noted that physical trajectories are dictated not necessarily by equation~(\ref{GR geodesic}) nor the autoparallel equation~(\ref{Autoparallel}), but by the dynamical equations of motion of the theory in question. This eliminates the need to `choose' between the metrical length-minimising geodesics and autoparallel geodesics.
Instead, nature prescribes the paths that particles will travel along, and whether or not this coincides with these two notions of `straightest' or `shortest' path depends on the theory itself. On the other hand, if we a priori decide that we want the motion predicted by our theory to coincide with these geometric notions of geodesics, this can tell us how to construct our theory. However, it is only observation and self-consistency, not aesthetics, that can tell us if our theory is correct.

\subsubsection{Decomposition of affine connection}
A general affine connection can be decomposed into a sum of its Levi-Civita part and contributions from torsion and non-metricity. It is convenient to introduce the so-called contortion tensor, defined as
\begin{equation} \label{contortion}
K_{\mu \nu}{}^{\lambda} = \frac{1}{2} g^{\lambda \rho}(- T_{\{\nu \mu \rho \}} + Q_{\{ \mu \rho \nu \}}) \, ,
\end{equation}
where the \textit{Schouten bracket} is defined as the following permutation of indices 
\begin{equation}
S_{\{ \mu \nu \lambda \} } = S_{\mu \nu \lambda} - S_{\nu \lambda \mu} + S_{\lambda \mu \nu}  \, .
\end{equation} 
Note that in some places what we have called contortion is sometimes called \textit{distortion}, whilst the permutation of torsion is also sometimes called the contortion (or contorsion). Moreover, the permutation of the non-metricity tensor is also sometimes known as the \textit{disformation} tensor. We will stick to just calling $K_{\mu \nu}{}^{\lambda}$ the contortion tensor.

Using these tensors, it is a well-known result~\cite{JS1954,Hehl:1994ue} that the general metric-affine connection can be written as
\begin{equation} \label{affine decomp}
\bar{\Gamma}^{\lambda}_{\mu \nu} = \Gamma^{\lambda}_{\mu \nu} + K_{\mu \nu}{}^{\lambda} \, .
\end{equation}
From the definition of contortion~(\ref{contortion}) we know that $K_{\mu \nu}{}^{\lambda}$ is a tensor, whilst the affine and Levi-Civita connections of course are not. This can be understood from the difference of two connections being a tensor itself, due to the cancellation of the inhomogeneous terms (as happens with the torsion tensor).

This decomposition will be extremely useful and will allow us to easily compare Levi-Civita objects with torsion and non-metricity. One result using this decomposition is that the affine curvature tensor~(\ref{Riemann tensor}) can be written as a sum of the Levi-Civita curvature and contortion terms
\begin{equation} \label{Riemann_decomp} 
\bar{R}_{\mu \nu \gamma}{}^{\lambda} = R_{\mu \nu \gamma}{}^{\lambda} + 2 K_{[\mu|\sigma}{}^{\lambda} K_{\nu] \gamma}{}^{\sigma} + 2\nabla_{[\mu} K_{\nu] \gamma}{}^{\lambda} \, ,
\end{equation}
where $\nabla$ is the covariant derivative with respect to the Levi-Civita connection.
Taking contractions gives the decompositions of the Ricci tensor and Ricci scalar
\begin{align} \label{Ricci_decomp}
\bar{R}_{\nu \gamma} &=  R_{\nu \gamma}+ 2 K_{[\lambda|\sigma}{}^{\lambda} K_{\nu] \gamma}{}^{\sigma} + 2\nabla_{[\lambda} K_{\nu] \gamma}{}^{\lambda} \, , 
\\
\bar{R} &=  R +  K_{\lambda \sigma}{}^{\lambda} K_{\nu}{}^{\nu \sigma} - K_{\nu \sigma}{}^{\lambda} K_{\lambda}{}^{\nu \sigma}  + \nabla_{ \lambda} K_{\nu}{}^{\nu \lambda} - \nabla_{\nu} K_{\lambda}{}^{\nu \lambda} 
\, .  \label{R_decomp}
\end{align}

Another nice result is that the geodesic equation found from minimising the length interval, given in terms of the Levi-Civita connection, can be expressed in terms of a general affine connection
\begin{align} \label{geodesic contortion}
\frac{d^2 x^{\mu}}{ d \tau^2} + \Gamma^{\mu}_{\nu \gamma} \frac{d x^{\nu}}{d \tau} \frac{d x^{\rho}}{d \tau} &= 0 \nonumber \\
\frac{d^2 x^{\mu}}{ d \tau^2}  + \bar{\Gamma}^{\mu}_{\nu \gamma} \frac{d x^{\nu}}{d \tau} \frac{d x^{\gamma}}{d \tau} &=  K_{\nu \gamma}{}^{\mu}  \frac{d x^{\nu}}{d \tau} \frac{d x^{\gamma}}{d \tau}  \, ,
\end{align}
where the contortion term appears in the equation as a Lorentz-like force term~\cite{Bahamonde:2021gfp}. Of course, the dynamics on the first and second lines are still totally equivalent.  

On the other hand, autoparallel geodesics of $\bar{\nabla}_{\dot{x}} \dot{x}^{\mu} =0$ take the explicit form
\begin{align} \label{geodesic contortion2}
\frac{d^2 x^{\mu}}{ d \tau^2}  + \Gamma^{\mu}_{\nu \gamma} \frac{d x^{\nu}}{d \tau} \frac{d x^{\gamma}}{d \tau} =  -K_{\nu \gamma}{}^{\mu}  \frac{d x^{\nu}}{d \tau} \frac{d x^{\gamma}}{d \tau}  \, ,
\end{align}
which are different from~(\ref{geodesic contortion}) unless the symmetric part vanishes $K_{(\mu \nu)}{}^{\lambda}=0$. Writing this condition explicitly in terms of torsion and non-metricity 
\begin{equation}
K_{(\mu \nu)}{}^{\lambda} = -T_{(\mu \nu)}{}^{\lambda} + Q_{(\mu \nu)}{}^{\lambda} - \frac{1}{2} Q^{\lambda}{}_{\mu \nu} \, ,
\end{equation}
where we have used the symmetry of non-metricity and skew-symmetry of torsion over their last two indices. We also therefore see that torsion and non-metricity do contribute to the symmetric part of the affine connection, and hence geodesics.

As a final point, it should be repeated that a spacetime manifold (endowed with a metric structure) does not itself possess curvature, torsion or non-metricity; this is only associated with a given affine connection on the space. However, we can always talk about the Levi-Civita curvature of a manifold with respect to a metric, independently of the connection.

\subsection{Tetrad formalism}
\label{section2.1.2}

In the previous sections we worked exclusively in the natural coordinate basis with $T_{p}(\mathcal{M})$ spanned by $\mathbf{e}_{\mu} = \partial_{\mu}$ and $T^*_{p} (\mathcal{M})$ by the basis one-forms $\mathbf{e}^{\mu} = dx^{\mu}$.
 If we instead work in an anholonomic basis we can begin to see a richer structure, and the link to gauge theories will be clearer. Moreover, non-coordinate bases are needed for coupling gravity to matter such as Dirac spinors. Note that we mainly follow~\cite{Nakahara:2003nw} and~\cite{Frankel:1997ec} in general, but follow Schouten~\cite{JS1954} for index conventions for Cartan's symbolic method. For a modern introduction, see~\cite{yepez2011einstein}.
 
A basis $\mathbf{e}_{a}$ (otherwise known as a \textit{frame} of vector fields) can be related to the coordinate basis (coordinate frame) through $\mathbf{e}_{a} = e_{a}{}^{\mu} \partial_{\mu}$. The dual basis of one-forms $\mathbf{e}^{a}$ are related to the coordinate one-forms by $\mathbf{e}^{a} = e^a{}_{\mu} dx^{\mu}$ with
\begin{equation}
\langle \mathbf{e}_{a} , \mathbf{e}^{b} \rangle =  \delta^{a}_{b} \, ,
\end{equation}
which we recognise to be the inner product~(\ref{innerproduct}).
The components $e_{a}{}^{\mu}$ are known as the \textit{tetrad}, or \textit{veirbein} in four dimensions, and are elements of the general linear group $GL(n,\mathbb{R})$. The inverse components $e^{a}{}_{\mu}$ are known as the \textit{co-tetrad}, and they satisfy the relations
\begin{equation}
e_{a}{}^{\mu} e^a{}_{\nu} = \delta^{\mu}_{\nu} \quad \quad , \quad \quad  e_{a}{}^{\mu} e^b{}_{\mu} =   \delta^{a}_{b} \, .
\end{equation}
We will often simply drop the `co' when referring to the co-tetrad.

 A spacetime vector field in an anholonomic basis is simply $v = v^{a} \mathbf{e}_{a}$ where $v^a$ are the components in the arbitrary basis, and similarly for a one-form $w = w_{a} \mathbf{e}^a$. As described in footnote~\ref{footnote:Lie}, the Lie bracket of frame fields is
 \begin{equation} \label{frame commutator}
 [\mathbf{e}_{a},  \mathbf{e}_{b} ] = \gamma^c{}_{a b} \mathbf{e}_{c} \, ,
 \end{equation}
 with $\gamma^c{}_{ab}$ the object of anholonomy. It follows that this object can be expressed as
 \begin{equation} \label{anholonomy}
 \gamma^{c}{}_{ab} = 2 e_{[a}{}^{\mu} e_{b]}{}^{\nu} \partial_{\nu} e^{c}{}_{\mu} \, .
 \end{equation}
The vanishing of the commutator is the definition of the coordinate frame, and then the basis is said to be integrable or holonomic.

The non-coordinate basis that will be used is the \textit{orthonormal basis} (ONB)
\begin{equation}
g(\mathbf{e}_{a},\mathbf{e}_{b}) = g_{\mu \nu} e_{a}{}^{\mu} e_{b}{}^{\nu} = \eta_{a b} \, , 
\end{equation}
where $\eta_{a b} = \textrm{diag}(-1,1,...,1)$ is the flat Minkowski metric of the tangent space. Note that if the coordinate basis is also orthonormal $\gamma^{c}{}_{ab}=0$, the spacetime is flat $g_{\mu \nu} = \eta_{\mu \nu}$.

 In the ONB we also have the relations
\begin{equation} \label{ONB relations}
g_{\mu \nu} = e^{a}{}_{\mu} e^{b}{}_{\nu}  \eta_{a b} \quad \quad , \quad \quad g^{\mu \nu} = e_{a}{}^{\mu} e_{b}{}^{\nu} \eta^{a b} \, .
\end{equation}
In general, Greek indices, ${\mu}, \nu$, etc., refer to the spacetime coordinates whilst Latin indices, $a,b$, etc., refer to the local tangent space. Hence, these are labelled spacetime indices and Lorentz (or flat-space, or tangent space) indices respectively. The flat Minkowski metric $\eta_{ab}$ and its inverse $\eta^{ab}$ are used to raise and lower Lorentz indices.
The tetrad then acts as a bridge between the spacetime coordinates and the local Lorentz coordinates at each point. Any geometric object in (holonomic) spacetime coordinates can be translated into its Lorentz coordinate counterpart using the tetrad.

\subsubsection{Local Lorentz transformations}
In the local Lorentz space a change of basis, or frame transformation, is represented by a \textit{local Lorentz transformation}. These are the transformations of flat Minkowski space, but applied to the tangent space at each point on the manifold.
This acts on the basis one-forms as
\begin{equation} \label{Lorentz1}
\mathbf{e}^{a} \rightarrow \hat{\mathbf{e}}^{a} = \Lambda^{a}{}_{b} (x) \mathbf{e}^{b} \, ,
\end{equation}
where $\Lambda^{a}{}_{b}(x)$ is an element of the Lorentz group $SO(1,3)$
\begin{equation}
 \Lambda^{a}{}_{b} \eta_{ac}  \Lambda^{c}{}_{d}= \eta_{bd} \, .
\end{equation}
Again, we will often omit the dependence on the spacetime points for brevity. For the basis vectors we have
\begin{equation} \label{Lorentz2}
\mathbf{e}_{a} \rightarrow \hat{\mathbf{e}}_{a} =  \Lambda_{a}{}^{b}(x) \mathbf{e}_{b}
\end{equation}
where $\Lambda_{a}{}^{b} =  (\Lambda^{-1})^{b}{}_{a}$ are the inverse Lorentz transformations satisfying $\Lambda_{a}{}^{b} \Lambda^{a}{}_{c} = \delta^{b}_{c}$.
The tetrad components then transform as
\begin{equation}
e^{a}{}_{\mu} \rightarrow \hat{e}{}^{a}{}_{\mu} = \Lambda^{a}{}_{b} e^{b}{}_{\mu}  \quad \quad , \quad \quad e_{a}{}^{\mu} \rightarrow \hat{e}_{a}{}^{\mu} = \Lambda_{a}{}^{b} e_{b}{}^{\mu} \, .
\end{equation}

Under a simultaneous general coordinate transformation~(\ref{Tensor transformation}) and local Lorentz transformation, a tensor picks up a Lorentz matrix term $\Lambda^{a}{}_{b}(x)$ for each of its Lorentz indices and a coordinate Jacobian term for each of its spacetime indices
\begin{equation} \label{Tensor Lorentz}
T^{\mu a}{}_{\nu b} \rightarrow \hat{T}^{\mu a}{}_{\nu b} = \frac{\partial \hat{x}^{\mu}}{\partial x^{\lambda}} 
\Lambda^{a}{}_{c}  \frac{\partial x^{\rho}}{\partial \hat{x}^{\nu}} \Lambda_{b}{}^{d}
T^{\lambda c}{}_{\rho d} \, .
\end{equation}
We can talk about `local Lorentz covariance' in the same way as we previously spoke about spacetime covariance: i.e., if an object transforms covariantly according to equations~(\ref{Lorentz1}) and~(\ref{Lorentz2}). 

\subsubsection{Spin structure}
The affine connection in the non-coordinate basis is
\begin{equation}
\bar{\nabla}_{a} \mathbf{e}_{b} = \mathbf{e}_{c} \bar{\Gamma}^c_{ab} \, .
\end{equation}
We can  relate the connection in the non-coordinate basis to the coordinate one by simply expanding the components of the covariant derivative 
\begin{equation} \label{connection mixed}
\bar{\Gamma}^{c}_{ab} = e^{c}{}_{\mu} e_{a}{}^{\nu} (\partial_{\nu} e_{b}{}^{\mu} + e_{b}{}^{\lambda} \bar{\Gamma}^{\mu}_{\nu \lambda}) \, .
\end{equation}

We then define the matrix-valued \textit{spin connection} one-form as
\begin{equation} \label{spin form}
\omega^{a}{}_{b} := \bar{\Gamma}^{a}_{cb} \mathbf{e}^{c} = \bar{\Gamma}^{a}_{cb} e^{c}{}_{\mu} dx^{\mu}  \, ,
\end{equation}
where we make use of the differential form notation and suppress the spacetime indices of forms
\begin{equation}
\omega^{a}{}_b \equiv \omega_{\mu}{}^{a}{}_b dx^{\mu} \, .
\end{equation}
Note that under a local Lorentz transformation, the spin connection transforms inhomogeneously 
\begin{equation} \label{spin_LLT}
\omega^{a}{}_{b} \rightarrow \hat{\omega}^{a}{}_{b} = \Lambda^{a}{}_{c} \Lambda_{b}{}^{d} \omega^{c}{}_{d} + \Lambda^{a}{}_{c} d \Lambda_{b}{}^{c} \, ,
\end{equation}
but under a general coordinate transformation it transforms as a differential form (covariantly)\footnote{Contrast this with the spacetime connection, which is a Lorentz scalar but inhomogeneous under coordinate transformations. These transformations can be seen by explicit computations using equation~(\ref{spin and affine}).}.

Comparison of $\omega{}^{a}{}_{b}$ with~(\ref{connection mixed}) leads to the following relations between the spin connection and the spacetime connection
\begin{equation} \label{spin and affine}
\omega_{\mu}{}^{a}{}_{b} = e^a{}_{\nu} \partial_{\mu} e_{b}{}^{\nu} + e^{a}{}_{\nu} e_{b}{}^{\lambda} \bar{\Gamma}^{\nu}_{\mu \lambda}  \quad \quad , \quad \quad \bar{\Gamma}^{\nu}_{\mu \lambda} = e_{a}{}^{\nu} \partial_{\mu} e^{a}{}_{\lambda} + e_{a}{}^{\nu} e^{b}{}_{\lambda} \omega_{\mu}{}^{a}{}_{b} 
\, ,
\end{equation}
where we used $e^{a}{}_{\mu} \partial_{\nu} e_{a}{}^{\lambda} = - e_{a}{}^{\lambda} \partial_{\nu}  e^{a}{}_{\mu}$.
From the above definitions it should be clear that the affine connection and the spin connection represent the same fundamental object with respect to different spaces.

Making use of the spin connection, we can think of the \textit{total} spacetime covariant derivative of Lorentz-index tensors as giving rise to spin connection terms such as
\begin{equation}
\bar{\nabla}_{\mu} v^{a} = \partial_{\mu} v^{a} + \omega_{\mu}{}^{a}{}_{b} v^{b} \, ,
\end{equation}
and for a mixed spacetime-Lorentz index object
\begin{equation} \label{total derivative}
\bar{\nabla}_{\lambda} T^{\mu a}{}_{\nu b} = \partial_{\lambda} T^{\mu a}{}_{\nu b} + \bar{\Gamma}^{\mu}_{\lambda \gamma} T^{\gamma a}{}_{\nu b} + \omega_{\lambda}{}^{a}{}_{c} T^{\mu c}{}_{\nu b}   - \bar{\Gamma}^{\gamma}_{\lambda \nu} T^{\mu a}{}_{\gamma b} - \omega_{\lambda}{}^{c}{}_{b} T^{\mu a}{}_{\nu c} \, .
\end{equation}
A consequence is that the total covariant derivative of the tetrad vanishes identically
\begin{equation}
\bar{\nabla}_{\mu} e^{a}{}_{\nu} = \partial_{\mu}  e^{a}_{\nu} + \omega_{\mu}{}^{a}{}_{b} e^{b}{}_{\nu} -  \bar{\Gamma}_{\mu \nu}^{\lambda} e^{a}{}_{\lambda} = 0 \, .
\end{equation}
By taking the commutator of covariant derivatives acting on a Lorentz vector, we can write the Riemann and torsion tensors in terms of the spin connection
\begin{equation} \label{Riemann mixed2}
[\bar{\nabla}_{\mu}, \bar{\nabla}_{\nu}] v^{a} = \bar{R}_{\mu \nu b}{}^{a} v^{b} - T^{b}{}_{\mu \nu} e_{b}{}^{\lambda} \bar{\nabla}_{\lambda} v^{a} \, ,
\end{equation}
where the components with mixed spacetime-Lorentz indices are
\begin{align} \label{Riemann mixed}
\bar{R}_{\mu \nu a}{}^{b} &= 2\partial_{[\mu} \omega_{\nu]}{}^{b}{}_{a} + 2\omega_{[\mu}{}^{b}{}_{|c|} \, \omega_{\nu]}{}^{c}{}_{a} \, , \\ \label{Torsion mixed}
 T^{a}{}_{\mu \nu} &=2 \partial_{[\mu} e^{a}{}_{\nu]} +2 e^{b}{}_{[\nu} \omega_{\mu]}{}^{a}{}_{b} \,. 
\end{align}
In the coordinate basis where the Lie bracket of the frame fields vanishes~(\ref{frame commutator}), this matches our previous definitions of torsion~(\ref{Torsion}) and curvature~(\ref{Riemann tensor}). We could also write the torsion tensor in terms of the objects of anholonomy by expressing it fully in tangent space indices
\begin{equation} \label{torsion_ONB}
T^{a}{}_{bc} = e_{b}{}^{\mu} e_{c}{}^{\nu} T^{a}{}_{\mu \nu} = 2 e_{b}{}^{\mu} \omega_{[\mu}{}^{a}{}_{c]} - \gamma^{a}{}_{bc} \, ,
\end{equation}
which again matches our abstract notation for torsion in a general basis~(\ref{Torsion v0}).

Using the differential form notation, such as with the spin connection one-form~(\ref{spin form}), we introduce the \textit{torsion two-form} $T^{a}$ and \textit{curvature two-form} $\bar{R}_a{}^{b}$, which we write with reference to the coordinate basis for clarity
\begin{align}
T^{a} & \equiv \frac{1}{2} T^{a}{}_{bc} \, \mathbf{e}^{b} \wedge  \mathbf{e}^{c} = \frac{1}{2} T^{a}{}_{\mu \nu} dx^{\mu} \wedge dx^{\nu} \, ,   \\
\bar{R}_a{}^{b} & \equiv \frac{1}{2} \bar{R}_{cda}{}^{b} \, \mathbf{e}^{c} \wedge  \mathbf{e}^{d} =\frac{1}{2}   \bar{R}_{\mu \nu a}{}^{b} dx^{\mu} \wedge dx^{\nu} \, .
\end{align}
The connection satisfies \textit{Cartan's structure equations}
\begin{align} \label{Cartan_torsion}
d \mathbf{e}^a + \omega^a{}_{b} \wedge \mathbf{e}^b &= T^a \, , \\
d \omega^{a}{}_{b} + \omega^a{}_{c} \wedge \omega^{c}{}_{b} &= \bar{R}_{b}{}^{a} \, ,
\end{align}
where $d$ is the exterior derivative~(\ref{exterior}). It is not hard to see the equivalence with the tensors~(\ref{Riemann mixed}) and~(\ref{Torsion mixed}). 

It is also useful to introduce the \textit{covariant exterior derivative} $D$ which generalises the exterior derivative $d$ in a covariant manner with respect to the spin connection
\begin{equation} \label{total_d}
D \mathbf{e}^{a} := d \mathbf{e}^{a} + \omega^{a}{}_{b} \wedge  \mathbf{e}^{b} \, . 
\end{equation}
With this notation, Cartan's equation for torsion can be written elegantly as $D\mathbf{e}^a = T^a$, and one can also define the non-metricity one-form via $Q_{ab} = -D \eta_{ab}$. The Bianchi identities in differential form notation can be written compactly as  
\begin{align}
D T^{a} &= \bar{R}_{b}{}^{a} \wedge \mathbf{e}^b \, , \\
D \bar{R}_{b}{}^{a} &= 0 \, ,
\end{align}
see Schouten~\cite{JS1954} for a derivation in both notations.

\subsubsection{Levi-Civita spin connection}
Lastly, we mention the properties of the Levi-Civita spin connection, denoted by $\mathring{\omega}^a{}_{b}$, with vanishing torsion and non-metricity. From the vanishing of the non-metricity one-form $Q_{ab} =0$ we find
\begin{align}
\nabla_{\mu} \eta_{ab} = 0 = \partial_{\mu} \eta_{ab} -  \mathring{\omega}_{\mu}{}^{c}{}_{a}  \eta_{cb} - \mathring{\omega}_{\mu}{}^{c}{}_{b}  \eta_{ac}   
=  -2\mathring{\omega}_{\mu (ab)}  \, ,
\end{align}
therefore the Levi-Civita spin connection $ \mathring{\omega}_{ab}$ is antisymmetric. Vanishing torsion means that the spin connection can be written totally in terms of the tetrads using
\begin{equation}
D \mathbf{e}^{a} = d \mathbf{e}^{a} +  \mathring{\omega}^{a}{}_{b} \wedge \mathbf{e}^b = 0 \, .
\end{equation}

Using the objects of anholonomy introduced in equation~(\ref{frame commutator}), the Levi-Civita spin connection in the orthonormal basis is 
\begin{equation} \label{spin anholonomy}
\mathring{\omega}_{abc} = \frac{1}{2} \gamma_{\{bac\}} = \frac{1}{2} \big(\gamma_{bac} - \gamma_{acb} + \gamma_{cba}\big)   \, , 
\end{equation}
where $\omega_{abc} = e_{a}{}^{\mu} \eta_{bd} \omega_{\mu}{}^{d}{}_{c}$ and $\gamma^c{}_{ab}$ is defined in~(\ref{anholonomy}). These are sometimes called the Ricci rotation coefficients.
Alternatively, one can take the expression for the spin connection in terms of the spacetime affine connection in~(\ref{spin and affine}) and fix this to be the Levi-Civita one~(\ref{Levi-Civita}). The result, equivalent to~(\ref{spin anholonomy}), is just an equation in terms of the tetrad field and its first partial derivatives. One nicely sees that the torsion tensor of the Levi-Civita geometry in the anholonomic basis vanishes by use of~(\ref{torsion_ONB}) in equation~(\ref{spin anholonomy}).

One could also go on to express the general affine spin connection in terms of its Levi-Civita part, torsion, non-metricity and anholonomy, analogous to the spacetime decomposition~(\ref{affine decomp}), again see~\cite{JS1954} for details. However, we will primarily be working in the metric-affine framework only in the natural coordinate basis, unless otherwise stated, so this won't be necessary.

\section{General Relativity}
\label{section2.2}

In this section we take a step back from the more abstract mathematics and examine Einstein's theory of General Relativity. We discuss the postulates that led Einstein to develop GR from a physical and intuitive perspective, before studying the Einstein field equations and the Einstein-Hilbert action. The Gibbons-Hawking-York term is also introduced, as we look closely at the boundary conditions required in our variations. Then, the action using the tetrad formalism will be briefly examined. But in order to build up to the theory of General Relativity we must first begin with Special Relativity, the theory of spacetime in the absence of gravitation.

\subsubsection{Special Relativity}
 In Special Relativity, the absolute three dimensions of space and one dimension of time in pre-relativistic physics are replaced by the unified notion of \textit{spacetime}. Four-dimensional spacetime is a continuum where each point represents a specific event in space and time. Though a drastic change from the fixed space and time of Newton and Galilei, the four-dimensional spacetime picture has been shown to be consistent with the other laws of physics. Moreover, and very importantly, Special Relativity is compatible with Maxwell's theory of electromagnetism.

The two fundamental postulates of Special Relativity are the following:
\begin{enumerate}[(i)]
  \item \textit{Physical laws should remain the same in inertial frames of reference.}
  \item \textit{The speed of light in a vacuum is constant in any inertial frame}.
\end{enumerate}
An inertial frame of reference is defined as one undergoing no acceleration. 

These postulates led to the realisation that the old notions of simultaneity must be discarded. Observers in different inertial frames of reference will no longer agree on the simultaneity of events because neither time intervals nor space intervals alone are invariant. Instead, one has the invariant \textit{spacetime interval}
\begin{equation} \label{Minkowski}
ds^2 = - c^2 dt^2 + dx^2 + dy^2 + dz ^2 = \eta_{\mu \nu} d x^{\mu} dx^{\nu} \, ,
\end{equation}
where $\eta_{\mu \nu} = \textrm{diag}(-1,+1,+1,+1)$ is the \textit{Minkowski metric}. The interval $ds^2$ is observer-independent, and it describes the causal light-cone structure of spacetime.

The set of all transformations which leave the spacetime interval invariant is the \textit{Poincar\'{e} group}, or the inhomogeneous Lorentz group. It is the semi-direct product of the translations group $T(4)$ and the (homogeneous) Lorentz group $SO(1,3)$, which includes boosts and spatial rotations. A general Poincar\'{e} transformation can be written in index notation as
\begin{equation} \label{Poincare}
x^{\mu} \rightarrow \hat{x}^{\mu} = \Lambda^{\mu}{}_{\nu} x^{\nu} + a^{\mu} \,  ,
\end{equation}
where $\Lambda$ is a Lorentz matrix $\Lambda^{\mu}{}_{\nu} \in SO(1,3)$ and $a^{\mu}$ is a constant spacetime vector. The interval~(\ref{Minkowski}) is invariant under these transformations.
 These are then isometries of spacetime and represent the physical symmetries of the theory, which will be covered in more detail in Sec.~\ref{section2.3}. Invariance under the Poincar\'{e} group can be seen as the foundation of Special Relativity\footnote{Note that in the tetrad formulation we introduced \textit{local} Lorentz transformations within the tangent space at every spacetime point. These were not Poincar\'{e} transformations, which include translations. Local and global symmetries should not be confused.}.
 
Using mathematical language, we see that Special Relativity is described by a four-dimensional differentiable manifold with metric tensor $\eta_{\mu \nu}(x)$ given in~(\ref{Minkowski}). Further, using the Minkowski metric we see that the manifold of Special Relativity is flat, known as Minkowski space $\mathbb{R}^{(1,3)}$. 

As an aside, it is interesting to note that the theory can be studied in terms of the frame fields (tetrads). If one does this, it can then be shown that the spin connection represents the inertial effects in Lorentz rotated frames, and is equal to just the Levi-Civita connection~(\ref{spin anholonomy}) of a flat spacetime, see \cite{Krssak:2018ywd,Pereira:2019woq}. These lessons are useful when studying Teleparallel gravity, the translational gauge theory of gravity, see Sec.~\ref{section3.3}.

Now we transition from Special Relativity to General Relativity, in order to accommodate the gravitational interaction. 
 The principles of Special Relativity, that apply in the \textit{special} frames of reference of inertial observers, must be generalised to accommodate all frames: \textit{``The laws of physics must be of such a nature that they apply to systems of reference in any kind of motion''}~\cite{Einstein:1916vd}. In the new framework, the inability to construct globally inertial frames will lead to the conclusion that spacetime is not described by the flat Minkowski space $\mathbb{R}^{(1,3)}$.

Some of the guiding principles for deriving such a theory came in the way of the fundamental postulates of General Relativity\footnote{It should be kept in mind that Einstein was familiar with Riemannian geometry, not the full metric-affine structure laid out in the previous section \cite{Schrodinger:2011gqa}.}. The two main postulates are the principle of equivalence and the principle of covariance, though they are not completely independent. A related concept is the minimal coupling principle. We will review these principles and see how they lead to the theory of General Relativity.

A very crucial point is that in the following section, we will \textit{not} go beyond the Riemannian geometry described by the Levi-Civita connection. We will assume that the affine connection of spacetime is the unique Levi-Civita one. This is for two reasons: firstly, to understand the development of General Relativity from a historical perspective, and secondly, because many of the already qualitative and imprecise postulates become even less clear in the metric-affine setting. 
In Chapter~\ref{chapter3} we relax this assumption and explore gravitational theories with an independent affine connection.

\subsection{Principles of General Relativity}
\label{section2.2.1}
\subsubsection{Principle of equivalence}
Perhaps the most important postulate is the \textit{equivalence principle}. There are various ways to state this principle, with subtle differences between them\footnote{For a review of the various versions of the equivalence principle and their physical interpretations see~\cite{Will:2014kxa}. For a historical review see~\cite{NORTON1985203}.}. Different conclusions can be drawn depending on the precise formulation of the statement. Because our aim in introducing the equivalence principle is to give some physical intuition of how one arrives at General Relativity from Special Relativity, we will use the definitions which make this leap easiest.

Weinberg states the equivalence principle as the following~\cite{Weinberg:1972kfs}:
\begin{quote}
\textit{
``At every space-time point in an arbitrary gravitational field it is possible to choose a ``locally inertial coordinate system'' such that, within a sufficiently small regime of the point in question, the laws of nature take the same form as in unaccelerated Cartesian coordinate systems in the absence of gravitation.''}
\end{quote}
He then goes on to clarify that ``the same form as in unaccelerated Cartesian coordinate systems'' means the form of laws in Special Relativity. The `sufficiently small regime' condition means that tidal forces can safely be ignored. To paraphrase, our working definition is that in a small enough region of spacetime, the laws of physics will reduce to those of flat Minkowski space. Hence we see that the equivalence principle, stated in the form above, is directly related to the notion that spacetime is locally Minkowski $\mathbb{R}^{(1,3)}$.

The principle is often also split up into its weaker and stronger statements, which differ by asserting \textit{which} physical laws must obey the principle of equivalence.
\\
 \textit{\textbf{Weak Equivalence Principle (WEP):}}
 \vspace{-0.5mm}
 \begin{quote}
 At every spacetime point, in a small enough region, it is possible to choose a locally inertial frame (LIF) such that the \textbf{\textit{laws of motion}} for a freely falling test particle take the same form as in Special Relativity.
\end{quote}
  \textit{\textbf{Einstein Equivalence Principle (EEP):}}
 \vspace{-0.5mm}
 \begin{quote}
 At every spacetime point, in a small enough region, it is possible to choose a locally inertial frame (LIF) such that the \textbf{\textit{non-gravitational laws of nature}} for a freely falling test particle take the same form as in Special Relativity.
\end{quote}
\clearpage
  \textit{\textbf{Strong Equivalence Principle (SEP):}}
 \vspace{-0.5mm}
 \begin{quote}
 At every spacetime point, in a small enough region, it is possible to choose a locally inertial frame (LIF) such that the \textbf{\textit{laws of nature}} for a freely falling self-gravitating particle take the same form as in Special Relativity.
\end{quote}
The WEP, applicable to the laws of motion for non-gravitational objects, is generalised in the EEP to apply to all physical laws for non-gravitational objects. This is generalised once again in the SEP, which applies to objects that have appreciable self-gravity. The SEP stated above is equivalent to the original quotation of the equivalence principle.

The WEP says that the motion of freely falling test particles is independent of their internal structure and composition, known as the \textit{universality of free fall}~\cite{Will:2014kxa}. This is equivalent with the statement from Newtonian mechanics that the inertial and gravitational masses of an object are the same. It can be argued that the WEP implies that test particles must follow along geodesics\footnote{According to Will~\cite{Will:2014kxa}, geodesic motion is a consequence of the EEP and not the WEP; this comes down to slightly different definitions and interpretations. Because of these disagreements, many authors agree that the equivalence principle(s) are of little practical value due to their qualitative nature, and instead construct quantitative postulates such as the `metric postulates'~\cite{Thorne:1971iat,Will:2014kxa}. However, these appear to be representation dependent~\cite{Sotiriou:2007zu}.}. Weinberg presents this argument in~\cite{Weinberg:1972kfs}, which we will sketch here. 

For a test particle in free-fall, we can define local inertial coordinates at a point $y^{\mu}(p)$,  such that the particle experiences no acceleration 
\begin{equation} \label{straightline}
\frac{d^2 y^{\mu}}{d\lambda^2}= 0 \, .
\end{equation}
Upon a change of coordinates to the system $\{x^{\mu}\}$ where $y^{\mu} = y^{\mu}(x^{\nu})$, the equation can be rewritten as
\begin{equation}
\frac{\partial y^{\mu}}{\partial x^{\nu}} \frac{ d^2 x^{\nu}}{d \lambda^2} + \frac{\partial^2 y^{\mu}}{\partial x^{\nu} \partial x^{\rho}} \frac{d x^{\nu} }{ d \lambda} \frac{d x^{\rho} }{ d \lambda}  = 0 \, .
\end{equation}
From here, one uses the definition of the affine connection transformed from the locally flat coordinates $\Gamma^{\rho}_{\nu \lambda}(y^{\mu})=0$ to the new coordinates $x^{\mu}$ to obtain the geodesic equation
\begin{equation} \label{prince.geo}
\frac{ d^2 x^{\mu}}{d \lambda^2} + \Gamma^{\mu}_{\nu \rho} \frac{d x^{\nu} }{ d \lambda} \frac{d x^{\rho} }{ d \lambda}  =0 \, .
\end{equation}
It is quite clear that this equation is the covariant version of the coordinate-dependent equation of motion~(\ref{straightline}), analogous to how the covariant derivative is introduced to fix the coordinate dependence of the partial derivative of tensors.

The EEP is what forces our theory (not just the laws of motion) to be geometric; our earlier assumption of a four-dimensional Lorentzian manifold realises this principle. The inertial frames of Special Relativity are now `promoted' to \textit{locally} inertial frames at each point in spacetime, due to the tangent space of our manifold looking like $\mathbb{R}^{(1,3)}$. By geometrising spacetime, the EEP enforces the local Lorentz invariance and coordinate invariance of our theory. Blagojevic also argues that the equivalence principle can be expressed as a principle of local symmetry~\cite{Blagojevic:2002du}. We should note that it has been conjectured that consistent theories satisfying the WEP will also satisfy the EEP~\cite{schiff1960experimental}, and it seems difficult to construct theories that only violate the latter.

Let us give two examples, one that satisfies the (weak) equivalence principle and the other that does not. The action for a test particle of mass $M$ in General Relativity is given by
\begin{equation}
S_{1} = - M \int \sqrt{-g_{\mu \nu} \dot{x}^{\mu} \dot{x}^{\nu}} d \tau \, .
\end{equation}
Solutions to extremising the action are the geodesics~(\ref{GR geodesic}) with $\nabla_{\dot{x}} \dot{x}^{\mu}=0$. This theory therefore satisfies the WEP.
Now imagine a theory where the action is instead given by
\begin{equation}
S_{2} = - M \int e^{\phi(x)} \sqrt{-g_{\mu \nu} \dot{x}^{\mu} \dot{x}^{\nu}} d \tau \, ,
\end{equation}
where $\phi(x) = \phi(x(\tau))$ is a new scalar field and a dynamical variable of the theory. The solutions obtained from extremising this action now include the scalar field $\phi(x)$, and we no longer have the solution $\nabla_{\dot{x}} \dot{x}^{\mu}=0$. The WEP is therefore violated. This is similar in presentation to the scalar-tensor theories in the Einstein frame, as covered in the introduction. 

Returning to the SEP, we now must take into account the gravitational self-energy of an object. Even for a theory that is minimally coupled to matter and obeys the EEP, such as the scalar-tensor theories in the Jordan frame representation~(\ref{ST_Jordan}), it is not necessarily the case that the SEP will be satisfied. This is because objects with different gravitational binding energies may take different trajectories through spacetime. This effect, referred to as the Nordtvedt effect~\cite{PhysRev.169.1014,PhysRev.169.1017}, does not play a role in General Relativity but is known to apply to extended massive objects in other gravitational theories, such as scalar-tensor theories~\cite{PhysRevD.80.104002}.

The SEP can be viewed as the assumption that gravity is \textit{entirely} geometric, or that the metric tensor is the only fundamental gravitational variable\footnote{Here we have again implicitly assumed that the metric is the only geometric variable, and ignored the independent affine connection $\bar{\Gamma}$. The equivalence principle in non-Riemannian theories $(\mathcal{M},g,\bar{\Gamma})$ becomes even more subtle, and claims can be made on both sides about whether metric-affine theories adhere to the principle or not (see, for example, \cite{Capozziello:2022zzh,Bajardi:2020fxh}). In the following chapter this should become more clear when looking at concrete examples.}. For theories that satisfy the SEP, the gravitational action will be $S_{\textrm{grav}} = S_{\textrm{grav}}[g]$, again see~\cite{Will:2014kxa,Sotiriou:2007zu}. It has also been claimed that General Relativity is the only viable gravitational theory adhering to the SEP~\cite{1993tegp.book.....W}. 

To summarise, the principle of equivalence tells us that the force of gravity is universal, and that spacetime is geometric. Hence General Relativity should be described geometrically by a Lorentzian manifold. We further postulate that the fixed Minkowski metric of Special Relativity~(\ref{Minkowski}) is generalised to a non-fixed dynamical metric $g_{\mu \nu}$ which represents geometry, and this allows us to interpret gravity as the \textit{curvature of spacetime}. This can be intuited from the fact that gravity is universal and that it does not behave like a force, due to the ability to find locally inertial coordinates for a freely falling particle in an arbitrary gravitational field. Given that we already have a differentiable manifold to describe spacetime, this makes curvature a well-motivated choice for the source of gravity. This `leap' will be further motivated in the following\footnote{It is historically interesting that, at least in the early years of General Relativity (1918), Einstein saw his gravitational theory as following necessarily from the principle of equivalence: ``from this [the principle of equivalence] and from the special theory of relativity it follows necessarily that the symmetric ``fundamental tensor” $g_{\mu \nu}$ determines the metric properties of space, the inertial behaviour of bodies in this space, as well as the gravitational effects"~\cite{janssen2002collected}. In hindsight, however, we see that more assumptions are needed.}.

\subsubsection{Principle of covariance}
Another postulate of General Relativity, which was important for its development, but arguably less important from a modern perspective, is the \textit{principle of general covariance}. This states that the form of the laws of physics should not depend on the coordinate system or frame of reference. To quote Einstein's 1916 paper `The foundation of the general theory of relativity'~\cite{Einstein:1916vd},
\begin{quote}
\textit{
``The general laws of nature are to be expressed by equations which hold good for all systems of coordinates, that is, are covariant with respect to any substitutions whatever (generally covariant).''}
\end{quote}
This is the generalisation of the Lorentz covariance between inertial frames of reference of Special Relativity to the general covariance of any basis or frame related by general coordinate transformations. Hence, \textit{all} frames are equally valid. In essence this means that we want our theory to be described by spacetime tensorial expressions, which again leads to our geometric treatment. Einstein also referred to this principle exclusively as the \textit{principle of relativity}~\cite{janssen2002collected}.

In more mathematical language, one speaks of the \textit{diffeomorphism invariance} of GR. Under a diffeomorphism (smooth, invertible map between manifolds) from the spacetime to itself $\psi: \mathcal{M} \rightarrow \mathcal{M}$, the set $(\mathcal{M},g,\varPhi)$ is mapped to $(\mathcal{M},\psi^*g,\psi^* \varPhi)$, where $g$ and $\varPhi$ are the metric and matter fields respectively. The two sets are physically indistinguishable with the same solutions, hence the theory is diffeomorphism invariant. Diffeomorphisms can be thought of simply as `active coordinate transformations', i.e. moving points around on the manifold, or as `passive coordinate transformations', relabelling spacetime points\footnote{Here terminology could become an issue as many authors disagree on what they mean by diffeomorphism invariance and use `passive' vs `active' differently: passive usually being equivalent to coordinate transformations, with `active' being a statement about the theory itself. For example, Rovelli~\cite{Rovelli:1990ph} says that ``Passive diffeomorphism invariance is a property of the formulation of a dynamical theory, while active diffeomorphism invariance is a property of the dynamical theory itself". In this interpretation, only background-independent theories (explained below) can be actively diffeomorphism invariant, whilst other theories formulated in a covariant manner such as QED or QCD would not be~\cite{Gaul:1999ys}. In Section~\ref{section2.3.2} we make our definitions clear so as to avoid any confusion.}. Both interpretations are computationally equivalent and lead to the same transformation rules (\ref{Tensor transformation}). In Section~\ref{section2.3} we cover diffeomorphism invariance in greater detail.
 
It should be noted that almost all physical theories, including pre-relativistic theories, can be made covariant or diffeomorphism invariant by formulating them in a way that makes no explicit reference to coordinates (e.g., in the language of differential geometry, with the possible addition of fixed tensor fields). Many authors argue then that the principle of covariance is a vacuous statement~\cite{Gaul:1999ys,norton1993general,Carroll:2004st}. 

A related concept that \textit{does} set General Relativity apart from other physical theories is \textit{background independence}, also sometimes called `\textit{no prior geometry}'~\cite{Misner:1973prb,Wald:1984,norton1993general}. This is the notion that spacetime geometry, described by the metric tensor, is a dynamical variable of the theory and not fixed a priori. The metric is both acting on, and being acted upon by matter and energy in the universe~\cite{brown2005physical}.
Theories with a fixed metric, such as the Minkowski metric of Special Relativity, are not background independent. Background independence is what sets General Relativity apart from other gravitational theories with fixed metrics or additional non-dynamical structures, Nordstr\"{o}m theory being a prime example of the latter~\cite{Misner:1973prb}. Having relaxed the assumption of a fixed, flat spacetime metric in favour of a dynamical one, it becomes natural to consider the implications of spacetime \textit{curvature}.

\subsubsection{Minimal-coupling principle}
A final heuristic that is useful when transitioning from the flat, Minkowski space of Special Relativity, to the curved manifolds of General Relativity is the \textit{minimal-coupling principle}. This involves transforming the laws of physics of Minkowski spacetime, valid in inertial coordinates, to covariant laws that are valid in any set of coordinates and remain true in curved spacetime. Clearly this replacement is a necessity for a covariant description.

In the context of GR, minimal coupling entails the following replacements
\begin{equation}
\eta_{\mu \nu}(x) \rightarrow g_{\mu \nu}(x)  \quad \quad , \quad \quad \partial_{\mu} \rightarrow \nabla_{\mu} \, .
\end{equation}
The similarities with the gauge coupling procedure in gauge theories is immediately apparent, and will be explored further in Chapter~\ref{chapter3}. 

This replacement is exactly what we performed when going from the equation of motion of the freely-falling particle in~(\ref{straightline}) to the covariant geodesic equation~(\ref{prince.geo}), which is easily verified using the chain rule
\begin{align}
\frac{d^2 x^{\mu}}{d \lambda^2} =  \frac{d x^{\nu}}{d \lambda} \partial_{\nu} \Big( \frac{d x^{\mu}}{d \lambda}\Big) \nonumber
&\rightarrow  \frac{d x^{\nu}}{d \lambda} \nabla_{\nu} \Big( \frac{d x^{\mu}}{d \lambda}\Big)  \\
&= \frac{d^2 x^{\mu}}{d \lambda^2} + \Gamma^{\mu}_{\nu \rho}  \frac{d x^{\nu}}{d \lambda}  \frac{d x^{\rho}}{d \lambda}  \, .
\end{align}
This shows the simplicity of this principle.

There is some ambiguity in the procedure when second derivatives are present, due to the non-commutativity of covariant derivatives.
Take Maxwell's theory of electromagnetism in Minkowski spacetime as an example, described by the Lagrangian
\begin{align}
L_{\textrm{EM}} = -\frac{1}{4} F_{\mu \nu} F^{\mu \nu} - A_{\mu} J^{\mu} \, ,
\end{align}
where 
\begin{equation}
F_{\mu \nu} = \partial_{\mu} A_{\nu} - \partial_{\nu} A_{\mu} \, , \label{field strength}
\end{equation}
is the two-form field strength tensor, $A^{\mu}$ is the electromagnetic potential and $J^{\mu}$ is the vector current. Maxwell's equations are
\begin{align}
\partial_{\nu} F^{\mu \nu} &= J^{\mu} \label{Maxwell1} \\
\partial_{[\mu} F_{\nu \lambda]} &= 0 \, .  \label{Maxwell2}
\end{align}
Due to the commutativity of partial derivatives, the left-hand side of~(\ref{Maxwell1}) could be rewritten equally as 
\begin{equation} \label{ambiguity}
\partial_{\nu} \partial^{\mu} A^{\nu} - \partial_{\nu}  \partial^{\nu} A^{\mu} =  \partial^{\mu}  \partial_{\nu}A^{\nu} - \partial_{\nu}\partial^{\nu}   A^{\mu}  \, .
\end{equation}
These differ by a curvature term when the minimal coupling procedure is applied
\begin{align} \label{ambiguity2}
\partial_{\nu} \partial^{\mu} A^{\nu} - \partial_{\nu}  \partial^{\nu} A^{\mu} &\rightarrow \nabla_{\nu} \nabla^{\mu} A^{\nu} - \nabla_{\nu}  \nabla^{\nu} A^{\mu}  \nonumber \\
&= R^{\mu}{}_{\nu} A^{\nu} + \nabla^{\mu} \nabla_{\nu} A^{\nu} - \nabla_{\nu} \nabla^{\nu} A^{\mu} \, ,
\end{align}
whereas the Ricci tensor term would be absent if we started with the right-hand side of~(\ref{ambiguity}).

It turns out that writing the field strength in terms of covariant derivatives 
\begin{equation}
F_{\mu \nu} = \nabla_{\mu} A_{\nu} - \nabla_{\nu} A_{\mu} \, , \label{field strength2}
\end{equation}
which is equivalent to~(\ref{field strength}) due to the symmetry of the Christoffel symbols,
and directly applying the procedure on equations~(\ref{Maxwell1}) and~(\ref{Maxwell2}) leads to the correct formulation in curved spacetime
\begin{align}
\nabla_{\nu} F^{\mu \nu} &= J^{\mu} \, , \\
\nabla_{[\mu} F_{\nu \lambda]} &= 0 \, .
\end{align}
This matches the first choice~(\ref{ambiguity2}). 
Maxwell's equations can actually be seen much more clearly in the differential form notation, where the equations take the form $d F = 0$ and $d \star F = J$, where $\star$ is the Hodge star~\cite{Frankel:1997ec}. This holds in both flat and curved spacetimes, and it immediately follows that $d J =0$ by Poincar\'{e}'s lemma.

Lastly, we give a short example of the minimal coupling principle for Dirac theory, which requires the use of the tetrad and spin connection to couple fermions. In flat spacetime $e_{a}{}^{\mu} e_{b}{}^{\nu} \eta^{ab} = \eta^{\mu \nu}$ the Dirac equation is
\begin{equation} 
(i \gamma^{\mu} \partial_{\mu} - m) \Psi = 0 \, ,
\end{equation}
where $\gamma^{\mu} = \gamma^{a} e_{a}{}^{\mu}$ are the gamma matrices satisfying $\{ \gamma^{a} , \gamma^{b} \} = 2 \eta^{a b}$ and $\Psi(x)$ is a spinor field. The minimal coupling procedure in this case entails the replacement
\begin{equation}
\partial_{\mu} \rightarrow \mathfrak{D}_{\mu} \, ,
\end{equation}
where the (Fock-Ivaneko) covariant derivative is defined as acting on spinors as
\begin{equation}
\mathfrak{D}_{\mu} \Psi = \partial_{\mu} \Psi + \frac{i}{2} \mathring{\omega}_{\mu}{}^{ab} S_{ab} \, ,
\end{equation}
with $S_{ab}$ the Lorentz generator in the spinor representation~\cite{Aldrovandi:2013wha}
\begin{equation}\label{Lorentz generator}
S_{ab} = \frac{i}{4} [\gamma_{a} , \gamma_{b} ] \, .
\end{equation}
The Dirac equation in curved spacetime is then just
\begin{equation} 
(i \gamma^{\mu} \mathfrak{D}_{\mu} - m) \Psi = 0 \, ,
\end{equation}
which again highlights the simplicity of the procedure, even in the case of spinors. This concludes our examples of applying the minimal coupling principle.

Covariance, background independence, the equivalence principle and the minimal-coupling procedure do not, however, uniquely lead to General Relativity. One still needs to find the form of the equations of motion, or, equivalently, the action of General Relativity. Some intuition for what these laws should look like can be gained by revisiting Newton's theory of gravitation.

\subsubsection{Gravity and matter}
We now turn to the dynamical equations of motion that govern gravity. In Newtonian physics, the gravitational force between two masses $m_1$ and $m_2$ separated by the vector $\mathbf{r}$ is 
\begin{equation}
\mathbf{F} = -\frac{G m_1 m_2}{|\mathbf{r}|^2} \mathbf{\hat{r}} \, ,
\end{equation}
where $\mathbf{\hat{r}} = \mathbf{r} / |\mathbf{r}|$ is the unit vector and $G$ is Newton's gravitational constant. From Newton's second law $\mathbf{F} = m \mathbf{a}$ and the equivalence of gravitational and inertial mass, we see that gravitating objects feel an attractive acceleration in the direction of the unit vector.

In terms of the gravitational field $\mathbf{g}(\mathbf{r})$ defined by
\begin{equation}
\mathbf{g} = - G \frac{m_1}{|\mathbf{r}|^2}  \mathbf{\hat{r}} \, ,
\end{equation}
we have $\mathbf{F} = m \mathbf{g} = m \mathbf{a}$, where the gravitational field is simply the acceleration felt by an object under the influence of gravity.
The gravitational field is conservative and can hence be written in terms of the gradient of a scalar potential $\Phi(r)$
\begin{equation} \label{Newt1}
\mathbf{g} = -  \boldsymbol{\nabla} \Phi \, ,
\end{equation}
where here $\boldsymbol{\nabla}$ refers to the spatial gradient operator.
For a gravitating object with mass density $\rho$, Gauss's law can be written as
\begin{equation}  \label{Newt2}
\boldsymbol{\nabla} \cdot \mathbf{g} = - 4 \pi G \rho \, ,
\end{equation}
which leads to Poisson's equation
\begin{equation} \label{Newt3}
\boldsymbol{\nabla}^{2} \Phi = 4 \pi G \rho \, ,
\end{equation}
where $\boldsymbol{\nabla}^{2} = \delta^{ij} \partial_{i} \partial_{j}$ is the spatial Laplacian.

This is the type of equation we are looking to generalise in order to describe gravity in General Relativity. We are also required to reproduce the dynamics of Newtonian gravity in the so-called `Newtonian limit', which corresponds to slow speeds $v \ll c$ and weak, slowly evolving gravitational fields~\cite{Carroll:2004st}.
Taking the geodesic equation in this limit, we find
\begin{equation} \label{geodesic limit1}
\frac{d^2 x^{\mu}}{d \tau^2} + \Gamma_{tt}^{\mu} \frac{d t}{d \tau}  \frac{d t}{d \tau}= 0 \, ,
\end{equation} 
where we have used that $dx^{i}/d \tau \ll dt/d \tau$ from $v \ll c$. For a slowly evolving gravitational field, the time derivative of the metric will be approximately zero $\partial_{t} g_{\mu \nu} \approx 0$, whilst the condition of weak gravitational fields implies that the metric can be written as a perturbation around flat Minkowski space 
\begin{equation}
g_{\mu \nu} = \eta_{\mu \nu} + h_{\mu \nu} \, ,
\end{equation}
with $|h_{\mu \nu}| \ll 1$ (in Cartesian coordinates). Plugging these into~(\ref{geodesic limit1}) and using the definition of the Levi-Civita connection leads to
\begin{equation} \label{geodesic limit2}
\frac{d^2 x^{\mu}}{d \tau^2} - \frac{1}{2} \eta^{\mu \nu} \partial_{\nu} h_{tt}  \frac{d t}{d \tau}  \frac{d t}{d \tau} = 0 \, .
\end{equation}
The time component $\mu=t$ of the equation is simply
\begin{equation}
\frac{d^2 t}{d \tau^2} = 0 \, ,
\end{equation}
whilst the spatial components can be written as
\begin{equation}
\frac{d^2 x^{i}}{d t^2} - \frac{1}{2} \partial^{i} h_{tt} = 0 \, ,
\end{equation}
by dividing through by $(d t/d \tau)^2$ or simply using the solution $t(\tau) = a \tau + b$.

Looking back at our Newtonian equations~(\ref{Newt1})-(\ref{Newt3}), we can make the identification between the gravitational acceleration $\mathbf{g}(\mathbf{r})$ and our acceleration term $d^2 x^{i} / d t^2$, and the gravitational potential $\Phi$ and the time-component of the metric perturbations
\begin{equation}
 h_{tt} = - 2 \Phi \, .
\end{equation}
We therefore recover the Newtonian limit if the metric takes the form $g_{tt} = -(1+2 \Phi)$. 

Now we wish to find a set of fully covariant gravitational field equations that reduce to the Poisson equation $\boldsymbol{\nabla}^2 \Phi = 4 \pi G \rho $ in the Newtonian limit studied above. 
 Recall that in this limit, the connection has components
\begin{equation}
\Gamma^{i}_{tt} = - \frac{1}{2} \partial^{i} h_{tt} = \partial^{i} \Phi \, .
\end{equation}
It then follows that the Ricci tensor has components
\begin{equation}
R_{t t} = \partial_{i} \partial^{i} \Phi = \boldsymbol{\nabla}^{2} \Phi \, .
\end{equation}
Next, we note that the energy-momentum tensor for a perfect fluid is
\begin{equation}
T_{\mu \nu} = (\rho + p) U_{\mu} U_{\nu} + p g_{\mu \nu} \, ,
\end{equation}
where $U^{\mu}$ is the timelike fluid four-velocity satisfying $U^{\mu} U_{\mu} =-1$. In the Newtonian limit, where $\rho$ is small, we have the non-vanishing component $T_{tt} = \rho$. It then follows that the Poisson equation can be written as
\begin{equation}
R_{tt} = 4 \pi G T_{tt}    \, .
\end{equation}

This gives us a good starting point from which to guess the form of the covariant gravitational equations. One extra requirement is that the conservation of energy in flat spacetime should be generalised via the minimal coupling procedure to a covariant conservation law
\begin{equation}
\partial_{\mu} T^{\mu \nu} =0 \rightarrow \nabla_{\mu} T^{\mu \nu} = 0 \,.
\end{equation}

The first logical guess would be $R_{\mu \nu} = 4\pi G T_{\mu \nu}$. However, the covariant conservation of energy-momentum implies $\nabla_{\mu} R^{\mu \nu} = 0$, from which it follows from the twice-contracted Bianchi identities~(\ref{Bianchi2}) that $ \nabla_{\mu} R=  \nabla_{\mu} T =0$, where $T = g^{\mu \nu} T_{\mu \nu}$ is the energy-momentum trace. Clearly this implies that $T$ is constant at all points on the manifold, which is not a very satisfactory theory to describe both the vacuum and matter sources.

The next choice is the Einstein tensor
\begin{equation} \label{EFE1}
 R_{\mu \nu} -\frac{1}{2} g_{\mu \nu} R =  \kappa T _{\mu \nu} \, ,
\end{equation}
where $\kappa$ is a constant related to $4 \pi G$. Following the same arguments as before, in the Newtonian limit one finds $R = 2 \boldsymbol{\nabla}^2 \Phi$ and the coupling constant is determined to be $\kappa = 8 \pi G$. Due to the twice-contracted Bianchi identity in the Levi-Civita framework~(\ref{Bianchi2}), the covariant derivative of the Einstein tensor vanishes identically, implying the covariant conservation of matter. These are the field equations of General Relativity, known as the \textit{Einstein field equations}.

Recall from Lovelock's theorem that we are also free to add a constant term to the field equations
\begin{equation}
G_{\mu \nu} + g_{\mu \nu} \Lambda = \kappa T_{\mu \nu} \, .
\end{equation}
The cosmological constant $\Lambda$ is necessary to account for the observed accelerating expansion of the universe, though as discussed in the introduction, this is not without problems. The Einstein field equations with the inclusion of the cosmological constant are then the geometric laws which govern the gravitational interaction.

To summarise, in General Relativity the Lorentzian manifold of Special Relativity $\mathbb{R}^{(1,3)}$ is no longer assumed to be flat, leading to a general curved manifold. Due to this manifold structure, spacetime locally looks like Minkowski space, and we recover the physics of Special Relativity on small enough scales, in accordance with the principle of equivalence. The Einstein field equations are geometric in nature and agree with Newton's law of gravitation in the appropriate limits. 

The connection is chosen to be the Levi-Civita one, and the motions of particles are described by these Levi-Civita geodesics. The fundamental object of importance is the metric tensor alone $g_{\mu \nu}(x)$, from which all relevant tensorial quantities are derived. The minimal coupling procedure gives us a simple way of incorporating the non-gravitational laws of physics into Einstein's geometric theory of General Relativity.

\subsection{Einstein-Hilbert action}
\label{section2.2.2}

Einstein's theory can be derived from a Lagrangian approach, which is noticeably more direct and elegant.
Once the Riemannian setting has been specified, simplicity and aesthetics can serve as a guide to choosing the form of the scalar action. And indeed, the simplest geometric scalar derived from the metric tensor in four dimensions is the Ricci scalar~(\ref{Ricci scalar}). To be more formal, one can propose that the action should satisfy the following constraints:
\begin{enumerate}
\item Be a scalar constructed from just the metric tensor $S_{\textrm{grav}} = S_{\textrm{grav}}[g,\partial g, ...]$
\item Contain no higher than second derivatives of the metric $S_{\textrm{grav}} = S_{\textrm{grav}}[g,\partial g, \partial^2 g]$
\item Be at most linear in those second derivatives $\partial^2 g$
\item Vanish in flat spacetime $g_{\mu \nu} = \eta_{\mu \nu}$
\end{enumerate}
The only choice is Ricci scalar $R$, which is also the simplest non-trivial choice of scalar constructed from the metric tensor and its derivates. There are of course alternative axiomatisations that lead to General Relativity (see for example section 17.5 of MTW for six distinct routes~\cite{Misner:1973prb}), but the resulting action and equations are the same.

The action formed from the Ricci scalar is the Einstein-Hilbert action
\begin{equation} \label{EH}
S_{\textrm{EH}} = \frac{1}{2 \kappa} \int R  \sqrt{-g}  d^4x \, ,
\end{equation}
where $\sqrt{-g}  d^4x$ is the invariant volume measure and $g$ is the metric determinant. A simple calculation verifies that the measure is a coordinate scalar, such that the action $S_{\textrm{EH}}$ is invariant. The replacement $d^4x \rightarrow \sqrt{-g}  d^4x$ can also be seen as part of the minimal coupling procedure when applied to actions.

Let us sketch out the derivation of the field equations from this action, which can be found in most textbooks\cite{Misner:1973prb,Carroll:2004st,Wald:1984}. From the stationary action principle, small variations in the field variables will leave the action functional unchanged
\begin{equation}
\delta S_{\textrm{EH}}[g] = 0 \, ,
\end{equation} 
where the fields are the metric tensor $g_{\mu \nu}$. A few useful relations in the derivation are
\begin{equation}
\delta g^{\mu \nu} = - \delta g_{\alpha \beta} g^{\alpha \mu} g^{\beta \nu} \, , \quad \delta \sqrt{-g} = \frac{1}{2} \sqrt{-g}  g^{\mu \nu} \delta g_{\mu \nu} \, .
\end{equation}

It can then be shown using standard definitions that the variation of the Levi-Civita Ricci scalar is
\begin{equation}
\delta R = \big( \nabla^{\mu} \nabla^{\nu} - g^{\mu \nu} \Box - R^{\mu \nu} \big) \delta g_{\mu \nu} \, ,
\end{equation}
where $\Box = \nabla^{\mu} \nabla_{\mu}$ is the Levi-Civita d'Alembert operator. This can also be derived from the so-called Palatini identity 
\begin{equation}
\delta R_{\mu \nu} = \nabla_{\lambda} \delta \Gamma^{\lambda}_{\nu \mu} - \nabla_{\nu} \delta \Gamma^{\lambda}_{\lambda \mu} \, .
\end{equation}
Plugging this into the variation of the Einstein-Hilbert action therefore leads to
\begin{equation}
\begin{multlined}
\delta S_{\textrm{EH}} = \frac{1}{2 \kappa} \int \sqrt{-g} \delta g^{\mu \nu} \big(R_{\mu \nu} - \frac{1}{2}g_{\mu \nu} R) d^4 x  \\ \quad \quad \quad - \frac{1}{2 \kappa}  \int \sqrt{-g} \nabla_{\lambda} \Big( \nabla_{\nu} \delta g^{\lambda \nu} - g_{\mu \nu} \nabla^{\lambda} \delta g^{\mu \nu}  \Big) d^4 x \, .
\end{multlined}
\end{equation}
The first term gives the Einstein field equations, whilst the final term takes the form of a total derivative and will vanish on a manifold without a boundary. If the manifold is not compact, we must do a little more work to make sure our variation procedure is well-posed. This is because the boundary term includes variations of the metric and their first derivatives, and Dirichlet boundary conditions fix just the metric at the boundary $\delta g_{\mu \nu} |_{\partial \mathcal{M}} = 0$. For an overview, and the approach we will follow, see Wald~\cite{Wald:1984}. For a nice modern review that focuses on this subject in General Relativity and modified theories, see~\cite{dyer2009boundary}.

Firstly we state the generalised Stoke's theorem, which will be important for many of the action variations throughout the thesis. In the language of differential forms, for an orientable manifold $\mathcal{M}$ with boundary $\partial \mathcal{M}$, the theorem states that\footnote{For a more detailed treatment see~\cite{Lee00}.}
\begin{equation}
\int_{\partial \mathcal{M}} \omega = \int_{ \mathcal{M}} d \omega \, ,
\end{equation}
where $\omega$ is a form and $d$ the exterior derivative~(\ref{exterior}). For a manifold without a boundary this clearly vanishes.
In tensorial language we can write this for the divergence of vectors (i.e., duals of one-forms) on our four-dimensional spacetime manifold
\begin{equation}
\int_{ \mathcal{M}} \sqrt{-g} \nabla_{\mu} V^{\mu} d^4x  = \int_{ \mathcal{M}} \partial_{\mu} (\sqrt{-g} V^{\mu}) d^4x = \int_{\partial \mathcal{M}}  V^{\mu} n_{\mu} \sqrt{-\gamma} d^3 x \, ,
\end{equation}
where $\gamma_{\mu \nu} = g_{\mu \nu} \pm n_{\mu} n_{\nu}$ is the induced metric on the boundary. In what follows we take $n_{\nu}$ to be the (spacelike) outwards unit normal vector.

Applying this to our boundary term yields
\begin{align} \label{EHbound_var}
\delta S_{\textrm{EH}}^{\textrm{boundary}}  &= - \frac{1}{2 \kappa}  \int_{ \mathcal{M}} \partial_{\lambda} \Big( \sqrt{-g} (\nabla_{\nu} \delta g^{\lambda \nu} - g_{\mu \nu} \nabla^{\lambda} \delta g^{\mu \nu}  ) \Big) d^4 x \nonumber \\
&= - \frac{1}{2 \kappa}  \int_{\partial \mathcal{M}} n_{\lambda} \big(  \nabla_{\nu} \delta g^{\lambda \nu} - g_{\mu \nu} \nabla^{\lambda} \delta g^{\mu \nu} \big) \sqrt{-\gamma} d^3 x \nonumber \\
&= \frac{1}{2 \kappa}  \int_{\partial \mathcal{M}}  n^{\lambda} g^{\mu \nu} \big( \nabla_{\mu} \delta g_{\nu \lambda} - \nabla_{\lambda}  \delta g_{\mu \nu} \big) \sqrt{-\gamma} d^3 x 
\, .
\end{align}
Using the decomposition $\gamma_{\mu \nu} = g_{\mu \nu} - n_{\mu} n_{\nu}$ gives
\begin{align} 
\delta S_{\textrm{EH}}^{\textrm{boundary}}  &= \frac{1}{2 \kappa}  \int_{\partial \mathcal{M}}  n^{\lambda} \gamma^{\mu \nu} \big( \nabla_{\mu} \delta g_{\nu \lambda} -\nabla_{\lambda}  \delta g_{\mu \nu}\big) \sqrt{-\gamma} d^3 x \nonumber \\
&= - \frac{1}{2 \kappa}  \int_{\partial \mathcal{M}}  n^{\lambda} \gamma^{\mu \nu} \nabla_{\lambda}  \delta g_{\mu \nu}\sqrt{-\gamma} d^3 x 
\, , \label{EHbound}
\end{align}
where we have used that the tangential derivatives vanish on the boundary $\gamma^{\mu \nu} \nabla_{\mu} \delta g_{\nu \lambda} =0$ from $\delta g^{\mu \nu} = 0$ on $\partial \mathcal{M}$.

Ensuring this remaining term vanishes is usually solved via the introduction of the Gibbons-Hawking-York counterterm~\cite{PhysRevD.15.2752}, constructed from the trace of the intrinsic curvature on the boundary $K= \gamma^{\mu \nu} K_{\mu \nu} = \gamma^{\mu}{}_{\nu} \nabla_{\mu} n^{\nu}$ 
\begin{equation} \label{GHY}
S_{\textrm{GHY}} = \frac{1}{\kappa} \int_{\partial \mathcal{M}} K \sqrt{-\gamma}d^3 x \, .
\end{equation}
Varying the intrinsic curvature leads to~\cite{Wald:1984}
\begin{equation}
\delta K = \frac{1}{2} n^{\lambda} \gamma^{\mu \nu} \nabla_{\lambda} \delta g_{\mu \nu} \, ,
\end{equation}
which exactly cancels the boundary term variation~(\ref{EHbound}). Hence, the total variations in the presence of a non-trivial boundary lead to the Einstein field equations
\begin{equation} \label{EH variation}
\delta S_{\textrm{EH}} + \delta S_{\textrm{GHY}} =  \frac{1}{2 \kappa} \int_{\mathcal{M}} \sqrt{-g} \delta g^{\mu \nu} \big(R_{\mu \nu} - \frac{1}{2}g_{\mu \nu} R) d^4 x  \, .
\end{equation}
Note that this additional counter term is not strictly covariant because it depends on the choice of foliation~\cite{chakraborty2017boundary}. We will revisit this issue in Chapter~\ref{chapter4}.

For modified theories, analogous counter-terms need to be found and subtracted from the action. The calculations are far from simple in generalised theories, but this has been done, for example, in $f(R)$ gravity~\cite{dyer2009boundary} and more general theories of the form $f(L_{\textrm{Lovelock}})$~\cite{Bueno:2016dol} and $f(R_{\gamma \mu \nu}{}^{\lambda})$~\cite{Deruelle:2009zk}. These counter terms play a prominent role in the ADM formalism~\cite{arnowitt1959dynamical,Deruelle:2009zk} and in topics relating to the entropy of black holes~\cite{gibbons1977action,Brown:1992bq,TP:2010,briscese2008black}.
It should also be noted that if one wants to make the action finite, additional non-dynamical boundary terms may need to be added. See~\cite{barrow1988action} for an interesting discussion on this topic in the context of cosmology.

Returning to our action, let us now include minimally coupled matter $S_{\textrm{M}}[g,\varPhi^A]$, where $\varPhi^A$ refers to any matter fields (e.g., scalars, tensors, spinors, etc). Working in the Levi-Civita framework, any derivatives $\nabla$ of the fields can be written in terms of the metric. The condition of minimal coupling means that the metric and its derivatives only enter into the matter Lagrangian via the Levi-Civita covariant derivative and not, say, through any direct couplings such as $R \varPhi$. 

Dropping the GHY counter term for simplicity, the variation of the total action with respect to the metric is
\begin{align}
\delta S_{\textrm{total}} &= \delta S_{\textrm{EH}} + \delta S_{\textrm{M}}  \nonumber \\ 
&=\frac{1}{2\kappa} \int \frac{S_{\textrm{EH}} }{\delta g^{\mu \nu}} \delta g^{\mu \nu}  d^4x + \int \frac{ \delta S_{\textrm{M}} }{\delta g^{\mu \nu } } \delta g^{\mu \nu} d^4x  \nonumber \\
&= \frac{1}{2\kappa} \int  \delta g^{\mu \nu} \big(R_{\mu \nu} - \frac{1}{2}g_{\mu \nu} R) \sqrt{-g} d^4 x - \frac{1}{2} \int \delta g^{\mu \nu} T_{\mu \nu}  \sqrt{-g}  d^4x \, ,
\end{align}
where we define the (Hilbert) metric energy-momentum tensor as
\begin{equation} \label{energy-momentum tensor}
T_{\mu \nu} = -\frac{2}{\sqrt{-g}} \frac{\delta S_{\textrm{M}} }{\delta g^{\mu \nu}} \, .
\end{equation}
The gravitational field equations are then
\begin{equation} \label{EFE}
R_{\mu \nu} - \frac{1}{2}g_{\mu \nu} R = \kappa T_{\mu \nu} \, ,
\end{equation}
from which it follows that $\nabla_{\mu} T^{\mu \nu} = 0$.
The variations of $S_{\textrm{M}}[g,\varPhi^{A}]$ with respect to the fields $\varPhi^{A}$ lead to the standard matter equations of motion.

\subsubsection{Tetradic action}
In the above derivation we worked exclusively in the coordinate basis. But we can instead regard the tetrad field $\mathbf{e}_{a}$ as the fundamental variable of the theory, writing our action in differential form notation as
\begin{equation} \label{EH tetrad0}
S_{\textrm{EH}}[e] = \frac{1}{2 \kappa} \int \epsilon_{abcd} \big( \frac{1}{4} \mathbf{e}^{a} \wedge \mathbf{e}^{b} \wedge R^{cd} \big) \, .
\end{equation}
Here one can interpret the tetrad as opposed to the metric as representing the gravitational field~\cite{Rovelli:2004tv}.
Equivalently, in full index notation we have
\begin{equation} \label{EH tetrad}
S_{\textrm{EH}}[e] = \frac{1}{2 \kappa} \int R(e) e d^4x \, ,
\end{equation}
where $R(e) = e_{a}{}^{\mu} e_{b}{}^{\nu} R_{\nu \mu}{}^{ab}(\mathring{\omega}(e))$ from~(\ref{Riemann mixed}) and $e = \sqrt{-g}$ is the tetrad determinant. This is why the tetrad is sometimes called the `square root of the metric'. Clearly the Ricci scalar is both a coordinate and local Lorentz scalar, and so the $R(e)$ in~(\ref{EH tetrad}) and $R(g) = g^{\mu \nu} R_{\lambda \mu \nu}{}^{\lambda}(\Gamma(g))$ are completely equivalent. Hence the action~(\ref{EH tetrad}) is also equivalent to the metric formulation, as we will now prove.

 From $g_{\mu \nu} = e^{a}{}_{\mu} e^{b}{}_{\nu} \eta_{ab}$ we can relate the variations of the metric to variations of the tetrad
\begin{align}
\delta g_{\mu \nu} &=   \delta e^{a}{}_{\mu} e^{b}{}_{\nu} \eta_{ab} + e^{a}{}_{\mu} \delta e^{b}{}_{\nu} \eta_{ab}
\nonumber \\
&= -\delta e_{a}{}^{\lambda} \big(g_{\mu \lambda} e^{a}{}_{\nu} + g_{\nu \lambda} e^{a}{}_{\mu} \big) \, .
\end{align}
The variation of the action~(\ref{EH tetrad}) with respect to the tetrad is then
\begin{equation} \label{EH tetrad var}
\delta S_{\textrm{EH}}[e] = \frac{1}{\kappa} \int \delta e_{a}{}^{\nu} e^{a}{}_{\mu}  \big(R^{\mu}{}_{\nu} - \frac{1}{2}\delta^{\mu}_{\nu} R\big)  e \, d^4x \, ,
\end{equation}
leading to the total tetrad field equations 
\begin{equation} \label{EFE tetrad}
\big(R^{\mu}{}_{\nu} - \frac{1}{2} \delta^{\mu}_{\nu} R) e^{a}{}_{\mu} = \kappa e^{a}{}_{\mu} T^{\mu}{}_{\nu} \, .
\end{equation}
Contracting with $\eta_{ab} e^{b}{}_{\lambda}$ leads to the standard Einstein field equations~(\ref{EFE}). Hence these are field equations for the \textit{metric}, only determining the tetrad up to local Lorentz transformations~(\ref{Lorentz1})~\cite{Weinberg:1972kfs}. This is what we expected because we started with a Lorentz-invariant action. We delay the discussion of the energy-momentum tensor $T^{a}{}_{\nu}$ appearing on the RHS of~(\ref{EFE tetrad}) until Chapter~\ref{chapter3}.

In the following section we will look more closely at the implications of local Lorentz invariance, but it is already clear from the previous calculations what the conclusion will be. If the action is a local Lorentz scalar then the equations of motion will share this Lorentz covariance. This implies that the field equations, like~(\ref{EFE tetrad}), will determine the metric and not the tetrad. Therefore the equations of motion will be symmetric, from the symmetry of $g_{\mu \nu}$. In other words, the tetrad only contributes to the dynamical equations through its relation to the metric $g_{\mu \nu} = e^{a}{}_{\mu} e^{b}{}_{\nu} \eta_{ab}$, up to local Lorentz transformations.

\subsubsection{Gravitons}
The postulates in the previous section give us an intuitive and physical justification for the Einstein field equations, and the variation procedure above gives us an elegant mathematical derivation in the setting of Riemannian geometry. However, from a more modern perspective, it is illustrative to use field-theoretic language. Beginning with spin-2 particles in the linearised theory, one can build up to Einstein's full non-linear theory. Nowadays these types of arguments are known as bootstrapping. We will briefly sketch the idea here, following Feynman's lectures on gravitation~\cite{feynman2018feynman,preskill1995foreword}.

The quantum mechanical massless spin-2 graviton has two helicity states, implying there should be two degrees of freedom in the classical gravitational theory. A spin-2 particle is described by a symmetric tensor $h_{\mu \nu}$, which propagates in flat Minkowski space. It has ten components, with the six $h_{ij}$ being dynamical~\cite{feynman2018feynman}. To remove the additional degrees of freedom we will require the theory to have some gauge invariance.

The equations of motion for a free massless spin-2 field on a Minkowski background are derived from the (massless) Fierz-Pauli Lagrangian~\cite{Fierz:1939ix}
\begin{equation} \label{FP}
\mathcal{L}_{\textrm{FP}} = \frac{1}{2} \partial_{\lambda} h_{\mu \nu} \partial^{\lambda} h^{\mu \nu} - \partial_{\lambda} h^{\lambda \mu} \partial_{\nu} h^{\nu}{}_{\mu} + \partial_{\lambda} h^{\lambda}{}_{\nu} \partial^{\nu} h - \frac{1}{2} \partial_{\lambda} h \partial^{\lambda} h \, ,
\end{equation}
where $h_{\mu \nu}$ is the symmetric tensor field describing the massless spin-2 field, and $h = h^{\mu}{}_{\mu}$ is the trace. The theory is invariant under the gauge transformations 
\begin{equation} \label{gauge}
h_{\mu \nu} \rightarrow h_{\mu \nu} + 2 \partial_{(\mu} \xi_{\nu)} \, ,
\end{equation}
ensuring that the theory has the correct number of degrees of freedom.
We will later see that these are just linearised diffeomorphisms. It is not difficult to check that~(\ref{FP}) is the most general, invariant, quadratic Lagrangian up to second derivatives, modulo integration by parts~\cite{zee2013einstein}.

Next we linearly couple our theory to matter
\begin{equation} \label{coupled}
\mathcal{L}_{\textrm{coupled}}  = \mathcal{L}_{\textrm{FP}} - \frac{\lambda}{2} h_{\mu \nu}T^{\mu \nu} \, ,
\end{equation}
where $\lambda$ is the coupling constant and $T^{\mu \nu}$ the matter energy-momentum tensor (independent of $h_{\mu \nu}$).
 The variations with respect to $h_{\mu \nu}$ lead to the gravitational equations of motion. From the field equations and 
 the gauge invariance\footnote{In Sec.~\ref{section2.3.2} we show explicitly how conservation laws are a consequence of gauge invariance, see equation~(\ref{EH diff}).} of the Fierz-Pauli Lagrangian~(\ref{gauge}) we have that $\partial_{\mu} T^{\mu \nu}=0$. However, this presents a problem, because the matter equations of motion, usually satisfying $\partial_{\mu} T^{\mu \nu}=0$, will no longer satisfy this equation due to the new coupling in the action $h_{\mu \nu}T^{\mu \nu}$. Hence there is an inconsistency between the gravitational and matter field equations.

The resolution to this problem is to assume that the graviton couples not only to matter but also to itself, in the form of its own gravitational energy-momentum tensor. Therefore higher order, non-linear terms need to be added to $\mathcal{L}_{\textrm{coupled}}$. The procedure itself is not so simple, and usually involves iterating an infinite number of times to obtain the full non-linear theory~\cite{preskill1995foreword}. To put it qualitatively, the argument as laid out by Gupta~\cite{Gupta:1954zz} is that the source term for the spin-2 field equations should be modified by introducing a new energy-momentum tensor (which includes the energy-momentum of the graviton itself). If the source term is chosen to be the original matter tensor $T_{\mu \nu}$ of the free theory, then coupling this to $h_{\mu \nu}$ leads to cubic terms in the action. Repeating this process, again requiring the divergence to vanish, leads to quartic terms, and so on. The aim is that the infinite series can be summed to the full non-linear theory, though this was not fully carried out by Gupta and others like Feynman. Kraichnan gave the first full derivation~\cite{kraichnan1955special}, while Thirring obtained the EFE from arguments focussing on gauge invariance~\cite{thirring1961alternative}. For a different but very elegant approach leading to the same conclusions, without needing an infinite number of iterations, see Deser~\cite{Deser:1969wk}.

This highlights why gravity must be described by a non-linear theory: the graviton carries energy-momentum, and energy-momentum is the source of gravity, and hence the graviton itself must gravitate. This is exactly the same as in non-Abelian gauge theories such as Yang-Mills, where the gauge fields couple to their own gauge currents (unlike in Abelian theories such as electromagnetism). The key difference between Yang-Mills theory and gravity is that, for Yang-Mills, the non-linear coupling only includes \textit{quartic terms}; in gravity, this goes on forever~\cite{coleman2019quantum}. 

Either way, the resulting theory, with the full metric tensor expanded around flat Minkowski space as $g_{\mu \nu} = \eta_{\mu \nu} + \lambda h_{\mu \nu}$ and $\lambda \propto \sqrt{G}$, leads to the Einstein field equations.
Working at the level of action, however, there is the caveat that none of these iterative procedures beginning with the massless spin-2 theory should actually obtain the full Einstein-Hilbert action~\cite{Padmanabhan:2004xk}. At least, not without further assumptions such as diffeomorphism invariance.
 This can be seen schematically by noting that the Einstein-Hilbert action $R \sim \partial \Gamma + \Gamma^2$ to lowest order is
\begin{equation}
\mathcal{L}_{\textrm{EH}} \propto \frac{1}{\lambda^2} R \propto (\partial h)^2 + \frac{1}{\lambda} \partial^2 h \, ,
\end{equation}
with the second term being non-analytic in the coupling term $\lambda$. This result is proven in~\cite{Padmanabhan:2004xk}, where it is shown that only the quadratic part of the Ricci scalar can be obtained via this method, and not its boundary term (see equation~(\ref{RGB}) for this decomposition into quadratic and boundary parts). This is quite interesting for our work, as it is this quadratic part that will be the starting point for the modified theories in the following chapters. 

More recent bootstrapping calculations that do indeed compute the full iterative process (though with a slightly different starting point than the Fauz-Pauli action~(\ref{FP})) have been calculated by Butcher et. al.~\cite{Butcher:2009ta}. However, they also acknowledge the arguments of Padmanabhan that the Einstein-Hilbert action can only be obtained \textit{modulo surface terms}, i.e., without the non-analytic boundary part. This can alternatively be viewed as the statement that the full diffeomorphism invariance of the Lagrangian cannot be obtained by this procedure and must instead be assumed. These topics will be discussed in the following section as well as in Chapter~\ref{chapter4}. Another more recent bootstrapping approach has been carried out in the fully metric-affine setting~\cite{Delhom:2022zbi}, as well as the symmetric teleparallel geometries~\cite{BeltranJimenez:2018vdo}. The latter of these reproduces a `covariant version' of the quadratic part of the Ricci scalar $\Gamma^2$, which we discuss in Sec.~\ref{section4.3}.

\section{Symmetries}
\label{section2.3}

In this section, the concepts of symmetry and covariance in the context of General Relativity are covered. Importantly, we distinguish between \textit{physical symmetries}, or \textit{isometries} of the spacetime metric, and \textit{fundamental symmetries}, which are built into our theories. Isometries describe particular symmetries of a given spacetime geometry, whereas fundamental symmetries can be likened to gauge symmetries. We then study how these invariances give rise to conservation laws. Some of these topics we have already encountered, such as the gauge-invariance of the linearised theory~(\ref{FP}). 

Another type of symmetry that will be especially useful in the next chapter is \textit{projective invariance}. This is related to transformations of the affine connection. For further details on projective transformations, such as how it applies to the affine Ricci scalar, see Appendix~\ref{appendixA}. 
 In the remainder of this section, we focus on isometries and fundamental symmetries. For reference, we mostly follow the conventions and presentations used in the textbooks of Wald~\cite{Wald:1984}, Hawking \& Ellis~\cite{Hawking:1973uf} or Frankel~\cite{Frankel:1997ec}.

\subsection{Isometries}
\label{section2.3.1}
To study physical symmetries of the spacetime, we look at symmetry transformations of the metric $\psi^{*}g(\psi(p))= g(p)$ where $\psi$ is a diffeomorphism (smooth, invertible map from the manifold to itself, see Sec.~\ref{section2.3.2}). The operation $\psi^*$ represents the pullback~\cite{Wald:1984,Frankel:1997ec}, and $g(\psi(p))$ simply refers to the metric evaluated at the point $\psi(p)$~\cite{Nakahara:2003nw}. The notation will become clear in the following.

 It is useful to introduce the Lie derivative, which measures the change under an infinitesimal diffeomorphism generated by a vector field $\xi$
\begin{equation} \label{Lie derivative diff} 
\mathcal{L}_{\xi} T(p) = \lim_{t \to 0} \frac{T(p) - \psi^{*}_{t} T(\psi_{t} (p))}{t} \, ,
\end{equation}
where $T$ is an arbitrary tensor and $t$ parameterises the flow along $\xi$. To be explicit, $T(p)$ is the tensor evaluated at the point $p$, whilst the term $\psi^{*}_{t} T(\psi_{t} (p))$ is the tensor at $\psi(p)$ pulled back to $p$, so that both terms are evaluated at the same spacetime point. This is necessary to have a meaningful derivative. It is then clear why in the symmetry transformation we have the object $\psi^{*}g(\psi(p))$ as opposed to $\psi^{*}g(p)$. Strictly speaking, these are one-parameter families of diffeomorphisms.

One can work with the more familiar coordinate transformations instead of abstract diffeomorphisms, which are computationally equivalent (for an explicit verification see~\cite{Frankel:1997ec,Wald:1984}). In this case, we look at the infinitesimal coordinate transformation $x^{\mu} \rightarrow \hat{x}^{\mu} = x^{\mu} + \epsilon \xi^{\mu}(x)$ where $|\epsilon| \ll 1$ is small. The Lie derivative can then be restated in local coordinates as
\begin{equation} \label{Lie derivative}
\mathcal{L}_{\xi} T(x) = \lim_{\epsilon \to 0} \frac{T(x) - \hat{T}(x)}{\epsilon} \, ,
\end{equation}
where $\hat{T}(x)$ is the transformed tensor evaluated in the original coordinates $x^{\mu}$. We then can use the usual transformation rules for tensors (\ref{Tensor transformation}), whilst remembering that we want $\hat{T}(x)$ not $\hat{T}(\hat{x})$, to get the Lie derivative of a tensor in local coordinates~\cite{Wald:1984}
\begin{equation} \label{Lie derivative coords}
\begin{split}
\mathcal{L}_{\xi} T^{\mu_1 ... \mu_k}{}_{\nu_1... \nu_l} =  \xi^{\alpha} \partial_{\alpha} T^{\mu_1 ... \mu_k}{}_{\nu_1... \nu_l} - \sum^{k}_{i=1} T^{... \mu_{i-1} \alpha \mu_{i+1} ...}{}_{\nu_1... \nu_l} \partial_{\alpha} \xi^{\mu_i} \\ +
\sum^{l}_{i=1} T^{\mu_1 ... \mu_k}{}_{... \nu_{i-1} \alpha \nu_{i+1}...} \partial_{\nu_i} \xi^{\alpha} \, .
\end{split}
\end{equation}

To see how we obtain an equation of the form (\ref{Lie derivative coords}), let us take the simple example of a vector $v^{\mu}$. From (\ref{Tensor transformation}) we know that under a general coordinate transformation 
\begin{equation} \label{vector example}
v^{\mu}(x) \rightarrow \hat{v}^{\mu}(\hat{x}) = \frac{\partial \hat{x}^{\mu}}{\partial x^{\nu}} v^{\nu}(x) \, .
\end{equation}
For the infinitesimal coordinate transformation  $x^{\mu} \rightarrow \hat{x}^{\mu} = x^{\mu} + \epsilon \xi^{\mu}(x)$ this becomes
\begin{equation} \label{vector example 2}
\hat{v}^{\mu}(\hat{x}) = (\delta^{\mu}_{\nu} + \epsilon \partial_{\nu} \xi^{\mu}) v^{\nu}(x) = v^{\mu}(x) + \epsilon v^{\nu}(x) \partial_{\nu} \xi^{\mu} \, .
\end{equation}
Taylor expanding the left hand side gives
\begin{align}
\hat{v}^{\mu}(x^{\nu} + \epsilon \xi^{\nu}(x)) &= \hat{v}^{\mu}(x) + \epsilon \partial_{\nu} \hat{v}^{\mu}(x) \xi^{\nu} + \mathcal{O} (\epsilon^2) \nonumber  \\
&=  \hat{v}^{\mu}(x) + \epsilon \xi^{\nu} \partial_{\nu} \big( v^{\mu}(x) + \mathcal{O}(\epsilon) \big) + \mathcal{O} (\epsilon^2) \nonumber \\
&= \hat{v}^{\mu}(x) +  \epsilon \xi^{\nu} \partial_{\nu} v^{\mu}(x) + \mathcal{O} (\epsilon^2)  \, .
\end{align}
 Putting this back into the (\ref{vector example 2}) we have
 \begin{align}
 \hat{v}^{\mu} +  \epsilon \xi^{\nu} \partial_{\nu} v^{\mu}  &= v^{\mu} + \epsilon v^{\nu} \partial_{\nu} \xi^{\mu}  + \mathcal{O} (\epsilon^2) \, ,
 \end{align}
 where everything is now a function of the original coordinates $x^{\mu}$. We can now rearrange these terms to match the form of the Lie derivative expression (\ref{Lie derivative})
\begin{equation}
v^{\mu}(x) - \hat{v}^{\mu}(x) = \epsilon \big(\xi^{\nu} \partial_{\nu}v^{\mu} - v^{\nu} \partial_{\nu} \xi^{\mu}) + \mathcal{O}(\epsilon^2) \, .
 \end{equation}
 We therefore find using (\ref{Lie derivative}) that
 \begin{align}
 \mathcal{L}_{\xi} v^{\mu}(x) =  \lim_{\epsilon \to 0} \frac{v^{\mu}(x) - \hat{v}^{\mu}(x)}{\epsilon} &= \lim_{\epsilon \to 0} \frac{  \epsilon \big(\xi^{\nu} \partial_{\nu}v^{\mu} - v^{\nu} \partial_{\nu} \xi^{\mu}) + \mathcal{O}(\epsilon^2) }{\epsilon} \nonumber \\
 &= \xi^{\nu} \partial_{\nu}v^{\mu} - v^{\nu} \partial_{\nu} \xi^{\mu} \ .
 \end{align}
 This is the rank $(1,0)$ tensor version of the formula given in~(\ref{Lie derivative coords}). The explicit method used here will also be used when looking at objects which are not tensors, when we cannot use the standard formula~(\ref{Lie derivative coords}).

The Lie derivative shares similar properties to the derivative operators previously introduced~(\ref{exterior}) and~(\ref{Covariant derivative}). It obeys a Leibniz-like rule
\begin{equation}
\mathcal{L}_{\xi} ( T \otimes S) = (\mathcal{L}_{\xi} T) \otimes S + T \otimes (\mathcal{L}_{\xi} S) \, ,
\end{equation}
and commutes with the exterior derivative on differential forms $[\mathcal{L}_{\xi}, d] \omega =0$, see~\cite{Hawking:1973uf}. The Lie derivative of a vector field is just the Lie bracket
\begin{equation}
\mathcal{L}_{\xi} v = \Big(\xi^{\mu} \frac{\partial v^{\nu}}{\partial x^{\mu}} - v^{\mu}  \frac{\partial \xi^{\nu}}{\partial x^{\mu}} \Big) \partial_{\nu} =  [\xi,v]  \, ,
\end{equation}
which we saw in footnote~\ref{footnote:Lie}. Lastly, if the two sets $(\mathcal{M}, g)$ and $(\mathcal{M}, \psi^*g)$ are related by an isometry $\psi^* g = g$ then the sets are equivalent. Hence spacetime is defined up to an equivalence class of pairs $(\mathcal{M}, \psi^*g)$ related by isomorphisms~\cite{Hawking:1973uf}.
 
 Returning to spacetime symmetries, the Lie derivative of the metric is given by
   \begin{align} \label{Lie metric}
 \mathcal{L}_{\xi} g_{\mu \nu} &=  \xi^{\lambda} \partial_{\lambda}g_{\mu \nu} + g_{\mu \lambda} \partial_{\nu}\xi^{\lambda} + g_{\lambda \nu} \partial_{\mu}\xi^{\lambda} \nonumber \\
 &= 2\nabla_{(\mu} \xi_{\nu)} \ ,
 \end{align}
 where we have made use of the metric-compatibility and symmetry of the Levi-Civita covariant derivative. Note that this always holds true, as we can always define the Levi-Civita derivative for a given metric tensor. If the Lie derivative of the metric vanishes we have a spacetime isometry
  \begin{equation} \label{Lie metric vanish}
 \mathcal{L}_{\xi} g_{\mu \nu} = 2\nabla_{(\mu} \xi_{\nu)} = 0 \ .
 \end{equation}
 The vector fields $\xi$ are then known as \textit{Killing vector fields}. The flows generated by Killing vector fields $\xi^{\mu}$ are continuous isometries of spacetime. Using (\ref{Lie metric}) one can categorise spacetimes based on its symmetries. For example, a spacetime is \textit{stationary} if it has a timelike Killing vector, i.e., there exists the Killing vector $\xi = \partial_t$ with $\partial_{t} g_{\mu \nu}=0$ and $g_{tt} <0$ in local coordinates $(t,x^i)$. A spacetime is \textit{static} if there are coordinates $(t,x^i)$ for which there exists a Killing vector $\xi = \partial_t$ with $\partial_{t} g_{\mu \nu}=0$, $g_{tt} <0$ and $g_{ti}=0$.

Working again in the metric-affine setting, we calculate the Lie derivative for the affine connection to be
\begin{align}
\mathcal{L}_{\xi} \bar{\Gamma}^{\lambda}_{\mu \nu} &=
\xi^{\rho} \partial_{\rho}  \bar{\Gamma}^{\lambda}_{\mu \nu} - \partial_{\rho} \xi^{\lambda}  \bar{\Gamma}^{\rho}_{\mu \nu} + \partial_{\mu} \xi^{\rho}  \bar{\Gamma}^{\lambda}_{\rho \nu} + \partial_{\nu} \xi^{\rho}  \bar{\Gamma}^{\lambda}_{\mu \rho} + \partial_{\mu} \partial_{\nu} \xi^{\lambda}  \nonumber \\ 
\label{Lie connection}
&= \bar{\nabla}_{\mu} \bar{\nabla}_{\nu} \xi^{\lambda} + \xi^{\rho} \bar{R}_{\rho \mu \nu}{}^{\lambda} + \bar{\nabla}_{\mu}(T^{\lambda}{}_{\nu \rho} \xi^{\rho}) \, ,
\end{align}
which matches the textbook results~\cite{JS1954,yano2020theory}. This can be found by simply evaluating $\bar{\Gamma}(x) - \hat{\bar{\Gamma}}(x)$. In going from the first to the second line we use the standard properties of the curvature and torsion tensors. It is interesting to note that despite the connection not being a tensor, clearly its Lie derivative is~(\ref{Lie connection}). This can again be understood from the cancelling of the inhomogeneous terms in $\bar{\Gamma}(x)$ and $\hat{\bar{\Gamma}}(x)$.

In the Levi-Civita case this reduces to
\begin{align}
\mathcal{L}_{\xi} \Gamma^{\lambda}_{\mu \nu} &=
\xi^{\rho} \partial_{\rho}  \Gamma^{\lambda}_{\mu \nu} - \partial_{\rho} \xi^{\lambda}  \Gamma^{\rho}_{\mu \nu} + \partial_{\mu} \xi^{\rho}  \Gamma^{\lambda}_{\rho \nu} + \partial_{\nu} \xi^{\rho}  \Gamma^{\lambda}_{\mu \rho} + \partial_{\mu} \partial_{\nu} \xi^{\lambda}  \nonumber \\ 
\label{Lie connection2}
&= \nabla_{\mu} \nabla_{\nu} \xi^{\lambda} + \xi^{\rho} R_{\rho \mu \nu}{}^{\lambda} \, .
\end{align}
Moreover, it can be shown that the above equation can be written in terms of the Lie derivatives of the metric tensor~\cite{yano2020theory}
\begin{equation}
\mathcal{L}_{\xi}  \Gamma^{\lambda}_{\mu \nu} = \frac{1}{2} g^{\lambda \rho} \Big( \nabla_{\mu} \mathcal{L}_{\xi} g_{\rho \nu} + \nabla_{\nu} \mathcal{L}_{\xi} g_{\rho \mu} - \nabla_{\rho} \mathcal{L}_{\xi} g_{\mu \nu} \Big) \, .
\end{equation}
From this it follows that the Lie derivative of the Levi-Civita connection vanishes for Killing vector fields of the metric~(\ref{Lie metric vanish}).

Killing vectors give rise to conserved quantities along spacetime geodesics, such as $u^{\mu} \xi_{\mu}$ where $\xi$ is Killing and $u$ satisfies the (Levi-Civita) geodesic equation. It follows straightforwardly from equations~(\ref{Parallel}) and~(\ref{Lie metric vanish}) that this quantity is covariantly conserved
\begin{equation}
\nabla_{u}(u^{\mu} \xi_{\mu}) = \xi_{\mu} \nabla_{u} u^{\mu} +  u^{\mu} u^{\nu} \nabla_{(\nu} \xi_{\mu)} =0 \, .
\end{equation}
We can similarly construct the conserved quantity $T^{\mu \nu} \xi_{\mu}$ which has $\nabla_{\nu} (T^{\mu \nu} \xi_{\mu})=0$, where $T^{\mu \nu}$ is the covariantly conserved energy-momentum tensor. This again follows from the Killing equation~(\ref{Lie metric vanish}) along with the covariant conservation of energy-momentum.

For theories with an independent affine connection, the concepts of isometries, Killing vectors and conserved quantities can of course be generalised. For example, Killing vectors can be defined as isometries of the metric \textit{and} of the affine connection~\cite{Hecht:1992xn,Hohmann:2015pva}. In the teleparallel theories (with vanishing affine curvature) these topics are slightly less clear~\cite{Coley:2019zld,Hohmann:2019nat} and we will touch upon this in the next chapter.

\subsection{Fundamental symmetries}
\label{section2.3.2} 
\subsubsection{Diffeomorphisms}
As previously mentioned, the diffeomorphism invariance of General Relativity is the invariance under the smooth $C^{\infty}(\mathcal{M})$, invertible mapping $\psi:\mathcal{M} \rightarrow \mathcal{M}$. In other words, the two sets $\{\mathcal{M}, g, \varPhi \}$ and $\{\mathcal{M}, \psi ^* g, \psi ^* \varPhi \}$ are physically indistinguishable, describing the same physics. Here it should be understood that by $\psi ^* g$ we mean $\psi ^* g(p)$ and not $\psi ^* g(\psi(p))$. This is the standard invariance under a change of coordinates, which will be shown below. 

One can think of this as a gauge symmetry and a redundancy in our description of the physical theory: specifically, invariance under the group of diffeomorphisms $\textrm{Diff}(\mathcal{M})$.
It should be noted that though there are similarities with the gauge symmetries of gauge theories, such as Yang-Mills theory, the diffeomorphism invariance of GR is not a true gauge symmetry in the equivalent sense. This is primarily because in gauge theories, the symmetry is \textit{internal}, whereas for GR we have the \textit{external} spacetime gauge symmetry~\cite{Blagojevic:2002du}. Moreover, for standard gauge theories, their Lie groups are compact, but this is not the case for $\textrm{Diff}(\mathcal{M})$. For more on the gauge approach to gravitation, see~\cite{Ivanenko:1983fts,Blagojevic:2002du, Blagojevic:2013xpa}.

\begin{figure}[!hbt]
\centering

\begin{subfigure}[b]{0.9\textwidth} 
\centering
\begin{tikzpicture}
\draw (2,2) ellipse (2cm and 1cm);
\shadedraw[inner color=black!0,outer color=black!0, draw=black] (2,2) ellipse (2.5cm and 1.5cm) node[anchor=south east]
 {$$}; 
\draw [->] (1,1) .. controls (1,3) and (3,1) .. (3,3);
\draw [->]  (1.4,0.6) .. controls (1.4,2.6) and (3.4,0.6) .. (3.4,2.6);
\draw [black,->] (3.3,3.4) arc (120:58:3cm);
\put(5,10){$\mathcal{M}$};
\put(135,94){$\psi$};
\put(66,32){$T(x^{\mu})$};
\put(60,48.5){$p$};
\filldraw [black] (2.3,1.59) circle (1pt);

\shadedraw[inner color=gray!0,outer color=black!0, draw=black] (8,2) ellipse (2.5cm and 1.5cm) node[anchor=south east] {$$};
\draw [black!90,->] (7.5,1.1) .. controls (7,3) and (9,1) .. (8.5,3);
\draw [black!90,->] (7.9,0.7) .. controls (7.4,2.6) and (9.4,0.2) .. (9,2.6);
\put(180,10){$\mathcal{M}$};
\put(250,30){$\hat{T}(\hat{x}^{\mu})$};
\put(230,46){${\color{red}\psi(p)}$};
\filldraw [red] (8.8,1.45) circle (1pt);

    \foreach \x in {0.6,1.2,...,3.6}{
        \draw[gray, dashed] (\x,0.8) -- (\x,3.4);
    }
    \foreach \y in {1.0,1.4,...,3.0}{
        \draw[gray, dashed] (0.2,\y) -- (3.8,\y);
    }
    
    \foreach \x in {6.6,7.2,...,9.6}{
        \draw[gray, dashed] (\x,0.8)..controls + (.075*22,-0) and +(-.075*24,0) .. (\x,3.4);
    }
    \foreach \y in {1.0,1.4,...,3.0}{
        \draw[gray, dashed] (6.2,\y) -- (9.8,\y);
    }        
\end{tikzpicture}
\caption{Active diffeomorphism where the point $p$ is transformed to $\psi(p)$ inducing a change in coordinates. Tensor fields are transformed according to~(\ref{Tensor transformation}).}\label{fig.diff1}
\end{subfigure}
\vspace{4mm}

\begin{subfigure}[b]{0.9\textwidth} 
\centering
\begin{tikzpicture}
\draw (2,2) ellipse (2cm and 1cm);
\shadedraw[inner color=red!0,outer color=red!0, draw=black] (2,2) ellipse (2.5cm and 1.5cm) node[anchor=south east] {$$};
\draw [->] (1,1) .. controls (1,3) and (3,1) .. (3,3);
\draw [->]  (1.4,0.6) .. controls (1.4,2.6) and (3.4,0.6) .. (3.4,2.6);
\draw [black,->] (3.3,3.4) arc (120:58:3cm);
\put(5,10){$\mathcal{M}$};
\put(135,94){$\hat{x}^{\mu}$};
\put(66,32){$T(x^{\mu})$};
\put(60,48.5){$p$};
\filldraw [black] (2.3,1.59) circle (1pt);

\shadedraw[inner color=blue!0,outer color=blue!0, draw=black] (8,2) ellipse (2.5cm and 1.5cm) node[anchor=south east] {$$};
\draw [->] (7,1) .. controls (7,3) and (9,1) .. (9,3);
\draw [->]  (7.4,0.6) .. controls (7.4,2.6) and (9.4,0.6) .. (9.4,2.6);
\put(180,10){$\mathcal{M}$};
\put(230,30){$\hat{T}(\hat{x}^{\mu})$};
\put(230,48.5){$p$};
\filldraw [black] (8.3,1.59) circle (1pt);

    \foreach \x in {0.6,1.2,...,3.6}{
        \draw[gray, dashed] (\x,0.8) -- (\x,3.4);
    }
    \foreach \y in {1.0,1.4,...,3.0}{
        \draw[gray, dashed] (0.2,\y) -- (3.8,\y);
    }
    
    \foreach \x in {6.6,7.2,...,9.6}{
        \draw[gray, dashed] (\x,0.8)..controls + (-.075*22,-0) and +(+.075*24,0) .. (\x,3.4);
    }
    \foreach \y in {1.0,1.4,...,3.0}{
        \draw[gray, dashed] (6.2,\y) -- (9.8,\y);
    }  
\end{tikzpicture}
\vspace{4mm}
\caption{Passive diffeomorphism where points are left alone but changing coordinates induces a change in tensor fields according to~(\ref{Tensor transformation}).} \label{fig.diff2} 
\end{subfigure}

\vspace{4mm}

\begin{subfigure}[b]{0.9\textwidth} 
\centering
\begin{tikzpicture}
\draw (2,2) ellipse (2cm and 1cm);
\shadedraw[inner color=red!0,outer color=red!0, draw=black] (2,2) ellipse (2.5cm and 1.5cm) node[anchor=south east] {$$};
\draw [->] (1,1) .. controls (1,3) and (3,1) .. (3,3);
\draw [->]  (1.4,0.6) .. controls (1.4,2.6) and (3.4,0.6) .. (3.4,2.6);
\draw [black,->] (3.3,3.4) arc (120:58:3cm);
\put(5,10){$\mathcal{M}$};
\put(125,94){$ \hat{x}^{\mu} \circ \psi $};
\put(66,32){$T(x^{\mu})$};
\put(60,48.5){$p$};
\filldraw [black] (2.3,1.59) circle (1pt);

\shadedraw[inner color=blue!0,outer color=blue!0, draw=black] (8,2) ellipse (2.5cm and 1.5cm) node[anchor=south east] {$$};
\draw [black!90,->] (7.5,1.1) .. controls (7,3) and (9,1) .. (8.5,3);
\draw [black!90,->] (7.9,0.7) .. controls (7.4,2.6) and (9.4,0.2) .. (9,2.6);
\put(180,10){$\mathcal{M}$};
\put(230,30){$\hat{T}(x^{\mu})$};
\put(230,47.5){${\color{red}\psi(p)}$};
\filldraw [red] (8.5,1.445) circle (1pt);

    \foreach \x in {0.6,1.2,...,3.6}{
        \draw[gray, dashed] (\x,0.8) -- (\x,3.4);
    }
    \foreach \y in {1.0,1.4,...,3.0}{
        \draw[gray, dashed] (0.2,\y) -- (3.8,\y);
    }
    
    \foreach \x in {6.6,7.2,...,9.6}{
        \draw[gray, dashed] (\x,0.8) -- (\x,3.4);
    }
    \foreach \y in {1.0,1.4,...,3.0}{
        \draw[gray, dashed] (6.2,\y) -- (9.8,\y);
    }        
\end{tikzpicture}
\vspace{3mm}
\caption{Active diffeomorphism followed by a transformation back to original coordinates. Geometric quantities change by their Lie derivative~(\ref{Lie derivative}).} \label{fig.diff3}
\end{subfigure}
\vspace{2mm}

\caption{Diffeomorphism from spacetime manifold $\mathcal{M}$ to itself. Transformations (a) and (b) lead to the standard coordinate transformation law $\hat{T}(\hat{x})$~(\ref{Tensor transformation}) whilst (c) leads to the Lie derivative $T(\hat{x})$~(\ref{Lie derivative}).}
\label{fig.diff}
\end{figure}
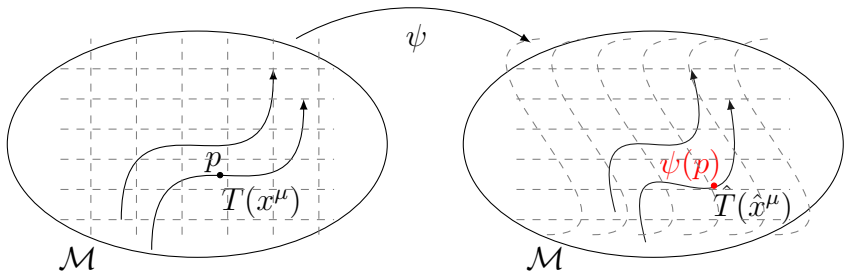
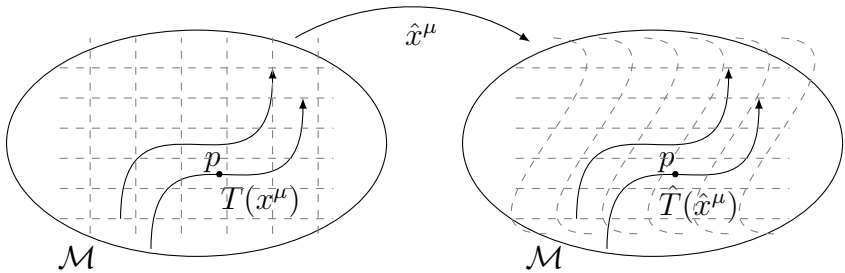
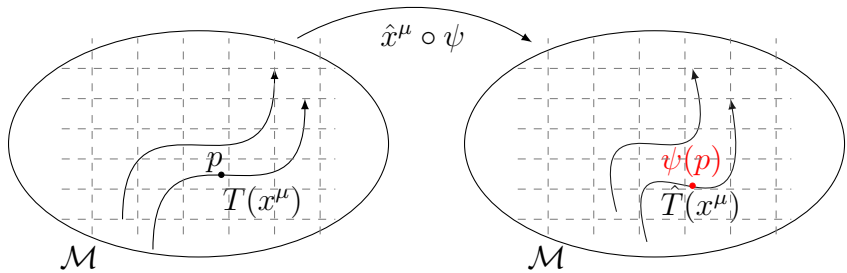

In the context of GR, for all intents and purposes, diffeomorphisms can be equally viewed as active or passive coordinate transformations (with appropriate assumptions regarding the overlapping of coordinate charts~\cite{Nakahara:2003nw}). This equivalence between active and passive points of view relies on the fact that GR is background independent (\textit{no prior structure}), as discussed previously. Any theory can be made diffeomorphism invariant in the passive sense by introducing additional background structures (e.g., preferred foliations~\cite{Visser:2011mf}) but not all theories will be actively diffeomorphism invariant. The active view can be thought of as moving points around on the manifold, which induces a change in coordinates as the natural coordinate basis at $\psi(p)$ differs from the original coordinates $x^{\mu}$ at $p$~\cite{Wald:1984}, see Fig.~\ref{fig.diff1} for an illustration. The passive view is just the relabelling of coordinates of spacetime points, see Fig.~\ref{fig.diff2}. Here, these are equivalent, the result being that objects change by their standard coordinate transformation rules. 

We can define a different transformation, however, which actively moves points on the manifold and then makes a coordinate transformation back to the original coordinates. In other words, looking at the difference between an object at the spacetime point $p$ and its value at $\psi_t (p)$ pulled back to the point $p$ and evaluated in the \textit{original coordinates}. This is also sometimes called a \textit{symmetry transformation}\footnote{This is what some authors mean computationally when they refer to an (active) diffeomorphism~\cite{Rovelli:1990ph}, whilst older works on GR only use diffeomorphism to refer to this $\psi^{*}_{t} T(\psi_{t} (p))$ type of transformation~\cite{Hawking:1973uf}.}. This is illustrated in Fig.~\ref{fig.diff3}, where we see the coordinate system on the right-hand side is the same as the original system on the left.
What makes this distinction different is that we are evaluating the difference between $T(x)$ and $\hat{T}(x)$, as opposed to the standard coordinate transformation which maps $T(x)$ to $\hat{T}(\hat{x})$. But this transformation to $\hat{T}(x)$ is just the Lie derivative, defined in equation~(\ref{Lie derivative}).

To see how this is useful, let us think about actions. Any generally covariant, scalar action is invariant under passive diffeomorphisms or GTCs, so $S(x) - \hat{S}(\hat{x}) = 0$. Let us instead consider the symmetry transformation of the type in Fig.~\ref{fig.diff3},
generated be the infinitesimal coordinate transformation $x^{\mu} \rightarrow \hat{x}^{\mu} = x^{\mu} + \epsilon \xi^{\mu}(x)$. For the Lagrangian density $\mathcal{L} = \sqrt{-g}L(x)$, where $L(x)$ is a coordinate scalar, this leads to
\begin{align}
\delta_{\xi} (L \sqrt{-g}) = \mathcal{L}_{\xi}(L \sqrt{-g}) &= \sqrt{-g} \mathcal{L}_{\xi} (L) + L \, \mathcal{L}_{\xi}(\sqrt{-g}) \nonumber \\
& = \sqrt{-g}  \xi^{\mu} \partial_{\mu} L + \frac{1}{2} L \sqrt{-g} g^{\mu \nu} \mathcal{L}_{\xi}(g_{\mu \nu})
\nonumber \\ 
&= \sqrt{-g}  \xi^{\mu}  \partial_{\mu}L + L \sqrt{-g} \nabla_{\mu} \xi^{\mu} \nonumber \\
&= \sqrt{-g} \nabla_{\mu}(\xi^{\mu} L) \, , \label{Lagrangian diff}
\end{align}
where we have used the Leibniz rule for the Lie derivative and the metric's transformation~(\ref{Lie metric}). There is no need to include the coordinates $d^4 x$ because we are evaluating the action in the original coordinate system $x^{\mu}$~\cite{blau2011lecture,Weinberg:1972kfs}. The final term is just a total derivative and therefore vanishes in an action when $\xi$ is held fixed on the boundary. 

 Applying this to the Einstein action (\ref{EH}) leads to
 \begin{equation}
 \delta_{\xi} S_{\textrm{EH}} = 0 \, ,
 \end{equation}
 as expected.
 However, if we compare this to the variation with respect to the metric (\ref{EH variation}) and let the variation in $g_{\mu \nu}$ be from an arbitrary diffeomorphism (symmetry transformation) generated by $\xi$ we find
 \begin{align} \label{EH diff}
 \delta_{\xi} S_{\textrm{EH}} = -\frac{1}{2\kappa} \int G^{\mu \nu} \delta_{\xi}( g_{\mu \nu} )  \sqrt{-g} d^4x &= 0 \nonumber \\
-\frac{1}{2\kappa} \int G^{\mu \nu} \mathcal{L}_{\xi}(g_{\mu \nu})   \sqrt{-g} d^4x &= 0  \nonumber \\
-\frac{1}{\kappa} \int G^{\mu \nu} \nabla_{\mu}\xi_{\nu} \sqrt{-g}  d^4 x &= 0 \, .
 \end{align}
 Next we integrate by parts and require $\xi$ to vanish on the boundary
 \begin{align}
  \delta_{\xi} S_{\textrm{EH}} &=  \frac{1}{\kappa} \int \xi_{\nu} \nabla_{\mu}(\sqrt{-g} G^{\mu \nu}) d^4 x = 0 \, ,
 \end{align}
 from which we can conclude
 \begin{equation} \label{Bianchi identity}
 \nabla_{\mu} G^{\mu \nu} = 0 \, .
 \end{equation}
This is the twice-contracted Bianchi identity, which we can interpret as a consequence of the diffeomorphism invariance of GR~\cite{Weinberg:1972kfs,Wald:1984}. This is an application of Noether's theorem, which states that invariances of the action are associated with conserved quantities~\cite{DN:2017}. 

We also note that this result, which can be found in most GR textbooks~\cite{Wald:1984,Carroll:2004st,TP:2010}, is often stated without explicitly indicating that it is the symmetry transformation of Fig.~\ref{fig.diff3} being performed, as opposed to a naive coordinate transformation. Clearly for a standard infinitesimal coordinate transformation, the variation of the metric is
\begin{equation}
\delta_{\xi} g_{\mu \nu} =  g_{\mu \nu} (x) -  \hat{g}_{\mu \nu} (\hat{x}) = 2 g_{\lambda (\mu} \partial_{\nu)} \xi^{\lambda}  \, ,
\end{equation}
which is not what we are after. This is sometimes called the \textit{total variation}, with symmetry transformations called \textit{form variations}~\cite{Blagojevic:2002du}, the latter being related to isometries and the Lie derivative. We apply the form variation because in the action $x^{\mu}$ is just a variable of integration, hence $d^4x$ is left alone. This is nicely explained by Weinberg~\cite{Weinberg:1972kfs} or Ort\'{i}n~\cite{ortin2004gravity}.
With these distinctions made, and it now being clear the types of transformations we are considering for the action~(\ref{Lagrangian diff}), we will continue to use the word diffeomorphism to describe any of the passive, active or symmetry transformations. If the context is not clear, we will describe the transformation explicitly.

Looking now at the matter action $S_{\textrm{M}}[g_{\mu \nu},\varPhi^{A}]$, which is also diffeomorphism invariant\footnote{This follows from us assuming the total action $S_{\textrm{total}} = S_{\textrm{EH}} + S_{\textrm{M}}$ is diffeomorphism invariant, because we showed already that $S_{\textrm{EH}}$ is diffeomorphism invariant. This would also follow from us assuming $S_{\textrm{M}}$ is a coordinate scalar by construction.}, the variation is given by 
\begin{align} \label{matter action}
\delta S_{\textrm{M}} &= \int \frac{\delta S_{\textrm{M}}}{\delta g^{\mu \nu} }\delta g^{\mu \nu} d^4 x + \int \frac{\delta S_{\textrm{M}}}{\delta \varPhi^{A} } \delta \varPhi^A d^4 x  \nonumber \\
&=  \frac{1}{2} \int \sqrt{-g} T^{\mu \nu} \delta g_{\mu \nu} d^4 x  \, ,
\end{align} 
where we have assumed the matter fields $\varPhi^{A}$ satisfy their equations of motion (on-shell). We will continue to make this assumption throughout without explicitly stating the on-shell condition, but see~\cite{Jacobson:2011cc} for a nice discussion regarding off-shell generalised Bianchi identities.

 If we let this variation be generated by an infinitesimal diffeomorphism just as we did for the Einstein-Hilbert action (\ref{EH diff}), we find
\begin{align} \label{matter action2}
\delta_{\xi} S_{\textrm{M}}  &=  \frac{1}{2} \int \sqrt{-g} T^{\mu \nu} \mathcal{L}_{\xi} g_{\mu \nu}  \, d^4 x \nonumber \\
&= \int \sqrt{-g} T^{\mu \nu} \nabla_{\mu} \xi_{\nu}  d^4 x \nonumber \\
&= - \int \sqrt{-g} \nabla_{\mu}(T^{\mu \nu}) \xi_{\nu} d^4 x \, ,
\end{align}
where in the last line we have integrated by parts. We have therefore derived the covariant conservation of energy-momentum by considering the diffeomorphism invariance of the matter action
\begin{equation}
\nabla_{\mu} T^{\mu \nu} = 0 \ .
\end{equation}

We can also consider the diffeomorphism invariance of total action $S_{\textrm{total}}$ comprised of both the gravitational and matter actions
\begin{align} \label{matter conservation}
\delta_{\xi}S_{\textrm{total}} = \delta_{\xi} S_{\textrm{EH}} + \delta_{\xi} S_{\textrm{M}} &= 0 \nonumber \\
  \nabla_{\mu}G^{\mu \nu} - \nabla_{\mu} T^{\mu \nu} &= 0 \nonumber \\
 \Rightarrow \nabla_{\mu} T^{\mu \nu} &= 0 \ ,
\end{align}
where again the outcome is the same by virtue of (\ref{Bianchi identity}). However, in this form we see a relationship between the covariant conservation of energy-momentum and the contracted Bianchi identity, and we did not need to assume the invariance of the individual gravity and matter actions. In the modified theories studied later in the thesis, it will be the total action that will be of importance, and from it we will derive our analogous conservation equations.

 Lastly, note that in flat space we obtain the expected energy-momentum conservation law $\partial_{\mu}T^{\mu \nu} =0$ that results from Noether's theorem. This is exactly what we claimed to be the case when working with the spin-2 graviton action~(\ref{coupled}), with $\partial_{\mu}T^{\mu \nu} =0$ a consequence of the linearised diffeomorphism invariance. It is not difficult to see that plugging $\mathcal{L}_{\xi} h_{\mu \nu}$ into~(\ref{matter action2}) yields this result.

\subsubsection{Local Lorentz invariance}
Local Lorentz invariance is also built into our theories. This is the invariance under the choice of local Lorentz frame $\mathbf{e}_{a}$. This is more closely related to gauge symmetries because these can be viewed as \textit{internal}; for a thorough discussion see~\cite{Blagojevic:2002du, Blagojevic:2013xpa}.
As we have been working in the coordinate basis, we have neglected using Lorentz indices, but we could equally well make the same statements about Lorentz covariance that we have made about spacetime covariance. This naturally leads to the historic works of Sciama~\cite{sciama1962analogy} and Kibble~\cite{Kibble:1961ba}, who studied the role of global Lorentz transformations, and then subsequently the gauging of these symmetries and \textit{local} Lorentz invariance.

It is illustrative to briefly recall the consequences of the \textit{global} Lorentz invariance of Special Relativity. In flat Minkowski space $\mathbb{R}^{(1,3)}$, we have invariance under (global) Poincar\'{e} transformations~(\ref{Poincare}). The infinitesimal transformations are
\begin{equation}
\delta x^{\mu} = \epsilon^{\mu}{}_{\nu} x^{\nu} + \epsilon^{\mu} \, ,
\end{equation}
where  $\epsilon_{\mu \nu} = - \epsilon_{\nu \mu}$ parameterises Lorentz rotations and $ \epsilon^{\mu}$ is a vector for translations. It is understood that these parameters are taken to be small and we have omitted the infinitesimal $\epsilon$ term factors for clarity. The fields transform according to
\begin{equation}
\delta \varPhi^{A}(x) = \frac{1}{2} \epsilon^{\mu \nu} J_{\mu \nu} \varPhi^{A}(x) \, ,
\end{equation}
with $J_{\mu \nu}=-J_{\nu \mu}$ being the Lorentz generators~\cite{Kibble:1961ba} satisfying
\begin{equation}
[J_{\mu \nu},J_{\rho \sigma}] = \eta_{\nu \rho} J_{\mu \sigma} + \eta_{\mu \sigma} J_{\nu \rho} - \eta_{\nu \sigma} J_{\mu \rho} - \eta_{\mu \rho} J_{\nu \sigma} \, .
\end{equation}
To be clear, scalar fields are invariant (trivial representation), spinors have their spinorial representation~(\ref{Lorentz generator}), and so on~\cite{Weinberg:1995mt}.

It is well known that for any Lorentz scalar Lagrangian (density) $\mathcal{L}(x)$, invariance leads to the set of ten conservation laws~\cite{Kibble:1961ba}
\begin{align} \label{SRc1}
\partial_{\mu} \Sigma^{\mu \nu} &= 0 \, , \\
\partial_{\mu}(S^{\mu}{}_{\rho \sigma} - x_{\rho} \Sigma^{\mu}{}_{\sigma} + x_{\sigma} \Sigma^{\mu}{}_{\rho}) &= 0 \, ,
\end{align}
where here we define the canonical energy-momentum tensor
\begin{equation} \label{EM_canonical}
\Sigma^{\mu}{}_{\nu} := \frac{\partial \mathcal{L}}{\partial ( \partial_{\mu} \varPhi^{A})}  \partial_{\nu} \varPhi^{A} - \delta^{\mu}_{\nu} \mathcal{L} \, ,
\end{equation}
and the tensor of intrinsic angular momentum (spin)
\begin{equation} \label{SR_canonical}
S^{\mu}{}_{\rho \sigma} := - \frac{\partial \mathcal{L}}{\partial ( \partial_{\mu} \varPhi^{A})}  J_{\rho \sigma}  \varPhi^{A}\, ,
\end{equation}
with $S^{\mu}{}_{\rho \sigma} = -S^{\mu}{}_{\sigma \rho}$.
These equations then represent the conservation of energy-momentum and angular momentum. For scalar fields, where the Lorentz generators vanish, we have $S^{\mu}{}_{\rho \sigma} =0$ and retrieve the more familiar form of angular momentum conservation. Note that the angular momentum equation can be rewritten as~\cite{sciama1962analogy}
\begin{equation} \label{SRc2}
\partial_{\mu} S^{\mu \rho \sigma} = \Sigma^{\rho \sigma} - \Sigma^{\sigma \rho} \, ,
\end{equation}
hence $\Sigma^{\mu \nu}$ need not be symmetric. Lastly, the Belinfante~\cite{belinfante1940current} and Rosenfeld~\cite{rosenfeld1940energy} procedure can be used to construct the symmetric and divergence-free energy-momentum tensor
\begin{equation} \label{SR_BR}
\mathcal{T}^{\mu \nu} = \Sigma^{\mu \nu} + \frac{1}{2} \partial_{\rho}  \big(S^{\mu \rho \nu} + S^{\nu \rho \mu} + S^{\rho \nu 
\mu} \big) \, ,
\end{equation}
which will be useful for later comparison.

Moving back to General Relativity, let us work with the \textit{local} Lorentz transformations~(\ref{Lorentz1}), which infinitesimally take the form
\begin{equation} \label{LLT_infinitesimal}
\Lambda^{a}{}_{b}(x) = \delta^{a}{}_{b} + \epsilon^{a}{}_{b}(x) \, ,
\end{equation}
where $\epsilon_{ab} = - \epsilon_{ba}$. We no longer include translations because we are working with the homogeneous Lorentz group $SO(1,3)$. In the foundational works by Utiyama~\cite{utiyama1956invariant}, Sciama~\cite{sciama1962analogy} and Kibble~\cite{Kibble:1961ba}, they look at the consequences of starting with flat Minkowski space $\mathbb{R}^{(1,3)}$ and gauging the Lorentz group (or Poincar\'{e} groups in the case of Sciama and Kibble). They then show that this directly leads to a theory that looks very much like General Relativity. However, the geometry is not Levi-Civita but instead Einstein-Cartan (containing torsion). In Chapter~\ref{chapter3} we introduce these theories, before studying their modifications in the following chapters. For now, we stick with Riemannian General Relativity. 

Under an infinitesimal local Lorentz transformation the tetrad field transforms as
\begin{equation}
\delta_{\Lambda} e_{a}{}^{\mu} = \epsilon_{a}{}^{b}(x) e_{b}{}^{\mu} \, .
\end{equation}
Note that the coordinates remain unchanged under LLTs $\mathbf{e}_{a}(x) \rightarrow \hat{\mathbf{e}}_{a}(x)$, so the situation differs from the calculation using general coordinate transformations\footnote{Hence, we do not have the same notion of the Lie derivative for objects taking values in the local tangent space as we did for spacetime coordinate objects. For more details see~\cite{Ortin:2002qb}.}.

Now we let the tetrad variation in the Einstein-Hilbert action~(\ref{EH tetrad}) be generated by this infinitesimal local Lorentz transformation, see for example~\cite{Weinberg:1972kfs,Aldrovandi:2013wha}. The resulting variations~(\ref{EH tetrad var}) lead to
\begin{align} \label{EH_tetrad_LLT}
\delta_{\Lambda} S_{\textrm{EH}}[e] &= \frac{1}{ \kappa} \int (\delta_{\Lambda} e_{a}{}^{\nu}) e^{a}{}_{\mu}  \big(R^{\mu}{}_{\nu} - \frac{1}{2}\delta^{\mu}_{\nu} R\big)  d^4x \nonumber \\
&= \frac{1}{ \kappa} \int \epsilon_{a}{}^{b} e_{b}{}^{\nu} e^{a}{}_{\mu}  \big(R^{\mu}{}_{\nu} - \frac{1}{2}\delta^{\mu}_{\nu} R\big)  e \, d^4x  \nonumber \\
&= \frac{1}{\kappa} \int \epsilon^{\mu \nu}  \big(R_{\mu \nu} - \frac{1}{2} g_{\mu \nu} R\big)  e \, d^4x = 0 \, ,
\end{align}
with the final line vanishing due to the antisymmetry of $\epsilon$. This could be applied to any Lorentz scalar action $\mathcal{L}$, where the infinitesimal variation will vanish. This can also be seen as a consequence of Noether's theorem~\cite{Blagojevic:2013xpa}.

Moving to matter, let us for now assume the action takes the form $S_{\textrm{M}}[e,\varPhi^{A}]$, and that the gravitational dynamical variable is the tetrad $\mathbf{e}_a$. Local Lorentz invariance is equivalent to the requirement that the field equations be symmetric
\begin{equation} \label{EH_tetrad_sym}
\delta_{\Lambda} S[e, \varPhi^{A}] = \int \Big( \epsilon_{a}{}^{b} e_{b}{}^{\mu} \frac{\delta S}{\delta e_{a}{}^{\mu}}  +  \frac{\delta S}{\delta \varPhi^{A}} \delta \varPhi^{A}\Big) e d^4x  = - \int \epsilon^{ab} \Theta_{ab} e d^4 x = 0  \, ,
\end{equation}
where we have assumed that matter is on-shell and defined the (tetradic) energy-momentum tensor\footnote{\label{footnote_tensor}It is important to distinguish between the different types of energy-momentum tensor. Here, in the Riemannian setting (which we will soon see is equivalent to vanishing hypermomentum~\cite{FH1976}, see equation~(\ref{Belinfante})), it turns out that~(\ref{tetradic_EM}) coincides with both the symmetric (Hilbert) energy-momentum tensor \textit{and} the \textit{canonical} (Noether) energy-momentum tensor~\cite{sciama1962analogy,belinfante1940current}. This coincidence happens purely because we are working with a matter action that we assume depends only on the tetrad and not the spin connection; this rules out working with fermionic matter. This will not be true in general, which we clarify shortly in the following chapter. For a review of how these tensors relate to one another see~\cite{leclerc2006canonical}.}
\begin{equation}  \label{tetradic_EM}
\Theta^{a}{}_{\mu} := - \frac{1}{e} \frac{\delta S_{\textrm{M}}[e, \varPhi^{A}]}{\delta e_{a}{}^{\mu}} \, .
\end{equation}
It follows that $\Theta_{[ab]}=0$ holds \textit{on-shell}. We also have an equivalence with the metrical energy-momentum tensor $e^{a}{}_{\mu} e^{b}{}_{\nu} \Theta_{ab} = T_{\mu \nu}$, following from the implicit assumption that matter couples to the tetrad only via the metric. On top of this, for matter that does not couple to the spin connection (e.g., bosonic matter), the tetradic energy-momentum tensor is also equivalent to the \textit{canonical} energy-momentum tensor~(\ref{EM_canonical})~\cite{leclerc2006canonical,belinfante1940current}. This is explained in more detail in~\cite{Hehl:1994ue}, or for a more introductory and modern treatment see section 4 of~\cite{ortin2004gravity}.

It is not difficult to see the similarities between $\nabla_{\mu} T^{\mu}{}_{\nu}=0$ and $T_{[\mu \nu]}=0$ and the Special Relativistic conservation laws~(\ref{SRc1}) and~(\ref{SRc2}). However, in standard GR there is no analogous object for $S^{\mu}{}_{
\rho \sigma}$, which corresponds to spin~\cite{sciama1962analogy,FH1976}. This necessitates a framework beyond Riemannian geometry, which will be the focus of the next chapter.

In Chapter~\ref{chapter5} and~\ref{chapter6}, we look at the consequences of breaking these symmetries, where our modified actions are not manifestly invariant under diffeomorphisms or Local Lorentz transformations. For the Einstein-Hilbert action, Noether's theorem for diffeomorphisms and Lorentz transformations lead to the equations $\nabla_{\mu} G^{\mu \nu} =0$ and $G_{[\mu \nu]}=0$, but these were already identically true. For the modified theories we will study, if we assume that these fundamental symmetries hold, then Noether's theorem will lead to new constraints that are not identities a priori.

\begin{savequote}[70mm]
Symmetry is a vast subject, significant in art and nature. Mathematics lies at its root, and it would be hard to find a better one on which to demonstrate the working of the mathematical intellect.
\qauthor{`Symmetry' (1952) \\
Hermann Weyl}
\end{savequote}

\chapter{Metric-affine theories}
\label{chapter3}
In the previous chapter, we introduced the metric-affine geometric framework as well as the standard formulation of General Relativity and the Einstein-Hilbert action. One of the assumptions when constructing General Relativity was that the affine connection was chosen to be the unique, torsion-free, metric-compatible Levi-Civita connection. In this chapter we will drop that assumption and consider the wider framework of gravitational theories with different choices of connection, the most general being the full metric-affine framework. All of the necessary mathematical tools were introduced in the previous chapter, in particular Sec.~\ref{section2.1.1}.

Within these broader geometric frameworks, the choice of gravitational action is less obvious. For example, in Poincar\'{e} gravity, with both curvature and torsion, the most general quadratic Lagrangian contains many additional terms beyond the Einstein-Hilbert term~\cite{baekler2011beyond}. More generally, in metric-affine theories, still more combinations are possible~\cite{Hehl:1994ue}. As a guiding criterion, it is often chosen that the gravitational Lagrangian be quadratic in curvature, torsion and non-metricity; the reason for this is to preserve the \textit{quasilinearity} of the field equations~\cite{Hehl:1999sb}. Even so, the theory becomes much more complicated than GR, see for instance~\cite{vitagliano2011dynamics,baldazzi2022metric}.

One immediate consequence of adopting the metric-affine framework is that gravity can now be described in a \textit{first-order} way, i.e., containing no higher than first-derivatives of the fundamental variables. This makes the similarities with gauge theories much more apparent, as first explored by Weyl~\cite{Weyl:1929fm} and then later by Yang and Mills~\cite{Yang:1954ek} and Utiyama~\cite{utiyama1956invariant}, followed by Sciama~\cite{sciama1962analogy} and Kibble~\cite{Kibble:1961ba}. In fact, the inclusion of torsion alongside curvature, resulting in an Einstein-Cartan geometry, is a direct consequence of consistently gauging the Poincar\'{e} group $P(1,3)$~\cite{sciama1962analogy,Kibble:1961ba}. Extending this to the general linear group $GL(4,\mathbb{R})$ leads to metric-affine theories, with an independent connection possessing curvature, torsion and non-metricity~\cite{lord1978metric,Blagojevic:2013xpa}.

We will first look at the Palatini approach to General Relativity, which treats the metric and connection as independent. We then move on to the fully metric-affine theories, including Einstein-Cartan theory. As mentioned, the rise of gauge theoretic ideas in the 1950s-60s led to the rediscovery of non-Riemannian geometries, and they play an important role in the motivation of metric-affine theories. We will briefly talk about the gauging of the Lorentz and Poincar\'{e} groups, and show how this relates to gravity, before concluding with the teleparallel theories.
For further details on metric-affine theories of gravity, and our main sources for this chapter, we refer to many of the works by Hehl and others~\cite{Hehl:1994ue,hehl1978metric,hehl1981metric,FH1976,HehlKerlickHeyde+1976+111+114,HehlKerlickHeyde+1976+524+527,HehlKerlickHeyde+1976+823+827,hehl1976new}, and the textbooks~\cite{ortin2004gravity,Blagojevic:2013xpa,Aldrovandi:2013wha}. For more on the gauge-theoretic approach to gravity specifically, see~\cite{Blagojevic:2013xpa,Hehl:1994ue,lord1978metric}.

\section{Palatini formalism}
\label{section3.1}
Before studying metric-affine theories, its useful to study the so-called `Palatini formalism'\footnote{Incidentally, the modern formulation of treating the metric and connection independently, often attributed to Attilio Palatini, was in fact first used by Einstein~\cite{ferraris1982variational}. We will, however, continue to incorrectly use this naming convention to match with the literature.}. In this approach, though the metric  and connection are taken to be independent and varied simultaneously, it is assumed that matter does not couple to the connection $S_{\textrm{M}} = S_{\textrm{M}}[g,\varPhi^A]$. This assumption on the matter sector is used in many Palatini  approaches in modified theories, see~\cite{Sotiriou:2008rp} and references therein. In these scenarios, the affine connection does not play a physical role. For example, it is not associated with parallel transport because we have matter satisfying \textit{Levi-Civita} geodesics $\nabla_{\dot{\gamma}} \dot{\gamma} =0$ as opposed to autoparallel geodesics $\bar{\nabla}_{\dot{\gamma}} \dot{\gamma} =0$. The connection can then be seen to be merely an auxiliary field~\cite{Sotiriou:2006qn}, and the geometry is  really just the standard (pseudo-)Riemannian one of GR~\cite{Sotiriou:2008rp}. In this section we do not yet consider modified theories, but show that the Palatini prescription leads back to General Relativity.

The Palatini theories will be of less interest than the fully metric-affine ones, mainly because we want the connection to play a non-trivial geometric role. However, the calculations on the gravitational side will be largely the same. Moreover, they are equivalent to the metric-affine theories in vacuum. We will also see that they give us another perspective from which to view GR. This approach is sometimes called the first-order formalism because the action $\bar{R} = g^{\mu \nu} \bar{R}_{\mu \nu}(\bar{\Gamma})$ is now only first-order in derivatives. This also begins to make the connection with other formulations of gravity (such as supergravity~\cite{ortin2004gravity} or gauge-gravity~\cite{Blagojevic:2013xpa}) more clear, but we will return to this point later.

\subsection{Einstein-Palatini action}
\label{section3.1.1}
Let us first look at the Einstein-Hilbert action in the Palatini formalism, where we treat the affine connection as independent. The action is given by
\begin{equation} \label{S_Palatini}
S_{\textrm{Palatini}}[g,\bar{\Gamma}] = \frac{1}{2 \kappa} \int \bar{R} \sqrt{-g} d^4 x \, ,
\end{equation}
where the affine Ricci scalar $\bar{R} = g^{\mu \nu} \bar{R}_{\mu \nu}(\bar{\Gamma})$ is defined in~(\ref{Ricci scalar}). We assume our matter action to be \textit{independent} from the affine connection $S_{\textrm{M}}[g,\varPhi^A]$. For example, if any covariant derivatives acting on indexed objects were to appear in the matter action, they would necessarily be Levi-Civita ones $S_{\textrm{M}}[g,\varPhi^A, \nabla \varPhi^A]$, which can subsequently be rewritten in terms of the metric and its derivatives $S_{\textrm{M}}[g, \partial g, \varPhi^A, \partial \varPhi^A]$. This also reveals some of the unnaturalness that comes with assuming the matter action does not couple to the affine connection: the gravitational action is first-order, but the same is not necessarily true for matter\footnote{This is precisely the reason why in this Palatini setting the energy-momentum tensors arising from metric/tetrad variations are not the usual canonical ones, again see the discussion and citations in footnote~\ref{footnote_tensor}.}.

Moving on to the field equations, variations with respect to the metric lead to the (symmetrised) affine Einstein tensor, whilst the connection variations lead to the \textit{Palatini tensor}~\cite{Hehl:1978a,Hehl:1978b} 
\begin{equation}
\delta S_{\textrm{Palatini}} =  \frac{1}{2 \kappa} \int  \Big( \delta g^{\mu \nu} \bar{G}_{\mu \nu} + \delta \bar{\Gamma}^{\lambda}_{\mu \nu} P^{\mu \nu}{}_{\lambda} \Big) \sqrt{-g} d^4 x \, .
\end{equation}
The Palatini tensor is defined as
\begin{multline} \label{Palatini0}
  P^{\mu \nu}{}_{\lambda} := -Q_{\lambda}{}^{\mu \nu} +  \frac{1}{2} g^{\mu \nu} Q_{\lambda \rho}{}^{\rho} + \delta_{\lambda}^{\mu} Q_{\rho}{}^{\rho \nu} -
  \frac{1}{2} \delta^{\mu}_{\lambda} Q^{\nu}{}_{\rho}{}^{\rho} \\ + T^{\mu}{}_{\lambda}{}^{\nu} + g^{\mu \nu} T^{\rho}{}_{\rho \lambda} +
  \delta^{\mu}_{\lambda} T^{\rho \nu}{}_{\rho}  \, , \qquad
\end{multline} 
and is algebraic in torsion $T^{\lambda}{}_{\mu \nu}$~(\ref{Torsion})  and non-metricity $Q_{\lambda \mu \nu}$~(\ref{nonmetricity}).
Note that we have again assumed all boundary terms vanish, which here only depend on first derivatives of the fields. This is in contrast to the second-order formulation of Einstein's theory studied in the previous sections.
For further details see, for example,~\cite{Sotiriou:2008rp,Sotiriou:2006qn}. We will also explicitly calculate these variations and rederive the Palatini tensor in Sec.~\ref{section4.4}.

It follows that the metric field equations are
\begin{equation} \label{Palatini_eq1}
\bar{G}_{(\mu \nu)} = \kappa T_{\mu \nu} \, ,
\end{equation}
where $T_{\mu \nu}$ is the standard metrical (Hilbert) energy-momentum tensor defined in~(\ref{energy-momentum tensor}). Because matter only couples to the metric, we do not yet need to make any redefinitions of the energy-momentum tensor. The connection equation of motion is simply
\begin{equation} \label{Palatini_eq2}
 P^{\mu \nu}{}_{\lambda}  = 0 \, ,
\end{equation}
because $S_{\textrm{M}}$ is independent of $\bar{\Gamma}$. 

By decomposing the affine connection into its Levi-Civita part and contributions from torsion and non-metricity~(\ref{affine decomp}), it is possible to solve the algebraic equation~(\ref{Palatini_eq2}) for the affine connection~\cite{Hehl:1978a}. We first note that the Palatini tensor is trace-free over its last two indices, implying $P^{\mu \nu}{}_{\nu} = 0$ identically. This is a consequence of the \textit{projective invariance} of the action, which we show explicitly in Appendix~\ref{appendixA}. In short, the affine Ricci scalar $\bar{R}$ is invariant under transformations of the connection of the form
\begin{equation}
\bar{\Gamma}^{\lambda}_{\mu \nu} \rightarrow \bar{\Gamma}^{\lambda}_{\mu \nu} + \delta^{\lambda}_{\nu} P_{\mu} \, ,
\end{equation}
where $P^{\mu}$ is an arbitrary vector, see equation~(\ref{Proj3}).
This issue will also be revisited in the following chapters when working in the metric-affine framework, but for a detailed discussion see~\cite{Hehl:1978a} or more recently~\cite{Afonso:2017bxr}.

This amounts to an indeterminacy in equation~(\ref{Palatini_eq2}), which can only fix 60 of the 64 independent components of the affine connection. The solution for the connection is
 \begin{align} \label{connection_sol}
  \bar{\Gamma}{}^{\lambda}_{\mu \nu} = \Gamma^{\lambda}_{\mu \nu} - \frac{1}{2} Q_{\mu} \delta^{\lambda}_{\nu}
  \quad \mbox{or} \quad
  \bar{\Gamma}{}^{\lambda}_{\mu \nu} = \Gamma^{\lambda}_{\mu \nu}+\frac{1}{3} T_{\mu} \delta^{\lambda}_{\nu} \,,
\end{align}
where $\Gamma^{\lambda}_{\mu \nu}$ is the Levi-Civita connection and the additional term is the non-metricity vector $Q_{\mu} = Q_{\mu}{}^{\rho}{}_{\rho}$ or the torsion vector  $T_{\mu} = T^{\rho}{}_{\mu \rho}$. 

Standard General Relativity is recovered when the solution for the connection~(\ref{connection_sol}) is substituted into the metric-affine Einstein equation~(\ref{Palatini_eq1}). This is because the additional torsion or non-metricity vector term arising in~(\ref{connection_sol}) does not contribute to the affine Einstein tensor when symmetrised~\cite{Dadhich:2012htv}.  One then arrives at
$G_{\mu \nu} = \kappa T_{\mu \nu}$.
This freedom represented by $T_{\mu}$ or $Q_{\mu}$ in the solution for the affine connection can be seen as a gauge freedom, with the projective transformation taking the form of a gauge transformation~\cite{Julia:2000er}. However, it is also possible to use Lagrange multipliers to fix this freedom
\begin{equation} \label{Lag_mult}
S_{\lambda} = \int \lambda^{\mu} T_{\mu} \sqrt{-g} d^4x \, ,
\end{equation}
and then vary the total action with respect to $g^{\mu \nu}$, $\bar{\Gamma}^{\rho}_{\mu \nu}$, and the Lagrange multiplier $\lambda^{\mu}$. Then the connection will be exactly the Levi-Civita one. We note that the gauge character of the projective invariance, first identified in~\cite{Julia:2000er}, is not widely known. It can then be argued~\cite{Dadhich:2012htv} that the gauge-fixing terms~(\ref{Lag_mult}) are not in fact necessary, as they are not dynamically relevant.

\subsection{Tetradic Palatini action}
\label{section3.1.2}
Let us now work with the Palatini prescription but in the tetrad formalism.  In this case, the fundamental variables are the independent tetrad $\mathbf{e}_{a}$ and spin connection $\omega_{ab}$. Using the tetrad formalism is essential for coupling fermions to gravity, since fermions require the use of the spin connection. However, for now we continue to assume matter couples only to the tetrad, such that it takes the form $S_{\textrm{M}}[e, \varPhi^{A}, \nabla \varPhi^{A} ] = S_{\textrm{M}}[e, \mathring{\omega}, \varPhi^{A}, \partial \varPhi^{A}]$ where $\mathring{\omega}_{ab}$ is the Levi-Civita spin connection. Hence the matter action is not actually assumed to be first-order but may in fact contain derivatives of the tetrad $S_{\textrm{M}}[e,\partial e,  \varPhi^{A}, \partial \varPhi^{A} ]$. 

The gravitational action is defined as
\begin{equation} 
S_{\textrm{Palatini}}[e,\omega] = \frac{1}{2\kappa} \int \frac{1}{4} \epsilon_{abcd} \mathbf{e}^{a} \wedge \mathbf{e}^{b} \wedge \bar{R}^{cd}(\omega) \, ,
\end{equation}
or in full index notation
\begin{equation} \label{S_Palatini_tetrad}
S_{\textrm{Palatini}}[e,\omega] = \frac{1}{2\kappa} \int e_{a}{}^{\mu} e_{b}{}^{\nu} \bar{R}_{\mu \nu}{}^{ab}(\omega) e d^4x \, ,
\end{equation}
where $\bar{R}_{\mu \nu}{}^{ab}$ is defined in terms of the independent spin connection in~(\ref{Riemann mixed}). Variations of the gravitational action with respect to the tetrad lead to the mixed-index, non-symmetric, affine Einstein tensor $\bar{G}^{a}{}_{\mu}$. The tetrad field equations are then just
\begin{equation}
\bar{G}^{a}{}_{\mu} = \kappa \Theta^{a}{}_{\mu} \, ,
\end{equation}
where here $\Theta^{a}{}_{\mu}$ is the tetradic energy-momentum tensor~(\ref{tetradic_EM}).

Local Lorentz invariance of the matter action enforces the on-shell symmetry of the energy-momentum tensor $\Theta_{[ab]}=0$, and hence the symmetry of the affine Einstein tensor via the gravitational field equations $\bar{G}_{[ab]}=0$. It is perhaps important to elaborate on this point: because our matter action, as formulated here $S_{\textrm{M}}[e,\partial e,  \varPhi^{A}, \partial \varPhi^{A}]$, couples \textit{only} to the tetrad and its derivatives, an infinitesimal local Lorentz transformation leads to the variations presented in the previous section~(\ref{EH_tetrad_sym}). On-shell the matter field variations are zero and the only remaining term is $\epsilon^{ab} \Theta_{ab}$, with $\epsilon^{ab} = - \epsilon^{ba}$. It follows that local Lorentz invariance $\delta_{\Lambda} S_{\textrm{M}} =0$ implies the symmetry of the energy-momentum tensor $\Theta_{[ab]}=0$.
Note that the Local Lorentz invariance of the gravitational action $S_{\textrm{Palatini}}[e,\omega]$ does not directly imply $\bar{G}_{[ab]}=0$, because the variations with respect to the spin connection must also be considered.

Let us now compute the variations with respect to the independent spin connection. Note that in most works in Palatini gravity, the metric compatibility of the spin connection is often assumed $\omega_{ab} = \omega_{ba}$, see for example~\cite{VanNieuwenhuizen:1981ae,ortin2004gravity}. In that case, one is quickly led to an equation of the form $D \mathbf{e}^{a} = 0$, where $D$ is the covariant exterior derivative defined in~(\ref{total_d}). Notice that this is exactly just the Cartan structure equation for torsion~(\ref{Cartan_torsion}), and hence one retrieves standard GR~\cite{Clifton:2011jh}.

To be fully general, and to make the discussion as clear as possible, let us calculate this derivation for a fully general spin connection. One route is to decompose the affine spin connection into its Levi-Civita and contortion components, but we will use some methods that will be employed regularly in the next chapter. These are the relations between the partial derivatives of the tetrad and the Levi-Civita spin connection, which hold independent of the geometry.  

Variations of the Riemann tensor with respect to the spin connection are 
\begin{equation}
\delta_{\omega} \bar{R}_{\mu \nu a}{}^{b} = 2 \partial_{[\mu} \delta \omega_{\nu]}{}^{a}{}_{b} + 2 \delta \omega_{\mu}{}^{b}{}_{|c|} \omega_{\nu]}{}^{c}{}_{a} +  2 \omega_{\mu}{}^{b}{}_{|c|} \delta \omega_{\nu]}{}^{c}{}_{a} \, .
\end{equation}
Substituting this into the Ricci scalar density $e e_{a}{}^{\mu} e_{b}{}^{\nu} \bar{R}_{\mu \nu}{}^{ab} $, and performing an integration by parts on the derivative terms, leads to an expression of the form
\begin{equation} \label{spin_var}
2 \delta \omega_{\alpha}{}^{i}{}_{j} \Big[ - \partial_{\beta}\big(e e_{a}{}^{[\beta} e_{i}{}^{\alpha]} \eta^{ja} \big) + e \omega_{\beta}{}^{j}{}_{a} e_{b}{}^{[\alpha} e_{i}{}^{\beta]} \eta^{ab} + e \omega_{\beta}{}^{b}{}_{i} e_{a}{}^{[\beta} e{}_{b}{}^{\alpha]} \eta^{ja} \Big] \, .
\end{equation}
It should be understood that the boundary term has been discarded as it vanishes under the integral, and we have made some algebraic simplifications in the above expression. Now we use that the partial derivatives of the tetrad terms can be rewritten in terms of Levi-Civita affine and spin connection contributions\footnote{This follows from the definition of the Levi-Civita spin connection, which holds even whilst the affine spin connection and tetrad are assumed to be completely independent. Here we make use of equations~(\ref{e_formulae}) from the following chapter.}. Using these relations the first term can be expanded as
\begin{align}
\frac{1}{e} \partial_{\beta}\big(e e_{a}{}^{[\beta} e_{i}{}^{\alpha]} \eta^{ja} \big) & 
\begin{multlined}[t] = 
\Gamma_{\beta \lambda}^{\lambda} e_{a}{}^{[\beta} e_{i}{}^{\alpha]} \eta^{ja} +  \mathring{\omega}_{\beta}{}^{b}{}_{a} e_{b}{}^{[\beta} e_{i}{}^{\alpha]} \eta^{ja} + \mathring{\omega}_{\beta}{}^{b}{}_{i} e_{b}{}^{[\alpha} e_{a}{}^{\beta]} \eta^{ja}  \\ 
 - \Gamma^{[\beta}_{\beta \mu} e_{i}{}^{\alpha]} e_{a}{}^{\mu} \eta^{ja} - \Gamma^{[\alpha}_{\beta \mu} e_{a}{}^{\beta]} e_{i}{}^{\mu} \eta^{ja} 
\end{multlined} \nonumber \\
&=  \mathring{\omega}_{\beta}{}^{b}{}_{a} e_{b}{}^{[\beta} e_{i}{}^{\alpha]} \eta^{ja} + \mathring{\omega}_{\beta}{}^{b}{}_{i} e_{b}{}^{[\alpha} e_{a}{}^{\beta]} \eta^{ja} \, ,
\end{align}
where all of the Levi-Civita spacetime connection terms cancel with one another. The details of calculations such as this one are covered in Sec.~\ref{section4.2}, see for example equations~(\ref{e_formulae})-(\ref{boundary_tetrad_e2}).

Substitution of this result back into~(\ref{spin_var}) leads to the final expression
\begin{multline} \label{spin_var2}
2e  \delta \omega_{\alpha}{}^{i}{}_{j} \Big[ \omega_{i}{}^{j \alpha} - \omega^{a j}{}_{a} e_{i}{}^{\alpha} + \omega^{j \alpha}{}_{i} - \omega_{a}{}^{a}{}_{i} e_{b}{}^{\alpha} \eta^{jb}- \mathring{\omega}_{a}{}^{a j} e_{i}{}^{\alpha} \\
+ \mathring{\omega}_{i}{}^{\alpha j} - \mathring{\omega}^{j \alpha}{}_{i} + \mathring{\omega}{}_{a}{}^{a}{}_{i} e_{b}{}^{\alpha} \eta^{j b} \Big] \, ,
\end{multline}
where we have taken liberties with exchanging spacetime and tangent space indices on the spin connection terms. For arbitrary variations in the spin connection, this expression must vanish as we are dealing with matter that couples only to the tetrad. Therefore the terms in the bracket of~(\ref{spin_var2}) must equal zero. Making use of the skew-symmetry of the Levi-Civita spin connection $\mathring{\omega}_{\mu a b} = - \mathring{\omega}_{\mu b a}$, which follows from metric compatibility, we can group the terms in~(\ref{spin_var2}) to give the expected result 
\begin{multline} \label{spin_Palatini}
\Big[ (\omega^{j \alpha}{}_{i} - \mathring{\omega}^{j \alpha}{}_{i}) + (\omega_{i}{}^{j \alpha} - \mathring{\omega}_{i}{}^{j \alpha}) - (\omega^{a j}{}_{a} e_{i}{}^{\alpha} - \mathring{\omega}^{a j}{}_{a} e_{i}{}^{\alpha} )  \\
 - ( \omega_{a}{}^{a}{}_{i} e_{b}{}^{\alpha} \eta^{jb} - \mathring{\omega}_{a}{}^{a}{}_{i} e_{b}{}^{\alpha} \eta^{jb} ) \Big] = 0 \, .
\end{multline}

Solving this equation leads to an analogous result to the corresponding one in the coordinate basis case: the affine spin connection is simply equal to the Levi-Civita one $\mathring{\omega}_{\mu}{}^{a}{}_{b}$ with an additional undetermined vector component. See also~\cite{Dadhich:2012htv}, where a different method is used to come to the same conclusion. The undetermined contribution comes from the vanishing of the trace over the last two indices of equation~(\ref{spin_Palatini}).
This is again just a manifestation of the projective invariance of the action~(\ref{S_Palatini_tetrad}), this time written in terms of the orthonormal basis and spin connection
\begin{equation}
\omega_{\mu}{}^{a}{}_{b} \rightarrow \omega_{\mu}{}^{a}{}_{b} + \delta_{b}^{a} P_{\mu} \, .
\end{equation}
It is also not difficult to see that the LHS of~(\ref{spin_Palatini}) is indeed tensorial, as the difference between the connection and the Levi-Civita connection is simply the contortion tensor. In fact, the expression~(\ref{spin_Palatini}) is simply the Palatini tensor~(\ref{Palatini0}) written in terms of the mixed spacetime-Lorentz indices.

Perhaps it is no surprise that these first-order formulations still lead to equivalent results as the second-order formulations of GR. This is precisely because of our assumptions about the geometric quantities to which matter couples, and the unequal treatment of $S_{\textrm{grav}}$ and $S_{\textrm{M}}$. We will now study the case where matter couples to geometry via both the metric (or tetrad) and the affine connection, in accordance with the prescription used for the first-order formulation of gravity.

\section{Metric-affine gravitation}
\label{section3.2}

For matter that does couple to the connection, the crucial difference is that the connection variations no longer vanish in general. Moreover, the definitions of the energy-momentum tensors coming from the variations with respect to the metric or tetrad need to be modified. The result of considering a fully metric-affine theory is that the geometry is no longer the Riemannian (Levi-Civita) one of GR. The effects of curvature, torsion and non-metricity now all play a prominent role. We will, however, see the limits in which one can reinterpret the geometry as Riemannian but with additional modified matter contributions. 

Within this framework, we assume our gravitational action to be no higher than first-order in derivatives of the metric (or tetrad) and affine connection, e.g., $S_{\textrm{grav}}[g,\partial g, \bar{\Gamma}, \partial \bar{\Gamma}]$. The requirement of coordinate invariance implies that this can be written as $S_{\textrm{grav}}[g,\bar{R},T,Q]$, with the last three symbols standing for curvature, torsion and non-metricity. The whole gravitational theory can then be consistently called first-order, with the first-order matter action\footnote{Here we are assuming minimally coupled matter with up to first derivatives of the matter fields $S_{\textrm{M}}[\varPhi^{A}, \bar{\nabla} \varPhi^{A}]$. We then work with the set $(g, \bar{\Gamma}, \varPhi^A)$ or $(e, \omega, \varPhi^A)$, which both include the first partial derivatives of $\varPhi^{A}$. Note that one can instead work with torsion and non-metricity explicitly, see~\cite{Hehl:1994ue}.} taking the form $S_{\textrm{M}}[g,\bar{\Gamma},\varPhi^{A}]$ or $S_{\textrm{M}}[e,\omega,\varPhi^{A}]$. This also seems more elegant and natural from a mathematical standpoint. In the metric framework, variations of the matter action lead to the  \textit{metric (Hilbert) energy-momentum tensor}
\begin{equation} \label{affine_EM}
{}^{(\bar{\Gamma})} T_{\mu \nu} := -\frac{2}{\sqrt{-g}} \frac{\delta S_{\textrm{M}}[g,\bar{\Gamma},\varPhi^{A}]}{\delta g^{\mu \nu}} \, ,
\end{equation}
where the superscript $\bar{\Gamma}$ makes explicit that the metric and connection are
treated as completely independent variables~\cite{Boehmer:2023fyl}. This tensor is again symmetric by definition.

In the first-order tetrad formalism, the variation of the matter action with respect to the tetrad field leads to the \textit{canonical (Noether) energy-momentum tensor}
\begin{equation} \label{canonical}
\Sigma^{a}{}_{\mu} := -\frac{1}{e} \frac{\delta S_{\textrm{M}}[e,\omega,\varPhi^{A}]}{\delta e_{a}{}^{\mu}} \, ,
\end{equation}
which is no longer symmetric in general, see~\cite{ortin2004gravity,Hehl:1978b}. As we commented previously, this is equivalent to the (symmetric) tetradic energy-momentum tensor in the case where matter does not couple to the spin-connection, e.g., bosonic matter~\cite{leclerc2006canonical}.

For further details on the different matter tensors we again refer to the review by Hehl et. al.~\cite{Hehl:1994ue}. In particular, they study the corresponding Noether currents of matter in the first-order formalism and give more details on the role of the metrical (Hilbert) versus canonical (Noether) energy-momentum tensors. For the precise relationship between the matter tensors $\Theta^{a}{}_{\mu}$ and $\Sigma^{a}{}_{\mu}$ in equations~(\ref{tetradic_EM}) and~(\ref{canonical}) respectively, see~\cite{leclerc2006canonical,Blagojevic:2013xpa,Hehl:1994ue,Hehl:1978b}.

Let us now look at the affine connection. Here we will stick to working in the metric formalism, to save writing all equations twice. 
For the variations with respect to the connection, let us define the \textit{hypermomentum} current as
\begin{equation} \label{hypermomentum}
\Delta^{\mu \nu}{}_{\lambda} := - \frac{1}{\sqrt{-g}} \frac{\delta S_{\textrm{M}}[g,\bar{\Gamma},\varPhi^{A}]}{\delta \bar{\Gamma}^{\lambda}_{\mu \nu}} \, ,
\end{equation}
which should be regarded as an intrinsic property of matter. Note the index placement of $\Delta^{\mu \nu}{}_{\lambda}$, which often differs in the literature. 
For further important details such as how it decomposes into spin, dilation and shear currents, along with the physical interpretation of this tensor, see~\cite{HehlKerlickHeyde+1976+111+114,HehlKerlickHeyde+1976+524+527,HehlKerlickHeyde+1976+823+827,Obukhov:1993pt,Babourova:1998mgh}. 
 Note that one can instead decompose the affine connection into its torsion and non-metricity components and take the fundamental variables to be $(g,T,Q)$, see~\cite{HehlKerlickHeyde+1976+823+827}.
However, care must be taken when reformulating in terms of these variables, especially with regard to the metric/tetrad matter currents, because the order of derivatives in the Lagrangian will change.

For the simplest choice of action, the affine Ricci scalar, we are led to the field equations~\cite{Hehl:1978a,Hehl:1978b} 
\begin{align} \label{MA_1}
\bar{G}_{(\mu \nu)} &= \kappa {}^{(\bar{\Gamma})} T_{\mu \nu} \, , \\
P^{\mu \nu}{}_{\lambda} &= 2\kappa \Delta^{\mu \nu}{}_{\lambda} \, . \label{MA_2}
\end{align}
Clearly for vanishing hypermomentum, we retrieve GR, just like in the Palatini case. %
In the first work on metric-affine gravity, the proposed action contained an additional term $L = \bar{R} + \beta Q_{i} Q^{i}$, where $Q_{i}$ is the non-metricity vector~\cite{hehl1976new}. The purpose of the additional term is to fix the indeterminacy presented in the Palatini tensor as a result of the projective invariance of the affine Ricci scalar~\cite{Hehl:1978a}. For more general Lagrangians and studies of their solutions, see~\cite{Blagojevic:2013xpa} and references therein. We also refer to~\cite{Obukhov:2014nja} for details of conservation laws in metric-affine theories.

Because the connection equation of motion~(\ref{MA_2}) is algebraic instead of dynamical, it is possible to remove the connection altogether. We will show this explicitly in the simple case of Einstein-Cartan theory, but this has been shown more generally for a wide class of metric-affine actions that lead to algebraic connection field equations~\cite{vitagliano2011dynamics}. The procedure involves rewriting the torsion and non-metricity of the affine connection algebraically in terms of the matter fields (through the hypermomentum equation). This can then be substituted into the metric field equation, leading to an Einstein-like equation of motion
\begin{equation}
\bar{G}_{(\mu \nu)} = \kappa \mathcal{T}_{\mu \nu} \, ,
\end{equation}
where $\mathcal{T}_{\mu \nu}$ is a modified rank 2 energy-momentum tensor depending on ${}^{(\bar{\Gamma})} T_{\mu \nu}$ and $\Delta^{\mu \nu}{}_{\lambda}$. In this sense, these theories can be interpreted as General Relativity with modified matter interactions. However, this is somewhat unnatural\footnote{For further details on why torsion and non-metricity should not simply be regarded as additional tensor fields on top of the Riemannian structure of GR, see Chapter 7 of~\cite{Blagojevic:2013xpa}. This is intimately linked with the gauge structure of metric-affine theories.} and spoils the minimal coupling interpretation~\cite{FH1976}.

In order to have the affine structure play a more prominent role, the independent components of the connection can be made to be dynamical. This is precisely what we propose in Section~\ref{section5.3}, where we study solutions with propagating torsion. This is accomplished by considering actions that lead to field equations which are dynamical in torsion and non-metricity. It is useful to first study Einstein-Cartan theory in some detail, outlining how the different approaches are equivalent.

\subsection{Einstein–Cartan–Sciama–Kibble theory}
\label{section3.2.1}
Some of the earliest explorations of gravitational theories beyond Riemannian geometry can be seen in Cartan's work~\cite{cartan2001riemannian}, where the connection is allowed to possess torsion as well as curvature. This geometry is more generally known as a Riemann-Cartan spacetime. The prototypical example of a consistent gravitational theory within this framework is \textit{Einstein–Cartan–Sciama–Kibble theory}~\cite{sciama1962analogy,Kibble:1961ba}, or Einstein-Cartan for short. The action is simply the Ricci scalar of a connection containing both curvature and torsion but without non-metricity. We can therefore choose to describe the theory by the set $(g,T)$ or by $(g,\tilde{\Gamma})$ where $\tilde{\Gamma}$ is the \textit{Cartan connection} with vanishing non-metricity. Equally one can work in the orthonormal basis with the tetrad and spin connection instead, which we will outline below. In Sec.~\ref{section3.2.2} we use the Lagrange multiplier method to eliminate non-metricity from our theory, and show how this compares to the other formulations.

The standard presentation, working with the tetrad and Cartan spin connection $\tilde{\omega}$, leads to equations of motion~\cite{sciama1962analogy,Kibble:1961ba}
\begin{align}
\tilde{G}_{a}{}^{\mu} &= \kappa \Sigma_{a}{}^{\mu} \, ,\\
T^{\mu}{}_{ab} + e_{a}{}^{\mu} T^{\nu}{}_{b \nu} - e_{b}{}^{\mu} T^{\nu}{}_{a \nu}  &= 2\kappa S^{\mu}{}_{ab}
 \, ,
 \end{align}
where $\Sigma_{a}{}^{\mu}$ is the canonical energy-momentum tensor~(\ref{canonical}), here with respect to the Cartan connection $\tilde{\omega}_{\mu a b} = - \tilde{\omega}_{\mu ba}$. We also define the \textit{intrinsic spin} tensor as
\begin{equation} \label{spin_current}
S^{\mu}{}_{a}{}^{b} := -\frac{1}{e} \frac{\delta S_{\textrm{M}}[e, \tilde{\omega},\varPhi^{A}]}{\delta \tilde{\omega}_{\mu}{}^{a}{}_{b}} \, .
\end{equation}
 This is the gravitational analogue of the spin tensor of Special Relativity~(\ref{SR_canonical}). It follows that the connection equations are also antisymmetric.
 Writing this purely in terms of spacetime indices by contracting with the tetrad leads to
\begin{align}\label{EC_2a}
\tilde{G}_{\mu \nu} &= \kappa \Sigma_{\mu \nu} \, , \\
T^{\mu}{}_{\rho \sigma} + \delta^{\mu}_{\rho} T^{\nu}{}_{\sigma \nu} - \delta^{\mu}_{\sigma} T^{\nu}{}_{\rho \nu}  &= 2\kappa S^{\mu}{}_{\rho \sigma} \label{EC_2b} \, .
\end{align}
For more details on these equations of motion, see~\cite{trautman1972einstein,FH1976,Shapiro:2001rz}. We will briefly discuss solutions and the deviations from GR shortly.

In Riemann-Cartan spaces, the spin connection equation of motion is equivalent to the Palatini tensor $P^{\mu}{}_{\rho \sigma}$~(\ref{Palatini0}) when non-metricity is set to zero. This follows from the fact that the hypermomentum can be decomposed into spin, sourced by torsion, and shear and dilation sourced by non-metricity~\cite{Hehl:1994ue}. In a Riemann-Cartan space, only the spin contribution survives. It follows that~(\ref{EC_2b}) can be written compactly as $\tilde{P}^{\mu}{}_{\sigma \rho} = 2 \kappa S^{\mu}{}_{\rho \sigma}$, where $\tilde{P}^{\mu}{}_{\rho \sigma}$ is the Palatini tensor with vanishing non-metricity. (Note the ordering of indices on the Palatini tensor, with an overall minus sign coming from the definitions of hypermomentum and spin.) It is important to note that the metric-affine equations of motion cannot be directly compared to the Cartan ones here, as the matter currents differ. This will become more clear when using the Lagrange multiplier method.

By decomposing the field equations into their symmetric and antisymmetric parts, and making use of the geometric identity $(\tilde{\nabla}_{\lambda} + T^{\sigma}{}_{\lambda \sigma}) \tilde{P}^{\lambda}{}_{\nu \mu} =2 \tilde{R}_{[\mu \nu]}$~\cite{JS1954}, it is possible to retrieve a conservation equation of the form
\begin{equation} \label{spin_skew}
(\tilde{\nabla}_{\lambda} + T^{\sigma}{}_{\lambda \sigma}) S^{\lambda}{}_{\mu \nu} = \Sigma_{[\mu \nu]} \, .
\end{equation}
This is the Riemann-Cartan generalisation of the Special Relativistic conservation law of total angular momentum~\cite{Kibble:1961ba}, given in equation~(\ref{SRc2}). One can then construct a symmetric Belinfante-Rosenfeld energy-momentum tensor in an analogous way to~(\ref{SR_BR}), see~\cite{FH1976}. There is also of course a similar Noether identity (covariant conservation law) for the canonical matter current $\Sigma_{\mu \nu}$, which can be found in~\cite{Hehl:1994ue}, but we will not need the explicit form here.

We have therefore answered the question posed at the end of Section~\ref{section2.3.2}, which was to search for the General Relativistic analogue of spin-angular momentum~(\ref{SR_canonical}). Of course, this is now localised and viewed as an \textit{internal} property of matter. This should be no surprise as Einstein-Cartan theory can be derived from gauging the local Lorentz group~\cite{sciama1962analogy,Kibble:1961ba}. Before discussing more of the gauge theory motivations, let us compare the above formulation with the metric-affine one (in the coordinate basis) by making use of Lagrange multipliers.

\subsection{Lagrange multiplier approach}
\label{section3.2.2}
Let us now approach Einstein-Cartan theory using the Lagrange multiplier method. This, in our opinion, is the clearest and most direct way to implement constraints on the affine connection within the metric-affine framework. This also shows the equivalence between the different variational methods, see~\cite{kichenassamy1986lagrange}. 
Here we follow our work~\cite{Boehmer:2023fyl}. We also refer back to the works on variation principles in metric-affine theories by Hehl~\cite{Hehl:1978a,Hehl:1978b}. Do note, however, that the indices of many of the geometric objects, such as torsion, the Palatini tensor and hypermomentum, are permuted with respect to their definitions in Hehl's work. We stay consistent with our own definitions given throughout the thesis.

Let us supplement our action $L_{\textrm{grav}} = \bar{R}$ with the following Lagrange multiplier term
\begin{align} \label{EC_lambda}
  S_Q = -\int \frac{1}{2 }\lambda^{\mu \nu \rho} Q_{\mu \nu \rho} \sqrt{-g}\, d^4 x =
  \int \frac{1}{2 }\lambda^{\mu \nu \rho} \bar{\nabla}_{\mu} g_{\nu \rho} \sqrt{-g}\, d^4 x\,,
\end{align}
which ensures that non-metricity vanishes. The Lagrange multiplier shares the symmetries with the non-metricity tensor $\lambda^{\mu \nu \rho} = \lambda^{\mu \rho \nu}$, and we now vary with respect to the set $(g,\bar{\Gamma},\lambda)$. The constraint equation from the $\lambda^{\mu \nu \rho}$ variations leads to the condition
\begin{equation}
\bar{\nabla}_{\mu} g_{\nu \rho} = 0 \, .
\end{equation}
After implementing vanishing non-metricity, the inclusion of the metric and connection variations in the full affine field equations~(\ref{MA_1})-(\ref{MA_2}) yields
\begin{align}
  \label{ECl1}
  \tilde{G}_{(\mu \nu)} + \kappa (\tilde{\nabla}_{\rho} + T^{\sigma}{}_{\rho \sigma}) \lambda^{\rho}{}_{\mu \nu} &=
  \kappa {}^{(\overline{\Gamma})}T_{\mu \nu} \,, \\
  \label{ECl2}
  \tilde{P}_{\mu \nu \rho} - 2\kappa \lambda_{\mu \nu \rho} &= 2\kappa \Delta_{\mu \nu \rho} \,,
\end{align}
which can also be found in~\cite{Hehl:1978b,Boehmer:2023fyl}.

The symmetric part of~(\ref{ECl2}) singles out the Lagrange multiplier $\lambda_{\mu \nu \rho} = - \Delta_{\mu(\nu \rho)}$, whilst the skew-symmetric part of this equation gives $\tilde{P}_{\mu \nu \rho} = 2\kappa \Delta_{\mu[\nu \rho]}$. Substituting these results back into the first field equation~(\ref{ECl1}) gives the desired field equations 
\begin{align}
  \label{EEl1b}
  \tilde{G}_{(\mu \nu)} &=
  \kappa {}^{(\overline{\Gamma})} T_{\mu \nu} + \kappa (\tilde{\nabla}_{\rho} + T^{\sigma}{}_{\rho \sigma}) \Delta^{\rho}{}_{(\mu \nu)}\,, \\
  \label{EC2b}
  \tilde{P}_{\mu \nu \rho} &= 2\kappa \Delta_{\mu[\nu \rho]} \, .
\end{align}
Now that the Lagrange multipliers have been eliminated, these two final equations can be taken as the equations for Einstein-Cartan theory.

Let us compare this with our previous calculations involving the tetrad and Cartan spin connection, which we saw led to the canonical energy-momentum tensor on the right-hand side of the Einstein equation~(\ref{EC_2a}). Symmeterising that equation and equating it to our (symmetric) Einstein equation~(\ref{EEl1b}) leads to
\begin{equation}
\Sigma_{(\mu \nu)} =   \kappa {}^{(\overline{\Gamma})} T_{\mu \nu} + \kappa (\tilde{\nabla}_{\rho} + T^{\sigma}{}_{\rho \sigma}) \Delta^{\rho}{}_{(\mu \nu)} \, .
\end{equation}
The skew-symmetric part of the connection equations lets us relate the hypermomentum current with the intrinsic spin tensor 
\begin{equation}
\Delta_{\lambda [\mu \nu]} = S_{\lambda \nu \mu} \, .
\end{equation}
Lastly, using the conservation equation for the skew-symmetric part of the canonical energy-momentum tensor in~(\ref{spin_skew}) allows us to obtain the following equation
\begin{align} \nonumber
\Sigma_{\mu \nu} &= \Sigma_{(\mu \nu)} +\Sigma_{[\mu \nu]} =  
  \kappa {}^{(\overline{\Gamma})} T_{\mu \nu} + \kappa (\tilde{\nabla}_{\rho} + T^{\sigma}{}_{\rho \sigma}) \Delta^{\rho}{}_{(\mu \nu)} +\kappa (\tilde{\nabla}_{\rho} + T^{\sigma}{}_{\rho \sigma}) \Delta^{\rho}{}_{[\nu \mu]} \\
  &=  \kappa {}^{(\overline{\Gamma})} T_{\mu \nu} + \kappa (\tilde{\nabla}_{\rho} + T^{\sigma}{}_{\rho \sigma}) \Delta^{\rho}{}_{\nu \mu} \, . \label{Belinfante}
\end{align}
This is the Belinfante-Rosenfeld relation between the Hilbert and canonical energy-momentum tensors in this first-order formalism (for reference, this exact equation can be found in (47) of~\cite{Hehl:1978b}). It follows that we can rewrite the field equations~(\ref{EEl1b}) and~(\ref{EC2b}) exactly in the form given previously in~(\ref{EC_2a}) and~(\ref{EC_2b}).

Let us briefly comment on the solutions of Einstein-Cartan theory, and how they differ from General Relativity. Because the connection equation of motion~(\ref{EC_2b}) is algebraic in torsion, one can solve for torsion in terms of the matter fields. Upon substitution into the first field equation~(\ref{EEl1b}), it is possible to decompose the Einstein tensor into its Levi-Civita part and additional matter terms. The result is a modified Einstein field equation with a modified energy-momentum tensor on the right-hand side, with contributions from the intrinsic spin of matter. One can then ask how much of a role these additional spin terms play. The answer is not much, except at very high spin-densities~\cite{FH1976}. For example, the critical density for electrons, where the spin contributions are comparable to their mass density, is around $\sim 10^{48} \textrm{g} /\textrm{cm}^{3}$~\cite{Hehl:1973qn}. However, in extreme environments, such as in the early universe, these effects could dominate. For more details on the properties and solutions of Einstein–Cartan–Sciama–Kibble theory we again refer to~\cite{trautman1972einstein,FH1976,Shapiro:2001rz}.

\subsection{Gauge theoretic approach}
\label{section3.2.3}

Einstein–Cartan–Sciama–Kibble gravity is a particular case of \textit{Poincar\'{e} Gauge theory}~\cite{Hayashi:1979wj,hehl1980four}, the class of gravitational theories within the Riemann-Cartan geometry. As the name suggests, it is constructed from the gauging of the Poincar\'{e} group~\cite{Blagojevic:2013xpa,Ivanenko:1983fts}. In Utiyama's work~\cite{utiyama1956invariant} he looked at deriving General Relativity by gauging the (homogeneous) Lorentz group $SO(1,3)$. However, the Levi-Civita connection was assumed a priori. Kibble instead studied the (inhomogeneous) Poincar\'{e} group~\cite{Kibble:1961ba}, with the translation potentials represented by $\mathbf{e}_{a}$ and the Lorentz potentials represented by $\mathbf{\omega}^{a}{}_{b}$. The use of the Poincar\'{e} group is essential because it includes the \textit{translational} subgroup, which is linked with the conservation of energy-momentum~\cite{Blagojevic:2002du}. This was shown to lead to a consistent gravitational framework with both torsion and curvature.

Historically, gauge theories arose from Weyl's study of electrodynamics and the unitary group $U(1)$. The invariance under \textit{local} phase transformations
\begin{equation}
\phi(x) \rightarrow e^{i \alpha(x)} \phi(x) \, , \qquad \qquad A_{\mu} \rightarrow A_{\mu} + \partial_{\mu} \alpha(x) \, ,
\end{equation}
necessitate the use of the gauge covariant derivative
\begin{equation}
D_{\mu} = \partial_{\mu} - i A_{\mu} \, .
\end{equation}
Gauging the $U(1)$ group implies that the parameters $\alpha(x)$ are functions of spacetime, as opposed to constants like in the global setting.
Hence, if we wish our Lagrangian to remain invariant, we see the appearance of the gauge potentials $A_{\mu}(x)$, which in this case represent the electromagnetic potential field. 

In general, the gauging of global (or rigid) symmetries involves promoting these invariances to local ones; to preserve the invariance of the Lagrangian, compensating gauge fields (connections) are then introduced. The rigid symmetries that are associated with conserved currents by Noether's theorem $dJ = 0$ then couple to the gauge fields $dJ = 0 \rightarrow D J = 0$, where $D$ is the covariant derivative operator associated with the gauge connection. Thus, the conservation laws are promoted to local, covariant ones.

 The gauge invariance of electromagnetism was generalised to the non-Abelian group $SU(2)$ by Yang and Mills~\cite{Yang:1954ek}, and then to semi-simple Lie groups including the Lorentz group by Utiyama~\cite{utiyama1956invariant}. As mentioned, Kibble~\cite{Kibble:1961ba} then extended this to the Poincar\'{e} group\footnote{This is then further extended to the full affine group, which is associated with the metric-affine theories studied above, see~\cite{lord1978metric,Blagojevic:2013xpa}.}. The principles remain the same throughout, and the appearance of the gauge potentials $\mathbf{e}_{a}$ and $\mathbf{\omega}^{a}{}_{b}$ in transitioning to the local Poincar\'{e} group is quite remarkable. 

In Poincar\'{e} gauge theory, the Lagrangian can generally be taken to be of the form
\begin{equation} \label{Poincare_L}
L_{\textrm{total}} = L_{\textrm{grav}}(e,T,\tilde{R}) + L_{\textrm{M}}(e,\varPhi^{A}, \tilde{\nabla} \varPhi^{A}) \, ,
\end{equation}
see for example~\cite{Blagojevic:2013xpa}.
Einstein-Cartan is a degenerate case, in the sense that the translational excitation (the variations with respect to torsion) vanishes trivially. GR is then contained within Einstein-Cartan when the intrinsic spin of matter vanishes. Another important degenerate case, which we will study in Sec.~\ref{section3.3}, is \textit{teleparallel gravity}, where the variations with respect to the curvature vanishes. In Chapters~\ref{chapter4} and~\ref{chapter5} we will go beyond this framework and study modified theories where the gravitational action is even more general than the one in~(\ref{Poincare_L}). All of the previous work on non-Riemannian theories serves as motivation for investigating models in these generalised geometric spaces.

Let us conclude this subsection with a relevant quote on torsion and Einstein-Cartan theory from Sciama\footnote{Private communication from Sciama to Hehl~\cite{hehl1980four}.}: 
\begin{quote}
\textit{
``The idea that
spin gives rise to torsion should not be regarded as an ad hoc modification of general
relativity. On the contrary, it has a deep group theoretical and geometric basis. If history
had been reversed and the spin of the electron discovered before 1915, I have little doubt
that Einstein would have wanted to include torsion in his original formulation of general
relativity. On the other hand, the numerical differences which arise are normally very
small, so that the advantages of including torsion are entirely theoretical.''}
\end{quote}

\section{Teleparallel gravity}
\label{section3.3}
Teleparallel gravity is the geometric framework where curvature vanishes but torsion and non-metricity do not. The geometry was first introduced by Weitzenb\"{o}ck and Cartan with the so-called \textit{Weitzenb\"{o}ck connection} $\overset{\mathcal{W}}{\Gamma}{}_{\mu \nu}^{\lambda}$~\cite{weitzenbock1923invariantentheorie}. A subset of Riemann-Cartan spaces, the Weitzenb\"{o}ck space has non-zero torsion but is globally flat
\begin{equation}
\bar{R}_{\mu \nu a}{}^{b}(\mathcal{W}) = 0 \, , \qquad \overset{\mathcal{W}}{\nabla}_{\mu} g_{\nu \lambda} = 0 \, .
\end{equation}
We will later refer to this as \textit{metric teleparallel} gravity, to distinguish it from the \textit{symmetric teleparallel} theories with vanishing torsion but containing non-metricity instead, as studied in Sec.~\ref{section3.3.2}.

This (metric) teleparallel framework was used by Einstein in 1928~\cite{einstein1928riemann,Unzicker:2005in}, under the name \textit{Fernparallelismus}, or \textit{teleparallelism} or \textit{absolute parallelism}. A manifold and connection $(\mathcal{M}, \bar{\Gamma})$ is said to possess absolute parallelism if a quantity can be parallel transported between any two points and the end result is independent from the choice of curve between those points~\cite{JS1954}. Requiring teleparallelism in a Riemannian space leads to flat (Minkowski) spacetime, but this is not the case in a Riemann-Cartan geometry (or in a Weyl geometry). The Weitzenb\"{o}ck connection is then \textit{defined by} having this property $\overset{\mathcal{W}}{\nabla} \mathbf{e}_{a} = 0$, where
\begin{equation} \label{Weitzenbock}
 \overset{\mathcal{W}}{\nabla}_{\mu} e_{a}{}^{\lambda} := \partial_{\mu} e_{a}{}^{\lambda} + \overset{\mathcal{W}}{\Gamma}{}_{\mu \nu}^{\lambda} e_{a}{}^{\nu} = 0 \, .
\end{equation}
It follows that the Weitzenb\"{o}ck connection can be written explicitly as $ \overset{\mathcal{W}}{\Gamma}{}_{\mu \nu}^{\lambda} = -e^{a}{}_{\nu} \partial_{\mu} e_{a}{}^{\lambda}$, and the associated torsion tensor is $T^{\lambda}{}_{\mu \nu} =2 e^{a}{}_{[\mu} \partial_{\nu]} e_{a}{}^{\lambda}$.
Note that the covariant derivative above is defined to act only on the spacetime indices, as opposed to the \textit{total} covariant derivative introduced in~(\ref{total derivative}). However, from this same definition it follows that the Weitzenb\"{o}ck spin connection is zero $\overset{\mathcal{W}}{\omega}{}_{\mu}{}^{a}{}_{b} = 0$. We will revisit this shortly, and see how this relates to local Lorentz transformations and the choice of frame $\mathbf{e}_{a}$.

Just like Weyl ten years prior, Einstein's goal was to unify gravity and electromagnetism~\cite{Goenner:2004se,Goenner:2014mka}. Despite being unsuccessful, the ideas were revived in the sixties by M\o{}ller~\cite{moller1961conservation,Moller:1962vis} with a different aim; M\o{}ller was instead interested in conservation laws and the energy-momentum of the gravitational field, and found the tetrad formalism and Einstein's teleparallelism to be useful. We will revisit these topics, and M\o{}ller's tetradic constructions in particular, in the following chapter. 

Not long after M\o{}ller's work, teleparallel gravity was given a Lagrangian formulation~\cite{pellegrini1963tetrad}, and then reformulated as a gauge theory~\cite{Hayashi:1967se,hayashi1977gauge}. From here, Cho showed that the action of teleparallel gravity was equivalent to the Einstein-Hilbert action, up to boundary terms (see Sec.~\ref{chapter4})~\cite{Cho:1975dh}. This was extended by Hayashi and Shirafuji to a more general action, known as \textit{New General Relativity}~\cite{Hayashi:1979qx}. Modified theories of teleparallel gravity have now become popular in recent years~\cite{Bahamonde:2021gfp}, and we study these in Chapter~\ref{chapter5}. For a review of the teleparallel equivalent of General Relativity see~\cite{Maluf:2013gaa}, and for an introduction and detailed treatment of teleparallel gravity as a whole, see~\cite{Aldrovandi:2013wha}.

 Just as Einstein-Cartan theory is a natural result of gauging the inhomogeneous Poincar\'{e} group, the (metric) teleparallel theories arise from gauging the group of translations, see~\cite{Hayashi:1967se,hayashi1977gauge,blagojevic2000gauge}. An interesting consequence of viewing teleparallel gravity as a gauge theory is that the equivalence principle, and the universality of free-fall, is not actually a necessary requirement~\cite{Aldrovandi:2003pa,Aldrovandi:2013wha}. It should be noted, however, that the gauge structure of the translation group $T(4)$ is quite different to that of the Lorentz group $SO(1,3)$. Hence, some extra care must be taken in the interpretation of teleparallel gravity as a gauge theory\footnote{These issues primarily stem from the fact that gauge theory was developed with  \textit{internal} symmetry groups in mind, such as $U(1)$, $SU(2)$, $SO(1,3)$, etc.}, especially with regards to the roles of spacetime coordinate transformations~\cite{nester1984gravity}. 
 For further discussions on the gauge theory approach to teleparallel theories, see~\cite{Pereira:2019woq,blagojevic2000gauge,Blagojevic:2013xpa,Aldrovandi:2013wha}.
 
Teleparallel gravity can also be approached from the metric-affine framework~\cite{Obukhov:2002tm,BeltranJimenez:2018vdo,BeltranJimenez:2019esp}. This will be the approach taken here. We will also now use the term `teleparallelism' to refer to theories with vanishing affine curvature, `metric teleparallelism' for theories with vanishing curvature and non-metricity, and `symmetric teleparallelism' for theories with vanishing curvature and torsion. The Weitzenb\"{o}ck connection is therefore associated with metric teleparallism.

\subsubsection{General teleparallel gravity}

Here we will focus on studying teleparallel gravity geometrically, by beginning with the metric-affine structure endowed by an independent metric and connection $(g, \bar{\Gamma})$, or tetrad and spin connection pair $(e, \omega)$. Working in the coordinate basis, the teleparallel constraint of vanishing curvature can be written explicitly as
\begin{equation} \label{Teleparallel_constraint}
\bar{R}_{\mu \nu \lambda}{}^{\gamma} = 2 \bar{\Gamma}^{\lambda}_{[\mu| \rho} \bar{\Gamma}^{\rho}_{\nu] \gamma} + 2 \partial_{[\mu} \bar{\Gamma}^{\lambda}_{\nu] \gamma} = 0  \, .
\end{equation}
In the language of differential forms and in an orthonormal basis this is 
\begin{equation} \label{Teleparallel_constraint2}
\bar{R}_{b}{}^{a} = d \omega^a{}_{b} + \omega^{a}{}_{c} \wedge \omega^{c}{}_{b} = 0 \, .
\end{equation}

As mentioned previously, the Riemann tensor serves as an integrability condition for the connection. With the teleparallel constraint, we can then solve~(\ref{Teleparallel_constraint}) for the affine connection
\begin{equation} \label{Teleparallel_sol1}
\bar{\Gamma}^{\lambda}_{\mu \nu} = \Omega^{\lambda}{}_{a} \partial_{\mu} (\Omega^{-1})^{a}{}_{\nu} \, ,
\end{equation}
where the arbitrary matrix $\Omega^{a}{}_{\mu}(x) \in GL(4,\mathbb{R})$ is to be later determined~\cite{BeltranJimenez:2019esp,Krssak:2018ywd}. Similarly, in the differential form calculus we can solve~(\ref{Teleparallel_constraint2}) to obtain
\begin{equation} \label{Teleparallel_sol2}
\omega^{a}{}_{b} =  \Omega^{a}{}_{c} d (\Omega^{-1})^{c}{}_{b} \, ,
\end{equation}
which can also be obtained from~(\ref{Teleparallel_sol1}).
We will use these solutions in the next parts.

If we use the decomposition of the affine Ricci scalar into its Levi-Civita part and contortion~(\ref{R_decomp}), in the teleparallel limit we have 
\begin{equation} \label{teleparallel_decomp}
0 = R +  K_{\lambda \sigma}{}^{\lambda} K_{\nu}{}^{\nu \sigma} - K_{\nu \sigma}{}^{\lambda} K_{\lambda}{}^{\nu \sigma}  + \nabla_{ \lambda} K_{\nu}{}^{\nu \lambda} - \nabla_{\nu} K_{\lambda}{}^{\nu \lambda}  \, .
\end{equation}
The first term is the Einstein-Hilbert Lagrangian, and the final two terms are total derivatives that do not affect equations of motion. From this equation, we therefore see that the specific contractions of the contortion tensors $K_{\lambda \sigma}{}^{\lambda} K_{\nu}{}^{\nu \sigma} - K_{\nu \sigma}{}^{\lambda} K_{\lambda}{}^{\nu \sigma}$ can be equated with the action of General Relativity, up to boundary terms. This serves as the basis and starting point of the \textit{teleparallel equivalents of General Relativity}. It is then not hard to be convinced that the general teleparallel action, supplemented with a Lagrange multiplier term $r^{\mu \nu \lambda}{}_{\gamma}$ enforcing the vanishing curvature constraint
\begin{multline} \label{teleparallel_general}
S[g,\bar{\Gamma},r] = -\frac{1}{2\kappa} \int \big( K_{\lambda \sigma}{}^{\lambda} K_{\nu}{}^{\nu \sigma} - K_{\nu \sigma}{}^{\lambda} K_{\lambda}{}^{\nu \sigma} \big) \sqrt{-g} d^4 x  \\ 
+ \int \bar{R}_{\mu \nu \lambda}{}^{\gamma} r^{\mu \nu \lambda}{}_{\gamma} \sqrt{-g} d^4 x \, ,
\end{multline}
leads to equivalent equations of motion as General Relativity. To see this explicitly, we refer to~\cite{BeltranJimenez:2019odq,Hohmann:2021fpr}.

Note that other approaches besides the Lagrange multiplier method can also be taken, such as the restricted variations~\cite{BeltranJimenez:2019odq,Bahamonde:2021gfp}. We will talk about how these are implemented in the next part. Moreover, in the next two chapters we will look specifically at the derivations of these types of equations of motion. 

\subsection{Metric teleparallel gravity}
\label{section3.3.1}

Let us focus on the teleparallel framework with torsion only, where non-metricity has also been set to zero. Historically, the connection was chosen to be the Weitzenb\"{o}ck one, as originally studied by Einstein and others. In this case, the (metric) teleparallel equivalent of General Relativity, which we will refer to as TEGR, can be written completely in terms of the tetrad and its first derivatives~\cite{Maluf:2013gaa}. The contortion tensors in equation~(\ref{teleparallel_decomp}) are simply permutations of the torsion tensor $T^{\lambda}{}_{\mu \nu} =2 e^{a}{}_{[\mu} \partial_{\nu]} e_{a}{}^{\lambda}$,

However, note that this object is not a local Lorentz scalar. Moreover, the torsion tensor $T^{a}{}_{\mu \nu} =2 \partial_{[\mu} e^{a}{}_{\nu]}$ does not even transform covariantly with respect to local Lorentz transformations. For this reason, this original formulation of TEGR does not inherit the full local Lorentz symmetry that is usual to GR~\cite{Krssak:2018ywd}\footnote{Instead, the theory is \textit{pseudo-invariant}, as shown in the early work~\cite{Cho:1975dh}. We will cover this topic in much more detail shortly, so will delay these discussions until then.}. However, the resulting field equations (and hence the dynamics of the theory as a whole) are equivalent to Einstein's General Relativity. 
For more details on TEGR we refer to~\cite{Hohmann:2021fpr,BeltranJimenez:2019esp,Nester:1998mp,Aldrovandi:2013wha,Maluf:2013gaa}.

Let us sketch how one arrives at the metric teleparallel action (for explicit details, see any of the previous references). First, take the decomposition of the affine Ricci scalar~(\ref{teleparallel_decomp}), set non-metricity to zero and then explicitly expand the contortion tensor into torsion terms. For the covariant derivative terms, it is not difficult to see that they can be written as $2\nabla_{[\lambda} K_{\nu]}{}^{\nu \lambda} = 2\nabla_{\lambda} T^{\nu \lambda}{}_{\nu}  = \frac{2}{e} \partial_{\mu}(e T^{\lambda \mu}{}_{\lambda})$. Including a similar calculation for the quadratic torsion terms in~(\ref{teleparallel_decomp}) leads to~\cite{Aldrovandi:2013wha}
\begin{equation} \label{R_torsion_scalar}
0 = R + \Big( \frac{1}{4} T^{abc} T_{abc} + \frac{1}{2} T^{abc} T_{bac} - T^{b}{}_{b}{}^{a} T^{c}{}_{c a} \Big) + \frac{2}{e} \partial_{\mu}(e T^{\lambda \mu}{}_{\lambda}) \, .
\end{equation}
It is useful to introduce the \textit{torsion superpotential}, defined as
\begin{align} 
 \mathscr{S}^{\mu \nu \lambda} &:= \frac{1}{2} K^{\mu \lambda \nu} + g^{\mu [\nu} T^{|\rho|}{}_{\rho}{}^{\lambda]} \nonumber \\
 &=  \frac{1}{4} \big( T^{\mu \nu \lambda} + T^{\nu \mu \lambda} + T^{\lambda \nu \mu}  \big) +  \frac{1}{2} \big( g^{\mu \nu} T^{\rho}{}_{\rho}{}^{\lambda} -g^{\mu \lambda} T^{\rho}{}_{\rho}{}^{\nu} \big) \, ,\label{T_superpotential}
\end{align}
which is skew-symmetric $\mathscr{S}^{\mu \nu \lambda} = - \mathscr{S}^{\mu \lambda \nu}$. 
We now define the quadratic part in the brackets of~(\ref{R_torsion_scalar}) as the \textit{torsion scalar}, and the total derivative term (with a minus sign) as the \textit{torsion boundary term}
\begin{align} \label{T_scalar}
T &:=  \mathscr{S}^{abc} T_{abc} = \frac{1}{4} T^{abc} T_{abc} + \frac{1}{2} T^{abc}{}_{bac} - T^{b}{}_{b}{}^{a} T^{c}{}_{c a} \\
B_T &:= - \frac{2}{e} \partial_{\mu}(e T^{\lambda \mu}{}_{\lambda}) \, . \label{B_T}
\end{align}
It follows that we can equate the Levi-Civita Ricci scalar with these two terms
\begin{equation} \label{R_T_B}
R(e) = -T + B_T \, .
\end{equation}
Here the use of $R(e)$ is a reminder that the Ricci scalar is defined solely in terms of the tetrad and its derivatives. It should be understood that the torsion terms are defined in terms of the Weitzenb\"{o}ck connection, and hence just the tetrad as well.

The boundary term will not contribute to the equations of motion, so the action for TEGR\footnote{Note that here we are looking at the so-called `restricted' approach, where have explicitly used the solutions of the metric teleparallel constraints $\bar{R}_{\mu \nu \lambda}{}^{\gamma}=0$, $Q_{\mu \nu \lambda}=0$ in our connection solution $\overset{\mathcal{W}}{\Gamma}{}_{\mu \nu}^{\lambda} = -e^{a}{}_{\nu} \partial_{\mu} e_{a}{}^{\lambda}$. This differs from the Lagrange multiplier method, but the results are equivalent.} is defined to be the torsion scalar
\begin{equation}
S_{\textrm{TEGR}}[e] = -\frac{1}{2 \kappa} \int T e \, d^4x \, .
\end{equation}
The equations of motion from the variations with respect to the tetrad will be equivalent to General Relativity in the tetrad formalism, see~(\ref{EFE tetrad}). We will study this action and its properties explicitly in Sec.~\ref{section4.2}, but we can already see from the definition of the Weitzenb\"{o}ck connection that the torsion scalar~(\ref{T_scalar}) does not in fact transform as a local Lorentz scalar. However, the equations of motion (Einstein's field equations) are covariant. This is because the action is instead \textit{pseudo-invariant}, meaning it is invariant up to boundary terms~\cite{Cho:1975dh}. 
This fact caused some confusion and disagreement in the early literature, especially when considering modifications.

A modern approach, which indeed fixes this issue of non-invariance, is to not a priori choose the Weitzenb\"{o}ck connection. Instead, recall the teleparallel constraint for the spin connection~(\ref{Teleparallel_sol2}). This should be taken as the general solution for the spin connection 
\begin{equation} \label{spin_inertial}
\omega^{a}{}_{b}(\Lambda) = \Lambda^{a}{}_{c} d (\Lambda^{-1})^{c}{}_{b}  \, ,
\end{equation}
along with the vanishing non-metricity constraint $\omega_{(ab)}=0$. Here we have chosen to use the symbols $\Lambda^{a}{}_{b}(x)$ for the matrices. The reason is that they can equally be interpreted as the local Lorentz matrices arising from performing LLTs. Applying a local Lorentz transformation on the vanishing Weitzenb\"{o}ck spin connection leads to
\begin{align}
\overset{\mathcal{W}}{\omega}{}^{a}{}_{b} &\rightarrow  \Lambda^{a}{}_{c} \Lambda_{b}{}^{d}\overset{\mathcal{W}}{\omega}{}^{c}{}_{d} + \Lambda^{a}{}_{c} d \Lambda_{b}{}^{c} \nonumber \\
&= \Lambda^{a}{}_{c} d \Lambda_{b}{}^{c} \, ,
\end{align}
where we have, as usual, used the notation $\Lambda_{a}{}^{b} = (\Lambda^{-1})^{b}{}_{a}$ for the inverse Lorentz matrices. We now see that this is exactly just the form of the general teleparallel connection in~(\ref{spin_inertial}), and we have a natural interpretation for the matrices.

The vanishing Weitzenb\"{o}ck spin connection $\overset{\mathcal{W}}{\omega}{}^{a}{}_{b} = 0$ can now be seen as a special choice of frame $\Lambda^{a}{}_{b} = \delta^{a}_{b}$ of the general solution~\cite{krvsvsak2015spin,krvsvsak2016covariant,Krssak:2018ywd}. This is sometimes called the `purely inertial spin connection' or `pure gauge spin connection'. In other words, by performing an LLT on the purely inertial spin connection $\omega(\Lambda)$, we can always set it to zero. This specific class of frames are known as `proper frames'~\cite{Lucas:2009nq}.
 We therefore see that all calculations in the \textit{Weitzenb\"{o}ck gauge} are totally legitimate, from a covariant point of view, and also describe the exact same physics. 
Variations with respect to either the Weitzenb\"{o}ck-gauge-fixed connection or the purely inertial spin connection lead to equivalent equations of motion, as long as these constraints are implemented consistently~\cite{Golovnev:2017dox}.

The formulation of (metric) teleparallel gravity with the purely inertial spin connection~(\ref{spin_inertial}) is often called the `covariant formulation'~\cite{krvsvsak2016covariant}, whilst the original approach using the fixed Weitzenb\"{o}ck connection $\overset{\mathcal{W}}{\omega}{}^{a}{}_{b} = 0$ is called the `pure tetrad' approach. Initially, these different formulations led to some disagreements~\cite{Maluf:2018coz}, especially for the modified $f(T)$ theories, and we will discuss this in Chapter~\ref{chapter5}. However, it now seems that the general consensus is that both formulations are equally valid, and the main differences have become philosophical. (An analogy can be made with the two formulations of unimodular gravity studied in the introduction, though for the modified $f(T)$ theories there are additional considerations.) Crucially, the dynamics and predictions of both formulations are equivalent.

Using the covariant approach, all the previous objects can be rewritten using the purely inertial spin connection and have their symmetries restored~\cite{Krssak:2015rqa,Krssak:2018ywd}. The torsion tensor takes its usual form~(\ref{Torsion mixed}),
\begin{equation}
 T^{a}{}_{\mu \nu}(e,\omega(\Lambda)) =2 \partial_{[\mu} e^{a}{}_{\nu]} +2 e^{b}{}_{[\nu} \omega_{\mu]}{}^{a}{}_{b} \, ,
\end{equation}
which is now covariant as expected. We explicitly include the dependence on the parameters $\Lambda$ to show that this is the purely inertial spin connection. Note that we now need to transform the tetrad and inertial spin connection \textit{simultaneously}, see~\cite{Blixt:2022rpl} for some interesting comments on this point.
 The torsion scalar in particular can be decomposed into terms depending just on the tetrad and just on the connection
\begin{equation} \label{T_decomp}
T(e,\omega(\Lambda)) = T(e,0) + b_{\Lambda} \, ,
\end{equation}
where $T(e,0)$ is equivalent to the torsion scalar in the Weitzenb\"{o}ck gauge. Additionally, we have defined the \textit{purely gauge boundary term}\footnote{We use the term `purely gauge' to mean a term that exhibits none of its own dynamics and can always be set to zero via a suitable frame/coordinate transformation. This is somewhat nonstandard terminology, but seems appropriate in this context.} as~\cite{Krssak:2015rqa,krvsvsak2016covariant}
\begin{equation} \label{b_Lambda}
 b_{\Lambda} :=  -\frac{2}{e} \partial_{\mu}\big(e \omega^{\mu}) =  \frac{2}{e} \partial_{\mu}\big(e \omega_{\nu}{}^{a}{}_{b} e_{a}{}^{\nu} e^b{}_{\lambda} g^{\mu \lambda} ) \, .
\end{equation}
 The final term is a boundary term, but also is `purely gauge', in the sense that it can always be set to zero via a local Lorentz transformation (to the Weitzenb\"{o}ck frame where $\omega^{a}{}_{b} = 0$). Therefore it is clear that it plays no physical role. This is also in line with our earlier comments about the pseudo-invariance of the (Weitzenb\"{o}ck) TEGR action, as $T(e,0)$ still differs from a Lorentz scalar by a boundary term. 
 
 For completeness, note that we can also rewrite the teleparallel boundary term $B_T$ in terms of its tetradic part and this purely gauge boundary term
 \begin{equation} \label{B_T_decomp}
B_T(e,\omega(\Lambda)) = B_T(e,0) + b_{\Lambda} \, .
\end{equation}
We then see that for the unique combination $-T(e,\omega(\Lambda)) + B_T(e,\omega(\Lambda)) = R$, the gauge boundary terms $b_{\Lambda}$ cancel with one another. This must have been the case, because the Levi-Civita Ricci scalar $R$ is completely independent from the affine spin connection by definition. It also follows that $-T(e,0) + B_T(e,0) = R$ is locally Lorentz invariant too.
Clearly there are many boundary terms that arise here, and we refer to~\cite{Boehmer:2021aji} for more details. We will also study these boundary terms in great detail in the next chapter.

Let us give a few more general remarks before concluding the discussion on metric teleparallel gravity. There are at least three distinct approaches to deriving the field equations of TEGR: using the fixed Weitzenb\"{o}ck connection and varying with respect to the tetrad; using the purely inertial spin connection $\omega(\Lambda)$ and varying with respect to the tetrad and $\Lambda(x)$ matrices; and using an independent tetrad and spin connection along with appropriate Lagrange multiplier terms. These all lead to the same results, which is trivial from the decomposition~(\ref{T_decomp}). The equivalence of these different approaches will be less trivial in the modified $f(T)$ case.

We should also add that (metric) teleparallel gravity is not free from criticism. For example, the torsion tensor is not fully determined by the field equations~\cite{Blagojevic:2013xpa,kopczynski1982problems}. However, if we only care about determining the metric and not this gauge redundancy\footnote{Again, here we do not mean `gauge' in the gauge-theoretic sense of being related to physical gauge fields. The inertial spin connection can always be removed via an LLT and should not be viewed as a physical field. This is made clear in~\cite{Pereira:2019woq}, where they point out that local Lorentz invariance is a \textit{kinematic} symmetry of teleparallel gravity.} -- and in TEGR, it is the \textit{metric} that is physical -- then this is not an issue~\cite{Golovnev:2020zpv}. Another point to consider is the coupling of matter. Fermions couple to the connection, and if this is chosen to be the Weitzenb\"{o}ck connection (or its covariant, inertial version), this leads to inconsistencies~\cite{Obukhov:2004hv}. To remedy this issue, either matter must couple to the \textit{Levi-Civita} spin connection or a more general geometric framework (such as Einstein-Cartan) must be used, see for example~\cite{BeltranJimenez:2020sih}.

\subsection{Symmetric teleparallel gravity}
\label{section3.3.2}

Moving on to the symmetric teleparallel framework, we now have vanishing curvature and torsion but non-metricity is present. The `symmetric' part comes from the symmetry of the spacetime connection, following from the zero torsion constraint. Interestingly, this geometric framework only gained interest relatively recently~\cite{Nester:1998mp}, long after its metric counterpart. This is reflected in the extensive review by Hehl~\cite{Hehl:1994ue}, where this particular case of vanishing torsion and curvature is curiously absent. Hence, symmetric teleparallel gravity came late to the game.

However, we now know that an equivalent of General Relativity also exists in the purely non-metric framework~\cite{Adak:2006rx,Adak:2005cd}, the symmetric teleparallel equivalent of General Relativity (STEGR). This is also clear by looking at the decomposition of the affine Ricci scalar in the teleparallel limit~(\ref{teleparallel_decomp}). The calculation in STEGR follows the same logic as in the metric teleparallel case described above.

Taking the teleparallel solution for the connection~(\ref{Teleparallel_sol1}) and imposing the symmetry of the connection restricts the matrices to be $(\Omega^{-1})^{\mu}{}_{\nu}(x) = \partial_{\nu} \xi^{\mu}(x)$, for some arbitrary $\xi^{\mu}(x)$ that parameterise the connection. This leads to the symmetric teleparallel connection 
\begin{equation} \label{STG_connection}
\bar{\Gamma}^{\lambda}_{\mu \nu} = \frac{\partial x^{\lambda}}{\partial \xi^{\rho}} \partial_{\mu} \partial_{\nu} \xi^{\rho} \, .
\end{equation}
In analogy with the purely inertial spin connection, which can be obtained by performing a local Lorentz transformation from the vanishing Weitzenb\"{o}ck spin connection, the symmetric teleparallel connection is obtained by performing a coordinate transformation from the trivial (vanishing) spacetime connection. If the coordinates where $\bar{\Gamma}$ vanishes are given by $\{ \xi^{\mu} \}$, then the transformation~(\ref{Christoffel transformation}) to a general coordinate system $\{ x^{\mu} \}$ leads to
\begin{align}
 \bar{\Gamma}^{\gamma}_{\mu \nu}(\xi) \rightarrow \hat{\bar{\Gamma}}^{\gamma}_{\mu \nu}(x) &=  \frac{\partial \xi^{\alpha}}{\partial x^{\mu}} \frac{\partial \xi^{\beta}}{\partial x^{\nu}} \frac{\partial x^{\gamma}}{\partial \xi^{\sigma}} \bar{\Gamma}^{\sigma}_{\alpha \beta}(\xi) + \frac{\partial^2 \xi^{\sigma}}{\partial x^{\mu} \partial x^{\nu} } \frac{\partial x^{\gamma}}{\partial \xi^{\sigma}} \nonumber \\
 &=\frac{\partial x^{\gamma}}{\partial \xi^{\sigma}} \partial_{\mu} \partial_{\nu} \xi^{\rho} \, .
\end{align}
Hence, $\xi^{\mu}$ are associated with coordinate transformations.
The gauge where the connection vanishes $\bar{\Gamma}=0$ is known as the `coincident gauge' or `unitary gauge'~\cite{Nester:1998mp,BeltranJimenez:2017tkd,Adak:2008gd}. It can always be obtained by performing a suitable coordinate transformation $\xi^{\mu} = x^{\mu}$.

The primary object is the non-metricity tensor, which we can now write explicitly in terms of the parameters $\xi^{\mu}$ as
\begin{align} \label{Q_xi}
Q_{\lambda \mu \nu} = - \bar{\nabla}_{\lambda} g_{\mu \nu} &= -\partial_{\lambda} g_{\mu \nu} + 2 \bar{\Gamma}^{\rho}_{\lambda (\mu} g_{\nu) \rho} \, \nonumber \\
&= -\partial_{\lambda} g_{\mu \nu} + 2 \frac{\partial x^{\rho}}{\partial \xi^{\sigma}} g_{\rho (\nu} \partial_{\mu)} \partial_{\lambda} \xi^{\sigma} \, .
\end{align}
In the coincident gauge $\xi^{\mu} = x^{\mu}$ the final terms vanish and the non-metricity tensor coincides with the partial derivative of the metric. Clearly any calculations made in the coincident gauge will be as equally valid in any other choice of coordinates. However, the choice of coordinates that are usually adapted to a given problem may not be compatible with this gauge choice, so some care must be taken here. We will revisit these topics in the following chapters.

Repeating the process of taking the decomposition of the affine Ricci scalar in the teleparallel limit~(\ref{teleparallel_decomp}), this time with torsion set to zero, we obtain
\begin{equation}
R(g) = -Q - B_Q \, ,
\end{equation}
where we have defined the non-metricity scalar and boundary terms as~\cite{BeltranJimenez:2019esp}
\begin{align} \label{Q_scalar}
Q &:= \frac{1}{4} Q_{\mu \nu \lambda} Q^{\mu \nu \lambda} - \frac{1}{2} Q_{\mu \nu \lambda} Q^{\nu \lambda \mu} - \frac{1}{4} Q_{\mu}{}^{\nu}{}_{\nu} Q^{\mu \lambda}{}_{\lambda} + \frac{1}{2} Q_{\mu}{}^{\nu}{}_{\nu} Q_{\lambda}{}^{\mu \lambda} \, ,  \\
B_Q & := \frac{2}{\sqrt{-g}} \partial_{\mu} \big(\sqrt{-g} Q^{[\lambda \mu]}{}_{\lambda} \big) \, . \label{B_Q}
\end{align}
The action for STEGR is then the non-metricity scalar~\cite{BeltranJimenez:2017tkd}
\begin{equation}
S_{\textrm{STEGR}}[g,\xi] = -\frac{1}{2\kappa} \int Q \sqrt{-g} d^4x \, .
\end{equation}
Note that we can vary with respect to the metric and parameters $\xi$, or use the Lagrange multiplier method. Again, they lead to equivalent results $G_{\mu \nu} = \kappa T_{\mu \nu}$, provided the constraints are implemented properly. For further details on solutions, see~\cite{Adak:2008gd,Adak:2005cd}. Note that if we fix the coincident gauge so that the action only depends on the metric, we also derive the standard Einstein field equations. Just like in the metric teleparallel case, this can be seen as a result of the connection only entering the action by a boundary term.

Let us show this explicitly by decomposing the non-metricity scalar into parts depending purely on the metric and its derivatives, and another \textit{purely gauge boundary term} depending on the parameters $\xi^{\mu}$. We then have
\begin{equation}  \label{Q_decomp}
Q(g,\bar{\Gamma}(\xi)) = Q(g,0) - b_{\xi} \, ,
\end{equation}
where we define~\cite{Boehmer:2021aji}
\begin{equation} \label{b_xi}
b_{\xi} := \frac{2}{\sqrt{-g}} \partial_{\nu} \Big(\sqrt{-g}  \frac{\partial x^{[\nu}}{\partial \xi^{\sigma}} g^{\lambda] \mu} \partial_{\mu} \partial_{\lambda} \xi^{\sigma} \Big) \,.
\end{equation}
A similar decomposition for the non-metricity boundary term gives
\begin{equation}  \label{B_Q_decomp}
B_Q(g,\bar{\Gamma}(\xi)) = B_Q(g,0) + b_{\xi} \, .
\end{equation}
We will discuss further properties of both of these purely gauge boundary terms $b_{\Lambda}$ and $b_{\xi}$ in the next chapter. There, we will also study the gauge-fixed scalars $T(e,0)$ and $Q(g,0)$ and show how they can be derived in a non-geometric setting.

Lastly, let us mention that there has also been some confusion in the literature concerning the properties of symmetric teleparallel gravity. A prime example is that it was thought to suffer from the \textit{second clock effect}~\cite{quiros2022nonmetricity,Delhom:2020vpe}. This phenomenon is related to the effects of non-metricity on the speed at which clocks `tick' due to their past motion through spacetime. In short, two initially synchronised clocks will tick at different rates after having followed different timelike worldlines and being brought back together at rest~\cite{Hobson:2021iwy}. 
In the context of symmetric teleparallel gravity, this was subsequently shown not to be the case~\cite{BeltranJimenez:2020sih}. Other confusion seems to arise on the role of the coincident gauge, which, as we mentioned, does not play a physical role. The lack of physical significance of the coincident gauge is also made clear in~\cite{BeltranJimenez:2022azb}. We will return to this point in Chapter~\ref{chapter4}.

\begin{figure}[!hbt]
\centering
\begin{tikzpicture}[xscale=3,yscale=3,baseline={(0,0)}]
  \def\is{7.0} 
  \def\scal{1.75}
  
  \node[inner sep=\is,align=center] (MA) at (1/2,0) {\large \textbf{Metric-affine}};
  \node[inner sep=\is] (A) at ( -{sqrt(3)/2 *\scal}+1/2, \scal/2) {\textbf{Riemann-Cartan}};
  \node[inner sep=\is] (B) at ( +{sqrt(3)/2 *\scal}+1/2, \scal/2) {\textbf{Weyl}};
  \node[inner sep=\is] (C) at (1/2, -\scal) {\textbf{Teleparallel}};
  
    \node[inner sep=\is] (D) at ( 1/2, +\scal) {Riemann};
    \node[inner sep=\is,align=center] (E) at (  -{sqrt(3)/2 *\scal}+1/2 , \scal/2-\scal*0.95) {Metric \\ teleparallel};
    \node[inner sep=\is,align=center] (F) at (  +{sqrt(3)/2 *\scal}+1/2 ,  \scal/2-\scal*0.95) {Symmetric\\ teleparallel};

  \draw[->] (MA) -- (C) node[pos=0.45,left=-1,scale=0.9] {$\bar{R}=0$};
  \draw[->] (MA) -- (A) node[pos=0.6,below=8,scale=0.9] {$Q=0$};
  \draw[->] (MA) -- (B) node[pos=0.6,below=8,scale=0.9] {$T=0$};
  
  

  \draw[->] (A) -- (D) node[pos=0.6,left=10,scale=0.9] {$T=0$};
  \draw[->] (B) -- (D) node[pos=0.6,right=10,scale=0.9] {$Q=0$};
  
  \draw[->] (A) -- (E) node[pos=0.45,left=3,scale=0.9] {$\bar{R}=0$};
  \draw[->] (C) -- (E) node[pos=0.45,left=10,scale=0.9] {$Q=0$};
  
  \draw[->] (B) -- (F) node[pos=0.45,right=3,scale=0.9] {$\bar{R}=0$};
  \draw[->] (C) -- (F) node[pos=0.45,right=10,scale=0.9] {$T=0$};
  
\end{tikzpicture}
\caption{Relationship between different geometries. Metric-affine is the most general, from which we obtain Riemann-Cartan, Weyl or teleparallel by imposing vanishing non-metricity, torsion or curvature respectively. Riemann, symmetric teleparallel and metric teleparallel geometries possess only curvature, non-metricity or torsion respectively.}
\label{fig:mag}
\end{figure}
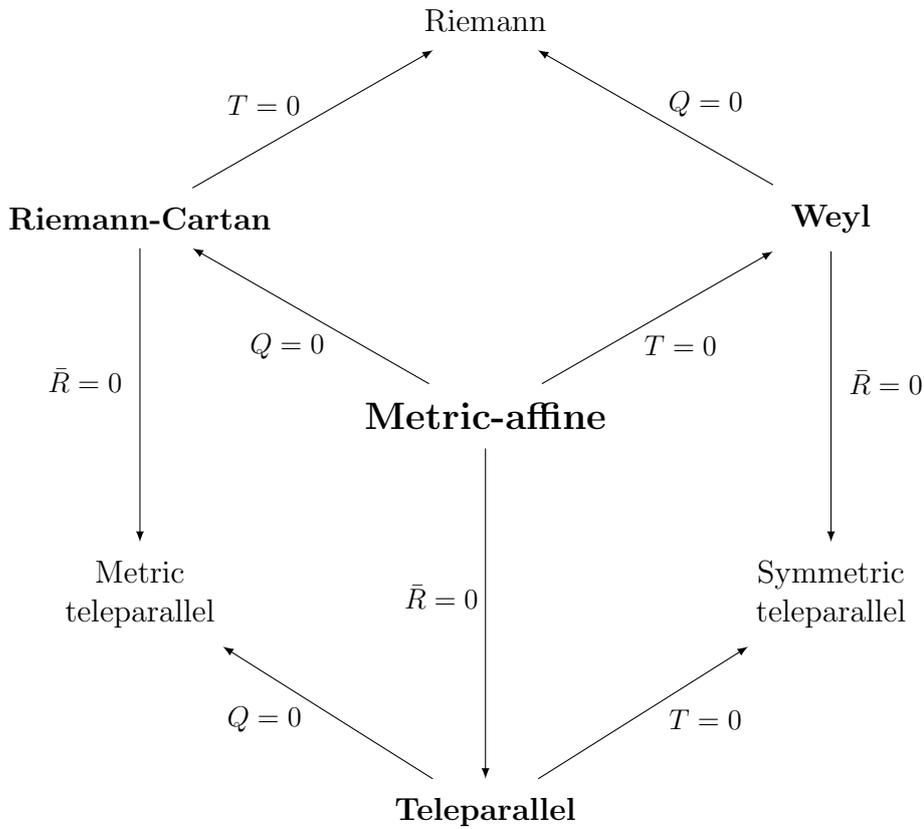

To summarise this chapter, the geometric frameworks of all of these non-Riemannian theories can be nicely visualised based on the presence of curvature, torsion and non-metricity. This is illustrated in Fig.~\ref{fig:mag}, an adaptation of a figure from Hehl's review~\cite{Hehl:1994ue}. Beginning with the most general, metric-affine theories at the centre, we obtain the other geometries by setting some combination of $\bar{R}_{\mu \nu \lambda}{}^{\gamma}$, $T^{\mu}{}_{\nu \lambda}$, or $Q_{\mu \nu \lambda}$ to zero. Arrows pointing downwards imply vanishing curvature, to the right, vanishing torsion, and to the left, vanishing non-metricity. (Visualising Figure~\ref{fig:mag} as a cube, the eighth (hidden) vertex would represent Minkowski space $\mathbb{R}^{(1,3)}$, which is reached by travelling in all three directions.) We have shown that the geometries of Riemann, metric teleparallel and symmetric teleparallel all share a formulation of General Relativity. These are the Einstein-Hilbert, TEGR and STEGR actions respectively. This has since been dubbed the `geometrical trinity of gravity'~\cite{BeltranJimenez:2019esp}.

\begin{savequote}[70mm]
Thus, Beauty is neither an appearance nor a being, but a relationship: the transformation of being into appearance.
\qauthor{`Saint Genet, Actor and Martyr' (1952) \\
Jean-Paul Sartre}
\end{savequote}
\chapter{Decomposing gravity with boundary terms}
\label{chapter4}

In this chapter we look at decomposing the action of General Relativity into bulk and boundary terms. The boundary terms take the form of a total derivative, whilst the bulk part is only first-order in derivatives and gives rise to all of the dynamical equations of motion.
These decompositions naturally introduce new, non-tensorial objects, which we will study in great detail. Theories of gravity can then be constructed and built from these newly derived terms. Interestingly, this naturally leads to modified frameworks that break the fundamental symmetries studied in Section~\ref{section2.3.2}, namely, diffeomorphism invariance and local Lorentz invariance.

The first decomposition we will study is the Levi-Civita Ricci scalar in the natural coordinate basis. This gives rise to the so-called Einstein action, which is fairly well-known and well-studied, though our approach introduces some new elements. Diffeomorphism invariance and gravitational pseudotensors are discussed for this action. We then move on to the Ricci scalar decomposition in anholonomic frames, using the tetrad formalism. There, we discuss local Lorentz invariance. A direct comparison between the two distinct decompositions gives rise to new boundary terms.

A highly interesting result is that these decompositions are naturally related to the teleparallel theories from Sec.~\ref{section3.3}. In fact, it can be shown that the bulk and boundary parts are completely equivalent to the teleparallel scalars and boundary terms, up to the additional `purely gauge' boundary terms identified in equations~(\ref{b_Lambda}) and~(\ref{b_xi}). As we previously saw, the auxiliary fields associated with the inertial connections can always be set to zero via a coordinate or frame transformation~\cite{BeltranJimenez:2019tjy}. It follows that these decompositions can be interpreted as `gauge-fixed' versions of the teleparallel theories. Alternatively, making no reference to non-Riemannian geometries, one can begin from these decomposed bulk and boundary terms and restore their invariance by introducing new gauge/auxiliary fields. We show this explicitly by making use of the Stueckelberg mechanism. It then turns out that these newly covariantised scalars are exactly the scalars of the teleparallel theories. We elucidate these results in Sec.~\ref{section4.3} and comment on the differences between these constructions.

 Finally, in Section~\ref{section4.4} we take a look at decomposing the metric-affine Ricci scalar into its bulk and boundary parts, which has not been previously investigated~\cite{Boehmer:2023fyl}. Moreover, we look at the properties of this decomposition and the pseudo-diffeomorphism invariance of the action. In this case, there is no longer any direct analogy between the teleparallel theories, which requires the (gauge) fields to be related to the teleparallel connection. For the metric-affine decomposition, we focus on the coordinate basis representation. We also investigate the projective invariance of these new terms.
 
Although we focus on the Ricci scalar, this type of decomposition can also be applied to any of the Lovelock scalars~(\ref{LovelockI}), see for example~\cite{Mukhopadhyay:2006vu,Padmanabhan:2013xyr}. There, the authors investigate the interesting holographic and thermodynamic interpretations of these theories, and we will briefly comment on this holographic relation later on. However, for the other Lovelock scalars besides the Ricci curvature, the decomposed bulk terms are not first-order in derivatives, which is our aim in this work. Hence, non-linear modifications will generically lead to fourth-order equations of motion.

\section{Einstein action}
\label{section4.1}

The Einstein-Hilbert action~(\ref{EH}) is given by the Levi-Civita Ricci scalar. It includes derivatives of the Christoffel symbols and, hence, second derivatives of the metric tensor. However, the Einstein field equations are themselves only second-order in derivatives of the metric. This result follows from the fact that the Ricci scalar can be decomposed into a first-order `bulk' term, and a second-order boundary term. The second derivative terms appearing in the Ricci scalar are linear, meaning they can be removed via integration by parts.

In fact, this decomposition was originally written down by Einstein in 1916, and the bulk term can be found explicitly in his work~\cite{Einstein:1916cd}. For this reason, it is sometimes known as the \textit{Einstein action}\footnote{Other common names include the \textit{Gamma-squared action}, the \textit{Schr\"{o}dinger action}, or the \textit{Dirac action}, see below.}. However, it suffers the property of no longer being an invariant scalar, which is why the usual Einstein-Hilbert action $R$ is more commonly used instead. We will discuss these invariance properties in depth, as well as showing the explicit calculations for the equations of motion.

After its original introduction by Einstein, the action has appeared in many other contexts and was notably studied by Weyl~\cite{Weyl1919}, Schr\"{o}dinger~\cite{Schrodinger:2011gqa} and Dirac~\cite{Dirac:1958sc}. In the 1950s, M\o{}ller used the action in relation to pseudotensors and conserved quantities~\cite{MOLLER1958347}. The use of non-tensorial objects, such as the energy-momentum complex, is necessary for localised descriptions of gravitational energy, and so the non-covariant Einstein action naturally arises in this context. In Einstein's original work~\cite{Einstein:1916cd}, he also looks at conserved pseudotensor quantities. We will show this in Sec.~\ref{section4.1.4}, where we also examine the interesting relationship with the Bianchi identities.

 In more recent times, the decomposition has been studied extensively by Padmanabhan~\cite{TP:2010}, who has also explored its thermodynamic interpretations~\cite{Mukhopadhyay:2006vu}. As mentioned in the introduction of this chapter, this is connected to the holographic relationship between the bulk and boundary terms of this decomposition. Here we will focus on the coordinate basis form of the Einstein action, whilst in the following sections we study its form in different bases.

\subsection{Ricci scalar decomposition}
\label{section4.1.1}
Moving on to the decomposition, the Levi-Civita Ricci scalar density is 
\begin{equation} \label{E1}
    \sqrt{-g} R = \sqrt{-g} \Big( g^{\mu \nu} \partial_{\lambda} \Gamma^{\lambda}_{\mu \nu} -  g^{\mu \nu} \partial_{\mu} \Gamma^{\lambda}_{\lambda \nu} + g^{\mu \nu} \Gamma^{\lambda}_{\lambda \sigma} \Gamma^{\sigma}_{\mu \nu} - g^{\mu \nu} \Gamma^{\lambda}_{\mu \sigma} \Gamma^{\sigma}_{\lambda \nu} \Big) \, .
\end{equation}
The partial derivative of the metric can be written in terms of the Levi-Civita connection
\begin{equation} \label{PDrelation}
 \partial_{\rho} g^{\mu \nu} = - \Gamma_{\rho \lambda}^{\mu} g^{\lambda \nu} -  \Gamma_{\rho \lambda}^{\nu} g^{\mu \lambda} \, , \quad   \partial_{\rho} g_{\mu \nu} =  \Gamma_{\rho \mu}^{\lambda} g_{\lambda \nu} +  \Gamma_{\rho \nu}^{\lambda} g^{\mu \lambda}  \, , 
\end{equation}
and similarly for the square root of the determinant 
\begin{equation}
\partial_{\rho} \sqrt{-g} = \frac{1}{2} \sqrt{-g} g^{\mu \nu} \partial_{\rho} g_{\mu \nu} = \sqrt{-g} \Gamma^{\lambda}_{\rho \lambda} \, .
\end{equation}
Using these relations, one can use the chain rule to write the partial derivative of the connection as
\begin{align} \label{E2}
  \sqrt{-g} g^{\mu \nu} \partial_{\rho} (\Gamma^{\rho}_{\mu \nu}) &= \partial_{\rho} ( \sqrt{-g} g^{\mu \nu} \Gamma^{\rho}_{\mu \nu}) -  
    \sqrt{-g} g^{\mu \nu} \Gamma^{\lambda}_{\rho \lambda}  \Gamma^{\rho}_{\mu \nu} + 2\sqrt{-g} g^{\sigma \nu}  \Gamma^{\mu}_{\rho \sigma}
    \Gamma^{\rho}_{\mu \nu} \, , \\
     \sqrt{-g} g^{\mu \nu} \partial_{\mu} (\Gamma^{\lambda}_{\lambda \nu}) &= 
    \partial_{\mu} (\sqrt{-g} g^{\mu \nu} \Gamma^{\lambda}_{\lambda \nu}) - 
    \sqrt{-g} g^{\mu \nu} \Gamma^{\sigma}_{\mu \sigma} \Gamma^{\lambda}_{\lambda \nu}  + 2
    \sqrt{-g} g^{\rho (\mu} \Gamma^{\nu)}_{\mu \rho} \Gamma^{\lambda}_{\lambda \nu}  \, .
\end{align}
Substituting these into the Ricci scalar~(\ref{E1}) gives
\begin{equation} \label{E3}
   \sqrt{-g}R =\sqrt{-g} g^{\mu \nu} \Big( \Gamma^{\lambda}_{\mu \sigma} \Gamma^{\sigma}_{\lambda \nu} - \Gamma^{\sigma}_{\mu \nu} \Gamma^{\lambda}_{\lambda \sigma} \Big) + \partial_\sigma \Big( \sqrt{-g} \big( g^{\mu \nu} \Gamma^{\sigma}_{\mu \nu} - g^{\sigma \nu} \Gamma^{\lambda}_{\lambda \nu} \big) \Big) \, .
\end{equation}
We therefore see that the Ricci scalar can be written as a part which is quadratic in the connection and a term taking the form of a total derivative. We write this as
\begin{equation} \label{RGB}
\sqrt{-g} R = \sqrt{-g}( \ourG + \ourB ) \, ,
\end{equation}
where we introduce the notation $\ourG$ for the \textit{bulk term} and $\ourB$ for the \textit{boundary term}
\begin{align}\label{G}
\ourG &:= g^{\mu \nu} \Big( \Gamma^{\lambda}_{\mu \sigma} \Gamma^{\sigma}_{\lambda \nu} - \Gamma^{\sigma}_{\mu \nu} \Gamma^{\lambda}_{\lambda \sigma} \Big)  \, , \\ \label{B}
\ourB &:= \frac{1}{\sqrt{-g}} \partial_\sigma \Big( \sqrt{-g} \big( g^{\mu \nu} \Gamma^{\sigma}_{\mu \nu} - g^{\sigma \nu} \Gamma^{\lambda}_{\lambda \nu} \big) \Big) \, .
\end{align}

An additional useful piece of notation is to label the trace of the connection terms in $\ourB$ as the \textit{boundary vector} 
\begin{equation} \label{Bvector}
B^{\sigma} := g^{\mu \nu} \Gamma^{\sigma}_{\mu \nu} - g^{\sigma \nu} \Gamma^{\lambda}_{\lambda \nu} \, ,
\end{equation}
so that the boundary term is
\begin{equation} \label{B2}
\ourB = \frac{1}{\sqrt{-g}} \partial_\sigma (\sqrt{-g} B^{\sigma}) = \nabla_{\sigma} B^{\sigma} \, .
\end{equation}
In the final equality it should be understood that the covariant derivative of the `boundary vector' is meant to imply the usual rule of $\nabla_{\sigma} B^{\sigma} = \partial_{\sigma} B^{\sigma} + \Gamma^{\sigma}_{\sigma \lambda} B^{\lambda}$.
Note that $B^{\sigma}$ is \textit{not} a vector, so the use of the covariant derivative here should be understood in the previously defined way. In general we will be careful not to use this sort of notation, as the covariant derivative is defined as only acting on tensorial objects. 
Similarly, the bulk and boundary terms $\ourG$ and $\ourB$ are not scalars despite having no free indices, which we will show explicitly shortly. 

Lastly, the boundary term can be written compactly in terms of the metric and its second derivatives as
\begin{equation}\label{B3}
\ourB = \frac{1}{\sqrt{-g}} \partial_{\nu} \Big( \frac{\partial_{\mu}(g g^{\mu \nu})}{\sqrt{-g}}\Big)  \, .
\end{equation}
An even more interesting and somewhat surprising form of the boundary term can be found by relating it to the variation of the bulk term \begin{equation}\label{B4}
\ourB = - \frac{1}{\sqrt{-g}} \partial_{\lambda} \Big(g_{\mu \nu} \frac{\partial(\sqrt{-g} \ourG)}{\partial(\partial_{\lambda}g_{\mu \nu}) }\Big) \, ,
\end{equation}
which was first noted by Padmanabhan~\cite{Padmanabhan:2002jr,Padmanabhan:2004fq}. This is directly related to the fact that gravity can be given a `holographic interpretation', see \cite{Padmanabhan:2004fq,Kolekar:2010dm} for further details. The use of holographic here simply means that the bulk and boundary can be directly related, such that the information of one can be defined fully in terms of the other. Also note that the constants in this final relation are specifically for $n=4$ dimensions. The equivalence between the different expressions for the boundary term~(\ref{B}),~(\ref{B3}) and~(\ref{B4}) is not immediately clear, so we will show this explicitly later in this section.

Let us first briefly pause here to comment on this decomposition~(\ref{RGB}) and the nature of the bulk and boundary terms. From the geodesic equation~(\ref{GR geodesic}), the connection coefficients can be thought of in some sense as representing the gravitational force.
It is then interesting to note that $\ourG$, being quadratic in the Christoffel symbols, can be seen as an analogue of the Lagrangians in gauge theories, such as electromagnetism or Yang-Mills\footnote{Yang-Mills theory perhaps provides a closer analogy than electromagnetism, with the theory also being non-linear, and the gauge bosons being charged under their currents in a similar way to gravitons carrying energy in GR~\cite{coleman2019quantum}.}. In those contexts, it is natural to look for an action quadratic in the field strength terms. Moreover, linearising the bulk term leads to the Lagrangian of the spin-2 field~\cite{TP:2010}.

However, $\ourG$ has the unusual property of not being invariant under general coordinate transformations (see Sec.~\ref{section4.1.2}). This means that it will take a different value in different sets of coordinates.
Despite this fact, we know that the combination that gives the invariant Ricci curvature is $\ourG + \ourB =R$, which is of course a scalar. We will therefore refer to $\ourG$ as a \textit{pseudo-scalar} with respect to coordinate transformations; we define a pseudo-scalar as a geometric object differing from an invariant scalar by a boundary term. A very important point is that because this decomposition~(\ref{RGB}) is not spacetime covariant, it is also not unique~\cite{Aldrovandi:2013wha}. As we shall see in Section~\ref{section4.2}, using a different basis leads to a different form of decomposition, with the bulk and boundary terms being inequivalent to our $\ourG$ and $\ourB$.

It should also be noted that $\ourB$ takes the form of a total derivative, but whether or not it vanishes under an integral depends on whether the manifold $\mathcal{M}$ has a boundary (and if it does, it depends on the nature of the boundary conditions). In general, a manifold with a boundary will need to be supplemented by appropriate additional boundary terms, such as the Gibbons-Hawking-York term, for the variation procedure to be well-posed. This will be discussed in more detail in Sec.~\ref{section4.1.3}.

Let us introduce two important objects that are naturally related to $\ourG$ and $\ourB$, and will also appear in the transformation properties and field equations throughout the thesis. Let us define the non-tensorial, rank-three object 
\begin{equation}
  \label{M}
  M^{\mu \nu}{}_{\lambda}:=
  \frac{\partial \ourG}{\partial \Gamma^{\lambda}_{\mu \nu}} =
  2 g^{\rho (\nu} \Gamma^{\mu)}_{\lambda \rho} - g^{\mu \nu} \Gamma^{\rho}_{\rho \lambda} -
  g^{\rho \sigma}  \delta^{(\nu}_{\lambda} \Gamma^{\mu)}_{\rho \sigma} \, ,
\end{equation}
which is simply the variation of the bulk term with respect to the Levi-Civita connection
\begin{align} \label{Mderiv}
\delta_{\Gamma} \ourG &= g^{\mu \nu} \big( \delta \Gamma^{\sigma}_{\lambda \mu} \Gamma^{\lambda}_{\sigma \nu} + \delta \Gamma^{\lambda}_{\sigma \nu} \Gamma^{\sigma}_{\lambda \mu} - \delta \Gamma^{\sigma}_{\mu \nu} \Gamma^{\lambda}_{\sigma \lambda} - \delta \Gamma^{\lambda}_{\sigma \lambda} \Gamma^{\sigma}_{\mu \nu} \big)\nonumber \\
    &=  g^{\mu \nu} \big(\delta^{\beta}_{\mu} \Gamma^{\alpha}_{\gamma \nu} + \delta^{\beta}_{\nu} \Gamma^{\alpha}_{\gamma \mu} - \delta^{\alpha}_{\mu} \delta^{\beta}_{\nu} \Gamma^{\lambda}_{\gamma \lambda} - \delta^{\beta}_{\gamma} \Gamma^{\alpha}_{\mu \nu}) \delta \Gamma^{\gamma}_{\alpha \beta} \nonumber \\
    &=  \big( 2 g^{\beta \nu}\Gamma^{\alpha}_{\gamma \nu} - g^{\alpha \beta} \Gamma^{\lambda}_{\gamma \lambda} - g^{\mu \nu} \delta^{\beta}_{\gamma} \Gamma^{\alpha}_{\mu \nu}    \big) \delta \Gamma^{\gamma}_{\alpha \beta} \nonumber \\
    &= M^{\alpha \beta}{}_{\gamma}  \delta \Gamma^{\gamma}_{\alpha \beta} 
    \, .
\end{align}
Note that this object is symmetric over its first two indices by construction, because its skew part will not contribute to~(\ref{Mderiv}).
It also immediately follows that one can write
\begin{equation} \label{G2}
\ourG = \frac{1}{2} M^{\mu \nu}{}_{\lambda} \Gamma^{\lambda}_{\mu \nu} \, .
\end{equation}

If we want to instead relate $M^{\mu \nu}{}_{\lambda}$ to the variation of $\ourG$ with respect to the partial derivative of the metric, then we simply use that $2\Gamma^{\lambda}_{\mu \nu} = g^{\lambda \rho} g_{\{\rho \nu , \mu \}}$ where $\{ ...\}$ is the Schouten bracket. Applying this to~(\ref{Mderiv}) one straightforwardly obtains
\begin{equation}\label{MG1}
\frac{\partial \ourG}{\partial g_{\{\lambda \nu , \mu\}} } = \frac{1}{2} M^{\mu \nu \lambda} \, ,
\end{equation}
or equivalently\footnote{Note the subtle difference between the index ordering in $\partial_{\{\mu} g_{\lambda \nu\}}$ compared with $g_{\{\lambda \nu ,  \mu \}}$. The key point to remember is that in~(\ref{MG1}) the \textit{first} index in $M^{\mu \nu \lambda}$ is associated with the partial derivative of the metric, whilst later in $E^{\lambda \mu \nu}$ in~(\ref{EG1}) it is the \textit{last} index.}
\begin{equation}
\frac{\partial \ourG}{\partial(\partial_{\mu} g_{\nu \lambda})} = \frac{1}{2} M^{ \{ \mu \nu \lambda \} }  = \frac{1}{2} \Big( M^{\mu \nu \lambda} - M^{\lambda \nu \mu} + M^{\nu \mu \lambda} \Big) \, .
\end{equation}
It then becomes convenient to introduce the permutation 
\begin{align} \label{E}
E^{\mu \nu \lambda}  &:= M^{\{ \lambda \mu \nu \}} =
  M^{\lambda \mu \nu} + M^{\nu \lambda \mu} - M^{\mu \nu \lambda} \nonumber \\
  &= 2 g^{\rho \mu} g^{\nu \sigma} \Gamma^{\lambda}_{\rho \sigma} -
  2 g^{\lambda (\mu} g^{\nu) \sigma} \Gamma^{\rho}_{\rho \sigma} + g^{\mu \nu} g^{\lambda \rho} 
  \Gamma^{\sigma}_{\sigma \rho} - g^{\mu \nu} g^{\rho \sigma} \Gamma^{\lambda}_{\rho \sigma} \, .
\end{align}
Then this object is naturally associated to the variation with respect to the partial derivative of the metric
\begin{equation} \label{EG1}
E^{\mu \nu \lambda} = 2 \frac{\partial \ourG}{\partial (g_{\mu \nu,\lambda})}  \, .
\end{equation}
It follows that the object $E^{\mu \nu \lambda}$ is also symmetric over its first two indices.
 We will call $E^{\mu \nu \lambda}$ the \textit{superpotential} because it resembles the canonical field momentum\footnote{\label{footnoteE}Typically, canonical field momenta take the form $p_i = \frac{\partial L }{\partial \dot{q}_i}$. For gravity we have
 $\pi^{\mu \nu \lambda} = \frac{ \partial \mathcal{L}}{\partial_{\mu}g_{\nu \lambda}}$ and the Lagrangian density $\mathcal{L} = \sqrt{-g} \ourG$, which implies that $\pi^{\mu \nu \lambda} = \frac{1}{2} \sqrt{-g} E^{\nu \lambda \mu}$.} with respect to $\ourG$~\cite{Padmanabhan:2002jr}. As such, we can also write
\begin{equation} \label{G3}
\ourG  =  \frac{1}{4} E^{\mu \nu \lambda} \partial_{\lambda} g_{\mu \nu} \, .
\end{equation}
One further useful relation between these objects, that can be seen by symmetrising the last two indices in equation~(\ref{E}), is
\begin{equation} \label{M to E}
M^{\mu \nu \lambda} = E^{\lambda (\mu \nu)} \, .
\end{equation}

The boundary vector $B^{\sigma}$ in equation~(\ref{Bvector}) can be obtained by contracting $E^{\mu \nu \lambda}$ over specific indices
\begin{equation} \label{Bvector2}
E_{\mu}{}^{\mu \sigma}  = (2-n) B^{\sigma}  \, ,
\end{equation}
where $n$ is the number of spacetime dimensions, which we will now take to be fixed to $n=4$.
The full boundary term can then be written as
\begin{align} \label{b_holographic}
\ourB =  \frac{1}{\sqrt{-g}} \partial_\sigma (\sqrt{-g} B^{\sigma}) &= -\frac{1}{2\sqrt{-g}} \partial_\sigma (\sqrt{-g} E_{\mu}{}^{\mu \sigma} )  \nonumber \\
&= - \frac{1}{\sqrt{-g}} \partial_\sigma \Big( g_{\mu \nu} \frac{\partial( \sqrt{-g} \ourG)}{\partial( \partial_{\sigma} g_{\mu \nu})} \Big)\, ,
\end{align}
which proves our assertion in~(\ref{B4}). Therefore in four dimensions the Ricci scalar density can be written as 
\begin{equation}
\sqrt{-g} R = \sqrt{-g} \ourG -  \partial_\sigma \Big( g_{\mu \nu} \frac{\partial( \sqrt{-g} \ourG)}{\partial( \partial_{\sigma} g_{\mu \nu})} \Big)\, ,
\end{equation}
or in terms of our superpotential
\begin{equation} \label{R_super}
\sqrt{-g} R = \sqrt{-g} \frac{1}{4} E^{\mu \nu \lambda} \partial_{\lambda} g_{ \mu \nu} -\frac{1}{2} \partial_\sigma (\sqrt{-g} E_{\mu}{}^{\mu \sigma} )  \, .
\end{equation} 

Note that these holographic relations between the bulk and boundary terms~(\ref{b_holographic}) exist not only for the Einstein-Hilbert action but for all of the Lovelock scalars, see~\cite{Kolekar:2010dm,Padmanabhan:2013xyr}. This is a consequence of the fact that the scalars can be decomposed into bulk and boundary terms like in equation~(\ref{RGB}). However, as we previously stated, for the other Lovelock scalars these decompositions do not give rise to a \textit{first-order} bulk term, see~\cite{Kolekar:2010dm,Padmanabhan:2013xyr,Mukhopadhyay:2006vu}. To our knowledge, a general decomposition of the Lovelock invariants (other than the Ricci scalar) into first-order bulk terms and second-order boundary terms cannot be found\footnote{This is likely due to Lovelock invariants beyond the Ricci scalar having at least quadratic second-order terms in the action. A decomposition of the Gauss-Bonnet term into first-order and boundary terms within the ADM framework can be found in~\cite{guilleminot2021first}. More generally, for Lovelock gravity, the ADM formalism has been completed in~\cite{Teitelboim:1987zz}, with the action purged of boundary terms being only first-order. However, we are not here working in the ADM formalism. In~\cite{Mukhopadhyay:2006vu} the authors claim that of the Lovelock invariants, only the Ricci scalar can be decomposed into a first-order bulk term.}. 

To conclude, we emphasise that none of the objects $\ourG$, $\ourB$, $M^{\mu \nu}{}_{\lambda}$ or $E^{\mu \nu \lambda}$ are spacetime tensors. In fact, they can be directly related to many of the pseudotensors that arise when studying the quasi-local energy-momentum associated with gravity~\cite{ChenNA}. In footnote~\ref{footnoteE} we hinted at this by writing $E^{\mu \nu \lambda}$ as the field momenta for the Lagrangian $L = \ourG$. This also works well with our terminology for $\ourG$ as a pseudo-scalar.
 We will revisit these ideas in Sec.~\ref{section4.1.4}. First we will study the transformation properties of these objects, before looking at the Einstein action and deriving the Einstein field equations.

\subsection{Coordinate transformations}
\label{section4.1.2}

An important property of the bulk and boundary terms $\ourG$ and $\ourB$ is their transformation properties under a general coordinate transformation. In practice, we only really need to know their infinitesimal transformation properties, given by the Lie derivative~(\ref{Lie derivative}), as this will give enough information for working with action principles. In Appendix~\ref{appendixB} we compute these transformations.

Recall that an infinitesimal coordinate transformation $x^{\mu} \rightarrow \hat{x}^{\mu} = x^{\mu} + \xi^{\mu} (x)$ leads to the standard relations
\begin{align}
  \frac{\partial \hat{x}^{\mu}}{\partial x^{\nu}} = \delta^{\mu}_{\nu} + \partial_{\nu} \xi^{\mu} \,,
  \qquad
  \frac{\partial x^{\mu}}{\partial \hat{x}^{\nu}} = \delta^{\mu}_{\nu} - \partial_{\nu} \xi^{\mu} \, ,
\end{align}
where $\xi$ is now assumed to be small $|\xi^{\mu}| \ll 1$ and we discard terms of order $\mathcal{O} (\xi^2)$.
For the metric, its inverse, and the Christoffel symbols we then have
\begin{align}
  \hat{g}_{\mu \nu}(\hat{x}) &= g_{\mu \nu} - \partial_{\mu} \xi^{\lambda} g_{\lambda \nu} -
  \partial_{\nu} \xi^{\lambda} g_{\mu \lambda}  \,, \\
  \hat{g}^{\mu \nu}(\hat{x})  &= g^{\mu \nu} + \partial_{\lambda} \xi^{\mu} g^{\lambda \nu} +
  \partial_{\lambda} \xi^{\nu} g^{\mu \lambda} \,, \\
  \hat{\Gamma}^{\gamma}_{\mu \nu}(\hat{x})  &= \Gamma^{\gamma}_{\mu \nu} + \partial_{\lambda} \xi^{\gamma} \Gamma^{\lambda}_{\mu \nu} - \partial_{\mu} \xi^{\lambda} \Gamma^{\gamma}_{\nu \lambda} - \partial_{\nu} \xi^{\lambda} \Gamma^{\gamma}_{\mu \lambda} -\partial_{\mu} \partial_{\nu} \xi^{\gamma}   \,,
\end{align}
with the terms on the right-hand side functions of the original coordinates $x^{\mu}$.

 It can then be shown that the bulk term $\ourG$~(\ref{G}) transforms as
\begin{align}
  \label{G_infinitesimal}
  \hat{\ourG}(\hat{x})  &=
  \hat{g}^{\mu \nu}\big( \hat{\Gamma}^{\lambda}_{\mu \sigma} \hat{\Gamma}^{\sigma}_{\lambda \nu} -
  \hat{\Gamma}^{\sigma}_{\mu \nu} \hat{\Gamma}^{\lambda}_{\lambda \sigma} \big) \nonumber \\
  &= g^{\mu \nu}\big( \Gamma^{\lambda}_{\mu \sigma} \Gamma^{\sigma}_{\lambda \nu} -
  \Gamma^{\sigma}_{\mu \nu} \Gamma^{\lambda}_{\lambda \sigma} \big) - \big(2 g^{\mu (\alpha} \Gamma^{\beta)}_{\mu \gamma} - g^{\alpha \beta} \Gamma^{\lambda}_{\lambda \gamma} - g^{\mu \nu} \delta^{(\beta}_{\gamma} \Gamma^{\alpha)}_{\mu \nu} \big) \partial_{\alpha} \partial_{\beta} \xi^{\gamma}  \nonumber \\
  &= \ourG - M^{\alpha \beta}{}_{\gamma} \partial_{\alpha} \partial_{\beta} \xi^{\gamma} \,,
  \end{align}
with $M^{\alpha \beta}{}_{\gamma}$ defined in~(\ref{M}). For the full details of these derivations, see Appendix~\ref{appendixB.1}. Recall that $M^{\mu \nu}{}_{\lambda}$ can be related to the superpotential via equation~(\ref{M to E}), so this inhomogeneous term can equally be written as
\begin{equation}
M^{\alpha \beta}{}_{\gamma} \partial_{\alpha} \partial_{\beta} \xi^{\gamma} = E_{\gamma}{}^{\alpha \beta} \partial_{\alpha} \partial_{\beta} \xi^{\gamma} \, ,
\end{equation}
due to the commuting of partial derivatives.

A similar calculation for the boundary vector~(\ref{Bvector}) leads to
\begin{align}
\hat{B}^{\sigma}(\hat{x})  =  B^{\sigma} + \partial_{\eta} \xi^{\sigma} B^{\eta} - \partial^{\nu} \partial_{\nu} \xi^{\sigma} + \partial^{\sigma} \partial_{\lambda} \xi^{\lambda} \,,
\end{align}
and for the full boundary term~(\ref{B}) we find
\begin{align}
  \label{B_infinitesimal} 
  \hat{\ourB}(\hat{x})  &=
  \hat{\Gamma}^{\lambda}_{\lambda \sigma} \hat{B}^{\sigma} + \hat{\partial}_{\sigma} \hat{B}^{\sigma}
  \nonumber \\
  &= \ourB +  \big(2 g^{\mu (\alpha} \Gamma^{\beta)}_{\mu \gamma} -
  g^{\alpha \beta} \Gamma^{\lambda}_{\lambda \gamma} - g^{\mu \nu}
  \delta^{(\beta}_{\gamma} \Gamma^{\alpha)}_{\mu \nu} \big) \partial_{\alpha} \partial_{\beta} \xi^{\gamma}
  \nonumber \\
  &=  \ourB + M^{\alpha \beta}_{}{\gamma} \partial_{\alpha} \partial_{\beta} \xi^{\gamma} \,,
\end{align}
see Appendix~\ref{appendixB}.
This explicitly verifies that $\ourG$ and $\ourB$ are not coordinate scalars.
Equations~(\ref{G_infinitesimal}) and~(\ref{B_infinitesimal}) of course show that the combination $\hat{\ourG} + \hat{\ourB} = \ourG + \ourB$ is invariant, with the $\xi$ terms cancelling.

The inhomogeneous terms appearing in the transformations~(\ref{G_infinitesimal}) and~(\ref{B_infinitesimal}) may look slightly suspect, because they should not contribute to any equations of motion. In other words, the additional term should be a total derivative, which it does not immediately appear to be. However, we can rewrite it using the standard product rule
\begin{align}
M^{\alpha \beta}{}_{\gamma} \partial_{\alpha} \partial_{\beta} \xi^{\gamma}  &
\begin{multlined}[t]
= \frac{1}{\sqrt{-g}} \Big( \partial_{\alpha}\partial_{\beta} (\sqrt{-g}M^{\alpha \beta}{}_{\gamma} \xi^{\gamma}) 
- \partial_{\alpha} \partial_{\beta} (\sqrt{-g} M^{\alpha \beta}{}_{\gamma}) \xi^{\gamma} \\
- 2 \partial_{\alpha} (\sqrt{-g} M^{\alpha \beta}{}_{\gamma}) \partial_{\beta} \xi^{\gamma}  \Big)
\end{multlined} \nonumber \\
& \begin{multlined}[b]
= \textrm{`surface term'} - \frac{1}{\sqrt{-g}} \partial_{\alpha} \partial_{\beta} (\sqrt{-g} M^{\alpha \beta}{}_{\gamma}) \xi^{\gamma}  \\
-  \frac{2}{\sqrt{-g}}  \partial_{\alpha} (\sqrt{-g} M^{\alpha \beta}{}_{\gamma}) \partial_{\beta} \xi^{\gamma}  \, ,
\end{multlined}
\end{align}
where we have used the symmetry of $M^{\alpha \beta}{}_{\gamma} = M^{(\alpha \beta)}{}_{\gamma}$. The first term has been identified as a surface term, taking the form $ \partial_{\mu}( \sqrt{-g}v^{\mu})$ in an action.
Doing the same again with the last term gives
\begin{align}
-\frac{2}{\sqrt{-g}}  \partial_a (\sqrt{-g} M^{\alpha \beta}{}_{\gamma}) \partial_{\beta} \xi^{\gamma} &
\begin{multlined}[t]
=  -\frac{2}{\sqrt{-g}} \partial_{\alpha} \big( \partial_{\beta} (\sqrt{-g} M^{\alpha \beta}{}_{\gamma} ) \xi^{\gamma} \big)
  \\
 + \frac{2}{\sqrt{-g}} \partial_{\alpha} \partial_{\beta} (\sqrt{-g} M^{\alpha \beta}{}_{\gamma}) \xi^{\gamma}  
\end{multlined} \nonumber \\
&= \textrm{`surface term'}  + \frac{2}{\sqrt{-g}} \partial_{a} \partial_{\beta} (\sqrt{-g} M^{\alpha \beta}{}_{\gamma}) \xi^{\gamma} \, .
\end{align}
Putting these together we have 
\begin{align} \label{M surface}
M^{\alpha \beta}{}_{\gamma} \partial_{\alpha} \partial_{\beta} \xi^{\gamma} = \textrm{`surface terms'} +  \frac{1}{\sqrt{-g}} \partial_{\alpha} \partial_{\beta} (\sqrt{-g} M^{\alpha \beta}{}_{\gamma}) \xi^{\gamma} \, .
\end{align}
The final remaining term, which \textit{is not} a surface term, is particularly interesting. We will show in Section~\ref{section4.1.4} that it actually vanishes identically, and is therefore fully covariant, despite its appearance. Overall we see that $\ourG$ and $\ourB$ transform covariantly up to surface terms\footnote{In Appendix~\ref{appendixD} we also show that $\ourG$ differs from an invariant scalar by a boundary term, by instead considering (finite) GCTs and applying the Stueckelberg trick.}.

Lastly, we also include the infinitesimal transformations for $M^{\alpha\beta}{}_{\gamma}$ and $E_{\rho \sigma}{}^{\gamma}$. For the former we find
\begin{multline}
  \label{M_infinitesimal}
  \hat{M}^{\alpha \beta }{}_{\gamma}  =
  M^{\alpha \beta }{}_{\gamma} + \partial_{\lambda} \xi^{\alpha } M^{\lambda \beta }{}_{\gamma} +
  \partial_{\lambda} \xi^{\beta } M^{\alpha  \lambda}{}_{\gamma} - \partial_{\gamma} \xi^{\lambda} M^{\alpha  \beta }{}_{\lambda} \\
  - 2 \partial^{(\alpha }\partial_{\gamma} \xi^{\beta )} + g^{\alpha \beta } \partial_{\lambda} \partial_{\gamma} \xi^{\lambda} +
  g^{\mu \nu} \delta^{(\beta }_{\gamma} \partial_{\mu} \partial_{\nu} \xi^{\alpha )} \,,
\end{multline}
while the latter gives
\begin{multline}
  \label{E_infinitesimal}
  \hat{E}_{\rho \sigma}{}^{\gamma}  =  E_{\rho \sigma}{}^{\gamma} -
  E_{\rho \eta}{}^{\gamma} \partial_{\sigma} \xi^{\eta} -
  E_{\sigma \eta}{}^{\gamma} \partial_{\rho} \xi^{\eta} +
  E_{\rho \sigma}{}^{\eta} \partial_{\eta} \xi^{\gamma} \\ 
  - 2 \partial_{\rho} \partial_{\sigma} \xi^{\gamma} +
  2 \delta_{(\rho}^{\gamma} \partial_{\sigma)} \partial_{\lambda} \xi^{\lambda} -
  g_{\rho \sigma} \partial^{\gamma} \partial_{\lambda} \xi^{\lambda} +
  g_{\rho \sigma} \partial^{\lambda} \partial_{\lambda} \xi^{\gamma} \,.
\end{multline}

Using all of the infinitesimal coordinate transformations calculated above, we can simply state their Lie derivatives~(\ref{Lie derivative}). Note that we have assumed our transformations $\xi^{\mu}(x)$ to be small, implicitly absorbing the parameter $\epsilon$ into their definition. For the Lie derivatives, it should be understood that we have taken the limit $\epsilon \rightarrow 0$, see Sec.~\ref{section2.3.1} for details.

For the bulk term $\ourG$ we have
\begin{align}
  \label{Lie_G}
  \mathcal{L}_{\xi} \ourG = \xi^{\mu}  \partial_{\mu} \ourG +
  M^{\alpha \beta}{}_{\gamma} \partial_{\alpha} \partial_{\beta} \xi^{\gamma} \,,
\end{align}
and for the boundary term $\ourB$ 
\begin{align} \label{Lie_B}
\mathcal{L}_{\xi} \ourB = \xi^{\mu} \partial_{\mu} \ourB - M^{\alpha \beta}{}_{\gamma} \partial_{\alpha} \partial_{\beta} \xi^{\gamma} \,.
\end{align}
From our previous calculation in equation~(\ref{M surface}), we see that the Lie derivatives of $\ourG$ and $\ourB$ give rise to boundary terms in $\xi$ (plus a term that will be shown to be identically zero).
The Lie derivatives of the three-index objects are
\begin{multline} 
  \label{Lie_M}
  \mathcal{L}_{\xi} M^{\alpha \beta }{}_{\gamma} = \xi^{\mu} \partial_{\mu} M^{\alpha \beta }{}_\gamma -
  2 \partial_{\lambda} \xi^{(\alpha } M^{\beta ) \lambda}{}_{\gamma}  + \partial_{\gamma} \xi^{\lambda} M^{\alpha \beta }{}_{\lambda} \\
  + 2 \partial^{(\alpha }\partial_{\gamma} \xi^{\beta )} - g^{\alpha \beta } \partial_{\lambda} \partial_{\gamma} \xi^{\lambda} -
  g^{\mu \nu} \delta^{(\beta }_{\gamma} \partial_{\mu} \partial_{\nu} \xi^{\alpha )} \,,
\end{multline}
and for the superpotential
\begin{multline}
  \label{Lie_E}
  \mathcal{L}_{\xi} {E}_{\rho \sigma}{}^{\gamma} =  \xi^{\mu} \partial_{\mu} E_{\rho \sigma}{}^{\gamma} +
  2 E_{\eta (\rho}{}^{\gamma} \partial_{\sigma)} \xi^{\eta} -
  E_{\rho \sigma}{}^{\eta} \partial_{\eta} \xi^{\gamma} \\ 
  + 2 \partial_{\rho} \partial_{\sigma} \xi^{\gamma} - 2 \delta_{(\rho}^{\gamma} \partial_{\sigma)}
  \partial_{\lambda} \xi^{\lambda} + g_{\rho \sigma} \partial^{\gamma} \partial_{\lambda} \xi^{\lambda} -
  g_{\rho \sigma} \partial^{\lambda} \partial_{\lambda} \xi^{\gamma} \,.
\end{multline}

The right-hand sides of equations~(\ref{Lie_G})--(\ref{Lie_E}) are all non-tensorial. However, this is not immediately obvious as the expression for $\mathcal{L}_{\xi} \Gamma$~(\ref{Lie connection2}) also appears non-covariant at first. Intuitively, one can see that an object like $\ourG$, being quadratic in the connection, will transform like $\mathcal{L}_{\xi} \Gamma^2 \sim \Gamma \mathcal{L}_{\xi} \Gamma$ which cannot be a tensor as it is a product of a tensor and a connection. To be more concrete, using the calculations in Appendix~\ref{appendixB.1}, it can be shown that under a (different) infinitesimal coordinate transformation $\hat{x}^{\mu} = x^{\mu} + \epsilon \zeta^{\mu}(x)$ the Lie derivative of $\ourG$ is not invariant
\begin{equation} \label{LieLieG}
(\widehat{\mathcal{L}_{\xi} \ourG}) = \mathcal{L}_{\xi} \ourG - (\mathcal{L}_{\xi} M^{\alpha \beta}{}_{\gamma} ) \partial_{\alpha} \partial_{\beta} \zeta^{\gamma} \, .
\end{equation}
The same holds for equations~(\ref{Lie_B})--(\ref{Lie_E}). With all of the relevant transformation properties concluded, let us move on to studying gravitational actions.

\subsection{Action and field equations}
\label{section4.1.3}

Given the preceding sections, we now have a good understanding of the bulk and boundary terms $\ourG$ and $\ourB$. Applying this to gravity, using our decomposition~(\ref{RGB}) the Einstein-Hilbert action can be written as
\begin{equation}
S_{\textrm{EH}} = \frac{1}{2\kappa} \int R \sqrt{-g} d^4 x =  \frac{1}{2\kappa} \int \ourG \sqrt{-g} d^4x +  \frac{1}{2\kappa} \int \ourB \sqrt{-g} d^4x \, .
\end{equation}
On a manifold without a boundary, variations of the final term will vanish leaving only the bulk contribution.
We therefore define an action from just the bulk term as the \textit{Einstein action} 
\begin{equation} \label{Einstein_action}
S_{\textrm{E}} :=  \frac{1}{2\kappa} \int \ourG \sqrt{-g} d^4x \, ,
\end{equation}
which is first-order in metric derivatives. Because the action is first-order, the requirement of the metric variations vanishing on the boundary is enough to have a well-posed variation principle. It is quite interesting that the non-covariant action $S_{\textrm{E}}$ does not need any additional terms in the case of manifolds with a boundary.

For said manifolds with a boundary, it is well known that the Einstein-Hilbert action must be supplemented with additional counter terms, see Sec.~\ref{section2.2.2}. This can also be seen directly from varying the boundary term~(\ref{B2}) with respect to the metric 
\begin{align}
\frac{1}{2\kappa}  \int_{\mathcal{M}} \delta( \ourB \sqrt{-g} )d^4x  = \frac{1}{2\kappa} \int_{\mathcal{M}} \partial_{\sigma} \big( \sqrt{-g} \delta B^{\sigma} -\frac{1}{2} \sqrt{-g} g_{\mu \nu} \delta g^{\mu \nu} B^{\sigma}  \big) d^4x \, ,
\end{align}
with $\delta B^{\sigma}$ containing terms proportional to derivatives of the metric $\delta \Gamma \propto \nabla \delta g$. This was shown explicitly in equation~(\ref{EHbound_var}).
The counter term needed is the Gibbons-Hawking-York term~(\ref{GHY}), so that we can express the full Einstein-Hilbert action (with counter terms) as
\begin{align}
S_{\textrm{EH'}} = S_{\textrm{EH}}  + S_{\textrm{GHY}}  &=  \frac{1}{2\kappa} \int_{\mathcal{M}} R \sqrt{-g} d^4 x  + \frac{1}{\kappa} \int_{\mathcal{\partial M}} K \sqrt{h} d^3 x \nonumber \\
&= S_{\textrm{E}} +  \frac{1}{2\kappa} \int_{\mathcal{M}} \ourB \sqrt{-g} d^4x + \frac{1}{\kappa} \int_{\mathcal{\partial M}} K \sqrt{h} d^3 x \, .
\end{align}
Using the decomposition $g_{\mu \nu} = h_{\mu \nu} - n_{\mu} n_{\nu}$ and following the same steps as in Sec.~\ref{section2.2.2}, it can be shown that the difference between these two boundary terms vanishes under variations provided $\delta g_{\mu \nu}|_{\partial \mathcal{M}}=0$~\cite{TP:2010}
\begin{equation} \label{EH equiv}
\delta S_{\textrm{EH'}}  = \delta S_{\textrm{E}} \, .
\end{equation}
Hence the resulting field equations are the same. In general though $S_{\textrm{EH'}}  \neq S_{\textrm{E}}$, so quantities depending on the total value of the action need to be carefully considered.

It may appear that $S_{\textrm{EH'}}$ is a more natural choice for an action compared to the non-covariant Einstein action, with the former taking a covariant form~\cite{TP:2010}
\begin{equation}
L_{\textrm{EH'}} = R - 2 \nabla_{\mu}(n^{\mu} \nabla_{\nu} n^{\nu})\, .
\end{equation}
But this Lagrangian depends on additional, fixed vector fields. Any non-covariant expression can be rewritten in a covariant form with the inclusion of fixed vector fields~\cite{Gao:2014soa}, Einstein-Aether theories~\cite{Eling:2004dk} and Horava gravity~\cite{Jacobson:2010mx} being key examples. There is therefore little \textit{conceptual} advantage of choosing the Einstein-Hilbert action with the Gibbons-Hawking-York term over the non-covariant Einstein action. Moreover, it should be mentioned that the Einstein action is known to have technical advantages when it comes to topics such as quantization and graviton vertices~\cite{BeltranJimenez:2018vdo, Tomboulis:2017fim}.

Moving on to the equations of motion, we know from~(\ref{EH equiv}) that the field equations derived from the Einstein action will be the same field equations as from $S_{\textrm{EH}}$. For completeness, we show these explicit computations in Appendix~\ref{appendixC.1.1} following our work in~\cite{Boehmer:2021aji}. We therefore have
\begin{equation}
\delta S_{\textrm{E}} = \frac{1}{2\kappa} \int \delta g^{\mu \nu} \big(R_{\mu \nu} - \frac{1}{2} g_{\mu \nu} R\big) \sqrt{-g} d^4 x \, .
\end{equation}
The inclusion of minimally coupled matter leads to the standard Einstein field equations 
\begin{equation}
G_{\mu \nu} = \kappa T_{\mu \nu} \, ,
\end{equation}
with $T_{\mu \nu}$ the metric energy-momentum tensor~(\ref{energy-momentum tensor}). We reiterate that no additional boundary terms are needed in this derivation.

\subsection{Diffeomorphism invariance and pseudotensors}
\label{section4.1.4}

Let us look at the diffeomorphism invariance of the Einstein action. Under an infinitesimal symmetry transformation the action will change by its Lie derivative~(\ref{Lagrangian diff}), and hence
\begin{align}
\delta_{\xi} S_{\textrm{E}} &= \frac{1}{2 \kappa} \int \mathcal{L}_{\xi}(\sqrt{-g} \ourG ) d^4 x \nonumber \\
&= \frac{1}{2 \kappa} \int \Big(\frac{1}{2} g^{\mu \nu} \ourG \mathcal{L}_{\xi} g_{\mu \nu} + \mathcal{L}_{\xi} \ourG \Big) \sqrt{-g} d^4 x \nonumber \\ 
&= \frac{1}{2 \kappa} \int \Big( \ourG \nabla_{\mu} \xi^{\mu} + \xi^{\mu} \partial_{\mu} \ourG + M^{\mu \nu}{}_{\gamma} \partial_{\mu} \partial_{\nu} \xi^{\gamma} \Big) \sqrt{-g} d^4 x \nonumber \\
&=  \frac{1}{2 \kappa} \int \partial_{\mu} (\sqrt{-g} \xi^{\mu} \ourG) d^4x +  \frac{1}{2 \kappa} \int M^{\mu \nu}{}_{\gamma} \partial_{\mu} \partial_{\nu} \xi^{\gamma}  \sqrt{-g} d^4 x  \, , \label{E_diff}
\end{align}
where we have simply inserted the definition of the Lie derivative~(\ref{Lie_G}). Assuming that $\xi$ is held fixed on the boundary, the first term is a total derivative and vanishes. In equation~(\ref{M surface}) we showed that the remaining term could be written as the sum of surface terms\footnote{Note that the surface terms that we have now discarded from~(\ref{M surface}) depend on $\xi$ and its first derivatives, which we assume to vanish on the boundary.} and a term that should  vanish 
\begin{equation}
\delta_{\xi} S_{\textrm{E}} =  \textrm{`surface terms'}+  \frac{1}{2 \kappa} \int \partial_{\mu} \partial_{\nu} \big( \sqrt{-g} M^{\mu \nu}{}_{\gamma}\big) \xi^{\gamma} d^4 x \, .
\end{equation}
And in fact, this term is identically zero 
\begin{equation} \label{Bianchi_M}
\partial_{\mu} \partial_{\nu} (\sqrt{-g} M^{\mu \nu}{}_{\lambda}) \equiv  0 \, ,
\end{equation}
therefore $\delta_{\xi} S_{\textrm{E}} =0$ as expected.

This identity~(\ref{Bianchi_M}) has long been known, and was originally considered an identity by Einstein as a result of Noether's theorem~\cite{Einstein:1916cd,janssen2002collected,Chen:2015vya}, which he presented in the form
\begin{equation}
\partial_{\mu} \partial_{\nu} \Big( g^{\mu \rho} \frac{\partial (\sqrt{-g} \ourG) }{ \partial( \partial_{\nu} g^{\lambda \rho})} \Big) \equiv 0 \, .
\end{equation}
We see that this is equivalent to our equation~(\ref{Bianchi_M}) by using our definitions~(\ref{EG1}) and standard relations~(\ref{M to E}). The geometric meaning behind this identity is that it is in fact just the
twice contracted Bianchi identity~(\ref{Bianchi2}), albeit written in a form that does not appear covariant~\cite{MOLLER1958347,Duan:2018fke}. Moreover, this was not immediately known to Einstein and others at the time~\cite{Chen:2015vya}.

We therefore see that invariance of the Einstein action under infinitesimal diffeomorphisms is recovered due to the contracted Bianchi identity. This is perhaps not so surprising because the same thing happened with the Einstein-Hilbert action~(\ref{EH diff}). Alternatively, we can see the contracted Bianchi identity as a consequence of the (quasi)-diffeomorphism invariance of the Einstein action.

The simple calculation presented here~(\ref{E_diff}), which could also be obtained by taking the metric variations and plugging in the Lie derivative of the metric, gives us a concrete way of obtaining conservation laws associated with Noether's theorem. The benefit of this explicit method is that it can be applied to objects which are not tensorial. It also means that we can \textit{enforce} (infinitesimal) diffeomorphism invariance by requiring these conservation laws to be satisfied.

Lastly, let us briefly mention the relationship between our approach and the pseudotensors that often occur in General Relativity. Defining the canonical energy-momentum (pseudo-)tensor~\cite{Einstein:1916cd,Chen:2015vya} for the Einstein action 
\begin{equation} \label{grav_pseudo}
 \sqrt{-g} \, t_{\textrm{E}}^{\mu}{}_{\nu} :=   \delta^{\mu}_{\nu} \mathcal{L}_{\textrm{E}} - \frac{\partial \mathcal{L}_{\textrm{E}}}{\partial (\partial_{\mu} g_{\alpha \beta})} \partial_{\nu} g_{\alpha \beta} \, ,
\end{equation}
with $2\kappa \mathcal{L}_{\textrm{E}} = \sqrt{-g} \ourG$, leads to 
\begin{equation}
2 \kappa  t_{\textrm{E}}^{\mu}{}_{\nu} =   \delta^{\mu}_{\nu} \ourG - \frac{1}{2} E^{\alpha \beta \mu} \partial_{\nu} g_{\alpha \beta}  = \frac{1}{4} \big( \delta^{\mu}_{\nu} E^{\alpha \beta \gamma} \partial_{\gamma} g_{\alpha \beta} - 2  E^{\alpha \beta \mu} \partial_{\nu} g_{\alpha \beta} \big) \, .
\end{equation}
We then have via Noether's theorem the conserved quantity
\begin{equation} \label{Noether conserved}
\partial_{\mu} \big(\sqrt{-g}( t_{\textrm{E}}^{\mu}{}_{\nu}  +  T^{\mu}{}_{\nu}) \big) = 0 \, ,
\end{equation}
where $T_{\mu \nu}$ is the energy-momentum tensor. This can be seen explicitly by calculating\footnote{To verify equation~(\ref{d_pseudo}) is straightforward but a long calculation. Here we resort to using the Mathematica xAct package, which indeed verifies the left and right-hand sides are equivalent~\cite{martin2002xact}.}
\begin{equation} \label{d_pseudo}
2\kappa \partial_{\mu}(\sqrt{-g}  t_{\textrm{E}}^{\mu}{}_{\nu} ) \equiv -2\kappa \frac{\delta \mathcal{L}_{\textrm{E}}}{\delta g_{\alpha \beta}} \partial_{\nu} g_{\alpha \beta} =- \sqrt{-g} G^{\alpha \beta} \partial_{\nu} g_{\alpha \beta}  \, .
\end{equation}
From here one simply uses the contracted Bianchi identity 
\begin{align}
\nabla_{\mu} G^{\mu}{}_{\nu}  &= \frac{1}{\sqrt{-g} }\partial_{\mu}(\sqrt{-g} G^{\mu}{}_{\nu}) - \frac{1}{2} G^{\mu \sigma} \partial_{\nu} g_{\sigma \mu} = 0 \nonumber \\
&\implies \sqrt{-g} G^{\alpha \beta} \partial_{\nu} g_{\alpha \beta}  = 2 \partial_{\mu}(\sqrt{-g} G^{\mu}{}_{\nu})  \, ,
\end{align} 
and the field equations $G_{\mu \nu} = \kappa T_{\mu \nu}$ to show that
\begin{equation}
\partial_{\mu} \big(\sqrt{-g}(  t_{\textrm{E}}^{\mu}{}_{\nu}  +  T^{\mu}{}_{\nu}) \big)  = -\frac{1}{2\kappa}   \sqrt{-g} G^{\alpha \beta} \partial_{\nu} g_{\alpha \beta} + \frac{1}{\kappa} \partial_{\mu}(\sqrt{-g} G^{\mu}{}_{\nu}) = 0 \, .
\end{equation}

Alternatively, it can be shown directly that the Einstein tensor can be rewritten in terms of the superpotential 
\begin{align}
\sqrt{-g} G^{\mu}{}_{\nu} &= \frac{1}{2} \partial_{\lambda} (\sqrt{-g} E_{\nu}{}^{\mu \lambda}) - \frac{ \sqrt{-g}}{8} \big( \delta^{\mu}_{\nu} E^{\alpha \beta \gamma} \partial_{\gamma} g_{\alpha \beta} - 2  E^{\alpha \beta \mu} \partial_{\nu} g_{\alpha \beta} \big) \nonumber \\
&=  \frac{1}{2} \partial_{\lambda} (\sqrt{-g} E_{\nu}{}^{\mu \lambda})  - \kappa   \sqrt{-g} t_{\textrm{E}}^{\mu}{}_{\nu} \, .
\end{align}
To obtain the first line we follow similar calculations as those leading to the Ricci scalar in terms of the superpotential~(\ref{R_super}). See also~\cite{Boehmer:2021aji,Boehmer:2023fyl} for technical details. The conservation law~(\ref{Noether conserved}) then follows from taking the divergence and using the Bianchi identity~(\ref{Bianchi_M}) along with the field equations
\begin{align}
\frac{1}{2\kappa} \partial_{\mu} \partial_{\lambda} (\sqrt{-g} E_{\nu}{}^{\mu \lambda}) &= \partial_{\mu}\Big(   \sqrt{-g} t_{\textrm{E}}^{\mu}{}_{\nu} + \frac{1}{\kappa} \sqrt{-g} G^{\mu}{}_{\nu}    \Big) \nonumber \\
&= \partial_{\mu} \Big( \sqrt{-g} ( t_{\textrm{E}}^{\mu}{}_{\nu} + T^{\mu}{}_{\nu} ) \Big) = 0 \, .
\end{align}
(We do not write the final line as an identity because it only holds on-shell, as opposed to the first line which vanishes identically off-shell). For further discussion, see~\cite{DeHaro:2021gdv} and~\cite{Chen:2015vya}.

\section{Tetradic Einstein action}
\label{section4.2}

In this section we repeat the decomposition of the Ricci scalar into its bulk and boundary parts, but in the orthonormal basis, making use of the tetrad and spin connection. Note that we are still working in the purely Levi-Civita framework, with vanishing torsion and non-metricity. This leads to a tetradic version of the Einstein action, which we inventively label the \textit{tetradic Einstein action}.

This action can once again be found in M\o{}ller's work~\cite{moller1961conservation,Moller:1962vis}, extending his previous studies on localised energy complexes and conservation laws. Here, making use of the tetrad framework, his arguments take a covariant form. Because the tetrad and spin connection both transform homogenously with respect to transformations of the spacetime coordinates, the resulting pseudotensors take a more satisfactory form. The trade-off comes in the way of non-invariance with respect to frame transformations (in the local tangent space). We will cover these topics in detail in this section. For a modern exposition of M\o{}ller's ideas on energy-momentum complexes and pseudotensors, see~\cite{mikhail1993energy}. Also see the extensive review by Szabados, particularly Chapter 3, which focuses on the topic of quasi-local mass-energy-momentum~\cite{Szabados:2009eka}.

\subsection{Tetradic Ricci scalar decomposition}
\label{section4.2.1}

Let us now follow the same arguments as above, with the decomposition of the Ricci scalar, but working with the frame fields in the orthonormal basis $\mathbf{e}_{a}$. There are two ways one could approach this. The first is to simply take the coordinate basis expressions for $\ourG$ and $\ourB$ and use the relations between the spacetime metric $g_{\mu \nu}$ and tetrads $e_{a}{}^{\mu}$ in~(\ref{ONB relations}), and between the spacetime and spin connections in~(\ref{spin and affine}). The second, and more illustrative way, is to follow the steps of integrating terms in the tetradic Ricci scalar by parts to obtain boundary terms. Though both lead to the same equations, the second way is considerably more straightforward because we do not need to write the Levi-Civita spin connection explicitly in terms of the tetrad.

The Levi-Civita Ricci scalar in terms of the tetrad and spin connection is
\begin{equation} \label{R spin}
R = e_{a}{}^{\mu} e_{b}{}^{\nu} R_{\nu \mu}{}^{a b} = 2\mathring{\omega}_{[\nu}{}^{\nu}{}_{|c|} \mathring{\omega}_{\mu]}{}^{c \mu} + 2 e_{a}{}^{\mu} e_{b}{}^{\nu} \partial_{[\nu} \mathring{\omega}_{\mu]}{}^{ba}  \, ,
\end{equation}
where we remind the reader that $\mathring{\omega}_{\mu ab} =- \mathring{\omega}_{\mu ba}$, so the last antisymmetry brackets can be dropped. We also take liberties with the spin connection and somewhat abuse notation by allowing any of the indices to be Greek, Latin, raised or lowered. We reiterate that all objects in~(\ref{R spin}) are contractions of the spin connection~(\ref{spin form}), despite the use of spacetime indices.

A few useful formulae for these calculations are 
\begin{equation} \label{e_formulae}
\partial_{\mu} e^a{}_{\nu} = - \mathring{\omega}_{\mu}{}^{a}{}_{b} e^{b}{}_{\nu} + \Gamma_{\mu \nu}^{\lambda} e^a{}_{\lambda} \quad \, , \quad  \partial_{\mu} e_{a}{}^{\nu} = \mathring{\omega}_{\mu}{}^{b}{}_{a} e_{b}{}^{\nu} - \Gamma_{\mu \lambda}^{\nu} e_{a}{}^{\lambda} \, ,
\end{equation}
where $ \Gamma_{\mu \lambda}^{\nu} $ is the spacetime Levi-Civita connection. Both of these relations follow directly from the definition of the spin connection~(\ref{spin and affine}). Using the chain rule we have
\begin{multline} \label{boundary_tetrad_e1}
\partial_{\nu} \big(e e_{a}{}^{\mu} e_{b}{}^{\nu} \mathring{\omega}_{\mu}{}^{ba} \big) = e e_{a}{}^{\mu} e_{b}{}^{\nu} \partial_{\nu} \mathring{\omega}_{\mu}{}^{ba} + e \Gamma_{\nu \lambda}^{\lambda} \mathring{\omega}_{\mu}{}^{\nu \mu} + e \mathring{\omega}_{\mu}{}^{\nu a} \mathring{\omega}_{\nu}{}^{\mu}{}_{a} + e \mathring{\omega}_{\mu}{}^{a \mu} \mathring{\omega}_{\nu}{}^{\nu}{}_{a} \\
- e \mathring{\omega}_{\mu}{}^{\nu \lambda} \Gamma_{\nu \lambda}^{\mu} - e \mathring{\omega}_{\mu}{}^{\lambda \mu} \Gamma^{\nu}_{\nu \lambda} \, .
\end{multline}
Due to the antisymmetry of the spin connection, along with the symmetry of the Levi-Civita spacetime connection, the final term on the second line cancels with the second term on the RHS of the first line. Similarly, the remaining term on the second line vanishes due to the contraction between symmetric and antisymmetric indices. We then have
 \begin{equation}  \label{boundary_tetrad_e2}
\partial_{\nu} \big(e e_{a}{}^{\mu} e_{b}{}^{\nu} \mathring{\omega}_{\mu}{}^{ba} \big) = e e_{a}{}^{\mu} e_{b}{}^{\nu} \partial_{\nu} \mathring{\omega}_{\mu}{}^{ba} + 2 e \mathring{\omega}_{[\nu}{}^{a \nu} \mathring{\omega}_{\mu]}{}^{\mu}{}_{a} \, ,
\end{equation}
which we can simply rearrange to match the derivative terms in the Ricci scalar~(\ref{R spin})
\begin{equation}
2 e e_{a}{}^{\mu} e_{b}{}^{\nu} \partial_{\nu} \mathring{\omega}_{\mu}{}^{ba} = 2 \partial_{\nu} \big(e e_{a}{}^{\mu} e_{b}{}^{\nu} \mathring{\omega}_{\mu}{}^{ba} \big)  - 4 e \mathring{\omega}_{[\nu}{}^{a \nu} \mathring{\omega}_{\mu]}{}^{\mu}{}_{a} \, .
\end{equation}

Substituting this into our Lagrangian $R(e)$~(\ref{R spin}) leads to
\begin{equation}
R(e) =  2\mathring{\omega}_{[\nu}{}^{\nu}{}_{|c|} \mathring{\omega}_{\mu]}{}^{\mu c}   + \frac{2}{e}\partial_{\nu} \big(e e_{a}{}^{\mu} e_{b}{}^{\nu} \mathring{\omega}_{\mu}{}^{ba} \big) \, .
\end{equation}
Note the quadratic part in the equation above is simply minus the quadratic combination occurring in the original expression~(\ref{R spin}) due to the swapped last indices.
Introducing the following notation for the tetradic bulk and boundary terms we can write
\begin{equation}
R(e) = \mathfrak{G} + \mathfrak{B} \, ,
\end{equation}
where we define
\begin{align} \label{G_e}
\mathfrak{G} &:=  2\mathring{\omega}_{[\nu}{}^{\nu}{}_{|c|} \mathring{\omega}_{\mu]}{}^{\mu c} = 2 \eta^{cd} e_{a}{}^{\nu} e_{b}{}^{\mu} \mathring{\omega}_{[\nu}{}^{a}{}_{|c|} \mathring{\omega}_{\mu]}{}^{b}{}_{d} \, , \\
\mathfrak{B} &:= \frac{2}{e} \partial_{\nu} (e \mathring{\omega}^{\nu}) = \frac{2}{e} \partial_{\nu} (e \mathring{\omega}_{\mu}{}^{\nu \mu})  \, . \label{B_e}
\end{align}

Recall that the spin connection transforms like a tensor under general coordinate transformations, but inhomogeneously under local Lorentz transformations~(\ref{spin_LLT}). Indices are raised and lowered by the spacetime and Minkowski metrics, which are invariant or covariant with respect to coordinate and local Lorentz transformations. Lastly, the tetrad transforms covariantly under both general coordinate and local Lorentz transformations. It then follows that the objects $\mathfrak{G}$ and $\mathfrak{B}$ are coordinate scalars but not invariant under local Lorentz transformations. In the section below, we will explicitly show that they are pseudo-invariant under local Lorentz transformations.

The bulk part of this action was first studied by M\o{}ller in~\cite{moller1961conservation}, and later in \cite{moller1961further,moller1978crisis}, where he writes the term as
\begin{equation} \label{G_Moller}
\mathfrak{G} = 2 \eta^{ab} \mathring{\nabla}_{[\nu} e_{|b|}{}^{\nu} \mathring{\nabla}_{\mu]} e_{a}{}^{\mu} \, ,
\end{equation}
with $\mathring{\nabla}$ representing the Levi-Civita covariant derivative acting on spacetime indices only. Note that this differs to our use of the \textit{total} covariant derivative, which has $\nabla_{\mu} e_{a}{}^{\nu}=0$. By definition~(\ref{spin and affine}) it follows that this covariant derivative acting on the tetrad is simply the spin connection $\mathring{\nabla}_{\mu} e_{a}{}^{\nu} = \mathring{\omega}_{\mu}{}^{\nu}{}_{a} = e_{b}{}^{\mu} \mathring{\omega}_{\mu}{}^{b}{}_{a}$. It is then not hard to see the equivalence between the two forms of $\mathfrak{G}$ in~(\ref{G_e}) and~(\ref{G_Moller}).

This action is also closely related to the action of metric teleparallel gravity, and this has been noted by some other authors. See, for example, Section 9.2 of the textbook \textit{Teleparallel Gravity}~\cite{Aldrovandi:2013wha} or the work~\cite{Maluf:2018coz}. However, these facts seem worth reiterating as there is some confusion regarding the direct equivalence between this tetradic bulk term~(\ref{G_e}) and the torsion scalar $T$ of equation~(\ref{T_scalar}). This is not surprising, given that here we are working strictly in the torsion-free Levi-Civita framework, whereas the teleparallel setting uses a different connection which is curvature-free. In Section~\ref{section4.3} we aim to clarify these relations and show the formal equivalence between the two different geometric approaches.

\subsection{Local Lorentz transformations}
\label{section4.2.2}
In analogy with the pseudo-invariance under diffeomorphisms of the coordinate basis bulk and boundary terms $\ourG$ and $\ourB$, we also have pseudo-local Lorentz invariance for $\mathfrak{G}$ and $\mathfrak{B}$. It is again convenient to introduce the variation of the bulk term with respect to the Levi-Civita spin connection 
\begin{equation} \label{M_tetrad}
\mathfrak{M}^{\mu}{}_{a}{}^{b} := \frac{\partial \mathfrak{G}}{\partial \mathring{\omega}_{\mu}{}^{a}{}_{b} } =  4 \eta^{b c} e_{[d}{}^{\nu} e_{a]}{}^{\mu} \mathring{\omega}_{\nu}{}^{d}{}_{c}  = 4 e_{a}{}^{[\mu} \mathring{\omega}_{\nu}{}^{\nu] b} \, .
\end{equation}

Recall that under a local Lorentz transformation~(\ref{Lorentz1}) the tetrad and spin connection transform as 
\begin{align} 
e^{a}{}_{\mu} &\rightarrow \hat{e}_{a}{}^{\mu} = \Lambda^{a}{}_{b} e^{b}{}_{\mu} \, , \\
e_{a}{}^{\mu} &\rightarrow \hat{e}_{a}{}^{\mu} = \Lambda_{a}{}^{b} e_{b}{}^{\mu} \, , \\
\omega_{\mu}{}^{a}{}_{b} &\rightarrow \hat{\omega}_{\mu}{}^{a}{}_{b} = \Lambda^{a}{}_{c} \Lambda_{b}{}^{d}\omega_{\mu}{}^{c}{}_{d} + \Lambda^{a}{}_{c} \partial_{\mu} \Lambda_{b}{}^{c} \, ,
\end{align} 
and for an infinitesimal transformation we have
\begin{equation}
\Lambda^{a}{}_{b}(x) = \delta^{a}_{b} + \epsilon^{a}{}_{b}(x) \, ,
\end{equation}
where it is understood that the parameter $\epsilon_{ab} = - \epsilon_{ba}$ is infinitesimal, i.e., we will discard terms of order $\mathcal{O}(\epsilon^2)$. Spacetime tensors, such as the metric or affine connection, as well as the tetrad determinant $e = \sqrt{-g}$ are invariant under LLTs.

Applying this to our tetradic bulk term, for a (finite) local Lorentz transformation we have
\begin{equation}
\mathfrak{G} \rightarrow \hat{\mathfrak{G}} = 2  \hat{e}{}_{a}{}^{\nu} \hat{e}_{b}{}^{\mu} \hat{\eta}^{dc} \hat{\mathring{\omega}}_{[\nu}{}^{a}{}_{|c|} \hat{\mathring{\omega}}_{\mu]}{}^{b}{}_{d} \, ,
\end{equation}
which leads to the following expression
\begin{multline}
 \hat{\mathfrak{G}}  =2  \Lambda_{a}{}^{e} \Lambda_{b}{}^{f} \Lambda^{d}{}_{g} \Lambda^{c}{}_{h}  e_{e}{}^{\nu} e_{f}{}^{\mu} \eta^{gh}\big(\Lambda^{a}{}_{i} \Lambda_{c}{}^{j} \mathring{\omega}_{[\nu}{}^{i}{}_{|j|} + \Lambda^{a}{}_{i} \partial_{[\nu} \Lambda_{|c|}{}^{i} \big) \\
\times \big( \Lambda^{b}{}_{|k|} \Lambda_{|d|}{}^{l} \mathring{\omega}_{\mu]}{}^{k}{}_{l} + \Lambda^{b}{}_{|k|} \partial_{\mu]} \Lambda_{d}{}^{k} \big) \, .
\end{multline}
The antisymmetry brackets in the expression above are acting on the spacetime indices $\nu$ and $\mu$ only. A straightforward algebraic manipulation leads to the following result
\begin{equation}
 \hat{\mathfrak{G}}  =  \mathfrak{G} + 4 \Lambda^{a}{}_{b} \mathring{\omega}_{[\nu}{}^{\nu b} \partial_{k]} \Lambda_{a}{}^{k}  +2  \eta^{ab} \partial_{[i} \Lambda_{|a|}{}^{i} \partial_{k]} \Lambda_{b}{}^{k} \, ,
\end{equation}
where it is understood that we use the shorthand $\partial_{a} = e_{a}{}^{\mu} \partial_{\mu}$.

For the infinitesimal transformation the quadratic part vanishes and the expression reduces to
\begin{align}
 \hat{\mathfrak{G}} = \mathfrak{G} + 4 \mathring{\omega}_{[\nu]}{}^{\nu a} \partial_{k]} \epsilon_{a}{}^{k} \, .
\end{align}
Written more formally, returning to standard notation, we have our final result
\begin{equation}
\delta_{\Lambda}\mathfrak{G} := \hat{\mathfrak{G}}  -  \mathfrak{G} = 4\eta^{ad} e_{c}{}^{\nu} e_{b}{}^{\mu} \mathring{\omega}_{[\nu}{}^{c}{}_{|d|} 
\partial_{\mu]} \epsilon_{a}{}^{b}(x) \, . 
\end{equation}
Introducing the superpotential $\mathfrak{M}^{\mu}{}_{a}{}^{b}$ in equation~(\ref{M_tetrad}),
it is then possible to rewrite the transformation above as 
\begin{equation}
\delta_{\Lambda}\mathfrak{G} = \mathfrak{M}^{\mu}{}_{a}{}^{b} \partial_{\mu} \epsilon_{b}{}^{a} (x) \, .
\end{equation}

It is not too difficult to show that the inhomogeneous transformation term can be rewritten as
\begin{equation} \label{M_tetrad_boundary}
 \mathfrak{M}^{\mu}{}_{a}{}^{b} \partial_{\mu} \epsilon_{b}{}^{a}(x) =  \frac{2}{e} \partial_{\mu} \big(e e_a{}^{\nu} e_{b}{}^{\mu} \partial_{\nu} \epsilon^{ba}(x) \big) \, ,
\end{equation}
which follows from the symmetry of partial derivatives and antisymmetry of $\epsilon_{ab}(x)$. The calculation follows the same steps as in equations~(\ref{boundary_tetrad_e1}) and~(\ref{boundary_tetrad_e2}), but it is more straightforward to work with the RHS first
\begin{align} \label{M_tetrad_boundary2}
  \frac{2}{e} \partial_{\mu} \big(e e_a{}^{\nu} e_{b}{}^{\mu} \partial_{\nu} \epsilon^{ba}(x) \big)  &= 2 \Big(\Gamma^{\lambda}_{\mu \lambda} e_a{}^{\nu} e_{b}{}^{\mu} + \mathring{\omega}_{\mu}{}^{c}{}_{a} e_{c}{}^{\nu} e_{b}{}^{\mu} + \mathring{\omega}_{\mu}{}^{c}{}_{b} e_{c}{}^{\mu} e_{a}{}^{\nu} \nonumber \\
 & \qquad \qquad - \Gamma^{\nu}_{\mu \lambda} e_{a}{}^{\lambda} e_{b}{}^{\mu} - \Gamma^{\mu}_{\mu \lambda} e_{b}{}^{\lambda} e_{a}{}^{\nu} \Big) \partial_{\nu} \epsilon^{ba} \nonumber \\
 &=2 \big(  \mathring{\omega}_{\mu}{}^{c}{}_{a} e_{c}{}^{\nu} e_{b}{}^{\mu} + \mathring{\omega}_{\mu}{}^{c}{}_{b} e_{c}{}^{\mu} e_{a}{}^{\nu} \big)  \partial_{\nu} \epsilon^{ba} \nonumber \\
 &= 2 \big(\eta^{bd} \mathring{\omega}_{\mu}{}^{c}{}_{d} e_{c}{}^{\mu}  e_{a}{}^{\nu}  - \eta^{bd} \mathring{\omega}_{\mu}{}^{c}{}_{d} e_{c}{}^{\nu} e_{a}{}^{\mu} \big)  \partial_{\nu} \epsilon_{b}{}^{a} \nonumber \\
 &= 4 \eta^{bd} e_{[c}{}^{\mu} e_{a]}{}^{\nu} \mathring{\omega}_{\mu}{}^{c}{}_{d}  \partial_{\nu} \epsilon_{b}{}^{a}\nonumber  \\
 &= \mathfrak{M}^{\nu}{}_{a}{}^{b} \partial_{\nu} \epsilon_{b}{}^{a}\, .
\end{align}
Note that in transitioning from the second to the third line we have used the antisymmetry under the indices $a, b$ to flip the sign of the first term, and then relabelled the dummy indices. The second partial derivative of $\epsilon^{ab}$ also vanishes due to symmetry.

Importantly, this term now takes the form of a total derivative. This is in complete agreement with the coordinate basis transformation, where the inhomogeneous term also could be written as a boundary term depending on the infinitesimal vector $\xi^{\mu}$~(\ref{M surface}). The similarities in the form of~(\ref{M_tetrad_boundary}) with the tetradic boundary term~(\ref{B_e}) should also be noted, and this will be revisited below.

Moving on to the transformation for the boundary term $\mathfrak{B}$ we have
\begin{align}
\mathfrak{B} \rightarrow \hat{\mathfrak{B}} &=\frac{2}{e} \partial_{\nu} (e \hat{\eta}^{cb} \hat{e}_{a}{}^{\nu}  \hat{e}_{b}{}^{\mu} \hat{\mathring{\omega}}_{\mu}{}^{a}{}_{c}) \nonumber \\
 &= \mathfrak{B} + \frac{2}{e} \partial_{\nu} \big( e \eta^{ab} e_{c}{}^{\nu} e_{b}{}^{\mu} \Lambda^{d}{}_{a} \partial_{\mu} \Lambda_{d}{}^{c} \big) \, .
\end{align}
For the infinitesimal local Lorentz transformation the change is given by
\begin{align}
\delta_{\Lambda}\mathfrak{B} := \hat{\mathfrak{B}}  -  \mathfrak{B} =  \frac{2}{e} \partial_{\nu} \big( e e_{c}{}^{\nu} e_{b}{}^{\mu} \partial_{\mu} \epsilon^{bc} (x) \big)  \, .
\end{align}
It is then useful to recall the relation given in~(\ref{M_tetrad_boundary}) and proven in equation~(\ref{M_tetrad_boundary2}), which relates the superpotential to a boundary term. That boundary term is exactly the transformation given above, with an opposite sign. It follows that $\mathfrak{B}$ transforms under an infinitesimal local Lorentz transformation as
\begin{equation}
\delta_{\Lambda}\mathfrak{B} = -  \mathfrak{M}^{\mu}{}_{a}{}^{b} \partial_{\mu} \epsilon_{b}{}^{a}(x) \, ,
\end{equation}
where we have used the skew symmetry of $\epsilon^{ab}$ to obtain the minus sign in the above equation. Note that both $\delta_{\Lambda}\mathfrak{G}$ and $\delta_{\Lambda}\mathfrak{B}$ can be written in a way that is manifestly first-order in derivatives of the tetrad, and also in the form of a total derivative.

In summary, we have the following infinitesimal transformation properties for the tetradic bulk and boundary terms
\begin{align} \label{G_LLT}
\mathfrak{G} &\rightarrow  \mathfrak{G} + \mathfrak{M}^{\mu}{}_{a}{}^{b} \partial_{\mu} \epsilon_{b}{}^{a}(x) \, ,
\\ \label{B_LLT} 
\mathfrak{B} &\rightarrow \mathfrak{B} - \mathfrak{M}^{\mu}{}_{a}{}^{b} \partial_{\mu} \epsilon_{b}{}^{a}(x) \, ,
\end{align}
such that $\hat{R} = \hat{\mathfrak{G}}+\hat{\mathfrak{B}} = \mathfrak{G} + \mathfrak{B} = R$ is invariant. Importantly, as already stressed, the inhomogeneous term can be written as a total derivative. This will be crucial in the following section when considering the tetradic action.

\subsection{Action and local Lorentz invariance}
\label{section4.2.3}

Following the same arguments as in the previous sections in the coordinate basis, let us begin by using our decomposition of the tetradic Ricci scalar in the Einstein-Hilbert action
\begin{equation} \label{Einstein_tetrad_decomp}
S_{\textrm{EH}}[e] = \frac{1}{2\kappa} \int R(e) e d^4x =  \frac{1}{2\kappa} \int \mathfrak{G} e d^4x +  \frac{1}{2\kappa} \int \mathfrak{B} e d^4x \, .
\end{equation}
In analogy with~(\ref{Einstein_action}), we define the first-order \textit{tetradic Einstein action} as
\begin{equation} \label{Einstein_tetrad}
S_{\textrm{E}}[e] = \frac{1}{2 \kappa} \int \mathfrak{G} e d^4 x \, .
\end{equation}
As should now be quite clear, the resulting equations of motion from this first-order action will be equivalent to those from the Einstein-Hilbert action due to the vanishing contribution from the boundary term in~(\ref{Einstein_tetrad_decomp}). This could be shown by explicitly varying with respect to the tetrad, but here we will simply make use of our knowledge that boundary terms do not contribute. Hence, one way or another, we arrive at
\begin{equation}
G_{\mu \nu} = \kappa \Theta_{\mu \nu} \, .
\end{equation}

In Chapter~\ref{chapter2} we discussed the derivation of the Einstein field equations from this tetradic action, see equations~(\ref{EH tetrad var})-(\ref{EFE tetrad}). We also explicitly showed that it was the metric and not the tetrad that led to the dynamics of the theory, see equations~(\ref{EH_tetrad_LLT})-(\ref{tetradic_EM}) and the surrounding discussion. This was demonstrated by assuming that matter coupled to geometry only via the metric tensor, and using the local Lorentz invariance of the theory.

In the case of the tetradic Einstein action, if one did not a priori know that the field equations were given by the symmetric Einstein tensor, the local Lorentz invariance of the theory may not be obvious. This is especially true when one notes that $\mathfrak{G}$ is not a local Lorentz scalar, transforming inhomogeneously under LLTs~(\ref{G_LLT}). Let us look explicitly at how the action transforms under an infinitesimal local Lorentz transformation
\begin{equation}
\delta_{\Lambda} S_{\textrm{E}}[e] = \frac{1}{2 \kappa} \int \delta_{\Lambda} \mathfrak{G} e d^4 x = \frac{1}{2 \kappa} \int ( \mathfrak{M}^{\mu}{}_{a}{}^{b} \partial_{\mu} \epsilon_{b}{}^{a} ) e d^4 x \, , 
\end{equation}
where $\epsilon_{ab} = -\epsilon_{ba}$ and $\mathfrak{M}^{\mu}{}_{a}{}^{b}$ is defined in~(\ref{M_tetrad}). Next, we make use of the relation given in~(\ref{M_tetrad_boundary}) that shows that this inhomogeneous term can be written as a surface term
\begin{equation}
\delta_{\Lambda} S_{\textrm{E}}[e]  = \frac{1}{\kappa} \int \partial_{\mu} \big(e e_a{}^{\nu} e_{b}{}^{\mu} \partial_{\nu} \epsilon^{ba} \big) d^4 x \, .
\end{equation}
It then immediately follows via Stokes' theorem that this integral is zero and the action is locally Lorentz invariant up to boundary terms $\delta_{\Lambda} S_{\textrm{E}}[e] =0$. Consequently, the equations of motion will be symmetric, and this was shown in the previous chapters, again see equation~(\ref{EH_tetrad_sym}). Indeed this is the case as $G_{[\mu \nu]} =0$ identically. For modified theories, where the action is no longer pseudo-invariant, this will not hold identically. We also refer back to Chapter~\ref{chapter3}, where we look at other formulations where the variations with respect to the tetrad were more subtle (such as in the first-order Palatini formalism). In these cases, one must be careful about the precise statements being made with regards to local Lorentz invariance. In the following chapters we will make sure to keep these subtleties in mind and be explicit with regards to the independent dynamical variables of the theory. Here, we simply have the tetrad $S_{\textrm{E}}[e]$ and so the outcome is clear.

In Section~\ref{section4.1.4} we looked at the pseudotensors in connection with the (metric) Einstein action. That discussion can also be applied here in the same way, defining the gravitational canonical energy-momentum pseudotensor for the tetradic Einstein action using~(\ref{grav_pseudo}). This leads to an equivalent conserved Noether current~(\ref{Noether conserved}), and the analysis largely follows in the same way. The key difference is that the pseudotensors are now non-covariant with respect to  local Lorentz transformations instead of spacetime diffeomorphisms. For further details, see~\cite{moller1961conservation}.

\subsection{Comparison between coordinate and orthonormal bases}
\label{section4.2.4}
To conclude this section, let us now look at comparing the tetradic bulk and boundary terms with their coordinate basis counterparts. This can be achieved by simply subtracting the boundary terms. Writing $\mathfrak{B}$ in terms of the spacetime connection
\begin{align}
\mathfrak{B} &= \frac{2}{e} \partial_{\nu} \big(e g^{\lambda [\mu } e_{b}{}^{\nu]} e^{c}{}_{\lambda} \mathring{\omega}_{\mu}{}^{b}{}_{c}  \big) \nonumber \\
&=  \frac{2}{e} \partial_{\nu} \big(e e^{c}{}_{\lambda} g^{\lambda [\mu}  \partial_{\mu} e_{c}{}^{\nu]} \big) +  \frac{2}{\sqrt{-g}} \partial_{\nu} \big(\sqrt{-g} g^{\lambda [\mu} \Gamma^{\nu]}_{\mu \lambda} \big) \, ,
\end{align}
where we have used the standard definitions of the spin connection. Now we immediately notice the final term as the metrical boundary term $\ourB$~(\ref{B}), so we can define the difference as
\begin{equation}
\ourB- \mathfrak{B} =  - \frac{2}{e} \partial_{\nu} \big(e e^{c}{}_{\lambda} g^{\lambda [\mu}  \partial_{\mu} e_{c}{}^{\nu]} \big) =: \mathbb{B} \, .
\end{equation}

This term is itself a new boundary term, which we identified in our work~\cite{Boehmer:2021aji}. There we also showed that the term in fact only depends on first derivatives of the tetrad, despite appearing to be second-order. This is not too difficult to show by expanding out the derivative and using the inherent symmetry properties
\begin{align} \label{Bnew2}
\mathbb{B} &= \frac{2}{e} \partial_{\nu} \big(e g^{\lambda [\mu} e_{c}{}^{\nu]} \partial_{\mu} e^{c}{}_{\lambda} \big) \\
&= \frac{2}{e} \partial_{\nu} \big(e g^{\lambda [\mu} e_{c}{}^{\nu]} \big)  \partial_{\mu}e^{c}{}_{\lambda} +  2 g^{\lambda [\mu} e_{c}{}^{\nu]}(\partial_{\nu} \partial_{\mu}  e^{c}{}_{\lambda} ) \\ 
&=  \frac{2}{e} \partial_{\nu} \big(e g^{\lambda [\mu} e_{c}{}^{\nu]} \big)  \partial_{\mu}e^{c}{}_{\lambda}  \, ,
\end{align}
with the last term on the middle line vanishing due to the contraction of antisymmetric and symmetric terms. We note that the term was also identified in a much earlier work\footnote{At the time of our work~\cite{Boehmer:2021aji} we were unaware of the previous appearance of this boundary term, mainly because we were working in the context of modified teleparallel theories. In Sec.~\ref{section4.3} we clarify how these different areas overlap.}~\cite{Charap:1982kn}, where the authors studied the role of surface terms in General Relativity.

The fact that this term can be written both as a total derivative and as a first-order term is just a logical necessity. The difference between two boundary terms will be a boundary term, but here it is also the difference between the two bulk terms $\ourG$ and $\mathfrak{G}$. This can be calculated explicitly, but it is easier just to note that by definition
\begin{align}
R = \ourG + \ourB &= \mathfrak{G} + \mathfrak{B} \\
\implies \ourB - \mathfrak{B} &=  \mathfrak{G} - \ourG \, ,
\end{align}
which must be true because the Ricci scalar is invariant.
Therefore it  also follows that the difference must only depend on first derivatives of the fields by consistency, and we have
\begin{equation} \label{Bnew}
\mathfrak{G} - \ourG = \mathbb{B} \, .
\end{equation}
This is similar to the previous discussions on the local Lorentz transformations, where the inhomogeneous term took a similar form. We also explicitly showed that it was both first-order in derivatives and could be written as a total derivative~(\ref{M_tetrad_boundary}), completely analogous to~(\ref{Bnew2}).

Another interesting feature of this term is that it is neither a coordinate scalar nor a local Lorentz scalar. This again follows as a logical necessity, because $\ourG$ and $\mathfrak{G}$ are only pseudo-invariant under these transformations respectively. In fact, it follows from~(\ref{Bnew}) that it transforms with exactly the sum of the inhomogeneous terms of both bulk terms
\begin{align}
\delta_{\xi} \mathbb{B} &= M^{\mu \nu}{}_{\gamma} \partial_{\mu} \partial_{\nu} \xi^{\gamma} \, \\
\delta_{\Lambda} \mathbb{B} &= \mathfrak{M}^{\mu}{}_{a}{}^{b} \partial_{\mu} \epsilon_{b}{}^{a} \, ,
\end{align}
where $\delta_{\xi}$ and $\delta_{\Lambda}$ represent infinitesimal coordinate and infinitesimal local Lorentz transformations respectively. In our work~\cite{Boehmer:2021aji} we further discuss the dependence of this term on both the chosen coordinates and tetrad frame fields. In the following chapter we revisit this topic in greater detail.

As a last point, one could then go on to write a more general action including this new boundary term as 
\begin{equation} \label{Einstein_general}
S_{\textrm{general}}[e] = \frac{1}{2 \kappa} \int \big( \ourG + \alpha \mathbb{B}) \sqrt{-g} d^4 x =  \frac{1}{2 \kappa} \int \big( \mathfrak{G} + (\alpha-1)\mathbb{B} \big) \sqrt{-g} d^4 x \, ,
\end{equation}
where $\alpha$ is some constant, see~\cite{Charap:1982kn}. For $\alpha = 0$ we retrieve the (metrical) Einstein action~(\ref{Einstein_action}), whereas for $\alpha = 1$ we obtain the tetradic version~(\ref{Einstein_tetrad}). Because $\mathbb{B}$ can be written as a total derivative, clearly the equations of motion for~(\ref{Einstein_general}) will also be the Einstein field equations.

 Similarly, all of the symmetries and (pseudo-)invariances will also be shared by this action, namely, diffeomorphism invariance and local Lorentz invariance. However, the action is not a true scalar for any choice of $\alpha$. In the case where $\alpha = 1$ it is a coordinate scalar, and for $\alpha =0$ it is a local Lorentz scalar. For $\alpha \neq 0,1$ it is neither. The peculiarities of this action would manifest when considering non-linear modifications, and we will discuss this briefly in Sec.~\ref{section5.2.2}. 
 
 To close, we briefly mention an alternative approach that could be taken by instead working with the Ricci scalar in a general anholonomic basis and separating off a boundary term. The Levi-Civita spin connection is then written in terms of the objects of anholonomy~(\ref{spin anholonomy}), and the same steps are repeated as above. The decomposition can be found in equation (A.8) of~\cite{Hehl:1979gp}. It is then possible to retrieve the decomposition in the coordinate and orthonormal bases from this general anholonomic treatment.

\section{Equivalence with teleparallel theories}
\label{section4.3}

Let us now put to use the decompositions of the Ricci scalar to bring to light an equivalence with the teleparallel theories of gravity. First, recall the decomposition of the metric-affine Ricci scalar into its Levi-Civita and non-Riemannian parts~(\ref{R_decomp}). Using the definitions in Sec.~\ref{section3.3} for the metric and symmetric teleparallel scalars, we can write the following fundamental equation in the coordinate and orthonormal bases 
\begin{align} \label{decomp_scalars1}
\bar{R} &= \ourG + \ourB + T - B_T + Q + B_Q + \ourC \, ,\\
\bar{R} &= \mathfrak{G} + \mathfrak{B} + T - B_T + Q + B_Q + \ourC \, . \label{decomp_scalars2}
\end{align}
These equations show all of the contributions to the affine Ricci scalar, decomposed into Levi-Civita, torsion, and non-metricity bulk and boundary terms.
The cross terms, represented by $\ourC$, consist of contractions between the torsion and non-metricity terms and will not be needed, but see~\cite{Boehmer:2021aji} for their explicit form. Also note that only our bulk and boundary terms are not true scalars, so all other terms are independent of the choice of basis.

In the metric teleparallel setting, using the decompositions $T(e,\omega(\Lambda)) = T(e,0) + b_{\Lambda}$ and $B_T(e,\omega(\Lambda)) = B_T(e,0) + b_{\Lambda}$ given in~(\ref{T_decomp}) and~(\ref{B_T_decomp}), the tetradic equation reduces to
\begin{align} \label{TEGR_G_decomp}
0 &= \mathfrak{G} + \mathfrak{B} + T(e,0) + b_{\Lambda} -  B_T(e,0) - b_{\Lambda} \nonumber \\
&= \mathfrak{G} + \mathfrak{B} + T(e,0) -  B_T(e,0) \, .
\end{align}
As we showed before, the terms depending on the teleparallel connection $\omega(\Lambda)$ parameterised by the fields $\Lambda$ cancel with one another, as one would expect for the invariant Levi-Civita Ricci scalar. The fact that the final line of~(\ref{TEGR_G_decomp}) does not contain any terms depending on $\omega(\Lambda$) may be part of the reason that the `covariant approach' was not studied until long after the Weitzenb\"{o}ck one.

An explicit calculation, which can be found in our work\footnote{Actually, in our work we show that $\ourG + \mathbb{B} = -T(e,0)$, which is equivalent to the statement here.}~\cite{Boehmer:2021aji}, verifies that
\begin{equation} \label{G_T_rel}
\mathfrak{G} = - T(e,0) \, , \qquad \mathfrak{B} = B_T(e,0) \, ,
\end{equation}
which we could also write in terms of the non-gauge-fixed torsion terms
\begin{equation} \label{G_T_rel2}
\mathfrak{G} = - T + b_{\Lambda} \, , \qquad \mathfrak{B} = B_T - b_{\Lambda} \, .
\end{equation} 
A simple consistency check can be made by writing the torsional boundary term $B_T$ in terms of our tetradic one $\mathfrak{B}$ and the spin connection terms in $b_{\Lambda}$~(\ref{b_Lambda})
\begin{align}
B_T(e,\omega) &= \mathfrak{B} + b_{\Lambda} \nonumber \\
&= \frac{2}{e} \partial_{\mu}(e \mathring{\omega}^{\mu}) - \frac{2}{e} \partial_{\mu}(e \omega^{\mu}) \, .
\end{align}
Here $\omega^{\mu} = \omega_{\nu}{}^{\mu \nu}$ is a general spin connection. Now, in the Levi-Civita limit we have $\omega_{\mu}{}^{a}{}_{b} = \mathring{\omega}_{\mu}{}^{a}{}_{b}$ and the above term vanishes, as we expect.

Returning to equations~(\ref{G_T_rel})-(\ref{G_T_rel2}), we see that the $\mathfrak{G}$ theories are equivalent to the teleparallel theories with $T(e,0)$, and equivalent with $T(e,\omega(\Lambda))$ up to the \textit{gauge} boundary term $b_{\Lambda}$. 
As we mentioned in the previous chapter, in the original formulations of teleparallel gravity, the Weitzenb\"{o}ck connection (or gauge) is used, and so the equivalence is absolute. It is also known that the term $b_{\Lambda}$ has no dynamics, and can be removed via a frame transformation~\cite{RA2013}; its only role is to restore the covariance of the theory~\cite{Krssak:2018ywd}. Also recall that in TEGR, local Lorentz invariance was a kinematic symmetry~\cite{Pereira:2019woq}. Hence TEGR and the tetradic Einstein action are fully equivalent (at least classically).

For the symmetric teleparallel theories, using~(\ref{Q_decomp}) and~(\ref{B_Q_decomp}), we have the decomposition in the metric framework
\begin{align}
0 &= \ourG + \ourB + Q(g,0) - b_{\xi} + B_Q(g,0) + b_{\xi} \nonumber \\
&= \ourG + \ourB + Q(g,0) + B_Q(g,0) \, .
\end{align}
This leads to a similar relation between the decomposed Einstein terms and the gauge-fixed non-metricity terms~\cite{Boehmer:2021aji}
\begin{equation}
\ourG = - Q(g,0) \, , \qquad \ourB = - B_Q(g,0) \, ,
\end{equation}
or equivalently in an arbitrary gauge
\begin{equation}  \label{G_Q_B}
\ourG = - Q - b_{\xi} \, , \qquad \ourB = - B_Q + b_{\xi} \, ,
\end{equation}
see also~\cite{BeltranJimenez:2017tkd}.
Again, the gauge boundary term $b_{\xi}$ can be set to zero via an appropriate coordinate transformation to the coincident gauge~\cite{BeltranJimenez:2017tkd}. It follows that the Einstein action and STEGR are equivalent. This is also well known, and was the reason for first considering symmetric teleparallel theories~\cite{Nester:1998mp}. The fact that the coincident gauge carries no physical significance has been well documented\footnote{We also thank Tomi Koivisto for insightful discussions on this topic.}~\cite{BeltranJimenez:2022azb}. However, we should also point out that this fact is sometimes misunderstood in the literature too. (E.g., some authors assume that by fixing the coincident gauge, some physical solutions may be lost, which of course is not the case. See our previous discussions, and also~\cite{Hohmann:2021ast,Bahamonde:2022zgj} for the correct implementation of this gauge in some modified $f(Q)$ gravity models.) 

Let us also stress that the discussion here presents nothing fundamentally new, see~\cite{BeltranJimenez:2018vdo,BeltranJimenez:2019odq,BeltranJimenez:2017tkd,Golovnev:2021lki}. For example, the relationship between TEGR and the decomposed Ricci scalar in terms of tetrads has long been known. M\o{}ller describes this in~\cite{moller1961conservation}, where he says [about the tetradic formulation using $\mathfrak{G}$]:
\begin{quote}
 \textit{``Space-time is here rather a space of the type considered first by Weitzenb\"{o}ck, although it may always be pictured as a Riemannian space with a built-in tetrad lattice.
This circumstance puts one in mind of Einstein’s old idea of `Fernparallelismus'.''}
\end{quote}

 Conversely, we may interpret the teleparallel theories not in a purely geometric way but as related to these Einstein actions.
For example, we can think of teleparallel gravity as being \textit{defined by} its pseudo-invariance under the fundamental symmetries, diffeomorphisms and local Lorentz transformations. Symmetric teleparallel gravity comes from choosing to work in the natural coordinate basis, whilst metric teleparallel gravity comes from choosing an orthonormal basis. This viewpoint will be useful in the following chapter when we look at taking modifications of these actions.

Moreover, the concept of restoring (strict) symmetries by introducing compensating quantities is a useful tool. In this case, these compensating fields are ultimately related to an affine connection of a teleparallel geometry, $\bar{\Gamma}^{\lambda}_{\mu \nu}(\xi)$ and $\omega_{\mu}{}^{a}{}_{b}(\Lambda)$.
Hence, we naturally obtain a geometric interpretation of these theories by requiring invariance under symmetry groups. More precisely, it has been shown that the Stueckelberg covariantisation of a theory constructed from partial derivatives of the metric $\partial_{\lambda} g^{\mu \nu}$ directly leads to the symmetric teleparallel connection $\partial_{\mu} g^{\mu \nu} \rightarrow \bar{\nabla}_{\lambda} g^{\mu \nu}$, see~\cite{BeltranJimenez:2022azb,Milgrom:2019rtd}. This is very similar to the Stueckelberg covariantisation of the unimodular action, see equation~(\ref{unimodular2}) and the surrounding discussion.

In fact, the result is straightforward to see: under a coordinate transformation (diffeomorphism), derivatives of the metric transform as
\begin{equation}
\partial_{\lambda} g_{\mu \nu} \rightarrow \hat{\partial}_{\lambda} \hat{g}_{\mu \nu} = \frac{\partial x^{\gamma}}{\partial \hat{x}^{\lambda}} \partial_{\gamma} \Big( \frac{\partial x^{\alpha}}{\partial \hat{x}^{\mu}} \frac{\partial x^{\beta}}{\partial \hat{x}^{\nu}}  g_{\alpha \beta} \Big) \, .
\end{equation}
Promoting the gauge parameters to Stueckelberg fields (which transform as scalars) $\hat{x}^{\mu} \rightarrow \xi^{\mu}(x)$ would render the term invariant. Instead, we define the \textit{covariant} (rank-three tensor) object
\begin{equation}
q_{\lambda \mu \nu} =- \frac{\partial \xi^{\gamma}}{\partial x^{\mu}} \frac{\partial \xi^{\kappa}}{ \partial x^{\nu}} \frac{\partial}{\partial x^{\lambda}} \Big(  \frac{\partial x^{\alpha}}{\partial \xi^{\gamma}} \frac{\partial x^{\beta}}{\partial \xi^{\kappa}} g_{\alpha \beta} \Big) \, .
\end{equation}
One can easily verify that under a general coordinate transformation, again with $\xi^{\mu}(x) \rightarrow \hat{\xi}^{\mu}(\hat{x}) = \xi^{\mu}(x)$, this indeed transforms as expected $q_{\lambda \mu \nu}  \rightarrow \frac{\partial x^{\gamma}}{\partial \hat{x}^{\lambda}}  \frac{\partial x^{\alpha}}{\partial \hat{x}^{\mu}}  \frac{\partial x^{\beta}}{\partial \hat{x}^{\nu}}q_{\gamma \alpha \beta} $. Moreover, it is not difficult to see that this object is simply the non-metricity tensor~(\ref{Q_xi}) with the symmetric teleparallel connection
\begin{equation}
q_{\lambda \mu \nu}  = - \partial_{\lambda} g_{\mu \nu} + 2 \frac{\partial x^{\gamma}}{\partial \xi^{\sigma}} g_{\gamma (\nu} \partial_{\mu)} \partial_{\lambda} \xi^{\sigma}  = Q_{\lambda \mu \nu} \, .
\end{equation}
Hence the $\xi$ fields that parameterise the connection in symmetric teleparallel gravity play the role of Stueckelberg fields associated with coordinate invariance~\cite{BeltranJimenez:2019tjy}.

One can instead begin with the Einstein action and apply the Stueckelberg procedure to this directly. This takes a little more work, but follows the exact same principles. An advantage is that there is less ambiguity in how the Stueckelberg procedure should be applied, because we are aiming for an invariant rather than a covariant object. This result exactly gives the non-metricity scalar $Q$ for a symmetric teleparallel connection
\begin{align} \label{Stueck_G}
\ourG \rightarrow \hat{\ourG}(\hat{x}) &
\begin{multlined}[t]
= \ourG + M^{\rho \sigma}{}_{\gamma} \Big(\frac{\partial^2 x^{\gamma}}{\partial \xi^{\mu} \partial  \xi^{\nu}} \frac{\partial  \xi^{\mu} }{\partial x^{\rho }}\frac{\partial  \xi^{ \nu} }{\partial x^{\sigma }} \Big)  
 \\
+2 g^{\alpha \beta} \frac{\partial^2 x^{\kappa}}{\partial  \xi^{\mu} \partial  \xi^{\lambda}} \frac{\partial^2 x^{\gamma}}{\partial  \xi^{\nu} \partial  \xi^{\rho}}  \frac{\partial  \xi^{\mu } }{\partial x^{\alpha}} \frac{\partial \xi^{ \rho} }{\partial x^{\kappa }} \Big(  \frac{\partial  \xi^{[\nu} }{\partial x^{\beta}} \frac{\partial  \xi^{ \lambda]} }{\partial x^{\gamma}} \Big)   
\end{multlined} \nonumber \\
 &= \ourG - M^{\mu \nu}{}_{\lambda} \bar{\Gamma}^{\lambda}_{\mu \nu} + 2 g^{\mu \nu} \bar{\Gamma}^{\lambda}_{\mu [\rho} \bar{\Gamma}_{\nu] \lambda}^{\rho} \nonumber \\
&= \ourG + b_\xi \nonumber \\
 &= - Q(g,\xi) \, ,
\end{align}
where we have applied the Stueckelberg trick in the first line, rewritten this explicitly in terms of the symmetric teleparallel connection on the second line, and then equated this with the boundary term $b_Q$ in the following line.
For the full details of this calculation, see Appendix~\ref{appendixD}. The terms depending on the Stueckelberg fields $\xi^{\mu}$ take the form of the boundary term $b_{\xi}$, hence, they do not exhibit any dynamics. This is a result of $\ourG$ being pseudo-invariant. However, we also see that they actually can be written quadratically in terms of the affine connection, see Appendix~\ref{appendixD} for the proof. In conclusion, we confirm that the covariantisation of the bulk term $\ourG$ leads to exactly the non-metricity scalar $Q$ of a symmetric teleparallel connection\footnote{Note that a very similar covariantisation of the Einstein action has been studied in~\cite{Tomboulis:2017fim}, using an additional reference connection and metric.}. Again, note that the invariance now depends on the presence of the additional $\xi$ fields.

 In the metric teleparallel case we have a similar process for restoring the invariance of the partial derivatives of the tetrad $\partial_{\nu} e_{a}{}^{\mu} \rightarrow \bar{\nabla}_{\nu} e_{a}{}^{\mu}$, see~\cite{Blixt:2022rpl}. Again, the same process can be carried out for the non-covariant tetradic Einstein action $\mathfrak{G}$. When the invariance is restored, this produces the torsion scalar for the teleparallel inertial spin connection. This should not come as any surprise, given the previous discussions.

As an aside, let us mention that we could perform a similar covariantisation of the more general Einstein action defined in~(\ref{Einstein_general}), which was neither a coordinate scalar nor a local Lorentz scalar. This could be accomplished by using a mixed anholonomic basis. The analogous theory in the general teleparallel framework (with both torsion and non-metricity) was recently considered in~\cite{Adak:2023ymc}, also making use of mixed anholonomic frames. The authors also discuss the gauge freedom present in teleparallel theories, which is quite relevant to our discussions above. 

Finally, we want to emphasise that the work here shows that it is possible to formulate the teleparallel theories \textit{completely without referring to alternative geometries}. We saw that the $\ourG$ and $\mathfrak{G}$ theories can be covariantized to \textit{exactly} give the invariant teleparallel scalars $Q$ and $T$, with no boundary terms needed. This relationship between the decomposition of the standard (Levi-Civita) Einstein-Hilbert action and the geometric teleparallel theories seems quite remarkable to us.  In their modified version, $f(\ourG)$ and $f(\mathfrak{G})$, there will also be an equivalence with the modified $f(Q)$ and $f(T)$ theories. However, this is less obvious due to the non-linear functions of boundary terms, which no longer vanish trivially. We study these modifications in Sec.~\ref{section5.2.2} and make sure to pay special attention to the equations of motion associated with the affine connection of the teleparallel theories.

\section{Metric-affine Einstein action}
\label{section4.4}

In Sections~\ref{section4.2} and~\ref{section4.3} we have worked exclusively in the Levi-Civita framework. Here we will abandon this assumption and consider the fully metric-affine Ricci scalar, see~(\ref{Ricci scalar}). It is important to remember that in metric-affine or Palatini formalism, where the metric and affine connection are considered to be independent, \textit{all} terms are at most first-order in derivatives of the fundamental variables $g$ and $\Gamma$, see Chapter~\ref{chapter3}. Hence there is a priori no special first-order bulk term that can be identified. However, we can still identify a natural boundary term (with respect to some basis). 

Moreover, in the case that matter does not couple to the connection (such as the Palatini formalism of Sec.~\ref{section3.1}), we know that the connection equation of motion can be solved algebraically. The solution is that the connection is related to the first derivatives of the metric~(\ref{connection_sol}), so derivatives of the connection are proportional to \textit{second} derivatives in $g$. We can therefore preemptively decompose the metric-affine Ricci scalar into a bulk term that depends linearly on the connection and a boundary term that depends on its first derivatives.
The added complexity of working in a fully metric-affine geometry will become apparent, but the essence of the calculations are the same. 

Another interesting point to note is that the form of the bulk term will not simply be $\ourG$~(\ref{G}) with Levi-Civita connections replaced by affine connections, as one might naively assume. There would be a number of problems with such a theory, the first of which is the ambiguity in the lower indices of the connection terms. If one were to study such an action of the form
\begin{equation*}
g^{\mu \nu} \big(\bar{\Gamma}^{\lambda}_{\mu \sigma} \bar{\Gamma}^{\sigma}_{\lambda \nu} - \bar{\Gamma}_{\mu \nu}^{\sigma} \bar{\Gamma}_{\lambda \sigma}^{\lambda} \big) \, ,
\end{equation*}
the theory would not reproduce either the Levi-Civita or the metric-affine Einstein field equations. Moreover, the action would not differ from a scalar quantity by a boundary term and hence would lead to non-covariant equations of motion. This was noted long ago in~\cite{el1973so}.

From the previous sections it should be intuitive that we instead want to obtain a pseudo-invariant quantity that gives rise to the same equations of motion as $\bar{R}$. The decomposition will also be different in the coordinate basis to the orthonormal basis, the former being related to diffeomorphisms and the latter to local Lorentz transformations. Here we will focus on the coordinate basis formulation. We follow our recent work~\cite{Boehmer:2023fyl}, beginning with the decomposition into bulk and boundary parts.

\subsection{Metric-affine Ricci scalar decomposition}
\label{section4.4.1}

The Ricci scalar density for a general metric-affine connection $\bar{\Gamma}$ is given by
\begin{align}
  \sqrt{-g}\, \bar{R} = \sqrt{-g}g^{\mu\lambda}
  \bigl(\partial_\kappa \bar{\Gamma}_{\mu\lambda}^\kappa -
  \partial_\mu \bar{\Gamma}_{\kappa\lambda}^\kappa \bigr) + \sqrt{-g}g^{\mu\lambda}
  \bigl(\bar{\Gamma}_{\kappa\rho}^\kappa \bar{\Gamma}_{\mu\lambda}^\rho -
  \bar{\Gamma}_{\mu\rho}^\kappa \bar{\Gamma}_{\kappa\lambda}^\rho
  \bigr) \,.
\end{align}
In order to identify a suitable boundary term, we again apply integration by parts to each of the terms containing a first derivative of the connection. Written out explicitly, we have the two relations
\begin{align}
&\begin{multlined}[b]  \sqrt{-g}g^{\mu\lambda} \partial_\kappa \bar{\Gamma}_{\mu\lambda}^\kappa =
  \partial_\kappa \bigl(\sqrt{-g}g^{\mu\lambda} \bar{\Gamma}_{\mu\lambda}^\kappa\bigr) -
  \sqrt{-g}\, \bar{\Gamma}_{\mu\lambda}^\kappa \partial_\kappa g^{\mu\lambda}  \\
  + \frac{1}{2} \sqrt{-g} g_{\alpha\beta} g^{\mu\lambda} \bar{\Gamma}_{\mu\lambda}^\kappa
  \partial_\kappa g^{\alpha\beta} \,,
\end{multlined} \\
&\begin{multlined}[b] 
\sqrt{-g}g^{\mu\lambda} \partial_\mu \bar{\Gamma}_{\kappa\lambda}^\kappa =
  \partial_\mu \bigl(\sqrt{-g}g^{\mu\lambda} \bar{\Gamma}_{\kappa\lambda}^\kappa\bigr) -
  \sqrt{-g}\, \bar{\Gamma}_{\kappa\lambda}^\kappa \partial_\mu g^{\mu\lambda}
    \\
 +\frac{1}{2} \sqrt{-g} g_{\alpha\beta} g^{\mu\lambda} \bar{\Gamma}_{\kappa\lambda}^\kappa
  \partial_\mu g^{\alpha\beta} \,,
\end{multlined} 
\end{align}
where the first term in each is a boundary term. Note that now we no longer have a metric compatible connection $\bar{\nabla}_{\mu}g_{\nu \lambda} \neq 0$, we do not rewrite derivatives of the metric in terms of connection terms. In fact, doing so would change the fundamental variables of the theory from the set $\{g,\bar{\Gamma} \}$ to $\{g, T,Q\}$ with $T$ and $Q$ representing torsion and non-metricity. Later we will look at formulating the theory with the latter representation.

 Next, we can separate off these boundary terms and arrive at
\begin{multline}
  \sqrt{-g}\, \bar{R} =
  \partial_\kappa \bigl(\sqrt{-g}g^{\mu\lambda} \bar{\Gamma}_{\mu\lambda}^\kappa\bigr) -
  \partial_\mu \bigl(\sqrt{-g}g^{\mu\lambda} \bar{\Gamma}_{\kappa\lambda}^\kappa\bigr) \\ +
  \sqrt{-g}g^{\mu\lambda}
  \bigl(\bar{\Gamma}_{\kappa\rho}^\kappa \bar{\Gamma}_{\mu\lambda}^\rho -
  \bar{\Gamma}_{\mu\rho}^\kappa \bar{\Gamma}_{\kappa\lambda}^\rho\bigr) -
  \sqrt{-g}\, \bar{\Gamma}_{\mu\lambda}^\kappa \partial_\kappa g^{\mu\lambda} +
  \sqrt{-g}\, \bar{\Gamma}_{\kappa\lambda}^\kappa \partial_\mu g^{\mu\lambda} \\+
  \frac{1}{2} \sqrt{-g} g_{\alpha\beta} g^{\mu\lambda} \bar{\Gamma}_{\mu\lambda}^\kappa
  \partial_\kappa g^{\alpha\beta} -
  \frac{1}{2} \sqrt{-g} g_{\alpha\beta} g^{\mu\lambda} \bar{\Gamma}_{\kappa\lambda}^\kappa
  \partial_\mu g^{\alpha\beta} \,.
  \label{eqn:ac1}
\end{multline}
Introducing the notation $\bar{\ourB}$ to denote the new boundary term and $\bar{\ourG}$ to denote the remaining bulk terms, we have the following decomposition
\begin{align} \label{RGB_affine}
   \sqrt{-g} \bar{R} =  \sqrt{-g} \big( \bar{\ourG} + \bar{\ourB} \big) \, ,
\end{align}
with $\bar{\mathbf{G}}$ and $\mathbf{\bar{B}}$ given by
\begin{align}
  \label{Gbar}
  \bar{\mathbf{G}} &:= g^{\mu \lambda} \big( \bar{\Gamma}^{\kappa}_{\kappa \rho} \bar{\Gamma}^{\rho}_{\mu \lambda} - \bar{\Gamma}^{\kappa}_{\mu \rho} \bar{\Gamma}^{\rho}_{\kappa \lambda} \big)
  + \big(\bar{\Gamma}^{\mu}_{\mu \lambda} \delta^{\nu}_{\kappa} - \bar{\Gamma}^{\nu}_{\kappa \lambda} \big) \big( \partial_{\nu} g^{\kappa \lambda} - \frac{1}{2} g_{\alpha \beta} g^{\kappa \lambda} \partial_{\nu} g^{\alpha \beta} \big)\,, \\
  \label{Bbar}
  \mathbf{\bar{B}} &:= \frac{1}{\sqrt{-g}} \partial_{\kappa} \big(\sqrt{-g}(g^{\mu \lambda}\bar{\Gamma}^{\kappa}_{\mu \lambda} - g^{\kappa \lambda} \bar{\Gamma}^{\mu}_{\mu \lambda})\big)\,.
\end{align}
The similarities with the Levi-Civita formulation in~(\ref{G}) and~(\ref{B}) are quite apparent: the boundary term is form equivalent with $\Gamma$ replaced by $\bar{\Gamma}$; the same is not true for the bulk term. Notice that the index combination on the quadratic part of $\bar{\ourG}$~(\ref{Gbar}) differs from $\ourG$~(\ref{G}), and the quadratic parts actually have opposite signs. 

The partial derivative terms in $\bar{\ourG}$ can be written in a more succinct way which makes the comparison more clear
\begin{align} \label{G2bar}
\bar{\ourG} = g^{\mu \lambda} \big( \bar{\Gamma}^{\kappa}_{\kappa \rho} \bar{\Gamma}^{\rho}_{\mu \lambda} - \bar{\Gamma}^{\kappa}_{\mu \rho} \bar{\Gamma}^{\rho}_{\kappa \lambda} \big)  - \frac{1}{2} \partial_{\lambda}g^{\mu \nu} \bar{E}_{\mu \nu}{}^{\lambda} \, ,
\end{align}
where the object $\bar{E}{}_{\mu \nu}{}^{\lambda}$ is defined by
\begin{align}
  \label{Ebar}
  \bar{E}_{(\mu \nu)}{}^{\lambda} := -2 \frac{\partial \ourG}{\partial_{\lambda} g^{\mu \nu} } = 2 \bar{\Gamma}^{\lambda}_{(\mu \nu)} -
  2 \delta^{\lambda}_{(\mu} \bar{\Gamma}^{\rho}_{| \rho |\nu)} - g_{\mu \nu} g^{\kappa \rho} \bar{\Gamma}^{\lambda}_{\rho \kappa} +
  g_{\mu \nu} g^{\kappa \lambda} \bar{\Gamma}^{\rho}_{\rho \kappa} \,.
\end{align}
This is just the metric-affine version of the superpotential used in the Levi-Civita case, see equations~(\ref{E}) and~(\ref{EG1}). Let us also note that equation~(\ref{G2bar}) only determines the symmetric part of $\bar{E}_{\mu \nu}{}^{\lambda}$, hence the explicit symmeterization brackets on~(\ref{Ebar}). Our first field equation is obtained by varying with respect to the metric and will only contain the symmetric part of $E_{\mu \nu}{}^{\lambda}$. However, when considering a generalised Einstein-Cartan type theory, we will be able to also choose its skew-symmetric part in a particular way to make the equations of motion consistent. This will be made more clear in the following.

For the unique torsion-free, metric-compatible, Levi-Civita connection the bulk terms simplify to
\begin{align}
  \bar{\ourG}|_{\bar{\Gamma}\rightarrow \Gamma} = g^{\mu \lambda} \bigl(\Gamma^{\kappa}_{\mu \rho} \Gamma^{\rho}_{\kappa \lambda} -
  \Gamma^{\kappa}_{\kappa \rho}\Gamma^{\rho}_{\mu \lambda}  \bigr) = \ourG  \,,
  \label{oldG}
\end{align}
which is just the Levi-Civita bulk term~(\ref{G}). In the equation above the partial derivatives of the metric in~(\ref{Gbar}) have been rewritten as connection terms. This result is not immediately obvious but can be seen by noting that in this limit
\begin{equation}
 \partial_{\lambda}g^{\mu \nu} \bar{E}_{\mu \nu}{}^{\lambda}|_{\bar{\Gamma} \rightarrow \Gamma}=  \partial_{\lambda}g^{\mu \nu} E_{\mu \nu}{}^{\lambda} = -4 \ourG \, ,
\end{equation}
with the final equality following from equation~(\ref{G3}). This result, along with the knowledge that the quadratic part of $\bar{\ourG}$ has an opposite sign to $\ourG$, leads to the total limit as being $-\ourG + 2 \ourG = \ourG$. 

\subsubsection{The Palatini tensor}

The above decomposition of the Ricci scalar is unique in the sense that there is only one canonical boundary term. However, one can write $\bar{\ourG}$ in a variety of different equivalent ways, as was already done in~(\ref{Gbar}) and~(\ref{G2bar}).
Formulation~(\ref{G2bar}) turns out to be useful as the field equations will contain the object $\bar{E}_{\mu \nu}{}^{\lambda}$.
Another previously used form is given by
\begin{align} \label{G3bar}
  \bar{\mathbf{G}} = g^{\mu \lambda}
  \bigl(\bar{\Gamma}^{\kappa}_{\rho \lambda} \bar{\Gamma}^{\rho}_{\mu \kappa} -
  \bar{\Gamma}^{\kappa}_{\mu \lambda} \bar{\Gamma}^{\rho}_{\rho \kappa} \bigr) +
  \bar{\Gamma}^{\kappa}_{\mu \lambda}  P^{\mu \lambda}{}_{\kappa} \,,
\end{align}
where one has to pay special attention to the index positions in the quadratic part compared to~(\ref{Gbar}) and~(\ref{G3bar}). This formulation can be found in many of the works by Hehl~\cite{Hehl:1978a,Hehl:1978b,HehlKerlickHeyde+1976+111+114,HehlKerlickHeyde+1976+524+527,HehlKerlickHeyde+1976+823+827}. Here $P^{\mu \lambda}{}_{\kappa}$ is the \textit{Palatini tensor} 
\begin{multline} \label{Palatini}
  P^{\mu \nu}{}_{\lambda} = -Q_{\lambda}{}^{\mu \nu} +  \frac{1}{2} g^{\mu \nu} Q_{\lambda \rho}{}^{\rho} + \delta_{\lambda}^{\mu} Q_{\rho}{}^{\rho \nu} -
  \frac{1}{2} \delta^{\mu}_{\lambda} Q^{\nu}{}_{\rho}{}^{\rho} \\  + T^{\mu}{}_{\lambda}{}^{\nu} + g^{\mu \nu} T^{\rho}{}_{\rho \lambda} +
  \delta^{\mu}_{\lambda} T^{\rho \nu}{}_{\rho} \, ,
\end{multline} 
which we originally introduced in equation~(\ref{Palatini0}). This object is a tensor, as can be seen from its definition in terms of other tensors, but clearly its contraction with the affine connection leads to a non-covariant term.

Let us show how one arrives at~(\ref{G3bar}) starting from our original affine bulk term $\bar{\ourG}$. From the definitions of torsion and non-metricity we have the relations
\begin{equation}
\partial_{\lambda} g^{\mu \nu} = Q_{\lambda}{}^{\mu \nu} - 2 \bar{\Gamma}{}^{(\mu}_{\lambda \eta} g^{\nu) \eta} \, , \qquad \bar{\Gamma}{}^{\mu}_{\nu \lambda} = T^{\mu}{}_{\nu \lambda} +  \bar{\Gamma}{}^{\mu}_{\lambda \nu} \, ,
\end{equation}
from which we obtain
\begin{align}
\bar{\ourG} &= g^{\mu \lambda} \big( \bar{\Gamma}{}^{\kappa}_{\kappa \rho} \bar{\Gamma}{}^{\rho}_{\mu \lambda} - \bar{\Gamma}{}^{\kappa}_{\mu \rho} \bar{\Gamma}{}^{\rho}_{\kappa \lambda} \big) + \big(\bar{\Gamma}{}^{\mu}_{\mu \lambda} \delta^{\nu}_{\kappa} - \bar{\Gamma}{}^{\nu}_{\kappa \lambda} \big) \big( \partial_{\nu} g^{\kappa \lambda} - \frac{1}{2} g_{\alpha \beta} g^{\kappa \lambda} \partial_{\nu} g^{\alpha \beta} \big)  \nonumber \\
&= g^{\mu \lambda} \big( \bar{\Gamma}{}^{\kappa}_{\kappa \rho} \bar{\Gamma}{}^{\rho}_{\mu \lambda} - \bar{\Gamma}{}^{\kappa}_{\mu \rho} \bar{\Gamma}{}^{\rho}_{\kappa \lambda} \big) + 2 g^{\mu \lambda} \big( \bar{\Gamma}{}^{\kappa}_{\rho \lambda} \bar{\Gamma}{}^{\rho}_{\mu \kappa} - \bar{\Gamma}{}^{\kappa}_{\mu \lambda} \bar{\Gamma}{}^{\rho}_{\rho \kappa}  \big)  \nonumber \\
&+ \bar{\Gamma}{}^{\kappa}_{\mu \lambda} \big(\delta^{\mu}_{\kappa} Q_{\rho}{}^{\rho \lambda} -Q_{\kappa}{}^{\mu \lambda}  + \frac{1}{2} g^{\mu \lambda} Q_{\kappa \rho}{}^{\rho} - \frac{1}{2} \delta^{\mu}_{\kappa} Q^{\lambda \rho}{}_{\rho} + T^{\mu}{}_{\kappa}{}^{\lambda} - g^{\mu \lambda} T^{\rho}{}_{\kappa \rho} + \delta^{\mu}_{\kappa} T^{\rho \lambda}{}_{\rho} \big) \nonumber \nonumber \\
&= g^{\mu \lambda} \big( \bar{\Gamma}{}^{\kappa}_{\rho \lambda} \bar{\Gamma}{}^{\rho}_{\mu \kappa} - \bar{\Gamma}{}^{\kappa}_{\mu \lambda} \bar{\Gamma}{}^{\rho}_{\rho \kappa}  \big) + \bar{\Gamma}{}^{\kappa}_{\mu \lambda}  P^{\mu \lambda}{}_{\kappa} \,.
\label{Gbar-nice1}
\end{align}
Note that the quadratic terms have flipped sign in going from the first to the final line. Throughout this section it is important to be aware of the index combinations on the quadratic part: in terms of the metric and affine connection only, such as in~(\ref{Gbar}) and~(\ref{G2bar}), the traced connection terms come first. In terms of the Palatini tensor~(\ref{G3bar}), the traced parts come last. For later use, we will define the quadratic part of the final line of~(\ref{Gbar-nice1}) as
\begin{equation} \label{Gbarquad}
\bar{\ourG}_{\textrm{quad}} := g^{\mu \lambda} \big( \bar{\Gamma}{}^{\kappa}_{\rho \lambda} \bar{\Gamma}{}^{\rho}_{\mu \kappa} - \bar{\Gamma}{}^{\kappa}_{\mu \lambda} \bar{\Gamma}{}^{\rho}_{\rho \kappa}  \big) \, .
\end{equation}
It follows that the quadratic part of the original definition~(\ref{Gbar}) is just minus $\bar{\ourG}_{\textrm{quad}}$, but this additional definition will be useful during the following calculations.

It is also important to be aware that this Palatini form of the bulk term~(\ref{G3bar}) should not be used in variation procedures, because the quadratic part depends on the affine connection whilst the Palatini tensor is written in terms of torsion and non-metricity. We instead will always work directly with the set $\{g,\bar{\Gamma}\}$ where $g$ and $\bar{\Gamma}$ are independent. This removes any confusion when making variations of an action that could include, for example, both the non-metricity tensor $Q_{\lambda \mu \nu}$ and derivatives of the metric, which are not independent.

It turns out that the Palatini tensor can be defined by the variation of $\bar{\ourG}$ with respect to the connection, which we will also show explicitly. Splitting the calculation into two parts, the variation of the quadratic part of $\bar{\ourG}$ and the part containing derivatives of the metric\footnote{Here we are looking at the variation of $\bar{\ourG}$ in the form~(\ref{Gbar}). We noted that the quadratic part has an opposite sign compared with the Levi-Civita bulk term $\ourG$ and the Palatini version in~(\ref{G3bar}). This is the reason for the minus sign in equation~(\ref{Gquad}).}, leads to
\begin{align} \label{Gquad}
-\delta_{\bar{\Gamma}} \bar{\ourG}_{\textrm{quad}} &= \delta \bar{\Gamma}{}^{\gamma}_{\alpha \beta} \Big( \delta^{\alpha}_{\gamma} g^{\mu \lambda} \bar{\Gamma}{}^{\beta}_{\mu \lambda} 
+ g^{\alpha \beta} \bar{\Gamma}{}^{\kappa}_{\kappa \gamma} - g^{\alpha \kappa}\bar{\Gamma}{}^{\beta}_{\gamma \kappa} - g^{\kappa \beta}\bar{\Gamma}{}^{\alpha}_{\kappa \gamma}
\Big ) \,,  \\
\delta_{\bar{\Gamma}} \bar{\ourG}_{\textrm{other}} &=  \delta \bar{\Gamma}{}^{\gamma}_{\alpha \beta} \Big(\delta^{\alpha}_{\gamma} \partial_{\mu}g^{\mu \beta} - \partial_{\gamma} g^{\alpha \beta} +\frac{1}{2} g_{\mu \nu} g^{\alpha \beta} \partial_{\gamma}g^{\mu \nu} - \frac{1}{2} \delta^{\alpha}_{\gamma} g_{\mu \nu} g^{\kappa \beta} \partial_{\kappa} g^{\mu \nu} \Big ) \,. \label{Gother}
\end{align}
Rewriting the partial derivatives of the metric in terms of non-metricity and connection terms, along with some cancellations, allows us to write the total variation as 
\begin{align}
\delta_{\bar{\Gamma}} \bar{\ourG} &=  -\delta_{\bar{\Gamma}} \bar{\ourG}_{\textrm{quad}} + \delta_{\bar{\Gamma}} \bar{\ourG}_{\textrm{other}}  \nonumber  \\
&\begin{multlined}= \delta \bar{\Gamma}{}^{\gamma}_{\alpha \beta} \Big[ \delta^{\alpha}_{\gamma} Q_{\lambda}{}^{\lambda \beta} - Q_{\gamma}{}^{\alpha \beta} + \frac{1}{2}g^{\alpha \beta} Q_{\gamma}{}^{\lambda}{}_{\lambda} - \frac{1}{2} \delta^{\alpha}_{\gamma} Q^{\beta \lambda}{}_{\lambda}  \\
+ \delta^{\alpha}_{\gamma} g^{\mu \beta} \big( \bar{\Gamma}{}^{\lambda}_{\mu \lambda} -  \bar{\Gamma}{}^{\lambda}_{\lambda \mu} \big) + g^{\alpha \beta}\big( \bar{\Gamma}{}^{\lambda}_{\lambda \gamma} - \bar{\Gamma}{}^{\lambda}_{\gamma \lambda}\big) + g^{\mu \beta} \big( \bar{\Gamma}{}^{\alpha}_{\gamma \mu} -  \bar{\Gamma}{}^{\alpha}_{\mu \gamma} \big)
\Big]
\end{multlined} \nonumber \\
&\begin{multlined}[b]= \delta \bar{\Gamma}{}^{\gamma}_{\alpha \beta} \Big[ \delta^{\alpha}_{\gamma} Q_{\lambda}{}^{\lambda \beta} - Q_{\gamma}{}^{\alpha \beta} + \frac{1}{2}g^{\alpha \beta} Q_{\gamma}{}^{\lambda}{}_{\lambda} - \frac{1}{2} \delta^{\alpha}_{\gamma} Q^{\beta \lambda}{}_{\lambda}  \\
+ \delta^{\alpha}_{\gamma} T^{\lambda \beta}{}_{\lambda} + g^{\alpha \beta} T^{\lambda}{}_{\lambda \gamma} + T^{\alpha}{}_{\gamma}{}^{\beta}  \Big] \,, \end{multlined} 
\end{align}
where in the final line we used the definition of the torsion tensor.
We recognise this object in the brackets to be the Palatini tensor~(\ref{Palatini}), hence
\begin{equation} \label{GbarGammavar}
\frac{\partial \bar{\ourG}}{\partial \bar \Gamma_{\mu \nu}^{\lambda}} = P^{\mu \nu}{}_{\lambda} \, .
\end{equation}

The fact that this is tensorial, unlike our previously introduced objects, deserves some comments. It can be understood for a few reasons: firstly, the metric and connection are treated independently; next, the bulk term is \textit{linear} in the connection; lastly, the connection equations of motion coming from $\delta / \delta \bar{\Gamma}$ (which will be tensorial\footnote{The equations of motion will be tensorial because the action is (quasi-)diffeomorphism invariant. This is proven shortly, where we see that $\bar{\ourG}$ is a pseudo-scalar.}) are derived solely from the bulk term. It therefore follows that in this case $\partial / \partial \bar{\Gamma} = \delta / \delta \bar{\Gamma}$, and the resulting equation~(\ref{GbarGammavar}) is a tensor. In the Levi-Civita case with $\ourG$ the connection was not independent $\delta \bar{\Gamma} \propto \delta \nabla g$, so these arguments did not hold. As for the affine superpotential $\bar{E}$ in~(\ref{Ebar}), we varied with respect to the partial derivative of the metric, holding the undifferentiated metric fixed. Hence this object also is not a tensor. This short discussion should clarify some of the differences between the metric-affine and the purely metric formulations, as well as the importance of the Palatini tensor.

Lastly, let us look at the limit of vanishing non-metricity. In this case we can swap the indices of the connection terms to introduce torsion tensors that cancel terms in the Palatini tensor
\begin{align}
  \bar{\ourG} &= g^{\mu \lambda} \big( \bar{\Gamma}{}^{\kappa}_{\rho \lambda} \bar{\Gamma}{}^{\rho}_{\mu \kappa} - \bar{\Gamma}{}^{\kappa}_{\mu \lambda} \bar{\Gamma}{}^{\rho}_{\rho \kappa}  \big) + \bar{\Gamma}{}^{\kappa}_{\mu \lambda}  P^{\mu \lambda}{}_{\kappa} \nonumber \\
& \stackrel{Q=0}{=} g^{\mu \lambda} \big( \bar{\Gamma}{}^{\kappa}_{\rho \lambda} \bar{\Gamma}{}^{\rho}_{\kappa \mu} - \bar{\Gamma}{}^{\kappa}_{\mu \lambda} \bar{\Gamma}{}^{\rho}_{\kappa \rho}  \big) - \bar{\Gamma}{}^{\lambda}_{\lambda \mu} T^{\rho}{}_{\rho}{}^{\mu} \,.
\end{align}
This is the form of the bulk term in an Einstein-Cartan space, see for example~\cite{FH1976}. When torsion also vanishes the final term is zero and we obtain $\ourG$.

\subsection{Coordinate and projective transformations}
\label{section4.4.2}
\subsubsection{Coordinate transformations}
Let us again study the coordinate transformations for these metric-affine bulk and boundary terms. Here we will go directly to stating the Lie derivatives, with the calculations detailed in Appendix~\ref{appendixB.2}. For clarity, we restate the Lie derivatives of the metric and affine connection here
\begin{align} \label{Lie_g_P}
 \mathcal{L}_{\xi} g_{\mu \nu} &= \xi^{\lambda} \partial_{\lambda} g_{\mu \nu} + \partial_{\mu} \xi^{\lambda} g_{\lambda \nu} + \partial_{\nu} \xi^{\lambda} g_{\lambda \mu} = 2\nabla_{(\mu} \xi_{\nu)} \, ,
 \\
 \mathcal{L}_{\xi} \bar{\Gamma}^{\lambda}_{\mu \nu} &=
\xi^{\rho} \partial_{\rho}  \bar{\Gamma}^{\lambda}_{\mu \nu} - \partial_{\rho} \xi^{\lambda}  \bar{\Gamma}^{\rho}_{\mu \nu} + \partial_{\mu} \xi^{\rho}  \bar{\Gamma}^{\lambda}_{\rho \nu} + \partial_{\nu} \xi^{\rho}  \bar{\Gamma}^{\lambda}_{\mu \rho} + \partial_{\mu} \partial_{\nu} \xi^{\lambda} \nonumber \\
&=\bar{\nabla}_{\mu} \bar{\nabla}_{\nu} \xi^{\lambda} + \xi^{\rho} \bar{R}_{\rho \mu \nu}{}^{\lambda} - \bar{\nabla}_{\mu}(T^{\lambda}{}_{\nu \rho} \xi^{\rho}) \, .   \label{Lie_Gamma_P}
\end{align}
 
It is most direct to work with the Palatini form of $\ourG$~(\ref{G3bar}), which is valid for these calculations. Using the relations above, following the same methods as in the previous sections, we find
\begin{align} \label{inf_Gbar}
\mathcal{L}_{\xi} \bar{\ourG} &=  \xi^{\mu} \partial_{\mu} \bar{\ourG} +\partial_{\mu} \partial_{\nu} \xi^{\lambda} \big(\bar{M}^{\mu \nu}{}_{\lambda} + P^{\mu \nu}{}_{\lambda}\big) \, , \\
 \mathcal{L}_{\xi} \bar{\ourB} &= \xi^{\mu} \partial_{\mu} \bar{\ourB} -\partial_{\mu} \partial_{\nu} \xi^{\lambda} \big(\bar{M}^{\mu \nu}{}_{\lambda} + P^{\mu \nu}{}_{\lambda}\big) \, , \label{inf_Bbar}
\end{align}
where $\bar{M}$ is defined by the variation of the quadratic part of $\bar{\ourG}$~(\ref{Gbarquad}) with respect to the connection
\begin{align} \label{Mbar}
\bar{M}^{\mu \nu}{}_{\lambda} := \frac{\partial \bar{\ourG}_{\textrm{quad}} }{\partial \bar{\Gamma}_{\mu \nu}^{\lambda} } =   g^{\rho \mu} \bar{\Gamma}^{\nu}_{\lambda \rho} +  g^{\rho \nu} \bar{\Gamma}^{\mu}_{\rho \lambda } -  g^{\mu \nu} \bar{\Gamma}^{\rho}_{\rho \lambda }- g^{\rho \sigma} \delta^{\mu}_{\lambda} \bar{\Gamma}^{\nu}_{\rho \sigma} \,.
\end{align}
The explicit calculations are given in Appendix~\ref{appendixB.2}.

It turns out that the sum of this new object and the Palatini tensor is independent from the affine connection altogether. In fact, it is just equal to the unsymmetrised Levi-Civita object $M^{\mu \nu}{}_{\lambda}$, with all of the torsion and non-metricity terms in $\bar{\Gamma}$ cancelling 
\begin{equation}
\bar{M}^{\mu \nu}{}_{\lambda} + P^{\mu \nu}{}_{\lambda} = g^{\rho \mu} \Gamma^{\nu}_{\lambda \rho} +  g^{\rho \nu} \Gamma^{\mu}_{\rho \lambda } -  g^{\mu \nu} \Gamma^{\rho}_{\rho \lambda }- g^{\rho \sigma} \delta^{\mu}_{\lambda} \Gamma^{\nu}_{\rho \sigma} \, ,
\end{equation}
again, see Appendix~\ref{appendixB.2} for the proof.
Note that the final term is not symmetric under $\mu \leftrightarrow \nu$.
In the transformations above, however, this combination appears contracted with the partial derivatives of $\xi$ and is symmetrised over its first two indices. It follows that we can rewrite the Lie derivatives in terms of the Levi-Civita version $M^{\mu \nu}{}_{\lambda}$ defined in~(\ref{M})
\begin{align} \label{Lie_Gbar}
\mathcal{L}_{\xi} \bar{\ourG} &=  \xi^{\mu} \partial_{\mu} \bar{\ourG} + M^{\mu \nu}{}_{\lambda}  \partial_{\mu} \partial_{\nu} \xi^{\lambda}  \, , \\ 
 \mathcal{L}_{\xi} \bar{\ourB} &= \xi^{\mu} \partial_{\mu} \bar{\ourB} - M^{\mu \nu}{}_{\lambda}  \partial_{\mu} \partial_{\nu} \xi^{\lambda} \, . \label{Lie_Bbar}
\end{align}
The affine connection no longer appears in the inhomogeneous parts of the pseudo-scalar transformation. This rather odd result reflects the fact that this decomposition, despite being in the metric-affine framework, solely depends on the choice of coordinates only. In fact, in Sec.~\ref{section4.4.4} we decompose the metric-affine bulk and boundary terms into their Levi-Civita and contortion parts, and see that $\bar{\ourG}$ must necessarily transform in the same way as $\ourG$. The Lie derivatives given in equations~(\ref{Lie_Gbar})-(\ref{Lie_Bbar}) then immediately follow.

\subsubsection{Projective transformations}
In the previous Levi-Civita cases, the connection depends only on the metric. Now that we have an independent affine connection, it is important to study the projective properties of our quantities. In Appendix~\ref{appendixA} we give an introduction and overview of projective transformations. Moreover, in Sec.~\ref{section3.1} we saw that a consequence of the projective invariance of the Ricci scalar was that the Palatini tensor was trace free $P^{\mu \nu}{}_{\nu} =0$. It is therefore important to consider how the objects $\bar{\ourG}$ and $\bar{\ourB}$ transform. 

The generalised projective transformation~(\ref{Projective}) takes the form
\begin{align}
  \label{Projective2}
  \bar{\Gamma}{}^{\gamma}_{\alpha \beta} \rightarrow \bar{\Gamma}{}^{\gamma}_{\alpha \beta} + c_1 \delta_{\alpha}^{\gamma} P_{\beta} + c_2 \delta_{\beta}^{\gamma} P_{\alpha} \,,
\end{align}
with $c_1=0$ corresponding to the standard projective transformations preserving geodesic motion. In Appendix~\ref{appendixA} we show that the Ricci scalar transforms as
\begin{equation} \label{Proj_R}
\bar{R} \rightarrow \bar{R}  + 2 c_1 \bar{\Gamma}{}^{\gamma}_{[\gamma \mu]} P^{\mu} +  c_1 (1-n) \bar{\nabla}^{\mu} P_{\mu} + c_1^2 (n-1) P^{\mu} P_{\mu} \, ,
\end{equation}
where $n$ is the number of spacetime dimensions. This is invariant for projective transformations with $c_1=0$.

A direct calculation shows that the affine bulk term transforms as
\begin{multline}
\bar{\mathbf{G}} \rightarrow \bar{\mathbf{G}} + c_1 \Big[ 2 \bar{\Gamma}{}^{\kappa}_{[\kappa \mu]} P^{\mu} + (n-1) \big(g^{\mu \lambda}  \bar{\Gamma}{}^{\rho}_{\mu \lambda} P_{\rho} +P_{\lambda} \partial_{\nu} g^{\nu \lambda} - \frac{1}{2} P^{\nu} g_{\alpha \beta} \partial_{\nu} g^{\alpha \beta} \big)  \\ 
 + c_1 (n-1) P^{\lambda} P_{\lambda}
\Big] \, .
\end{multline}
This can also be rewritten explicitly in terms of non-metricity and torsion\footnote{Here if one works with the Levi-Civita connection, we see only the $P^{\lambda}P_{\lambda}$ term remains. This, along with the total divergence term $\nabla_{\kappa} P^{\kappa}$  in the $\mathbf{B}$ transformation below, immediately gives the Schouten result for the Levi-Civita Ricci scalar~(\ref{Projective_Levi3}).}
\begin{multline}
  \bar{\mathbf{G}} \rightarrow \bar{\mathbf{G}} + c_1 \Big[ (2-n) T^{\kappa}{}_{\kappa \mu} P^{\mu} + (n-1) P_{\lambda}\big(\bar{\nabla}_{\nu}g^{\nu \lambda} - \frac{1}{2} g_{\alpha \beta} \bar{\nabla}^{\lambda} g^{\alpha \beta}\big) \\
  + c_1 (n-1) P^{\lambda} P_{\lambda} \Big] \,.
\end{multline}
For the boundary term $\bar{\mathbf{B}}$ we find
\begin{align}
  \bar{\mathbf{B}} \rightarrow \bar{\mathbf{B}} + c_1 (1-n)\frac{1}{\sqrt{-g}} \partial_{\kappa}(\sqrt{-g} P^{\kappa}) \,.
\end{align}
Both of these affine quantities share the projective properties of the Ricci scalar, being invariant for transformations with $c_1=0$
\begin{align} \label{Gproj}
\bar{\mathbf{G}} & \rightarrow \bar{\mathbf{G}} \, ,  \\
\bar{\mathbf{B}} & \rightarrow \bar{\mathbf{B}} \, .
\label{Broj}
\end{align}
 From this we can deduce that any action comprised of these objects will share this invariance, see~\cite{Sotiriou:2006qn}.

For consistency, we show that the above transformations give the correct transformation for the Ricci scalar. First, expanding the boundary term transformation gives
\begin{align}
\bar{\mathbf{B}} \rightarrow \bar{\mathbf{B}} + c_1 (1-n)\big(\partial_{\kappa} P^{\kappa}-\frac{1}{2} g_{\alpha \beta} P^{\kappa} \partial_{\kappa} g^{\alpha \beta} ) \,,
\end{align}
which then leads to
\begin{align}
\bar{\mathbf{G}} + \bar{\mathbf{B}} &\rightarrow \bar{\mathbf{G}} + \bar{\mathbf{B}}   + c_1 \Big[ 2\bar{\Gamma}{}^{\kappa}_{[\kappa \mu]} P^{\mu} + c_1 (n-1) P^{\lambda} P_{\lambda} \nonumber
 \\
&+ (n-1) \big(g^{\mu \lambda} \bar{\Gamma}{}^{\rho}_{\mu \lambda} P_{\rho} + P_{\lambda} \partial_{\nu} g^{\nu \lambda} - \frac{1}{2} P^{\nu} g_{\alpha \beta} \partial_{\nu} g^{\alpha \beta} + \frac{1}{2} g_{\alpha \beta} P^{\kappa} \partial_{\kappa} g^{\alpha \beta} - \partial_{\kappa} P^{\kappa} ) 
\Big]  \nonumber \\
&= \bar{\mathbf{G}} +\bar{\mathbf{B}}  + c_1 \Big[ 2\Gamma^{\kappa}_{[\kappa \mu]} P^{\mu} + c_1 (n-1) P^{\lambda} P_{\lambda} + (n-1) \big(g^{\mu \lambda} \bar{\Gamma}{}^{\rho}_{\mu \lambda} P_{\rho} - \partial^{\kappa} P_{\kappa} ) 
\Big] \nonumber \\
&=  \bar{\mathbf{G}} +\bar{\mathbf{B}}  + 2 c_1 \bar{\Gamma}{}^{\kappa}_{[\kappa \mu]} P^{\mu} +  c_1^2 (n-1) P^{\lambda} P_{\lambda} + c_1 (1-n)  \bar{\nabla}^{\mu} P_{\mu} \,,
\end{align}
which matches the result for the Ricci scalar in equation~(\ref{Proj_R}).

The projective invariance of the action will be important when considering the field equations in both the unmodified and the modified theories. It also brings with it some potential problems relating to the uniqueness of solutions and indeterminacy of the affine connection, which we discussed in the previous chapter, see~\cite{hehl1978metric,hehl1981metric,Julia:2000er}. We will see the consequences of this symmetry later on.

\subsection{Action and diffeomorphism invariance}
\label{section4.4.3}

Beginning with the metric-affine Einstein-Hilbert action $\bar{R}$, we can once again use our new decomposition $\bar{R} = \bar{\ourG} + \bar{\ourB}$ to extract the bulk part of the action. We then define the \textit{metric-affine Einstein action} as \begin{align}
  \label{Einstein_actionP}
  S_{\bar{\textrm{E}}}[g,\bar{\Gamma}] = \frac{1}{2\kappa} \int \sqrt{-g}\, \bar{\ourG} \, d^4 x \,.
\end{align}
This of course differs from the metric-affine Einstein-Hilbert action~(\ref{S_Palatini_tetrad}) by the boundary term $\bar{\ourB}$, and so will lead to equivalent equations of motion. 
Variations with respect to the metric and the connection lead to 
\begin{align}
  \label{Einstein_var}
  \delta   S_{\bar{\textrm{E}}} = \frac{1}{2\kappa} \int \sqrt{-g} \Big[
    \delta g^{\mu \nu} \bar{G}_{\mu \nu} + \delta \bar{\Gamma}^{\lambda}_{\mu \nu} P^{\mu \nu}{}_{\lambda}
    \Big] d^4x = 0 \,,
\end{align}
where $\bar{G}_{\mu \nu}$ is the Einstein tensor of a general affine connection, and $P^{\mu \nu}{}_{\lambda}$ is the Palatini tensor. The full details of this derivation can be found in Appendix~\ref{appendixC.1.2}. These calculations will be especially useful in Chapter~\ref{chapter5}, when we consider modified actions.

As was discussed in detail in Sec~\ref{section3.1.1}, the Palatini tensor is trace-free over $P^{\mu \nu}{}_{\nu} = 0$. This was a consequence of the projective invariance of the affine Ricci scalar $\bar{R}$. The invariance of the bulk term $\bar{\ourG}$ leads to the same conclusion, and the connection equation of motion therefore only fixes 60 of the 64 components of the affine connection. In the case of vanishing hypermomentum we obtain the solutions for the connection given in~(\ref{connection_sol}), and we return to GR in the Levi-Civita geometry.
The general field equations are thus
\begin{align}
\bar{G}_{(\mu \nu)} &= \kappa {}^{(\bar{\Gamma})} T_{\mu \nu} \, ,\\
P^{\mu \nu}{}_{\lambda} &= 2 \kappa \Delta^{\mu \nu}{}_{\lambda} \, , \, 
\end{align}
with the symmetric energy-momentum tensor and hypermomentum defined in equations~(\ref{affine_EM}) and~(\ref{hypermomentum}). The dynamics of this theory are then completely equivalent to the metric-affine formulation of the Einstein-Hilbert action studied in the previous chapter.

Let us now discuss the invariances of the metric-affine Einstein action. Firstly, the action is clearly invariant under local Lorentz transformations, as we have made no reference to the tangent space and tetrads whatsoever. Similarly, we showed the projective invariance of the affine bulk term $\bar{\ourG}$ and hence the invariance of the action as a whole. With regards to diffeomorphisms, the metric-affine bulk term transforms infinitesimally as
\begin{equation}
\mathcal{L}_{\xi} \bar{\ourG} = \xi^{\mu} \partial_{\mu} \bar{\ourG} + M^{\mu \nu}{}_{\lambda} \partial_{\mu} \partial_{\nu} \xi^{\lambda} \, .
\end{equation}
This transformation is exactly the same as in the Levi-Civita case, and the final term can be written as a total derivative. 
Applying this to the action leads to
\begin{align}
\delta_{\xi}   S_{\bar{\textrm{E}}}[g,\bar{\Gamma}] &= \frac{1}{2\kappa} \int  \delta_{\xi} (\sqrt{-g}\, \bar{\ourG}) \, d^4 x \nonumber \\
&=  \frac{1}{2 \kappa} \int \partial_{\mu} \partial_{\nu} \big( \sqrt{-g} M^{\mu \nu}{}_{\gamma}\big) \xi^{\gamma} d^4 x  = 0 \, ,
\end{align}
where we have discarded boundary terms depending on $\xi$ and its derivatives and made use of the contracted Bianchi identity~(\ref{Bianchi_M}) in the final equality.

Again, this result should not be surprising given that the action differs from a diffeomorphism invariant one by a boundary term. However, it is perhaps unexpected that this method does not give rise to any new identities besides the standard (Levi-Civita) contracted Bianchi identity. To highlight this, let us take the traditional approach of taking our standard field variations and letting them be generated by an infinitesimal diffeomorphism
\begin{align} \label{diffeo2}
  \delta_{\xi} S_{\bar{\textrm{E}}}[g,\bar{\Gamma}]  &= \frac{1}{2\kappa} \int \Big[
    \delta_{\xi} g^{\mu \nu} \bar{G}_{\mu \nu} + \delta_{\xi} \bar{\Gamma}{}^{\lambda}_{\mu \nu} P^{\mu \nu}{}_{\lambda}
    \Big]  \sqrt{-g} d^4x \nonumber \\
    =  \frac{1}{2\kappa} \int  \Big[ -2 & \nabla_{(\mu}  \xi_{\nu)} \bar{G}{}^{\mu \nu} 
    + P^{\mu \nu}{}_{\lambda} \Big( \bar{\nabla}_{\mu} \bar{\nabla}_{\nu} \xi^{\lambda} + \xi^{\rho} \bar{R}_{\rho \mu \nu}{}^{\lambda} 
    - \bar{\nabla}_{\mu}(T^{\lambda}{}_{\nu \rho } \xi^{\rho}) \Big)  \Big] \sqrt{-g} d^4 x  \nonumber \\
    =  & \  \textrm{`surface terms'} + \frac{1}{2\kappa} \int \xi^{\mu} \Big[ 2\nabla_{\nu} \bar{G}{}^{(\nu}{}_{\mu)}  
    + \frac{1}{\sqrt{-g}} \bar{\nabla}_{\nu} \bar{\nabla}_{\lambda} (\sqrt{-g} P^{\lambda \nu}{}_{\mu})
    \nonumber \\
    + &  \frac{1}{\sqrt{-g}}  \bar{\nabla}_{\nu}(\sqrt{-g} P^{\nu \lambda}{}_{\rho}) T^{\rho}{}_{\lambda \mu} + \bar{R}_{\mu \nu \lambda}{}^{\rho} P^{\nu \lambda}{}_{\rho} \Big] \sqrt{-g}  d^4 x = 0 \, ,
\end{align}
where we have simply inserted the Lie derivative definitions in~(\ref{Lie_g_P}) and~(\ref{Lie_Gamma_P}).
Discarding boundary terms, we are led to the identity 
\begin{multline} \label{identity2}
2 \nabla_{\nu}  \bar{G}{}^{(\nu}{}_{\mu )}   
    + \frac{1}{\sqrt{-g}} \bar{\nabla}_{\nu} \bar{\nabla}_{\lambda} (\sqrt{-g} P^{\lambda \nu}{}_{\mu})
    + \frac{1}{\sqrt{-g}}  \bar{\nabla}_{\nu}(\sqrt{-g} P^{\nu \lambda}{}_{\rho}) T^{\rho}{}_{\lambda \mu} \\ + \bar{R}_{\mu \nu \lambda}{}^{\rho} P^{\nu \lambda}{}_{\rho} \equiv 0 \,,
\end{multline}
which is manifestly covariant\footnote{We hazard the guess that this final identity~(\ref{identity2}) could be equated with the twice-contracted affine Bianchi identity~(\ref{Bianchi}), such that Noether's theorem again leads to a geometric identity as it did in the Levi-Civita case. However, the required algebraic manipulations to verify this equivalence appear tediously long.}. A similar result is obtained in the teleparallel case in~\cite{Hohmann:2021fpr}.

As a last note, we mention that the above analysis could be performed in the orthonormal basis working instead with the tetrad and affine spin connection $\{e,\omega \}$, see for instance~\cite{Maluf:2018coz}. In that case, we would discuss local Lorentz pseudo-invariance. The resulting equations of motion would still be the same, due to the action differing from the metric-affine Einstein-Hilbert action by a boundary term. However, the boundary term would of course be different to the $\bar{\ourB}$ studied here, just as the Levi-Civita boundary terms $\ourB$ and $\mathfrak{B}$ had a non-trivial difference. 

\subsection{Relation to teleparallel gravity}
\label{section4.4.4}

The affine Riemann tensor and its contractions can be decomposed into a sum of Levi-Civita terms and contortion terms, see equations~(\ref{Riemann_decomp})--(\ref{R_decomp}). For the affine Ricci scalar, these contracted contortion terms can be expressed in terms of the geometric scalars of torsion and non-metricity. We made use of this, along with the decomposition of the Ricci scalar into its bulk and boundary terms, in equations~(\ref{decomp_scalars1})-(\ref{decomp_scalars2}). Let us restate the first of those fundamental equations (in the coordinate basis) here for clarity
\begin{align} \label{fundamental}
\bar{R} = \overbrace{\ourG + \ourB}^{R} + \, T - B_T + Q + B_Q + \ourC \,.
\end{align}
Recall that $T$ and $Q$ are the torsion and non-metricity scalars, given in~(\ref{T_scalar}) and~(\ref{Q_scalar}) respectively. The torsion and non-metricity boundary terms $B_T$ and $B_Q$ are given in~(\ref{B_T}) and~(\ref{B_Q}). The cross terms $\ourC$ can be found in~\cite{Boehmer:2021aji}.

One can then show that the affine quantities $\bar{\ourG}$ and $\bar{\ourB}$ are related to the geometric scalars by~\cite{Boehmer:2023fyl}
\begin{align} \label{G_scalars}
  \bar{\ourG} &= \ourG + T + Q + \ourC \,, \\
  \label{B_scalars}
  \bar{\ourB} &= \ourB - B_T + B_Q \,.
\end{align}
The affine bulk term is simply the sum of the Levi-Civita, torsion, and non-metricity bulk terms, and similarly for the affine boundary term.

It is important to remember that the equivalence between the scalars $Q$ and $T$ and the bulk terms $\ourG$ and $\mathfrak{G}$ only holds in the teleparallel framework, where vanishing curvature $\bar{R}_{\lambda \mu \nu}{}^{\gamma}(\bar{\Gamma})=0$ has been enforced. Here we allow for connections with affine curvature $\bar{R}_{\lambda \mu \nu}{}^{\gamma} \neq 0$ and so this equivalence is broken. Hence, a combination like $\ourG + Q$ does not vanish in general in any specific set of coordinates or in any gauge. This marks a significant departure from the analysis in Sec.~\ref{section4.3}.

We also remark that even in the teleparallel case, the affine bulk and boundary terms are not forced to vanish individually. Instead, only their sum needs to vanish $\bar{\ourG} + \bar{\ourB} = 0$. Moreover, they will actually reduce to the \textit{gauge boundary terms} identified in~(\ref{b_Lambda}) and~(\ref{b_xi}). Let us show this in the simple case of a symmetric teleparallel geometry $T=\ourC=0$, in which case equations~(\ref{G_scalars}) and~(\ref{B_scalars}) can be written as
\begin{align} \label{G_scalars_2}
  \bar{\ourG} &= \ourG + Q = - b_{\xi} \,, \\
  \label{B_scalars_2}
  \bar{\ourB} &= \ourB  + B_Q = b_{\xi} \,,
\end{align}
where we have made use of the teleparallel relations~(\ref{G_Q_B}). We therefore see that the requirement for $\bar{\ourG}$ or $\bar{\ourB}$ to vanish is exactly the same as the conditions for $b_{\xi}$ to vanish. This is satisfied in the coincident gauge. 

When we study the modified metric-affine theories based on $\bar{\ourG}$ and $\bar{\ourB}$, we will be working in a geometric setting with non-vanishing curvature, and so the torsion and non-metricity scalars are no longer teleparallel. This opens up a new avenue of gravitational theories to study that will differ from GR more significantly. Instead, these theories will more closely resemble the metric-affine theories with a truly independent and dynamical connection, see Sec.~\ref{section3.2}.

This concludes this section and this chapter, where we have focussed on the different possible decompositions of the Ricci scalar within different geometric frameworks. All of the resulting actions were shown to be dynamically equivalent to their Einstein-Hilbert counterparts and exhibit the same invariant properties. In the Levi-Civita cases, we saw an interesting relationship with the teleparallel equivalents of General Relativity. In the next chapter, we look at modified theories of gravity based on non-linear modifications of these pseudo-invariant actions.

\begin{savequote}[75mm]
There are some scientists who go about preaching that Nature always takes on the simplest solutions. Yet the simplest solution by far would be nothing at all in the Universe. Nature is far more inventive than that, so I refuse to go along thinking it always has to be simple.
\qauthor{`Lectures on Gravitation' (1963) \\ Richard P. Feynman}
\end{savequote}
\chapter{Modified theories of gravity}
\label{chapter5}

In this chapter we study modified theories of gravity based on the non-covariant decompositions of the Einstein-Hilbert action. 
The form of these modifications follows the route taken in $f(R)$ gravity~(\ref{f(R)_action}), allowing our generalised action to depend on an arbitrary function of our decomposed terms. We do not introduce any additional fields and work in $n=4$ spacetime dimensions. Moreover, we only consider minimally coupled matter. However, diffeomorphism invariance or local Lorentz invariance is explicitly broken in these models, leading to a number of interesting consequences. 

In the first section we work in the standard Riemannian framework of General Relativity. This is based on our work~\cite{Boehmer:2021aji}. We take a detailed look at the role of diffeomorphism invariance and the conservation of energy-momentum in these theories. Our modifications can also be shown to be equivalent to the modified teleparallel theories, which we prove in Sec.~\ref{section5.2}. We then propose a unified framework from which one can study all of the previously introduced modified theories. In the final section we move on to considering our non-covariant modifications in the more general metric-affine setting, taking a close look at a modified Einstein-Cartan type theory, following our work~\cite{Boehmer:2023fyl}.

\section{Modifications of the Einstein action}
\label{section5.1}
Inspired by the modifications of $f(R)$, $f(T)$ and $f(Q)$ gravity, we now consider non-linear modifications of the Einstein action~(\ref{Einstein_action}). These take the form $f(\ourG)$, with $\ourG$ corresponding to the bulk part of the Ricci scalar defined in~(\ref{G}), and $f$ representing an arbitrary function. To be even more general, and to make contact with limiting cases such as $f(R)$ gravity, let us propose our action take the following form~\cite{Boehmer:2021aji}
\begin{equation} \label{S_f(g,b)}
S_{\textrm{grav}}[g] = \frac{1}{2\kappa} \int f(\ourG, \ourB) \sqrt{-g} d^4 x \, ,
\end{equation}
where $\ourB$ is the boundary term~(\ref{B}) in the decomposition $R = \ourG + \ourB$. These more general modifications are motivated by the analogous treatment in teleparallel $f(T,B_T)$ gravity~\cite{Bahamonde:2015zma}, where the similar relation  $R= -T+B_T$ holds.

The $f(\ourG,\ourB)$ class of theories encompass both General Relativity and $f(R)$ gravity for specific forms of the function $f$. However, they also give room for deviations by breaking the diffeomorphism symmetry present in both of those two cases. As shown in the previous chapter, the Einstein action given by $f = \ourG$ was diffeomorphism invariant up to boundary terms~(\ref{E_diff}). Similarly, the bulk and boundary terms transformed as pseudo-scalars, see~(\ref{Lie_G}) and~(\ref{Lie_B}).
However, a \textit{non-linear function of a boundary term is no longer a boundary term}. Hence, non-linear modifications will not share the pseudo-diffeomorphism invariance of the bulk and boundary terms $\ourG$ and $\ourB$.

Moving on to the dynamical equations of motion, variations of the action~(\ref{S_f(g,b)}) lead to
\begin{equation}
\delta S_{\textrm{grav}} = \frac{1}{2\kappa} \int \Big( \delta f(\ourG, \ourB) - \frac{1}{2} f(\ourG, \ourB) g_{\rho \sigma} \delta g^{\rho \sigma} \Big) \sqrt{-g} d^4 x \, ,
\end{equation}
where 
\begin{equation}
 \delta f(\ourG, \ourB)  = \frac{\partial f (\ourG, \ourB)}{\partial \ourG} \delta \ourG +  \frac{\partial f (\ourG, \ourB)}{\partial \ourB} \delta \ourB \, .
\end{equation}
Following our work in~\cite{Boehmer:2021aji}, the field equations are calculated to be
\begin{equation} \label{f(g,b)_EoM}
H_{\rho \sigma} = \kappa T_{\rho \sigma} \, ,
\end{equation}
where we define
\begin{multline}
  \label{f(g,b)_EoM2}
  H_{\rho \sigma} :=
     \frac{\partial f}{\partial \ourG} \Big[G_{\rho \sigma} + \frac{1}{2} g_{\rho \sigma} \ourG \Big] + \frac{1}{2}  E_{\rho \sigma}{}^{\gamma} \partial_{\gamma} \Big( \frac{\partial f}{\partial \ourG} \Big)  - \frac{1}{2} g_{\rho \sigma}  f(\ourG,\ourB) \ 
      \\
    + \frac{1}{2} \frac{\partial f}{\partial \ourB}  \ g_{\rho \sigma} \ourB  + g_{\rho \sigma}\partial^{\mu}  \partial_{\mu}\Big( \frac{\partial f}{\partial \ourB}\Big)
      - \partial_{\rho} \partial_{\sigma} \Big( \frac{\partial f}{\partial \ourB}\Big)
    \\
    + \frac{1}{2} g_{\rho \sigma} \partial_{\mu}(g^{\mu \nu}) \partial_{\nu}\Big(\frac{\partial f}{\partial \ourB}\Big) + \frac{1}{\sqrt{-g}} \partial_{(\rho}(\sqrt{-g})\partial_{\sigma)} \Big(\frac{\partial f}{\partial \ourB} \Big)  \, .
\end{multline} 
The superpotential $E_{\rho \sigma}{}^{\gamma}$ is defined in equation~(\ref{E}) and $T_{\rho \sigma}$ is the energy-momentum tensor of a minimally coupled matter action $S_{\textrm{M}}[g,\varPhi^A]$. For the full details of the derivation of these equations of motion, see Appendix~\ref{appendixC.2.1}. Also observe that for $f = \ourG + \ourB$ or $f = \ourG$, the field equations reduce to $G_{\rho \sigma} = \kappa T_{\rho \sigma}$ and GR is recovered as expected.

Let us make a few preliminary comments on these field equations. Firstly, they are of course symmetric $H_{[\rho \sigma]}=0$ because they were computed by varying with respect to the symmetric metric tensor. They are also generically fourth-order in derivatives of the metric, as there are terms with second derivatives of $\ourB$. A key departure from theories such as $f(R)$ gravity is that the gravitational equations~(\ref{f(g,b)_EoM2}) are no longer covariant, following from the transformation properties\footnote{Note that Levi-Civita connection and all of the non-covariant objects $\ourG$, $\ourB$ and $E_{\rho \sigma}{}^{\gamma}$ are still invariant under \textit{global} Poincar\'{e} transformations~(\ref{Poincare}).} of $\ourG$, $\ourB$ and $E_{\rho \sigma}{}^{\gamma}$. We also do not make any assumptions about the invariant properties of the matter action, which we will return to shortly.

The lack of diffeomorphism invariance at the level of the action~(\ref{S_f(g,b)}) implies that we do not automatically have a Bianchi-like identity. This is in contrast to the Einstein action~(\ref{Einstein_action}), and covariant theories such as $f(R)$ gravity~\cite{Koivisto:2005yk} or even $f(T)$ gravity~\cite{Golovnev:2020las}. As we do not (yet) assume our matter action $S_{\textrm{M}}[g,\varPhi^A]$ to be invariant under diffeomorphisms, we cannot infer the usual covariant conservation law $\nabla^{\mu} T_{\mu \nu}=0$. Furthermore, if $S_{\textrm{M}}[g,\varPhi^A]$ is not a scalar then the energy-momentum $T_{\mu \nu}$ may not be a true rank-two tensor, and the equation $\nabla^{\mu} T_{\mu \nu}$ would not be well-defined.

Having said this, let us now make the following observation. If we divide~(\ref{f(g,b)_EoM}) by $\partial f/\partial\ourG$ we can isolate the Einstein tensor $G_{\rho\sigma}$ and move all remaining terms to the right-hand side. This would yield an equation of the form
\begin{align} \label{Einstein_tensor_eq}
  G_{\rho\sigma} = \frac{\kappa}{f_{,\ourG}} T_{\rho\sigma} + T^{(f)}_{\rho\sigma} \,,
\end{align}
where $f_{,\ourG} = \partial f/\partial\ourG$ and $T^{(f)}_{\rho\sigma}$ stands for the collection of all the remaining terms. In this sense, the right-hand side can be thought of as modifications to the effective energy-momentum tensor.
This formulation has two implications: first, mathematical consistency requires that the right-hand side of~(\ref{Einstein_tensor_eq}) must behave as a rank 2 tensor, possibly under some additional conditions; second, the twice contracted Bianchi identities imply that the right-hand side must also be covariantly conserved
\begin{align}
  \nabla^\rho G_{\rho\sigma} = 0 \qquad \Rightarrow \qquad
  \nabla^\rho \Bigl[\frac{\kappa}{f_{,\ourG}} T_{\rho\sigma} +
    T^{(f)}_{\rho\sigma}\Bigr] = 0\,.
\end{align}
Perhaps unexpectedly, we arrive at a conservation equation.

 Note that we did not explicitly assume the diffeomorphism invariance of either of the actions, but in the next section we will show how such a condition has in fact been applied \textit{a posteriori}. We will also clarify, in a more fundamental way, how this conservation equation has emerged.
 
\subsection{Diffeomorphism invariance and conservation equations}
\label{section5.1.1}

Let us take a closer look at the diffeomorphisms of action~(\ref{S_f(g,b)}). As previously noted, the Einstein action was invariant under diffeomorphisms, owing to the fact that the action differed from a coordinate scalar by a boundary term. Clearly the modified action~(\ref{S_f(g,b)}) does not share this property, but it is illustrative to see this explicitly.

Let us proceed to calculate how the modified action transforms under an infinitesimal symmetry transformation. We make use of our previous expressions for the Lie derivatives of the bulk and boundary terms,~(\ref{Lie_G}) and~(\ref{Lie_B}), and find
\begin{align}
  \delta_{\xi} S_{\textrm{grav}}  &= \frac{1}{2\kappa}
  \int \mathcal{L}_{\xi} \big( \sqrt{-g} f(\ourG,\ourB) \big) d^4x
  \nonumber \\ 
  &\begin{multlined}[b]
  = \frac{1}{2\kappa}  \int \biggl[\nabla_{\mu} \xi^{\mu} f(\ourG,\ourB)  \\
 + \Bigl(\frac{\partial f(\ourG,\ourB) }{\partial \ourG} \mathcal{L}_{\xi} \ourG +
  \frac{\partial f(\ourG,\ourB) }{\partial \ourB} \mathcal{L}_{\xi} \ourB \Bigr)\biggr]
  \sqrt{-g}\, d^4x \,.
    \end{multlined}
\end{align}
This leads to
\begin{multline}
  \delta_{\xi} S_{\textrm{grav}}=
  \frac{1}{2\kappa} \int
  \biggl[ \nabla_{\mu} \xi^{\mu} f(\ourG,\ourB) +
    \Bigl(\frac{\partial f(\ourG,\ourB)}{\partial \ourG} \partial_{\mu}\ourG \xi^{\mu} +
    \frac{\partial f(\ourG,\ourB) }{\partial \ourB} \partial_{\mu} \ourB \xi^{\mu}  \Bigr) \\
  + M^{\alpha\beta}{}_{\gamma} \partial_\alpha \partial_\beta \xi^\gamma \Big( \frac{\partial f(\ourG,\ourB) }{\partial \ourG} - \frac{\partial f(\ourG,\ourB) }{\partial \ourB} \Big)\biggr] \sqrt{-g}\, d^4x
  \\
 = \textrm{`surface term'} + \frac{1}{2\kappa}  \int M^{\alpha\beta}{}_{\gamma} \partial_\alpha \partial_\beta \xi^\gamma \bigg( \frac{\partial f(\ourG,\ourB) }{\partial \ourG} - \frac{\partial f(\ourG,\ourB) }{\partial \ourB} \bigg) \sqrt{-g}\, d^4x \,,
 \end{multline} 
 which is analogous to equation~(\ref{E_diff}). Discarding the surface term leaves the following non-trivial contribution
\begin{equation}
  \delta_{\xi} S_{\textrm{grav}} =   \frac{1}{2\kappa}  \int \sqrt{-g} \big(f_{,\ourG}- f_{,\ourB} \big) M^{\alpha\beta}{}_{\gamma} \partial_\alpha \partial_\beta \xi^{\gamma}\, d^4x \,.
\end{equation}
As before, we integrate by parts twice and discard the boundary terms, which leads to
\begin{equation}
  \label{f(G,B)_conservation}
  \delta_{\xi} S_{\textrm{grav}} = \frac{1}{2\kappa} \int \partial_{\alpha} \partial_{\beta}
  \Bigl(\sqrt{-g} M^{\alpha\beta}{}_{\gamma} (f_{,\ourG}- f_{,\ourB} ) \Bigr) \xi^{\gamma}\, d^4 x \,.
\end{equation}
If we also include the matter action $S_{\textrm{M}}$, with matter fields assumed to be on-shell, the total action transforms as
\begin{align} 
  \delta_\xi S_{\textrm{ total}} &= \delta_{\xi} S_{\textrm{grav}} + \delta_\xi S_{\textrm{M}}
  \nonumber \\ &= 
  \int \biggl[
    \frac{1}{2 \kappa}\partial_{\alpha} \partial_{\beta}
    \Bigl(\sqrt{-g} M^{\alpha\beta}{}_{\gamma} (f_{,\ourG}- f_{,\ourB} ) \Bigr)
    - \nabla_\alpha (\sqrt{-g}T^{\alpha}{}_{\gamma})\biggr]\xi^{\gamma}\, d^4 x \,,
  \label{eqn:cons0}
\end{align}
where the covariant derivative of the energy-momentum `tensor' should be understood as implying its usual relation in terms of partial derivatives and connection terms $\partial_{\alpha} T^{\alpha}{}_{\gamma} + \Gamma^{\alpha}_{\alpha \beta} T^{\beta}{}_{\gamma} - \Gamma_{\alpha \gamma}^{\beta} T^{\alpha}{}_{\beta}$. This follows directly from the definition of the Lie derivative of the metric, but we use the notation $\nabla_{\alpha}$ as shorthand.

If at this point one requires the total action to be invariant under coordinate transformations, one finds a conservation equation, namely the vanishing of the integrand of~(\ref{eqn:cons0}). In the GR case, the integrand vanished identically~(\ref{Bianchi_M}). One can again regard this equation as a consequence of Noether's theorem, as it relates to the invariance of the action under transformations of coordinates.

Let us emphasise an important mathematical point that is required for the consistency of the theory. For the variational formulation to be well-defined, the total action must be a pseudo-scalar (a coordinate scalar up to boundary terms). Without this requirement, it seems difficult to construct a well-defined Lagrangian field theory\footnote{For a more formal and geometric discussion regarding the Lagrangian formulation see~\cite{giachetta1997new}. For our purposes, we simply require that the Euler-Lagrange equations be self-consistent.}. We will therefore assume that~(\ref{eqn:cons0}) is forced to vanish $ \delta_\xi S_{\textrm{ total}}=0$ and  show that this is in fact required by consistency.

The implementation of the vanishing of the conservation equation~(\ref{eqn:cons0}) leads to various interesting points to be addressed. First, let us assume that the matter action is itself based on a true scalar Lagrangian, which then implies that the matter energy-momentum tensor is independently conserved $\nabla_\alpha T^{\alpha}{}_{\gamma} =0$. Requiring $\delta_\xi S_{\textrm total} = 0$ then implies the additional equation
\begin{align}
  \partial_{\alpha} \partial_{\beta}
    \Bigl(\sqrt{-g} M^{\alpha\beta}{}_{\gamma} (f_{,\ourG}- f_{,\ourB} )\Bigr) = 0\,.
    \label{eqn:cons1}
\end{align}
 It is possible to show that this conservation equation~(\ref{eqn:cons1}) is in fact completely equivalent to the vanishing `covariant derivative' of the gravitational field equations
\begin{equation} \label{eqn:cons2}
\nabla_{\alpha} H^{\alpha}{}_{\gamma} := \partial_{\alpha} H^{\alpha}{}_{\gamma} + \Gamma^{\alpha}_{\alpha \beta} H^{\beta}{}_{\gamma} - \Gamma^{\beta}_{\alpha \gamma} H^{\alpha}{}_{\beta} = 0 \, ,
\end{equation}
where $H_{\rho \sigma}$ is defined in~(\ref{f(g,b)_EoM2}). A direct but long-winded calculation and expansion of both equations~(\ref{eqn:cons1}) and~(\ref{eqn:cons2}) verifies this fact. Alternatively, one can work with the metric variations of the action directly and let this variation be generated by an infinitesimal diffeomorphism to arrive at equation~(\ref{eqn:cons2}). (This is exactly how we first derived the contracted Bianchi identity for the Einstein-Hilbert action~(\ref{EH diff})).

Once the equivalence between both forms of conservation equation has been established, it is straightforward to see that these equations are exactly what we found in~(\ref{Einstein_tensor_eq}) following from $\nabla_{\alpha} G^{\alpha}{}_{\beta} = 0$. In other words, for our field equations to be mathematically consistent we require the vanishing diffeomorphism transformation of the total action $\delta_\xi S_{\textrm{ total}} = 0$. This can also be seen from the fact that the variations with respect to the dynamical field variables $g_{\mu \nu}$ are required to vanish $\delta S_{\textrm{ total}} = 0$, and the fields themselves are covariant. 

This is a similar conclusion to what is found in many other symmetry-breaking models. In Sec.~\ref{section1.2.4}, when looking at unimodular gravity, the action becomes diffeomorphism invariant on-shell when applying the field equations. In the non-covariant Chern-Simons modifications studied by Jackiw~\cite{RJackiw_2006,RJackiw:2007br}, it is found that diffeomorphism invariance is \textit{dynamically} reinstated as well. This is reflected in our modifications here also.

Moving on to the solutions of~(\ref{eqn:cons1}), we first note that the constraint is satisfied for all possible geometries if $f_{,\ourG}- f_{,\ourB} = 0$. This equation is easily integrated and one finds that all functions of the form $f(\ourG+\ourB)$ are general solutions, assuming a sufficiently regular function $f$. Recalling that this combination is in fact the Ricci scalar, the coordinate scalar constructed from $\ourG$ and $\ourB$, we find the expected result that within the family of $f(\ourG,\ourB)$ theories $f(R)$ gravity is the unique diffeomorphism invariant theory. We will show this result explicitly in Sec.~\ref{section5.2.1}.

Alternatively, let us now consider the case when $f(\ourG,\ourB) \neq f(\ourG + \ourB)$, i.e., non-$f(R)$ and non-covariant theories. Moreover, we will maintain the assumption that the matter action yields an independent conservation equation $\nabla_\alpha T^{\alpha}{}_{\gamma} =0$. As before, for the total action to be invariant, equation~(\ref{eqn:cons1}) has to be satisfied. It is now important that we note that the conservation equation~(\ref{eqn:cons1}) in fact depends on the coordinates used. As such, solutions for a given geometry $g_{\mu \nu}(x)$ will depend explicitly on the coordinate system $\{ x^{\mu} \}$.  
The simplest non-trivial example is when $\ourG$ and $\ourB$ are both constants with respect to some set of coordinates.

In the case when the conservation equation vanishes, the gravitational action will naively \emph{appear} to be invariant under diffeomorphisms. This is therefore associated with a specific set of preferred coordinates. 
This situation is similar to that in $f(T)$ gravity when so-called `good tetrads' are considered~\cite{Ferraro:2011us,Tamanini:2012hg,Bahamonde:2015zma}. This is now understood as a manifestation of the Weitzenb\"{o}ck gauge in $f(T)$ gravity~\cite{Krssak:2018ywd}. In our case here, these `good' coordinates are in fact just those associated with \textit{coincident gauge} in symmetric teleparallel gravity. Moreover, the guaranteed existence of the coincident gauge in the symmetric teleparallel theories~\cite{Adak:2008gd} also guarantees the existence of non-trivial solutions to~(\ref{eqn:cons1}). We will return to this point in Sec.~\ref{section5.2.2}, when we extend the analysis between the bulk and boundary terms $\ourG$ and $\ourB$ and the symmetric teleparallel scalars $Q$ and $B_Q$ to the modified theories $f(\ourG,\ourB)$ and $f(Q,B_Q)$.

Lastly, if we drop the assumption that matter be described by an invariant action, and hence the covariant conservation of the energy-momentum tensor, only the integrand of~(\ref{f(G,B)_conservation}) has to vanish to achieve diffeomorphism invariance of the total action. We then simply have an equation of the form
\begin{equation}
\nabla_{\alpha} \big(H^{\alpha}{}_{\gamma} - T^{\alpha}{}_{\gamma} \big) = 0 \, ,
\end{equation}
where we again understand the covariant derivative here to be shorthand for the expression in equation~(\ref{eqn:cons2}).
 This would allow us to study more exotic theories where energy-momentum is not conserved. An interesting example is models where the dark matter energy density can decrease over time whilst the dark energy density, here modelled through the contributions of $f(\ourG,\ourB)$, can increase over time. For more on these types of models, see the review~\cite{Bolotin:2013jpa}.

\subsection{Second-order $f(\ourG)$ gravity}
\label{section5.1.2}

Now that we have introduced the fourth-order $f(\ourG,\ourB)$ gravity class of theories, let us focus on the second-order models. This is achieved by simply considering functions of the bulk term only, or more specifically, $f(\ourG,\ourB) = f (\ourG) + c_1 \ourB$ for some constant $c_1$. Clearly the linear boundary term does not contribute to the classical dynamics, but we mention it briefly here for completeness\footnote{Recall in Sec.~\ref{section2.2.2} we made reference to the role of the Gibbons-Hawking-York counter term in relation to quantum gravity and black hole entropy. This same counter term can be included alongside the boundary term $\ourB$ if necessary.}. In what follows, we set $c_1=0$ as it plays no role in the discussion. 

In general, these theories will be non-covariant and differ from GR when $f(\ourG) \neq c_2  \ourG$. They also possess the attractive quality that they remain second-order, introduce no additional field content, are four-dimensional, and still reside in the standard Levi-Civita framework. Lovelock's theorem is instead violated due to the non-covariance of the theory.

Defining the second-order modified action as
\begin{equation} \label{f(G)}
S_{\textrm{grav}}[g] = \frac{1}{2 \kappa} \int f(\ourG) \sqrt{-g} d^4 x \, ,
\end{equation}
the equations of motion follow directly from~(\ref{f(g,b)_EoM2})
\begin{align}
  \label{f(G)field_equation}
  f'(\ourG) \Big[G_{\rho \sigma} + \frac{1}{2} g_{\rho \sigma} \ourG \Big] + \frac{1}{2} f''(\ourG) E_{\rho \sigma}{}^{\gamma} \partial_{\gamma} \ourG - \frac{1}{2} g_{\rho \sigma} f(\ourG) =
  \kappa T_{\rho \sigma} \,. 
\end{align}
The conservation equation for the gravitational sector~(\ref{eqn:cons1}) becomes
\begin{equation} \label{f(G)cons}
\partial_{\alpha} \partial_{\beta} \big( \sqrt{-g} M^{\alpha \beta}{}_{\gamma} f'(\ourG) \big) = 0 \, .
\end{equation}
Clearly this equation is only zero (for all geometries) in the case that $f''(\ourG)=0$, which leads back to GR as expected. We will again require coordinates to satisfy this constraint for all of the solutions that we study.

In the introduction we mentioned that diffeomorphism-breaking theories with a minimal length scale can be motivated by appeals to quantum gravity. We also mentioned that this had been studied in $f(T)$ gravity (in relation to breaking local Lorentz invariance) by using the Born-Infeld approach~\cite{Ferraro:2006jd,Fiorini:2013kba}. The Born-Infeld Lagrangian for our $f(\ourG)$ theory reads
\begin{equation} \label{BI_Lagrangian}
    L_{BI} = f(\ourG) = \lambda \Big(\sqrt{1+\frac{2 \ourG}{\lambda}} -1\Big) \ .
\end{equation}
This is simply $\ourG$ to leading order as $\lambda$ tends to infinity. In natural units, the parameter $\lambda$ has dimensions of inverse length squared. Taking $1/\sqrt{|\lambda|} \propto l$ to be the associated length scale of the theory, we recover GR in the limit that $l \rightarrow 0$ and $|\lambda| \rightarrow \infty$. One can then think of $l$ as the length scale related to the non-invariance of the theory under general coordinate transformations. Alternatively, this can be interpreted as some fundamental length scale of spacetime (e.g., the Planck length).

An interesting motivation for the Born-Infeld scheme comes from transitioning from the classical free particle Lagrangian to the relativistic one, see~\cite{Bohmer:2019vff,Ferraro:2006jd}. The free particle Lagrangian of Newtonian mechanics is $L_{p} = m \dot{x}^2/2$. In the Born-Infeld scheme this becomes $L_{BI} = \sqrt{1+L_{p} /(mc^2)}-1$, where the new scale $mc^2$ has been introduced. This new Lagrangian is simply that of a relativistic particle, and the low energy, non-relativistic limit $|\dot{x}| \ll c$ reduces to the Newtonian mechanics action.
We will study these modified Born-Infeld models in the cosmological setting in Sec.~\ref{section6.1.1}.

A different choice of function $f$ leads to an important comparison with the theories known as \textit{modified Newtonian dynamics} (MOND)~\cite{Milgrom:1983ca}. It was recently shown that the $f(\ourG)$ equations of motion~(\ref{f(G)field_equation}) can give rise to the MOND phenomenology~\cite{Milgrom:2019rtd}, see also \cite{Milgrom:2009gv,Milgrom:2022nyl}. The so-called `deep-MOND limit' is obtained for the choice of function $f \propto \ourG^{2/3}$.
In fact, Milgrom recently proposed the theories given by $\mathcal{F}(\ell_{M}^2 \mathcal{R})$, where $\ell_M$ is the MOND length scale and $\mathcal{R}$ is the quadratic part of the Ricci scalar~\cite{Milgrom:2019rtd}. This turns out to be exactly the $f(\ourG)$ theory presented here\footnote{We thank Moti Milgrom for pointing this out in response to our work~\cite{Boehmer:2021aji}.}. We will return to this topic when discussing covariantisation in Sec.~\ref{section5.2}. This shows yet another example of how the $f(\ourG)$ models can be chosen to lead to physically interesting scenarios.

Returning to the equations of motion, let us rewrite the field equations by dividing through by $f'$ and isolating the divergence-free Einstein tensor
\begin{align} \label{f(G)_2}
  G_{\rho\sigma} = \frac{\kappa}{f'} T_{\rho\sigma} + \frac{1}{2}\frac{1}{f'} \Bigl(
  (f - f' \ourG) g_{\rho\sigma} -
  f'' E_{\rho \sigma}{}^{\gamma} \partial_{\gamma} \ourG \Bigr) \,.
\end{align}
The vanishing covariant derivative of the right-hand side can be formally implemented by using the gravitational constraint equation~(\ref{f(G)cons}), which we can write as
\begin{align}
  \label{f(G)cons2}
  \frac{1}{2\kappa}\partial_{\alpha} \partial_{\beta}
  \bigl(\sqrt{-g} M^{\alpha\beta}{}_{\gamma} f'(\ourG)\bigr) - \nabla_\alpha T^\alpha{}_\gamma \sqrt{-g} = 0 \,.
\end{align}
For conventional matter based on a scalar action, the covariant divergence of the energy-momentum tensor vanishes and only the initial term survives. 
As we mentioned, the non-covariance of the gravitational theory means that certain systems of coordinates are picked out by the equation above. To illustrate what this means in practice, let us briefly consider the metrics for spherically symmetric and FLRW spacetimes, the latter of which will be studied in greater detail in Chapter~\ref{chapter6}.

The static and spherically symmetric metric is often written in spherical coordinates $(t,r,\theta,\phi)$, taking the form
\begin{equation}
ds^2 = - A(r)^2 dt^2 + B(r)^2 dr^2 + r^2 d \theta^2 + r^2 \sin^2 \theta d \phi^2 \, .
\end{equation}
However, if one uses this metric in this form in the conservation equation~(\ref{f(G)cons}), one is led to the constraint $f''(\ourG)=0$. This constraint also comes from the off-diagonal field equations. This then trivially leads back to GR. However, this is not a reflection of the theory or the physical symmetries of the spacetime, but instead a reflection of the choice of coordinates. In isotropic coordinates $(t,x,y,z)$ the metric takes the form
\begin{align} 
  ds^2 = -{A(R)}^2 dt^2 + {B(R)}^2 \bigl(dx^2 + dy^2 + dz^2 \bigr)\,, \qquad 
  R = \sqrt{x^2 + y^2 + z^2} \,,
\end{align}
and it is then found that~(\ref{f(G)cons}) vanishes independently of the function $f$. Hence, these are the appropriate coordinates to study models beyond GR. The same results hold in the more general $f(\ourG,\ourB)$ theories. Further discussion of spherically symmetric spacetimes is given in our work~\cite{Boehmer:2021aji}.

For the FLRW metric a similar situation occurs. Here we will consider the spatially flat $k=0$ case only\footnote{Spatially curved cases can also be considered, but the explicit form of the metric in the `good coordinates' is extremely complex.}. Working in spherical coordinates with the metric
\begin{equation}
ds^2 = - N(t)^2 dt^2 + a(t)^2 \big( dr^2 + r^2 d\theta^2 + \sin^2 \theta d \phi^2 \big) \, ,
\end{equation}
one again is lead to an equation of the form $f''(\ourG)=0$, forcing the theory back to GR. Working instead in Cartesian coordinates
\begin{equation} \label{FLRW_1}
ds^2 = - N(t)^2 dt^2 + a(t)^2 \big( dx^2 + dy^2 + dz^2 \big ) \, ,
\end{equation}
the constraint equation is satisfied for all choices of $f$. This again represents the correct choice of coordinates if studying theories beyond GR. As we pointed out previously, this is similar to the `good' and `bad' tetrads in modified teleparallel $f(T)$ gravity~\cite{Tamanini:2012hg}. It should now be clear from the previous discussions that these are related to the coincident and Weitzenb\"{o}ck gauges. We examine this more closely in Sec.~\ref{section5.2.2}. This will explain this unusual, seemingly coordinate-dependent behaviour of the theory presented here, showing that it is in fact just an artefact of an implicit gauge choice.
 
An interesting and related topic, which we unfortunately will not explore here, is the issue of the number of physical degrees of freedom of these theories. Without the full diffeomorphism symmetry, one would expect the number to be greater than the two polarisations of the gravitational field in GR. But to determine the precise number requires a full Hamiltonian analysis, which is far from straightforward. Moreover, though second-order theories are free from the Ostrogradsky instabilities described in Sec.~\ref{section1.2}, it still remains to be checked whether the additional degrees of freedom are healthy. For example, if additional propagating degrees of freedom exist but do not show up at lower perturbation orders, this may indicate a possible strong coupling problem\footnote{This is when a given background solution becomes strongly coupled at very low (IR) energies. At the classical level, the predictions from linear perturbations about this background then become unreliable. This is often hinted at when known degrees of freedom are absent at lower perturbation orders around specific background, e.g., around Minkowski and FLRW backgrounds in $f(T)$ gravity~\cite{BeltranJimenez:2020fvy,BeltranJimenez:2019nns,Golovnev:2018wbh}.}, see~\cite{Arkani-Hamed:2002bjr,Deffayet:2005ys,Charmousis:2009tc}. For details on these issues in $f(T)$ and $f(Q)$ gravity, see~\cite{Golovnev:2018wbh,BeltranJimenez:2019tme,BeltranJimenez:2020lee,Hu:2023juh}.

\subsubsection{Tetradic modifications}

We have focused on the modifications stemming from the coordinate basis decomposition of the Ricci scalar $R = \ourG + \ourB$, but we could instead work in the orthonormal basis $R(e) = \mathfrak{G} + \mathfrak{B}$. Modifications based on these tetradic terms, such as $f(\mathfrak{G})$ and $f(\mathfrak{G},\mathfrak{B})$, would lead to a different theory than those studied above. We will not go through the explicit details of the calculation, but the formulation follows analogously to the $f(\ourG,\ourB)$ theories. The main computational difference is that we instead vary with respect to the tetrad, and a conservation equation would follow from requiring the total action to be locally Lorentz invariant. In Section~\ref{section5.2.2} we show the resulting equations of motion, by making use of the aforementioned equivalence with the teleparallel theories. See equation~(\ref{f(T)_EoM}) for the explicit equations of motion of the $f(\mathfrak{G})$ theories.

We also note that $\mathfrak{G}$ can be obtained by making use of the boundary term $\mathbb{B}$ in equation~(\ref{Bnew2}). There, we showed it could be related to the metric bulk term by $\mathfrak{G} = \ourG + \mathbb{B}$. One could instead derive the $f(\mathfrak{G})$ theories by taking the tetradic variations of $f(\ourG + \mathbb{B})$. To be even more general, the mixed Einstein action~(\ref{Einstein_general}) could be used as the starting point for modifications. In Sec.~\ref{section5.2.3} we explore this possibility, showing how this leads to a unified approach.

\section{Relation to other modified theories}
\label{section5.2}
\subsection{$f(R)$ gravity}
\label{section5.2.1}

If we take $f(\ourG,\ourB)$ to be a function of the Ricci scalar only, that is $f(\ourG,\ourB) = f(\ourG + \ourB) = f(R)$, we readily recover the $f(R)$ field equations.  This also serves as a good consistency check. First we use that
\begin{equation}
    \frac{\partial f(\ourG+\ourB)}{\partial \ourG} = \frac{\partial f(\ourG+\ourB)}{\partial \ourB} \ = \frac{\partial f(R)}{\partial R} \,.
\end{equation}
With this simplification we can rewrite the left-hand side of equation~(\ref{f(g,b)_EoM2}) as follows
\begin{multline}
  \label{f(GB)variation_expansion1}
  \frac{\partial f}{\partial R} \Big[ G_{\rho \sigma} + \frac{1}{2} g_{\rho \sigma} \ourG +
    \frac{1}{2} g_{\rho \sigma} \ourB \Big] - \frac{1}{2} g_{\rho \sigma} f(R) +
  g_{\rho \sigma} \partial^{\mu} \partial_{\mu} \Big(\frac{\partial f}{\partial R} \Big) -
  \partial_{\rho} \partial_{\sigma} \Big( \frac{\partial f}{\partial R} \Big) 
  \\
  + \partial_{\gamma} \Big( \frac{\partial f}{\partial R} \Big)
  \Big[
    \frac{1}{2} E_{\rho \sigma}{}^{\gamma} + \frac{1}{2} g_{\rho \sigma} \partial_{\mu}g^{\mu \gamma} +
    \frac{1}{\sqrt{-g}} \partial_{(\rho}(\sqrt{-g}) \delta^{\gamma}_{\sigma)}  
    \Big] \,.
\end{multline}
Expanding the definition of $E_{\rho \sigma}{}^{\gamma}$~(\ref{E}), writing the partial derivatives of the metric in terms of Christoffel symbols and using $\ourG + \ourB = R$ gives
\begin{multline}
  \label{f(GB)variation_expansion2}
    \frac{\partial f}{\partial R} \Big[ G_{\rho \sigma} + \frac{1}{2} g_{\rho \sigma}R \Big]
    - \frac{1}{2} g_{\rho \sigma} f(R) + g_{\rho \sigma} \partial^{\mu} \partial_{\mu} \Big(\frac{\partial f}{\partial R} \Big)  - \partial_{\rho} \partial_{\sigma} \Big( \frac{\partial f}{\partial R} \Big) 
    \\
    + \partial_{\gamma} \Big( \frac{\partial f}{\partial R} \Big)
    \Big[ \Gamma^{\gamma}_{\rho \sigma} - g_{\rho \sigma} g^{\mu \nu} \Gamma^{\gamma}_{\mu \nu}
    \Big]  \,.
\end{multline}
The term in the first square bracket is the Ricci tensor $R_{\rho \sigma}$. The other connection and partial derivative terms are simply the covariant derivatives acting on $\partial f/\partial R$, which, unlike $\ourG$ and $\ourB$, is a true coordinate scalar.

To see the equivalence more clearly, take the $f(R)$ field equations~(\ref{f(R)field_equation})
\begin{equation}
    \frac{\partial f}{\partial R} R_{\rho \sigma} + \big[g_{\rho \sigma} \Box - \nabla_{\rho} \nabla_{\sigma}\big]\frac{\partial f}{\partial R} - \frac{1}{2} g_{\rho \sigma} f(R) \,,
\end{equation}
and expand the derivative terms
\begin{align}
    \big[g_{\rho \sigma} \Box - \nabla_{\rho} \nabla_{\sigma}\big] f_{,R} &=
    g_{\rho \sigma} g^{\mu \nu} \big(\partial_{\mu} \nabla_{\nu} - \Gamma^{\gamma}_{\mu \nu} \nabla_{\gamma} \big) f_{,R} - 
     \nabla_{\rho} \partial_{\sigma} f_{,R} \\
    &= \Big( g_{\rho \sigma} g^{\mu \nu} \partial_{\mu} \partial_{\nu} 
    - g_{\rho \sigma} g^{\mu \nu} \Gamma^{\gamma}_{\mu \nu} \partial_{\gamma} - \partial_{\rho} \partial_{\sigma} + \Gamma^{\gamma}_{\rho \sigma} \partial_{\gamma} \Big) f_{,R} \,.
\end{align}
Comparing~~(\ref{f(GB)variation_expansion2}) with the expressions above shows that these are indeed the same, such that $f(\ourG+\ourB)$ yields the standard $f(R)$ field equations. These are generally fourth-order, due to the second-order terms in the Ricci scalar (which are confined to the boundary in GR) contributing to the equations of motion when $f(R)$ is a non-linear function.

In Fig.~\ref{fig:f(G,B)} we illustrate the relationship between $f(R)$ gravity and our $f(\ourG,\ourB)$ theories. The most general model is $f(\ourG,\ourB)$, which reduces to $f(R)$ by requiring covariance. This is represented by the vanishing of the conservation equation~(\ref{eqn:cons1}).
 Alternatively, the most general second-order theory is $f(\ourG)$ gravity. Requiring the field equations to be both covariant and second-order leads back to General Relativity.

\begin{figure}[!hbt]
\centering
\begin{tikzpicture}
  \matrix (m) [matrix of math nodes,row sep=5em,column sep=10em,minimum width=2em]
  {
    f(\ourG,\ourB) & f(\ourG) \\
    f(\ourG+\ourB) & \text{GR} \\
  };
  \path[-stealth]
  (m-1-1) edge node [above] {second-order} (m-1-2)
  (m-1-1) edge node [left] {diff.~invariant} (m-2-1)
  (m-2-1) edge node [above] {second-order} (m-2-2)
  (m-1-2) edge node [right] {diff.~invariant}(m-2-2);
\end{tikzpicture}
\caption{Relationship between the modified theories of gravity and General Relativity.}
\label{fig:f(G,B)}
\end{figure}
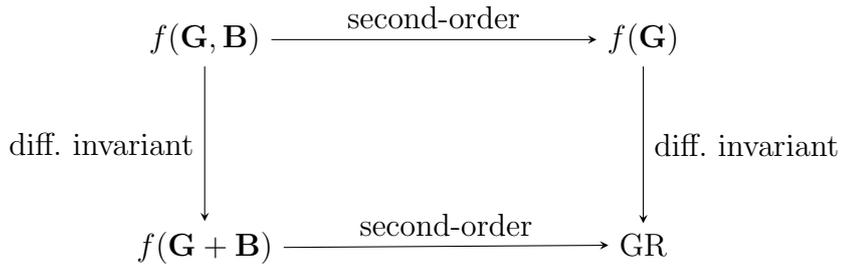

\subsection{Teleparallel $f(T)$ and $f(Q)$ gravity}
\label{section5.2.2}

The modified teleparallel theories $f(T)$ and $f(Q)$ gravity take a very similar form to our $f(\ourG)$ theories~\cite{Boehmer:2021aji}. They too are both second-order in derivatives of the metric or tetrad, though they are based in different geometric settings. In Section~\ref{section4.3} we showed the equivalence of the Einstein actions and the teleparallel equivalents of General Relativity, TEGR and STEGR. Here we extend this discussion to their modified versions as well.
One may then wonder if the equivalence between $f(\ourG)$ and $f(Q)$, and between $f(\mathfrak{G})$ and $f(T)$, is broken for these modified theories. Such an equivalence between the Ricci scalar and the bulk terms is clearly broken when transitioning to non-linear modifications. However, quite remarkably, the theories are still completely equivalent. 

The variations with respect to the inertial connections for the teleparallel theories lead to non-trivial equations, and the fields that parameterise the connection become dynamical. Evidence for this can be seen in the increased number of degrees of freedom\footnote{We note that this issue is not fully settled, but for $f(Q)$ gravity it is clear there are \textit{at least} two additional propagating degrees of freedom~\cite{Jimenez:2019ovq}.} in $f(T)$ or $f(Q)$ gravity~\cite{DAmbrosio:2023asf,Blixt:2020ekl}. But these equations are exactly the same as the ones imposed by requiring the non-covariant theories $f(\ourG)$ and $f(\mathfrak{G})$ to be invariant under diffeomorphisms and local Lorentz transformations. And, as we saw in the previous sections, these constraints were required by mathematical consistency. Hence, the modified theories are in fact dynamically equivalent, which we show below. 

\subsubsection{Metric teleparallel gravity}

The action of modified metric teleparallel gravity is given by
\begin{equation} \label{f(T)}
S_{f(T)}[e,...] = -\frac{1}{2 \kappa} \int f(T) e d^4 x \, .
\end{equation}
We now have the choice of what additional fields to consider as dynamical, indicated by the ellipsis in $S_{f(T)}[e,...]$. Recall that in TEGR, the purely inertial spin connection did not play a physical role, and the fields that parameterised the connection $\Lambda$ were not dynamical. In fact, they could be separated off in a term that we labelled the `purely gauge boundary term', see~(\ref{T_decomp}). For non-linear modifications this is no longer the case.

For a consistent variational procedure in the modified action~(\ref{f(T)}), we have the following three choices: work in the fixed Weitzenb\"{o}ck gauge $\omega(\Lambda)=0$; use the purely inertial spin connection parameterised by the fields $\Lambda$; or vary with respect to an independent spin connection $\omega$ with Lagrange multipliers enforcing vanishing curvature and non-metricity. The former of these approaches breaks local Lorentz invariance, which will be discussed below. The latter two methods are covariant but introduce additional dynamical fields. These various approaches are compared in detail in~\cite{Golovnev:2017dox}, so we refer to that work, as well as~\cite{Aldrovandi:2013wha}, for the explicit calculations.

In the fixed Weitzenb\"{o}ck gauge, known as `pure $f(T)$ gravity', the equations of motion are
\begin{equation} \label{f(T)_EoM}
    f'(T) \Big[G_{\rho \sigma} - \frac{1}{2} g_{\rho \sigma} T \Big] + 2 f''(T) \mathscr{S}_{\rho \sigma}{}^{\gamma} \partial_{\gamma} T + \frac{1}{2} g_{\rho \sigma} f(T) = \kappa \Theta_{\rho \sigma} \,,
\end{equation}
where $\mathscr{S}^{\mu \nu \lambda}$ is the teleparallel superpotential~(\ref{T_superpotential}) and $\Theta_{\rho \sigma}$ is the symmetric\footnote{We assume the matter action to be a local Lorentz scalar and to only couple to the \textit{Levi-Civita} spin connection, hence $S_{\textrm{M}} = S_{\textrm{M}}[e,\varPhi^{A}]$ and $\Theta_{[\rho \sigma]}=0$. See the discussion at the end of Chapter~\ref{chapter2}.} (tetradic) energy-momentum tensor~(\ref{tetradic_EM}), see~\cite{Li:2011wu,li2011degrees}. Importantly, the antisymmetric part of the field equations do \textit{not} vanish in general, leading to an additional set of equations that must be satisfied
\begin{equation} \label{f(T)_skew}
 f''(T) \mathscr{S}_{[\rho \sigma]}{}^{\gamma} \partial_{\gamma} T  = 0 \, .
\end{equation}
This equation is a consequence of the non-invariance under local Lorentz transformations. For TEGR, $f(T) = T$ and $f''(T) = 0$ and this equation vanishes identically, representing the local Lorentz (pseudo-)invariance.  A similar situation arises in the $f(T,B_T)$ theories, where the teleparallel boundary term is also included. The combination $f(-T+B_T)$ leads to the locally Lorentz invariant $f(R)$ theories, and the antisymmetric field equation also vanishes identically~\cite{Bahamonde:2015zma}.

In the fixed, Weitzenb\"{o}ck framework, the torsion tensor is not covariant under LLTs, and hence the field equations take different values in different choices of frame $\{\mathbf{e}_{a}\}$. Frames in which the antisymmetric equation~(\ref{f(T)_skew}) forces $f''(T) =0$ were known as `bad tetrads', because this trivially leads back to GR. On the other hand, if the equation is satisfied for all $f$, then these were known as `good tetrads'~\cite{Tamanini:2012hg}. A typical example is when the torsion scalar is a constant. The similarities with the choices of coordinates that lead to $f''(\ourG)=0$ in our theories is immediately apparent.

Let us now consider the approach with the inclusion of the purely inertial spin connection. The action then takes the form $f(T(e,\omega(\Lambda)))$, and is invariant under local Lorentz transformations performed on the tetrad and spin connection simultaneously. The explicit form of the field equations can be found in~\cite{krvsvsak2016covariant}, but the conclusion is that the tetradic field equations are simply~(\ref{f(T)_EoM}) in an arbitrary gauge. Transforming to the Weitzenb\"{o}ck gauge $\omega =0$ exactly gives the pure $f(T)$ field equations. 

The equations of motion obtained from the variations of the spin connection end up being identical to the skew-symmetric part~(\ref{f(T)_skew}), again in an arbitrary gauge. Therefore the covariant $f(T)$ equations are simply equivalent to the Weitzenb\"{o}ck ones with gauge-invariance restored. Importantly, this means that all possible solutions of the theory in either approach are the same. Any solution of the covariant $f(T)$ equations can be transformed via a suitable local Lorentz transformation to the Weitzenb\"{o}ck gauge, and hence exists in the pure approach. 
For more discussions on this topic, see~\cite{Golovnev:2020zpv,Golovnev:2021lki,Krssak:2018ywd,Blixt:2022rpl}.

There have been many studies on $f(T)$ gravity, working with a plethora of different models, see the reviews~\cite{Cai:2015emx,Bahamonde:2021gfp} and references therein. A recent focus has been on constraining such models with cosmological observations, for example~\cite{Cai:2018rzd,Nunes:2018xbm,Kumar:2022nvf}. Notably, the EFT techniques mentioned in Sec.~\ref{section1.1.2} have been applied to $f(T)$ gravity in this cosmological context~\cite{Li:2018ixg,Mylova:2022ljr,Ren:2022aeo}. This appears to be a promising avenue to study these theories in a model-independent way. Moreover, problems such as strong coupling have been approached using an EFT perspective~\cite{Hu:2023juh}.

Moving on to the comparison with our theories, recall that $\mathfrak{G}$ and $T(e,0)$ were equivalent~(\ref{G_T_rel}), modulo  sign. It follows immediately that their modifications are also equivalent, and so $f(\mathfrak{G})$ gravity is simply pure $f(T)$ gravity\footnote{Again, we assume that matter couples only to the Levi-Civita connection, as is standard in teleparallel gravity. In our $f(\mathfrak{G})$ theories, this is the only natural coupling.}. All of the previous discussions on the equivalence of the covariant vs pure approach now apply to the non-covariant $f(\mathfrak{G})$ models. This seems quite natural, as $f(\mathfrak{G})$ gravity is not invariant under local Lorentz transformations, while the presence of additional degrees of freedom can be seen explicitly in its covariant form. It is the $\Lambda$ matrices parameterising the purely inertial spin connection $\omega(\Lambda)$ that restore the covariance of the theory, and these fields become dynamical when we consider non-linear modifications. 
In the next part, when we study symmetric teleparallel gravity, we will show this equivalence explicitly.

Lastly, we note all of these discussions extend to the modified theories including boundary terms, namely $f(\mathfrak{G},\mathfrak{B})$ and the teleparallel $f(T,B_T)$ models~\cite{Bahamonde:2015zma}. The equivalence follows from the discussion of the boundary terms in Sec.~\ref{section4.2.3}, and the relation $B_T = \mathfrak{B} + b_{\Lambda}$. In Sec.~\ref{section5.2.3} we look closer at these possible extensions.

\subsubsection{Symmetric teleparallel gravity}

Moving on to modified symmetric teleparallel gravity, the action is given by\footnote{Note that in some of the $f(Q)$ literature, the non-metricity scalar is defined with an overall minus sign different from ours. To match up with these formulations, appropriate field redefinitions are needed. Similar care must also be taken with non-equivalent definitions of the non-metricity tensor. This will be discussed below when comparing with $f(\ourG)$ gravity.\label{footnotef(Q)}}~\cite{BeltranJimenez:2017tkd}
\begin{equation} \label{f(Q)}
S_{f(Q)}[g,...] = -\frac{1}{2\kappa} \int f(Q) \sqrt{-g} d^ 4x \, .
\end{equation}
Again, we must decide which fields to consider independent and dynamical. Fixing the coincident gauge breaks the diffeomorphism invariance of the action and immediately gives the $f(\ourG)$ action~(\ref{f(G)}). In this case, only the metric tensor enters the action. Alternatively, a covariant formulation can be used by considering the fields $\xi$ that parameterise the symmetric teleparallel connection~(\ref{STG_connection}), or using the Lagrange multiplier approach with an affine connection $\bar{\Gamma}^{\lambda}_{\mu \nu}$. Either of these latter two approaches lead to the same equations of motion, and we will show that they are dynamically equivalent to the gauge-fixed formulation shortly.

The field equations from varying the action $S_{f(Q)}[g,\xi]$ with respect to the metric are 
\begin{equation}
f'(Q) \big( G_{\rho \sigma} - \frac{1}{2} g_{\rho \sigma} Q \big) - 2 f''(Q) \mathscr{P}^{\lambda}{}_{\rho \sigma} \partial_{\lambda} Q + \frac{1}{2} g_{\rho \sigma} f(Q) = \kappa T_{\rho \sigma} \, ,
\end{equation}
see~\cite{BeltranJimenez:2019tme,Zhao:2021zab,DAmbrosio:2021pnd}. The \textit{non-metricity conjugate}, or superpotential, is defined as
\begin{equation}
\mathscr{P}^{\lambda \rho \sigma}  := \frac{1}{2} \frac{\partial Q}{\partial Q_{\lambda \rho \sigma}} = \frac{1}{4}Q^{\lambda \rho \sigma} - \frac{1}{2} Q^{(\rho \sigma)\lambda} - \frac{1}{4} g^{\rho \sigma}( Q^{\lambda \mu}{}_{\mu} - Q^{\mu}{}_{\mu}{}^{\lambda}) + \frac{1}{4} g^{\lambda (\rho} Q^{\sigma) \mu}{}_{\mu} \, ,
\end{equation}
from which we can obtain the non-metricity scalar via the contraction $\mathscr{P}^{\mu \nu \lambda} Q_{\mu \nu \lambda} = Q$.
The connection equation of motion is given by
\begin{equation} \label{f(Q)_connection}
\bar{\nabla}_{\mu} \bar{\nabla}_{\nu} \big( \sqrt{-g} \mathscr{P}^{\mu \nu}{}_{\lambda} f'(Q) \big) = 0 \, ,
\end{equation}
where $\bar{\nabla}_{\mu}$ is the covariant derivative associated with the symmetric teleparallel connection. Note that we have assumed that hypermomentum vanishes, as is standard in these teleparallel models. For further details on these equations of motion we refer to~\cite{BeltranJimenez:2018vdo,BeltranJimenez:2019tme}.

Let us now show that these equations in the coincident gauge are exactly equivalent to our $f(\ourG)$ ones. Firstly, we of course have that $\ourG = -Q$. The superpotentials of both theories are also easily shown to coincide
\begin{align} \label{PtoE}
\mathscr{P}_{\lambda \rho \sigma}(g,0) &= \frac{1}{4}\Big( - \partial_{\lambda} g_{\rho \sigma} + 2 \partial_{(\rho} g_{\sigma) \lambda} + 2 g_{\rho \sigma} g^{\alpha \beta} \partial_{[\lambda} g_{\alpha] \beta} - g^{\alpha \beta} g_{\lambda(\rho} \partial_{\sigma)} g_{\alpha \beta} \Big) \nonumber \\
&= \frac{1}{4}\Big( 2 g_{\gamma \lambda} \Gamma_{\rho \sigma}^{\gamma} - 2 g_{\lambda (\rho} \Gamma_{\sigma) \gamma}^{\gamma} + g_{\rho \sigma} \Gamma_{\lambda \gamma}^{\gamma} - g_{\rho \sigma} g_{\gamma \lambda} g^{\alpha \beta}  \Gamma_{\alpha \beta}^{\gamma}  \Big) \nonumber \\\
&= \frac{1}{4} E_{\rho \sigma \lambda} \, ,
\end{align}
where we have used that the non-metricity tensor in the coincident gauge is simply the partial derivative of the metric~(\ref{Q_xi}). We have then expressed the derivatives of the metric in terms of the Levi-Civita Christoffel symbols, which gives an equivalent expression to the superpotential $E^{\rho \sigma \lambda}$~(\ref{E}).

 Lastly, we must make appropriate field redefinitions such that the original $f(Q)$ action~(\ref{f(Q)}) aligns with our $f(\ourG)$ action~(\ref{f(G)}). These redefinitions\footnote{Alternatively, we can begin with the $f(\ourG)$ equations and make the transformation $f(\ourG) \rightarrow -f(-\ourG)$. Note again that there are different definitions of $Q_{\mu \nu \lambda}$, $Q$ and $P^{mu \nu \lambda}$ in the literature, see footnote~\ref{footnotef(Q)}. Conveniently, in works where the non-metricity scalar differs from ours by a minus sign, the equivalence with our $f(\ourG)$ equations is present immediately without any redefinitions needed, see equation (18) of~\cite{Zhao:2021zab}.} are $f(Q) \rightarrow -f(-Q) = -f (\ourG)$, from which it follows that $f'(Q) \rightarrow f'(\ourG)$ and $f''(Q) \rightarrow -f''(\ourG)$.
Along with equation~(\ref{PtoE}), we see that the coincident gauge $f(Q)$ metric field equations are the same as the $f(\ourG)$ ones~(\ref{f(G)field_equation}).

For the connection equation of motion~(\ref{f(Q)_connection}), in the coincident gauge we have $\bar{\nabla}_{\mu} \rightarrow \partial_{\mu}$. Using the equivalence of the superpotentials~(\ref{PtoE}), we obtain
\begin{equation}
\bar{\nabla}_{\mu} \bar{\nabla}_{\nu} \big( \sqrt{-g} \mathscr{P}^{\mu \nu}{}_{\lambda} f'(Q) \big) = \frac{1}{4} \partial_{\mu} \partial_{\nu} \big( \sqrt{-g} E_{\lambda}{}^{\mu \nu} f'(\ourG) \big) = 0 \, .
\end{equation}
With the identity $E^{\lambda (\mu \nu)} = M^{\mu \nu \lambda}$~(\ref{M to E}), we see the above equation is exactly our constraint coming from diffeomorphism invariance~(\ref{f(G)cons}). Further details on the relationship between the connection variations and infinitesimal diffeomorphisms can be found in~\cite{BeltranJimenez:2019tme,BeltranJimenez:2018vdo}. 

Just like in the $f(T)$ case, the fields $\xi^{\mu}$ that parameterise the connection act to restore the invariance of the theory: their equations of motion are exactly those found from the symmetry transformations of the gauge-fixed theories, as we expected. Performing the covariantisation \'{a} la Stueckelberg on the non-covariant $f(\ourG)$ theories leads uniquely to $f(Q)$ gravity. This has also been shown by Milgrom in~\cite{Milgrom:2019rtd}.

The discussion can again be extended to arbitrary functions of the boundary terms as well. In our work~\cite{Boehmer:2021aji} (Sec.~\ref{section5.1}), we gave the first formulation of the $f(\ourG,\ourB)$ theories along with their equations of motion. More recently, the extended $f(Q,B_Q)$ modified theories have been considered~\cite{Capozziello:2023vne,De:2023xua}. However, it is clear that the dynamics, and hence the theories themselves, are completely equivalent to our non-covariant theories. The symmetric teleparallel $f(Q,B_Q)$ theories are simply the covariantised versions of our $f(\ourG,\ourB)$ models. Likewise, it is not difficult to show that fixing the coincident gauge for the $f(Q,B_Q)$ theories leads exactly to $f(\ourG,\ourB)$ gravity.

As a final observation, recall our brief discussion of Ho\v{r}ava gravity in Sec.~\ref{section1.2.4}. A further comparison can be made between the situation there and the $f(\ourG)$ and covariant $f(Q)$ theories. In Ho\v{r}ava gravity, when working in the (gauge-fixed) ADM formalism, one finds the dynamical equations do not in fact provide enough information to determine the (dynamical) free lapse function $N$. This is then fixed by imposing the Hamiltonian constraint~\cite{Blas:2009yd}. However, on certain backgrounds the Hamiltonian constraint vanishes and leaves the lapse undetermined. In other words, it exhibits some residual reparameterisation invariance. When covariance is restored via the Stueckelberg trick, this constraint equation becomes the equation for the Stueckelberg fields. On these backgrounds, this equation constrains an otherwise propagating degree of freedom, and so the additional mode becomes non-dynamical. This is then related to the aforementioned strong-coupling issue~\cite{Blas:2009yd}.

The same phenomena occur in $f(\ourG)$ and $f(Q)$. The diffeomorphism conservation constraint $\partial_{\mu} \partial_{\nu} \big(\sqrt{-g} M^{\mu \nu}{}_{\lambda} f'(\ourG) \big)= 0$ becomes the equation of motion for the Stueckelberg fields (the $\xi$'s parameterising the symmetric teleparallel connection). 
Similarly, for some backgrounds, like FLRW in Cartesian coordinates~(\ref{FLRW_1}), the theory retains a residual reparameterisation  invariance~\cite{BeltranJimenez:2019esp,BeltranJimenez:2019tme}. 
This allows us to generically set the lapse to one $N(t)=1$. The authors of~\cite{BeltranJimenez:2019esp,BeltranJimenez:2019tme} confirm that in the minisuperspace action, the lapse indeed becomes non-dynamical, and the theory remains time-reparameterisation invariant.

 In comparison to other symmetry-breaking models, the situation in the $f(\ourG)$, $f(Q)$, $f(\mathfrak{G})$ and $f(T)$ modified theories is even more complicated, with the full diffeomorphism or local Lorentz symmetries effectively being broken. In $f(T)$ gravity, this has resulted in many investigations into the so-called `remnant symmetry'~\cite{Ferraro:2014owa,Chen:2014qtl,Golovnev:2020nln,Bejarano:2019fii}. This is the residual local Lorentz symmetry of the tetrad about a given solution. The transformation properties of $\mathfrak{G}$, given in~(\ref{G_LLT}), are no doubt related to this residual symmetry. However, we will not explore this issue further here.

\subsubsection{Matter and symmetry}

In the modified teleparallel theories, we have assumed that the matter action does not depend on the teleparallel connection $S_{\textrm{M}}[e,\varPhi^{A}]$. This is the usual approach when working with these theories, but some further comments on this  topic are warranted. In the metric teleparallel framework, it is well known that the use of the Weitzenb\"{o}ck connection is inconsistent with Dirac spinor matter~\cite{Arcos:2004tzt}. This is because the Dirac action couples to the connection, but the Weitzenb\"{o}ck connection is totally inertial and can be set to zero via a frame transformation. In this case, one finds that the matter energy-momentum tensor is not necessarily conserved, leading to inconsistencies. To solve this issue, matter must instead couple to the Levi-Civita connection of GR rather than the teleparallel one, which is slightly at odds with the standard gauge theory procedure.

On the other hand, it has been pointed out that in the symmetric teleparallel geometry, the Dirac Lagrangian only couples to the Levi-Civita part of the connection~\cite{BeltranJimenez:2018vdo,BeltranJimenez:2020sih}. Therefore none of the problems of the metric teleparallel framework are present in the symmetric case. Moreover, the natural coupling procedure of promoting partial derivatives to covariant derivatives is less unambiguous than in theories with torsion.
For further details on matter couplings, see~\cite{Harko:2018gxr,Xu:2019sbp,BeltranJimenez:2018vdo,BeltranJimenez:2020sih}.

Another important topic is spacetime symmetries. In teleparallel theories beyond the equivalents of General Relativity, one needs to also take account of the additional geometric structure. This means ensuring that the metric and affine connection pair, or tetrad and spin connection pair, all respect the symmetries of the spacetime. 
In metric teleparallel gravity, this can be achieved by generalising isometries of the metric to \textit{affine frame symmetries}, where the Lie derivatives of the tetrad $\mathcal{L}_{\xi} \mathbf{e}_{a}$ and spin connection $\mathcal{L}_{\xi} \omega_{ab}$ must both vanish in a given frame~\cite{Coley:2019zld}. Alternatively, a frame independent notion of spacetime symmetry can be adapted from Cartan geometries, which has been employed in~\cite{Hohmann:2019nat,Hohmann:2015pva}. The Weitzenb\"{o}ck gauge can then also be enforced, see for example~\cite{Pfeifer:2022txm}. This makes the method applicable to the pure $f(T)$ formulation as well, and hence also $f(\mathfrak{G})$ gravity.

In symmetric teleparallel gravity, one requires the Lie derivative of both the metric~(\ref{Lie metric}) and affine connection~(\ref{Lie connection}) to vanish~\cite{Hohmann:2021ast}. In this natural coordinate basis, these concepts are easier to implement. This is a purely covariant approach, where the affine connection is not assumed to vanish. However, we are also still free to transform our system to the coincident gauge via a suitable coordinate transformation. For the spatially flat FLRW metric~(\ref{FLRW_1}), this happens to take the simple form of Cartesian coordinates. For spatially curved cosmologies it has been noted that the coincident gauge is not ideal, due to the inability to retain the usual spherical coordinate form of the metric tensor~\cite{Zhao:2021zab,Hohmann:2021ast}. However, coincident gauge coordinates can still be found and used, leading to the exact same equations of motion, albeit in a more complicated form. This is the choice of coordinates that would be appropriate for the gauge-fixed $f(\ourG)$ theories.

\subsection{A unified framework}
\label{section5.2.3}

Putting together all of the modifications studied so far, we can construct a unified framework that encompasses all of these theories. To do so, we make use of the relationship between the metric bulk and boundary terms $\ourG$, $\ourB$, and the tetradic bulk and boundary terms $\mathfrak{G}$, $\mathfrak{B}$. These were given by
\begin{equation} \label{GBrels}
\mathfrak{G} - \ourG = \mathbb{B} \, , \qquad \qquad \ourB - \mathfrak{B} = \mathbb{B} \, ,
\end{equation}
with $\mathbb{B}$ defined in~(\ref{Bnew2}). We also use that the metrical decomposed terms are equivalent to the gauge-fixed symmetric teleparallel scalars, and the tetradic decomposed terms are equivalent to the gauge-fixed metric teleparallel scalars. And as emphasised in the previous sections, the covariant theories are dynamically and physically equivalent.

We then propose the following generalised modified theory in the \textit{purely Riemannian framework}, described by the action
\begin{equation}
S_{\textrm{grav}}[g,e] = \frac{1}{2\kappa} \int f \big(\ourG(g), \mathbb{B}(e), \ourB(g) \big) e  d^4 x \, ,
\end{equation}
where the dynamical variables are the metric and tetrad\footnote{We could have written this just in terms of the tetrad, but we include the metric to emphasise that $\ourG$ and $\ourB$ can be written just in terms of the metric.}. This can be seen as a modification and generalisation of the mixed Einstein action in equation~(\ref{Einstein_general}). Crucially, the most general function $f(\ourG,\mathbb{B},\ourB)$ is neither locally Lorentz invariant nor diffeomorphism invariant, and gives rise to fourth-order equations of motion. 

In Fig.~\ref{fig:unified} we give a representation of the different theories based on this unified $f(\ourG,\mathbb{B},\ourB)$ action. Requiring Lorentz invariance leads to the metric modifications $f(\ourG,\ourB)$, which can be expressed in terms of the natural coordinate basis. On the other hand, requiring diffeomorphism invariance leads to the tetradic theories $f(\ourG + \mathbb{B},\ourB-\mathbb{B}) = f(\mathfrak{G},\mathfrak{B})$ in the orthonormal basis. We retrieve $f(R)$ gravity for $f(\ourG+\ourB) = f(\mathfrak{G} + \mathfrak{B})$, which is invariant under both coordinate and local Lorentz transformations.

\begin{landscape}
\centering
\vspace*{\fill}
\begin{figure}[!htpb]
\centering
\begin{tikzpicture}
  \matrix (m) [matrix of math nodes,row sep=6em,column sep=6em,minimum width=3em]
  {\mbox{}  & \mbox{} & f(\ourG,\mathbb{B},\ourB) & \mbox{} & \mbox{}  \\
   f(Q,B_Q) &  f(\ourG,\ourB) & f(R) & f(\mathfrak{G},\mathfrak{B})  & f(T,B_T) \\
   f(Q) & f(\ourG)& \text{GR} & f(\mathfrak{G}) & f(T)  \\};
  \path[-stealth]
  (m-1-3) edge node [left] { \parbox{2cm}{   \centering \text{Lorentz invar.} \text{(metrical)}} $\ \ $}  (m-2-2)
  (m-1-3) edge node [right] {  $\ \ $ \parbox{2cm}{   \centering \text{diff. invar.} \text{(tetradic)}}} (m-2-4)
  (m-2-2) edge node [above] {$\text{diff. invar}$} (m-2-3)
  (m-2-4) edge node [above] {$\text{Lorentz invar}$} (m-2-3)
   (m-2-3) edge node [left] {} (m-3-3)
    (m-2-2) edge node [right] {$\text{2nd-order}$} (m-3-2)
    (m-2-4) edge node [left] {$\text{2nd-order}$} (m-3-4)
     (m-2-4) edge node [left] {$\text{2nd-order}$} (m-3-4)
      (m-2-1) edge node [left] {$\text{2nd-order}$} (m-3-1)
       (m-2-5) edge node [right] {$\text{2nd-order}$} (m-3-5)
        (m-3-2) edge node [below] {$\text{diff. invar.}$} (m-3-3)
         (m-3-4) edge node [below] {$\text{Lorentz invar.}$} (m-3-3);
     \path[-stealth]
       (m-2-4)  edge[bend left=12] node [above] {$\text{covariantise}$} (m-2-5)
       (m-2-5)  edge[bend left=12] node [below] { \parbox{2cm}{   \centering \text{gauge fix} $ \omega =0$}} (m-2-4)
       (m-2-2)  edge[bend right=12] node [above] {$\text{covariantise}$} (m-2-1)
       (m-2-1)  edge[bend right=12] node [below] { \parbox{2cm}{   \centering \text{gauge fix} $ \bar{\Gamma} =0$}} (m-2-2)
      (m-3-1) edge[bend right=12] node[below]  { \parbox{2cm}{   \centering \text{gauge fix} $  \bar{\Gamma} =0$}}(m-3-2)
      (m-3-2) edge[bend right=12] node[above] {$\text{covariantise}$} (m-3-1)
      (m-3-4)  edge[bend left=12] node [above] {$\text{covariantise}$} (m-3-5)
       (m-3-5)  edge[bend left=12] node [below] { \parbox{2cm}{   \centering \text{gauge fix} $ \omega =0$}} (m-3-4);
\end{tikzpicture}
\caption{Relationships between the Riemannian $f(\ourG,\mathbb{B},\ourB)$ class of theories. The most general setup is $f(\ourG,\mathbb{B},\ourB)$, from which we obtain $f(\ourG,\ourB)$ or $f(\mathfrak{G},\mathfrak{B})$ in the metric and tetradic representations respectively. The former is locally Lorentz invariant while the latter is diffeomorphism invariant.
Covariantisation of these models leads to the modified symmetric and metric teleparallel theories. Considering only the bulk terms leads to the second-order modifications. From $f(\ourG)$ and $f(\mathfrak{G})$ we obtain GR by requiring invariance under diffeomorphisms and local Lorentz transformations.}
\label{fig:unified}
\end{figure}
\vfill
\end{landscape}

As we have discussed, covariantisation of these theories gives rise to the teleparallel modifications. Specifically, we retrieve $f(Q,B_Q)$ gravity from the $f(\ourG,\ourB)$ theories, and $f(T,B_T)$ from the $f(\mathfrak{G},\mathfrak{B})$ theories. Despite the teleparallel theories being fully covariant, the $f(R)$ limit still represents a special case possessing an `enhanced' symmetry: invariance under coordinate transformations of the metric tensor and local Lorentz transformations of the tetrad alone (without considering the affine connection). This is also special because the degrees of freedom in $f(R)$ gravity are fewer than in the modified teleparallel theories~\cite{DAmbrosio:2023asf}. This can be understood as the (dynamical) fields parameterising the teleparallel connection dropping out from the action entirely in the $f(R)$ limit. We then are led to GR in the second-order, fully invariant limit, as expected.

On the other hand, if one wanted to work in the other direction, one can begin in the teleparallel framework and simply fix the gauge using the formulae developed in this work. This could be accomplished by introducing Lagrange multipliers to the teleparallel actions. We will briefly comment on this approach in the next section, when working in the metric-affine geometries. 

We should also point out the similar modifications in general teleparallel gravity~\cite{Hohmann:2022mlc,BeltranJimenez:2019odq}, described by $f(\mathring{G})$. There, $\mathring{G}$ represents the quadratic contortion terms of a teleparallel geometry with both torsion and non-metricity. This is the generalisation of the teleparallel action given in~(\ref{teleparallel_general}). Clearly this is closely related to our non-covariant theory here. However, the contractions of the contortion tensor include torsion and non-metricity cross terms, which are not immediately present in our $f(\ourG,\mathbb{B},\ourB)$ model. The simplest realisation of this theory into our non-covariant scheme would be to work in a general anholonomic frame which is not orthonormal~\cite{Adak:2023ymc}. Then, the gauge-fixed theories would correspond to the vanishing affine connection in the mixed basis. However, this is beyond the scope of this thesis.

This unified framework also serves as a useful tool to compare and contrast the differences between the metric teleparallel and the symmetric teleparallel theories. It is curious that the equations of motion for these theories often coincide, leading to equivalent solutions\footnote{Of course, when perturbations are taken into account, metric vs tetrad perturbations may lead to differences.}. The case of FLRW cosmology is well-known and well-studied, and it can be attributed to the high amount of symmetry that these spacetimes possess. The same also occurs in spherically symmetric spacetimes, which we first pointed out in~\cite{Boehmer:2021aji}, and has later been studied in more detail~\cite{DAmbrosio:2021zpm}.

Clearly a necessary condition for the equivalence of these theories is for the boundary term $\mathbb{B} = \mathfrak{G} -\ourG$ to vanish. As we noted, $\mathbb{B}$ is neither a coordinate scalar nor a Lorentz scalar, hence depends on both the coordinates and frames. One would hope that studying a theory of the form $f(\ourG+\alpha \mathbb{B})$, and using the additional constraints that come from diffeomorphism invariance and local Lorentz invariance, would be useful in further investigating the relationship between the $f(\ourG)$ and $f(\mathfrak{G})$ theories. It would be interesting to prove (or disprove) whether a background solution of $f(Q)$ gravity is always a solution of $f(T)$ gravity, and vice-versa. Our gauge-fixed theories seem to be a useful approach for these types of questions, where we only have the metric and tetrad fields to deal with.

\section{Metric-affine generalisations} 
\label{section5.3}

Here we return to the metric-affine setting with an independent metric tensor and affine connection. This allows us to make contact with the other metric-affine theories directly, such as Einstein-Cartan theory. In particular, beginning with the affine decomposition studied in Sec.~\ref{section4.3}, we will assume that the total curvature $\bar{R}_{\mu \nu \gamma}{}^{\lambda}$ is non-vanishing. To set either non-metricity or torsion to zero, we will use the Lagrange multiplier method employed throughout the previous chapters.

Setting both torsion and non-metricity to zero will return to the Levi-Civita framework. But we will see that in some cases, such as when non-linear functions of boundary terms are considered, the modified theories do not reduce to their metric counterparts studied in the previous sections. This will be explained in more detail below.

\subsection{\texorpdfstring{$f(\bar{\ourG},\bar{\ourB})$}{f(G,B)} gravity}
\label{section5.3.1}

Let us now study the modifications of the metric-affine decomposition of the Ricci scalar $\bar{R} = \bar{\ourG} + \bar{\ourB}$, where we remind the reader that $\bar{\ourG}$ and $\bar{\ourB}$ are defined as
\begin{align}
  \nonumber
  \bar{\mathbf{G}} &:= g^{\mu \lambda} \big( \bar{\Gamma}^{\kappa}_{\kappa \rho} \bar{\Gamma}^{\rho}_{\mu \lambda} - \bar{\Gamma}^{\kappa}_{\mu \rho} \bar{\Gamma}^{\rho}_{\kappa \lambda} \big)
  + \big(\bar{\Gamma}^{\mu}_{\mu \lambda} \delta^{\nu}_{\kappa} - \bar{\Gamma}^{\nu}_{\kappa \lambda} \big) \big( \partial_{\nu} g^{\kappa \lambda} - \frac{1}{2} g_{\alpha \beta} g^{\kappa \lambda} \partial_{\nu} g^{\alpha \beta} \big)\,, \\
  \nonumber
  \mathbf{\bar{B}} &:= \frac{1}{\sqrt{-g}} \partial_{\kappa} \big(\sqrt{-g}(g^{\mu \lambda}\bar{\Gamma}^{\kappa}_{\mu \lambda} - g^{\kappa \lambda} \bar{\Gamma}^{\mu}_{\mu \lambda})\big)\,.
\end{align}
Recall that all terms with a bar now refer to their metric-affine versions, and we work in the natural coordinate basis.

We propose the modified gravitational action to be an arbitrary function of the affine bulk and boundary terms~\cite{Boehmer:2023fyl}
\begin{align} \label{affine_action}
  S_{\textrm{ grav}} [g, \bar{\Gamma}] =  \frac{1}{2\kappa} \int f(\bar{\ourG}, \bar{\ourB}) \sqrt{-g}\, d^4x \, .
\end{align}
Variations with respect to the independent metric and connection lead to
\begin{multline}
  \label{affine_EoM}
  \delta S_{\textrm{grav}} = \frac{1}{2\kappa} \int \biggl\{
  \delta g^{\mu \nu} \Big[
    - \frac{1}{2} g_{\mu \nu} f +
    f_{,\bar{\ourG}} \big(\bar{G}_{\mu \nu} + \frac{1}{2} g_{\mu \nu} \bar{\ourG}\big) +
    \frac{1}{2} \bar{E}_{\mu \nu}{}^{\lambda} \partial_{\lambda} f_{,\bar{\ourG}}  \\
    +
    \frac{1}{2} g_{\mu \nu} f_{,\bar{\ourB}} \bar{\ourB} -
    \frac{1}{2}  \bar{E}_{\mu \nu}{}^{\lambda}\partial_{\lambda}f_{,\bar{\ourB}}
    \Big] +
  \delta \bar{\Gamma}{}^{\lambda}_{\mu \nu} \Big[ P^{\mu \nu}{}_{\lambda} f_{,\bar{\ourG}} +
    2 \partial_{\sigma} f_{,\bar{\ourB}}\delta_{\lambda}^{[\mu} g^{\sigma] \nu}
    \Big]
  \biggr\} \sqrt{-g}\, d^4 x \,,
\end{multline}
where the superpotential $\bar{E}_{\mu \nu}{}^{\lambda}$ is defined in~(\ref{Ebar}). For the derivation of the field equations, see Appendix~\ref{appendixC.2.2}. Let us also include the matter action $S_{\textrm{M}}[g,\bar{\Gamma},\varPhi^{A}]$, which we assume to be minimally coupled to gravity. The full field equations are  
\begin{equation}
  \label{affine_field1}
  - \frac{1}{2} g_{\mu \nu} f +
  f_{,\bar{\ourG}} \big(\bar{G}_{(\mu \nu)} + \frac{1}{2} g_{\mu \nu} \bar{\ourG}\big) +
  \frac{1}{2} \bar{E}_{(\mu \nu)}{}^{\lambda} \partial_{\lambda} (f_{,\bar{\ourG}} - f_{,\bar{\ourB}}) +
  \frac{1}{2} g_{\mu \nu} f_{,\bar{\ourB}} \bar{\ourB} =
  \kappa {}^{(\bar{\Gamma})}T_{\mu \nu} \,,
  \end{equation}
  \begin{equation}
  \label{affine_field2}
  P^{\mu \nu}{}_{\lambda} f_{,\bar{\ourG}} +
  2 \partial_{\rho} f_{,\bar{\ourB}} \delta_{\lambda}^{[\mu} g^{\rho] \nu} =
  2\kappa \Delta^{\mu \nu}{}_{\lambda} \,,
  \end{equation}
where the metric (Hilbert) energy-momentum tensor is defined in~(\ref{affine_EM}) and the hypermomentum defined in~(\ref{hypermomentum}).

In the case that the action depends only on the bulk term $f(\bar{\ourG})$, all $ f_{,\bar{\ourB}} $ terms vanish. Then, the connection equation is proportional to Palatini tensor $P^{\mu \nu}{}_{\lambda}$~(\ref{Palatini}). Recall that $\bar{\ourG}$ is also algebraic in torsion and non-metricity~(\ref{Gbar}), therefore in the $f(\ourG)$ theories the connection field equation is only algebraic. It is well known that in Einstein-Cartan theory, torsion does not propagate, which implies the absence of torsional gravitational waves in vacuum. This is in stark contrast to metric perturbations, which can propagate through vacuum regions of spacetime.
In order to obtain dynamical behaviour in either of these geometric quantities, we must include the boundary term $\bar{\ourB}$. It is precisely those boundary terms that contain the derivatives of torsion and non-metricity. 

It is interesting to note that making the Palatini variations of $f(\bar{\ourG})$ and then choosing the Levi-Civita connection (consistent with solving the connection field equation with vanishing hypermomentum) leads back to the metric $f(\ourG)$ gravity theory~(\ref{f(G)field_equation}). However, performing the same procedure with $f(\bar{\ourG},\bar{\ourB})$ does not yield the metric $f(\ourG,\ourB)$ model. This is perhaps unsurprising as the same situation occurs in Palatini $f(R)$ gravity~\cite{Sotiriou:2006qn,Sotiriou:2008rp,DeFelice:2010aj}. The reason is clear from the form of~(\ref{affine_EoM}): if one assumes the Levi-Civita connection, the Palatini tensor $P^{\mu \nu}{}_{\lambda}$ vanishes identically. Therefore none of the content of the $f(\ourG)$ model is lost in the connection equations of motion. In contrast, the fourth-order terms coming from the $f(\ldots,\ourB)$ field equation are lost in the connection equation of motion. Specifically, the Levi-Civita connection includes derivatives of the metric, which means an integration by parts would be performed on the $\partial_{\lambda} f_{,\ourB}$ term, leading to the fourth-order terms in the Levi-Civita variation of the $f(\ourG,\ourB)$ action.

In order to make direct contact with the teleparallel theories, one would need to consider a more general action than~(\ref{affine_action}). Firstly, looking back at Fig.~\ref{fig:unified}, one can imagine a slightly more general setting with $f(\bar{\ourG},\mathbb{B},\bar{\ourB})$, where $\mathbb{B}$ relates the coordinate basis and orthonormal basis bulk and boundary terms. This extension follows readily from the work in this chapter and the previous one. In the Levi-Civita limit, we would obtain all of Fig.~\ref{fig:unified}. Decomposing the affine quantities into their different geometric parts~(\ref{fundamental}) yields an even more general type of theory $f(\ourG,T,Q,\ourC,\mathbb{B},\ourB,B_T,B_Q)$, where again $\mathbb{B}$ is used to relate the coordinate and orthonormal bases. Teleparallel theories are then directly included in this action. Moreover, one easily obtains the limit $f(\ourG+T+Q+\ourC, \ourB -B_T + B_Q) = f(\bar{\ourG},\bar{\ourB})$. 
In this section we will continue to work the $ f(\bar{\ourG},\bar{\ourB})$ theories, but this discussion should show how more general theories can be constructed, and how these relate with one another.

\subsection{Diffeomorphism and projective invariance}
\label{section5.3.2}

Let us again return to the topic of symmetry and invariance for our modified action.
In Sec.~\ref{section4.4.2} it was shown that the affine bulk and boundary terms were projectively invariant under the transformation
\begin{equation}
\bar{\Gamma}^{\gamma}_{\alpha \beta} \rightarrow \bar{\Gamma}^{\gamma}_{\alpha \beta} + \delta^{\gamma}_{\beta} P_{\alpha} \, .\nonumber
\end{equation}
The same was the case for the affine Ricci scalar $\bar{R}$, and in Sec.~\ref{section3.1.1} we showed that the consequence was that the Palatini tensor is trace-free $P^{\mu \nu}{}_{\nu} = 0$. Similarly, our modified action $f(\bar{\ourG}, \bar{\ourB})$ is then invariant under projective transformations, and this is reflected in the connection equation also being trace-free over the same indices. This is easily verified
\begin{equation}
  P^{\mu \nu}{}_{\nu} f_{,\bar{\ourG}} +2\partial_{\lambda} f_{, \bar{\ourB}} g^{\nu [\lambda} \delta^{\mu]}_{\nu} =
  \partial_{\lambda} f_{, \bar{\ourB}} g^{\mu \lambda} -  \partial_{\lambda} f_{, \bar{\ourB}} g^{\lambda \mu} = 0 \, .\end{equation}
The situation here is the same as in the standard metric-affine GR case, or in modifications such as metric-affine $f(\bar{R})$ gravity. There is little else to say on this topic, as these modifications neither help nor hinder any issues associated with projective invariance.

Moving on to diffeomorphisms, the action transforms under an infinitesimal symmetry transformation as
\begin{align}  \nonumber
\delta_{\xi} S_{\textrm{grav}} &= \frac{1}{2\kappa} \int \mathcal{L}_{\xi} \big( \sqrt{-g}  f(\bar{\ourG},\bar{\ourB}) \big) d^4x 
\\
&= \frac{1}{2\kappa} \int \xi^{\lambda}  \partial_{\mu} \partial_{\nu} \Big[ \sqrt{-g} M^{\mu \nu}{}_{\lambda} \big(  f_{,\bar{\ourG}}  -  f_{,\bar{\ourB}}  \big)  \Big] d^4 x  \,.
\end{align}
Here we have performed the same calculations shown explicitly in Sec.~\ref{section5.1.1}, and discarded any surface terms. We have also made use of the formulae for the Lie derivatives of these affine terms, see equations~(\ref{Lie_Gbar})-(\ref{Lie_Bbar}). For arbitrary $\xi$, this leaves the expression
\begin{align}  \label{affine_conservation}
 \partial_{\mu} \partial_{\nu} \Big[ \sqrt{-g} M^{\mu \nu}{}_{\lambda} \big(  f_{,\bar{\ourG}}  -  f_{,\bar{\ourB}}  \big)  \Big] \, .
\end{align}
This expression could also be easily obtained by using the decompositions of the affine objects
\begin{equation}
 \bar{\ourG} = \ourG + T + Q + C \, , \qquad \bar{\ourB} = \ourB - B_T + B_Q  \, ,
 \nonumber
\end{equation}
which we gave in equations~(\ref{G_scalars}) and~(\ref{B_scalars}). All terms on the right-hand side, except the bulk and boundary terms, are coordinate scalars. Hence, under a diffeomorphism, it is only the $\ourG$ and $\ourB$ term that transforms non-covariantly, which directly leads to the conservation equation~(\ref{affine_conservation}).

Unlike for an invariant scalar action, the above transformation need not vanish identically. The expression is also non-covariant and depends on the choice of coordinates, due to the non-tensorial nature of the bulk and boundary terms. One immediately sees that for the choice of function $f(\bar{\ourG} + \bar{\ourB})=f(\bar{R})$ the expression vanishes, reflecting the diffeomorphism invariance of the Palatini $f(\bar{R})$ theories.

If we insist that our gravitational action remain invariant under the above transformations, we obtain the conservation law
\begin{align} \label{affine_cons1}
 \partial_{\mu} \partial_{\nu} \Big[ \sqrt{-g} M^{\mu \nu}{}_{\lambda} \big(  f_{,\bar{\ourG}}  -  f_{,\bar{\ourB}}  \big)  \Big] = 0  \, .
\end{align}
We discussed this topic in detail in Sec.~\ref{section5.1.1}, so we refer to that section for further analysis.
If we also include minimally coupled matter in our total action, which we remind the reader depends on the metric and the connection, we have the following (on-shell) conservation law
\begin{multline} \label{affine_cons2}
\frac{1}{2\kappa} \partial_{\mu} \partial_{\nu} \Big[ \sqrt{-g} M^{\mu \nu}{}_{\lambda} \big(  f_{,\bar{\ourG}}  -  f_{,\bar{\ourB}}  \big)  \Big] -\sqrt{-g}\nabla_{\mu} {}^{(\bar{\Gamma})}T^{\mu}{}_{\lambda} 
\\
- \bar{\nabla}_{\nu} \bar{\nabla}_{\mu}(\sqrt{-g} \Delta^{\mu \nu}{}_{\lambda}) - \sqrt{-g} \bar{R}_{\lambda \mu \nu}{}^{\rho} \Delta^{\mu \nu}{}_{\rho} - \bar{\nabla}_{\mu}(\sqrt{-g} \Delta^{\mu \nu}{}_{\rho}) T^{\rho}{}_{\nu \lambda}
  = 0 \,.
\end{multline}
For a consistent variation principle of the total action $S_{\textrm{total}} = S_{\textrm{grav}} + S_{\textrm{M}}$ we require the above equation~(\ref{affine_cons2}) to be satisfied. This is the most general case, as we do not consider non-minimal couplings here. One sees that the equation~(\ref{affine_cons2}) reduces to the metric $f(\ourG,\ourB)$ conservation law given in equation~(\ref{eqn:cons0}) when the connection is Levi-Civita and hypermomentum is zero.

 If the matter action $S_{\textrm{M}}$ is a coordinate scalar then we must impose that the first of these equations~(\ref{affine_cons1}) is satisfied. In that case, one obtains the usual metric-affine matter-hypermomentum conservation laws
\begin{multline} \label{affine_cons3}
 \sqrt{-g}\nabla_{\mu} {}^{(\bar{\Gamma})}T^{\mu}{}_{\lambda} 
+ \bar{\nabla}_{\nu} \bar{\nabla}_{\mu}(\sqrt{-g} \Delta^{\mu \nu}{}_{\lambda}) + \sqrt{-g} \bar{R}_{\lambda \mu \nu}{}^{\rho} \Delta^{\mu \nu}{}_{\rho} \\ 
+ \bar{\nabla}_{\mu}(\sqrt{-g} \Delta^{\mu \nu}{}_{\rho}) T^{\rho}{}_{\nu \lambda}
  = 0 \, ,
\end{multline}
see for example~\cite{Hohmann:2021fpr}. Vanishing hypermomentum leads to the standard energy-momentum conservation law $\nabla_{\mu} T^{\mu}{}_{\lambda} =0$. In Chapter~\ref{chapter6}, when working in the cosmological setting, we will work with the standard perfect fluid matter representations, where hypermomentum vanishes and the matter action is a scalar. We will see that this forces the choice of coordinates to satisfy the gravitational conservation constraint~(\ref{affine_cons1}).

In our work~\cite{Boehmer:2023fyl}, where we first proposed this theory, we conjectured on the covariantisation of the $f(\bar{\ourG},\bar{\ourB})$ models. Our main questions were the following: how would such a procedure be implemented, and what would the resulting theory look like? We now feel to be in a better position to answer such questions. In Section~\ref{section4.3} we showed how the Stueckelberg covariantisation could be applied to the Levi-Civita bulk and boundary terms in both the coordinate and orthonormal bases. Importantly, this procedure made no reference to geometry, even though it turned out that the compensating Stueckelberg fields could be given a geometric interpretation. It follows that the same procedure could be applied here, i.e., using the decomposition of $\bar{\ourG}$ into $\ourG$ and the other affine scalars~(\ref{G_scalars}), and then applying the Stueckelberg trick to the (metric) bulk part $\ourG$. This appears to give a satisfactory answer to the first question.

With regards to the second question, on the interpretation of the resulting theory, we must note that because the connection is not flat $\bar{R}_{\mu \nu \lambda}{}^{\gamma} \neq 0$, the Stueckelberg fields cannot be related to the affine connection of the spacetime. In other words, we do not obtain a neat geometric picture like we did when performing the covariantisation of the metric $f(\ourG,\ourB)$ theories.

However, one answer to this second question is that they could be related to a \textit{different}, independent teleparallel connection. This seems reminiscent of the modified theories with two independent connections~\cite{Tamanini:2012mi}, as well as the so-called `hybrid-Palatini' theories~\cite{Harko:2011nh,Tamanini:2013ltp}. In our case, the second (auxiliary) connection will be in terms of the Stueckelberg fields, taking the form of the `purely gauge' symmetric teleparallel connection~(\ref{STG_connection}). This was actually shown to be consistent by Milgrom (in the context of the bimetric MOND theories~\cite{Milgrom:2009gv}), where the connection associated with the auxiliary metric was fixed to be the symmetric teleparallel one~\cite{Milgrom:2019rtd}. This led to the covariant $f(Q)$ theories. A similar approach was used for the non-covariant Einstein action by introducing an additional reference connection and metric~\cite{Tomboulis:2017fim}.
This would be an interesting avenue to explore further in the metric-affine context, which we believe has not been looked at before.

\subsection{Equivalence with metric-affine $f(R)$ gravity}
\label{section5.3.3}

Metric-affine $f(\bar{R})$ gravity is of course also included in our approach and achieved by simply setting $f(\bar{\ourG},\bar{\ourB}) = f(\bar{\ourG} + \bar{\ourB})$. This again follows from the decomposition of the affine Ricci scalar~(\ref{RGB_affine}).
For the metric field equations~(\ref{affine_field1}), the $\bar{E}$ terms cancel to give
\begin{align}
  - \frac{1}{2} g_{\mu \nu} f + f'(R) \bigl(\bar{G}_{(\mu \nu)} +
  \frac{1}{2} g_{\mu \nu} \bar{\ourG} + \frac{1}{2} g_{\mu \nu} \bar{\ourB} \bigr) =
  \kappa    {}^{(\bar{\Gamma})} T_{\mu \nu} \,,
\end{align}
which can be written as
\begin{align} \label{f(R)}
  f'(\bar{R}) \bar{R}_{(\mu \nu)} - \frac{1}{2} g_{\mu \nu} f(\bar{R}) =
  \kappa   {}^{(\bar{\Gamma})} T_{\mu \nu} \,,
\end{align}
where we have used the relations $\bar{G}_{\mu \nu} + g_{\mu \nu} \bar{R}/2 = \bar{R}_{\mu \nu}$ and $\bar{\ourG}+\bar{\ourB} = \bar{R}$. The connection field equation is 
\begin{align}
  \label{f(R)_connection}
  P^{\mu \nu}{}_{\lambda} f'(R) + 2 \partial_{\rho} f'(\bar{R}) \delta_{\lambda}^{[\mu} g^{\rho] \nu} =
  2\kappa \Delta^{\mu \nu}{}_{\lambda} \,.
\end{align}
This can be written in a more familiar form by noting that
\begin{multline}
  \label{covar_deriv}
  \frac{2}{\sqrt{-g}} \bar{\nabla}_{\rho} (f'(R) \sqrt{-g}  \delta_{\lambda}^{[\mu} g^{\rho] \nu}) =
  2 \partial_{\rho} f'(\bar{R}) \delta_{\lambda}^{[\mu} g^{\rho] \nu} \\
  + f'(\bar{R})
  \Big(\frac{1}{2} Q_{\lambda}{}^{\rho}{}_{\rho} g^{\mu \nu} -
  \frac{1}{2} Q^{\nu \rho}{}_{\rho} \delta^{\mu}_{\lambda} +
  Q_{\rho}{}^{\rho \nu} \delta^{\mu}_{\lambda} - Q_{\lambda}{}^{\mu \nu }\Big)
  \\ =
  2 \partial_{\rho} f'(\bar{R}) \delta_{\lambda}^{[\mu} g^{\rho] \nu}  +
  f'(\bar{R})  P^{\mu \nu}{}_{\lambda} -  f'(\bar{R})(T^{\mu}{}_{\lambda}{}^{\nu} +
  g^{\mu \nu} T^{\rho}{}_{\rho \lambda} + \delta^{\mu}_{\lambda} T^{\rho \nu}{}_{\rho}) \,,
\end{multline}
where in the first line we expanded out the covariant derivatives of the metric in terms of the non-metricity tensor, and in the second line we inserted the definition of $P^{\mu \nu}{}_{\lambda}$. Putting together~(\ref{f(R)_connection}) and~(\ref{covar_deriv}) we arrive at
\begin{align}
  \label{f(R)_connection_final}
  \frac{2}{\sqrt{-g}} \bar{\nabla}_{\rho} (f'(R) \sqrt{-g} \delta_{\lambda}^{[\mu} g^{\rho] \nu}) +
  f'(\bar{R})(T^{\mu}{}_{\lambda}{}^{\nu} + g^{\mu \nu} T^{\rho}{}_{\rho \lambda} +
  \delta^{\mu}_{\lambda} T^{\rho \nu}{}_{\rho} )  =2 \kappa \Delta^{\mu \nu}{}_{\lambda} \,.
\end{align}
This is the familiar form of the $f(\bar{R})$ connection equation of motion~\cite{Sotiriou:2006qn}. 

In the case that matter does not couple to the connection, \textit{\`{a} la} Palatini, it is known that this theory is equivalent to the scalar-tensor theory of Brans-Dicke~(\ref{ST_Jordan}) with parameter $\omega = -3/2$~\cite{Sotiriou:2006hs}. Interestingly, this differs from its metric $f(R)$ counterpart with $\omega=0$. As discussed earlier, this inequivalence is related to the boundary terms present in the affine Ricci scalar.

Taking the trace of the metric equation~(\ref{f(R)}) leads to the algebraic  relation $-2f(\bar{R}) + f'(\bar{R})\bar{R} = \kappa T$. This implies an algebraic relation between the Ricci scalar $\bar{R}$ and the trace of the energy-momentum tensor $T$~\cite{Sotiriou:2006hs}. Contrast this with the $f(\bar{\ourG},\bar{\ourB})$ equations~(\ref{affine_field1}), where the trace leads to a dynamical equation as opposed to an algebraic one. This leads to a number of interesting possibilities beyond what can be found in the metric-affine $f(\bar{R})$ theories.

\subsection{Modified Einstein-Cartan gravity}
\label{section5.3.4}

Einstein-Cartan theory is derived in the metric-affine setting by imposing metric-compatibility of the connection. This was achieved in Sec.~\ref{section3.2.2} by using the Lagrange multiplier action~(\ref{EC_lambda}). Once the Lagrange multiplier was eliminated from the equations of motion, they read 
\begin{align} \nonumber
  \tilde{G}_{(\mu \nu)} &=
  \kappa {}^{(\bar{\Gamma})} T_{\mu \nu} + \kappa (\tilde{\nabla}_{\rho} + T^{\sigma}{}_{\rho \sigma}) \Delta^{\rho}{}_{(\mu \nu)}\,, \\ \nonumber
  \tilde{P}_{\mu \nu \rho} &= 2\kappa \Delta_{\mu[\nu \rho]} \,,
\end{align}
see~(\ref{EEl1b})-(\ref{EC2b}). Also recall that we were able to make use of geometric identities in order to rewrite the first field equation in such a way that the symmetry brackets could be removed
\begin{equation}
  \tilde{G}_{\mu \nu} =
  \kappa {}^{(\bar{\Gamma})} T_{\mu \nu} + \kappa (\tilde{\nabla}_{\rho} + T^{\sigma}{}_{\rho \sigma}) \Delta^{\rho}{}_{\nu \mu} \,, \nonumber
\end{equation}
Its symmetric part gives the original first field equation, whilst its skew part coincides with the second equation. Moreover, the right-hand side was shown to be equivalent with the canonical energy-momentum tensor~(\ref{Belinfante}).

Following our work~\cite{Boehmer:2023fyl},  and using the same Lagrange multiplier term
\begin{align} \nonumber
  S_Q = -\int \frac{1}{2 }\lambda^{\mu \nu \rho} Q_{\mu \nu \rho} \sqrt{-g}\, d^4 x =
  \int \frac{1}{2 }\lambda^{\mu \nu \rho} \bar{\nabla}_{\mu} g_{\nu \rho} \sqrt{-g}\, d^4 x \,.
\end{align}
let us apply this to our modified theory. The complete modified metric field equation~(\ref{affine_field1}) now reads
\begin{multline}
  \label{fieldc1}
  - \frac{1}{2} g_{\mu \nu} f +
  f_{,\tilde{\ourG}} \big(\tilde{G}_{(\mu \nu)} + \frac{1}{2} g_{\mu \nu} \tilde{\ourG}\big) + 
  \frac{1}{2} \tilde{E}_{(\mu \nu)}{}^{\sigma} \partial_{\sigma} f_{,\tilde{\ourG}} \\ +
  \frac{1}{2} g_{\mu \nu} f_{,\tilde{\ourB}} \tilde{\ourB} -
  \frac{1}{2} \tilde{E}_{(\mu \nu)}{}^{\sigma} \partial_{\sigma}f_{,\tilde{\ourB}} +
  \kappa \tilde{\nabla}_{\sigma} \lambda^{\sigma}{}_{\mu \nu} + \kappa T^{\rho}{}_{\sigma \rho}\lambda^{\sigma}{}_{\mu \nu} =
  \kappa {}^{(\bar{\Gamma})}T_{\mu \nu} \,,
\end{multline}
while the connection~(\ref{affine_field2}) and constraint equation are given by
\begin{align}
  \label{fieldc2}
  \tilde{P}^{\mu \nu}{}_{\rho} f_{,\tilde{\ourG}} +
  2 \partial_{\sigma} f_{,\tilde{\ourB}} \delta_{\rho}^{[\mu} g^{\sigma] \nu} - 2 \kappa \lambda^{\mu \nu}{}_{\rho}
  &= 2\kappa \Delta^{\mu \nu}{}_{\rho} \,, \\
  \label{fieldc3}
  \bar{\nabla}_{\mu} g_{\nu \rho} &= 0 \,.
\end{align}
Note that we have already implemented the constraint equation to remove any non-metricity terms which would otherwise be present in the field equations. For this reason, affine quantities appearing in the field equations now are denoted with a tilde, which implies non-metricity has been set to zero.

We will now follow the same method as shown for the standard Einstein-Cartan theory of eliminating the Lagrange multipliers and writing the equations in a unified way. We begin with the $f(\tilde{\ourG})$ case, without boundary terms, and then look at the more general $f(\tilde{\ourG},\tilde{\ourB})$ theories.

\subsubsection{Modified Einstein-Cartan theory without boundary terms}

Let us now return to the field equations~(\ref{fieldc1})--(\ref{fieldc3}) and assume $f(\tilde{\ourG},\tilde{\ourB}) = f(\tilde{\ourG})$ so that $f_{,\tilde{\ourB}} = 0$. Moreover, we assume that non-metricity vanishes via the introduction of the Lagrange multiplier, in which case the field equations simplify to
\begin{multline}
  \label{fieldd1EC}
  - \frac{1}{2} g_{\mu \nu} f +
  f_{,\tilde{\ourG}} \big(\tilde{G}_{(\mu \nu)} + \frac{1}{2} g_{\mu \nu} \tilde{\ourG}\big) +
  \frac{1}{2} \tilde{E}_{(\mu \nu)}{}^{\rho} \partial_{\rho} f_{,\tilde{\ourG}}  \\ +
  \kappa \tilde{\nabla}_{\rho} \lambda^{\rho}{}_{\mu \nu} + \kappa T^{\sigma}{}_{\rho \sigma}\lambda^{\rho}{}_{\mu \nu} =
  \kappa {}^{(\bar{\Gamma})}T_{\mu \nu} \,,
  \end{multline}
  \begin{equation}
  \tilde{P}^{\mu \nu}{}_{\rho} f_{,\tilde{\ourG}} - 2 \kappa \lambda^{\mu \nu}{}_{\rho} = 2\kappa \Delta^{\mu \nu}{}_{\rho} \,.
  \end{equation}
As before, the Palatini tensor is now skew-symmetric in the last pair of indices and we find $-\lambda_{\mu \nu \rho} = \Delta_{\mu(\nu \rho)}$ and $\tilde{P}_{\mu \nu \rho} f_{,\tilde{\ourG}} = 2\kappa \Delta_{\mu[\nu \rho]}$. Hence, we can write the first and second field equations without any Lagrange multipliers in the form
\begin{multline}   \label{fieldd2EC}
 - \frac{1}{2} g_{\mu \nu} f +
  f_{,\tilde{\ourG}} \big(\tilde{G}_{(\mu \nu)} + \frac{1}{2} g_{\mu \nu} \tilde{\ourG}\big) +
  \frac{1}{2} \tilde{E}_{(\mu \nu)}{}^{\rho} \partial_{\rho} f_{,\tilde{\ourG}} \\ =
  \kappa \Bigl({}^{(\bar{\Gamma})}T_{\mu \nu} + (\tilde{\nabla}_{\rho} + T^{\sigma}{}_{\rho \sigma})\Delta^{\rho}{}_{(\nu \mu)}\Bigr)\,,
\end{multline}
\begin{equation}
  \tilde{P}_{\mu \nu \rho} f_{,\tilde{\ourG}} = 2\kappa \Delta_{\mu[\nu \rho]} \,.
  \label{fieldd2b}
\end{equation}
The right-hand side of equation~(\ref{fieldd2EC}) now takes the familiar form of the symmetrised canonical energy-momentum tensor. 

We now wish to follow the same approach that we used in the standard Einstein-Cartan theory, of removing the symmetry brackets and considering the full equation instead. Taking the skew-symmetric part of~(\ref{fieldd2EC}), we find
\begin{align}
  f_{,\tilde{\ourG}} \tilde{G}_{[\mu \nu]} +
  \frac{1}{2} \tilde{E}_{[\mu \nu]}{}^{\rho} \partial_{\rho} f_{,\tilde{\ourG}} &=
  \kappa (\tilde{\nabla}_{\rho} + T^{\sigma}{}_{\rho \sigma})\Delta^{\rho}{}_{[\nu \mu]} \\
  &=
  \frac{1}{2} (\tilde{\nabla}_{\rho} + T^{\sigma}{}_{\rho \sigma})(\tilde{P}^{\rho}{}_{\nu \mu } f_{,\tilde{\ourG}}) \,.
  \label{fieldd3}
\end{align}
After taking into account the geometric identity $\tilde{G}_{[\mu \nu]} = (\tilde{\nabla}_{\rho} + T^{\sigma}{}_{\rho \sigma}) \tilde{P}^{\rho}{}_{ \nu \mu }/2$ we arrive at the interesting equation\footnote{Note that here we have somewhat abused notation by assuming that $ f_{,\tilde{\ourG}}$ acts as a scalar when acted on by the covariant derivative. This is of course not well defined. However, our purpose here is only to rewrite the field equations in a more suggestive form. We will use the original forms~(\ref{fieldd2EC})-(\ref{fieldd2b}) when performing calculations.}
\begin{align}
  \frac{1}{2} \tilde{E}_{[\mu \nu]}{}^{\rho} \partial_{\rho} f_{,\tilde{\ourG}} =
  \frac{1}{2} \tilde{P}^{\rho}{}_{ \nu \mu} \partial_{\rho} f_{,\tilde{\ourG}} \,,
  \label{fieldd4}
\end{align}
which immediately gives $\tilde{E}_{[\mu \nu]}{}^{\rho} = \tilde{P}^{\rho}{}_{ \nu \mu}$. This equation can be seen as determining the skew-symmetric part of $\tilde{E}_{[\mu \nu]}{}^{\rho}$ such that, in complete analogy with Einstein-Cartan theory, the full first field equation contains the second field equation. This allows us to define a new object as follows
\begin{align}
  \mathcal{E}_{\mu \nu}{}^{\rho} &:= \tilde{E}_{(\mu \nu)}{}^{\rho} + \tilde{E}_{[\mu \nu]}{}^{\rho}
  \nonumber \\ &:=
  2 \tilde{\Gamma}^{\rho}_{(\mu \nu)} -
  2 \delta^{\rho}_{(\mu} \tilde{\Gamma}^{\lambda}_{|\lambda |\nu)} - g_{\mu \nu} (g^{\kappa \lambda} \tilde{\Gamma}^{\rho}_{\lambda \kappa} -
  g^{\kappa \rho} \tilde{\Gamma}^{\lambda}_{\lambda \kappa}) + \tilde{P}^{\rho}{}_{ \nu \mu } \nonumber \\
  &= 2 \tilde{\Gamma}{}^{\rho}_{  \mu \nu } + 2 \delta^{\rho}_{[ \mu } \tilde{\Gamma}^{\lambda}_{\nu ]\lambda} - 2 \delta^{\rho}_{ \mu } \tilde{\Gamma}{}^{\lambda}_{\lambda  \nu } - g_{\mu \nu} (g^{\kappa \lambda} \tilde{\Gamma}^{\rho}_{\lambda \kappa} -
  g^{\kappa \rho} \tilde{\Gamma}^{\lambda}_{\lambda \kappa})  \,.
  \label{fieldd5}
\end{align}
We are now able to re-write the complete first field equation
\begin{align}
  - \frac{1}{2} g_{\mu \nu} f +
  f_{,\tilde{\ourG}} \big(\tilde{G}_{\mu \nu} + \frac{1}{2} g_{\mu \nu} \tilde{\ourG}\big) +
  \frac{1}{2} \mathcal{E}_{\mu \nu}{}^{\lambda} \partial_{\lambda} f_{,\tilde{\ourG}} =
  \kappa \Bigl({}^{(\tilde{\Gamma})}T_{\mu \nu} + (\tilde{\nabla}_{\lambda} + T^{\sigma}{}_{\lambda \sigma})\Delta^{\lambda}{}_{ \nu \mu  }\Bigr)\,,
  \label{fieldd6}
\end{align}
so that its skew-symmetric part coincides with the second field equation. This result is somewhat surprising but illustrates the intricate relationship between the two field equations. It also reduces to standard Einstein-Cartan when $f(\bar{\ourG}) = \bar{\ourG}$ as expected.

\subsubsection{Modified Einstein-Cartan theory with boundary terms}

Let us now work with the full $f(\tilde{\ourG},\tilde{\ourB})$ theory, given by equations~(\ref{fieldc1})-(\ref{fieldc3}). The connection equation with lowered indices is
\begin{align}
  \tilde{P}_{\mu \nu \rho} f_{,\tilde{\ourG}} + 2 g_{\mu[\rho} \partial_{\nu]} f_{,\tilde{\ourB}} - 2 \kappa \lambda_{\mu \nu \rho} = 2\kappa \Delta_{\mu \nu \rho} \,,
  \label{ECmod1}
\end{align}
from which we immediately see the symmetric part $\lambda_{\mu \nu \rho} = - \Delta_{\mu(\nu \rho)}$ isolates the Lagrange multiplier, just like before. The skew-symmetric part now includes an additional dynamical term
\begin{align}
  P_{\mu \nu \rho} f_{,\tilde{\ourG}} + 2 g_{\mu[\rho} \partial_{\nu]} f_{,\tilde{\ourB}} = 2\kappa \Delta_{\mu[\nu \rho]}\,.
  \label{ECmod2}
\end{align}
The metric field equation is then
\begin{multline}
  - \frac{1}{2} g_{\mu \nu} f +
  f_{,\tilde{\ourG}} \big(\tilde{G}_{(\mu \nu)} + \frac{1}{2} g_{\mu \nu} \tilde{\ourG}\big) +
  \frac{1}{2} \tilde{E}_{(\mu \nu)}{}^{\lambda} \partial_{\lambda} (f_{,\tilde{\ourG}} - f_{,\tilde{\ourB}})  +
  \frac{1}{2} g_{\mu \nu} f_{,\tilde{\ourB}} \tilde{\ourB} \\
   =
  \kappa \Big( {}^{(\bar{\Gamma})}T_{\mu \nu} + ( \tilde{\nabla}_{\lambda} + T^{\sigma}{}_{\lambda \sigma}) \Delta^{\lambda}{}_{( \nu \mu)} \Big) \,.
  \label{ECmod3}
\end{multline}
Just as before, we now remove the symmetry brackets and consider the full field equation. First, the antisymmetric part can be written as
\begin{align}
  f_{,\tilde{\ourG}} \tilde{G}_{[\mu \nu]} +
  \frac{1}{2} \tilde{E}_{[\mu \nu]}{}^{\lambda} \partial_{\lambda} (f_{,\tilde{\ourG}} -f_{,\tilde{\ourB}} )
=
  \frac{1}{2} (\tilde{\nabla}_{\lambda} + T^{\sigma}{}_{\lambda \sigma})  \Big( \tilde{P}^{\lambda}{}_{  \nu \mu } f_{,\tilde{\ourG}} + 2 \delta^{\lambda}_{[ \mu } \partial_{  \nu ]} f_{,\tilde{\ourB}}  \Big)\,,
  \label{ECmod4}
 \end{align}
where on the right-hand side we have used the skew-symmetric part of the connection equation. By again using the geometric identity $\tilde{G}_{[\mu \nu]} = (\tilde{\nabla}_{\lambda} + T^{\sigma}{}_{\lambda \sigma}) \tilde{P}^{\lambda}{}_{ \nu \mu }/2$ we find 
\begin{align}
  \label{fGB1}
  \frac{1}{2}  \partial_{\lambda} f_{,\tilde{\ourG}}  \Big( \tilde{E}_{[\mu \nu]}{}^{\lambda} - \tilde{P}^{\lambda}{}_{  \nu \mu } \Big) - \frac{1}{2}\tilde{E}_{[\mu \nu]}{}^{\lambda} \partial_{\lambda} f_{,\tilde{\ourB}} = (\tilde{\nabla}_{\lambda} + T^{\sigma}{}_{\lambda \sigma}) \delta^{\lambda}_{[\mu } \partial_{\nu]}  f_{,\tilde{\ourB}} \,.
\end{align} 
The part in the first bracket was derived in the $f(\tilde{\ourG})$ case~(\ref{fieldd4}). We can expand the right-hand side, cancelling the second partial derivatives, to obtain
\begin{align}
 (\tilde{\nabla}_{\lambda} + T^{\sigma}{}_{\lambda \sigma}) \delta^{\lambda}_{[\mu} \partial_{ \nu]}  f_{,\tilde{\ourB}}  &=  \frac{1}{2}\Big( T^{\lambda}{}_{  \nu \mu} \partial_{\lambda}  f_{,\tilde{\ourB}} + T^{\sigma}{}_{ \mu \sigma} \partial_{  \nu}  f_{,\tilde{\ourB}} - T^{\sigma}{}_{ \nu \sigma} \partial_{ \mu}  f_{,\tilde{\ourB}} \Big) \nonumber \\
  &= - \frac{1}{2} \tilde{P}^{\lambda}{}_{  \nu \mu } \partial_{\lambda}    f_{,\tilde{\ourB}} \,,
  \label{ECmod5}
\end{align}
where the first torsion term comes from commutator of covariant derivatives. Putting this back into our skew-symmetric field equation~(\ref{fGB1}) we find
\begin{align}
  \Big(\tilde{E}_{[\mu \nu]}{}^{\lambda} - \tilde{P}^{\lambda}{}_{ \nu \mu} \Big)
  \Big(\partial_{\lambda} f_{,\tilde{\ourG}} - \partial_{\lambda} f_{,\tilde{\ourB}}\Big) = 0 \,,
  \label{ECmod6}
\end{align}
where we again have $\tilde{E}_{[\mu \nu]}{}^{\lambda} = \tilde{P}^{\lambda}{}_{ \nu \mu }$. If we use the object $\mathcal{E}_{\mu \nu}{}^{\lambda}$ introduced in~(\ref{fieldd5}), we can write the full field equation as 
\begin{multline}
  - \frac{1}{2} g_{\mu \nu} f +
  f_{,\tilde{\ourG}} \big(\tilde{G}_{\mu \nu} + \frac{1}{2} g_{\mu \nu} \tilde{\ourG}\big) +
  \frac{1}{2} \mathcal{E}_{\mu \nu}{}^{\lambda} \partial_{\lambda} \big(f_{,\tilde{\ourG}}-f_{,\tilde{\ourB}} \big) +
  \frac{1}{2} g_{\mu \nu} f_{,\tilde{\ourB}} \tilde{\ourB}  \\ =
  \kappa \Big( {}^{(\bar{\Gamma})}T_{\mu \nu} + (\tilde{\nabla}_{\lambda} + T^{\sigma}{}_{\lambda \sigma}) \Delta^{\lambda}{}_{  \nu \mu} \Big) \,.
  \label{ECmod7}
\end{multline}
The symmetric part contains the metric field equation~(\ref{ECmod3}) and the skew part coincides with the connection equation~(\ref{ECmod2}). Again, the right-hand side could be written using the canonical energy-momentum tensor.

The presence of the boundary terms in the form $f_{,\tilde{\ourB}}$ in the connection field equation makes this model distinctly different to the previous modified Einstein-Cartan type theories. In equations~(\ref{EC2b}) and~(\ref{fieldd2b}), the vanishing of the source term $\Delta_{\mu[\nu \lambda]}=0$ (which corresponds to the intrinsic spin of matter) implies the vanishing of the Palatini tensor. This in turn implies the vanishing of the torsion tensor, and one finds that torsion is non-dynamical. This is the case in standard Einstein-Cartan and the $f(\tilde{\ourG})$ modifications.

Let us now set the intrinsic spin to zero $\Delta_{\mu[\nu \lambda]}=0$ in our $f(\tilde{\ourG}, \tilde{\ourB})$ modifications~(\ref{ECmod2}). Using the expression for the Palatini tensor in terms of torsion, we can solve this equation for the torsion tensor
\begin{align}
  T^{\mu}{}_{\lambda \nu} = \frac{1}{f_{,\tilde{\ourG}}}
  \delta^{\mu}_{[\lambda} \partial_{\nu]} f_{,\tilde{\ourB}} \,.
  \label{ECmod8}
\end{align}
Therefore, in general, torsion does not vanish in source-free regions of spacetime. It should be recalled that the boundary term itself contains contributions from torsion, which means that $\partial_{\mu} f_{,\tilde{\ourB}}$ contains (second) derivatives of the torsion tensor. Consequently, Eq.~(\ref{ECmod8}) is a partial differential equation for torsion. Using the chain rule we have
\begin{align}
  \partial_{\mu} f_{,\tilde{\ourB}} = f_{,\tilde{\ourB}\,\tilde{\ourB}} \partial_{\mu}  \tilde{\ourB} +
  f_{,\tilde{\ourB}\,\tilde{\ourG}} \partial_{\mu}  \tilde{\ourG} \,,
  \label{ECmod9}
\end{align}
so that our previous expression~(\ref{ECmod8}) can also be written as
\begin{align}
  T^{\mu}{}_{\lambda \nu} = \frac{f_{,\tilde{\ourB}\,\tilde{\ourB}}}{f_{,\tilde{\ourG}}}
  \delta^{\mu}_{[\lambda} \partial_{\nu]} \tilde{\ourB} +
  \frac{f_{,\tilde{\ourB}\,\tilde{\ourG}}}{f_{,\tilde{\ourG}}}
  \delta^{\mu}_{[\lambda} \partial_{\nu]} \tilde{\ourG} \,.
  \label{ECmod10}
\end{align}
Neglecting the peculiar situations in which the partial derivative terms vanish, we can make the following general statements. Firstly, the $f(\tilde{\ourG},\tilde{\ourB})$ Einstein-Cartan type models have non-dynamical torsion if $f_{,\tilde{\ourB}} = \mathrm{const.}$, which corresponds to functions linear in the boundary term. In other words, this reduces to $f(\tilde{\ourG})$ as we expected. Secondly, if $f_{,\tilde{\ourB}} \neq \mathrm{const.}$, torsion becomes dynamical and has the potential to propagate.

As was already remarked on various occasions, these boundary terms play a crucial role in these modified theories of gravity. Models depending on non-linear functions of these boundary terms show distinctly different properties than models with a linear term. In the next chapter we will study the cosmological applications of these modified boundary models in the Einstein-Cartan framework.

Also note that just like in all the theories studied in this chapter, equation~(\ref{ECmod10}) has the peculiar property of having some tensor quantities (here the torsion tensor) equated with non-covariant objects. This must be understood in the context of having \textit{fixed} a system of coordinates, which is imposed by the conservation equations~(\ref{affine_cons1}). Once coordinates have been fixed, these equations have a well-defined and proper meaning. We also discussed how all of these theories can be seen as gauge-fixed versions of fully covariant theories, which gives a legitimate interpretation of `fixed coordinates'. How this is implemented in practice will be seen explicitly in the following chapter.

\begin{savequote}[70mm]
Infinities and indivisibles transcend our finite understanding, the former on account of their magnitude, the latter because of their smallness; Imagine what they are when combined.
\qauthor{`Two New Sciences' (1638) \\
Galileo Galilei}
\end{savequote}
\chapter{Cosmology}
\label{chapter6}

In this chapter we study the cosmological solutions for the modified theories introduced in Chapter~\ref{chapter5}. We take for granted the standard cosmology of General Relativity as background knowledge, see~\cite{Hawking:1973uf,Dodelson:2003ft} for reference. We also do not consider perturbations, working only at the background level. This makes the analysis fairly simple. For more detailed studies in relation to cosmological observations, such as Baryonic Acoustic Oscillations (BAO) data and CMB constraints, as well as gravitational wave propagation, we refer to the works~\cite{Bahamonde:2021gfp,Nunes:2018xbm,nunes2018new} for $f(T)$ gravity and~\cite{Frusciante:2021sio,Lazkoz:2019sjl,Anagnostopoulos:2021ydo} for $f(Q)$ gravity. These results also apply to our $f(\ourG)$ and $f(\mathfrak{G})$ theories by extension.

As demonstrated in the previous chapter, the dynamics of $f(\ourG)$ and $f(Q)$ gravity and of $f(\mathfrak{G})$ and $f(T)$ gravity are completely equivalent. In the first section we explicitly show that this equivalence also extends to the background cosmological equations of $f(\ourG)$ and $f(\mathfrak{G})$ gravity. As a concrete example, we then study the Born-Infeld Lagrangian in our $f(\ourG)$ models, which leads to early-time inflationary dynamics. This has also been reported in the $f(T)$ case~\cite{Ferraro:2006jd}.

In Sec.~\ref{section6.2} we apply a dynamical systems approach to all of the second-order modified theories. This is again possible because the cosmological field equations for both $f(\ourG)$ and $f(\mathfrak{G})$ coincide, provided we work in appropriate coordinates and tetrads. In our analysis, we focus on the particularly interesting solutions that describe accelerated (de Sitter) expansion. We derive a number of specific constraints that a model must satisfy in order for these solutions to exist. Importantly, these are model-independent results, making the analysis very broad. We then focus on the phase portraits for two recently proposed $f(Q)$ models, comparing their qualitative behaviour to the standard $\Lambda$CDM cosmologies.

In the last section we study the $f(\bar{\ourG},\bar{\ourB})$ theories, which depart from the teleparallel models by being genuine metric-affine theories. A simple quadratic boundary term model is chosen to study the cosmological dynamics in the Riemann-Cartan geometry. In this model, torsion is dynamical despite its source current (hypermomentum or intrinsic spin) being zero in cosmology. This propagating torsion field is found to give rise to early-time inflationary behaviour~\cite{Boehmer:2023fyl}.

\section{Second-order modified cosmology}
\label{section6.1}

Here we study the cosmological solutions of the Riemannian second-order modified theories of the previous chapter, namely $f(\ourG)$ and $f(\mathfrak{G})$ gravity. This of course extends to the $f(Q)$ and $f(T)$ theories via covariantisation. In our gauge-fixed theories, we must be careful that the choice of coordinates and tetrad lead to non-trivial dynamics, whilst also respecting the symmetries of the spacetime. This will be checked explicitly.

Throughout this chapter, we assume the validity of the cosmological principle, namely, that on sufficiently large scales the universe is homogeneous and isotropic\footnote{We do, however, note that there are good reasons to consider going beyond this assumption and studying models with anisotropy such as those of the Bianchi kind~\cite{Aluri:2022hzs}.}. Consequently, we can model the universe using the Friedmann-Lema{\^i}tre-Robertson-Walker (FLRW) line element. We also restrict our study to spatially flat models, in agreement with most current observational data~\cite{Efstathiou:2020wem}. This line element reads
\begin{align} \label{FLRW}
    ds^2=-N^2(t)dt^2+a^2(t) \big( dx^2+dy^2+dz^2 \big) \,,
\end{align}
where $a(t)$ is the scale factor and $N(t)$ is the lapse function.

 In some symmetry-breaking theories of modified gravity, it is not necessarily true that the lapse function can be set to one. On the other hand, in fully diffeomorphism invariant theories one can always rescale the time coordinate to absorb the lapse function into the new coordinates. However, in $f(\ourG)$ gravity for this metric~(\ref{FLRW}), there exists the aforementioned time reparameterisation invariance~\cite{BeltranJimenez:2019esp,BeltranJimenez:2019tme}. We therefore see that we can indeed set $N(t)=1$. This is also the case in $f(\mathfrak{G})$ gravity, where the full diffeomorphism symmetry under spacetime coordinate transformations is preserved.
  
We assume the energy-momentum tensor to be modelled by a single perfect fluid with pressure $p(t)$ and energy density $\rho(t)$, taking the form
\begin{equation}
T^{\mu \nu} = (\rho + p) U^{\mu} U^{\nu} + p g^{\mu \nu} \ ,
\end{equation}
where $U^{\mu}$ is the normalised timelike four-velocity $U^{\mu}U_{\mu}=-1$.
 We will later also assume a linear equation of state $p = w \rho$ to obtain closed-form solutions. The $f(\ourG)$ cosmological field equations are then given by~\cite{Boehmer:2021aji}
\begin{align}
  \label{f(G)FLRW1}
  f(\ourG) N^2 + 12 H^2 f'(\ourG)
  &= 2  \kappa \rho \,, \\
  \label{f(G)FLRW2}
  -f(\ourG) - f'(\ourG) \frac{4}{N^2} \Bigl(\dot{H} + 3H^2 - H\frac{\dot{N}}{N} \Bigr) 
  + f''(\ourG) \frac{48H^2}{N^4} \Bigl(\dot{H}-H\frac{\dot{N}}{N}\Bigr)
  &= 2  \kappa p \,,
\end{align}
where $H = \dot{a}/a$ is the Hubble function. In these coordinates $\ourG$ is related to the Hubble parameter by
\begin{align}
  \label{FLRW_G}
  \ourG = -6\frac{H^2}{N^2} \,.
\end{align}
The field equations are second-order in derivatives of the metric, or first-order in $\{H(t),N(t)\}$. The first equation~(\ref{f(G)FLRW1}) is a generalised Friedmann constraint equation, containing no derivatives of the Hubble and lapse functions, while the second~(\ref{f(G)FLRW2}) is a modified acceleration equation.

These equations satisfy the usual continuity equation 
\begin{equation} \label{continuity}
\dot{\rho} + 3H (\rho + p)=0 \,, 
\end{equation}
which can be verified explicitly from~(\ref{f(G)FLRW1}) and~(\ref{f(G)FLRW2}). This can also be seen by noting that the conservation equation~(\ref{f(G)cons}) vanishes for this choice of metric and coordinates, independent from the choice of $f$. If one works in the full $f(\ourG,\ourB)$ theory, the conservation equation~(\ref{eqn:cons1}) is also zero. This implies directly that the energy-momentum tensor is covariantly conserved $\nabla_{\mu} T^{\mu}{}_{\nu} = 0$, see equation~(\ref{f(G)cons2}).

In the language of symmetric teleparallel gravity, we see that the Cartesian coordinates are consistent with the coincident gauge (zero affine connection). The vanishing of the $f(\ourG)$ conservation equation implies that the $f(Q)$ connection equation of motion~(\ref{f(Q)_connection}) also vanishes.
The non-metricity scalar then takes the form $Q = - \ourG = 6 H^2 /N^2$, with the gauge boundary term being equal to zero $b_{\xi}=0$. Along with the appropriate field redefinitions, explained in Sec.~\ref{section5.2.2}, the cosmological equations for $f(Q)$ gravity in the coincident gauge are then simply~(\ref{f(G)FLRW1})-(\ref{f(G)FLRW2}).

Cosmological solutions in symmetric teleparallel gravity have been well studied, see for instance~\cite{BeltranJimenez:2019tme}. In~\cite{DAmbrosio:2021pnd,Hohmann:2021ast} the authors also consider solutions beyond the coincidence gauge. This is especially useful when studying non-flat $k \neq 0$ cosmologies\footnote{Note that $k \neq 0$ can still be studied in the coincident gauge, but the coordinates for the metric tensor become extremely complicated~\cite{Hohmann:2021ast}.}. We also note that for the $f(Q)$ theories, the geometry is non-Riemannian and so extra care must be taken with regard to the coupling of matter. In particular, the role of hypermomentum can become important, which has been studied in this cosmological context~\cite{Iosifidis:2020gth,Iosifidis:2020upr,Hohmann:2021ast}. In our $f(\ourG)$ framework, however, the geometry is Riemannian and no such considerations are needed.

Setting the lapse function to one, the field equations reduce to the simple form~\cite{Boehmer:2022wln}
\begin{align}
    6 f'  H^2 + \frac{1}{2}f &= \kappa \rho \,, 
    \label{Friedmann}\\
    (12 H^2 f'' - f' )\dot H &= \frac{\kappa}{2}(\rho+p) 
    \label{Acceleration} \, ,
\end{align}
with the bulk term given by $\ourG = - 6H^2$. This can be seen to match the $f(Q)$ field equations~\cite{BeltranJimenez:2019tme}, again making sure to take the correct redefinitions to match the various sign conventions throughout the literature. To be concrete and make contact with the cosmological equations shown in~\cite{BeltranJimenez:2019tme,Hohmann:2021ast}, whose non-metricity scalar matches our conventions, we can make the redefinition $f(\ourG) \rightarrow - f(-\ourG)$ and then apply the coincident gauge relation $-\ourG = Q$ to obtain
\begin{align}
6 f'(Q) H^2 - \frac{1}{2} f(Q) &= \kappa \rho \, ,
    \label{FriedmannQ}\\
    \Big(12 H^2 f''(Q) + f'(Q) \Big)\dot H &= -\frac{\kappa}{2}(\rho+p) 
    \label{AccelerationQ} \,.
\end{align}

 What is quite remarkable is that these equations are identical to the field equations of $f(T)$ gravity~\cite{Bengochea:2008gz,DAmbrosio:2021pnd} (or equivalently, $f(\mathfrak{G})$ gravity). 
To show this equivalence, we first must choose a frame field $\mathbf{e}_{a}$ to work with. Due to the invariance of the metric under local Lorentz transformations, there is no unique choice for the tetrad leading to the FLRW metric~(\ref{FLRW}). For the pure formulation of $f(T)$ gravity or $f(\mathfrak{G})$ gravity, we must ensure that our choice is compatible with the Weitzenb\"{o}ck gauge~\cite{Krssak:2018ywd}. A `proper tetrad' that satisfies this gauge choice~\cite{krvsvsak2016covariant} is given by
\begin{equation} \label{tetrad_FLRW}
e^{a}{}_{\mu} = {\textrm{diag}} \big( 1,a(t),a(t),a(t) \big) \, ,
\end{equation}
which leads to the FLRW metric~(\ref{FLRW}) with the lapse set to one. The tetradic bulk term is found to coincide with its metric counterpart $\mathfrak{G} = -6H(t)^2 = \ourG$, and hence the torsion scalar also coincides with the non-metricity scalar $T = Q = 6H(t)^2 $.

The equivalence of the metric and tetradic bulk terms can be seen as a consequence of the vanishing of the boundary term $\mathbb{B} = \ourG - \mathfrak{G}$ for this particular frame~(\ref{tetrad_FLRW}). Moreover, this choice of tetrad and metric can also be seen as a requirement imposed by the coincident and Weitzenb\"{o}ck gauges, or alternatively, the equations related to diffeomorphism and local Lorentz invariance. It is clear that further study of this boundary term $\mathbb{B}$ is useful in relation to these discussions. 

What is perhaps more surprising is that the superpotentials of both theories also coincide, leading to a complete equivalence between the $f(\ourG)$ and $f(\mathfrak{G})$ cosmological field equations, see~\cite{Boehmer:2021aji,Boehmer:2022wln}. It then follows that equations~(\ref{Friedmann})-(\ref{Acceleration}) describe the background dynamics for both metric and tetradic theories, as well as the teleparallel $f(Q)$ and $f(T)$ theories (in their respective unitary gauges). We will therefore use these equations to study the cosmologies of all of these second-order modified theories.

\subsection{Born-Infeld cosmology}
\label{section6.1.1}

As a brief example of cosmological dynamics beyond GR, let us look at the Born-Infeld Lagrangian~(\ref{BI_Lagrangian}) within the $f(\ourG)$ framework
\begin{equation} \label{BI_Lagrangian2}
    L_{BI} = f(\ourG) = \lambda \Bigg( \sqrt{1 + \frac{2 \ourG}{\lambda}} - 1\Bigg) \, ,
\end{equation}
where we now have $\ourG = -6 H^2$. The analysis largely follows the analogous $f(T)$ case, see~\cite{Ferraro:2006jd}.
We mentioned in Sec.~\ref{section5.1.2} that these models can be constructed to break diffeomorphism invariance around some characteristic length scale $l \propto 1/\sqrt{|\lambda|}$. We retrieve GR in the limit that $|\lambda| \rightarrow \infty$ and $l \rightarrow 0$. We are therefore interested in large but finite values of $\lambda$ such that $l$ is small but nonzero\footnote{Note that to make contact with the $f(Q)$ and $f(T)$ theories using our conventions, one can perform $\lambda \rightarrow - \lambda$, which is equivalent to $f(\ourG) \rightarrow - f(-\ourG)$.}.

For this action~(\ref{BI_Lagrangian2}), the cosmological field equations can be written in the following form
\begin{align} \label{BI1}
\frac{6 H(t)^2}{\sigma(t) + \sigma(t)^2} &=  \rho(t) \, \\
\frac{6 H(t)^2}{\sigma(t) + \sigma(t)^2} + \frac{2\dot{H}(t)}{\sigma(t)^3}&= - w \rho(t) \, ,  \label{BI2}
\end{align}
where we assumed a barotropic equation of state $w = p/\rho$ with $w > -1$ and set $\kappa=1$. We have also introduced the convenient function $\sigma(t)$ to simplify the field equations, defined by
\begin{equation}
\sigma(t) := \sqrt{1 + \frac{2 \ourG}{\lambda}} = \sqrt{1 - \frac{12 H(t)^2}{\lambda}} \,  .
\end{equation}
In the GR limit we simply have $\sigma(t) \rightarrow 1$, and one sees the cosmological equations~(\ref{BI1})-(\ref{BI2}) reduce to their usual form.

The field equations lead to the following differential equation for the Hubble factor
\begin{equation} \label{Hdiff}
\dot{H} = \frac{1}{4}(1+w)(\lambda - 12 H^2)(\sigma(t)-1) \, .
\end{equation}
In the GR limit $\lambda \rightarrow \infty$ the RHS simplifies to $-3 H^2 (1+w)/2$, yielding the standard solution for the scale factor $a(t) \propto (t-t_0)^{2/(3(1+w))}$. Equation~(\ref{Hdiff}) yields analytic solutions, but it turns out to be easier to eliminate the Hubble function and work directly with the energy density.

From the field equations we find that the matter energy density $\rho(t)$ satisfies the differential equation
\begin{equation} \label{rhodiff}
\dot{\rho} = \pm \sqrt{3}(1+w ) \frac{\sqrt{\lambda} \sqrt{ \rho^3 (\rho + \lambda)}}{\lambda + 2 \rho} \, ,
\end{equation}
where decreasing solutions $\dot{\rho}(t) <0$ corresponding to an expanding universe $H(t) > 0$ and increasing solutions $\dot{\rho}(t) >0$ corresponding to a contracting universe $H(t) < 0$. Here we will focus on the expanding $H(t) > 0$ solutions.

 In the limit $\lambda \rightarrow \infty$ we obtain $\rho \propto 1/t^2$ as expected. Somewhat surprisingly, equation~(\ref{rhodiff}) yields analytic solutions, without needing to make any series expansion for $\lambda$. This is in contrast to some previous results for Born-Infeld models in $f(T)$. For example, in~\cite{Jana:2014aca} the authors study the $f(T)$ Born-Infeld cosmological equations up to order $1/\lambda$. However, the results are qualitatively the same, and we refer to that work for further details.  
 
\begin{figure}[!htb]
     \centering
     \begin{subfigure}[b]{0.75\textwidth}
         \centering
         \includegraphics[width=\textwidth]{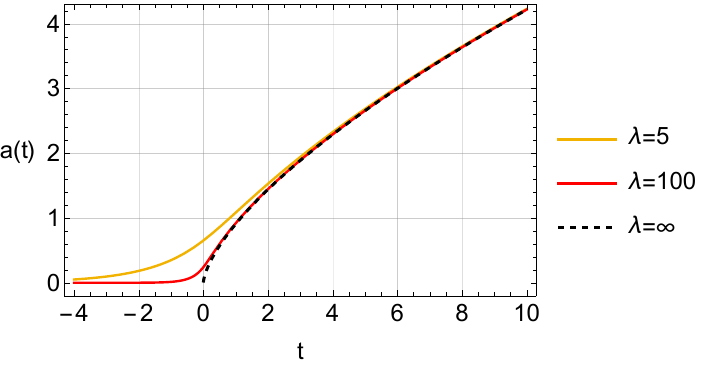}
         \caption{Solutions for scale factor $a(t)$ with equation of state $w =0$. The $\lambda = \infty$ solution corresponds to the GR limit with $a(t) \propto t^{2/3}$.}
         \label{fig:scalea}
         \vspace{6mm}
     \end{subfigure}
    \begin{subfigure}[b]{0.75\textwidth}
         \centering
     \includegraphics[width=\textwidth]{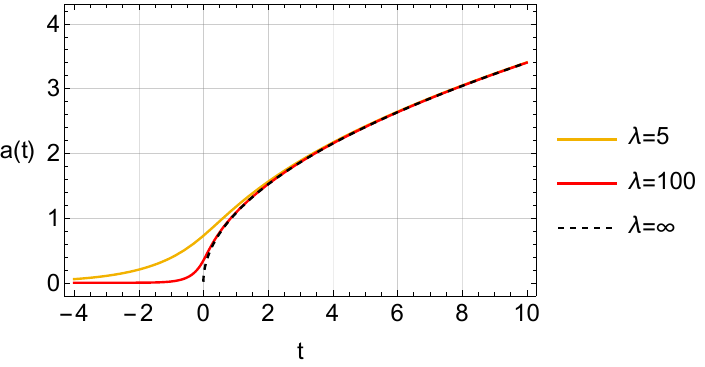}
         \caption{Solutions for scale factor $a(t)$ with equation of state $w =1/3$. The $\lambda = \infty$ solution corresponds to the GR limit with $a(t) \propto t^{1/2}$.}
         \label{fig:scaleb}
     \end{subfigure}
        \caption{Scale factor solutions for the Born-Infeld action~(\ref{BI_Lagrangian2}).}
        \label{fig:scale}
\end{figure}

Using the continuity equation~(\ref{continuity}) leads to the standard relationship between the energy density and the scale factor 
\begin{equation} \label{rho0}
\rho(t) = \rho_0  a(t)^{-3(1+w)} \,,
\end{equation}
where $\rho_0$ is a constant related to the present-day energy density and scale factor. Using this with the solutions to the differential equation~(\ref{rhodiff}), we can find exact solutions for the scale factor. In Fig.~\ref{fig:scale} we plot these solutions for the scale factor $a(t)$ for the values $\lambda = 5$, $\lambda = 100$ and $\lambda = \infty$ with equation of state $w =0$ and $w = 1/3$, corresponding to matter and radiation fluids respectively. The $\lambda = \infty$ case is simply the GR solution with $a(t) \propto t^{2/(3+3w)}$. For large values of $t$ the modified scale factor $a(t)$ tracks the GR solution. However, for finite $\lambda$ the initial time $t \rightarrow 0$ singularity of GR is removed, with $a(t)$ continuing smoothly in the negative $t$ direction.

\begin{figure}[!htb]
     \centering
     \begin{subfigure}[b]{0.8\textwidth}
         \centering
         \includegraphics[width=\textwidth]{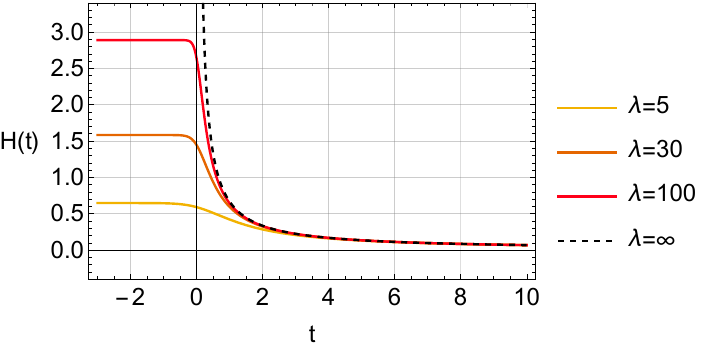}
         \caption{Hubble factor $H(t)$.}
         \label{subfig:Hubble}
                  \vspace{6mm}
     \end{subfigure} 
    \begin{subfigure}[b]{0.75\textwidth}
         \centering
     \includegraphics[width=\textwidth]{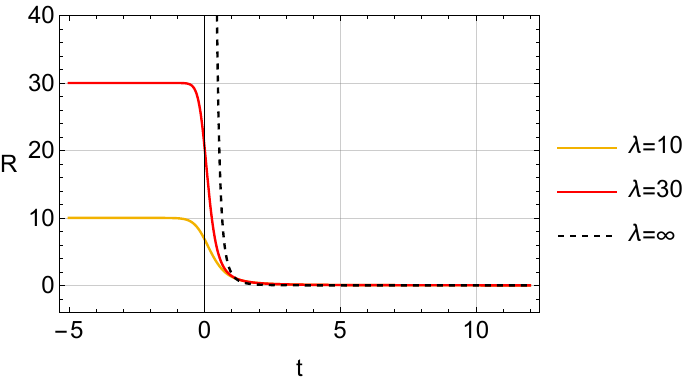}
         \caption{Ricci scalar $R$}
         \label{subfig:R}
                  \vspace{6mm}
     \end{subfigure}
     \begin{subfigure}[b]{0.8\textwidth}
         \centering
     \includegraphics[width=\textwidth]{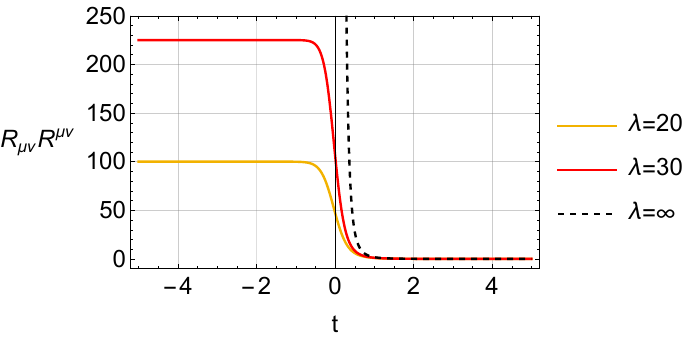}
         \caption{Ricci tensor squared $R_{\mu \nu}R^{\mu \nu}$.}
         \label{subfig:Rsquare}
     \end{subfigure}
        \caption{Hubble factor and curvature scalars for the Born-Infeld action~(\ref{BI_Lagrangian2}). Solutions give rise to constant curvature regimes as $t \rightarrow -\infty$.}
        \label{fig:hub}
\end{figure}

In fact, as $t \rightarrow - \infty$ the modified scale factor behaves like the de Sitter solution
\begin{equation} \label{de Sitter}
a(t) \approx \exp ( \alpha t ) \qquad , \qquad \alpha = \frac{\sqrt{\lambda}}{2 \sqrt{3}} \, .
\end{equation}
The Hubble function therefore approaches the constant value $H(t) = \alpha$ as $t \rightarrow - \infty$. This is shown in Fig.~\ref{subfig:Hubble}, where various values of $\lambda$ are plotted against the GR solution. To see more clearly that these regions represent  regimes of constant curvature, in Fig.~\ref{subfig:R} and~\ref{subfig:Rsquare} we also plot the curvature invariants  for the modified expanding solutions
\begin{equation}
R = 6 \frac{\dot{a}^2 + a \ddot{a}}{a^2} \, , \qquad R_{\mu \nu}R^{\mu \nu} = 3\frac{3 a^2 \ddot{a}^2 + (2 \dot{a}^2 + a \ddot{a})^2}{a^4} \, .
\end{equation}
We find that these two invariants take a maximum value of $R_{\textrm{max}} = \lambda$ and $(R_{\mu \nu}R^{\mu \nu})_{\textrm{max}} = \lambda^2/4$ in the limit $t\rightarrow - \infty$, as one would expect for the de Sitter solutions~(\ref{de Sitter}). They also remain finite for all $t$. These results are also reported in~\cite{Ferraro:2006jd,Fiorini:2013kba} for $f(T)$ gravity, which verifies our solutions.

In conclusion, we find the Born-Infeld extensions are a particularly interesting class of models to study within these modified theories. In particular, they lead to geodesically complete spacetimes without an initial big bang singularity~\cite{Ferraro:2006jd}. Instead, there is a geometrically sourced inflationary stage with de Sitter solutions for $t <0$. We will study similar solutions in the metric-affine setting in Sec.~\ref{section6.3}. In the next section we focus on modifications that lead to \textit{late-time} de Sitter solutions.

\section{Dynamical systems approach}
\label{section6.2}
Let us analyse the cosmological field equations for the second-order modifications, given in~(\ref{Friedmann})-(\ref{Acceleration}). We can choose to use any of the pseudo-scalars $\ourG$ and $\mathfrak{G}$ or teleparallel scalars $Q$ and $T$ for this discussion, with appropriate redefinitions of the function $f$, due to the aforementioned equivalence 
\begin{equation}
T = Q = -\ourG = - \mathfrak{G} = 6 H(t)^2 \, .
\end{equation}
In this section we will choose to use the non-metricity scalar $Q$.

At the background level, it makes sense to study all of these theories in a unified way. The theory of dynamical systems turns out to be extremely useful in this regard. We refer to the standard textbooks on dynamical systems~\cite{perko2013differential,wiggins2003introduction}, and the textbooks~\cite{wainwright1997dynamical,Coley:2003mj} for applications in cosmology. For a recent review of applications in modified gravity, see~\cite{Bahamonde:2017ize}. Here we will follow our work~\cite{Boehmer:2022wln} and outline some of the results of our analysis\footnote{Actually, our work~\cite{Boehmer:2022wln} can be applied to \textit{any} second-order system, provided the first constraint equation is some function of $H(t)$ and the second equation is a function of $H(t)$ and $\dot{H}(t)$.}. In Appendix~\ref{appendixE} we include supplementary details for the dynamical systems analysis, referenced where necessary. 

For a universe comprised of two fluids, matter $\rho_m$ ($w= 0$) and radiation $\rho_r$ ($w= 1/3$), the cosmological equations~(\ref{FriedmannQ})-(\ref{AccelerationQ}) can be written as 
\begin{align} \label{cosmo1}
-\frac{f}{6H^2} - \frac{\rho_m}{3H^2} - \frac{\rho_r}{3H^2}+ 2 f' &= 0 \, , \\
\label{cosmo2}
\rho_m + \frac{4}{3}  \rho_r + 2 \dot{H} (f' + 12 H^2 f') &= 0 \, .
\end{align}
We now introduce the dynamical variables 
 \begin{align}
    X_1 = \Omega_{m} = \frac{\rho_m}{3 H^2 } \,, \qquad X_2 = \Omega_{r} = \frac{\rho_r}{3 H^2} \,, \qquad Z = \frac{H^2}{H_0^2 +H^2} \, .
    \label{X1X2Z}
\end{align}
The first two variables are the standard density parameters for matter and radiation~\cite{wainwright1997dynamical}, whilst the $Z$ is related to the Hubble function and chosen to make the phase space compact~\cite{Hohmann:2017jao}. Note that the variables are independent from the function $f$, making this approach quite different from the usual dynamical systems studies~\cite{Bahamonde:2017ize}. It also follows that $X_1$ and $X_2$ are always non-negative, and $0 \leq Z \leq 1$.

The Friedmann constraint in terms of the new variables is 
\begin{align} \label{2fluidF}
    X_1 + X_2 =  \big(1-\frac{1}{Z}\big) \frac{f}{6 H_0^2} + 2 f' \ ,
\end{align}
where we treat $f = f({6 H_0^2 Z}/{(1-Z)})$ as a function of $Z$. Clearly the Friedmann constraint now depends on the form of $f$, which is the price to pay for using dynamical variables independent from the function $f$. Traditionally, additional variables are often introduced to completely eliminate $f$ from the Friedmann constraint, at the cost of increasing the dimensionality of the system, e.g.,~\cite{Khyllep:2022spx,BeltranJimenez:2019tme}. Our formulation has the advantage of having fewer variables with a lower dimensional phase space and is therefore computationally simpler to study. We provide a thorough discussion of this topic and a comparison of different approaches in~\cite{Boehmer:2022wln,Boehmer:2023knj}.

The three-dimensional phase space $\{X_1,X_2,Z\}$ is reduced to two dimensions by using the Friedmann constraint~(\ref{2fluidF}). It is most convenient to remove either $X_1$ or $X_2$, so let us choose to eliminate $X_1$ and consider the evolution of the system $\{X_2,Z\}$ described by the dynamical equations
\begin{align} \label{X2 eq long}
    \frac{d X_2}{d N} &= - \frac{X_2 \left((Z-1)^2 f + 2 H_0^2 Z \left( X_2 (Z-1) + 2(Z-1) f' + 48 H_0^2 Z f'' \right) \right) }{2 H_0^2 Z \big((1-Z) f' + 12 H_0^2 Z f'' \big)} \, , \\ \label{Z eq long}
    \frac{d Z}{d N} &=  -\frac{(Z-1)^2 \left((Z-1)f + 2 H_0^2 Z (X_2 + 6 f' ) \right)}{2 H_0^2 \big((1-Z)f' + 12 H_0^2 Z f'' \big)} \, ,
\end{align}
where $N = \log a(t)$. The acceleration equation~(\ref{cosmo2}) and Friedmann constraint~(\ref{2fluidF}) have both been used to remove $\dot{H}$ terms and $X_1$ terms respectively. The system is then closed when a function $f$ has been specified.
 Note that $f$ has dimensions $H_0^2$, $f'$ is dimensionless and $f''$ has units of $H_0^{-2}$. Consequently, equations~(\ref{X2 eq long})--(\ref{Z eq long}) are dimensionless.

The system (\ref{X2 eq long})--(\ref{Z eq long}) can be written more compactly by introducing functions related to $f$ and its derivatives
\begin{align} \label{m(Z)}
    m(Z) &:= \frac{2 Q f''(Q)}{f'(Q)} = \frac{12 H_0^2 Z f''}{(1-Z) f'} \, , \\ \label{n(Z)}
    n(Z) &:= \frac{f(Q)}{Q f'(Q)} = \frac{f (1-Z)}{6H_0^2 Z f'} \, ,
\end{align}
where the functions on the right-hand side should again be treated as $f({6 H_0^2 Z}/{(1-Z)})$.
The dynamical equations can then be brought into the following form
\begin{align} 
    \label{X2 mn}
    X_2' &= \frac{X_2 \left(2-3n-4m+ X_2/f' \right)}{1+m} \, , \\ 
    \label{Z mn}
    Z' &= \frac{Z(Z-1) \left(X_2 - 3(n-2)f' \right)}{(1+m)f'} \ .
\end{align}
It turns out that many properties of this system and its critical points are independent of $f$, and hence are valid for all models. This makes our approach very broad, as will be demonstrated below.

\subsection{Fixed points}

The fixed points of the system~(\ref{X2 mn})-(\ref{Z mn}) can be divided into two families of solutions: one with $X_2=0$, and the other with $X_2=(4m+3n-2) f' =: X_2^*$. For the case where $X_2=0$, the points with $Z=Z^*$ are solutions of the algebraic equation 
\begin{align} 
    \label{eq Z*}
    \frac{3Z(Z-1)(n(Z)-2)}{1+m(Z)}=0 \ .
\end{align}
For the $X_2=X_2^*$ case, the $Z$ coordinates take values $Z=0$ or $Z=1$. The $X_2$ coordinates must then be evaluated at $Z=0$ or $Z=1$ because $X_2^*$ is function of $Z$.

These results have been collected in Table~\ref{tab2}. The points A and B are solutions with $X_2=X_2^*$ evaluated at $Z=0$ and $Z=1$ respectively. The P's represent points along the $X_2=0$ line with $Z=Z^*$ satisfying~(\ref{eq Z*}). The special cases of $Z^*$ when $n(Z)=2$ and $m(Z)\rightarrow \infty$ are labelled as P$_n$ and P$_m$ respectively. Note that $m(Z)$ can diverge when $f' \rightarrow 0$. However, $f'$ appears in the denominator of $n(Z)$~(\ref{n(Z)}) and in both dynamical equations~(\ref{X2 mn}) and~(\ref{Z mn}). Therefore for $m(Z) \rightarrow \infty$ to correspond to a critical point, we must also require $n(Z)$ to remain finite. In other words, we require that $f'' \rightarrow \infty$ or that  both $f \rightarrow 0$ and $f' \rightarrow 0$, for some value(s) of $Z$. For further details, see~\cite{Boehmer:2022wln}.

\begin{table}[htb!]
    \centering
    \begin{tabular}{c|c|c|c|c}
         Point/Line & $X_2$ & $Z$ & $X_1$ & conditions \\ \hline\hline
         P$_1$ & $0$ & $0$ & $(2-n(0)) f'(0)$ & $m(0) \neq -1$, $n(0) \neq 2$  \\
           P$_2$ & $0$ & $1$ & $(2-n(1)) f'(\infty)$ & $m(1) \neq -1$, $n(1) \neq 2$  \\
          P$_n$ & $0$ & $Z^*$ & $ 0$ & $n(Z^*)=2$ \\
           P$_m$ & $0$ & $Z^{*}$ & $(2-n(Z^{*})) f'(Z^{*})$ & $m(Z^{*}) \rightarrow \infty$ \\
          A & $X_2^*|_{Z=0}$ & $0$ & $-4(m(0)+n(0)-1) f'(0)$ & \\ 
          B & $X_2^*|_{Z=1}$ & $1$ & $-4(m(1)+n(1)-1)f'(+\infty)$ & \\
           L1 & any & $Z_c$ & $ -X_2 +(2-n(Z_c)) f'(Z_c)$ & $m(Z_c) = -1$ \\
           L2 & any & $Z_d$ & $-X_2 + (2-n(Z_d)) f'(Z_d)$ & $n(Z_d) \rightarrow \infty$ \\
    \end{tabular}
    \caption{Table of fixed points of system~(\ref{X2 mn})--(\ref{Z mn}) for arbitrary function $f$. For existence, we assume that all variables and functions are finite at the critical point unless otherwise stated.}
    \label{tab2}
\end{table}

 The system also displays divergent behaviours along `singular lines'. The line L1 exists for values $Z=Z_c$ such that $m(Z_c)=-1$, while L2 exists if $n(Z) \rightarrow \infty$ with $m(Z)$ finite. This means that either $f \rightarrow \infty$ or that both $f' \rightarrow 0$ and $f'' \rightarrow 0$ such that $n(Z)$ diverges but $m(Z)$ does not, similar to the conditions for P$_m$.

The corresponding values of $X_1$ determined by the Friedmann constraint have also been included in Table~\ref{tab2}. The final column states the specific conditions for each point, and it is assumed that all variables $X_1$, $X_2$, $Z$ must be finite for existence. Note that the fixed points are not necessarily unique and in many cases they coincide with one another. Moreover, `asymptotic points' (e.g., at $X_2 \rightarrow \infty$) are not captured by the above analysis due to the assumptions of finiteness. A key example where this type of critical point exists will be shown in Sec.~\ref{section6.2.2}. For compact $f(T)$ or $f(Q)$ models, however, all critical points can simply be read off from the above table.

In Appendix~\ref{appendixE.1} we provide a detailed stability analysis for the critical points of the general system. The de Sitter point P$_{n}$ is found to be stable, with negative eigenvalues~(\ref{Pnstab}). For the other points, the stability depends on the specific form of the function $f$ and are model-dependent.

\subsubsection{Physical Parameters}

Using the formulation above (\ref{m(Z)})--(\ref{Z mn}) along with the cosmological field equations, the deceleration parameter can be expressed in terms of the new variables 
\begin{align}
    q := -\frac{\ddot{a}a}{\dot{a}^2} = -1-\frac{\dot{H}}{H^2} = -1 - \frac{(\frac{3}{2}X_1+2X_2)(n-2)}{(X_1+X_2)(1+m)} \, .
\end{align}
The value of $q$ for each of the fixed points is given in Table~\ref{tab3}. For the points P$_n$, P$_m$, A and B one obtains fixed values of $q$, which are model-independent. The points P$_1$ and P$_2$ depend on the functions $m$ and $n$ evaluated at their respective $Z$ values, and so a specific $f$ is needed to determine $q$. For the lines L1 and L2, the divergent behaviour of the system makes any concrete claims about physical parameters untrustworthy, so they have not been included.

The effective equation of state parameter $w_{\textrm{eff}}$ can be expressed in a similar form
\begin{align}
   w_{\textrm{eff}} := \frac{p_{\textrm{tot}}}{\rho_{\textrm{tot}}} = 
    -1 + \frac{(X_1+\frac{4}{3}X_2)(n-2)}{(X_1+X_2)(1+m)} \ ,
\end{align}
where the total energy density is $\rho_{\textrm{tot}}= \rho_{m}+\rho_{r} + \rho_{f}$ and $\rho_{f}$ represents the extra terms in (\ref{cosmo1}) that do not appear in the standard Friedmann equation of GR
\begin{equation}
    \rho_{f} := 3H^2 + \frac{1}{2} f - 6 H^2 f' \, .
\end{equation}
 Similarly, the total pressure is $p_{\textrm{tot}}= p_{m}+p_{r} + p_{f}$ with $p_{f}$ representing the additional terms in the acceleration equation~(\ref{cosmo2}). The values of $w_{\textrm{eff}}$ evaluated at the critical points are also given in Table~\ref{tab3}. Lastly, the values of the Hubble parameter $H(t)$ are also included, provided they are well-defined. 

\begin{table}[htb!]
    \centering
    \begin{tabular}{c|c|c|c}
         Point & $q$ & $w_{\textrm{eff}}$ & $H(t)$ \\ \hline\hline
         && \\[-8pt]
       P$_1$ & $\displaystyle\frac{4-2m(0)-3n(0)}{2+2m(0)}$ & $\displaystyle-\frac{-1+m(0)+n(0)}{1+m(0)}$ & $H(t) = 0$ \\[8pt]
         P$_2$ & $\displaystyle\frac{4-2m(1)-3n(1)}{2+2m(1)}$ & $\displaystyle-\frac{-1+m(1)+n(1)}{1+m(1)}$ & $H(t) \rightarrow \pm \infty$ \\
         P$_n$ & $-1$ & $-1$ & $H(t) =$ const. \\
         P$_m$ & $-1$ & $-1$ & $H(t) =$ const. \\
         A & $1$ & $1/3$ & $H(t) = 0$ \\
         B & $1$ & $1/3$ & $H(t) \rightarrow \pm \infty$ \\
    \end{tabular}
    \caption{Physical quantities at critical locations}
    \label{tab3}
\end{table}

The density parameter of the additional non-GR terms can be expressed in terms of our dynamical variables as
\begin{equation} \label{omega f}
    \Omega_{f} := \frac{\rho_{f}}{3H^2} = 1 - (2-n)f' \, ,
\end{equation}
which satisfies $\Omega_1 + \Omega_2 + \Omega_f = 1$ from the Friedmann equation. Hence we have obtained a very neat expression representing the contributions of the modified theory beyond GR. As we explain in~\cite{Boehmer:2023knj}, a more standard dynamical systems formulation would introduce a 
dynamical variable for $\Omega_f$, but here this is not necessary because $\Omega_f$ can be written totally in terms of $Z$. We therefore obtain a phase space with fewer dimensions at the expense of a more cumbersome Hubble constraint.

For the fixed points P$_n$ satisfying $n(Z)=2$ one immediately has $\Omega_f =1$. Similarly for points P$_m$ with $m(Z) \rightarrow \infty$, we see that if $f' \rightarrow 0$ with $n(Z)$ staying finite, we may also obtain $\Omega_f =1$. For these two solutions, the deceleration parameter and equation of state are fixed to be $q=w_{\textrm{eff}}=-1$, representing a de Sitter universe. In fact, this is a necessary requirement for any de Sitter solution\footnote{By inspection of Table~\ref{tab3}, we see that points P1 and P2 cannot have $q=-1$ or $w_{\textrm{eff}}=-1$ unless $n(Z) =2$ or $m(Z) \rightarrow \infty$. These are exactly the conditions for points P$_n$ and P$_m$}. Models where $n(Z) \neq2$ and $m(Z) \nrightarrow \infty$ for some $Z$ in the range $(0,1)$ cannot possess a de Sitter fixed point. This immediately rules out models such as $f(Q) \propto Q^{\alpha }$ for $\alpha  \neq 1/2$, or $f(Q) = Q + \beta Q^2$ for $\beta \geq 0$. This powerful result highlights the utility of the dynamical systems approach.

\subsubsection{Physical phase space}
For the physical phase space, we first have that $X_1, X_2 \geq 0$ and that $0 \leq Z \leq 1$, which holds for all models. In order to enforce further conditions we must use the Friedmann constraint~(\ref{2fluidF}) for a given $f$ to constrain $X_2$ and $Z$. Then one can simply read off the fixed points from Table~\ref{tab2} (that satisfy their appropriate existence conditions) and determine whether they are located within the physical phase space. 

\begin{figure}[!htb]
    \centering
    \includegraphics[width=0.7\textwidth]{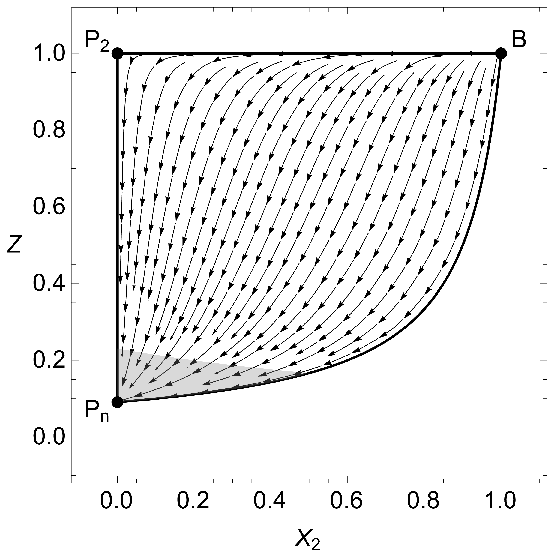}
    \caption{Phase space for General Relativity with $f(Q)=Q + 6 \Lambda H_0^2$ and $\Lambda >0$.}
    \label{fig:gr2fluid}
\end{figure}

For example, for pure General Relativity $f(Q)=Q$ equation~(\ref{2fluidF}) simplifies to $X_1+X_2 =1$ and the physical phase space in $\{X_2,Z\}$ is the unit square. On the other hand, for General Relativity with a positive cosmological constant $f(Q)=Q + 6 \Lambda H_0^2$ and $\Lambda > 0$ one instead has $X_1 + X_2 = 1 + \Lambda - \Lambda/Z$, and therefore $X_2 \leq 1$. The physical phase space is then the region satisfying $Z \geq \Lambda /(1-X_2+ \Lambda)$ and $0 \leq Z \leq 1$, see Fig.~\ref{fig:gr2fluid}. We emphasise that only the fixed points in Table~\ref{tab2} that lie within the region constrained by~(\ref{2fluidF}) are physical. In Fig.~\ref{fig:gr2fluid} we see the points B, P$_2$ and P$_n$, representing the radiation repeller, matter saddle and de Sitter attractor, respectively. The grey shaded region has $w_{\textrm{eff}} < - 1/3$, representing accelerated expansion. This is simply the typical $\Lambda$CDM dynamical systems phase space~\cite{Bahamonde:2017ize} written in slightly different coordinates.

\subsection{Beltr\'an et al. model}
\label{section6.2.2}

Let us now apply these dynamical systems techniques to some concrete modified gravity models. These models were proposed in the context of $f(Q)$ gravity, but as we have shown, are equally applicable in the cosmological context to $f(T)$ gravity, as well as our gauge-fixed $f(\ourG)$ and $f(\mathfrak{G})$ theories. 

The first model of interest is given by
\begin{align} \label{Beltran}
    f(Q)=Q-\lambda_p \frac{H_0^4}{Q} - 2\lambda H_0^2 \,,
\end{align}
which was suggested by Beltr\'an et al. in~\cite{BeltranJimenez:2019tme}, with cosmological solutions studied in~\cite{BeltranJimenez:2016wxw}. In the action~(\ref{Beltran}) the cosmological constant is represented by $\lambda$, and we have introduced factors of $H_0$ such that both parameters $\lambda_p$ and $\lambda$ are dimensionless. 

The authors~\cite{BeltranJimenez:2019tme} also applied a dynamical systems formulation for the $f(Q)$ cosmologies, though with a very different set of variables leading to a different system of equations. However, it appears that the proposed dynamical system is not closed, and so it is difficult to see whether the analysis is valid. We will study this model with our dynamical systems formalism outlined above, which will always lead to a closed system when $f$ is specified.

We give the explicit form of of dynamical equations for this model in~\cite{Boehmer:2022wln}. What is particularly useful to see are the functions $m(Z)$ and $n(Z)$
\begin{align} 
    \label{beltran m}
    m(Z) &= -\frac{4 \lambda_p (Z-1)^2}{\lambda_p - 2 Z \lambda_p + Z^2 (36 + \lambda_p)}, \\ 
    n(Z) &= \frac{Z^2(36+12 \lambda - \lambda_p) - \lambda_p + 2Z (\lambda_p- 6 \lambda)}{\lambda_p - 2 Z \lambda_p + Z^2 (36+\lambda_p)} \label{beltran n} \ .
\end{align}
We will assume that $\lambda_p \neq 0$ as this would reduce the model back to GR with a cosmological constant term. The sign of the parameter $\lambda_p$ determines whether the phase space is naturally compact or not, so we will look at these two cases separately. This follows from the modified Friedmann constraint~(\ref{2fluidF}), which now takes the form 
\begin{align} 
    \label{beltran friedmann}
    X_1 = \frac{Z(4\lambda-2\lambda_p) + Z^2(12-12 X_2 -4 \lambda + \lambda_p) + \lambda_p}{12 Z^2} \geq 0 \, ,
\end{align}
and must hold for any values of the parameters. 

\subsubsection{Compact case: $\lambda_p <0$}
\begin{figure}[!htb]
     \centering
     \begin{subfigure}[b]{0.48\textwidth}
         \centering
         \includegraphics[width=\textwidth]{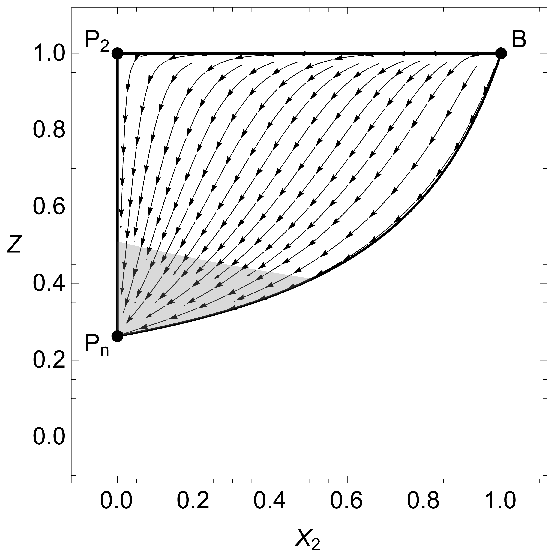}
         \caption{$\lambda_p < 0$ and $\lambda < 0$.}
         \label{fig:beltran1}
         \vspace{5mm}
     \end{subfigure} \hspace{1mm}
    \begin{subfigure}[b]{0.48\textwidth}
         \centering
     \includegraphics[width=\textwidth]{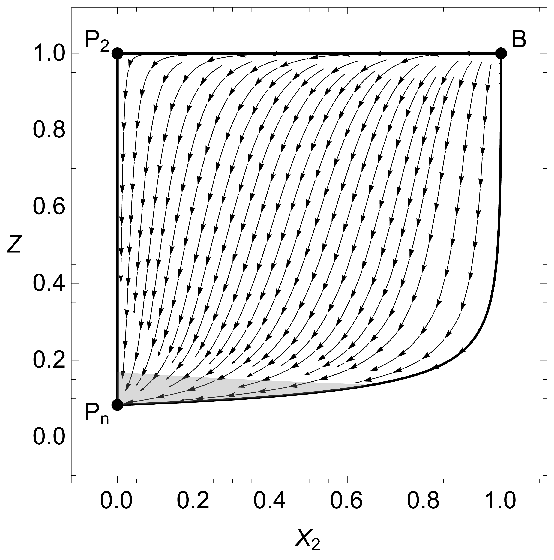}
         \caption{$\lambda_p < 0$ and $\lambda = 0$.}
         \label{fig:beltran1b}
         \vspace{5mm}
     \end{subfigure}
     \hfill
     \begin{subfigure}[b]{0.75\textwidth}
         \centering
         \includegraphics[width=\textwidth]{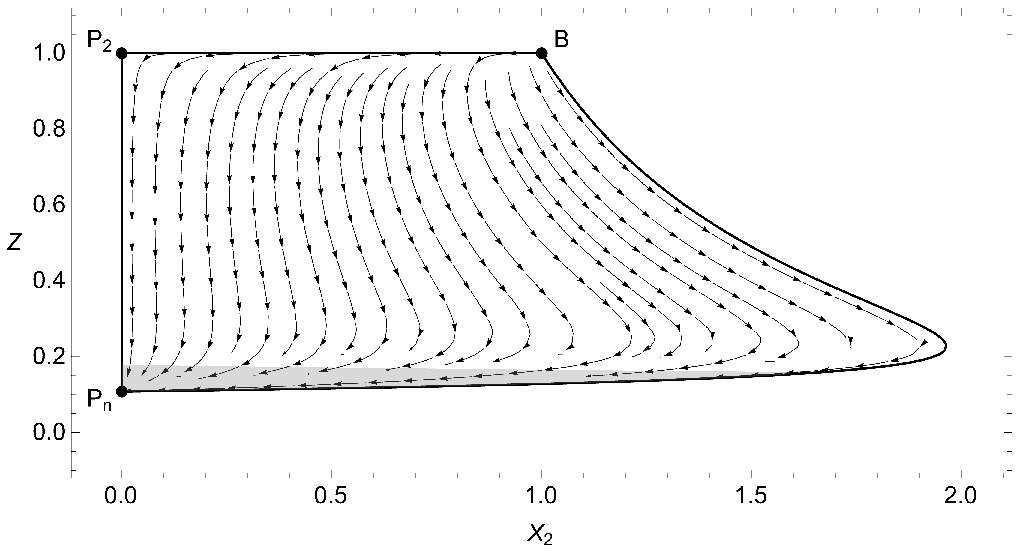}
         \caption{$\lambda_p < 0$ and $\lambda > 0$.}
         \label{fig:beltran2}
     \end{subfigure}
        \caption{Phase space for the Beltr\'an model~(\ref{Beltran}) with $\lambda_p <0$.}
        \label{fig:beltran sub1}
\end{figure}

First, assume $\lambda_p <0 $ and $\lambda \leq 0$. In this case, the model is compact and $X_1$ and $X_2$ are bounded between $[0,1]$, which can be deduced from~(\ref{beltran friedmann}).  
The phase portraits for $\lambda <0$ and $\lambda =0$ are given in Fig.~\ref{fig:beltran1} and Fig.~\ref{fig:beltran1b} respectively. The fixed points B, P$_2$ and P$_n$ represent the radiation dominated repeller $\Omega_{r}=1$, matter dominated saddle point $\Omega_{m}=1$ and de Sitter attractor $\Omega_{f}=1$ respectively. This can be deduced from the definitions of $X_2$ and $Z$ at the fixed points, and by using the Friedmann constraint to determine the value of $X_1$. From Table \ref{tab3} the effective equation of state parameter can be read off at the points B and P$_n$ as $w_{\textrm{eff}}=1/3$ and $w_{\textrm{eff}}=-1$. For the matter saddle point P$_2$ one easily finds $w_{\textrm{eff}}=0$ as expected. Again, the grey shaded region represents accelerated expansion with $q < 0$.

All orbits start at B and end at P$_n$, with trajectories attracted towards the matter saddle P$_2$. The qualitative features are very similar to those of GR with a positive cosmological constant, shown in Fig.~\ref{fig:gr2fluid}. Interestingly, in the absence of a cosmological constant term $\lambda =0$, Fig.~\ref{fig:beltran1b}, the phase space still remains the same.

When $\lambda_p$ is negative but $\lambda > 0$, the phase space is still compact but no longer bounded between $[0,1]$ in the variables $X_1$ and $X_2$, see Fig.~\ref{fig:beltran2}. The upper bound instead depends on the value of $\lambda$ through relations in~(\ref{beltran friedmann}). This can be understood physically by the modified density parameter $\Omega_f$~(\ref{omega f}) taking negative values within this parameter space, see~\cite{Boehmer:2023knj} for further discussion.
The late-time de Sitter point P$_n$ at $(0,Z^*)$ remains the only global late-time attractor of the physical phase space, and the other two fixed points are the same. However, trajectories beginning at the past attractor, point B, can instead follow trajectories in the positive $X_2$ direction before terminating at the future-time attractor P$_n$. This is clearly visible in the phase portraits, Fig.~\ref{fig:beltran2}.

\subsubsection{Noncompact case: $\lambda_p >0$}
For the case when $\lambda_p > 0$ the situation differs. Firstly, the variables $X_1$ and $X_2$ are no longer bounded from above, meaning the phase space is not compact. The phase space instead stretches to infinity in the positive $X_2$ direction as $Z$ tends to zero. 

\begin{figure}[!htp]
     \centering
     \begin{subfigure}[b]{0.48\textwidth}
         \centering
         \includegraphics[width=\textwidth]{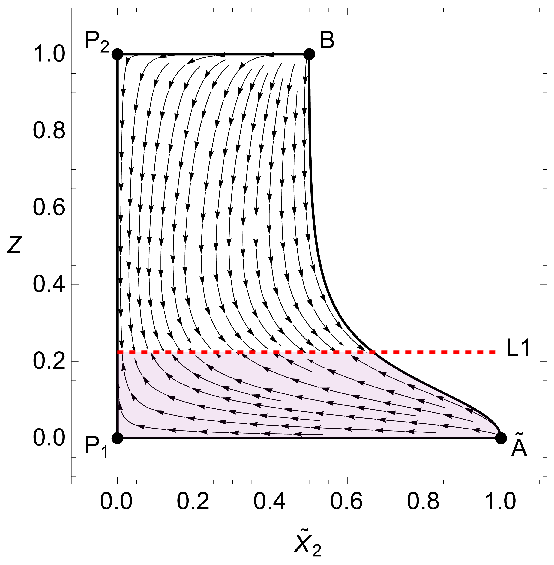}
         \caption{$\lambda_p > 0$ and $\lambda > - \sqrt{3 \lambda_p}$.}
         \label{fig:beltran3}
         \vspace{6mm}
     \end{subfigure}  \hspace{1mm}
     \begin{subfigure}[b]{0.48\textwidth}
         \centering
         \includegraphics[width=\textwidth]{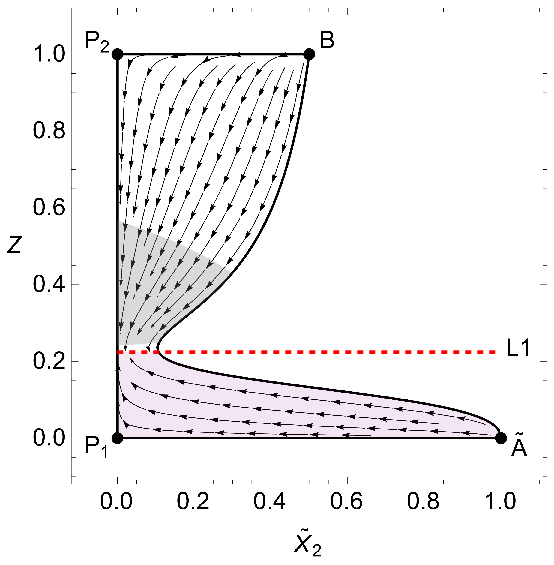}
         \caption{$\lambda_p > 0$ and $\lambda > - \sqrt{3 \lambda_p}$.}
         \label{fig:beltran4}
         \vspace{6mm}
     \end{subfigure}
     \begin{subfigure}[b]{0.48\textwidth}
         \centering
         \includegraphics[width=\textwidth]{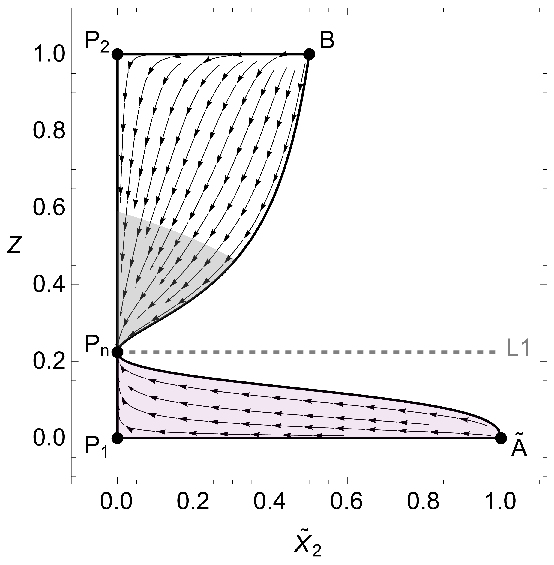}
         \caption{$\lambda_p > 0$ and $\lambda = - \sqrt{3 \lambda_p}$.}
         \label{fig:beltran5}
     \end{subfigure}  \hspace{1mm}
     \begin{subfigure}[b]{0.48\textwidth}
         \centering
         \includegraphics[width=\textwidth]{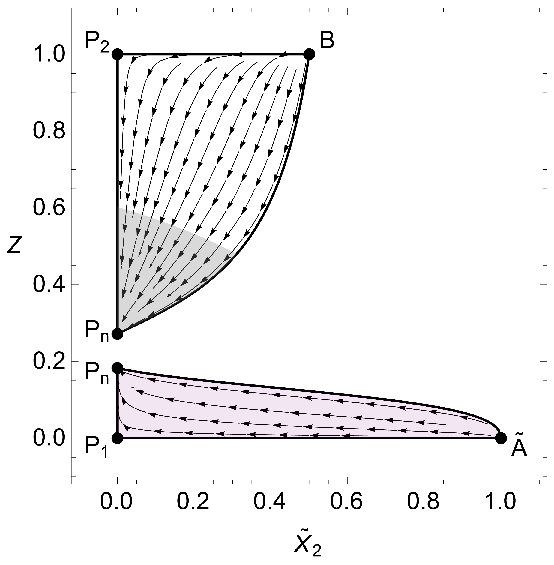}
         \caption{$\lambda_p > 0$ and $\lambda < - \sqrt{3 \lambda_p}$.}
         \label{fig:beltran6}
     \end{subfigure}
        \caption{Compactified phase space for the Beltr\'an model (\ref{Beltran}) with $\lambda_p >0$. Grey shading represents accelerated expansion $q <0$, while purple shading indicates a phantom equation of state $w_{\textrm{eff}} < -1$.}
        \label{fig:beltran sub2}
\end{figure}

The denominator of equations~(\ref{X2 mn}) and~(\ref{Z mn}) vanishes when the $Z$ coordinate takes the fixed value $Z=Z_c (\lambda_p)$ such that $m(Z_c)=-1$. This corresponds to the line L1 with $Z$ coordinate $Z_c = (-2 \sqrt{3 \lambda_p} + \lambda_p) / (\lambda_p -12)$. If $\lambda > - \sqrt{3 \lambda_p}$ the physical region is connected, whereas for $\lambda \leq - \sqrt{3 \lambda_p}$ the space bifurcates into two disconnected parts at the point $\{0,Z=Z_c\}$. However, for all $\lambda$, trajectories are always confined to regions either above or below the line L1 at $Z=Z_c$. This will be shown clearly in the compactified phase portraits, Fig.~\ref{fig:beltran sub2}.

In order to compactify the phase space we introduce the variable 
\begin{align}
    \tilde{X}_2 = \frac{X_2}{1+X_2} \ ,
\end{align}
such that $0 \leq \tilde{X}_2 \leq 1$. The new dynamical system in $\{\tilde{X}_2,Z\}$ can then be found by taking the derivative of $\tilde{X}_2$ with respect to $N$, see~\cite{Boehmer:2022wln}. Similarly, the Friedmann constraint is rewritten with $\tilde{X_2}$ leading to the compactified phase spaces given in Fig.~\ref{fig:beltran sub2}. Note that the fixed points in Table~\ref{tab2} and Table~\ref{tab3} are still valid in the original variables $X_2$ and $Z$, therefore they will also represent critical points in the compactified variables (at the new $\tilde{X}_2$ coordinates).

The two critical points B and P$_2$ are the same as in the $\lambda_p<0$ case, being the unstable radiation dominated repeller $\Omega_r=1$ and the matter dominated saddle $\Omega_m=1$ respectively. The line L1 splits the phase space vertically, as shown by  Figs.~\ref{fig:beltran3}--\ref{fig:beltran5}. Trajectories change direction above and below this line due to a sign change in the dynamical equations. On the lower part of the phase space we now have the critical point P$_1$ and the asymptotic point $\tilde{\textrm{A}}$, both along $Z=0$. At point $\tilde{\textrm{A}}$ one finds that $X_2 = \Omega_r \rightarrow \infty$ and $X_1 = \Omega_m \rightarrow \infty$, whilst at P$_1$ we have $X_2= \Omega_r = 0$ and $X_1 = \Omega_m \rightarrow \infty$. At both P$_1$ and $\tilde{\textrm{A}}$, and along the entirety of the $Z=0$ line, the effective equation of state parameter is $w_{\textrm{eff}}=-2$. Moreover, the effective equation of state is $w_{\textrm{eff}} < -1$ at all points below L1, representing a phantom regime. This is represented by the purple shading in Fig.~\ref{fig:beltran sub2}.

It has been argued that a phantom equation of state ($w_{\textrm{eff}} < -1$) may be necessary to address cosmological tensions~\cite{Heisenberg:2022gqk}, see also~\cite{Escamilla:2023oce}. More specifically, for late-time modifications of $\Lambda$CDM, in order to solve both the $H_0$ and $S_8$ tensions, the effective equation of state has to cross the phantom divide at $w_{\textrm{eff}} =-1$~\cite{Heisenberg:2022gqk}. However, trajectories for this model in Fig.~\ref{fig:beltran sub2} cannot cross this phantom divide due to the presence of the singular line L1.

Figs.~\ref{fig:beltran3} and~\ref{fig:beltran4} show the phase space for $\lambda > - \sqrt{3 \lambda_p}$, with the phase space beginning to pinch off in Fig.~\ref{fig:beltran4} as $\lambda$ decreases. All trajectories approach the line, and the relative magnitudes of the parameters determine the shape of the phase space.

For the case with $\lambda < -\sqrt{3 \lambda_p}$, the phase space splits into two disjoint parts along the line L1. Fig.~\ref{fig:beltran5} shows the limiting case with $\lambda = -\sqrt{3 \lambda_p}$. The late-time de Sitter points P$_n$ exist for $\lambda \leq -\sqrt{3 \lambda_p}$ and are solutions to the equation
\begin{align}
    n(Z)= \frac{Z^2(36+12 \lambda -\lambda_p) - \lambda_p + 2 Z(\lambda_p-6\lambda)}{\lambda_p - 2 \lambda_p Z + Z^2 (36+\lambda_p)} = 2 \ .
\end{align}
This has two solutions in the physical $Z$-range when the inequality is strict, see Fig.~\ref{fig:beltran6}. In the particular case  $\lambda = -\sqrt{3 \lambda_p}$, the two points merge into one, which coincides with the line L1 (Fig.~\ref{fig:beltran5}). As mentioned in the stability analysis, the points P$_n$ are always stable, and they attract all trajectories in their respective sections of the phase space. As $\lambda$ decreases relative to $\lambda_p$, the P$_n$ points move apart along the $Z$-axis. This is an interesting bifurcation rarely discussed before in cosmological models.

\subsection{Anagnostopoulos et al. model}
\label{section6.2.3}

Let us move on to our second and final example.
The power-exponential model proposed by Anagnostopoulos et al. in~\cite{Anagnostopoulos:2021ydo} displays a number of interesting features and was shown to pass a variety of observational tests~\cite{Anagnostopoulos:2021ydo,Anagnostopoulos:2022gej}. In particular, the authors studied the model against Supernovae type Ia (SNIa), BAO, cosmic chronometers (CC), and
Redshift Space Distortion (RSD) data and found that it is comparable, and for some datasets favourable, over the $\Lambda$CDM model. Moreover, it immediately passes early universe constraints. As such, it has been shown to be a genuine alternative to the $\Lambda$CDM concordance model and worthwhile studying from a dynamical systems perspective.

The model is given by the function
\begin{align} 
    \label{Saridakis lagrangian}
    f(Q)=Q e^{\lambda \frac{Q_0 }{Q}} \,,
\end{align}
with the single free parameter $\lambda$.
A dynamical systems analysis was recently performed for this model in~\cite{Khyllep:2022spx}. There the authors studied the background and perturbation equations of a universe with a single fluid matter component ($w=0)$, and the subsequent phase space was three-dimensional. It is interesting to then study this model in our reduced dimensionality formulation with an additional matter fluid component, which will turn out to be two-dimensional. Moreover, the reduced dimensions in our approach will turn out to make the stability analysis much simpler to compute.

In the limit that $\lambda$ vanishes the model~(\ref{Saridakis lagrangian}) reduces to GR without a cosmological constant. It does not however have a direct $\Lambda$CDM limit.
When the parameter $\lambda$ is small, to first order the function behaves like GR with a cosmological constant term $Q_0 \lambda$, and this behaviour will be observed in the phase space analysis. The sign of the parameter $\lambda$ leads to different phase spaces, and so we will investigate both cases. We will also assume that $\lambda \neq 0$, as this trivially leads back to GR.

\begin{figure}[!htb]
     \centering
     \begin{subfigure}[b]{0.48\textwidth}
         \centering
         \includegraphics[width=\textwidth]{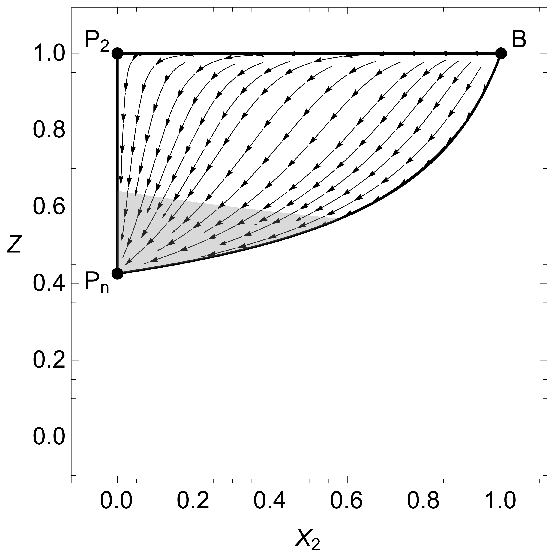}
         \caption{$\lambda > 0$.}
         \label{fig:ana1}
         \vspace{5mm}
     \end{subfigure} \hspace{1mm}
    \begin{subfigure}[b]{0.48\textwidth}
         \centering
     \includegraphics[width=\textwidth]{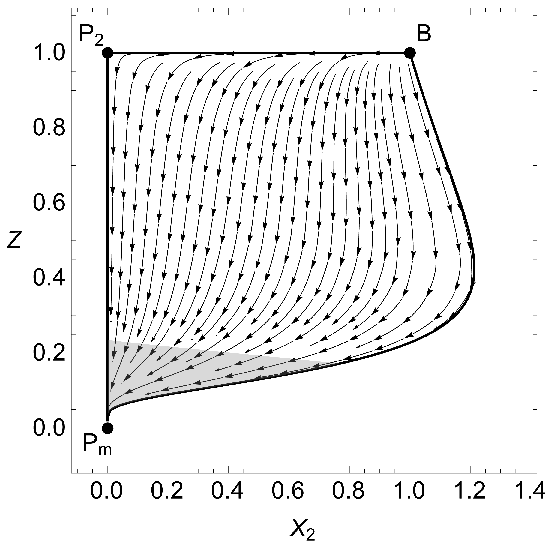}
         \caption{$\lambda < 0$.}
         \label{fig:ana2}
         \vspace{5mm}
     \end{subfigure}
        \caption{Phase space for the Anagnostopoulos model~(\ref{Saridakis lagrangian}).}
        \label{fig:ana}
\end{figure}

We now give an overview of our dynamical systems formulation applied to this model~\cite{Boehmer:2023knj}. The functions $m(Z)$ and $n(Z)$ take the form
\begin{align} \label{n Saridakis}
    m(Z) &= \frac{2(Z-1)^2 \lambda^2}{Z (Z + \lambda (Z -1))} \,, \\
    \label{m Saridakis}
      n(Z) &= \frac{Z}{Z + \lambda (Z -1)} \, .
\end{align}
It follows that the system is notably more simple than the Beltr\'an et al. model, and we list all of the fixed points in Table~\ref{tab Saridakis}. The stability has also been included, which is calculated in Appendix~\ref{appendixE.2}.
There are only two different possible phase configurations, one with $\lambda$ positive and the other with $\lambda$ negative. For $\lambda >0$, the phase space resembles GR with a positive cosmological constant, see Fig.~\ref{fig:ana1}. The radiation dominated past attractor, matter saddle, and late-time de Sitter attractor are given by points B, P$_2$ and P$_n$ respectively. The shaded region represents accelerated expansion.

\begin{table}[htb!]
    \centering
    \begin{tabular}{c|c|c|c|c|c|c|c}
         Point & $X_1$ & $X_2$ & $Z$ & $q$ & $w_{\textrm{eff}}$  & conditions & stability \\ \hline\hline 
         P$_2$ & $1$ & $0$ & $1$ & $1/2$ & $0$ & none & Saddle
         \\
         B & $0$ & $1$ & $1$ & $1$ & $1/3$  & none & Unstable \\
         P$_{\textrm{m}}$ & $0$ & $0$ & $0$ & $-1$ & $-1$ & $\lambda < 0$ & Nonhyperbolic \\
         P$_{\textrm{n}}$ & $0$ & $0$ &$2\lambda/(1+2 \lambda)$  & $-1$ & $-1$ & $\lambda > 0$ & Stable \\
    \end{tabular}
    \caption{Table of fixed points for the Anagnostopoulos et al. model.}
    \label{tab Saridakis}
\end{table}

 For $\lambda < 0$ the de Sitter point P$_n$ is replaced by the de Sitter point P$_m$, see Fig.~\ref{fig:ana2}. Importantly, all of the physical properties of the new de Sitter point P$_m$ are identical to P$_n$. This is proven in~\cite{Boehmer:2023knj}. We refer again to Appendix~\ref{appendixE.2} for further details on the stability of  P$_m$, which we determine to be an attractor within the physical phase space. We also note that the phase space now extends past $X_2 =1$, due to the modified density parameter $\Omega_f$ having a negative lower bound for $\lambda <0$.

\clearpage

\begin{figure}[!htb]
     \centering
     \begin{subfigure}[b]{0.75\textwidth}
         \centering
         \includegraphics[width=\textwidth]{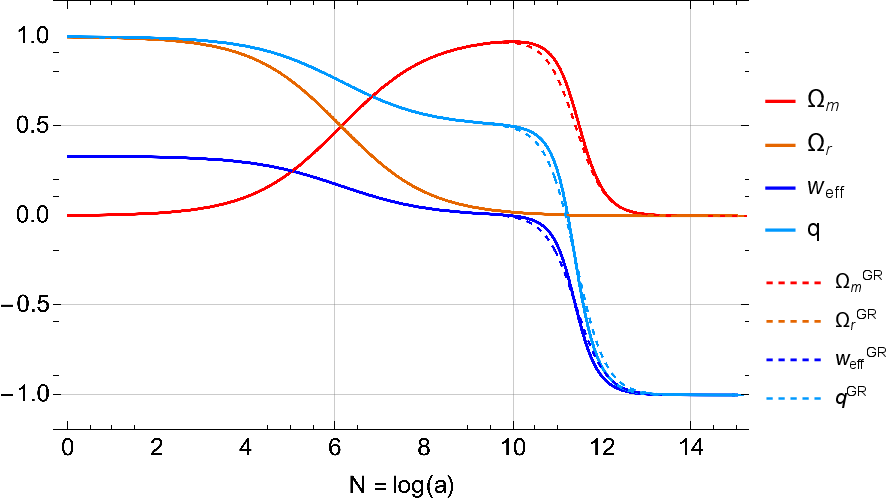}
         \caption{Evolution for positive $\lambda$.}
         \label{fig:3A}
     \end{subfigure}\\[2ex]
    \begin{subfigure}[b]{0.75\textwidth}
         \centering
     \includegraphics[width=\textwidth]{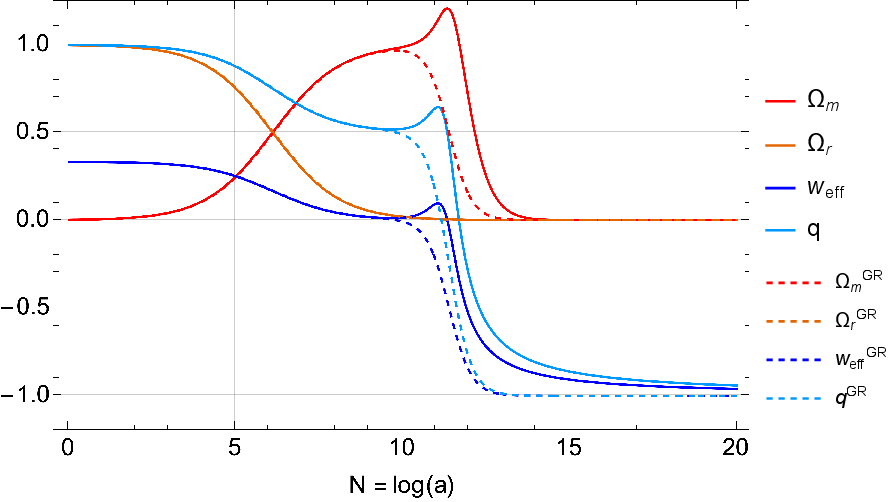}
         \caption{Evolution for negative $\lambda$.}
         \label{fig:3B}
     \end{subfigure}
        \caption{Evolution of density parameters $\Omega_m$, $\Omega_r$, effective equation of state $w_{\textrm{eff}}$, and deceleration parameter $q$ for GR with a positive cosmological constant (dashed) and Anagnostopoulos et al. model (solid line).}
        \label{fig:evo}
\end{figure}

To better see the quantitative differences between these models and GR, in Fig.~\ref{fig:evo} we plot the evolution of the matter and radiation density parameters, the deceleration parameter $q$ and the effective equation of state $w_{\textrm{eff}}$. Fig.~\ref{fig:3A} shows the evolution for $\lambda>0$ of a trajectory following a heteroclinic orbit from points B $\rightarrow$ P$_2$ $\rightarrow$ P$_{\textrm{n}}$, while Fig.~\ref{fig:3B} follows the heteroclinic orbit B $\rightarrow$ P$_2$ $\rightarrow$ P$_{\textrm{m}}$ for $\lambda<0$. The dashed lines represent the evolution of these parameters for GR with a positive cosmological constant $f(Q)=Q+ \Lambda  Q_0  = 6H^2 + 6 H_0^2 \Lambda$. The initial $N \rightarrow 0$ and final $N \rightarrow \infty$ states are the same for the modified model and for GR, as one would expect from the critical points.

\clearpage

The background evolution of the model with positive $\lambda$ can be seen to match very closely with its GR counterpart, given the same initial conditions. This is in agreement with the analysis of~\cite{Anagnostopoulos:2021ydo}. For the negative $\lambda$ case in Fig.~\ref{fig:3B}, a sharp spike in the matter density parameter can be noted before approaching the de Sitter point with $w_{\textrm{eff}}=-1$. It would be interesting to further investigate the differences between these parameter choices.

\section{Metric-affine cosmology}
\label{section6.3}

In this section we return to the metric-affine framework and the $f(\bar{\ourG},\bar{\ourB})$ theories of Sec.~\ref{section5.3}. We will focus on the modified Einstein-Cartan theories, again working in the FLRW framework with vanishing hypermomentum. The field equations were written compactly in equation~(\ref{ECmod7}). We begin by briefly reviewing the role of torsion in cosmological spacetimes and showing how this applies to our modified theories. We then study a model including a non-linear function of the boundary term $\bar{\ourB}$, leading to interesting early-time dynamics. This follows our work~\cite{Boehmer:2023fyl}.

\subsection{Cosmological torsion}
\label{section6.3.1}

It is well known that the cosmological principle reduces the number of independent components of the metric tensor to just one free function, namely the scale factor $a(t)$. As we saw in the previous sections, the lapse function for cosmological spacetimes could be set to one $N(t)=1$, due to the reparameterisation invariance of these modified theories. Moreover, the conservation equation coming from infinitesimal diffeomorphisms for the $f(\ourG,\ourB)$  theories vanished for the FLRW metric in Cartesian coordinates with an arbitrary lapse function~(\ref{FLRW}). One can also verify that the analogous metric-affine conservation equation~(\ref{affine_cons1}) is satisfied for a general lapse function $N(t)$ in these coordinates. We therefore again work with $N(t)=1$.

Recall that the spatially flat FLRW metric is
\begin{align}
  ds^2 = -dt^2 + a^2(t) (dx^2 + dy^2 +dz^2)\,.
  \label{cos1}
\end{align}
The torsion tensor is also strongly constrained when assuming isotropy and homogeneity~\cite{TSAMPARLIS197927}. The only allowed components of a cosmological torsion tensor are given by
\begin{align}
  T^1{}_{10} = T^2{}_{20} = T^3{}_{30} &=: h(t)/a^3(t) \,, \\
  T^1{}_{23} = T^2{}_{31} = T^3{}_{12} &=: k(t) \,, 
  \label{cos2}
\end{align}
where $h$ and $k$ are arbitrary functions of the time coordinate. It will become clear shortly why the factor of $a^{-3}$ was included.

 Let us now consider our equation relating the torsion tensor to our $f(\tilde{\ourG},\tilde{\ourB})$ action~(\ref{ECmod8}), which we rewrite here for clarity
 \begin{align}
  T^{\mu}{}_{\lambda \nu} = \frac{1}{f_{,\tilde{\ourG}}}
  \delta^{\mu}_{[\lambda} \partial_{\nu]} f_{,\tilde{\ourB}} \,. \label{ECmod8copy}
\end{align}
  In the cosmological setting all objects $\tilde{\ourG}$ and $\tilde{\ourB}$ are functions of time only, and so the right-hand side of~(\ref{ECmod8copy}) identically vanishes if all indices take spatial values $\{x,y,z\}$. Therefore, this source-free equation immediately yields $T^1{}_{23} = T^2{}_{31} = T^3{}_{12} = 0$, so that we can set $k(t)=0$, and one is left with only a vector torsion\footnote{Note that the torsion vector here $T^{\sigma}{}_{\sigma \mu}$ differs by a minus sign from the torsion vector of Chapter~\ref{chapter3}, which used the convention $T^{\sigma}{}_{\mu \sigma}$. This does not affect any calculations, but we point this out for consistency.} contribution 
\begin{align}
  T_\mu = T^\sigma{}_{\sigma\mu} = \{3h/a^3,0,0,0\} \,.
  \label{cos2a}
\end{align}
Equation~(\ref{ECmod8copy}) simplifies into a single equation
\begin{align}
  \frac{h}{a^3} = T^{1}{}_{10} = T^2{}_{20} = T^3{}_{30} =
  \frac{1}{2}\frac{1}{f_{,\tilde{\ourG}}}
  \partial_t f_{,\tilde{\ourB}} \,.
  \label{cos3}
\end{align}
This cosmological spacetime only contains the torsion vector and this torsion vector has only one non-trivial component. Therefore the norm of this vector, which is a scalar, contains all of the information about torsion in this spacetime
\begin{align}
  |T_{\textrm vec}| = \sqrt{-g_{\mu\nu} T^\mu T^\nu} = \frac{3h}{a^3} \,.
\end{align}
In what follows it will be most natural to discuss the properties of $h/a^3$.

The boundary term $\tilde{\ourB}$ with vanishing non-metricity contains a metric part $\ourB$ and a torsional contribution $B_T$, see Eq.~(\ref{B_scalars}). These are given by
\begin{align}
  \ourB &= \frac{1}{\sqrt{-g}}\partial_\nu
  \Bigl(\frac{1}{\sqrt{-g}} \partial_\mu(g g^{\mu\nu})\Bigr) = 18H^2 + 6\dot{H}  \, ,\\
  B_T &= \frac{2}{\sqrt{-g}}\partial_\mu(\sqrt{-g} T^\mu) = -\frac{3}{a^3} \dot{h}\,.
  \label{cos4}
\end{align}
Here $T^\mu = T_\lambda{}^{\lambda\mu}$ and $H(t)=\dot{a}/a$ is the Hubble function. It is the term $\sqrt{-g} T^\mu$ which motivated the above mentioned factor of $a^{-3}$ in~(\ref{cos2}). The torsion field equation now becomes
\begin{align}
  \frac{h}{a^3} = \frac{1}{2}\frac{1}{f_{,\tilde{\ourG}}}
  \bigl(f_{,\tilde{\ourB}\,\tilde{\ourB}} \partial_t \tilde{\ourB} +
  f_{,\tilde{\ourB}\,\tilde{\ourG}} \partial_t \tilde{\ourG} \bigr) \,.
  \label{cos5}
\end{align}
It is already clear at this point that torsion can be non-trivial despite the absence of source terms. This is similar to curvature in General Relativity, where one typically finds vacuum solutions with non-zero curvature. 

\subsection{Quadratic model and inflation}
\label{section6.3.2}

Let us propose to study specific models of the form
\begin{align}
  f(\tilde{\ourG}, \tilde{\ourB}) = \tilde{\ourG} + F(\tilde{\ourB}) \,,
\end{align}
with $F(\tilde{\ourB})$ being an arbitrary function of the boundary term satisfying $F''(\tilde{\ourB}) \neq 0$ such that torsion does not vanish. It is useful to make a redefinition of the torsion component $h(t) = \dot{a}a^2 + \mathfrak{h}(t)$. Then $h/a^3 = H + \mathfrak{h}/a^3$ which will simplify the subsequent field equations.

The metric cosmological field equations now take the following form
\begin{align}
  \frac{1}{2}F + 3\frac{\dot{\mathfrak{h}} }{a^3}F' + 3\frac{\mathfrak{h}^2}{a^6}  &= \rho \,,
  \label{cos5a} \\
  -\frac{1}{2} F - 3\frac{\dot{\mathfrak{h}}}{a^3} F' +
  \frac{12\mathfrak{h}}{a^6}\Bigl(\ddot{\mathfrak{h}}-3\frac{\dot{a}}{a}\dot{\mathfrak{h}}\Bigr) F'' + 2\frac{\dot{\mathfrak{h}}}{a^3} - \frac{\mathfrak{h}^2}{a^6} + 4\frac{\mathfrak{h}}{a^3}\frac{\dot{a}}{a} &= w \rho \,,
  \label{cos6}
\end{align}
where we have assumed a linear equation of state $p=w\rho$ and set $\kappa=1$. The connection equation is 
\begin{align}
  \label{cos7}
  6\Bigl(\frac{\ddot{\mathfrak{h}}}{a^3} -
  3 \frac{\dot{a}}{a}\frac{\dot{\mathfrak{h}}}{a^3}\Bigr) F'' -
  2\frac{\mathfrak{h}}{a^3} 2- \frac{\dot{a}}{a} = 0 \,,
\end{align}
with the right-hand side vanishing due to the absence of hypermomentum.

The dependence on $F(\tilde{\ourB})$ and its derivatives can then be eliminated, leading to a \textit{Riccati}\footnote{A Riccati equation is a non-linear ODE of the form $y'(x) = q_0(x) + q_1(x) y + q_2(x) y^2$. In the above we have $q_1 = 0$.} differential equation relating torsion, the scale factor and the matter content
\begin{align}
  \label{Ricatti}
  \dot{\mathfrak{h}} = \frac{1}{2}(1+w) a^3 \rho - 3\frac{\mathfrak{h}^2}{a^3} \,.
\end{align}
Because this equation does not depend explicitly on the function $F(\tilde{\ourB})$, torsion is uniquely determined once the scale factor and the matter evolution are known. However, these quantities do depend on the specific model in question and hence will affect the form of torsion. 

At this point, no further generic results can be extracted without specifying a concrete model. We choose one of the simplest possible settings, namely
\begin{align} \label{B_squared}
  F(\tilde{\ourB}) = -\beta \tilde{\ourB}^2 \,,
\end{align}
with $\beta > 0$. The chosen sign will become clear below. The task at hand is to solve the system of equations~(\ref{cos5a})--(\ref{cos6}), together with the Riccati equation~(\ref{Ricatti}) and continuity equation for matter. Since the boundary term depends on torsion and the scale factor, one naturally has a system of two equations in two unknowns. Note that matter again satisfies the standard conservation equation in our model so that $\rho = \rho_0/a^{3(1+w)}$, see~(\ref{rho0}). It turns out that one can eliminate torsion from the resulting equations and arrive at a single first-order equation in the scale factor. For simplicity, we will only state this equation for $w=0$, but the general equation is of similar form, 
\begin{align}
  \frac{\dot{a}}{a} = \frac{\sqrt{2}}{2\sqrt{3\beta}} \frac{Y\sqrt{Y^2+2Y-3}}{Y^2+4Y-1}\,,
  \qquad Y^2 = 1+\frac{36\beta\rho_0}{a^3} \,.
  \label{ODE}
\end{align}
The introduction of the variable $Y$ simplifies the presentation of this key equation substantially. The main observation is that the right-hand side is a function of the scale factor only. Therefore, this equation is, in principle, separable. Perhaps surprisingly, the resulting integral can be expressed in terms of elementary functions. This equation cannot be solved for $a(t)$ explicitly but gives a closed-form implicit solution. 

We begin by introducing the function 
\begin{align}
  \sinh(y(t)) = \sqrt{\frac{36\beta\rho_0}{a^3}}\,,
  \label{a2ODE}
\end{align}
which simplifies the ODE due to various hyperbolic identities. One then finds that~(\ref{ODE}) takes the form
\begin{align}
  \frac{dy}{dt} = -\frac{1}{\sqrt{3\beta}}\frac{\sqrt{3+\cosh(y)}\sech(y/2)}{2+8\coth(y)\csch(y)} \,.
  \label{a3ODE}
\end{align}
After separation of variables and integration, this leads to
\begin{multline} \label{ODEsol}
  \sqrt{3\beta/2}(t-t_0) = \sqrt{\frac{\cosh(y)+3}{\cosh(y)-1}} -
  2\arctan \Bigg(\frac{\sqrt{2}\sinh(y/2)}{\sqrt{3+\cosh(y)}} \Bigg) - \\
  2\arctanh \Bigg(\frac{1}{\sqrt{2}}\sinh(y/2) \Bigg) \,.
\end{multline}
Here $t_0$ is the constant of integration. Using the expression of $y$ in terms of $a$ gives an implicit formula for $a(t)$.

Despite the somewhat complicated form of~(\ref{ODE}) and the solution~(\ref{ODEsol}), we can make the following observations about limiting cases of this equation. First we make a series expansion assuming $a(t) \gg 1$, corresponding to the late universe, and arrive at
\begin{align}
  \frac{\dot{a}}{a} = \frac{\sqrt{\rho_0}}{\sqrt{3}} \frac{1}{a^{3/2}} + \mathcal{O} (a^{-5/2}) \,.
  \label{ODElate}
\end{align}
This is in agreement with a matter dominated universe, as we have used $w=0$. In particular, the late-time behaviour is independent of $\beta$ which strongly suggests that the boundary term affects the early-time dynamics of the universe only.

We now consider a series expansion of~(\ref{ODE}) assuming $a(t) \ll 1$, which gives
\begin{align}
  \frac{\dot{a}}{a} = \frac{\sqrt{2}}{3\sqrt{3}\sqrt{\beta}} + \mathcal{O} (a^{3/2}) \,,
  \label{ODEearly}
\end{align}
giving a constant Hubble function at early times. Writing $a = a_0 \exp(\lambda t)$ for the early universe, one finds the very neat result
\begin{align}
  \lambda = \frac{\sqrt{2}}{3\sqrt{3}\sqrt{\beta}} \,.
  \label{ODEearly2}
\end{align}
Our model can thus lead to a large amount of inflation for small values of $\beta$. Let us also note that the previous equation can easily be found for general $w$, which gives the expression
\begin{align}
  \lambda = \frac{\sqrt{2}}{\sqrt{3}\sqrt{\beta}}\frac{1}{3\sqrt{1-w^2}} \,.
  \label{ODEearly3}
\end{align}
Whilst the value of $\lambda$ is affected by the matter equation of state $w$, we can safely state that reasonable matter choices have no qualitative impact on this inflationary epoch.

This early-time inflation will nonetheless yield a late-time matter dominated universe whose expansion is independent of $\beta$. It is reasonable to expect that the introduction of the standard cosmological constant into this model will yield late-time accelerated expansion. Self-accelerating solutions with propagating torsion have been found in other theories with higher powers of curvature scalars even without spin/hypermomentum sources, e.g., in the context of massive gravity~\cite{Nikiforova:2016ngy,Nikiforova:2018pdk,Deffayet:2011uk}. However, given the simplicity of the model studied here, this is a most surprising result.

\begin{figure}[!htb] 
\centering
\includegraphics[width=.75\linewidth]{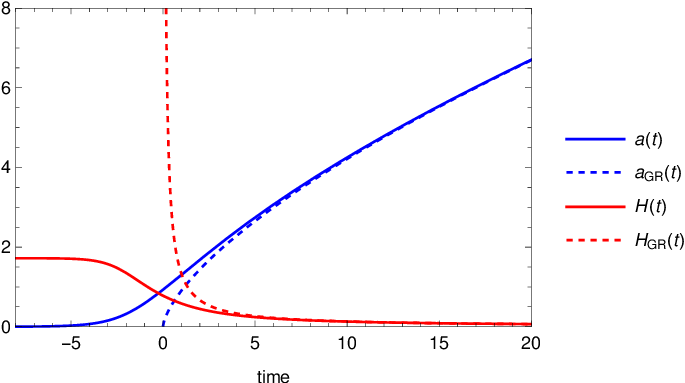}  
\caption{Solution of the field equations showing the scale factor $a(t)$ and Hubble factor $H(t)$ for a matter dominated universe. The dashed lines show the standard GR solutions. We note that the Hubble function was multiplied by a factor of 10 to improve the plot. The following numerical values were used: $w=0$, $\rho_0 =1$, $\kappa=1$, $\beta=1/10$. Other values give qualitatively similar results.}
\label{fig1Tora}
\end{figure}

Fig.~\ref{fig1Tora} shows the evolution of the scale factor and the Hubble parameter for a matter dominated universe, in agreement with the above asymptotic discussions. We also show the relevant GR solutions as dashed lines. Contrary to standard cosmology where the Hubble function diverges as $t \to 0$ we find that $H \to \lambda$ as $t \to -\infty$. This discussion can also be repeated for a radiation equation of state where the epoch of early-time inflation would then be followed by a radiation dominated epoch.

\begin{figure}[!hbt] 
\centering
\includegraphics[width=.80\linewidth]{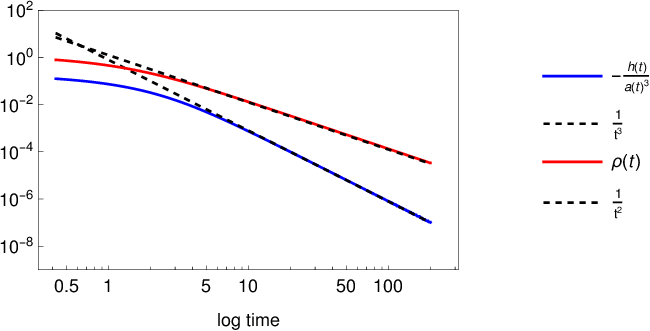}  
\caption{Log-log plot showing torsion $-h(t)/a^3$ (blue) and the energy density $\rho(t)$ (red) for a matter dominated universe together with the $1/t^3$ and $1/t^2$, shown as dashed lines. The following numerical values were used: $w=0$, $\rho_0 =1$, $\kappa=1$, $\beta=1/10$. Other values give qualitatively similar results.}
\label{fig2Tora}
\end{figure}

Next, let us turn to the behaviour of torsion. Since the late-time behaviour of the scale factor and matter agree with the GR results, we can assume $\rho \propto a^{-3} \propto t^{-2}$ and $a \propto t^{2/3}$ for late times, again assuming $w=0$. Using this input in~(\ref{Ricatti}) we can solve the Ricatti equation, explicitly or numerically. For the matter dominated epoch we find $-h/a^3 \propto 1/t^3 \propto \rho^{3/2}$ which, using the form of $a(t)$, also yields $-h \propto \sqrt{\rho}$. This is confirmed in Fig.~\ref{fig2Tora} where we display a log-log plot to emphasise the scaling behaviour of the late-time solution.

\clearpage

\begin{figure}[!htb] 
\centering
\includegraphics[width=.80\linewidth]{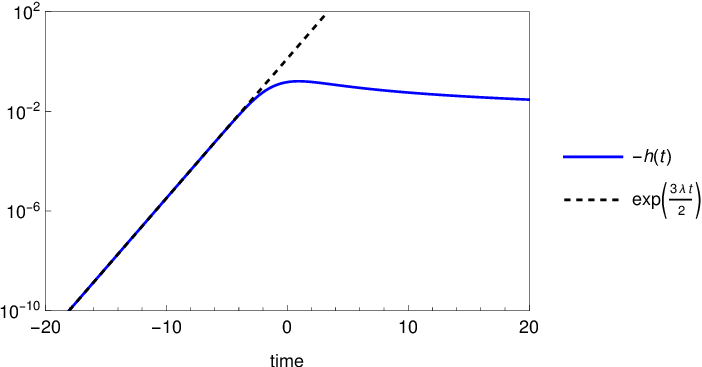}  
\caption{Log plot showing torsion $-h(t)$ and the solution $\exp(3\lambda t/2)$, again for a matter dominated universe. The following numerical values were used: $w=0$, $\rho_0 =1$, $\kappa=1$, $\beta=1/10$. Other values give qualitatively similar results.}
\label{fig2Torb}
\end{figure}

Likewise, we can also study the Ricatti equation assuming $a \propto \exp(\lambda t)$. As before one has $\rho \propto a^{-3}$ because the equation of state is unchanged. This gives $-h(t) \propto \exp(3\lambda t/2)$, which we can again verify using the numerical solutions, shown in Fig.~\ref{fig2Torb}. Rather interestingly, torsion again scales with the scale factor, and perhaps even more noteworthy, it thus scales with the matter. In particular one finds $-h(t) \propto 1/\sqrt{\rho}$ which is the inverse relation when compared to the late-time solution.

The dynamics of this model are very similar to the Born-Infeld $f(\ourG)$ models studied in Sec.~\ref{section6.1}. This should not come as much of a surprise, because a series expansion for the Born-Infeld Lagrangian leads to $\ourG - \ourG^2 /\lambda + \mathcal{O}(1/\lambda^2)$. The quadratic pseudo-scalar terms play a similar role in affecting the early-time dynamics of both models. This is reminiscent of the classic $R^2$ Starobinsky inflation~\cite{Starobinsky:1980te}, see Sec.~\ref{section1.2.1}. On the other hand, in the metric-affine setting, this can be attributed directly to the presence of torsion.

It should again be emphasised that this model, and those of the previous sections, can be seen as breaking diffeomorphism invariance. However, in the cosmological setting using Cartesian coordinates, the constraint equations are automatically satisfied. Consequently, the energy-momentum tensor is covariantly conserved, leading to the usual continuity equation. Another way of interpreting this result, using language \textit{\`{a} la} Stueckelberg, is that the equations for the auxiliary fields vanish identically for this background in these coordinates.

The cosmological solutions were studied in the Einstein-Cartan framework with vanishing non-metricity, but the full metric-affine theory could easily be incorporated into the study. Similarly, our general field equations and conservation equations allow for the possibility of connection source terms, the hypermomentum current. This may lead to a more interesting, albeit more complicated, set of solutions. One could also generalise the approach to allow for non-minimal couplings between the geometrical pseudo-scalars and the matter content, at which point the usual energy-momentum conservation laws would be modified. While non-minimal couplings between matter and geometry are straightforward to formulate at the level of the action, it is difficult to make qualitative statements about the effects of such terms without performing a detailed investigation. Moreover, we lack the knowledge of any guiding principles to select certain couplings over others, other than perhaps a preference for simpler models. Having said this, the minimally coupled theory presented here can be seen as the most obvious modification based on the natural decomposition of the Ricci scalar into its bulk and boundary parts within the metric-affine framework.

\label{chapterlabel6}

\begin{savequote}[70mm]
Even if there is only one possible unified theory, it is just a set of rules and equations. What is it that breathes fire into the equations and makes a universe for them to describe?
\qauthor{`A Brief History of Time' (1988) \\ Stephen Hawking}
\end{savequote}
\chapter{Outlook}
\label{chapter7}

In this thesis we have studied modified theories of gravity based on the decomposition of the Ricci scalar into its bulk and boundary terms. In the coordinate basis, using the metric tensor and its derivatives, this resulted in the Einstein action~(\ref{Einstein_action}), while in the orthonormal basis this led to the tetradic Einstein action~(\ref{Einstein_tetrad}). The former action was only pseudo-invariant under diffeomorphisms, due to the bulk term $\ourG$ transforming non-covariantly under coordinate transformations. The inhomogeneous part of the transformation could be written as a boundary term~(\ref{M surface}), so that the infinitesimal symmetry transformation of the total action vanished, see equation~(\ref{E_diff}). In the tetradic case, Sec.~\ref{section4.2}, the same argument could be applied but in terms of local Lorentz transformation. 

Modifications of these pseudo-invariant actions gave rise to second-order modified theories that break spacetime covariance or local Lorentz covariance. Crucially, a constraint equation could be obtained by requiring the total action (including matter) to be invariant under infinitesimal symmetry transformations. This was shown explicitly in the metrical case~(\ref{eqn:cons1}). These constraint equations ensure the consistency of our theory and simultaneously select out appropriate coordinates and frames to study solutions beyond GR.

We then went on to make comparisons with the modified teleparallel theories, known as $f(T)$ and $f(Q)$ gravity. Quite remarkably, these teleparallel theories in non-Riemannian geometries turn out to be equivalent to our non-covariant $f(\ourG)$ and $f(\mathfrak{G})$ theories. This was proven first for the unmodified theories in Sec.~\ref{section4.3}, where it was shown that covariantisation via the Stueckelberg procedure leads directly to the teleparallel scalars. In other words, the Stueckelberg-like fields associated with the invariance of these theories are exactly the degrees of freedom relating to the affine connection in teleparallel gravity.

In Sec.~\ref{section5.2.2} we extended this equivalence to the modified theories themselves. The exact relationship between all of these theories is illustrated in Fig.~\ref{fig:unified}, which summarises many of the results of this thesis. A dual interpretation is then possible: the non-covariant theories can be viewed as gauge-fixed versions of the fully covariant teleparallel theories, or the teleparallel theories can be seen as covariantised versions of the Riemannian modifications of the Einstein actions. To re-emphasise the points made in Sec.~\ref{section1.2.4}, breaking fundamental symmetries is synonymous with gauge-fixing. 

The question of coupling matter has always been somewhat complicated in the teleparallel theories and their modifications~\cite{Obukhov:2004hv,BeltranJimenez:2020sih}. However, in the approach laid out in this thesis, the question can be avoided altogether. This is because the background geometry for the non-covariant theories is simply Riemannian, just as in General Relativity. This sidesteps some of the more philosophical questions relating to matter couplings. Moreover, it exposes the intricate relationship between the decompositions of the Ricci scalar into first and second-order terms and the theories built from the teleparallel scalars.

We have also studied these decompositions in the non-Riemannian framework of metric-affine geometry. In this case, we worked exclusively in the coordinate basis, showing that the metric-affine bulk term $\ourG$ could be decomposed into a sum of its Levi-Civita and contortion parts. Modifications based on this metric-affine Einstein action~(\ref{Einstein_actionP}) gave rise to another non-covariant theory, see Sec.~\ref{section5.3}. There, the effects of torsion and non-metricity played a non-trivial role. Unlike in the previous Riemannian modifications, there is no direct link to the teleparallel theories.

In Chapter~\ref{chapter6} we looked at the cosmological applications of these modifications. In both the Riemannian setting, using the Born-Infeld Lagrangian~(\ref{BI_Lagrangian2}), and in the metric-affine setting, using a quadratic boundary model~(\ref{B_squared}), it was shown that an early inflationary period could be sourced geometrically. The initial big-bang singularity of GR was avoided without the need for additional fields, but the late-time modified solutions tracked their GR counterparts. We also utilised dynamical systems theory techniques to study the cosmologies of all of these theories. The critical points of the system indicated that de Sitter expansion is a generic feature of many models, and we derived the necessary criteria for the existence of these solutions.

The motivation for these modifications can be viewed in a number of different ways. Firstly, the decomposition of the Ricci scalar into bulk and boundary terms seems very natural, with the boundary term not contributing to the dynamics of GR. If one then uses the Einstein actions as a starting point, the $f(\ourG)$ and $f(\mathfrak{G})$ theories are logical choices for modifications. Returning to the theoretical motivations outlined in Sec.~\ref{section1.1.2}, if covariance is simply an emergent low-energy symmetry, then non-covariant theories based on extensions of $\ourG$ also seem plausible. Indeed, in the Born-Infeld models, diffeomorphism invariance was broken at a finite length scale, regularising the initial cosmological singularity of GR. Similarly, models giving rise to late-time accelerated expansion were studied in Sec.~\ref{section6.2}. In this way, many of the problems associated with General Relativity have the potential to be solved without requiring the introduction of additional fields or new matter content. These modified theories have the benefit of taking a simple and elegant form, but with enough freedom to address many of the obstacles that GR currently faces.

\section*{Future directions}

These modifications open up several new starting points for theories that break diffeomorphism or local Lorentz invariance in some way. For example, in the introduction we briefly mentioned the topological nature of the Ricci scalar in $n=2$ dimensions and the Gauss-Bonnet term in $n=4$ dimensions. However, by decomposing these scalars into their bulk and boundary parts, it is possible to make modifications that lead to non-trivial theories in these critical dimensions. In our work~\cite{Boehmer:2023lpb} we use this method to find a new $n \rightarrow 2$ limit of General Relativity, based on modifications of the bulk term $\ourG$. In the future, we hope to extend this analysis to the Gauss-Bonnet theories in $n = 4$ dimensions, see~\cite{fernandes20224d}. 

Other Gauss-Bonnet modifications can be approached by studying theories of the form $f(G_{\textrm{bulk}})$, where $G_{\textrm{bulk}}$ is its associated bulk term. Interestingly, this type of theory has been studied in the metric teleparallel framework~\cite{Kofinas:2014owa,Bahamonde:2016kba}, and very recently in symmetric teleparallel gravity too~\cite{Bajardi:2023gkd,Armaleo:2023rhj}. However, these modified Gauss-Bonnet theories lead to fourth-order equations of motion, due to the bulk part of the decomposition containing second-order terms. It would therefore be interesting to investigate whether it is possible to construct a second-order modified Gauss-Bonnet theory based on a new, first-order decomposition.

The unified approach of Section~\ref{section5.2.3}, making use of the boundary term $\mathbb{B} = \ourB - \mathfrak{B}$, opens up a new direction from which to compare and study the $f(T)$ and $f(Q)$ gravity theories. It is essentially the new boundary term $\mathbb{B}$ that contains all of the relevant dynamical information. Moreover, for a model of the form $f(\ourG + \alpha \mathbb{B})$, choices with $\alpha$ taking values of neither zero nor one constitute interesting new theories to be studied, breaking both diffeomorphism and local Lorentz invariance. In these cases, the geometric interpretations are less clear. Nonetheless, it would be interesting to explore this type of theory further.

In the fully metric-affine setting, we worked with the $f(\bar{\ourG}, \bar{\ourB})$ theories, but clearly this is only one possible choice of decomposition of $\bar{R}$. The next obvious choice is to consider the tetradic decomposition in the orthonormal basis, leading to $f(\bar{\mathfrak{G}},\bar{\mathfrak{B}})$ theories that break local Lorentz invariance. Again, with these theories the geometric interpretation is less clear. An area of future work that should be studied is explicitly restoring the invariance of these metric-affine modified theories by introducing auxiliary gauge fields. As we hypothesised at the end of Sec.~\ref{section5.3.2}, a natural interpretation may be a theory with two independent connections: the affine connection and a teleparallel connection. Alternatively, this can be seen as introducing an additional reference metric, such as in~\cite{Tomboulis:2017fim}.

In this work we have studied the background solutions in cosmology, but neglected cosmological perturbations. On the other hand, there has been much work studying exactly these topics in the equivalent $f(Q)$ or $f(T)$ theories. Though we expect the physical results to be the same, in our gauge-fixed theories we do not need to consider perturbations of the inertial affine or spin connection. This may offer some computational simplifications, and we expect to conduct research in this area in the future.

Though quite far removed from the classical work in this thesis, there are many questions relating to quantum theory that could be explored within this framework. Especially important are calculations that explicitly depend on the properties of boundary terms, such as the black hole entropy usually associated with the Gibbons-Hawking-York term in GR. It has been shown that the natural boundary term of TEGR in fact acts as a Gibbons-Hawking-York term, leading to the correct black hole Bekenstein entropy\footnote{See also~\cite{BeltranJimenez:2018vdo} for the STEGR case and further comments on quantization.}~\cite{Oshita:2017nhn}. This is of course important for the path integral formulations of quantum gravity. Moreover, as already pointed out, the Einstein action has technical advantages when it comes to studying perturbative quantum gravity~\cite{Tomboulis:2017fim}.

Lastly, and perhaps most importantly, the new degrees of freedom of these models must be investigated. In the modified  teleparallel theories this is known to be a difficult and complicated task~\cite{DAmbrosio:2023asf,Blixt:2020ekl}. In our case, we have new degrees of freedom associated with the breaking of symmetries. The number of propagating degrees of freedom around specific backgrounds should be studied, as well as any potential issues related to strong coupling or other pathologies. Perhaps the first step would again be to look at cosmological perturbations, which we plan to undertake in the future. Other techniques that may be useful include the EFT approach, which is beginning to see more of an application in modified theories of gravity. This would lead to a deeper understanding of these symmetry-breaking theories, and perhaps the role of diffeomorphism invariance and local Lorentz invariance in gravitational theories in general.

The modifications in this thesis serve as a novel approach to alternative theories of gravity. The close relationship with non-Riemannian theories should also not be understated. The breaking of diffeomorphism invariance and local Lorentz invariance can be seen as fundamental in these theories, leading to new phenomena beyond General Relativity. It is clear that there are many more directions that can be explored based on this work, and the close relationship with other theories of gravity naturally leads one to study these theories in greater detail. Theories that break fundamental symmetries are far less explored than other modifications of gravity, but our work shows the promising features of exploring exactly these types of modifications.

\renewcommand{\chapterheadstartvskip}{\vspace*{-6\baselineskip}}


\addcontentsline{toc}{chapter}{Appendices}

\appendix

 \addtocontents{toc}{\protect\setcounter{tocdepth}{1}}
 
\chapter{Projective transformations}
\label{appendixA}

\renewcommand{\theequation}{\thechapter.\arabic{equation}}

In this short appendix we give a brief introduction to projective transformations in gravitational theories.
Here we are working in the metric-affine setting, with an independent connection and metric. A \textit{projective transformation} is one that preserves the structure of autoparallel geodesics~(\ref{Autoparallel}), see~\cite{JS1954,ehlers1973geometry,eisenhart1997riemannian} for details. It is defined by the transformation of the affine connection
\begin{equation} \label{Projective_1}
\bar{\Gamma}^{\gamma}_{\alpha \beta} \rightarrow\bar{\Gamma}^{\gamma}_{\alpha \beta} + \delta_{\beta}^{\gamma} P_{\alpha} \, ,
\end{equation}
while leaving the metric and other fields unchanged. Here $P_{\alpha}(x)$ is an arbitrary spacetime one-form. Under this transformation the paths of geodesics remain unchanged (up to a different parameterisation), as does the parallelism of directions~\cite{JS1954}.

Let us consider a generalised version of the projective transformation, see for example~\cite{yano194460,hombu1941projective}, which we define as
\begin{align}
  \label{Projective}
  \bar{\Gamma}{}^{\gamma}_{\alpha \beta} \rightarrow \bar{\Gamma}{}^{\gamma}_{\alpha \beta} + c_1 \delta_{\alpha}^{\gamma} P_{\beta} + c_2 \delta_{\beta}^{\gamma} P_{\alpha} \,.
\end{align} 
When $c_{1}=0$ we retrieve~(\ref{Projective_1}).
Let us now study the effects on the curvature tensors. To our knowledge, these calculations for the generalised transformations have not appeared previously in the literature.

Following our calculations in~\cite{Boehmer:2023fyl}, we find the Riemann tensor transforms under the generalised projective transformation as
\begin{align}
\bar{R}_{\nu \mu \lambda}{}^{\gamma} &
\begin{multlined}[t]
\rightarrow \bar{R}_{\nu \mu \lambda}{}^{\gamma} + 2 c_2 \delta^{\gamma}_{\lambda} \partial_{[\nu} P_{\mu]} + 2 c_1  \delta_{[\mu}^{\gamma}  \partial_{\nu]}P_{\lambda} + 2 c_1 \bar{\Gamma}{}^{\gamma}_{[\nu \mu]} P_{\lambda} \\ +  2 c_1 \bar{\Gamma}{}^{\rho}_{[\mu| \lambda} \delta_{\nu]}^{\gamma} P_{\rho} + 2 c_1^2 \delta_{[\nu}^{\gamma}P_{\mu]} P_{\lambda} 
\end{multlined}
\nonumber \\
&= \bar{R}_{\nu \mu \lambda}{}^{\gamma} +  2 c_2 \delta^{\gamma}_{\lambda} \partial_{[\nu} P_{\mu]} + 2 c_1 \bar{\Gamma}{}^{\gamma}_{[\nu \mu]} P_{\lambda} + 2 c_1 \delta^{\gamma}_{[\mu} \bar{\nabla}_{\nu]} P_{\lambda} + 2 c_1^2 \delta_{[\nu}^{\gamma}P_{\mu]} P_{\lambda} \,,
\end{align}
from which we can calculate the Ricci tensor
\begin{align}
\bar{R}_{\mu \lambda} &\rightarrow \bar{R}_{\gamma \mu \lambda}{}^{\gamma} +  2 c_2 \delta^{\gamma}_{\lambda} \partial_{[\gamma} P_{\mu]} + 2 c_1 \bar{\Gamma}{}^{\gamma}_{[\gamma \mu]} P_{\lambda} + 2 c_1 \delta^{\gamma}_{[\mu} \bar{\nabla}_{\gamma]} P_{\lambda} + 2 c_1^2 \delta_{[\gamma}^{\gamma}P_{\mu]} P_{\lambda} \nonumber \\
&= \bar{R}_{\mu \lambda} +  2 c_2 \partial_{[\lambda} P_{\mu]} + 2 c_1 \, \bar{\Gamma}{}^{\gamma}_{[\gamma \mu]} P_{\lambda}   + c_1 (1-n) \bar{\nabla}_{\mu} P_{\lambda} + c_1^2 (n-1) P_{\mu} P_{\lambda} \, ,
\end{align}
and the Ricci scalar
\begin{align}
\bar{R} &\rightarrow g^{\mu \lambda} \bar{R} _{\mu \lambda} + g^{\mu \lambda} \big(2 c_2 \, \partial_{[\lambda} P_{\mu]} + 2 c_1 \bar{\Gamma}{}^{\gamma}_{[\gamma \mu]} P_{\lambda}   + c_1 (1-n) \bar{\nabla}_{\mu} P_{\lambda} + c_1^2 (n-1) P_{\mu} P_{\lambda} \big) \nonumber \\
&=\bar{R}  + 2 c_1 \bar{\Gamma}{}^{\gamma}_{[\gamma \mu]} P^{\mu} +  c_1 (1-n) \bar{\nabla}^{\mu} P_{\mu} + c_1^2 (n-1) P^{\mu} P_{\mu} \, .
\end{align}
Importantly, we see that for standard projective transformations of the form~(\ref{Projective_1}) with $c_1=0$ and $c_2 =1$ we obtain
\begin{align} \label{Proj1}
\bar{R}_{\nu \mu \lambda}{}^{\gamma} &\rightarrow \bar{R}_{\nu \mu \lambda}{}^{\gamma} +  2 \delta^{\gamma}_{\lambda} \partial_{[\nu} P_{\mu]}  \,, \\
\bar{R}_{\mu \lambda} &\rightarrow  \bar{R}_{\mu \lambda} +  2 \partial_{[\lambda} P_{\mu]} \, ,\\
\bar{R} & \rightarrow \bar{R} \, , \label{Proj3}
\end{align}
and the Ricci scalar is invariant. For more details on the properties and physical interpretations of equations~(\ref{Proj1})-(\ref{Proj3}) see~\cite{hehl1978metric,hehl1981metric}.

 Lastly, let us look at the case when torsion vanishes and the connection is symmetric~\cite{JS1954}. In this case, whether or not non-metricity is present makes no difference on the form of the equations, so we will work in the Levi-Civita case. For the Levi-Civita connection, the Riemann tensor transforms under the generalised projective transformation as
\begin{align} 
R_{\nu \mu \lambda}{}^{\gamma} &\rightarrow  R_{\nu \mu \lambda}{}^{\gamma}+  2 c_2 \delta^{\gamma}_{\lambda} \nabla_{[\nu} P_{\mu]}  + 2 c_1 \delta^{\gamma}_{[\mu} \nabla_{\nu]} P_{\lambda} + 2 c_1^2 \delta_{[\nu}^{\gamma}P_{\mu]} P_{\lambda} \,. 
\end{align}
Defining the tensor $P_{\mu \nu} := P_{\mu} P_{\nu} - \nabla_{\mu} P_{\nu}$ and setting $c_1 = c_2 = 1$, the above (Levi-Civita) transformation can be expressed as
\begin{align} \label{Projective_Levi1}
R_{\nu \mu \lambda}{}^{\gamma} \rightarrow R_{\nu \mu \lambda}{}^{\gamma} -2 \delta^{\gamma}_{\lambda} P_{[\nu \mu]} + 2 \delta^{\gamma}_{[\nu} P_{\mu] \lambda} \, ,
 \end{align}
which can be found in~\cite{JS1954}.
 The Levi-Civita Ricci tensor and Ricci scalar then have
\begin{align} \label{Projective_Levi2}
R_{\mu \lambda} &\rightarrow  R_{\mu \lambda}  - P_{\lambda \mu}  + n P_{\mu \lambda}  \,,\\
\label{Projective_Levi3} 
R &\rightarrow R + (n-1)P^{\nu}{}_{\nu} \,.
\end{align}

\chapter{Coordinate transformations}
\label{appendixB}

\renewcommand{\theequation}{\thesection.\arabic{equation}}

Here we include the explicit calculations for some of the more involved coordinate transformations used throughout the thesis, in particular those in Chapter~\ref{chapter4}. We refer to Sections~\ref{section2.2} and~\ref{section2.3} and references therein for further mathematical details regarding the role of coordinate transformations and their relation to diffeomorphisms and action principles. Some of the results of this appendix are also reported in our works~\cite{Boehmer:2021aji,Boehmer:2023fyl}, but here we show more details of these calculations.

\section{Levi-Civita framework}
\label{appendixB.1}

In this section we explicitly calculate the action of infinitesimal coordinate transformations. For the infinitesimal coordinate transformation $x^{\mu} \rightarrow \hat{x}^{\mu} = x^{\mu} + \xi^{\mu}(x)$, where $\xi^{\mu}$ is small, we have the following standard relations to first order in $\xi^{\mu}$, dropping terms of order $\mathcal{O} (\xi^2)$ as usual
\begin{align}
\frac{\partial \hat{x}^{\mu}}{\partial x^{\nu}} = \delta^{\mu}_{\nu} + \partial_{\nu} \xi^{\mu} \ , \quad
\frac{\partial x^{\mu}}{\partial \hat{x}^{\nu}} = \delta^{\mu}_{\nu} - \partial_{\nu} \xi^{\mu} \ .
\end{align}
The metric tensor and its inverse transform as
\begin{align} \label{append_inf_g1}
g_{\mu \nu}(x) \rightarrow
\hat{g}_{\mu \nu}(\hat{x}) &= g_{\mu \nu} - \partial_{\mu} \xi^{\lambda} g_{\lambda \nu} - \partial_{\nu} \xi^{\lambda} g_{\mu \lambda} \, , \\  
g^{\mu \nu}(x) \rightarrow
\hat{g}^{\mu \nu}(\hat{x}) &= g^{\mu \nu} + \partial_{\lambda} \xi^{\mu} g^{\lambda \nu} + \partial_{\lambda} \xi^{\nu} g^{\mu \lambda} \, , \label{append_inf_g2}
\end{align}
and the affine connection transforms as
\begin{align} \label{Chr transform}
 \bar{\Gamma}^{\gamma}_{\mu \nu} (x) \rightarrow \hat{\bar{\Gamma}}^{\gamma}_{\mu \nu}(\hat{x}) = \bar{\Gamma}^{\gamma}_{\mu \nu} + \partial_{\lambda} \xi^{\gamma} \bar{\Gamma}^{\lambda}_{\mu \nu} - \partial_{\mu} \xi^{\lambda} \bar{\Gamma}^{\gamma}_{\lambda \nu} - \partial_{\nu} \xi^{\lambda} \bar{\Gamma}^{\gamma}_{\mu \lambda} -\partial_{\mu} \partial_{\nu} \xi^{\gamma}  \, .
\end{align}
Almost all of the calculations in this section follow directly from the above expressions, though they are at times longwinded.

Let us now look at how the Levi-Civita objects of Sec.~\ref{section4.1} transform. In particular, we will prove the infinitesimal transformation equations for $\ourG$~(\ref{G_infinitesimal}), $\ourB$~(\ref{B_infinitesimal}), $M^{\mu \nu}{}_{\lambda}$~(\ref{M_infinitesimal}) and $E^{\mu \nu}{}_{\lambda}$~(\ref{E_infinitesimal}). 

\subsubsection{Bulk term $\ourG$}
Beginning with the bulk term $\ourG$, we have 
\begin{align}
\hat{\ourG} &= \hat{g}^{\mu \nu}\Big( \hat{\Gamma}^{\lambda}_{\mu \sigma} \hat{\Gamma}^{\sigma}_{\lambda \nu} -
  \hat{\Gamma}^{\sigma}_{\mu \nu} \hat{\Gamma}^{\lambda}_{\lambda \sigma} \Big) \nonumber \\
  & \begin{multlined}[b]
  =  \big(g^{\mu \nu} + \partial_{\gamma}\xi^{\mu} g^{\gamma \nu} + \partial_{\gamma}\xi^{\nu} g^{\gamma \mu}\big) \\
  \times
 \Bigg[ \Big( \Gamma^{\lambda}_{\mu \sigma} + \partial_{\eta} \xi^{\lambda} \Gamma^{\eta}_{\mu \sigma} - \partial_{\mu} \xi^{\eta} \Gamma^{\lambda}_{\eta \sigma} - \partial_{\sigma} \xi^{\eta} \Gamma^{\lambda}_{\mu \eta} - \partial_{\mu} \partial_{\sigma} \xi^{\lambda} \Big)   \\
\times \Big(\Gamma^{\sigma}_{\lambda \nu} + \partial_{\rho} \xi^{\sigma} \Gamma^{\rho}_{\lambda \nu} - \partial_{\lambda} \xi^{\rho} \Gamma^{\sigma}_{\rho \nu} - \partial_{\nu} \xi^{\rho} \Gamma^{\sigma}_{\lambda \rho} - \partial_{\lambda} \partial_{\nu} \xi^{\sigma} \Big)  \\
- \Big( \Gamma^{\sigma}_{\mu \nu} + \partial_{\eta} \xi^{\sigma} \Gamma^{\eta}_{\mu \nu} - \partial_{\mu} \xi^{\eta} \Gamma^{\sigma}_{\eta \nu}  - \partial_{\nu} \xi^{\eta} \Gamma^{\sigma}_{\eta \mu} - \partial_{\mu} \partial_{\nu} \xi^{\sigma} \Big) \\
\times \Big(   \Gamma^{\lambda}_{\lambda \sigma} - \partial_{\sigma} \xi^{\eta} \Gamma^{\lambda}_{\eta \lambda} - \partial_{\lambda} \partial_{\sigma} \xi^{\lambda} \Big)  \Bigg]  \, .
  \end{multlined}
\end{align}
With not too much work it can be seen that all terms proportional to $\partial \xi$ cancel with one another. Explicitly, after making some cancellations, we find 
\begin{subequations}
\begin{align}
 \hat{g}^{\mu \nu} \hat{\Gamma}^{\lambda}_{\mu \sigma} \hat{\Gamma}^{\sigma}_{\lambda \nu} &=  g^{\mu \nu} \big( \Gamma^{\lambda}_{\mu \sigma} \Gamma^{\sigma}_{\lambda \nu} - 2 \Gamma^{\lambda}_{\mu \sigma} \partial_{\lambda} \partial_{\nu} \xi^{\sigma}  \big) \, , \\
 \hat{g}^{\mu \nu}  \hat{\Gamma}^{\sigma}_{\mu \nu} \hat{\Gamma}^{\lambda}_{\lambda \sigma} &=  g^{\mu \nu} \big( \Gamma
 _{\mu \nu}^{\sigma} \Gamma^{\lambda}_{\lambda \sigma} - \Gamma^{\sigma}_{\mu \nu} \partial_{\lambda} \partial_{\sigma} \xi^{\lambda} - \Gamma^{\lambda}_{\lambda \sigma} \partial_{\mu} \partial_{\nu} \xi^{\sigma} \big) \, .
\end{align}
\end{subequations}
Note that we have made use of the symmetry of the Levi-Civita connection and discarded order $\mathcal{O} (\xi^2)$ terms. Combining these results we arrive at
\begin{align}
\hat{\ourG} &= \ourG - g^{\mu \nu} \big( 2 \Gamma^{\lambda}_{\mu \sigma} \partial_{\lambda} \partial_{\nu} \xi^{\sigma} - \Gamma^{\sigma}_{\mu \nu} \partial_{\lambda} \partial_{\sigma} \xi^{\lambda} - \Gamma^{\lambda}_{\lambda \sigma} \partial_{\mu} \partial_{\nu} \xi^{\sigma} \big) \nonumber
\\
&= \ourG - \partial_{\mu} \partial_{\nu} \xi^{\lambda} \big( 2g^{\rho (\nu} \Gamma^{\mu)}_{\lambda \rho} - g^{\mu \nu} \Gamma^{\rho}_{\rho \lambda} -
  g^{\rho \sigma}  \delta^{(\nu}_{\lambda} \Gamma^{\mu)}_{\rho \sigma}  \big)  \nonumber \\
&=  \ourG - M^{\mu \nu}{}_{\gamma} \partial_{\mu} \partial_{\nu} \xi^{\gamma} \, , \label{append_G_transform}
\end{align}
where we have inserted the definition of $M^{\mu \nu}{}_{\gamma}$ in~(\ref{M}).

\subsubsection{Boundary term $\ourB$}
For the boundary term $\ourB$, we have higher derivatives, so the result is slightly less obvious to see. We will therefore include more details of this calculation. Let us begin by writing $\ourB$ in the following form
\begin{align}
\ourB = \Gamma^{\lambda}_{\lambda \sigma} B^{\sigma} + \partial_{\sigma} B^{\sigma} \ .
\end{align}
where $B^{\sigma}$ is the boundary vector defined in~(\ref{Bvector})
\begin{equation}
 B^{\sigma} = g^{\mu \nu} \Gamma^{\sigma}_{\mu \nu} - g^{\sigma \nu} \Gamma^{\lambda}_{\lambda \nu} \, .
\end{equation}
The boundary vector transforms as
\begin{align}
\hat{B}^{\sigma}  &= \hat{g}^{\mu \nu} \hat{\Gamma}^{\sigma}_{\mu \nu} - \hat{g}^{\nu \sigma} \Gamma^{\lambda}_{\nu \lambda} \nonumber \\ 
&= \big( g^{\mu \nu} + \partial_{\eta} \xi^{\mu} g^{\eta \nu} + \partial_{\eta} \xi^{\nu} g^{\mu \eta} \big) \Big(\Gamma^{\sigma}_{\mu \nu} + \partial_{\lambda} \xi^{\sigma} \Gamma^{\lambda}_{\mu \nu} - \partial_{\mu} \xi^{\lambda} \Gamma^{\sigma}_{\lambda \nu} - \partial_{\nu} \xi^{\lambda} \Gamma^{\sigma}_{\mu \lambda} - \partial_{\mu} \partial_{\nu} \xi^{\sigma} \Big) \nonumber \\
&- \big( g^{\sigma \nu} + \partial_{\eta} \xi^{\sigma} g^{\eta \nu} + \partial_{\eta} \xi^{\nu} g^{\sigma \eta} \big) \Big(\Gamma^{\lambda}_{\lambda \nu} - \partial_{\nu} \xi^{\gamma} \Gamma^{\lambda}_{\gamma \lambda} - \partial_{\lambda} \partial_{\nu} \xi^{\lambda}\Big) \nonumber \\
&=  {g}^{\mu \nu} {\Gamma}^{\sigma}_{\mu \nu} - {g}^{\nu \sigma} \Gamma^{\lambda}_{\mu \lambda}  + \partial_{\eta} \xi^{\sigma} \big( g^{\mu \nu} \Gamma^{\eta}_{\mu \nu} - g^{\eta \nu} \Gamma^{\lambda}_{\lambda \nu}\big) - \partial^{\nu} \partial_{\nu} \xi^{\sigma} + \partial^{\sigma} \partial_{\lambda} \xi^{\lambda} 
 \nonumber \\
 &= B^{\sigma} + \partial_{\eta} \xi^{\sigma} B^{\eta} - \partial^{\nu} \partial_{\nu} \xi^{\sigma} + \partial^{\sigma} \partial_{\lambda} \xi^{\lambda} \, .
\end{align} 
The whole term then transforms as
\begin{align}
\hat{\ourB} &=  \hat{\Gamma}^{\lambda}_{\lambda \sigma} \hat{B}^{\sigma} + \hat{\partial}_{\sigma} \hat{B}^{\sigma}  \nonumber \\
& \begin{multlined}[b]
=  \big( \Gamma^{\lambda}_{\lambda \sigma} - \partial_{\sigma} \xi^{\eta} \Gamma^{\lambda}_{\eta \lambda} - \partial_{\lambda} \partial_{\sigma} \xi^{\lambda} \big) \big( B^{\sigma} + \partial_{\nu} \xi^{\sigma} B^{\nu} - \partial^{\nu} \partial_{\nu} \xi^{\sigma} + \partial^{\sigma} \partial_{\nu} \xi^{\nu} \big)  \\
+ \big( \partial_{\sigma} - \partial_{\sigma} \xi^{\beta} \partial_{\beta} \big) \big(B^{\sigma} + \partial_{\nu} \xi^{\sigma} B^{\nu} - \partial^{\nu} \partial_{\nu} \xi^{\sigma} + \partial^{\sigma} \partial_{\nu} \xi^{\nu}  \big) \, .
\end{multlined}
\end{align}
Expanding all the brackets and using the chain rule allows us to cancel most of the terms
\begin{multline}
\hat{\ourB} = \ourB - \overbrace{ \cancel{ \partial_{\sigma} \xi^{\eta} \Gamma^{\lambda}_{\eta \lambda}  B^{\sigma}}}^{\textrm{a}} -  
\overbrace{ \cancel{ \partial_{\lambda} \partial_{\sigma} \xi^{\lambda} B^{\sigma} }}^{\textrm{b}}  + \overbrace{\cancel{ \Gamma^{\lambda}_{\lambda \sigma}\partial_{\nu} \xi^{\sigma} B^{\nu}}}^{\textrm{a}} - \Gamma^{\lambda}_{\lambda \sigma}  \partial^{\nu} \partial_{\nu} \xi^{\sigma}  
 + \Gamma^{\lambda}_{\lambda \sigma} \partial^{\sigma} \partial_{\nu} \xi^{\nu}  \\
 + \overbrace{ \cancel{\partial_{\sigma} ( \partial_{\nu} \xi^{\sigma}) B^{\nu}} }^{\textrm{b}} + \overbrace{  \cancel{ \partial_{\sigma}(B^{\nu}) \partial_{\nu} \xi^{\sigma}} }^{\textrm{c}} 
 - \partial_{\sigma}  (\partial^{\nu} \partial_{\nu} \xi^{\sigma})   
 + \partial_{\sigma}( \partial^{\sigma} \partial_{\nu} \xi^{\nu}) - \overbrace{ \cancel{\partial_{\sigma} \xi^{\beta} \partial_{\beta} B^{\sigma}}}^{\textrm{c}} \, ,
\end{multline}
where we have indicated the terms that cancel with one another.

The third-order partial derivate terms can be rewritten as
\begin{align}
- \partial_{\sigma}  (\partial^{\nu} \partial_{\nu} \xi^{\sigma}) &= - \partial_{\sigma}(g^{\nu \gamma})\partial_{\gamma} \partial_{\nu} \xi^{\sigma}
- g^{\nu \gamma} \partial_{\sigma}\partial_{\gamma} \partial_{\nu} \xi^{\sigma} \nonumber \\
&= \left( \Gamma^{\nu}_{\sigma \eta} g^{\eta\gamma} + \Gamma^{\gamma}_{\sigma \eta} g^{\nu \eta} \right) \partial_{\gamma} \partial_{\nu} \xi^{\sigma} - \partial^{\nu} \partial_{\nu} \partial_{\sigma} \xi^{\sigma} \nonumber \\
&= 2 \Gamma^{\nu}_{\sigma \eta} \partial^{\eta} \partial_{\nu} \xi^{\sigma} - \partial^{\nu} \partial_{\nu} \partial_{\sigma} \xi^{\sigma} \ ,
\end{align}
\vspace{-12mm}
\begin{align}
\partial_{\sigma}( \partial^{\sigma} \partial_{\nu} \xi^{\nu}) &= \partial_{\sigma}( g^{\gamma \sigma}) \partial_{\gamma} \partial_{\nu} \xi^{\nu}
+ \partial^{\sigma} \partial_{\sigma} \partial_{\nu} \xi^{\nu} \nonumber \\
&= -(\Gamma^{\gamma}_{\sigma \eta} g^{\eta \sigma} + \Gamma^{\sigma}_{\sigma \eta} g^{\gamma \eta}) \partial_{\gamma} \partial_{\nu} \xi^{\nu} + \partial^{\sigma} \partial_{\sigma} \partial_{\nu} \xi^{\nu} \nonumber \\
&=  -\Gamma^{\gamma}_{\sigma \eta} g^{\eta \sigma} \partial_{\gamma} \partial_{\nu} \xi^{\nu} - \Gamma^{\sigma}_{\sigma \eta} \partial^{\eta} \partial_{\nu} \xi^{\nu} + \partial^{\sigma} \partial_{\sigma} \partial_{\nu} \xi^{\nu} \, ,
\end{align}
so that only the second-order derivatives remain.
In total we then have
\begin{align}
\hat{\ourB} &= \ourB -\Gamma^{\lambda}_{\lambda \sigma}  \partial^{\nu} \partial_{\nu} \xi^{\sigma}  + \cancel{ \Gamma^{\lambda}_{\lambda \sigma} \partial^{\sigma} \partial_{\nu} \xi^{\nu}} + 2 \Gamma^{\nu}_{\sigma \eta} \partial^{\eta} \partial_{\nu} \xi^{\sigma} - \cancel{\partial^{\nu} \partial_{\nu} \partial_{\sigma} \xi^{\sigma} } \nonumber \\ 
&- \Gamma^{\gamma}_{\sigma \eta} g^{\eta \sigma} \partial_{\gamma} \partial_{\nu} \xi^{\nu} - \cancel{ \Gamma^{\sigma}_{\sigma \eta} \partial^{\eta} \partial_{\nu} \xi^{\nu}} + \cancel{ \partial^{\sigma} \partial_{\sigma} \partial_{\nu} \xi^{\nu}} \nonumber \\
&= \ourB + \partial_{\alpha} \partial_{\beta} \xi^{\gamma} \big(2 g^{\mu (\alpha} \Gamma^{\beta)}_{\mu \gamma} - g^{\alpha \beta} \Gamma^{\lambda}_{\lambda \gamma} - g^{\mu \nu} \delta^{(\beta}_{\gamma} \Gamma^{\alpha)}_{\mu \nu} \big) \nonumber \\
 &= \ourB + M^{\alpha \beta}{}_{\gamma} \partial_{\alpha} \partial_{\beta} \xi^{\gamma} \, , \label{append_B_transform}
\end{align}
where we have again inserted the definition of $M^{\alpha \beta}{}_{\gamma}$ in the final line.
\subsubsection{Superpotential $E_{\mu \nu}{}^{\lambda}$}
Moving on to the superpotential, recall the form of $E_{\rho \sigma}{}^{\gamma}$~(\ref{E})
\begin{equation}
    E_{\rho \sigma}{}^{\gamma} = 2 \Gamma^{\gamma}_{\rho \sigma} - 2 \delta^{\gamma}_{(\rho} \Gamma^{\lambda}_{\sigma) \lambda } + g_{\rho \sigma} g^{\mu \gamma} \Gamma^{\lambda}_{\lambda \mu} - g_{\rho \sigma} g^{\mu \nu} \Gamma^{\gamma}_{\mu \nu} \ , \nonumber
\end{equation}
which transforms as 
\begin{align} \label{Appendix_inf_E1}
E_{\rho \sigma}{}^{\gamma} \rightarrow \hat{E}_{\rho \sigma}{}^{\gamma} &= 2 \hat{\Gamma}^{\gamma}_{\rho \sigma} - 2 \delta^{\gamma}_{(\rho} \hat{\Gamma}^{\lambda}_{\sigma)\lambda} + \hat{g}_{\rho \sigma} \hat{g}^{\mu \gamma} \hat{\Gamma}^{\lambda}_{\lambda \mu} - \hat{g}_{\rho \sigma} \hat{g}^{\mu \nu} \hat{\Gamma}^{\gamma}_{\mu \nu} \, .\end{align}
The individual terms of~(\ref{Appendix_inf_E1}) are
\begin{subequations}
\begin{align}
2 \hat{\Gamma}^{\gamma}_{\rho \sigma} &= 2 \Big( \Gamma^{\gamma}_{\rho \sigma} + \partial_{\eta} \xi^{\gamma} \Gamma^{\eta}_{\rho \sigma} - \partial_{\rho} \xi^{\eta} \Gamma^{\gamma}_{\sigma \eta} - \partial_{\sigma} \xi^{\eta} \Gamma^{\gamma}_{\rho \eta} - \partial_{\rho} \partial_{\sigma} \xi^{\gamma} \Big) \ ,
\\
2 \delta^{\gamma}_{(\rho} \hat{\Gamma}^{\lambda}_{\sigma)\lambda} &= 2 \delta^{\gamma}_{(\rho} \Big(\Gamma^{\lambda}_{\sigma)\lambda}  - \partial_{|\sigma)} \xi^{\eta} \Gamma^{\lambda}_{\lambda \eta} - \partial_{\lambda} \partial_{|\sigma)} \xi^{\lambda} \Big) \ ,
\\
 \hat{g}_{\rho \sigma} \hat{g}^{\mu \gamma} \hat{\Gamma}^{\lambda}_{\lambda \mu} &=
 \begin{multlined}[t]
  \Big(g_{\rho \sigma} g^{\mu \gamma} + 2 g_{\rho \sigma} \partial^{(\mu} \xi^{\gamma)} - 2 g^{\mu \gamma} g_{\eta (\sigma} \partial_{\rho)} \xi^{\eta} \Big) \Gamma^{\lambda}_{\lambda \mu} 
 \\ 
  - g_{\rho \sigma}g^{\mu \gamma} \Big( 
  \partial_{\mu}\xi^{\eta} \Gamma^{\lambda}_{\lambda \eta} + \partial_{\mu} \partial_{\lambda} \xi^{\lambda} \Big) \ , 
   \end{multlined}
  \\
  \hat{g}_{\rho \sigma} \hat{g}^{\mu \nu} \hat{\Gamma}^{\gamma}_{\mu \nu} & =
  \begin{multlined}[t]
 \Big(g_{\rho \sigma} g^{\mu \nu} + g_{\rho \sigma} \partial^{(\mu} \xi^{\nu)} - g^{\mu \nu} g_{\eta (\sigma} \partial_{\rho)} \xi^{\eta} \Big) \Gamma^{\gamma}_{\mu \nu} 
  \\
 + g_{\rho \sigma} g^{\mu \nu} \Big( \partial_{\eta} \xi^{\gamma} \Gamma_{\mu \nu}^{\eta}  - 2 \partial_{\nu}\xi^{\eta} \Gamma_{\mu \eta}^{\gamma} - \partial_{\mu} \partial_{\nu} \xi^{\gamma} \Big) \ .
   \end{multlined}
\end{align}
\end{subequations}
Collecting these terms together and performing a few cancellations leads to
\begin{multline}
 \hat{E}_{\rho \sigma}{}^{\gamma} =
2  \Gamma^{\gamma}_{\rho \sigma} -2 \delta^{\gamma}_{(\rho} \Gamma^{\lambda}_{\sigma) \lambda}  +  g_{\rho \sigma}g^{\mu \gamma} \Gamma^{\lambda}_{\lambda \mu} -  g_{\rho \sigma} g^{\mu \nu} \Gamma^{\gamma}_{\mu \nu} 
\\
+ \partial_{\rho} \xi^{\eta} \Big( -2\Gamma^{\gamma}_{\sigma \eta} + \delta^{\gamma}_{\sigma} \Gamma^{\lambda}_{\lambda \eta} - g_{\sigma\eta} g^{\mu \gamma} \Gamma^{\lambda}_{\lambda \mu} + g_{\sigma \eta} g^{\mu \nu} \Gamma^{\gamma}_{\mu \nu} \Big) 
\\
+  \partial_{\sigma} \xi^{\eta} \Big( -2  \Gamma^{\gamma}_{\rho \eta} +  \delta^{\gamma}_{\rho}  \Gamma^{\lambda}_{\lambda \eta} - g_{\rho \eta} g^{\mu \gamma} \Gamma^{\lambda}_{\lambda \mu} + g_{\rho \eta} g^{\mu \nu} \Gamma^{\gamma}_{\mu \nu} \Big) 
 \\
+ \partial_{\eta} \xi^{\gamma} \Big(2 \Gamma^{\eta}_{\rho \sigma} + g_{\rho \sigma} g^{\eta \mu} \Gamma^{\lambda}_{\lambda \mu} - g_{\rho \sigma} g^{\mu \nu} \Gamma^{\eta}_{\mu \nu} \Big)  
 \\
- 2 \partial_{\rho } \partial_{\sigma} \xi^{\gamma} + 2 \delta^{\gamma}_{(\rho} \partial_{\sigma)} \partial_{\lambda} \xi^{\lambda} - g_{\rho \sigma} g^{\mu \gamma} \partial_{\mu} \partial_{\lambda} \xi^{\lambda} + g_{\rho \sigma} g^{\mu \nu} \partial_{\mu} \partial_{\nu} \xi^{\gamma} \ .
\end{multline}
Inserting the definition of $E_{\rho \sigma}{}^{\gamma}$ we can simplify the above expression
\begin{align}
 \hat{E}_{\rho \sigma}{}^{\gamma} & \begin{multlined}[t]
   = E_{\rho \sigma}{}^{\gamma} +  \partial_{\rho} \xi^{\eta} ( -E_{\eta \sigma}{}^{\gamma} - \delta^{\gamma}_{\eta} \Gamma^{\lambda}_{\lambda \sigma} ) 
   \\
    \partial_{\sigma} \xi^{\eta} (-E_{\rho \eta}{}^{\gamma} - \delta^{\gamma}_{\eta} \Gamma^{\lambda}_{\lambda \sigma} )  + \partial_{\eta} \xi^{\gamma} (E_{\rho \sigma}{}^{\eta} + \delta_{\rho}^{\eta} \Gamma^{\lambda}_{\lambda \sigma} + \delta_{\sigma}^{\eta} \Gamma^{\lambda}_{\lambda \rho} )    \\ 
 - 2 \partial_{\rho } \partial_{\sigma} \xi^{\gamma} + 2 \delta^{\gamma}_{\rho} \partial_{\lambda} \partial_{\sigma} \xi^{\lambda} - g_{\rho \sigma} g^{\mu \gamma} \partial_{\mu} \partial_{\lambda} \xi^{\lambda} + g_{\rho \sigma} g^{\mu \nu} \partial_{\mu} \partial_{\nu} \xi^{\gamma} \nonumber 
 \end{multlined}
 \\
 &= E_{\rho \sigma}{}^{\gamma} - \partial_{\rho} \xi^{\eta} E_{\eta \sigma}{}^{\gamma} -  \partial_{\sigma} \xi^{\eta} E_{\rho \eta}{}^{\gamma} +  \partial_{\eta} \xi^{\gamma} E_{\rho \sigma}{}^{\eta} \nonumber 
 \\
 & \quad + 2 \partial_{\rho } \partial_{\sigma} \xi^{\gamma} - 2 \delta^{\gamma}_{(\rho} \partial_{\sigma)} \partial_{\lambda} \xi^{\lambda} + g_{\rho \sigma} g^{\mu \gamma} \partial_{\mu} \partial_{\lambda} \xi^{\lambda} - g_{\rho \sigma} g^{\mu \nu} \partial_{\mu} \partial_{\nu} \xi^{\gamma} \, .
 \end{align}
 In total we therefore find 
 \begin{multline}
\hat{E}_{\rho \sigma}{}^{\gamma} =  E_{\rho \sigma}{}^{\gamma} - E_{\rho \eta}{}^{\gamma} \partial_{\sigma} \xi^{\eta} - E_{\sigma \eta}{}^{\gamma} \partial_{\rho} \xi^{\eta} + E_{\rho \sigma}{}^{\eta} \partial_{\eta} \xi^{\gamma}  \\ 
- 2 \partial_{\rho} \partial_{\sigma} \xi^{\gamma} + 2 \delta_{(\rho}^{\gamma} \partial_{\sigma)} \partial_{\lambda} \xi^{\lambda} - g_{\rho \sigma} \partial^{\gamma} \partial_{\lambda} \xi^{\lambda} + g_{\rho \sigma} \partial^{\lambda} \partial_{\lambda} \xi^{\gamma} \, . \label{append_E_transform}
\end{multline}

By this point it should be quite clear how we arrive at the infinitesimal transformations for these Levi-Civita-based objects. Here we simply state the transformation for the object $M^{\mu \nu}{}_{\lambda}$, which follows the same rules as above
\begin{multline}
\hat{M}^{\alpha \beta}{}_{\gamma} =   M^{\alpha \beta }{}_{\gamma} + M^{\lambda \beta }{}_{\gamma}  \partial_{\lambda} \xi^{\alpha } +
M^{\alpha  \lambda}{}_{\gamma}    \partial_{\lambda} \xi^{\beta }  - M^{\alpha  \beta }{}_{\lambda}  \partial_{\gamma} \xi^{\lambda}  \\
  - 2 \partial^{(\alpha }\partial_{\gamma} \xi^{\beta )} + g^{\alpha \beta } \partial_{\lambda} \partial_{\gamma} \xi^{\lambda} +
  g^{\mu \nu} \delta^{(\beta }_{\gamma} \partial_{\mu} \partial_{\nu} \xi^{\alpha )} \, .
\label{append_M_transform}
\end{multline} 
The Lie derivatives~(\ref{Lie derivative}) follow directly from these transformations~(\ref{append_G_transform}), (\ref{append_B_transform}), (\ref{append_E_transform}) and~(\ref{append_M_transform}).

Lastly, in equation~(\ref{LieLieG}) we claim that performing an infinitesimal coordinate transformation with respect to $\zeta(x)$ on the object $\mathcal{L}_{\xi} \ourG$ leads to the following 
\begin{equation} 
(\widehat{\mathcal{L}_{\xi} \ourG}) = \mathcal{L}_{\xi} \ourG - (\mathcal{L}_{\xi} M^{\alpha \beta}{}_{\gamma} ) \partial_{\alpha} \partial_{\beta} \zeta^{\gamma} \, .
\end{equation}
The calculation follows the same steps as above but is much more long-winded. Here we sketch the main results, which are to perform an infinitesimal coordinate transformation on the object
\begin{equation}
\mathcal{L}_{\xi} \ourG =  \xi^{\mu}  \partial_{\mu} \ourG + M^{\alpha \beta}{}_{\gamma} \partial_{\alpha} \partial_{\beta} \xi^{\gamma} \, ,
\end{equation}
with the vector $\xi$ no longer treated as infinitesimal\footnote{This is because in taking the Lie derivative we have divided through by $\epsilon$, see Sec.~\ref{section2.3.1}.}. For the first term we find
\begin{equation}
\hat{\xi}^{\mu}  \hat{\partial}_{\mu} \hat{\ourG} =   \xi^{\mu} \partial_{\mu} \ourG - \xi^{\mu} \partial_{\mu} M^{\alpha \beta}{}_{\gamma} \partial_{\alpha} \partial_{\beta} \zeta^{\gamma} - \xi^{\mu} M^{\alpha \beta}{}_{\gamma} \partial_{\mu} \partial_{\alpha} \partial_{\beta} \zeta^{\gamma} \, ,
\end{equation}
while for the second term we have
\begin{multline}
 \hat{M}^{\alpha \beta}{}_{\gamma} \hat{\partial}_{\alpha} \hat{\partial}_{\beta} \hat{\xi}^{\gamma} =  M^{\alpha \beta}{}_{\gamma} \partial_{\alpha} \partial_{\beta} \xi^{\gamma} + \xi^{\mu} M^{\alpha \beta}{}_{\gamma}   \partial_{\mu}  \partial_{\alpha} \partial_{\beta}\zeta^{\gamma} 
 + \partial_{\alpha} \partial_{\beta} \zeta^{\gamma} \Big[ 2 M^{\beta \mu}{}_{\gamma} \partial_{\mu} \xi^{\alpha}
 \\ - M^{\alpha \beta}{}_{\mu} \partial_{\gamma} \xi^{\mu} - 2 g^{\alpha \mu} \partial_{\mu} \partial_{\gamma} \xi^{\beta} + \delta^{\alpha}_{\gamma} g^{\mu \nu} \partial_{\mu} \partial_{\nu} \xi^{\beta} + g^{\alpha \beta} \partial_{\gamma} \partial_{\mu} \xi^{\mu} \Big] \, ,
\end{multline}
which requires many cancellations and simplifications.

Finally, recall the Lie derivative of $M^{\alpha \beta}{}_{\gamma}$ in equation~(\ref{Lie_M}), which follows directly from~(\ref{append_M_transform}). We therefore see that the above transformation is in fact simply
\begin{equation}
\hat{\xi}^{\mu} \hat{\partial}_{\mu} \hat{\ourG} + \hat{M}^{\alpha \beta}{}_{\gamma} \hat{\partial}_{\alpha} \hat{\partial}_{\beta} \hat{\xi}^{\gamma}  = \xi^{\mu}   \partial_{\mu} \ourG  + M^{\alpha \beta}{}_{\gamma} \partial_{\alpha} \partial_{\beta} \xi^{\gamma}  - \big( \mathcal{L}_{\xi} M^{\alpha \beta}{}_{\gamma} \big) \partial_{\alpha} \partial_{\beta} \zeta^{\gamma} \, ,
\end{equation}
which verifies equation~(\ref{LieLieG}). This proves that the Lie derivative of the bulk term is not covariant (in contrast to the Lie derivative of the connection, which is in fact tensorial).

\section{Metric-affine framework}
\label{appendixB.2}

Working now in the metric-affine framework, making use of equations~(\ref{append_inf_g1})-(\ref{Chr transform}), we look at the infinitesimal coordinate transformations for the affine quantities of Sec.~\ref{section4.4}. These calculations are roughly the same as in the previous section, except that now we no longer have a symmetric connection. 

\subsubsection{Affine bulk term}

Recall that the affine bulk term could be written in the following two different forms
\begin{align*}
\bar{\ourG} &=  g^{\mu \lambda} \big( \bar{\Gamma}^{\kappa}_{\kappa \rho} \bar{\Gamma}^{\rho}_{\mu \lambda} - \bar{\Gamma}^{\kappa}_{\mu \rho} \bar{\Gamma}^{\rho}_{\kappa \lambda} \big)
  + \big(\bar{\Gamma}^{\mu}_{\mu \lambda} \delta^{\nu}_{\kappa} - \bar{\Gamma}^{\nu}_{\kappa \lambda} \big) \big( \partial_{\nu} g^{\kappa \lambda} - \frac{1}{2} g_{\alpha \beta} g^{\kappa \lambda} \partial_{\nu} g^{\alpha \beta} \big) \\ 
  &= g^{\mu \lambda}
  \bigl(\bar{\Gamma}^{\kappa}_{\rho \lambda} \bar{\Gamma}^{\rho}_{\mu \kappa} -
  \bar{\Gamma}^{\kappa}_{\mu \lambda} \bar{\Gamma}^{\rho}_{\rho \kappa} \bigr) +
  \bar{\Gamma}^{\kappa}_{\mu \lambda}  P^{\mu \lambda}{}_{\kappa}  \, ,
\end{align*}
where $P^{\mu \lambda}{}_{\kappa}$ is the Palatini tensor~(\ref{Palatini}). Either form can be used when working with coordinate transformations, but it turns out to be advantageous to use the Palatini form. This is because we can make use of the standard transformation rules for tensors, and the calculation is much quicker.

Performing another brute force calculation, beginning with the quadratic part of the bulk term~(\ref{Gbarquad}), we again find all terms proportional to $\partial \xi$ cancelling
\begin{multline}
 \hat{g}^{\mu \lambda}
  \bigl(\hat{\bar{\Gamma}}^{\kappa}_{\rho \lambda} \hat{\bar{\Gamma}}^{\rho}_{\mu \kappa} -
  \hat{\bar{\Gamma}}^{\kappa}_{\mu \lambda} \hat{\bar{\Gamma}}^{\rho}_{\rho \kappa} \bigr)  = g^{\mu \lambda}
  \bigl(\bar{\Gamma}^{\kappa}_{\rho \lambda} \bar{\Gamma}^{\rho}_{\mu \kappa} -
  \bar{\Gamma}^{\kappa}_{\mu \lambda} \bar{\Gamma}^{\rho}_{\rho \kappa} \bigr) \\
- \partial_{\alpha} \partial_{\beta} \xi^{\gamma} \big( 2 g^{\alpha \lambda} \bar{\Gamma}^{\beta}_{\lambda \gamma} - g^{\alpha \beta} \bar{\Gamma}^{\kappa}_{\kappa \gamma} - \delta^{\alpha}_{\gamma} g^{\mu \lambda} \bar{\Gamma}^{\beta}_{\mu \lambda} \big) \, .
\end{multline}
Making use of the object $\bar{M}^{\alpha \beta}{}_{\gamma}$ in~(\ref{Mbar}), we can write this as
\begin{equation}   \label{append_eq_m}
\hat{g}^{\mu \lambda} \big( \hat{\bar{\Gamma}}^{\kappa}_{\kappa \rho} \hat{\bar{\Gamma}}^{\rho}_{\mu \lambda}   - \hat{\bar{\Gamma}}^{\kappa}_{\mu \rho} \hat{\bar{\Gamma}}^{\rho}_{\kappa \lambda} \big) =  g^{\mu \lambda} \big( \bar{\Gamma}^{\kappa}_{\kappa \rho} \bar{\Gamma}^{\rho}_{\mu \lambda} - \bar{\Gamma}^{\kappa}_{\mu \rho} \bar{\Gamma}^{\rho}_{\kappa \lambda} \big) - \bar{M}^{\alpha \beta}{}_{\gamma}  \partial_{\alpha} \partial_{\beta} \xi^{\gamma} \, .
\end{equation}
For the Palatini part of $\bar{\ourG}$ we obtain the neat expression
\begin{align}
\hat{\bar{\Gamma}}^{\kappa}_{\mu \lambda}  \hat{P}^{\mu \lambda}{}_{\kappa}  & 
\begin{multlined}[t]
 = \big(\bar{\Gamma}^{\kappa}_{\mu \lambda} + \partial_{\eta} \xi^{\kappa} \bar{\Gamma}^{\eta}_{\mu \lambda} - \partial_{\mu} \xi^{\eta} \bar{\Gamma}^{\kappa}_{\eta \lambda} - \partial_{\lambda} \xi^{\eta} \bar{\Gamma}^{\kappa}_{\mu \eta} - \partial_{\mu} \partial_{\lambda} \xi^{\kappa} \big) 
\\  \nonumber
\times \big(P^{\mu \lambda}{}_{\kappa} + \partial_{\gamma} \xi^{\mu} P^{\gamma \lambda}{}_{\kappa} + \partial_{\gamma} \xi^{\lambda} P^{\mu \gamma}{}_{\kappa} -  \partial_{\kappa} \xi^{\gamma} P^{\mu \lambda}{}_{\gamma} \big) 
\end{multlined} \\
&=   \bar{\Gamma}^{\kappa}_{\mu \lambda}  P^{\mu \lambda}{}_{\kappa}  - P^{\mu \lambda}{}_{\kappa} \partial_{\mu} \partial_{\lambda} \xi^{\kappa} \, , \label{append_eq_pal}
\end{align}
where again all $\partial \xi$ terms cancel.
Putting together~(\ref{append_eq_m}) and~(\ref{append_eq_pal}) we arrive at
\begin{equation}
\hat{\bar{\ourG}} = \bar{\ourG}  - \partial_{\alpha} \partial_{\beta} \xi^{\gamma} \big(\bar{M}^{\alpha \beta}{}_{\gamma}  + P^{\alpha \beta}{}_{\gamma}  \big) \, ,
\end{equation}
as given in equation~(\ref{inf_Gbar}).

Lastly, it is possible to show that the symmetrised combination $\bar{M}^{(\alpha \beta)}{}_{\gamma}  + P^{(\alpha \beta)}{}_{\gamma}$ is equal to the Levi-Civita object $M^{\alpha \beta}{}_{\gamma}$. This can be seen directly by decomposing the affine connection into its Christoffel and contortion parts~(\ref{affine decomp}), from which we find
\begin{align}
\bar{M}^{\alpha \beta}{}_{\gamma}  &= g^{\rho \alpha} \bar{\Gamma}^{\beta}_{\gamma \rho} +  g^{\rho \beta} \bar{\Gamma}^{\alpha}_{\rho \gamma } -  g^{\alpha \beta} \bar{\Gamma}^{\rho}_{\rho \gamma }- g^{\rho \sigma} \delta^{\alpha}_{\gamma} \bar{\Gamma}^{\beta}_{\rho \sigma} \nonumber \\
& \begin{multlined}
= g^{\rho \alpha} \Gamma^{\beta}_{\gamma \rho} +  g^{\rho \beta} \Gamma^{\alpha}_{\rho \gamma } -  g^{\alpha \beta}  \Gamma^{\rho}_{\rho \gamma }- g^{\rho \sigma} \delta^{\alpha}_{\gamma} \Gamma^{\beta}_{\rho \sigma} 
 \\
+ T^{\alpha \beta}{}_{\gamma} + g^{\alpha \beta} T^{\lambda}{}_{\gamma \lambda} - \delta^{\alpha}_{\gamma} T^{\beta \lambda}{}_{\lambda} + Q_{\gamma}{}^{\alpha \beta} - \delta^{\alpha}_{\gamma} Q^{\lambda \beta}{}_{\lambda} + \frac{1}{2} \delta^{\alpha}_{\gamma} Q^{\beta \lambda}{}_{\lambda} - \frac{1}{2} g^{\alpha \beta} Q_{\gamma}{}^{\lambda}{}_{\lambda} \nonumber 
\end{multlined} \\ 
&= 2 g^{\rho (\alpha} \Gamma^{\beta)}_{\gamma \rho}  -  g^{\alpha \beta}  \Gamma^{\rho}_{\rho \gamma }- g^{\rho \sigma} \delta^{\alpha}_{\gamma} \Gamma^{\beta}_{\rho \sigma} - P^{\alpha \beta}{}_{\gamma} \, .
\end{align}
We then have 
\begin{equation} \label{MbarPM}
\bar{M}^{(\alpha \beta)}{}_{\gamma}  + P^{(\alpha \beta)}{}_{\gamma} = M^{\alpha \beta}{}_{\gamma} \, ,
\end{equation}
where the symmetry brackets are necessary due to $M^{\alpha \beta}{}_{\gamma}$ being symmetric by definition.
It follows that the Lie derivative of the affine bulk term can be written as 
\begin{equation}
\mathcal{L}_{\xi} \bar{\ourG} = \bar{\ourG}  + M^{\alpha \beta}{}_{\gamma} \partial_{\alpha} \partial_{\beta} \xi^{\gamma} \, .
\end{equation}

\subsubsection{Affine boundary term}

Moving on to the affine boundary term~(\ref{Bbar}), there are several different ways one can approach this calculation. It turns out to be surprisingly simple to follow the method of the previous section of using the boundary vector. Writing it in this form we have
\begin{equation}
\bar{\ourB} = \Gamma^{\lambda}_{\sigma \lambda} \bar{B}^{\sigma} + \partial_{\sigma} \bar{B}^{\sigma} \, ,
\end{equation}
where the affine boundary vector is defined as
\begin{equation}
\bar{B}^{\sigma} = g^{\mu \nu} \bar{\Gamma}^{\sigma}_{\mu \nu} - g^{\mu \sigma} \bar{\Gamma}^{\nu}_{\nu \sigma} \, .
\end{equation}
Care should be taken to distinguish between the Levi-Civita and affine connection terms. However, as we know our final transformation should give rise to the Levi-Civita object $M^{\mu \nu}{}_{\lambda}$, we will work with the Christoffel symbols directly.

The boundary vector transforms under an infinitesimal coordinate transformation as
\begin{equation}
\hat{\bar{B}}^{\sigma} = \bar{B}^{\sigma} + \partial_{\mu} \xi^{\sigma} \bar{B}^{\mu} + g^{\sigma \nu} \partial_{\mu} \partial_{\nu} \xi^{\mu} - g^{\mu \nu} \partial_{\mu} \partial_{\nu} \xi^{\sigma} \, ,
\end{equation}
and its derivative is then
\begin{equation}
\hat{\partial}_{\sigma} \hat{\bar{B}}^{\sigma} = \partial_{\sigma} \bar{B}^{\sigma} + \bar{B}^{\mu} \partial_{\nu} \partial_{\mu} \xi^{\nu} + \partial_{\rho} g^{\rho \nu}  \partial_{\mu} \partial_{\nu} \xi^{\mu} - \partial_{\rho} g^{\mu \nu} \partial_{\mu} \partial_{\nu} \xi^{\rho} \, .
\end{equation}
For the other term we find
\begin{equation}
\hat{\Gamma}^{\lambda}_{\sigma \lambda} \hat{\bar{B}}^{\sigma} = \Gamma^{\lambda}_{\sigma \lambda} \bar{B}^{\sigma} + \partial_{\mu} \partial_{\nu} \xi^{\gamma} \big( \delta^{\mu}_{\gamma} g^{\sigma \nu} \Gamma^{\lambda}_{\sigma \lambda} - g^{\mu \nu} \Gamma^{\lambda}_{\gamma \lambda} - \delta^{\mu}_{\gamma} \bar{B}^{\nu} \big) \, .
\end{equation}
In total, the boundary term transforms as
\begin{align}
\hat{\ourB} &= \ourB + \partial_{\mu} \partial_{\nu} \xi^{\gamma} \big( \delta_{\gamma}^{\mu} \bar{B}^{\nu}  + \delta^{\mu}_{\gamma} \partial_{\rho} g^{\rho \nu} - \partial_{\gamma} g^{\mu \nu} +
\delta^{\mu}_{\gamma} g^{\sigma \nu} \Gamma^{\lambda}_{\sigma \lambda} - g^{\mu \nu} \Gamma^{\lambda}_{\gamma \lambda} - \delta^{\mu}_{\gamma} \bar{B}^{\nu}  \big) \nonumber \\
&= \ourB + \partial_{\mu} \partial_{\nu} \xi^{\gamma} \big(\cancel{-\delta^{\mu}_{\gamma} g^{\eta \nu} \Gamma^{\rho}_{\rho \eta} }  - \delta^{\mu}_{\gamma}  g^{\rho \nu} \Gamma^{\nu}_{\rho \eta} + g^{\rho \nu} \Gamma^{\mu}_{\gamma \rho} + g^{\rho \mu} \Gamma^{\nu}_{\gamma \rho} + \cancel{\delta^{\mu}_{\gamma} g^{\sigma \nu} \Gamma^{\lambda}_{\sigma \lambda}} - g^{\mu \nu} \Gamma^{\lambda}_{\gamma \lambda} \big) \nonumber \\
&=   \ourB + \partial_{\mu} \partial_{\nu} \xi^{\gamma}  \big( 2 g^{\mu \rho} \Gamma^{\nu}_{\gamma \rho} - \delta^{\mu}_{\gamma} g^{\rho \sigma} \Gamma_{\rho\sigma}^{\nu} - g^{\mu \nu} \Gamma^{\lambda}_{\lambda \gamma} \big) \nonumber \\
&=  \ourB + \partial_{\mu} \partial_{\nu} \xi^{\gamma} M^{\mu \nu}{}_{\gamma} \, ,
\end{align}
where on the second line we have expanded the partial derivatives of the metric in terms of the Christoffel symbols. The remaining Levi-Civita terms give exactly the object $M^{\mu \nu}{}_{\gamma}$ defined in~(\ref{M}). This simple derivation shows that all of the affine quantities, contained in the vector $\bar{B}^{\sigma}$, drop out from the infinitesimal transformation. Using the equivalence~(\ref{MbarPM}), we can write the Lie derivative as
\begin{align}
\mathcal{L}_{\xi} \bar{\ourB} &= \xi^{\mu} \partial_{\mu} \ourB - \partial_{\mu} \partial_{\nu} \xi^{\gamma} M^{\mu \nu}{}_{\gamma} \nonumber  \\
&= \xi^{\mu} \partial_{\mu} \ourB - \partial_{\mu} \partial_{\nu} \xi^{\gamma} \big( \bar{M}^{\mu \nu}{}_{\gamma} + P^{\mu \nu}{}_{\gamma} \big) \, ,
\end{align}
as given in~(\ref{inf_Bbar}) and~(\ref{Lie_Bbar}).

\chapter{Equations of motion}
\label{appendixC}

In this appendix we derive the equations of motion for the various (coordinate basis) theories studied throughout the thesis. This begins with the Einstein actions, first for the Levi-Civita connection and then in the metric-affine setting. We then show the derivation for the modified field equations of the metrical $f(\ourG,\ourB)$ theories and the metric-affine $f(\bar{\ourG},\bar{\ourB})$ theories.

\section{Einstein actions}
\label{appendixC.1}

\subsection{Levi-Civita Einstein action}
\label{appendixC.1.1}

Here we calculate the field equations for the Levi-Civita Einstein action~(\ref{Einstein_action}), making use of the objects $M^{\mu \nu}{}_{\lambda}$~(\ref{M}) and $E^{\mu \nu \lambda}$~(\ref{E}). These calculations are originally to be found in our work~\cite{Boehmer:2021aji}.
The metric variation of the Einstein action $S_{\textrm{E}}$ and the matter action $S_{\textrm{M}}[g_{\mu \nu},\varPhi^A]$ can be written as
\begin{align}
  \delta S &= \delta S_{\textrm{E}}[g_{\mu \nu}] + \delta S_{\textrm{M}}[g_{\mu \nu},\varPhi^A] 
  \nonumber \\ &=
  \frac{1}{2 \kappa} \int d^4x \Big[ \delta \sqrt{-g} \ourG + \sqrt{-g} \delta \ourG +
    2 \kappa \frac{\delta S_{\textrm{M}}}{\delta g^{\mu \nu}}\delta g^{\mu \nu} \Big] \,,
  \label{Einstein_Action_Matter_Variation}
\end{align}
with $\ourG$ defined in~(\ref{G}). Focusing on the gravitational action, and temporarily dropping the factor of $2 \kappa$, we can write
\begin{multline}
  \label{Variation_of_Einstein_action}
  \delta S_{\textrm{E}} = \int \delta \sqrt{-g} g^{\mu \nu} (\Gamma_{\mu \nu})^2\, d^4x +
  \int \sqrt{-g} \ \delta g^{\mu \nu} (\Gamma_{\mu \nu})^2\, d^4x  \\ +
  \int \sqrt{-g} g^{\mu \nu} \delta (\Gamma_{\mu \nu})^2\, d^4x \,,
\end{multline}
where $(\Gamma_{\mu \nu})^2$ stands for $\Gamma^{\sigma}_{\lambda \mu} \Gamma^{\lambda}_{\nu \sigma} - \Gamma^{\sigma}_{\mu \nu} \Gamma^{\lambda}_{\sigma \lambda}$. The first two terms are
\begin{align}
  \label{metric_determinant_variation}
  \int \delta \sqrt{-g} g^{\mu \nu} (\Gamma_{\mu \nu})^2 d^4x &= -\frac{1}{2} \int \sqrt{-g} \ \delta g^{\rho \sigma} g_{\rho \sigma} g^{\mu \nu} (\Gamma^{\gamma}_{\lambda \mu} \Gamma^{\lambda}_{\gamma \nu} - \Gamma^{\gamma}_{\mu \nu} \Gamma^{\lambda}_{\gamma \lambda} ) d^4 x \,, \\
  \label{metric_tensor_variation}
  \int \sqrt{-g} \ \delta g^{\mu \nu} (\Gamma_{\mu \nu})^2 d^4x &= \int \sqrt{-g} \ \delta g^{\rho \sigma} (\Gamma^{\gamma}_{\lambda \rho} \Gamma^{\lambda}_{\gamma \sigma} - \Gamma^{\gamma}_{\rho \sigma} \Gamma^{\lambda}_{\gamma \lambda} ) d^4 x \,.
\end{align}
Next we expand $g^{\mu \nu} \delta (\Gamma_{\mu \nu})^2$ to arrive at
\begin{align}
    g^{\mu \nu} \delta (\Gamma_{\mu \nu})^2 &=g^{\mu \nu} \big( \delta \Gamma^{\sigma}_{\lambda \mu} \Gamma^{\lambda}_{\sigma \nu} + \delta \Gamma^{\lambda}_{\sigma \nu} \Gamma^{\sigma}_{\lambda \mu} - \delta \Gamma^{\sigma}_{\mu \nu} \Gamma^{\lambda}_{\sigma \lambda} - \delta \Gamma^{\lambda}_{\sigma \lambda} \Gamma^{\sigma}_{\mu \nu} \big) \nonumber \\
    &= g^{\mu \nu} (\delta^{\sigma}_{\mu} \Gamma^{\rho}_{\kappa \nu} + \delta^{\sigma}_{\nu} \Gamma^{\rho}_{\kappa \mu} - \delta^{\rho}_{\mu} \delta^{\sigma}_{\nu} \Gamma^{\lambda}_{\kappa \lambda} - \delta^{\sigma}_{\kappa} \Gamma^{\rho}_{\mu \nu}) \delta \Gamma^{\kappa}_{\rho \sigma} \nonumber \\
    &= M^{\rho \sigma}{}_{\kappa} \delta \Gamma^{\kappa}_{\rho \sigma} \,.
\end{align}
Here we have introduced the object 
\begin{align}
  \label{M_Appendix}
 M^{\alpha \beta}{}_{\gamma} :=   \frac{\partial \mathbf{G} }{\partial \Gamma^{\gamma}_{\alpha \beta}}  = 2 g^{\nu (\beta} \Gamma^{\alpha)}_{\gamma \nu} - g^{\alpha \beta} \Gamma^{\lambda}_{\gamma \lambda} - g^{\mu \nu}  \delta^{(\beta}_{\gamma} \Gamma^{\alpha)}_{\mu \nu} \,,
\end{align}
which is constructed to be symmetric over its first two indices to match the symmetry of the connection. This is also given in equation~(\ref{M}).

Expanding the variation of the Christoffel connection in terms of metric variations gives
\begin{align}
  \delta \Gamma^{\kappa }_{\rho \sigma } &= \frac{1}{2} \delta g^{\kappa \lambda } (g_{\rho \lambda ,\sigma } + g_{\sigma \lambda ,\rho } - g_{\rho \sigma ,\lambda } ) + \frac{1}{2} g^{\kappa \lambda } (\delta g_{\rho \lambda ,\sigma } + \delta g_{\sigma \lambda ,\rho } - \delta g_{\rho \sigma ,\lambda } )
  \nonumber \\
  &= \delta g^{\kappa \lambda } g_{\zeta \lambda } \Gamma^{\zeta }_{\rho \sigma } + \frac{1}{2} g^{\kappa \lambda } (\delta g_{\rho \lambda ,\sigma } + \delta g_{\sigma \lambda ,\rho } - \delta g_{\rho \sigma ,\lambda })
  \nonumber \\
  &= \delta g^{\kappa \lambda } g_{\zeta \lambda } \Gamma^{\zeta }_{\rho \sigma } + \frac{1}{2} g^{\kappa \lambda } (\delta g_{\alpha \beta , \gamma} \Delta^{\alpha \beta \gamma}_{\rho  \sigma  \lambda }) \,,
  \label{Chr_variation}
\end{align}
where we also define
\begin{align}
  \Delta^{\alpha \beta \gamma}_{\rho \sigma \lambda} =
  \delta^{\alpha}_{\{\lambda} \delta^{\beta}_{\rho} \delta^{\gamma}_{\sigma \}} =
  \delta^{\alpha}_{\lambda} \delta^{\beta}_{\rho} \delta^{\gamma}_{\sigma} +
  \delta^{\alpha}_{\sigma} \delta^{\beta}_{\lambda} \delta^{\gamma}_{\rho} -
  \delta^{\alpha}_{\rho} \delta^{\beta}_{\sigma} \delta^{\gamma}_{\lambda} \,,
\end{align}
which simply permutes indices using the \emph{Schouten bracket}, see~\cite{JS1954}. Again we note that only the symmetric part over the indices $\alpha \, \beta$ of $\Delta^{\alpha \beta \gamma}_{\rho \sigma \lambda}$ in~(\ref{Chr_variation}) contributes.

Putting the above equations back into the integrand, we can write the last term of~(\ref{Variation_of_Einstein_action}) as
\begin{multline}
  \int \sqrt{-g} g^{\mu \nu} \delta (\Gamma_{\mu \nu})^2\, d^4x =
  \int \sqrt{-g}  M^{\rho \sigma}{}_{\kappa} \delta \Gamma^{\kappa}_{\rho \sigma}\, d^4 x
   \\ =
  \int \sqrt{-g}  M^{\rho \sigma}{}_{\kappa} \big[  \delta g^{\kappa \lambda } g_{\zeta \lambda }
    \Gamma^{\zeta }_{\rho \sigma } + \frac{1}{2} g^{\kappa \lambda } (\delta g_{\alpha \beta , \gamma}
    \Delta^{\alpha \beta \gamma}_{\rho  \sigma  \lambda })\big]\, d^4 x \,.
\end{multline}
To simplify the second term, we make use of the object $E^{\mu \nu \lambda}$~(\ref{E}), which is the permutation of $M^{\rho \sigma \lambda}$ given by
\begin{multline}
  E^{\alpha \beta \gamma} := \Delta^{\alpha \beta \gamma}_{\rho \sigma \lambda} M^{\rho \sigma \lambda} = M^{\beta \gamma \alpha} + M^{\gamma \alpha \beta} - M^{\alpha \beta \gamma}
   \\ =
  2 g^{\nu \alpha} g^{\beta \mu} \Gamma^{\gamma}_{\mu \nu} -
  2 g^{\gamma (\alpha} g^{\beta) \mu} \Gamma^{\lambda}_{\lambda \mu} +
  g^{\alpha \beta} g^{\mu \gamma} \Gamma^{\lambda}_{\lambda \mu} -
  g^{\alpha \beta} g^{\mu \nu} \Gamma^{\gamma}_{\mu \nu} \,.
\end{multline}
In equation~(\ref{EG1}) we show this to be the variation of $\ourG$ with respect to the partial derivative of the metric. We then have
\begin{align}
  \label{Gammapre}
  \int \sqrt{-g} g^{\mu \nu} \delta (\Gamma_{\mu \nu})^2 d^4x = \int \sqrt{-g} \Big(\delta g^{\kappa \lambda } g_{\zeta \lambda } M^{\rho \sigma}{}_{\kappa} \Gamma^{\zeta }_{\rho \sigma }+ \frac{1}{2} E^{\alpha \beta \gamma} \partial_{\gamma}\delta g_{\alpha\beta} \Big)\, d^4 x \,.
\end{align}
Integrating the second term by parts and dropping the boundary terms, then rewriting the metric variation in terms of the inverse metric gives
\begin{multline}
  \label{Gamma_squared_variation}
  \int \sqrt{-g} g^{\mu \nu} \delta (\Gamma_{\mu \nu})^2\, d^4x =
  \int \delta g^{\rho \sigma} \Big[
    \sqrt{-g}  g_{\zeta \sigma}  M^{\mu \nu}{}_{\rho}
    \Gamma^{\zeta}_{\mu \nu} \\ 
    + \frac{1}{2} g_{\alpha \rho} g_{\beta \sigma}
    \partial_{\gamma} ( \sqrt{-g} E^{\alpha \beta \gamma}) \Big] \, d^4 x \,.
\end{multline}
As we wish to obtain the Einstein tensor, we will simply expand everything in terms of the connection and its derivatives.
The first term in the integrand of~(\ref{Gamma_squared_variation}) can be expanded as
\begin{multline}
        \delta g^{\rho \sigma} \sqrt{-g}\, g_{\zeta  \sigma}  M^{\mu \nu}{}_{\rho}
    \Gamma^{\zeta }_{\mu \nu}  =  \delta g^{\rho \sigma} \sqrt{-g} \Big( 2 g^{\nu  \eta} g_{\sigma \zeta } \Gamma^{\zeta }_{\mu  \nu } \Gamma^{\mu }_{\rho \eta} \\ 
    -  g^{\mu \nu } g_{\sigma \zeta } \Gamma^{\zeta }_{\mu \nu } \Gamma^{\lambda}_{\lambda \rho} - g^{\gamma \eta} g_{\sigma \zeta } \Gamma^{\zeta }_{\mu  \rho} \Gamma^{\mu }_{\gamma \eta} \Big) \,.
\end{multline}
The second term of~(\ref{Gamma_squared_variation}) involves taking the partial derivative of $E^{\alpha \beta \gamma}$, which leads to the following expression
\begin{multline} \label{DerivativeE_final}
    \frac{1}{2} \delta g^{\rho \sigma}  g_{\alpha \rho} g_{\beta \sigma} \partial_{\gamma} ( \sqrt{-g} E^{\alpha \beta \gamma}) =  \sqrt{-g} \delta g^{\rho \sigma} \Big( 
      2\Gamma^{\lambda}_{\kappa \lambda} \Gamma^{\kappa }_{\rho \sigma}
     -2\Gamma^{\gamma}_{\sigma \nu} \Gamma^{\nu}_{\gamma \rho}
     -g_{\sigma \rho} g^{\mu \nu} \Gamma^{\eta}_{\gamma \eta} \Gamma^{\gamma}_{\mu \nu}
     \\ -2 g_{\alpha \rho} g^{\epsilon \nu} \Gamma^{\gamma}_{\sigma \nu} \Gamma^{\alpha}_{\gamma \epsilon} 
     +  g_{\alpha \rho} g^{\eta \gamma} \Gamma^{\alpha}_{\gamma \eta} \Gamma^{\lambda}_{\sigma \lambda} 
     +g_{\alpha \rho} g^{\mu \nu} \Gamma^{\gamma}_{\mu \nu} \Gamma^{\alpha}_{\gamma \sigma} 
     + g_{\rho \sigma} g^{\eta \nu} \Gamma^{\gamma}_{\mu \nu} \Gamma^{\mu}_{\gamma \eta} \\
     + \partial_{\gamma} \Gamma^{\gamma}_{\rho \sigma} -  \partial_{\rho} \Gamma^{\lambda}_{\sigma \lambda} 
    + \frac{1}{2} g_{\sigma \rho} g^{\kappa  \gamma} \partial_{\gamma} \Gamma^{\lambda}_{\kappa  \lambda} - \frac{1}{2} g_{\rho \sigma} g^{\mu \nu} \partial_{\gamma} \Gamma^{\gamma}_{\mu \nu}
     \Big) \,.
\end{multline}
The full $(\Gamma_{\mu \nu})^2$ variation becomes
\begin{multline}
  \label{Final_gamma_squared_variation}
  \int  \sqrt{-g} g^{\mu \nu} \delta (\Gamma_{\mu \nu})^2\, d^4x =
  \int \delta g^{\rho \sigma} \sqrt{-g} \Big( 2 \Gamma^{\eta}_{\gamma \eta} \Gamma^{\gamma}_{\rho \sigma} 
    -2  \Gamma^{\gamma}_{\sigma \nu} \Gamma^{\nu}_{\gamma \rho}
    + g_{\rho \sigma} g^{\eta \nu} \Gamma^{\gamma}_{\mu \nu} \Gamma^{\mu}_{\gamma \eta} \\
    - g_{\sigma \rho} g^{\mu \nu} \Gamma^{\eta}_{\gamma \eta} \Gamma^{\gamma}_{\mu \nu}  
    + \partial_{\gamma} \Gamma^{\gamma}_{\rho \sigma}
    -  \partial_{\rho} \Gamma^{\lambda}_{\sigma \lambda} 
    + \frac{1}{2} g_{\sigma \rho} g^{\kappa \gamma} \partial_{\gamma} \Gamma^{\lambda}_{\kappa \lambda} - \frac{1}{2} g_{\rho \sigma} g^{\mu \nu} \partial_{\gamma} \Gamma^{\gamma}_{\mu \nu} \Big)\, d^4 x \,.
\end{multline}

Putting together equations~(\ref{metric_determinant_variation}),~(\ref{metric_tensor_variation}) and~(\ref{Final_gamma_squared_variation}), we get a final expression for the variation of the Einstein action
\begin{multline} \label{recovering_Einstein}
    \delta S_{\textrm{E}} = \frac{1}{2\kappa} \int \sqrt{-g} \delta g^{\rho \sigma} \Big[\Gamma^{\eta}_{\gamma \eta} \Gamma^{\gamma}_{\rho \sigma} -  \Gamma^{\gamma}_{\sigma \nu} \Gamma^{\nu}_{\gamma \rho} +
    \frac{1}{2} g_{\rho \sigma} g^{\eta \nu} \Gamma^{\gamma}_{\mu \nu} \Gamma^{\mu}_{\gamma \eta} -
    \frac{1}{2} g_{\sigma \rho} g^{\mu \nu} \Gamma^{\eta}_{\gamma \eta} \Gamma^{\gamma}_{\mu \nu}  \\
     + \partial_{\gamma} \Gamma^{\gamma}_{\rho \sigma}
    -  \partial_{\rho} \Gamma^{\lambda}_{\sigma \lambda} 
    + \frac{1}{2} g_{\sigma \rho} g^{\kappa \gamma} \partial_{\gamma} \Gamma^{\lambda}_{\kappa \lambda} - \frac{1}{2} g_{\rho \sigma} g^{\mu \nu} \partial_{\gamma} \Gamma^{\gamma}_{\mu \nu}
    \Big]\, d^4x \,.
\end{multline}
The terms in the square brackets are understood to be symmetrised over the indices $\rho \, \sigma$. We then recognise this to be the Einstein tensor,
\begin{align}
    \delta S_{\textrm{E}} = \frac{1}{2\kappa} \int \sqrt{-g} \delta g^{\rho \sigma} \big(R_{\rho \sigma} - \frac{1}{2} g_{\rho \sigma} R \big)\, d^4 x  =  \frac{1}{2\kappa}  \int \sqrt{-g} \delta g^{\rho \sigma} G_{\rho \sigma}\, d^4 x \,.
\end{align}
The metric variations of the full action~(\ref{Einstein_Action_Matter_Variation}) is then
\begin{align}
  \delta S = \delta S_{\textrm{E}}[g_{\mu \nu}] + \delta S_{\textrm{M}}[g_{\mu \nu},\varPhi] &= 0
  \nonumber \\
 \frac{1}{2 \kappa} \int \sqrt{-g}  \delta g^{\rho \sigma}
  \Big( G_{\rho \sigma} +
 \frac{2\kappa}{\sqrt{-g}} \frac{\delta S_{\textrm{M}}}{\delta g^{\rho \sigma}}\Big)\, d^4x &= 0 \, ,
\end{align}
where defining the metric energy-momentum tensor as
\begin{align}
  T_{\mu \nu} = -\frac{2}{\sqrt{-g}}
  \frac{\delta S_{\textrm{M}}[g_{\mu \nu}, \varPhi^A]}{\delta g^{\mu \nu}} \,,
\end{align}
leads to the Einstein field equations
\begin{equation}
G_{\rho \sigma} =\kappa T_{\rho \sigma} \, .
\end{equation}

\subsection{Metric-affine Einstein action}
\label{appendixC.1.2}

For the metric-affine Einstein action, or Einstein-Palatini action, given in equation~(\ref{Einstein_actionP}), we now have an independent metric and affine connection. Beginning with the metric variations, a direct calculation gives
\begin{align}
  \delta_g S_{\bar{\textrm{E}}} &= \frac{1}{2\kappa} \int \delta (\sqrt{-g} \bar{\ourG}) \, d^4 x
  \nonumber \\
  &\begin{multlined}
     =\frac{1}{2\kappa} \int \biggl\{- \frac{1}{2} \delta g^{\mu \nu} g_{\mu \nu} \bar{\ourG} +
     \Big[ \delta g^{\mu \nu}  \big( \bar{\Gamma}^{\kappa}_{\kappa \rho} \bar{\Gamma}^{\rho}_{\mu \nu} - \bar{\Gamma}^{\kappa}_{\mu \rho} \bar{\Gamma}^{\rho}_{\kappa \nu} \big)  \\
       + \big(\bar{\Gamma}^{\mu}_{\mu \lambda} \delta^{\nu}_{\kappa} - \bar{\Gamma}^{\nu}_{\kappa \lambda} \big)\big(\partial_{\nu} \delta g^{\kappa \lambda} - \frac{1}{2} \delta (g_{\alpha \beta} g^{\kappa \lambda} \partial_{\nu} g^{\alpha \beta})\big) \Big]\biggr\} \sqrt{-g} d^4x \, ,
   \end{multlined}
   \end{align}
 which then becomes
\begin{multline}
     =\frac{1}{2\kappa}  \int \sqrt{-g} \delta g^{\mu \nu} \Big( - \frac{1}{2} g_{\mu \nu} \bar{\ourG} + \big( \bar{\Gamma}^{\kappa}_{\kappa \rho} \bar{\Gamma}^{\rho}_{\mu \nu} - \bar{\Gamma}^{\kappa}_{\mu \rho} \bar{\Gamma}^{\rho}_{\kappa \nu} \big)  \Big) d^4x   \\
     +  \textrm{`surface terms'}  - \frac{1}{2\kappa} \int \delta g^{\kappa \lambda} \partial_{\nu} \Bigl(\sqrt{-g} \big(\bar{\Gamma}^{\mu}_{\mu \lambda} \delta^{\nu}_{\kappa} - \bar{\Gamma}^{\nu}_{\kappa \lambda} \big)\Bigr) d^4x
     \\
     - \frac{1}{4\kappa}\int  \sqrt{-g} \big(\bar{\Gamma}^{\mu}_{\mu \lambda} \delta^{\nu}_{\kappa} - \bar{\Gamma}^{\nu}_{\kappa \lambda} \big) 
     \big( \delta g_{\alpha \beta} g^{\kappa \lambda} \partial_{\nu} g^{\alpha \beta} + g_{\alpha \beta} \delta g^{\kappa \lambda} \partial_{\nu} g^{\alpha \beta} \big) d^4x
     \\
     - \textrm{`surface terms'}  + \frac{1}{4\kappa} \int \delta g^{\alpha \beta} \partial_{\nu}\Bigl(\sqrt{-g} \big(\bar{\Gamma}^{\mu}_{\mu \lambda} \delta^{\nu}_{\kappa} - \bar{\Gamma}^{\nu}_{\kappa \lambda} \big)  g_{\alpha \beta} g^{\kappa \lambda} \Bigr) d^4 x \,,
\end{multline}
where the surface terms come from integration by parts. Discarding these boundary terms leaves the expression
\begin{multline}
  \delta_g  S_{\bar{\textrm{E}}} = \frac{1}{2\kappa} \int \sqrt{-g} \delta g^{\sigma \gamma} \Big[ -\frac{1}{2} g_{\sigma \gamma} g^{\mu \lambda} \big( \bar{\Gamma}^{\kappa}_{\kappa \rho} \bar{\Gamma}^{\rho}_{\mu \lambda} - \bar{\Gamma}^{\kappa}_{\mu \rho} \bar{\Gamma}^{\rho}_{\kappa \lambda} \big) \\
-\frac{1}{2} g_{\sigma \gamma} \big(\bar{\Gamma}^{\mu}_{\mu \lambda} \delta^{\nu}_{\kappa} - \bar{\Gamma}^{\nu}_{\kappa \lambda} \big) 
 \big( \partial_{\nu} g^{\kappa \lambda} - \frac{1}{2} g_{\alpha \beta} g^{\kappa \lambda} \partial_{\nu} g^{\alpha \beta} \big) \\
+  \big( \bar{\Gamma}^{\kappa}_{\kappa \rho} \bar{\Gamma}^{\rho}_{\sigma \gamma} - \bar{\Gamma}^{\kappa}_{\sigma \rho} \bar{\Gamma}^{\rho}_{\kappa \gamma} \big) 
- \frac{1}{\sqrt{-g}} \partial_{\nu} \Big( \sqrt{-g} \big(\bar{\Gamma}^{\mu}_{\mu \gamma} \delta^{\nu}_{\sigma} - \bar{\Gamma}^{\nu}_{\sigma \gamma} \big) \Big)  \\
- \frac{1}{2} \big(\bar{\Gamma}^{\mu}_{\mu \lambda} \delta^{\nu}_{\kappa} - \bar{\Gamma}^{\nu}_{\kappa \lambda} \big) g^{\kappa \lambda} \partial_{\nu} g_{\sigma \gamma}   
-\frac{1}{2} (\bar{\Gamma}^{\mu}_{\mu \gamma} \delta^{\nu}_{\sigma} - \bar{\Gamma}^{\nu}_{\sigma \gamma} \big) g_{\alpha \beta} \partial_{\nu} g^{\alpha \beta}  \\
+ \frac{1}{2} \frac{1}{\sqrt{-g}} \partial_{\nu}\Big(\sqrt{-g} \big(\bar{\Gamma}^{\mu}_{\mu \lambda} \delta^{\nu}_{\kappa} - \bar{\Gamma}^{\nu}_{\kappa \lambda} \big)  g_{\sigma \gamma} g^{\kappa \lambda} \Big) 
 \Big] \, d^4 x \,,
\end{multline}
where we have used relations such as $\delta g_{\alpha \beta} \partial_{\nu} g^{\alpha \beta} = - \delta g^{\sigma \gamma} g_{\sigma \alpha} g_{\gamma \beta}  \partial_{\nu} g^{\alpha \beta} = \delta g^{\sigma \gamma} \partial_{\nu} g_{\sigma \gamma}$. Introducing the shorthand notation $\tilde{\Gamma}^{\nu}_{\kappa \lambda} := \bar{\Gamma}^{\mu}_{\mu \lambda} \delta^{\nu}_{\kappa} - \bar{\Gamma}^{{\nu}}_{\kappa \lambda}$ and expanding the partial derivatives, we have 
\begin{multline}
   \delta_g S_{\bar{\textrm{E}}} = \frac{1}{2\kappa} \int \sqrt{-g} \delta g^{\sigma \gamma} \bigg[ -\frac{1}{2} g_{\sigma \gamma} g^{\mu \lambda} \big( \bar{\Gamma}^{\kappa}_{\kappa \rho} \bar{\Gamma}^{\rho}_{\mu \lambda} - \bar{\Gamma}^{\kappa}_{\mu \rho} \bar{\Gamma}^{\rho}_{\kappa \lambda} \big) + \big( \bar{\Gamma}^{\kappa}_{\kappa \rho} \bar{\Gamma}^{\rho}_{\sigma \gamma} - \bar{\Gamma}^{\kappa}_{\sigma \rho} \bar{\Gamma}^{\rho}_{\kappa \gamma} \big) \\
    - \frac{1}{2} g_{\sigma \gamma} \tilde{\Gamma}^{\nu}_{\kappa \lambda} \partial_{\nu} g^{\kappa \lambda} + \frac{1}{4} g_{\sigma \gamma} \tilde{\Gamma}^{\nu}_{\kappa \lambda} g_{\alpha \beta} g^{\kappa \lambda} \partial_{\nu} g^{\alpha \beta} + \frac{1}{2} \tilde{\Gamma}^{\nu}_{\sigma \gamma} g_{\alpha \beta} \partial_{\nu} g^{\alpha \beta} \\ - \partial_{\nu} \tilde{\Gamma}^{\nu}_{\sigma \gamma} - \frac{1}{2}\tilde{\Gamma}^{\nu}_{\kappa \lambda} g^{\kappa \lambda} \partial_{\nu} g_{\sigma \gamma} - \frac{1}{2} \tilde{\Gamma}^{\nu}_{\sigma \gamma} g_{\alpha \beta} \partial_{\nu} g^{\alpha \beta} - \frac{1}{4} g_{\sigma \gamma} \tilde{\Gamma}^{\nu}_{\kappa \lambda} g_{\alpha \beta} g^{\kappa \lambda} \partial_{\nu} g^{\alpha \beta} \\ + \frac{1}{2} \partial_{\nu} \tilde{\Gamma}^{\nu}_{\kappa \lambda} g_{\sigma \gamma} g^{\kappa \lambda} + \frac{1}{2}\tilde{\Gamma}^{\nu}_{\kappa \lambda} g^{\kappa \lambda} \partial_{\nu} g_{\sigma \gamma} + \frac{1}{2} \tilde{\Gamma}^{\nu}_{\kappa \lambda} g_{\sigma \gamma} \partial_{\nu} g^{\kappa \lambda} \bigg]  d^4x \,.
\end{multline}
Lastly, making simplifications reduces the metric variations to
\begin{align}
  \delta_g S_{\bar{\textrm{E}}} &\begin{multlined}[t]
= \frac{1}{2\kappa}  \int \sqrt{-g} \delta g^{\sigma \gamma} \Big[ \big( \bar{\Gamma}^{\kappa}_{\kappa \rho} \bar{\Gamma}^{\rho}_{\sigma \gamma} - \bar{\Gamma}^{\kappa}_{\sigma \rho} \bar{\Gamma}^{\rho}_{\kappa \gamma} \big)  -\frac{1}{2} g_{\sigma \gamma} g^{\mu \lambda} \big( \bar{\Gamma}^{\kappa}_{\kappa \rho} \bar{\Gamma}^{\rho}_{\mu \lambda} - \bar{\Gamma}^{\kappa}_{\mu \rho} \bar{\Gamma}^{\rho}_{\kappa \lambda} \big) \nonumber  \\
+ \partial_{\nu} \bar{\Gamma}^{\nu}_{\sigma \gamma} - \partial_{\sigma} \bar{\Gamma}^{\lambda}_{\lambda \gamma} -\frac{1}{2} g_{\sigma \gamma} g^{\kappa \lambda}(\partial_{\nu} \bar{\Gamma}^{\nu}_{\kappa \lambda} - \partial_{\kappa} \bar{\Gamma}^{\mu}_{\mu \lambda}) \Big] d^4x  \quad
 \end{multlined}
  \\
  &= \frac{1}{2\kappa} \int \sqrt{-g} \delta g^{\sigma \gamma} \bar{G}_{\sigma \gamma} \, d^4x \,.
\end{align}
Hence we have obtained the metric-affine Einstein tensor from the metric variation of the bulk term. Also note that only the symmetric part contributes, due to the contraction with the (symmetric) metric variation.

The connection variations simply give rise to the Palatini tensor~(\ref{Palatini}), which was already shown explicitly in equation~(\ref{GbarGammavar}). Substituting this into the action immediately gives
\begin{align}
\delta_{\bar{\Gamma}} S_{\bar{\textrm{E}}} =  \frac{1}{2\kappa}  \int \sqrt{-g} \delta_{\bar{\Gamma}} (\bar{\ourG}) \, d^4x  = \frac{1}{2\kappa}  \int \sqrt{-g} \delta \bar{\Gamma}{}^{\lambda}_{\mu \nu} P^{\mu \nu}{}_{\lambda} d^4 x \,.
\end{align}

The total variations for the metric-affine Einstein action are therefore given by
\begin{equation}
\delta   S_{\bar{\textrm{E}}} = \frac{1}{2\kappa} \int \sqrt{-g} \Big[
    \delta g^{\mu \nu} \bar{G}_{\mu \nu} + \delta \bar{\Gamma}^{\lambda}_{\mu \nu} P^{\mu \nu}{}_{\lambda}
    \Big] d^4x \, ,
\end{equation}
as given in equation~(\ref{Einstein_var}).

\section{Modified actions}
\label{appendixC.2}
Here we calculate the equations of motion for the modified theories of Chapter~\ref{chapter5}. Many of the results from the previous appendix are used, especially for the variations of the Levi-Civita and metric-affine bulk and boundary terms. These results can also be found in our work~\cite{Boehmer:2021aji,Boehmer:2023fyl}.
\subsection{\texorpdfstring{$f(\ourG,\ourB)$}{f(G,B)} action}
\label{appendixC.2.1}
For the fourth-order $f(\ourG,\ourB)$ action~(\ref{S_f(g,b)}), the metric variations are
\begin{align} 
\delta S_{\textrm{grav}} &= \frac{1}{2\kappa }  \int \Big[ \sqrt{-g} \delta f(\ourG, \ourB) + \delta \sqrt{-g} f(\ourG, \ourB) \Big] d^4x \nonumber \\
& \begin{multlined}[b]
= \frac{1}{2\kappa}  \int \sqrt{-g} \Big[ \frac{\partial f(\ourG, \ourB)}{\partial \ourG}  \delta \ourG + \frac{\partial f(\ourG, \ourB)}{\partial \ourB}  \delta \ourB \\
-\frac{1}{2} g_{\mu \nu} \delta g^{\mu \nu} f(\ourG, \ourB)   \Big] d^4x \,.
\end{multlined} \label{Field_equation_derivation_Appendix_eq1}
\end{align}
We will split the calculation into two parts, focusing on the bulk terms and the boundary terms separately. The boundary term variations are notably more complicated, due to the presence of higher derivative terms. In this appendix we refrain from discussing boundary conditions, but see~\cite{Boehmer:2021aji} for further details.

\subsubsection{Bulk variations}
\label{Appendix_Bulk_term}

In Appendix~\ref{appendixC.1.1} we calculated the variation of the bulk term, so using equations~(\ref{metric_tensor_variation}) and~(\ref{Gammapre}) we can write
\begin{align}
    \delta \ourG = \delta g^{\rho \sigma} \Big(\Gamma^{\gamma}_{\lambda \rho} \Gamma^{\lambda}_{\gamma \sigma} - \Gamma^{\gamma}_{\rho \sigma} \Gamma^{\lambda}_{\gamma \lambda} + g_{\gamma \sigma}
    \Gamma^{\gamma}_{\mu \nu} M^{\mu \nu }{}_{\rho} \Big) + \frac{1}{2} E^{\alpha \beta \gamma} \partial_{\gamma} \delta g_{\alpha \beta} \,.
\end{align}
The first term of the integrand~(\ref{Field_equation_derivation_Appendix_eq1}) is then
\begin{multline}
  \label{deltaGintegral_Appendix}
  \sqrt{-g} f_{,\ourG} \delta \ourG = \sqrt{-g} f_{,\ourG}
  \Big[ \delta g^{\rho \sigma} \Big(\Gamma^{\gamma}_{\lambda \rho} \Gamma^{\lambda}_{\gamma \sigma} -
    \Gamma^{\gamma}_{\rho \sigma} \Gamma^{\lambda}_{\gamma \lambda} + g_{\gamma \sigma}
    \Gamma^{\gamma}_{\mu \nu} M^{\mu \nu}{}_{\rho} \Big) \\ +
    \frac{1}{2} E^{\alpha \beta \gamma} \partial_{\gamma} \delta g_{\alpha \beta} \Big] \,.
\end{multline}
Expanding the definition of $ M^{\mu \nu}{}_{\rho} $~(\ref{M_Appendix}), we can write the final term in the curly brackets above as
\begin{equation} \label{f_G_1_Appendix}
 \delta g^{\rho \sigma}  g_{\gamma \sigma}
    \Gamma^{\gamma}_{\mu \nu} M^{\mu \nu }{}_{\rho} 
 =
    \delta g^{\rho \sigma}
     \Big( 2 g^{\nu \lambda} g_{\sigma \gamma} \Gamma^{\gamma}_{\mu \nu} \Gamma^{\mu}_{\rho \lambda} -  g^{\mu \nu} g_{\sigma \gamma} \Gamma^{\gamma}_{\mu \nu} \Gamma^{\lambda}_{\lambda \rho} - g^{\gamma \lambda} g_{\sigma \nu} \Gamma^{\nu}_{\mu \rho} \Gamma^{\mu}_{\gamma \lambda} \Big) \,.
\end{equation}
Finally, the last term in~(\ref{deltaGintegral_Appendix}) requires integration by parts, leading to
\begin{multline}
  \frac{1}{2} \int \sqrt{-g} f_{,\ourG} E^{\alpha \beta \gamma} \delta g_{\alpha \beta , \gamma} d^4x =
  \textrm{`surface term'} - \frac{1}{2} \int \delta g_{\alpha \beta} \partial_{\gamma}
  \big( \sqrt{-g} f_{,\ourG} E^{\alpha \beta \gamma} \big) d^4x \\ =
  \frac{1}{2} \int \Big[ \sqrt{-g} \delta g^{\rho \sigma} \partial_{\gamma}(f_{,\ourG})
    E_{\rho \sigma}{}^{\gamma} + \delta g^{\alpha \beta} g_{\alpha \rho} g_{\beta \sigma}
    f_{, \ourG} \partial_{\gamma}(\sqrt{-g} E^{\alpha \beta \gamma}) \Big] d^4 x \,,
  \label{f_G_E_Appendix}
\end{multline}
where the surface term proportional to $\delta g_{\alpha \beta}$ will be dropped as usual. Note that we had already calculated $\partial_{\gamma}(\sqrt{-g} E^{\alpha \beta \gamma})$, so the second term in the final line of~(\ref{f_G_E_Appendix}) is just $f_{, \ourG}$ times our result in~(\ref{DerivativeE_final})
\begin{multline}
  \label{Derivative_E}
    \frac{1}{2} \delta g^{\rho \sigma}  g_{\alpha \rho} g_{\beta \sigma} \partial_{\gamma} f_{, \ourG} ( \sqrt{-g} E^{\alpha \beta \gamma}) =  \sqrt{-g} \delta g^{\rho \sigma} f_{, \ourG} \Bigg[ 
      2\Gamma^{\lambda}_{\kappa \lambda} \Gamma^{\kappa}_{\rho \sigma}
     -2\Gamma^{\gamma}_{\sigma \nu} \Gamma^{\nu}_{\gamma \rho} \\ 
     -g_{\sigma \rho} g^{\mu \nu} \Gamma^{\eta}_{\gamma \eta} \Gamma^{\gamma}_{\mu \nu}
     -2 g_{\alpha \rho} g^{\epsilon \nu} \Gamma^{\gamma}_{\sigma \nu} \Gamma^{\alpha}_{\gamma \epsilon} 
     +  g_{\alpha \rho} g^{\eta \gamma} \Gamma^{\alpha}_{\gamma \eta} \Gamma^{\lambda}_{\sigma \lambda} 
     +g_{\alpha \rho} g^{\mu \nu} \Gamma^{\gamma}_{\mu \nu} \Gamma^{\alpha}_{\gamma \sigma} 
    \\
      + g_{\rho \sigma} g^{\eta \nu} \Gamma^{\gamma}_{\mu \nu} \Gamma^{\mu}_{\gamma \eta}
     + \partial_{\gamma} \Gamma^{\gamma}_{\rho \sigma} -  \partial_{\rho} \Gamma^{\lambda}_{\sigma \lambda} 
    + \frac{1}{2} g_{\sigma \rho} g^{\kappa \gamma} \partial_{\gamma} \Gamma^{\lambda}_{\kappa \lambda} - \frac{1}{2} g_{\rho \sigma} g^{\mu \nu} \partial_{\gamma} \Gamma^{\gamma}_{\mu \nu}
     \Bigg] \,.
\end{multline}
From comparison with equation~(\ref{recovering_Einstein}), we can immediately notice that the terms inside the square brackets of equations~(\ref{f_G_1_Appendix}) and~(\ref{Derivative_E}) almost give the Einstein tensor, but they are missing the variation of the metric determinant multiplied by $\ourG$~(\ref{metric_determinant_variation}). We therefore have that the  variation of the bulk term is given by
\begin{align}
  \label{bulk_term_variation}
  \int \sqrt{-g} f_{, \ourG} \delta \ourG \, d^4x  = \int \sqrt{-g} \delta g^{\rho \sigma} \Big[ f_{, \ourG} \Big( G_{\rho \sigma} + \frac{1}{2} g_{\rho \sigma} \ourG \big) + \frac{1}{2} E_{\rho \sigma}{}^{\gamma} \partial_{\gamma} f_{, \ourG} \Big] d^4x \,.
\end{align}

\subsubsection{Boundary variations}
\label{Appendix_Boundary_term}

The variation of the boundary term is
\begin{align}  
    \delta \ourB = \delta \Big( \frac{1}{\sqrt{-g}}\Big) \partial_{\nu} \Big( \frac{\partial_{\mu}(g g^{\mu \nu})}{\sqrt{-g}} \Big) + \frac{1}{\sqrt{-g}} \delta\Bigg( \partial_{\nu} \Big( \frac{\partial_{\mu}(g g^{\mu \nu})}{\sqrt{-g}} \Big) \Bigg) \,.
\end{align}
The first term simply gives the boundary term
\begin{align}
    \frac{1}{2 \sqrt{-g}} g_{\alpha \beta} \delta g^{\alpha \beta}
    \partial_{\nu} \Big( \frac{\partial_{\mu}(g g^{\mu \nu})}{\sqrt{-g}} \Big) = \frac{1}{2} \delta g^{\alpha \beta} g_{\alpha \beta} \ourB \,.
\end{align}
The second term we split into three calculations
\begin{multline} 
  \frac{1}{\sqrt{-g}} \delta\Bigg( \partial_{\nu} \Big( \frac{\partial_{\mu}(g g^{\mu \nu})}{\sqrt{-g}} \Big) \Bigg) = \frac{1}{\sqrt{-g}} \partial_{\nu} \Big\{
    \overbrace{\frac{\partial_{\mu} (\delta g g^{\mu \nu}) }{\sqrt{-g}}}^{1}
    + \overbrace{\frac{\partial_{\mu}(g \delta g^{\mu \nu}) }{\sqrt{-g}}}^{2} \\
    + \overbrace{\partial_{\mu}(g g^{\mu \nu}) \delta \Big(\frac{1}{\sqrt{-g}} \Big)}^{3} \Big\} \,.
\end{multline}
These three terms are computed to be the following
\begin{multline}
 \qquad \quad \partial_{\nu} \Big[   \overbrace{\frac{\partial_{\mu} (\delta g g^{\mu \nu}) }{\sqrt{-g}}}^{1} \Big] 
 =  \partial_{\nu} 
    \Bigg\{\sqrt{-g} \delta g^{\alpha \beta} \Big[2 \Gamma^{\eta}_{\mu \eta} g_{\alpha \beta} g^{\mu \nu} 
    + 2  \Gamma^{\eta}_{\mu \alpha} g_{\eta \beta}g^{\mu \nu} \\
    -  g_{\alpha \beta} \big( \Gamma^{\mu}_{\mu \eta} g^{\eta \nu} + \Gamma^{\nu}_{\mu \eta} g^{\mu \eta} \big) \Big] + \sqrt{-g} g_{\alpha \beta} g^{\mu \nu} \partial_{\mu}( \delta g^{\alpha \beta}) \Bigg\} \, ,
\end{multline}
\vspace{-9mm}
\begin{equation}
\quad \partial_{\nu} \Big[\overbrace{\frac{\partial_{\mu}(g \delta g^{\mu \nu}) }{\sqrt{-g}} }^{2}\Big] = \partial_{\nu} \Bigg[ 
    \sqrt{-g} \Big[2 \delta g ^{\mu \nu} \Gamma^{\alpha}_{\mu \alpha} + \partial_{\mu}(\delta g^{\mu \nu}) \Big] \Bigg] \, , \qquad \quad
    \end{equation}
    \begin{multline}
     \partial_{\nu} \bigg[\overbrace{\partial_{\mu}(g g^{\mu \nu}) \delta \Big(\frac{1}{\sqrt{-g}} \Big)}^{3}  \bigg] 
    = \partial_{\nu} \Bigg[ \sqrt{-g} \delta g^{\alpha \beta} \Big[
    \frac{g_{\alpha \beta}}{2} \big( \Gamma^{\mu}_{\mu \eta} g^{\eta \nu} +  \Gamma^{\nu}_{\mu \eta} g^{\mu \eta} \big) 
    \\ 
    - \Gamma^{\eta}_{\mu \eta} g_{\alpha \beta} g^{\mu \nu}\Big] \Bigg] \,.
      \end{multline}
Collecting these together gives the expression
\begin{multline}
    \frac{1}{\sqrt{-g}} \partial_{\nu} \Bigg\{\sqrt{-g} \bigg[ \delta g^{\alpha \beta} 
    \Big( \frac{1}{2} g_{\alpha \beta} g^{\mu \nu} \Gamma^{\eta}_{\mu \eta} - \frac{1}{2} g_{\alpha \beta} g^{\mu \eta} \Gamma^{\nu}_{\mu \eta}  + 2 g_{\eta \beta} g^{\mu \nu} \Gamma^{\eta}_{\mu \alpha} 
    - 2 \delta_{\alpha}^{\nu} \Gamma^{\gamma}_{\beta \gamma} 
    \Big) \\
    +  g_{\alpha \beta} g^{\mu \nu} \partial_{\mu}(\delta g^{\alpha \beta}) - \partial_{\mu}(\delta g^{\mu \nu}) \bigg]
    \Bigg\} \,.
\end{multline}
The total variation of $\ourB$ can then be written as
\begin{multline}
  \label{Variation_B_Appendix}
    \delta \ourB =  \frac{1}{2} \delta g^{\alpha \beta} g_{\alpha \beta} \ourB  + \frac{1}{\sqrt{-g}} \partial_{\nu} \Bigg\{ \sqrt{-g} \bigg[ \delta g^{\alpha \beta} 
    \Big( \frac{1}{2} g_{\alpha \beta} g^{\mu \nu} \Gamma^{\eta}_{\mu \eta} - \frac{1}{2} g_{\alpha \beta} g^{\mu \eta} \Gamma^{\nu}_{\mu \eta}  +  \\ 2 g_{\eta \beta} g^{\mu \nu} \Gamma^{\eta}_{\mu \alpha} 
    - 2 \delta_{\alpha}^{\nu} \Gamma^{\gamma}_{\beta \gamma} 
    \Big) 
    +  g_{\alpha \beta} g^{\mu \nu} \partial_{\mu}(\delta g^{\alpha \beta}) - \partial_{\mu}(\delta g^{\mu \nu}) \bigg]
    \Bigg\} \,.
\end{multline}
Substituting $\delta \ourB$~(\ref{Variation_B_Appendix}) into the variational integral~(\ref{Field_equation_derivation_Appendix_eq1}) yields
\begin{multline}
  \label{variation_of_B_Int_by_Parts}
    \int \sqrt{-g} f_{, \ourB} \delta \ourB \, d^4 x= \int \frac{1}{2} \sqrt{-g} f_{, \ourB}   \delta g^{\alpha \beta} g_{\alpha \beta} \ourB  \, d^4x  \\ 
  + f_{, \ourB} \partial_{\nu} \Bigg\{\sqrt{-g} \bigg[ \overbrace{\delta g^{\alpha \beta} 
    \Big( \frac{1}{2} g_{\alpha \beta} g^{\mu \nu} \Gamma^{\eta}_{\mu \eta} - \frac{1}{2} g_{\alpha \beta} g^{\mu \eta} \Gamma^{\nu}_{\mu \eta}  +  2 g_{\eta \beta} g^{\mu \nu} \Gamma^{\eta}_{\mu \alpha} 
    - 2 \delta_{\alpha}^{\nu} \Gamma^{\gamma}_{\beta \gamma} 
    \Big) }^{A} \\
    +  \overbrace{g_{\alpha \beta} g^{\mu \nu} \partial_{\mu}(\delta g^{\alpha \beta}) - \partial_{\mu}(\delta g^{\mu \nu}) }^{B} \bigg]
    \Bigg\}\, d^4x \,,
\end{multline}
where we perform integration by parts once on the $A$ terms and twice on the $B$ terms. For the $A$ terms we have
\begin{multline}
  \label{Integration_by_parts_B_first_term_Appendix}
    \int f_{, \ourB}  \partial_{\nu} \Bigg[ \sqrt{-g} \delta g^{\alpha \beta} 
    \Big( \overbrace{\frac{1}{2} g_{\alpha \beta} g^{\mu \nu} \Gamma^{\eta}_{\mu \eta} - \frac{1}{2} g_{\alpha \beta} g^{\mu \eta} \Gamma^{\nu}_{\mu \eta}  +  2 g_{\eta \beta} g^{\mu \nu} \Gamma^{\eta}_{\mu \alpha} 
    - 2 \delta_{\alpha}^{\nu} \Gamma^{\gamma}_{\beta \gamma} 
     }^{A} \Big) \Bigg] d^4x \\
    = \textrm{`surface term'} - \int \delta g^{\alpha \beta} \partial_{\nu}(f_{, \ourB}) 
   \overbrace{\big(...\big)}^{A} \sqrt{-g}
    \, d^4 x \,,
\end{multline} 
where $\overbrace{\big(...\big)}^{A}$ represents the terms underneath the brace on the previous line. For the $B$ terms we have
\begin{multline}
    \int f_{, \ourB}  \partial_{\nu} \Bigg[ \sqrt{-g} \Big( \overbrace{g_{\alpha \beta} g^{\mu \nu} \partial_{\mu}(\delta g^{\alpha \beta}) - \partial_{\mu}(\delta g^{\mu \nu}) }^{B} \Big) \Bigg] d^4 x \\ =
    \textrm{`surface term'} - \int \partial_{\nu} (f_{, \ourB}) \sqrt{-g}   \overbrace{\big(...\big)}^{B}  \, d^4 x  \,,
\end{multline}
Performing integration by parts once more leads to
\begin{equation}   \label{integration_by_parts_B_Appendix}
    \textrm{`surface terms'} +
    \int \delta g^{\alpha \beta} \Big[ \partial_{\mu} \Big( \sqrt{-g} g_{\alpha \beta} g^{\mu \nu} \partial_{\nu} f_{, \ourB} \Big) - \partial_{\alpha} \big( \sqrt{-g} \partial_{\beta} f_{, \ourB} \big) \Big] d^4 x \, ,
\end{equation}
where we now assume all surface terms proportional to $\delta g_{\mu \nu}$ and $\partial \delta g_{\mu \nu}$ vanish. 

Expanding the partial derivatives of~(\ref{integration_by_parts_B_Appendix}) gives
\begin{multline}
  \label{integration_by_parts_final_Appendix}
  \int \delta g^{\alpha \beta} \Big[
    \partial_{\mu} \Big( \sqrt{-g} g_{\alpha \beta} g^{\mu \nu} \partial_{\nu} f_{, \ourB} \Big) -
    \partial_{\alpha} \big( \sqrt{-g} \partial_{\beta} f_{, \ourB} \big)
    \Big] d^4 x  \\ =
  \int \sqrt{-g} \delta g^{\rho \sigma} \bigg[
    \partial_{\nu}(f_{, \ourB}) \Big[2 g_{\eta \rho} g^{\mu \nu} \Gamma^{\eta}_{\mu \sigma} -
      g_{\rho \sigma} g^{\mu \eta} \Gamma^{\nu}_{\mu \eta}  \Big] -
    \partial_{\sigma}(f_{,\ourB}) \Gamma^{\eta}_{\rho \eta} \\ +
    \partial_{\mu}\partial_{\nu}(f_{,\ourB})g_{\rho \sigma} g^{\mu \nu} -
    \partial_{\rho} \partial_{\sigma}(f_{,\ourB}) \bigg]  d^4 x \,.
\end{multline}
Putting equations~(\ref{Integration_by_parts_B_first_term_Appendix}) and~(\ref{integration_by_parts_final_Appendix}) into our expression for the boundary variation~(\ref{variation_of_B_Int_by_Parts}), and cancelling off terms, we arrive at
\begin{multline} 
    \int \sqrt{-g} f_{, \ourB} \delta \ourB  \, d^4 x= \int \delta g^{\rho \sigma} \sqrt{-g} \bigg[ \frac{1}{2} f_{,\ourB}  g_{\rho \sigma} \ourB 
     - \partial_{\nu}(f_{,\ourB}) \Big[ \frac{1}{2} g_{\rho \sigma} g^{\mu \nu} \Gamma^{\eta}_{\mu \eta} \\
      + \frac{1}{2} g_{\rho \sigma} g^{\mu \eta} \Gamma^{\nu}_{\mu \eta} - \delta^{\nu}_{\sigma} \Gamma^{\gamma}_{\rho \gamma} \Big] 
    + \partial_{\mu}\partial_{\nu}(f_{,\ourB})g_{\rho \sigma} g^{\mu \nu} - \partial_{\rho} \partial_{\sigma}(f_{,\ourB})
     \bigg] d^4 x \,.
    \end{multline}
Lastly, let us rewrite the connection pieces in terms of partial derivatives of the metric
\begin{multline}
  \label{boundary_term_variation_final_appendix}
  \int \sqrt{-g} f_{, \ourB} \delta \ourB  \, d^4 x =
  \int \sqrt{-g}  \delta g^{\rho \sigma} \bigg[
    \frac{1}{2} f_{, \ourB}  \ g_{\rho \sigma} \ourB  + g_{\rho \sigma}\partial^{\mu}
    \partial_{\mu}  f_{, \ourB} - \partial_{\rho} \partial_{\sigma} f_{, \ourB} \\ +
    \frac{1}{2} g_{\rho \sigma} \partial_{\mu}(g^{\mu \nu}) \partial_{\nu} f_{, \ourB} +
    \frac{1}{\sqrt{-g}} \partial_{\rho}(\sqrt{-g})\partial_{\sigma} f_{, \ourB} \bigg]  d^4x \,.
\end{multline}

\subsubsection{Full variation} 

The full variation of the gravitational $f(\ourG,\ourB)$ action is the sum of the variation of the bulk term~(\ref{bulk_term_variation}), the boundary term~(\ref{boundary_term_variation_final_appendix}) and the metric determinant
\begin{multline}
  \delta S_{\textrm {grav}}  = \frac{1}{2\kappa} \int \sqrt{-g} \delta g^{\rho \sigma} \Bigg[ f_{, \ourG} \Big( G_{\rho \sigma} + \frac{1}{2} g_{\rho \sigma} \ourG \big) + \frac{1}{2} E_{\rho \sigma}{}^{\gamma} \partial_{\gamma} f_{, \ourG} - \frac{1}{2}   g_{\rho \sigma} f  \\
    + \frac{1}{2}g_{\rho \sigma}  f_{, \ourB}  \ourB  + g_{\rho \sigma}\partial^{\mu}  \partial_{\mu}  f_{, \ourB}
    - \partial_{\rho} \partial_{\sigma} f_{, \ourB} \\
    + \frac{1}{2} g_{\rho \sigma} \partial_{\mu}(g^{\mu \nu}) \partial_{\nu} f_{, \ourB}+ \frac{1}{\sqrt{-g}} \partial_{\rho}(\sqrt{-g})\partial_{\sigma} f_{, \ourB}   \Bigg] \, d^4 x \,.
\end{multline}
Note that the equations are symmetric over $\rho$ and $\sigma$. This is the result given in equation~(\ref{f(g,b)_EoM2}).

\subsection{\texorpdfstring{$f(\bar{\ourG},\bar{\ourB})$}{f(G,B)} action}
\label{appendixC.2.2}

Now let us derive the field equations for the modified metric-affine action of Sec.~\ref{section5.3}. Here we will compute the variations with respect to the metric and independent connection simultaneously. For the action~(\ref{affine_action}) the variations lead to
\begin{align} \label{Appendix_var}
\delta S_{\textrm{grav}} &= \frac{1}{2\kappa }  \int \Big[ \delta \sqrt{-g} f(\bar{\ourG}, \bar{\ourB}) + \sqrt{-g} \delta f(\bar{\ourG}, \bar{\ourB}) \Big] d^4x \nonumber \\
& \begin{multlined}[b]
= \frac{1}{2\kappa}  \int \sqrt{-g} \Big[  -\frac{1}{2} g_{\mu \nu} \delta g^{\mu \nu} f(\bar{\ourG}, \bar{\ourB}) \\
+ \frac{\partial f(\bar{\ourG}, \bar{\ourB})}{\partial \bar{\ourG}}  \delta  \bar{\ourG} + \frac{\partial f(\bar{\ourG}, \bar{\ourB})}{\partial \bar{\ourB}}  \delta  \bar{\ourB} \Big] d^4x \,.
\end{multlined}
\end{align}
Using the previous calculations from Appendix~\ref{appendixC.1.2}, and again using the shorthand notation $\tilde{\Gamma}^{\nu}_{\kappa \lambda} := \bar{\Gamma}^{\mu}_{\mu \lambda} \delta^{\nu}_{\kappa} - \bar{\Gamma}^{{\nu}}_{\kappa \lambda}$, the variation of the bulk term is
\begin{multline}
\delta \bar{\ourG} = \delta g^{\mu \lambda} \big( \bar{\Gamma}^{\kappa}_{\kappa \rho} \bar{\Gamma}^{\rho}_{\mu \lambda} - \bar{\Gamma}^{\kappa}_{\mu \rho} \bar{\Gamma}^{\rho}_{\kappa \lambda} \big) + \tilde{\Gamma}^{\nu}_{\kappa \lambda} \partial_{\nu} \delta g^{\lambda \kappa} 
 - \frac{1}{2} \tilde{\Gamma}^{\nu}_{\lambda \kappa} g_{\alpha \beta} g^{\kappa \lambda} \partial_{\nu} \delta g^{\alpha \beta} \\
 - \frac{1}{2} \delta g_{\alpha \beta} g^{\kappa \lambda} \partial_{\nu} g^{\alpha \beta} \tilde{\Gamma}^{\nu}_{\kappa \lambda} - \frac{1}{2} \delta g^{\kappa \lambda} g_{\alpha \beta} \partial_{\nu} g^{\alpha \beta} \tilde{\Gamma}^{\nu}_{\kappa \lambda} + \delta \bar{\Gamma}{}^{\lambda}_{\mu \nu} P^{\mu \nu}{}_{\lambda} \,,
\end{multline}
where the final two terms on the first line will require integration by parts. We can write this in a nicer form by recalling that for $f(\bar{\ourG}) = \bar{\ourG}$ the metric variations, along with the variation of the metric determinant, give the metric-affine Einstein tensor. The only additional terms for the modified action will come from integration by parts when the partial derivatives hit the $f_{, \bar{\ourG}}$ term in~(\ref{Appendix_var}). In other words, we have
\begin{multline}
  \label{A_G_final}
  \int \sqrt{-g}  f_{, \bar{\ourG}} \delta_g \bar{\ourG} \, d^4 x =  \int \sqrt{-g} \Big( \delta g^{\mu \nu} f_{, \bar{\ourG}}  \bar{G}_{\mu \nu} + \frac{1}{2} \delta g^{\mu \nu} f_{, \bar{\ourG}}  g_{\mu \nu} \bar{\ourG} \\+ \delta g^{\mu \nu} \partial_{\lambda} f_{, \bar{\ourG}} \big(
  \frac{1}{2} \tilde{\Gamma}^{\lambda}_{\rho \kappa} g_{\mu \nu} g^{\kappa \rho}
  -  \tilde{\Gamma}^{\lambda}_{\mu \nu}
   \big)
  \Big) d^4 x  \,.
\end{multline}
Also recall that the bulk term can be written in the form given in~(\ref{G2}) where all the derivative terms are included in $\frac{1}{2} \partial_{\lambda} g^{\mu \nu}  \bar{E}_{\mu \nu}{}^{\lambda}$. Therefore it must be the case that this final term is proportional to $ \bar{E}_{\mu \nu}{}^{\lambda}$, which is easily verified
 \begin{align} \label{E_verify}
\delta g^{\mu \nu} \partial_{\lambda} f_{, \bar{\ourG}} \Big(
  \frac{1}{2} & \tilde{\Gamma}^{\lambda}_{\rho \kappa} g_{\mu \nu} g^{\kappa \rho}
  -  \tilde{\Gamma}^{\lambda}_{\mu \nu} \Big) \nonumber \\
  &= \delta g^{\mu \nu} \partial_{\lambda} f_{, \bar{\ourG}} \Big( \frac{1}{2} g_{\mu \nu} g^{\kappa \lambda} \bar{\Gamma}{}^{\rho}_{\rho \kappa} - \frac{1}{2} g_{\mu \nu} g^{\kappa \lambda}  \bar{\Gamma}{}^{\lambda}_{\rho \kappa} +  \bar{\Gamma}{}^{\lambda}_{\mu  \nu} - \delta^{\lambda}_{\mu}  \bar{\Gamma}{}^{\rho}_{\rho \nu} \Big) \nonumber \\
  &= \frac{1}{2} \delta g^{\mu \nu} \partial_{\lambda} f_{, \bar{\ourG}}  \bar{E}_{\mu \nu}{}^{\lambda} \,.
\end{align}
In total, for the bulk term we have
 \begin{align}
\int \sqrt{-g}   & f_{, \bar{\ourG}} \delta \bar{\ourG} \, d^4 x \nonumber \\
 &=  \int \sqrt{-g}  \delta g^{\mu \nu}  \Big(f_{, \bar{\ourG}}  \bar{G}_{\mu \nu} + \frac{1}{2} f_{, \bar{\ourG}}  g_{\mu \nu} \bar{\ourG} +  \frac{1}{2} \partial_{\lambda} f_{, \bar{\ourG}} \bar{E}_{\mu \nu}{}^{\lambda}   \Big) d^4 x  \,.
\end{align}

For the boundary term~(\ref{B}) we have
\begin{multline}
\delta  \bar{\ourB} = \frac{1}{2} g_{\alpha \beta} \delta g^{\alpha \beta}  \bar{\ourB}  + \frac{1}{\sqrt{-g}} \partial_{\kappa} \Bigg[\sqrt{-g}  \Big(  -\frac{1}{2} g_{\alpha \beta} \delta g^{\alpha \beta} (g^{\mu \lambda} \bar{\Gamma}{}^{\kappa}_{\mu \lambda} - g^{\kappa \lambda} \bar{\Gamma}{}^{\mu}{}_{\mu \lambda}) \\
+ \delta g^{\mu \lambda} \bar{\Gamma}{}^{\kappa}_{\mu \lambda} - \delta g^{\kappa \lambda} \bar{\Gamma}{}^{\mu}_{\mu \lambda} \Big) \Bigg] 
+ \frac{1}{\sqrt{-g}} \partial_{\kappa} \Big( \sqrt{-g} g^{\mu \lambda} \delta \bar{\Gamma}{}^{\kappa}_{\mu \lambda} - \sqrt{-g} g^{\kappa \lambda} \delta \bar{\Gamma}{}^{\mu}_{\mu \lambda} \Big) \,,
\end{multline}
where all but the first term require integration by parts. It is clear that if $f$ is a linear function of $\bar{\ourB}$ then this term is a pure boundary term and does not contribute to the equations of motion. Plugging these into the variations~(\ref{Appendix_var}) and discarding boundary terms we obtain 
\begin{multline}
  \int \sqrt{-g} f_{, \bar{\ourB}} \delta  \bar{\ourB}  \, d^4x =
  \int \sqrt{-g} \Bigg\{ \delta g^{\alpha \beta} \Big[\frac{1}{2} g_{\alpha \beta}
    f_{, \bar{\ourB}} \bar{\ourB} + (\partial_{\kappa} f_{, \bar{\ourB}})
    \Big(\frac{1}{2} g_{\alpha \beta} (g^{\mu \lambda} \bar{\Gamma}{}^{\kappa}_{\mu \lambda} \\
     -
    g^{\kappa \lambda} \bar{\Gamma}{}^{\mu}_{\mu \lambda}) -
    \bar{\Gamma}{}^{\kappa}_{\alpha \beta} + \delta^{\kappa}_{\alpha}
    \bar{\Gamma}{}^{\mu}_{\mu \beta} \Big) \Big]   +
  \delta \bar{\Gamma}{}^{\kappa}_{\mu \nu} \partial_{\lambda}
  f_{,\bar{\ourB}}( \delta^{\mu}_{\kappa} g^{\lambda \nu} - \delta^{\lambda}_{\kappa} g^{\mu \nu}) \Bigg\} d^4x \,. 
\end{multline}
The final terms on the first line are exactly the terms found in~(\ref{E_verify}), with the opposite signs, so we can write this as
\begin{multline}
  \label{A_B_final}
  \int \sqrt{-g} f_{, \bar{\ourB}} \delta  \bar{\ourB}\, d^4x =
  \int \sqrt{-g} \Big[\frac{1}{2}\delta g^{\alpha \beta}
    \Big(g_{\alpha \beta} f_{, \bar{\ourB}} \bar{\ourB} -
    \partial_{\kappa} f_{, \bar{\ourB}}  \bar{E}_{\alpha \beta}{}^{\kappa} \Big)  \\
    + 2 \delta \bar{\Gamma}{}^{\kappa}_{\mu \nu} \partial_{\lambda}
    f_{, \bar{\ourB}}  g^{\nu [\lambda} \delta^{\mu]}_{\kappa} \Big] d^4x \,.
\end{multline}
Combining these results~(\ref{A_G_final}) and~(\ref{A_B_final}), the full variations of the modified action are 
\begin{multline}
  \delta S_{\textrm{grav}} = \frac{1}{2\kappa} \int \sqrt{-g} \bigg\{ \delta g^{\mu \nu}
  \Big[ f_{, \bar{\ourG}} (\bar{G}_{\mu \nu} +
    \frac{1}{2} g_{\mu \nu} \bar{\ourG}) -\frac{1}{2} g_{\mu \nu} f  +
    \frac{1}{2} \bar{E}_{\mu \nu}{}^{\lambda} (\partial_{\lambda}  f_{, \bar{\ourG}} -
    \partial_{\lambda}  f_{, \bar{\ourB}}) \\
     +
    \frac{1}{2} g_{\mu \nu}  f_{, \bar{\ourB}} \bar{\ourB} \Big] 
  + \delta \bar{\Gamma}{}^{\kappa}_{\mu \nu} \Big( P^{\mu \nu}{}_{\kappa}  f_{,\bar{\ourG}} +
  2 \partial_{\lambda} f_{, \bar{\ourB}} g^{\nu [\lambda} \delta^{\mu]}_{\kappa}  \Big) \bigg\} d^4x \,,
\end{multline}
which are given in~(\ref{affine_EoM}).

\chapter{Stueckelberg trick}
\label{appendixD}

\renewcommand{\theequation}{\thechapter.\arabic{equation}}

Here we show that the Stueckelberg trick applied to the Einstein action and metric bulk term $\ourG$ leads directly to the non-metricity scalar $Q$ for the symmetric teleparallel connection. This verifies the calculation in equation~(\ref{Stueck_G}), and also shows that the gauge boundary term $b_{\xi}$~(\ref{b_xi}) can be written in a form that is quadratic in the affine connection.

The bulk term transforms under an infinitesimal coordinate transformation as~(\ref{G_infinitesimal}), where the additional inhomogeneous $M^{\mu \nu}{}_{\lambda}$ term appears. In order to apply the Stueckelberg trick, however, we want the full general coordinate transformation. By a similar calculation it is not too difficult to see that $\ourG$ transforms as  
\begin{multline}
\ourG(x) \rightarrow \hat{\ourG}(\hat{x}) = \ourG(x) + 2 g^{\alpha \beta} \Gamma^{\kappa}_{\alpha \gamma}
\frac{\partial^2 x^{\gamma} }{\partial \hat{x}^{\nu} \partial \hat{x}^{\rho}} \frac{\partial \hat{x}^{\rho} }{\partial x^{\kappa }} \frac{\partial \hat{x}^{ \nu} }{\partial x^{ \beta}} 
 - g^{ \alpha \beta} \Gamma^{\lambda}_{\lambda \kappa} \frac{ \partial^2 x^{\kappa}}{\partial \hat{x}^{\mu} \partial \hat{x}^{\nu} }   \frac{\partial \hat{x}^{\mu } }{\partial x^{\alpha}} \frac{\partial \hat{x}^{\nu} }{\partial x^{ \beta}} \\
  - g^{\alpha \beta} \Gamma^{\kappa}_{\alpha \beta} \frac{\partial^2 x^{\gamma} }{\partial \hat{x}^{\lambda} \partial \hat{x}^{\rho} } \frac{\partial \hat{x}^{\rho} }{\partial x^{\kappa }}\frac{\partial \hat{x}^{ \lambda} }{\partial x^{\gamma }} 
+
g^{\alpha \beta} \frac{\partial^2 x^{\kappa}}{\partial \hat{x}^{\mu} \partial \hat{x}^{\lambda}} \frac{\partial^2 x^{\gamma}}{\partial \hat{x}^{\nu} \partial \hat{x}^{\rho}}  \frac{\partial \hat{x}^{\mu } }{\partial x^{\alpha}} \frac{\partial \hat{x}^{\nu} }{\partial x^{\beta}} \frac{\partial \hat{x}^{ \rho} }{\partial x^{\kappa }} \frac{\partial \hat{x}^{ \lambda} }{\partial x^{\gamma}} \\
- g^{\alpha \beta}  \frac{ \partial^2 x^{\kappa}}{\partial \hat{x}^{\mu} \partial \hat{x}^{\nu} }   \frac{\partial^2 x^{\gamma} }{\partial \hat{x}^{\lambda} \partial \hat{x}^{\rho} }    \frac{\partial \hat{x}^{\mu } }{\partial x^{\alpha}} \frac{\partial \hat{x}^{\nu} }{\partial x^{ \beta}} \frac{\partial \hat{x}^{\rho} }{\partial x^{\kappa}} \frac{\partial \hat{x}^{ \lambda} }{\partial x^{\gamma }}  \, .
\end{multline}
This can then be simplified as follows
\begin{multline} \label{G_GCT}
 \hat{\ourG}(\hat{x}) = \ourG + M^{\beta \kappa}{}_{\gamma} \Big(\frac{\partial^2 x^{\gamma}}{\partial \hat{x}^{\mu} \partial \hat{x}^{\nu}} \frac{\partial \hat{x}^{\mu} }{\partial x^{\kappa }}\frac{\partial \hat{x}^{ \nu} }{\partial x^{\beta}} \Big) 
\\ +2 g^{\alpha \beta} \frac{\partial^2 x^{\kappa}}{\partial \hat{x}^{\mu} \partial \hat{x}^{\lambda}} \frac{\partial^2 x^{\gamma}}{\partial \hat{x}^{\nu} \partial \hat{x}^{\rho}}  \frac{\partial \hat{x}^{\mu } }{\partial x^{\alpha}} \frac{\partial \hat{x}^{ \rho} }{\partial x^{\kappa }} \Big(  \frac{\partial \hat{x}^{[\nu} }{\partial x^{ \beta}} \frac{\partial \hat{x}^{ \lambda]} }{\partial x^{\gamma}} \Big) \ .
\end{multline}
As one would expect, the infinitesimal limit in~(\ref{G_infinitesimal}) is obtained immediately by using $\hat{x}^{\mu} = x^{\mu} + \xi^{\mu}(x)$.
Next we make various simplifications by applying the chain rule on the second-order Jacobian terms
\begin{align} 
 & \begin{multlined}[t]
  = \ourG - M^{\beta \kappa}{}_{\gamma} \Big( \frac{\partial \hat{x}^{\nu}}{ \partial x^{\beta}} \frac{\partial x^{\gamma}}{ \partial \hat{x}^{\mu}} \frac{\partial x^{\eta}}{\partial \hat{x}^{\nu}} \frac{\partial^2 \hat{x}^{\mu}}{\partial x^{\eta} \partial x^{\kappa} }
  \Big)   \\  
  + 2 g^{\alpha \beta} \frac{\partial x^{\kappa}}{\partial \hat{x}^{\mu}}  \frac{\partial x^{\eta}}{\partial \hat{x}^{\lambda}}  \frac{\partial x^{\gamma}}{\partial \hat{x}^{\rho}}  \frac{\partial x^{\sigma}}{\partial \hat{x}^{\nu}}  \frac{\partial^2 \hat{x}^{\mu}}{\partial x^{\eta} \partial x^{\alpha} }   \frac{\partial^2 \hat{x}^{\rho}}{\partial x^{\sigma} \partial x^{\kappa} }   \Big(  \frac{\partial \hat{x}^{[\nu} }{\partial x^{ \beta}} \frac{\partial \hat{x}^{ \lambda]} }{\partial x^{\gamma}} \Big)
 \nonumber
 \end{multlined}\\
 & \begin{multlined}[b]
 = \ourG - M^{\beta \kappa}{}_{\gamma} \Big( \frac{\partial x^{\gamma}}{ \partial \hat{x}^{\mu}} \frac{\partial^2 \hat{x}^{\mu}}{\partial x^{\beta} \partial x^{\kappa} }
  \Big)   \\ 
 + g^{\alpha \beta} \big(\delta^{\sigma}_{\beta} \delta^{\eta}_{\gamma} - \delta^{\eta}_{\beta} \delta^{\sigma}_{\gamma} \big) \frac{\partial x^{\kappa}}{\partial \hat{x}^{\mu}} \frac{\partial^2 \hat{x}^{\mu}}{\partial x^{\eta} \partial x^{\alpha} } \frac{\partial x^{\gamma}}{\partial \hat{x}^{\rho}}   \frac{\partial^2 \hat{x}^{\rho}}{\partial x^{\sigma} \partial x^{\kappa} } \, .
 \end{multlined} 
\end{align}

The next step is to promote the new coordinate to the Stueckelberg fields $\hat{x}^{\mu} \rightarrow \xi^{\mu}$, see for example~\cite{BeltranJimenez:2022azb} for details, which gives
\begin{equation}
 \hat{\ourG} = \ourG - M^{\beta \kappa}{}_{\gamma} \Big( \frac{\partial x^{\gamma}}{ \partial \xi^{\mu}} \frac{\partial^2 \xi^{\mu}}{\partial x^{\beta} \partial x^{\kappa} }
  \Big)  
 + g^{\alpha \beta} \big(\delta^{\sigma}_{\beta} \delta^{\eta}_{\gamma} - \delta^{\eta}_{\beta} \delta^{\sigma}_{\gamma} \big) \frac{\partial x^{\kappa}}{\partial \xi^{\mu}} \frac{\partial^2 \xi^{\mu}}{\partial x^{\eta} \partial x^{\alpha} } \frac{\partial x^{\gamma}}{\partial \xi^{\rho}}   \frac{\partial^2 \xi^{\rho}}{\partial x^{\sigma} \partial x^{\kappa} } \, .
\end{equation}
Now all that is left is to identify the terms above with the symmetric teleparallel connection~(\ref{STG_connection}), which we remind the reader is defined as
\begin{equation*}
\bar{\Gamma}^{\lambda}_{\mu \nu} = \frac{\partial x^{\lambda}}{\partial \xi^{\rho}} \frac{\partial^2 \xi^{\rho}}{\partial x^{\mu} \partial x^{\nu}}  \, .
\end{equation*}
We are therefore led to the final expression for the transformation of the bulk term
\begin{equation}
\hat{\ourG} = \ourG - M^{\beta \kappa}{}_{\gamma} \bar{\Gamma}^{\gamma}_{\beta \kappa} + 2 g^{\alpha \beta} \bar{\Gamma}^{\kappa}_{\alpha [\gamma} \bar{\Gamma}^{\gamma}_{\beta] \kappa} \, .
\end{equation}

Finally, we will equate these additional terms with the non-metricity gauge boundary term $b_{\xi}$ defined in~(\ref{b_xi}), which is given by
\begin{align}
b_{\xi} &= \frac{2}{\sqrt{-g}} \partial_{\kappa} \big( \sqrt{-g} g^{\lambda [\mu} \bar{\Gamma}^{\kappa]}_{\mu \lambda} \big)
 \nonumber \\
 & \begin{multlined}[b]
  = 2 \Gamma^{\rho}_{\kappa \rho} g^{\lambda [\mu} \bar{\Gamma}^{\kappa]}_{\mu \lambda}  - \big(\Gamma^{\lambda}_{\kappa \eta} g^{\eta \mu} + \Gamma^{\mu}_{\kappa \eta} g^{\lambda \eta} \big) \bar{\Gamma}^{\kappa}_{\mu \lambda}
 + \big(\Gamma^{\kappa}_{\kappa \eta} g^{\eta \lambda} + \Gamma^{\lambda}_{\kappa \eta} g^{\eta \kappa} \big) \bar{\Gamma}^{\mu}_{\mu \lambda} 
  \\
  + g^{\lambda \mu} \partial_{\kappa} \bar{\Gamma} ^{\kappa}_{\mu \lambda} - g^{\lambda \kappa} \partial_{\kappa} \bar{\Gamma}^{\mu}_{\mu \lambda} \, , \label{append_b_xi}
   \end{multlined} 
\end{align}
where we remind the reader that we are working with both the Levi-Civita connection $\Gamma$ and the symmetric teleparallel connection $\bar{\Gamma}$. There are two ways to deal with the derivative terms on the final line and rewrite them in terms of quadratic connection pieces. The first is to use the explicit form of the affine connection in terms of $\xi^{\mu}$, where one would easily find all the third derivatives of $\xi$ cancelling via symmetry, with quadratic terms in $\bar{\Gamma}$ remaining. A quicker route is to recall the teleparallel condition $\bar{R}=0$, such that derivatives of the affine connection can be related to quadratic connection terms. Either way, we obtain an expression for the derivative terms
\begin{align}
 g^{\lambda \mu} \partial_{\kappa} \bar{\Gamma} ^{\kappa}_{\mu \lambda} - g^{\lambda \kappa} \partial_{\kappa} \bar{\Gamma}^{\mu}_{\mu \lambda} = g^{\kappa \lambda} \bar{\Gamma}^{\mu}_{\eta \kappa} \bar{\Gamma}^{\eta}_{\mu \lambda} - g^{\mu \lambda} \bar{\Gamma}^{\kappa}_{\eta \kappa} \bar{\Gamma}^{\eta}_{\mu \lambda} \, .
\end{align}
Using this result along with the definition of $M^{\mu \nu}{}_{\lambda}$ shows that the boundary term~(\ref{append_b_xi}) can be written as
\begin{equation}
b_{\xi} = - M^{\beta \kappa}{}_{\gamma} \bar{\Gamma}^{\gamma}_{\beta \kappa} + 2 g^{\alpha \beta} \bar{\Gamma}^{\kappa}_{\alpha [\gamma} \bar{\Gamma}^{\gamma}_{\beta] \kappa}  \, .
\end{equation}

In total, we find that applying the Stueckelberg procedure to restore the invariance of the bulk term $\ourG$ leads exactly to the non-metricity scalar of symmetric teleparallel gravity
\begin{equation}
\hat{\ourG} = \ourG + b_{\xi} = -Q(g,\xi) \, ,
\end{equation}
as reported in equation~(\ref{Stueck_G}).

\chapter{Dynamical systems stability analysis}
\label{appendixE}

\renewcommand{\theequation}{\thesection.\arabic{equation}}

Here we include some further details for the stability analysis of the dynamical systems of Section~\ref{section6.3}. This material is adapted from our works~\cite{Boehmer:2022wln,Boehmer:2023knj}. Firstly, we perform the stability analysis for the general fixed points of the second-order modified cosmologies. We then study the stability analysis of the Anagnostopoulos et al. model Sec.~\ref{section6.2.3}, which has a nonhyperbolic fixed point. We must therefore use techniques beyond linear stability theory in this case.

\section{Stability analysis of the general system}
\label{appendixE.1}

The critical points of system~(\ref{X2 mn})--(\ref{Z mn}), as given in Table~\ref{tab2}, depend on the explicit form of $n(Z)$~(\ref{n(Z)}) and $m(z)$~(\ref{m(Z)}). These functions themselves depend on the choice of function $f$.
Nevertheless, it is possible to make certain general statements about the nature of the critical points. The stability matrix or Jacobian is given by
\begin{align} \label{Jacobian}
    J = \begin{pmatrix}
   \displaystyle \frac{\partial X_2'}{\partial X_2} &\displaystyle \frac{\partial X_2'}{\partial Z} \\[8pt]
 \displaystyle   \frac{\partial Z'}{\partial X_2} &\displaystyle \frac{\partial Z'}{\partial Z} 
    \end{pmatrix} \,.
\end{align}
We denote the two eigenvalues of $J$ by $\lambda_1$ and $\lambda_2$.

For the points in Table~\ref{tab2}, the necessary existence conditions shall be assumed, but care must be taken with the derivative of the function $m(Z)$, which will contain third derivatives of $f$ (the derivatives of $n(Z)$ can be rewritten in terms of $m$ and $n$). It is then often necessary to assume that $f^{(3)}$, which appears in the numerator of the elements of the Jacobian matrix, remains finite at the fixed point in order to determine the stability for a general $f$. We will make this assumption explicit when needed. 

For P$_1=(0,0)$ one obtains the eigenvalues
\begin{equation}
\lambda_1^{\textrm{P}_1} =  \frac{2-4m(0)-3n(0)}{1+m(0)} \qquad , \qquad  \lambda_2^{\textrm{P}_1} = \frac{ 3n(0)-6 }{ 1+m(0)} \, , 
 \label{P1eig}
 \end{equation}
  which has at least one negative eigenvalue. Conditions on $m(Z)$ and $n(Z)$ could then be imposed to determine its stability.
   The eigenvalues of point P$_2=(0,1)$ takes a similar form but with $m$ and $n$ evaluated at $Z=1$
\begin{equation}
\lambda_1^{\textrm{P}_2} =  \frac{2-4m(1)-3n(1)}{1+m(1)} \qquad , \qquad  \lambda_2^{\textrm{P}_2} = -\frac{ 3n(1)-6 }{ 1+m(1)} \, .
\end{equation}
Unsurprisingly, with $m$ and $n$ undefined for these critical points, there is little that can be said about their stability in full generality.

For the point P$_n=(0,Z^*)$ with $n(Z^*)=2$ one obtains constant eigenvalues 
\begin{equation} \label{Pnstab}
\lambda_1^{\textrm{P}_n} = -4  \qquad , \qquad  \lambda_2^{\textrm{P}_n} = -3 \, ,
\end{equation}
meaning this point is always stable when it exists. This is consistent with its interpretation as being a late-time de Sitter attractor with $q=-1$. The stability of the point P$_m=(0,Z^*)$ with $m(Z^*) \rightarrow \infty$ depends on the third derivatives of $f$, so one cannot make definitive claims about its stability. However, if one checks that both $n(Z^*)$ and $f^{(3)}(Z^*)$ remain finite while $m(Z^*) \rightarrow \infty$, the eigenvalues will also be $\lambda_1=-4$ and $\lambda_2=-3$, indicating stability. This would be consistent with the interpretation of P$_{m}$ as a stable, late-time, de Sitter point. Note that in other cases (such as for the Born-Infeld models of Sec.~\ref{section6.1}) the point P$_{m}$ is instead unstable, and acts as an early-time inflationary fixed point. It should therefore be emphasised that the properties and interpretation of P$_{m}$ is model dependent.

Point A $=(X_2^*,0)$, with $X_2^*$ defined previously, has eigenvalues 
\begin{equation}
\lambda_1^{\textrm{A}} = -4  \qquad , \qquad  \lambda_2^{\textrm{A}} = - \frac{2-4m(0)-3n(0)}{1+m(0)} \, . 
\end{equation}
The second eigenvalue is just minus the first of point P$_1$'s in~(\ref{P1eig}), therefore both points cannot be stable simultaneously. For point B the eigenvalues depend on the higher derivatives of $f$, so again, the stability cannot be determined without specifying a model $f$. If all derivatives of $f$ stay finite at $Z=1$ it is possible to at least conclude that this point must be unstable, as the eigenvalues cannot both be negative. Once again, this will be highly model-dependent.

\section{Stability analysis of the Anagnostopoulos et al. model}
\label{appendixE.2}

The fixed points for the model of Anagnostopoulos et al. are given in Table~\ref{tab Saridakis}. The stability has also been included, with each point having the following eigenvalues determined by the methods outlined above. Using the functions $m(Z)$ and $n(Z)$ for this model, given in equations~(\ref{n Saridakis}) and~(\ref{m Saridakis}) respectively, we find~\cite{Boehmer:2023knj}
\begin{itemize}
\item Point P$_2$ has eigenvalues $\lambda_1^{\textrm{P}_2}  = -1$, $\lambda_2^{\textrm{P}_2} = 3$ and is a saddle.
\item Point B has eigenvalues $\lambda_1^{\textrm{B}}  = 1$, $\lambda_2^{\textrm{B}} = 4$ and is unstable.
\item Point P$_n$ has eigenvalues $\lambda_1^{\textrm{P}_n}  = -4$, $\lambda_2^{\textrm{P}_n} = -3$ and is stable.
\item Point P$_m$ has eigenvalues $\lambda_1^{\textrm{P}_m}  = 0$, $\lambda_2^{\textrm{P}_m} = -4$ and is nonhyperbolic.
\end{itemize}

\begin{figure}[!htb]
		\includegraphics[width=0.65\textwidth]{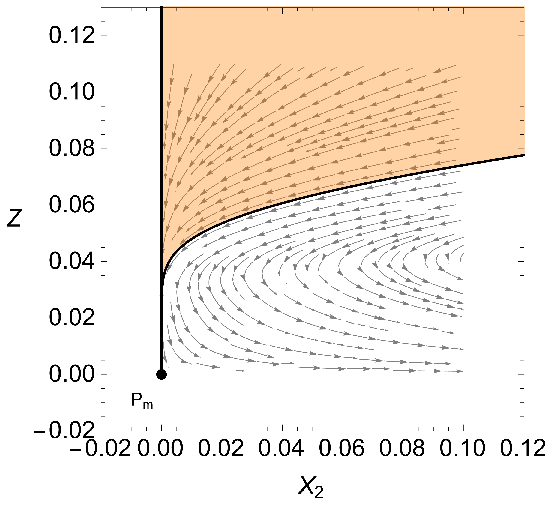}
		\centering
		\caption{Nonhyperbolic fixed point P$_{m}$.}
		\label{fig:Appendix1}
\end{figure}

	A closer look at the nonhyperbolic point is shown in Fig.~\ref{fig:Appendix1}, the orange region representing the physical phase space as determined from the Hubble constraint~(\ref{2fluidF}). Recall that we require $\lambda < 0 $ for the existence of this point. Despite the fact that the point appears to be mathematically unstable, with trajectories moving away from the point in Fig.~\ref{fig:Appendix1}, the Hubble constraint can be used to determine the fate of trajectories within the physical phase space.
	
	 The boundary of the physical phase space is described by the equations
	\begin{equation} \label{Append1}
		X_2 = e^{\lambda \big( \frac{1}{Z}-1 \big)} \Big(1 + \frac{2 \lambda(Z-1)}{Z} \Big) \quad , \quad \text{with} \ X_2 \geq 0 \ , \ 0 \leq Z \leq 1 \, \ , \ {\lambda < 0} \, ,
	\end{equation}
	see~\cite{Boehmer:2023knj} for details.
	As $Z$ approaches zero (from above) $X_2$ goes to zero. One can also show that for all $X_2 \geq 0$, trajectories always travel in the negative $Z$ direction whilst $Z$ is between $0$ and $1$. This can be most easily seen by substituting the expression for $X_2$ on the physical boundary~(\ref{Append1}) into the autonomous equation $dZ/dN$. This resulting equation is
	\begin{equation}
		\frac{dZ}{dN} = -\frac{4 Z^2(1-Z)\big(Z - 2\lambda(1-Z)\big) }{Z^2 + 2\lambda^2 (Z-1)^2 - \lambda Z(1-Z) } \, .
	\end{equation}
	All terms in the numerator and denominator are positive for $\lambda<0$, therefore $dZ/dN$ is negative and trajectories on the boundary approach the~origin.
	
	Following the same logic, the same result can be shown for the general equation $dZ/dN$ with $\lambda<0$. We can therefore conclude that all physical trajectories satisfying the Hubble constraint travel towards and terminate at the origin, point P$_{m}$. This is because trajectories are smooth and do not cross the boundary, and therefore terminate at $Z=0$ with $X_2=0$. This indeed matches what can be observed in the phase portraits, Fig.~\ref{fig:ana2} and~\ref{fig:Appendix1}.

\label{appendix3}


\renewcommand{\chapterheadstartvskip}{\vspace*{-2\baselineskip}} 

\cleardoublepage
\phantomsection

\end{document}